Tikhonov Moscow Institute of Electronics and Mathematics

NRU Higher School of Economics

(MIEM NRU HSE)

Kostinskiy Alexander Yulievich

# THE PLASMA STRUCTURES AND NETWORKS OF CHANNELS AS COMPONENTS OF THE SEQUENTIAL MECHANISM OF LIGHTNING INITIATION WITHIN THUNDER CLOUDS

25.00.29 - physics of the atmosphere and hydrosphere

Thesis for a degree Doctor of Physical and Mathematical Sciences
(PD Dr. Habilitation in Geophysics)

Moscow – 2021

# DISSERTATION ABSTRACT

## (in English)

In the physics of lightning, there is no more urgent and intricate problem than the problem of the mechanism of lightning initiation in thunderclouds.

The well-known lightning researcher Martin Uman, who consistently published (alone and with co-authors) the most cited review monographs on lightning physics for almost half a century [Uman, 1969, 1987, 2001], [Uman, 1972], [Rakov and Uman, 2003], [Dwyer & Uman, 2014], identified "The top ten questions in lightning research" [Dwyer & Uman, 2014, p.156]. The very first point in importance was the questions concerning the mechanism of initiation of lightning in thunderclouds: "{1} By what physical mechanism or mechanisms is lightning initiated in the thundercloud? What is the maximum cloud electric field magnitude and over what volume of the cloud? What, if any, high energy processes (runaway electrons, X-rays, gamma rays) are involved in lightning initiation and how? What is the role of various forms of ice and water in lightning initiation?". It is noteworthy that [Dwyer & Uman, 2014, p.156] suggested that the mechanism may consist of several physical phenomena, which involves the processes of electric field amplification, as well as runaway electrons, X-ray and gamma photons. Note that the possible key role of extensive cosmic ray air showers (EASs) in lightning initiation is also actively discussed (for example, [Gurevich et al., 1999], [Gurevich and Karashtin, 2013]]). Chapters 7 and 8 of the dissertation are devoted to an attempt to answer these questions and again, thanks to the proposed new mechanism, EAS play a key role in lightning initiation. The remaining chapters of the dissertation serve as necessary elements (links) of the general mechanism described in chapters 7 and 8.

The second most important issue [Dwyer & Uman, 2014, p.156] is the physics of lightning leader propagation: "{2} What physical mechanisms govern the propagation of the different types of lightning leaders (negative stepped, first positive, negative dart, negative dart-stepped, negative dart-chaotic) between cloud and ground and the leaders inside the cloud?"? Chapter 6 is devoted to the issues of stepwise propagation of the positive and negative leaders of a long spark, which is modeled by the development of lightning leaders. The fourth question singled out the problem of the physical nature of compact intracloud discharges (CID/NBE): "{4} What is the physics of compact intra-cloud discharges (CIDs) (that produce a narrow bipolar wideband electric field pulse, a narrow bipolar event or NBE, apparently multiple-reflecting propagating waves within 1 km height, and copious HF and VHF radiation)? How are CIDs related to other types of preliminary breakdown pulses? Are CIDs related to the Terrestrial Gamma-Ray Flashes (TGFs) observed on orbiting satellites or to the Transient Luminous Events (TLEs) photographed above cloud tops, particularly to so-called ''gigantic jets''?

[Rison et al., 2016] were the first to suggest that CID/NBE or their low-energy modification is the cause of the initiation of "all or almost all" lightning. That is, the first and fourth points of the "Uman-Dwyer research program" turned out to be closely related. As the physical mechanism of CID/NBE, they suggested the mechanism of fast positive breakdown (FPB), which, in their opinion, is a giant streamer flash that moves at a speed of $10^7$-$10^8$ m/s [Attanasio et al., 2019]. In our opinion, the idea of internal connection between CID/NBE (in most cases, a weak similar event) and the moment of lightning initiation is correct, but gas-discharge streamers at pressures of 0.1-1 atm cannot move at velocities of $10^7$-$10^8$ m/s in sub-breakdown electric fields [Les Renardières Group, 1972, 1974, 1977, 1981]. The issue of the close connection between lightning initiation and CID is the central part of the mechanism of lightning initiation proposed by us [Kostinskiy, Marshall & Stolzenburg, 2020a] (thesis chapters 7 and 8).

How relevant and complex the problem of lightning initiation in clouds can be seen by considering also the opinions of leading Russian researchers. The book "Lightning Physics and Lightning Protection" [Bazelyan and Raizer, 2000] contains a section that is called "Origination in Clouds", where the following view is expressed on the state of the problem at that time: "Although the process of propagation of a downward negative stepped leader is most familiar to lightning observers, the circumstances and mechanisms of its births are literally shrouded in mist. No one observed the start of lightning and did not follow the development of the process in the clouds. The origination process is not fully reproduced in the laboratory, although a negative step leader is obtained in experiments. But the conditions for its generation from a high-voltage metal electrode connected to a capacitor bank of a pulse generator have little in common with what happens in clouds. A cloud is not a lining of a capacitor and is not a conductor at all. The negative cloud charge is dispersed in a dielectric air medium on small hydrometeors. It is hard to imagine how this charge scattered in a huge volume, sitting on slow-moving particles, can gather and go into the plasma channel in milliseconds". In fact, lightning is considered by default to be a powerful stepped negative leader emerging from a thundercloud, as depicted in the now classic diagram [Uman, 1969], Figure B.1.

[Bazelyan and Raizer, 2000] also draw attention to the prospects for studying an artificially charged aerosol cloud, which was used in research in this dissertation: clouds. The fact that spark discharges sometimes occur in an environment with a dispersed charge is known thanks to investigations into the causes of explosions and fires in industrial premises with large volumes of electrostatically charged dust particles or droplets. Recently, reports have appeared on studies using steam jet generators that emit miniature electrically charged clouds into the atmosphere [Antsupov et al., 1991], [Верещагин и др., 1989]. Sometimes, elongated luminous formations with dimensions of the order of 10 cm appeared near the boundary of the charged aerosol; less often they turned into spark channels (up to 1 m long). Unfortunately, in the experiments it was not possible to measure the field at the start of the spark discharge, and the matter did not go further than stating the fact of spark excitation. Therefore, we can only guess about the mechanisms of initiation of lightning in clouds and sparks in laboratory aerosol mixtures."

Thus, the process of initiation of lightning in thunderclouds, as well as the process of initiation of compact intracloud discharges, are one of the most urgent problems in the physics of lightning and thunderstorms, and approaching the understanding of these processes is important both from fundamental and applied points of view, not only in a narrow but also in a broad sense, given that lightning and CID are discharge events in aerosol (multiphase) media, similar, for example, to discharges in volcanic ash clouds [McNutt and Thomas, 2015] or in the atmospheres of other planets [Brown et al., 2018].

**In the INTRODUCTION (Введение)**, the relevance of the research topic is substantiated, a detailed review of the problem of initiation of lightning and compact intracloud discharges (CID), as well as the lightning initiating event (IE), the initial electric field change (IEC) and the initial breakdown pulses (IBPs) is made.

**CHAPTER 1** is devoted to the discovery of a new class of electrical discharges in clouds of artificially charged droplets of water aerosol and the consequences of their discovery for the initiation of lightning in thunderclouds.

This chapter presents experiments where "unusual plasma formations" (UPFs) were observed for the first time within clouds of artificially charged water aerosol using a high-speed infrared camera, working in combination with a high-speed visible camera and measuring the current and glow of discharges in the optical range [Kostinskiy et al., 2015a]. The parameters of the infrared emission of the UPFs plasma channels were close to those of the positive leaders of long sparks observed in the same experiments, while the morphology of the UPFs channels differed from the parameters of any previously known leaders, so UPFs can be considered as a new, unusual type of intracloud and cloud discharges.

These plasma formations are probably the result of the collective processes of creating a complex hierarchical network of interacting channels at different stages of development (some of them are hot and are observed for milliseconds). We believe that similar phenomena can also occur in thunderclouds and may be one of the links in the process of lightning initiation inside thunderclouds. We were able to detect UPFs in the experiments described in this section, since the experiments used (a) a negative cloud of artificially charged water droplets with an average diameter of 0.5 μm and (b) an infrared IR camera, with a sensor sensitive in the wavelength range 2.7–5.5 μm (wavelengths an order of magnitude larger than the size of the droplets), which allowed us to "see" for the first time what is happening inside (within) the aerosol cloud.

*In section 1.1.* the experimental setup is described.

*Section 1.2.* the results of experiments are given.

*In section 1.2.1.* shown are two-frame image-enhanced camera frames and infrared frames of upward positive leaders and unusual plasma formations (UPFs) within the cloud.

*Section 1.2.2* shows images of UPFs located within a charged aerosol cloud below the top of the upward positive leader channel.

*Section 1.2.3* presents the results of IR sensing of the cloud and its surroundings in search of the location of UPFs initiation. The section shows that most UPFs occur at or near the visible boundary of the aerosol cloud within the aerosol cloud, but UPFs have also been found that occur near the grounded plane.

*Section 1.2.4* presents the results of observations of UPFs simultaneously, both in the IR and in the visible ranges. These experiments were important evidence that the images in the IR range correspond to the images in the visible range, and these images, discovered for the first time, are not artifacts of measuring instruments, but capture real physical phenomena.

*Section 1.2.5* shows IR images of events where multiple UPFs interact with each other and with positive leaders within a single event.

*Section 1.3* discusses the results obtained in Chapter 1.

*Section 1.4* discusses the results obtained in earlier works, which may have also recorded UPFs.

*Section 1.5* presents the conclusions of Chapter 1. The materials of the chapter are published in [Andreev et al., 2014], [Kostinskiy et al., 2015a].

**CHAPTER 2** describes the initiation of unusual plasma formations (UPFs) inside a positive streamer flash supported by the electric field of a negatively charged water aerosol. The purpose of this chapter was to establish the processes that lead to the occurrence of UPFs. To study the causes of UPFs, an experimental setup similar to that described in [Kostinskiy et al., 2015a] was supplemented with microwave diagnostics, which, together with other devices, made it possible to obtain experimental data that indicate a possible mechanism for the initiation of UPFs as hot plasma formations, formed inside a long positive streamer flash.

*Section 2.1* describes an experimental setup similar to that described in Chapter 1, supplemented by microwave diagnostics and a modified measurement scheme.

*Section 2.2* describes the obtained experimental results.

In *Section 2.3*, the experimental results of Chapter 2 are compared with those obtained in earlier studies, which may have also recorded UPFs [Анцупов и др., 1990]. The section shows that the new results are in good agreement with those obtained earlier and allow one to consistently explain the results of early experiments that did not have unambiguous interpretations.

*Section 2.4* discusses the results obtained in Chapter 2.

*Section 2.5* presents the conclusions of Chapter 2. The materials of the chapter are published in [Kostinskiy et al., 2021].

**CHAPTER 3** is devoted to detailed studies of the interaction between positive and negative leaders (breakthrough phase) in meter-scale electrical discharges generated by the electric fields of a negatively charged water aerosol [Kostinskiy et al., 2016]. For the first time, two images of the end-to-end phase of interaction between leaders are presented, showing a significant branching of the leader within the common streamer zone.

*Section 3.1* provides an introduction to the problem of the interaction of plasma channels (breakthrough phase).

*Section 3.2* describes an experimental setup similar to that described in chapters 1 and 2, modified for this scientific problem. The measurement scheme is supplemented with an optical high-speed camera in the visible range with a CMOS matrix.

*Section 3.3* presents the experimental results.

*Section 3.3.1* presents the breakthrough phase of the interaction of leaders, captured by a camera with image enhancement. Figure 3.1 shows two frames from the 4Picos camera showing the breakthrough phase of a negative discharge produced by a cloud of artificially charged water aerosol.

*Section 3.3.2* examines the brightness of the contact area of the leaders relative to the upper and lower parts of the resulting single channel. It has been experimentally shown that the IR brightness of the contact region (probably proportional to the energy input into the gas; see [Kostinskiy et. al., 2015a]) is 4–5 times higher than the brightness corresponding to a positive or negative leader above and below the contact region.

*Section 3.3.3* examines the contact point of the downward negative leader with the lateral surface of the upward positive leader. In the section, it was shown that not only interactions of leaders of the tip-to-tip (head-to-head) type take place, but also interaction when the leaders form a contact at angles close to 90 degrees to the direction of movement of one of the leaders. This phenomenon occurs when the leaders move in different planes in 3D space and the perpendicular to one of the channels is the shortest distance between the leaders.

*Section 3.3.4* analyzes the possible cause of an intracloud bidirectional leader. The early results of measurements [Анцупов и др., 1990] confirm the results of Chapter 2, where the cause of the appearance of hot plasma formations, which probably develop into bidirectional leaders, is the primary streamer flash rising from the grounded plane (see Chapter 2).

*In Section 3.3.5*, the upper positive part of the bidirectional intracloud leader is studied for the first time using IR measurements. For example, in Figures 3.2.I and 3.2.II (enlarged fragment) we see the central (1) and upper (2) part of the intracloud channel during the discharge, which ended with a quasi-return stroke (the lower part of this channel is a downward negative leader).

*In Section 3.4,* the results obtained are compared with those obtained in earlier studies ([Верещагин и др., 2003], [Temnikov et al. 2007], [Temnikov 2012a], [Temnikov et al. 2012b]), where contact of two leaders initiated in the electric field of an artificially charged aerosol cloud.

*Section 3.5* discusses the results of the experiments in Chapter 3 and compares them with the available experimental results for long sparks and lightning.

*Section 3.6* summarizes the conclusions of Chapter 3. The materials of Chapter 3 are partially published in [Kostinskiy et al., 2016].

**CHAPTER 4** investigates plasma formations, including bidirectional leaders, initiated in the electric field of a *positively charged* water aerosol. Plasma formations are open within the cloud, thanks to IR cameras that record in the range of 3-6 microns.

*Section 4.1* is an introduction to Chapter 4, which provides an overview of previous experiments with a positively charged aerosol cloud and plasma formations that have previously been detected [Верещагин и др. 1988], [Анцупов и др., 1990].

*Section 4.2* considers infrared images of plasma formations, including bidirectional leaders, initiated by the electric field of a positively charged water aerosol.

*Section 4.2.1* describes the experimental setup and features of a positively charged aerosol cloud.

*Section 4.2.2* describes the obtained experimental results.

*Section 4.2.2.1* considers the downward moving parts of the bidirectional leader and plasma formations.

*Section 4.2.2.2* describes the parts of the bidirectional leader and other plasma formations moving upward.

*Section 4.2.2.3* investigates the structures of intracloud discharges at the base of the cloud, near the grounded plane, and quasi-return strokes in the electric field of a positively charged cloud.

*Section 4.2.2.4* describes observations of discharges initiated by a positively charged cloud in the visible range. These studies agree well with IR measurements and previous experiments.

*Section 4.3* discusses the results obtained in Chapter 4.

*Section 4.4* presents the conclusions of Chapter 4. The materials of the chapter are partially published in [Kostinskiy et al., 2015b].

**CHAPTER 5** presents experimental data on modeling in laboratory experiments analogues of altitude-triggered lightning and "classical" triggered lightning in electric fields of negatively and positively charged water aerosol clouds.

*Section 5.1* is an introduction to Chapter 5, which provides a brief overview of the experimental data regarding triggered lightning.

*Section 5.2* describes an experimental setup similar to those described in Chapters 1-4 and the scheme of experiments, where a crossbow bolt with and without a grounded wire flies in the direction of a charged cloud.

*Section 5.3* describes the results of the experiments.

*Section 5.3.1* describes experiments where an ungrounded crossbow bolt flies in the direction of a charged cloud, which initiates a cloud-to-ground discharge. This type of discharge can be analogous to altitude-triggered lightning (ATL).

*Section 5.3.2* describes a discharge that can be analogous to a classical triggered lightning.

*Section 5.3.2.1* describes the "precursor pulses" of a positive upward leader initiated by a grounded bolt in the electric field of a negatively charged aerosol cloud.

*Section 5.3.2.2* presents, for the first time, unusual plasma formations (UPFs) detected with an IR camera, initiated by a crossbow bolt inside a negatively charged aerosol cloud.

*Section 5.3.2.3* describes upward positive leaders initiated from a grounded bolt before the bolt enters a negatively charged aerosol cloud.

*Section 5.3.2.4* shows for the first-time unusual plasma formations (UPFs) initiated by a crossbow bolt inside a *positively charged* aerosol cloud.

**Section 5.4** presents the conclusions of Chapter 5. The materials of the Chapter 5 are partially published in [Kostinskiy et al., 2015c].

**CHAPTER 6** describes the stepped development of a negative and positive leader leading to a powerful flash of a streamer corona: studies of long sparks initiated by Marx's high voltage generators. The purpose of the experiments is to physically simulate the stepped development of negative and positive lightning channels.

**Section 6.1** is an introduction to the problem of the motion of the negative stepped leader of a long spark and lightning.

**Section 6.2** is an introduction to the problem of the motion of the positive stepped leader of a long spark.

**Section 6.3** is an introduction to the problem of the motion of the lightning positive stepped leader.

**Section 6.4** describes the experimental setup.

**Section 6.5** shows recorded streamer flashes during the formation of steps of negative leaders and the fine structure of the streamer corona of the negative leader, including the space stem and the space leader.

**Section 6.6** investigates streamer flashes during the formation of steps of positive leaders.

**Section 6.7** discusses in detail the results of the experiments presented in Chapter 6 and their possible role in understanding the propagation of lightning and intracloud discharges.

**Section 6.8** contains the conclusions of Chapter 6. The materials of the chapter are published in [Kostinskiy et al., 2018].

**CHAPTER 7**. This chapter is the most important in the dissertation. It is written on the basis of all other chapters and is their synthesis and generalization. Based on experimental results of recent years, this chapter presents a qualitative description of a possible mechanism (termed the Mechanism) covering the main stages of lightning initiation, starting before and including the initiating event, followed by the initial electric field change (IEC), followed by the first few initial breakdown pulses (IBPs). The Mechanism assumes initiation occurs in a region of $\sim 0.1 - 1$ km$^3$ with average electric field E>0.3 MV/(m·atm), which contains, because of turbulence, numerous small "E$_{th}$ volumes" of $\sim 10^{-4}$-$10^{-3}$ m$^3$ with E $\geq 3$ MV/(m·atm). The Mechanism allows for lightning initiation by either of two observed types of events: a high‐power, very high frequency (VHF) event such as a Narrow Bipolar Event or a weak VHF event. According to the Mechanism, both types of initiating events are caused by a group of relativistic runaway electron avalanche particles (where the initial electrons are secondary particles of an extensive air shower) passing through many E$_{th}$ volumes, thereby causing the nearly simultaneous launching of many positive streamer flashes. Due to ionization-heating instability, unusual plasma formations (UPFs) appear along the streamers' trajectories. These UPFs combine into three-dimensional (3-D) networks of hot plasma channels during the IEC, resulting in its observed weak current flow. The subsequent development and combination of two (or more) of these 3-D networks of hot plasma channels then causes the first IBP. Each subsequent IBP is caused when another 3-D network of hot plasma channels combines with the chain of networks caused by earlier IBPs [Kostinskiy et al., 2020a].

In **CHAPTER 8**, we evaluate the dynamics of streamer burst initiation, which provide the spatiotemporal profile and propagation velocity of the phase wave of maximum VHF signals during the development of CID/NBE [Kostinskiy et al., 2020b].

Московский институт электроники и математики им. А.Н. Тихонова

Научно-исследовательского университета Высшая школа экономики

(МИЭМ НИУ ВШЭ)

На правах рукописи

Костинский Александр Юльевич

**ПЛАЗМЕННЫЕ СТРУКТУРЫ И ОБЪЕМНЫЕ СЕТИ КАНАЛОВ, КАК СОСТАВЛЯЮЩИЕ ПОСЛЕДОВАТЕЛЬНОГО-МЕХАНИЗМА ИНИЦИАЦИИ МОЛНИИ В ГРОЗОВЫХ ОБЛАКАХ**

25.00.29 – физика атмосферы и гидросферы

Диссертация на соискание учёной степени
доктора физико-математических наук

Москва - 2021

Моей любимой жене Оле,
без которой эта диссертация
никогда не была бы написана

Отличие молнии от искры состоит в том, что в лаборатории лавина зарядов, которая образует электрическую искру, создается с помощью высоковольтного генератора. Он поставляет энергию в канал искры. Молния же получает свою энергию от зарядов, находящихся на каплях и льдинках облака. Все эти частицы хорошо изолированы друг от друга. Расстояния между ними превышают космические <…>. Луна удалена от Земли на расстояние 300-400 тысяч км, что всего в 25-30 раз превышает диаметр Земли. Земля удалена от Солнца на расстоянии ста диаметров Солнца. Частицы же в облаке находятся на расстояниях, в сотни раз превышающие их диаметры. К тому же пространство между частицами заполнено плохо проводящим электрический ток воздухом. Как в этих условиях канал молнии собирает за несколько тысячных долей секунды заряды миллиарда миллиардов капель, распределенных по объему в несколько кубических километров? Это похоже на чудо! Причем чудо, которое повторяется при каждом ударе молнии, когда массы изолированных, «не знающих» друг друга капель, вдруг, как по сигналу, сливаются в единый коллектив с тем, чтобы через ничтожные доли секунды снова превратиться в массу суверенных частиц.

*И. Имянитов и Д. Тихий* [Имянитов и Тихий, 1980]

# Оглавление

# Введение[1]

> Понимание природы грозы и молнии существенно не только для метеорологов. Изучение электрических процессов в столь гигантских — по сравнению с масштабами лабораторий — объемах позволяет установить более общие физические закономерности природы высоковольтных разрядов, *разрядов в облаках аэрозолей.*
>
> *И.М. Имянитов,*
> *предисловие к книге [Мучник, 1973]*

Молния — одно из самых ярких и наиболее часто встречающееся опасное геофизическое явление, которое сопровождает человечество всю его историю и существовала задолго до него. Трудно найти человека, который никогда бы не видел вспышку молнии и не слышал бы раскаты грома. Каждую секунду на Земле происходит от 50 до 100 разрядов молний облако-земля (примерно 4-8 миллионов разрядов в день) [Rakov and Uman, 2003][2]. В среднем, на каждый квадратный километр поверхности Земли приходится 2-5 разряда молний в год [Базелян и Райзер, 2001].

Как самое частое опасное явление на Земле, грозы и молнии вызывают гибель людей и животных, нарушают работу линий электропередачи и связи, создают интенсивные радиопомехи, лесные пожары [Rakov and Uman, 2003, pp.642-655], [Dwyer & Uman, 2014]. Особую опасность молнии несут различным летательным аппаратам [Имянитов, 1970], [Laroche et al., 2012], [Rakov and Uman, 2003, pp.346-373]. Давно известно, что молния и другая внутриоблачная разрядная активность может быть важным диагностическим методом определения и предсказания грозовой активности, которая предупреждает о приближении ураганов, торнадо, сдвиге ветра, граде [Имянитов, 1960, 1961], [Мучник, 1974].

---

[1] Нумерация рисунков производится по главам. Во введении вместо номера главы первой ставится заглавная буква В, а в главах — номер главы. Рисунки данной диссертации пишутся с заглавной буквы и номером, чтобы их отличать от рисунков из других публикаций, которые пишутся со строчной буквы и номером.

[2] Все публикации в списке цитированной литературы расположены в алфавитном порядке (в начале русскоязычные) с добавлением к фамилии автора (авторов) года издания, чтобы различать публикации, написанные автором (авторами) в одном году, они маркируются буквами латинского алфавита по порядку (a, b, c…). Если в ссылке в квадратных скобках после фамилии два и более годов издания, то это различные публикации, где каждая публикация маркируется по фамилии и году издания.

Из-за частого наблюдения молнии и почти 270-летней истории ее исследований, начиная с Франклина, Далибара и Ломоносова, казалось, что молния уже достаточно хорошо изучена. Но в последние десятилетия были сделаны прорывные открытия, которые перевернули физику молнии, внутриоблачных разрядов и созданных ими других физических явлений. Первое по времени открытие, важнейшее для исследования молнии и атмосферных разрядов, сделал Дэвид Ле Вайн [Le Vine,1980]. Он обнаружил короткие (3-30 мкс) биполярные изменения электрического поля, которые впоследствии получили название «компактные внутриоблачные разряды» (compact intracloud discharge — CID) [Smith, 1999] или «узкие биполярные события» (narrow bipolar event — NBE) [Rison et al., 2016]. Однако CID вызвали сперва недоумение из-за своего резкого отличия от «привычных» проявлений молнии, а физика этого явления, по-видимому, тесно связанного с инициированием молнии, стала проясняться только сейчас (глава 7). Следующим выдающимся и неожиданным результатом стало открытие в 1989 году Джоном Рандольфом Винклером в средней атмосфере (50-70 км над Землей) гигантских, десятки километров в поперечнике, объемных электрических разрядов — спрайтов [Franz et al., 1990]. Спрайты оказались прямым следствием сильных ударов молнии. Оказалось, что это явление предсказал еще в 1925 году нобелевский лауреат, создатель «камеры Вильсона», Чарльз Томсон Риз Вильсон [Wilson, 1925a]. После открытия спрайтов были обнаружены еще несколько типов подобных разрядов, включая гигантские джеты, которые являлись разрядом, стартующим с верхней кромки грозовых облаков и достигающим нижней ионосферы [Pasko, 2010], [Sentman and Wescott, 1998], [Мареев, 2007]. Но на этом прорывные открытия не закончились. В 1994 году Джеральд Фишман с соавторами открыли, благодаря спутниковым измерениям, вспышки гамма-излучения, идущие с поверхности Земли [Fishman et al., 1994], которые были названы Terrestrial Gamma-Ray Flashes (TGFs). И это явление в том же 1925 году предсказал нобелевский лауреат Чарльз Вильсон [Wilson, 1925b]. Оказалось, что эти вспышки гамма-излучения тесно связаны с грозовой активностью и молнией [Dwyer & Uman, 2014, pp.187-223], но конкретный механизм генерации TGFs в настоящее время активно выясняется, хотя и здесь также наметился экспериментальный прорыв в понимании [Belz et al., 2020].

Несмотря на такое активное расширение тематики, относящееся к физике молнии и ее проявлениям, до сих остается не решенной, быть может самая главная проблема молнии: с помощью какого механизма (или механизмов) зарождается молния внутри грозовых облаков без присутствия протяженных проводящих объектов (типа самолетов и ракет). Этой проблеме посвящена данная диссертация и мы надеемся, что нам и нашим коллегам удалось сделать заметный и во многом неожиданный шаг в сторону решения этой проблемы.

## Актуальность темы исследования

В физике молнии нет более актуальной и запутанной проблемы, чем проблема механизма инициации молнии в грозовых облаках.

Известный исследователь молнии Мартин Юман, который последовательно в течение почти полувека публиковал (один и вместе с соавторами) наиболее цитируемые обзорные монографии по физике молнии [Uman, 1969, 1987, 2001], [Юман, 1972], [Rakov and Uman, 2003], [Dwyer & Uman, 2014], выделил «десять основных вопросов в исследовании молнии» [Dwyer & Uman, 2014, p.156]. Самым первым по важности пунктом были вопросы, касающиеся механизма инициирования молнии в грозовых облаках: «{1} Каким физическим механизмом или механизмами инициируется молния в грозовом облаке? Какова максимальная величина электрического поля внутри облака и в каком объеме облака? Какие процессы с высокой энергией (убегающие электроны, рентгеновские лучи, гамма-лучи) участвуют в возникновении молнии и как? Какова роль различных форм льда и воды в возникновении молнии?» Обращает на себя внимание, что [Dwyer & Uman, 2014, p.156] предположили, что механизм может состоять из нескольких физических явлений, в котором участвуют процессы усиления электрического поля, а также убегающие электроны, рентгеновские и гамма-фотоны. Отметим, что возможная ключевая роль широких атмосферных ливней космических лучей (ШАЛ) в инициации молнии, которая в эти годы активно обсуждается (например, [Gurevich et al., 1999], [Gurevich and Karashtin, 2013])), не включена в этот пункт из-за, по-видимому, отрицания Джозефом Двайером (Dwyer) этой роли. Главы диссертации 7 и 8 посвящены попытке ответить на эти вопросы и снова возвращают, благодаря предложенному новом механизму, ШАЛ ключевую роль в инициации молнии. Остальные главы диссертации служат необходимыми элементами (звеньями) общего механизма, описанного в главах 7 и 8.

Вторым по важности вопросом [Dwyer & Uman, 2014, p.156] считают вопросы физики распространения лидеров молнии: «{2} Какие физические механизмы управляют распространением различных типов лидеров молний (отрицательных ступенчатых, первых положительных, отрицательных стреловидных, отрицательных стреловидно-ступенчатых, отрицательных стреловидно-хаотических) распространяющихся между облаком и землей и внутриоблачными лидерами»? Вопросам ступенчатого распространения положительного и отрицательного лидеров длинной искры, которые моделирует развитие лидеров молнии, посвящена глава 6.

Третьим по важности [Dwyer & Uman, 2014, p.156] поставили вопрос взаимодействия нисходящего и восходящего лидеров молнии с наземными объектами, то есть, вопрос, актуальный для молниезащиты (который не рассматриваются в данной диссертации), а четвертым вопросом был отдельно выделена проблема физической природы компактных внутриоблачных разрядов — КВР (CID/NBE): «{4} Какова физика компактных внутриоблачных разрядов (CID) (которые производят узкий биполярный широкополосный импульс электрического поля, узкое биполярное событие (narrow bipolar event — NBE), сильно излучающий в HF и VHF-диапазоне? Как CID связаны с другими типами начальных импульсов пробоя (IBPs)? Связаны ли CID с земными гамма-вспышками (TGF), наблюдаемыми на спутниках, или с «переходными световыми явлениями» (TLE, то есть спрайтами, эльфами, джетами и гигантскими джетами – А.К.), зафиксированными над границами грозовых облаков, особенно с так называемыми «гигантскими джетами»?

[Rison et al., 2016] первыми предположили, что CID/NBE или их слабая по энергии модификация являются причиной инициации «всех или почти всех» молний. То есть, первый и четвертый пункт «программы исследований Юмана-Двайера» оказались тесно связанными. В качестве физического механизма CID/NBE они предположили механизм быстрого положительного пробоя – FPB, который, по их мнению, представляет собой гигантскую стримерную вспышку, которая движется со скоростью $10^7$-$10^8$ м/с [Attanasio et al., 2019]. На наш взгляд, идея внутренней связанности CID/NBE (в большинстве случаев слабого аналогичного события) и момента инициации молнии справедлива, но газоразрядные стримеры при давлениях 0.1-1 атм ни в коем случае не могут двигаться со скоростями $10^7$-$10^8$ м/с в подпробойных электрических полях [Les Renardières Group, 1972, 1974, 1977, 1981]. Ниже работы [Rison et al., 2016], [Attanasio et al., 2019] будут рассмотрены подробно. Вопрос тесной связи инициации молнии и КВР (CID) является центральной частью предложенного нами механизма инициации молнии (главы диссертации 7 и 8).

Насколько актуальна и сложна проблема инициации молнии в облаках, можно видеть, рассмотрев мнения ведущих российских исследователей. В книге «Физика молнии и молниезащиты» [Базелян и Райзер, 2001, стр.156] есть раздел, который так и называется — «Зарождение в облаках», где высказан следующий взгляд на состояние проблемы на тот момент: «Хотя процесс распространения нисходящего отрицательного ступенчатого лидера больше всего знаком наблюдателям молнии, обстоятельства и механизмы его зарождения в буквальном смысле слова покрыты туманом. Никто не наблюдал старта молнии и не следил за развитием процесса в облаках. Процесс зарождения не воспроизводится в полной мере в лаборатории, хотя в экспериментах получают отрицательный ступенчатый лидер. Но условия его зарождения от высоковольтного металлического электрода, присоединённого к конденсаторной батарее импульсного генератора, имеют мало общего с тем, что происходит в облаках. Облако — не обкладка конденсатора и вообще не проводник. Отрицательный облачный заряд рассеян в диэлектрической воздушной среде на малых по размерам гидрометеорах. Трудно вообразить, как может этот рассеянный в громадном объеме заряд, сидящий на малоподвижных частицах, собраться и за миллисекунды направиться в плазменный канал». Обратим внимание на точность оценок [Базелян и Райзер, 2001, стр.156]. Фактически молнией по умолчанию считается мощный ступенчатый отрицательный лидер, выходящий из грозового облака, как изображено на ставшей классической схеме-развертке из книги Мартина Юмана [Юман, 1972, стр.17], Рисунок В.1 Но в этом тексте есть и неточности, так как к началу 2000-х уже удалось надежно зафиксировать глубоко внутри грозового облака движение на расстояния нескольких километров отрицательного лидера с точностью не хуже 100 м (например, [Proctor et al, 1988], [Shao and Krehbiel,

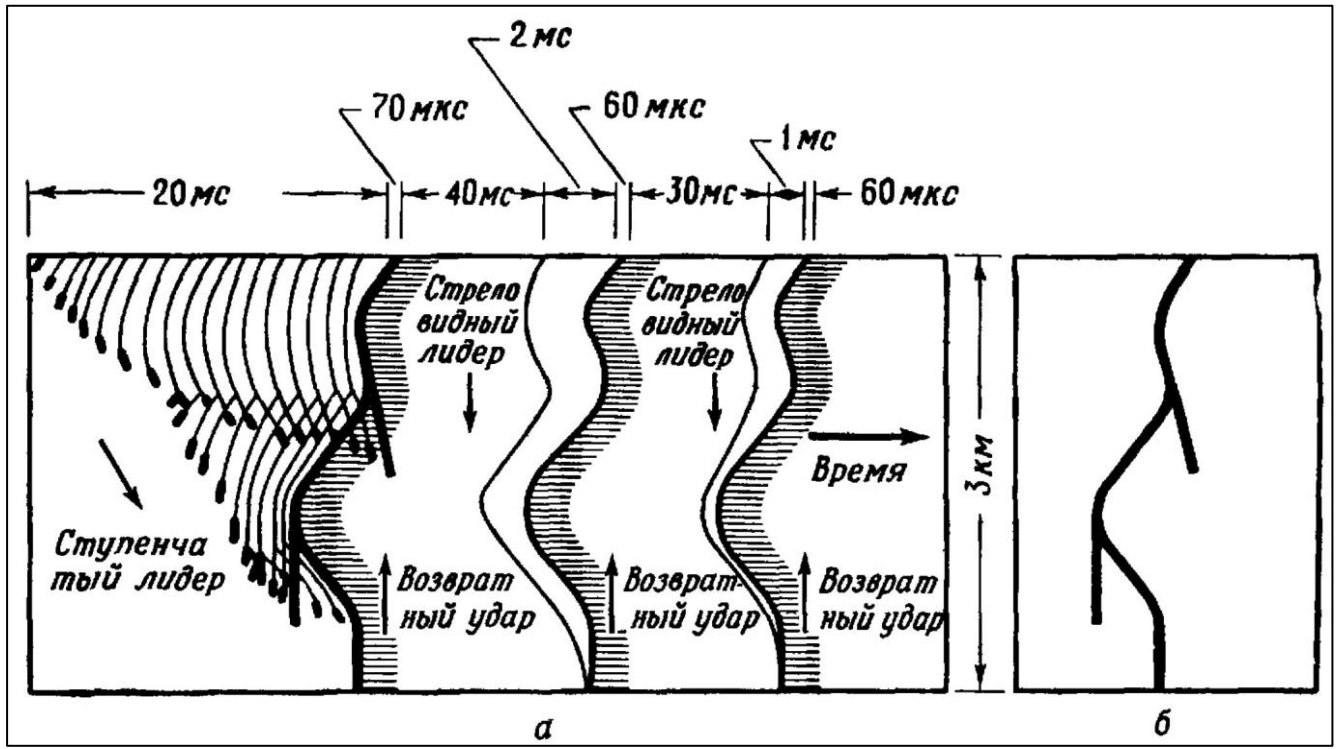

Рисунок В.1 (адаптировано из [Uman, 1969; Юман, 1972]). (а) Схема временной развертки трех ударов молнии, которые «записаны» камерой с неподвижным объективом и движущейся пленкой (подобно работе фотоэлектронных регистраторов или электронных стрик-камер). Время течет слева направо. Шкала времени для ясности изображения нелинейно. Ступени отрицательного лидера (слева) изображены жирными отрезками на кончиках тонких каналов лидера. Штриховка после обратных ударов подчеркивает, что ток через горячий плазменный канал течет и после прохождения обратного удара; (б) схема, подобная изображению обычным фотоаппаратом (с неподвижной пленкой или матрицей, без развертки изображения).

1996]), благодаря системам триангуляции радиоисточников по запаздыванию радиосигналов на разнесенных на несколько километров антеннах (VHF-time-of-arrival (ТОА)). Именно, опора одновременно на оптические (в видимом и ИК-диапазоне) и радиофизические методы исследования молнии привели в последние годы к прорыву в понимании внутриоблачных разрядов и процессов.

Далее [Базелян и Райзер, 2001, стр.156] обращают внимание на перспективы изучения искусственно заряженного аэрозольного облака, которое мы с коллегами впоследствии (надеюсь успешно) использовали в исследованиях, получивших отражение в данной диссертации: «Нельзя сказать, что в земной практике нет никаких намеков на что-либо сходное с зарождением искры в облаках. То, что в среде с дисперсно размещенным зарядом иногда возникают искровые разряды, известно благодаря расследованиям причин взрывов и пожаров в производственных помещениях с большими объемами электростатически заряженных пылевых частиц или капель. В последнее время появились отчеты об исследованиях при помощи пароструйных генераторов, выбрасывающих в атмосферу миниатюрные электрически заряженные облака ([4.3] — соответствует публикации [Antsupov et al., 1991], [4.4] — соответствует [Верещагин и др., 1989] – А.К.). Иногда у границы заряженного аэрозоля возникали вытянутые светящиеся образования с размерами порядка 10 см; реже они превращались в искровые каналы (длиной до 1 м). К сожалению, в экспериментах не удавалось измерить поле в месте старта искрового разряда и дальше констатации факта возбуждения искр дело не дошло. Поэтому о механизмах возбуждения молнии в облаках и искр в лабораторных аэрозольных смесях пока приходится только гадать». Обнаружению и подробному исследованию плазменных каналов и сетей, внутри этих «миниатюрных электрически заряженных облаков», которые получили название «необычных плазменных образований» (unusual plasma formations — UPFs) [Kostinskiy et al., 2015a, 2015b, 2015c], [Kostinskiy et al., 2016] посвящены главы 1-5 данной диссертации. Эти результаты активно использовались, как ключевые элементы (звенья) при построении механизма инициации молнии [Kostinskiy et al., 2020a, 2020b] (главы 7, 8).

Далее [Базелян и Райзер, 2001, стр.156] высказывают предположение, которому мы будем ниже следовать в наших оценках: «Умозрительные (а лишь такие сегодня и возможны) заключения по этому поводу приходится строить методом исключения. Облачную среду нельзя считать проводником, когда речь идет о снабжении током лидерного канала. Облачные заряды непосредственно в лидер не транспортируются и за время скоротечного лидерного процесса сами по себе из облака никуда не уходят. Стало быть, облачным зарядам уготована другая роль — быть источником электрического поля. Оно вызывает ионизацию молекул воздуха и рождение начальной плазмы, а потом поддерживает лидерный процесс. Для выполнения первой задачи поле в какой-то части заряженной области должно подняться выше порога ионизации ($E_i \approx$ 20-25 кВ/см на высоте заряда облака) или в облаке должны действовать включения, локально усиливающие поле своим поляризационным зарядом. По-видимому, ни того, ни другого нельзя исключать безоговорочно, хотя при зондировании облаков редко регистрировались поля выше нескольких киловольт на сантиметр. Эти результаты еще не говорят об отсутствии более сильных полей, поскольку чаще измерялись поля, усредненные на длинах в десятки метров, а сами измерения никогда не производились в момент рождения молнии (вероятность внести датчик в нужный момент в нужное место крайне мала). Но с другой стороны, условия для возбуждения лидерного процесса в облаке появляются не часто, иначе на квадратный километр поверхности земли приходилось бы не 2-5 ударов молний за грозовой сезон, а много больше». Мы также будем исходить в своих оценках в этой диссертации, что заряженные облачные гидрометеоры являются источником электрических полей, как локальных, так и крупномасштабных, а основные заряды, которые обеспечивают токи и заряды разрядов между плазменными каналами и сетями в облаке, сосредоточены в чехлах этих плазменных каналов. Чехлы плазменных каналов образованны стримерными коронами, которые создают при своем движении положительные и отрицательные лидеры [Базелян и Райзер, 1997, стр.229-230] и другие горячие высокопроводящие плазменные каналы [Les Renardieres Group, 1972, 1974, 1977, 1981], [Горин и Шкилев 1974, 1976]. То есть, ток и заряд таких масштабных разрядных явлений, как, обратные удары молнии или начальные импульсы пробоя (IBPs) черпается из чехлов двунаправленных лидеров и/или плазменных сетей, во время «возвратных разрядов» (не путать с обратными ударами молний) или по-другому этот процесс называют — «обратной короной» [Базелян и др., 1978, стр. 87-90], [Райзер, 1992, стр.451-452, 466]. Если токи и заряды велики, то и общая длина плазменных каналов также должна быть минимум несколько километров (примерно пропорциональной зафиксированным зарядам).

Таким образом процесс инициации молнии в грозовых облаках, как и процесс инициации компактных внутриоблачных разрядов являются одними из самых актуальных проблем физики молнии и грозы и приближение к пониманию этих процессов важно как с фундаментальной, так и с прикладной точек зрения не только в узком, но и широком смысле, учитывая, что молния и КВР (CID) являются разрядными событиями в аэрозольных (многофазных) средах, подобными, например, разрядам в облаках вулканического пепла [McNutt and Thomas, 2015] или в атмосферах других планет [Brown et al., 2018] и др.

## Степень разработанности проблемы инициации молнии и компактных внутриоблачных разрядов

### Что можно считать — молнией?

Для того, чтобы подробно обсуждать проблему инициации молнии в облаках необходимо более четко ответить на вопрос: что такое молния? Кажется, что на этот вопрос есть простой и однозначный ответ, так как это интуитивно понятно практически каждому человеку, а не только ученому. Выше мы приводили это общепринятое представление, в виде цитаты: «процесс распространения нисходящего отрицательного ступенчатого лидера» [Базелян и Райзер, 2001, стр.156] или Рисунка В.1 (слева), где схематически изображён ступенчатый лидер на общей схеме развития нисходящей отрицательной молнии [Юман, 1972]. Также, для наглядности, кадр скоростной съемки отрицательного ступенчатого лидера показан на Рисунке В.2 [Lu et al., 2016]. Конечно, земной поверхности достигают по крайней мере четыре типа молний (лидеров): отрицательные и положительные нисходящие лидеры (молнии) и отрицательные и положительные восходящие лидеры (Рисунок. В.3) [Rakov and Uman, 2003]. Но, так как основной темой нашего исследования является инициация молнии в облаках, то мы будем считать моментом появления молнии момент, когда возникнет настолько *большой* ступенчатый отрицательный лидер, что он будет хорошо различим на изображениях системы грозопеленгации молнии по сигналам источников VHF-излучения внутри облака (VHF-картографирования), например такой, как отрицательный лидер, изображенный на

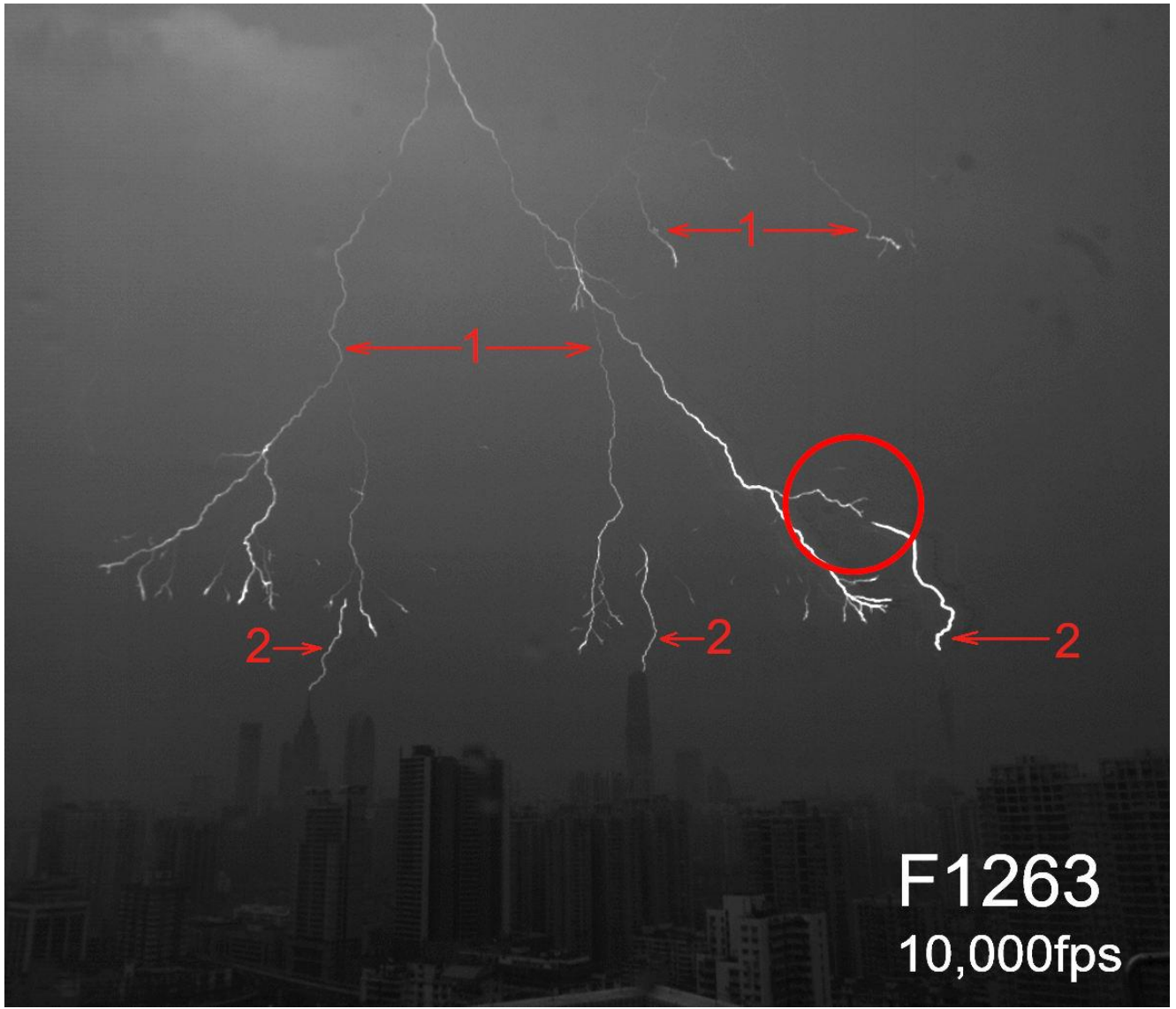

Рисунок В.2 (адаптировано из [Lu et al., 2016]). Ветви нисходящего отрицательного лидера (1), восходящие положительные лидеры (2). Красной окружностью показано место контакта ветви нисходящего отрицательного лидера и восходящего положительного лидера, которые перейдут в обратный удар. Снято скоростной камерой со скоростью 10'000 кадров в секунду в Гуанчжоу, Китай.

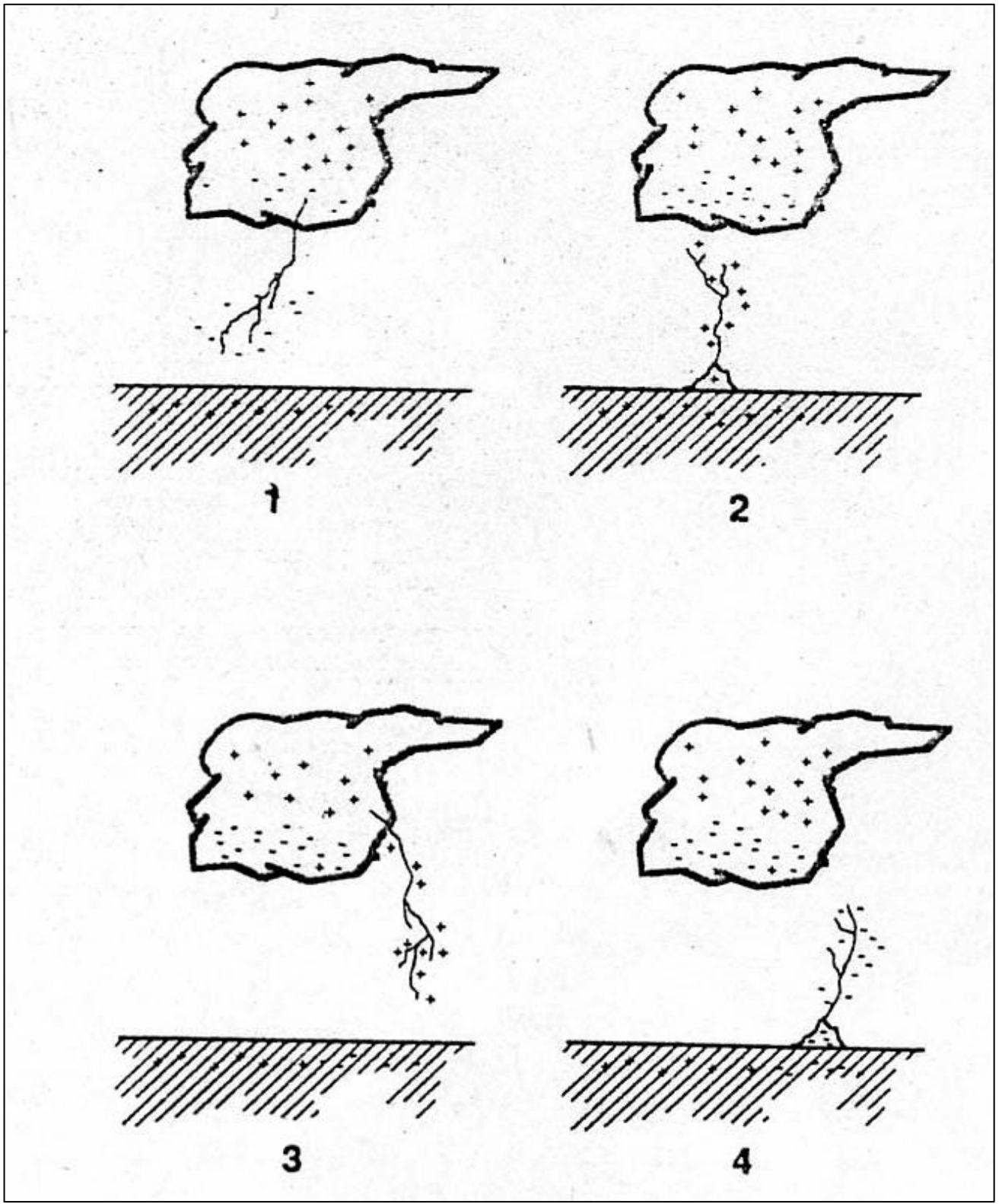

Рисунок В.3 (адаптировано из [Rakov and Uman, 2003]). Классификация типов четырех основных типов молний (типов лидеров), достигающих Земли. (1) — нисходящая отрицательная молния, (2) — восходящая положительная молния, (3) — нисходящая положительная молния, (4) — восходящая отрицательная молния.

Рисунке В.4 (пространственное разрешение не хуже 10 м) [Hare et al., 2019]. Мы будем считать моментом окончания инициации молнии именно момент появления *большого* отрицательного лидера, так как, именно его считают моментом образования молнии при исследовании внутриоблачных разрядов, как изображено на типичной LMA-карте распространения молнии (например, Рисунок В.5 [Rison et al., 2016]), так как с момента появления такого лидера распространение каналов молнии становится долгосрочным (десятки и сотни мс) и приводит с высокой вероятностью к внутриоблачным (IC) и облако-земля (CG) молниям. На Рисунке В.5 момент времени и пространственное расположение цепочки событий, инициирующих молнию показано красным кружком. Именно подобный период времени и промежуток пространства (внутри красного кружка на Рисунке В.5) будет предметом исследования данной диссертации. Красными квадратиками обозначен отрицательный лидер. Характерно, что изображение отрицательного лидера на карте появляется раньше (это типичная ситуация на LMA-картах), чем положительного, скорее всего из-за отчетливого радиоизлучения, которое характеризует большой отрицательный ступенчатый лидер. Исходя из многолетних исследований длинной искры и высотно инициированных триггерных молний (ATL), можно с высокой уверенностью предполагать, что положительный лидер должен возникнуть раньше, чем отрицательный, так как для этого требуется в 1.5-2 раза меньшее напряжение на головке лидера [Горин и Шкилев, 1974, 1976], [Les Renardières Group, 1977, 1981], [Rakov and Uman, 2003, pp.269-272]. Синими квадратиками на Рисунке В.5 обозначен положительный лидер, изображение которого появляется на LMA-карте через целых 25 мс после появления отрицательного лидера. Более того, опираясь на исследования длинной искры, создаваемой генераторами импульсного напряжения и искусственным аэрозольным облаком, мы можем с высокой вероятностью сказать, что небольшие плазменные образования и впоследствии лидеры должны появляться внутри области пространства и времени, ограниченного красным кружком на Рисунке В.5 (главы 1-5), но радиоизлучение этих начальных плазменных образований в большинстве случаев, вероятно, будет находиться за пределами чувствительности распространенных LMA-систем.

Таким образом, мы будем считать внутриоблачные плазменные образования молнией, когда возникнет *большой* отрицательный лидер, а моментом образования молнии (в обозначениях LMA-карт) будем считать момент, когда отрицательный лидер начнет «распространяться» за пределами красного кружка на Рисунке В.5. Из этого определения следует, что далеко не все плазменные образования и разряды внутри облака мы будем считать молнией.

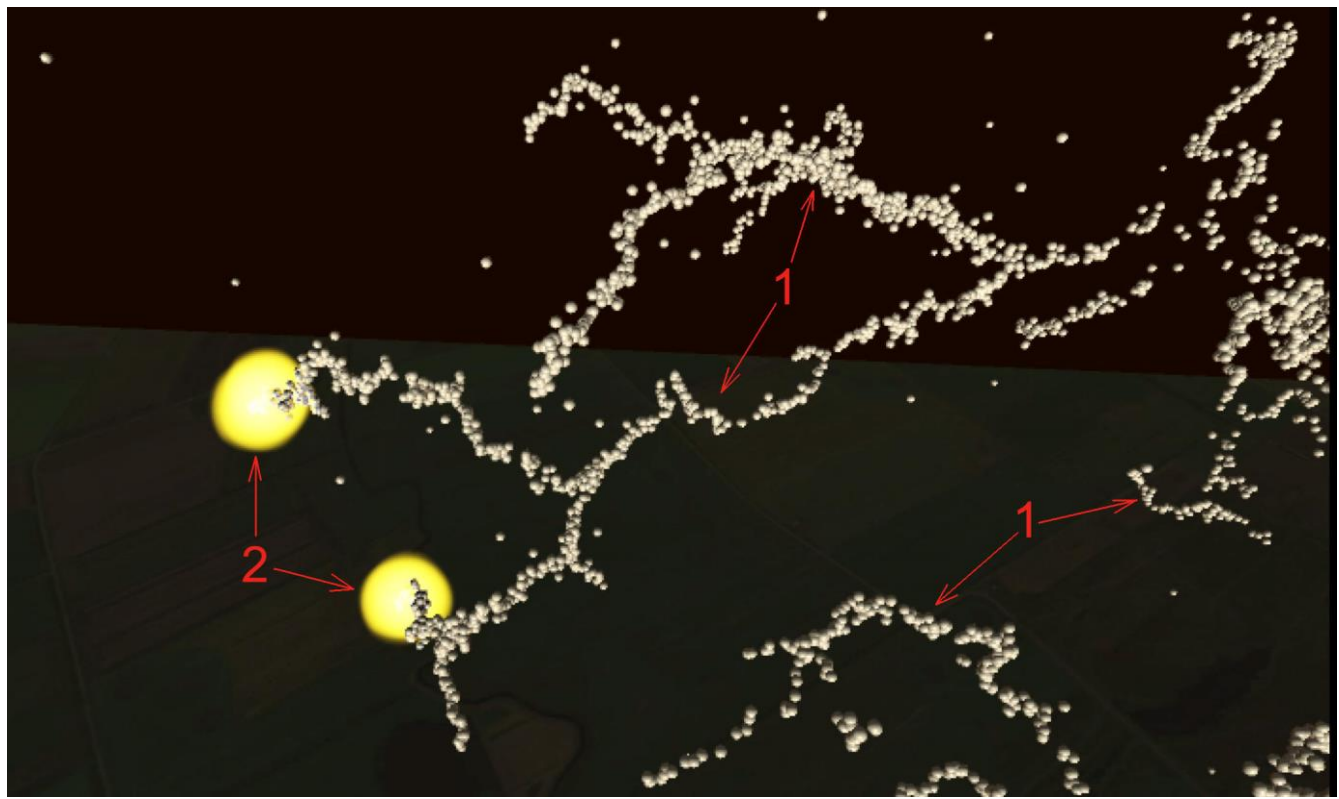

Рисунок В.4 (адаптированный кадр из видеоролика 3D-VHF-LMA-картографирования распространения радиоисточников каналов молнии внутри облака с пространственным разрешением не хуже 10 м [Hare et al., 2019]).
Ветви нисходящих отрицательных лидеров (1), момент образования ступени отрицательного лидера (2).
Ролик № 2019-41586_2019_1086_MOESM3_ESM-NEGATIVE_LEADER [Hare et al., 2019]

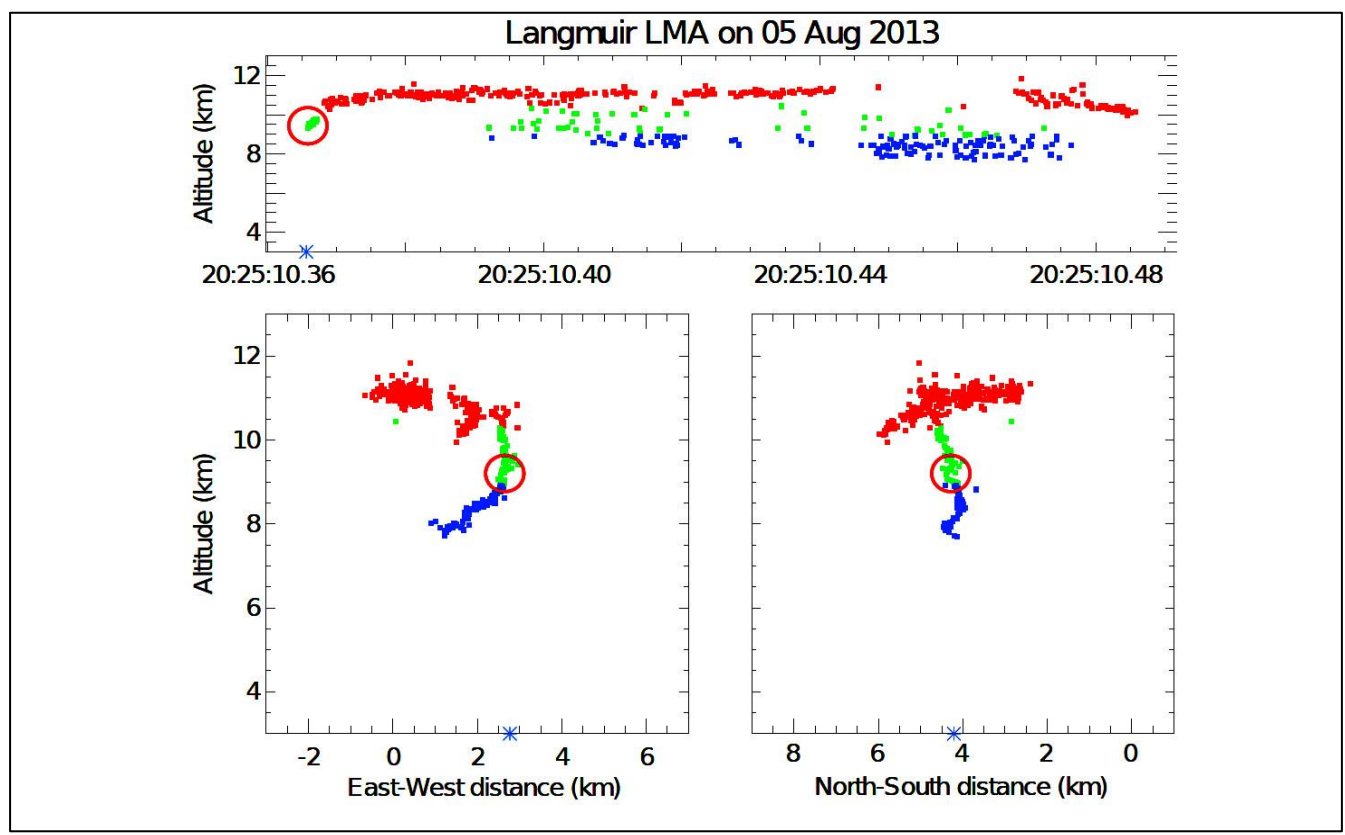

Рисунок В.5 (адаптировано из [Rison et al., 2016], это событие также рассматривается на Рисунках В.30-В.31). Трехмерная LMA-карта распространения радиоисточников каналов молнии внутри облака (внутриоблачная молния — IC), созданная, благодаря фиксированию импульсов излучения в VHF-диапазоне. Красным кружком показан момент и пространственное расположение события, инициирующего молнию. Красными квадратиками обозначен отрицательный лидер, синими квадратиками обозначен положительный лидер. Показаны первые 120 мс развития молнии. Все расстояния на панелях даны в км, а время на верхней панели изменяется в долях секунды от 0.36 до 0.48 с.

### Критерии инициации молнии

Канал нисходящей молнии, который удается зафиксировать на скоростных видеокамерах и стрик-камерах (фоторегистраторах с разверткой изображения) имеет размеры в несколько километров, токи и заряды при обратных ударах составляют десятки кА и несколько Кл, соответственно [Rakov and Uman, 2003]. Хорошо понятно, что плазменный канал с такими характеристиками не может возникнуть внезапно, а должен развиваться в какой-то последовательности. Изучая оптические записи лидеров молнии (включая восходящие лидеры) LMA-карты и карты, полученные на радиоинтерферометрах, подобные, изображенным на Рисунках В.2, В.4, В.5, исследователи приходят к выводу, что в первом приближении (через несколько миллисекунд или десятков миллисекунд после инициации) молнию можно описать, как большой двунаправленный ветвящийся канал. Поэтому задача исследования инициации молнии в грозовых облаках сводится к изучению первых нескольких миллисекунд после первого инициирующего события (an initiating event — IE) и построению на основе этих исследований механизма инициации и развития плазменных образований, которые превратятся в привычные ветвящиеся длинные лидеры. То есть, желательно построить цепочку последовательных преобразований плазмы, конечным итогом которых будут лидеры, подобные лидерам на Рисунках В.2, В.4.

Однако, построить цепочку этих первых шагов появления и преобразований плазмы оказалось чрезвычайно сложно, так как измеренные электрические поля в грозовых облаках были в несколько раз меньше, чем поля необходимые для создания газоразрядных электронных лавин [Rakov and Uman, 2003, pp.82-84], то есть условий, когда частота ионизации молекул воздуха электронами будет превышать частоту прилипания электронов к молекулам кислорода и воды (при атмосферном давлении этому критерию соответствует минимальное пороговое электрическое поле около 30 кВ/см или 3 МВ/м). Но электронные лавины в воздухе не возникнут даже если электрическое поле меньше порогового на 1% [Meek & Craggs, 1953], [Ретер, 1968]. Тем ни менее [Marshall et al., 2005] оценили, что область, где возникают CG-молнии, имела среднее электрическое поле $E > 284-350$ кВ/(м·атм) и занимала объем 1-4 км$^3$ с вертикальной и горизонтальной протяженностью 300-1000 м.

Таким образом задача инициирования молнии для большинства исследователей свелась к задаче получения первых электронных лавин или первого стримера. То есть до самого последнего времени, ввиду очень серьезных проблем практически не производились попытки построить механизм инициации молнии от первого инициирующего молнию события до большого канала отрицательного (или двунаправленного) лидера. Насколько нам известно, только [Petersen et al., 2008] впервые попытались описать возникновение молнии от инициации первого стримера до появления ступенчатого отрицательного спейс-лидера, то есть попытались создать непрерывную цепочку плазменных событий (Рисунок В.38). К сожалению, многообещающая попытка [Petersen et al., 2008] опиралась на мало обоснованную гипотезу [Griffiths and Phelps, 1976], которая описывала генерацию большой самоусиливающейся стримерной вспышки (подробный анализ существенных недостатков экспериментальных оснований гипотезы Гриффитса-Фелпса проведен в разделе «Компактные внутриоблачные разряды (КВР/CID/NBE) и их возможная роль в инициации молнии», стр.56). Кроме того, [Petersen et al., 2008] не очень хорошо разобрались в феноменологии спейс-стема длинной отрицательной искры, которую активно использовали (например, [Горин и Шкилев, 1976], [Les Renardières Group, 1981]), пользуясь устаревшими представлениями [Schonland, 1956]. Также можно отметить, что цепочка плазменных событий [Petersen et al., 2008] обрывалась почему-то на длинном спейс-лидере и не была доведена хотя бы до устойчиво развивающегося двунаправленного лидера.

Так как, как отмечалось выше, средние электрические поля в грозовом облаке малы, то они должны быть усилены. Первым и самым естественным кандидатом на усиление электрического поля являлись крупные гидрометеоры (крупа, град, капли), поляризованные в электрическом поле и/или имеющие большой заряд. Поэтому возникли гипотезы об инициации гидрометеорами коронного разряда, стримеров или стримерных вспышек (например, [Dawson and Duff, 1970], [Griffiths and Latham, 1974], [Phelps, 1974], [Griffiths and Phelps, 1976], [Petersen et al., 2015], [Babich et al., 2017]). Если, по мнению этих авторов, стримеры успешно стартовали с поверхности гидрометеоров, то вопрос инициации молнии решен. Но, как справедливо отмечали [Базелян и др., 1978]: «Лабораторные исследования показывают, что слабые ионизационные процессы, подобные короне на гидрометеорах, не могут возбудить лидерного процесса, поскольку возникновение канала лидера связано с термоионизацией воздуха и возможно лишь при достаточно большой плотности выделения энергии. Необходимая плотность энергии обеспечивается только стримерными процессами, которые всегда предшествуют лидеру и требуют для своего возбуждения полей напряженностью порядка 20 кВ/см на длине по крайней мере в несколько сантиметров [Стекольников, 1960]». Действительно, стримеры являются холодными плазменными волнами ионизации, после прохождения которых через электроотрицательный газ (воздух), электроны прилипают к молекулам кислорода за ~100 нс и плазма распадается. Для того, чтобы плазма «жила» хотя бы несколько микросекунд, она должна стать горячей настолько, чтобы пошла термическая ионизация воздуха. Хорошо известно из физики длинной искры, что на электроде раз за разом в течении многих минут могут возникать стримерные вспышки, но лидерный канал может так и не возникнуть («Если позволяет мощность лабораторного источника напряжения, стримерную корону тоже можно наблюдать часами. К короткому замыканию это не приводит» [Базелян и Райзер, 1997, стр.14], [Bazelyan & Raizer 1998, p.6]). Еще более показательный пример сложности инициации устойчиво распространяющегося положительного лидера даже при наличии мощных стримерных вспышек (а не одного стримера), которые сопровождаются токами 30-60 А и несут заряд в несколько десятков микрокулон [Lalande et al., 1998], [Rakov and Uman, 2003, стр.274-275] дают восходящие с кончика триггерной ракеты положительные лидеры триггерных молний. Отметим, что заряд даже самых сильно заряженных гидрометеоров не превосходит 1 нКл ([Marshall and Winn, 1982]; [Marshall and Marsh, 1993]; [Marsh and Marshall, 1993]; [Stolzenburg and Marshall, 1998]; [Bateman et al., 1999]), что на четыре порядка меньше, чем заряд стримерных вспышек с кончика триггерной ракеты. На Рисунке В.7 (панель С, красная стрелка [Hill et al., 2012, Fig. 7.3]) хорошо видна типичная ситуация, когда на протяжении целых 0.5 секунды возникают прекурсоры (стримерные вспышки и возможно начальные лидеры), которых больше 40, но не возникает устойчиво распространяющийся положительный лидер.

Таким образом, критерий возникновения одного стримера с гидрометеора, как критерий возникновения молнии имеет серьезные недостатки, несмотря на всю сложность проблемы инициации хотя бы одного стримера в грозовом облаке (см. подробнее разделы 8.1.1-8.1.2) и поэтому предложенный нами механизм почти одновременного

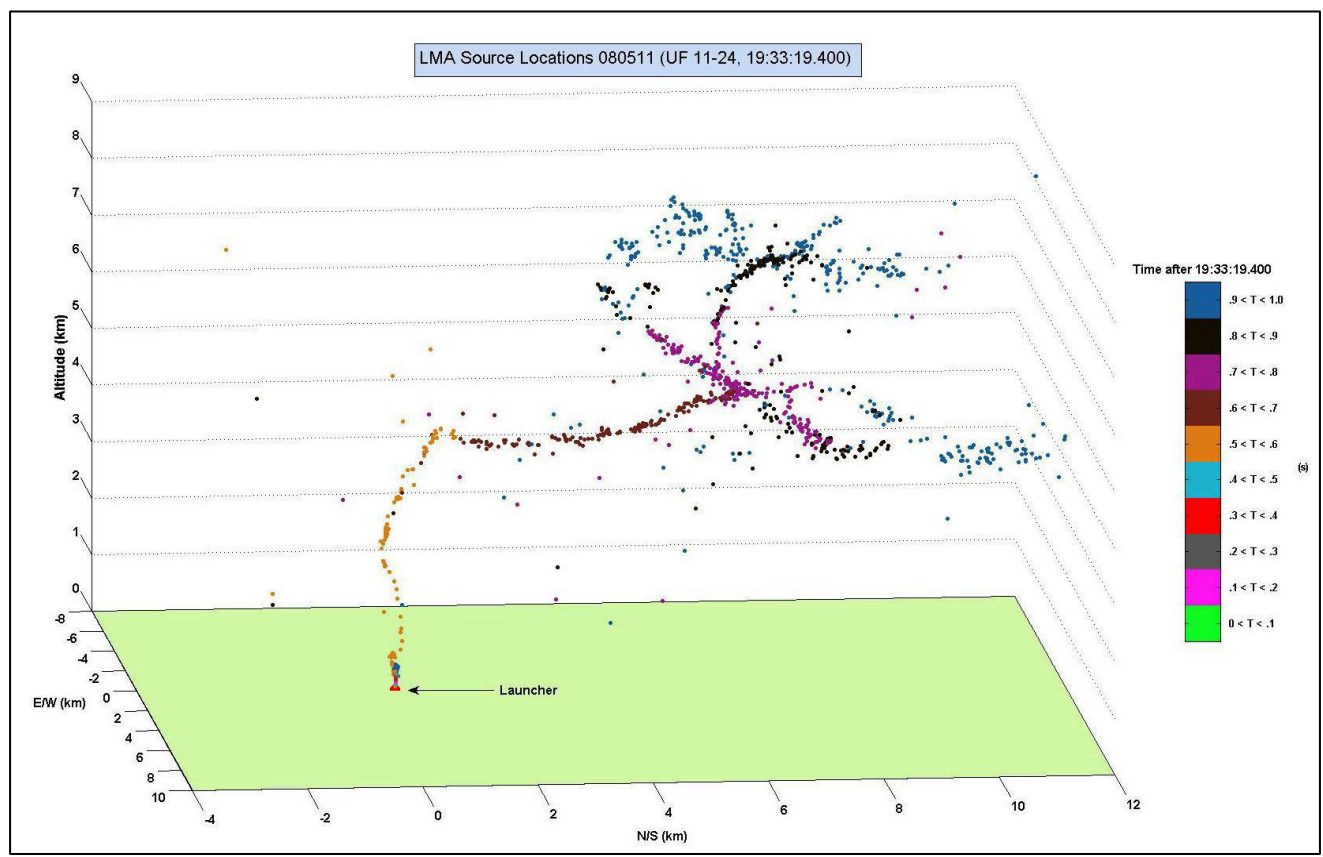

Рисунок В.6 (адаптировано из [Hill et al., 2012, Fig.7-2]) Трехмерная LMA-карта местоположения источников VHF-излучения восходящей триггерной молнии для вспышки UF 11-24 5 августа 2011 года [Hill et al., 2012]. Запись VHF-излучения соответствует общему времени — 1 с, начиная с 19:33: 19.400 (UT), и время на LMA-карте имеют цветовую кодировку в соответствии со столбцом справа, где каждый цвет соответствует промежутку времени 100 мс. Место стартового комплекса, откуда стартовала ракета с проводом обозначено стрелкой. Каждая точка на LMA-карте соответствует одному, самому сильному источнику VHF-излучения, в рамках временного окна 80 мкс. То есть, временное разрешение этой LMA-карты составляет 80 мкс. Эта же карта дана на Рисунке В.7, но в проекции на каждую ось и удобным графиком по времени.

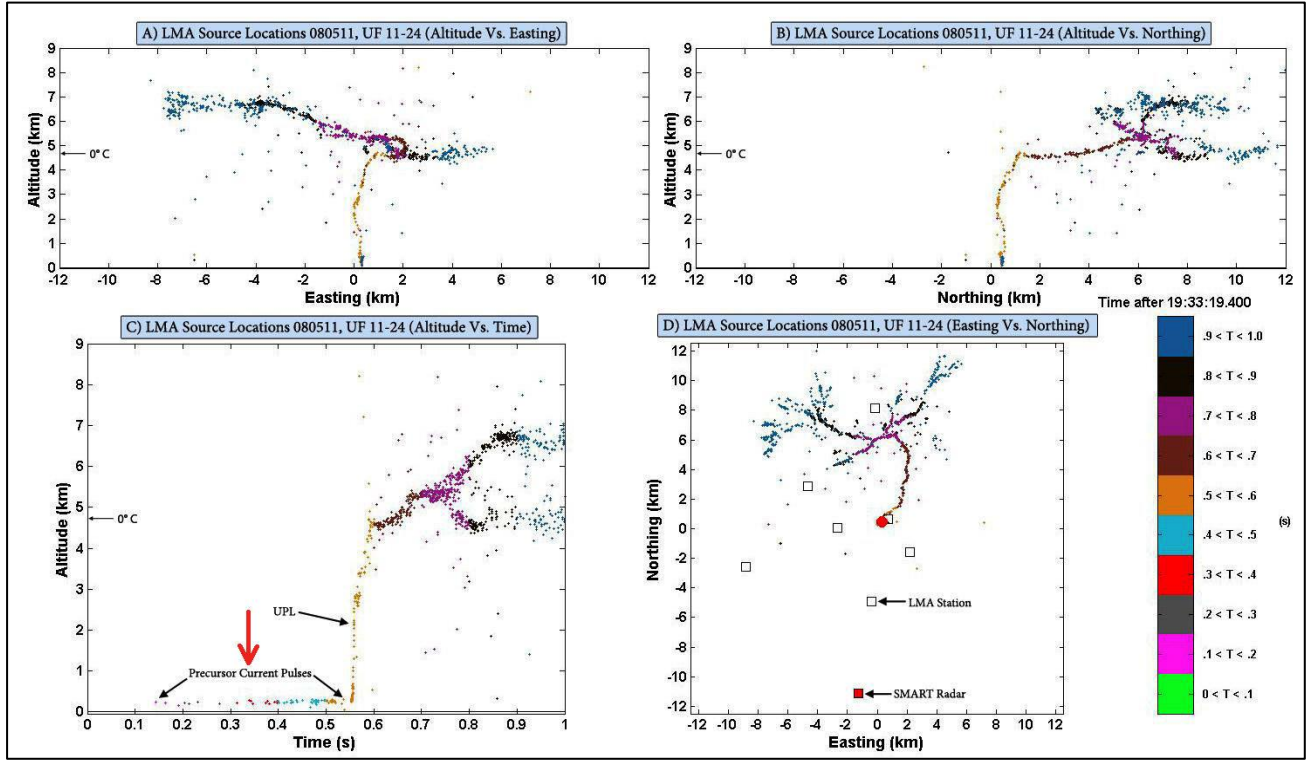

Рисунок B.7 (адаптировано из [Hill et al., 2012, Fig.7.3], см. это событие также на Рисунке B.6). Четыре проекции трехмерного изображения местоположения на LMA-карте источников VHF-излучения, которое возникает при инициации восходящей триггерной молнии для вспышки UF 11-24 5 августа 2011 года (трехмерное изображение этой же LMA-карты с цветовой маркировкой времени см. выше на Рисунке B.6). A) график источников VHF-излучения в зависимости восточного направления от высоты (вид на север), B) график источников VHF-излучения в зависимости северного положения от высоты (вид на запад), C) график зависимости источников VHF-излучения от высоты со временем и D) график зависимости источников VHF-излучения восточного направления от северного (вид сверху, стартовая площадка ракеты, несущей провод отмечена большим красным кружком). Запись VHF-излучения соответствует общему времени — 1 с, начиная с 19:33: 19.400 (UT), и время на всех LMA-проекциях также имеет цветовую кодировку в соответствии со столбцом в правом нижнем углу, где каждый цвет соответствует промежутку времени 100 мс. Черной стрелкой показана высота уровня температуры 0 ° С. Красным квадратом на панели D обозначен двухполяризационный радар. «Толстая» красная стрелка на панели (C) указывает на промежуток времени около 0.5 с, когда на кончике ракеты возникали стримерные вспышки (прекурсоры), но устойчиво распространяющийся положительный лидер не мог стартовать. Устойчиво распространяющийся положительный лидер стартовал в момент времени около 0.55 с (панель С). Каждая точка на LMA-карте соответствует одному, самому сильному источнику VHF-излучения, в рамках временного окна 80 мкс. То есть, временное разрешение этой LMA-карты составляет 80 мкс.

коллективного инициирования множества ($10^3$-$10^7$) стримерных вспышек в ограниченном объеме резко повышает вероятность инициации и поддержания развития лидера в такой трехмерной системе инициации молнии (см. подробнее главу 7).

В 1992 году Александр Викторович Гуревич [Gurevich et al., 1992], [Gurevich & Zybin, 2001] с соавторами предложил совершенно новый механизм инициации молнии, который они назвали «пробоем на убегающих электронах». [Gurevich et al., 1992] предположили, что космические лучи с энергией вторичных электронов больше, чем 300-500 кэВ могут порождать в сильных электрических полях грозового облака экспоненциально развивающиеся лавины убегающих электронов (не классические газоразрядные лавины). [Gurevich et al., 1992] предположили, что лавина убегающих электронов приведет непосредственно к разряду молнии, раз существует экспоненциальное увеличение числа электронов. К сожалению, убегающие электроны производят вдоль своего пути на высоте, например, 6 км, всего лишь 40 вторичных электронов на сантиметр, а для инициации стримера необходимы минимум $10^8$ электронов в размере меньше, чем миллиметр [Райзер, 2009, стр. 570]. В последствии [Gurevich et al., 1999] предложили модификацию своей гипотезы, в которой инициация стримера происходит непосредственно перед энергичной космической частицей с энергией большей, чем $10^{15}$-$10^{17}$ эВ, которая порождает широкий атмосферный ливень космических лучей (ШАЛ). К сожалению, даже ШАЛ в непосредственной близости от оси, не может обеспечить необходимое число свободных электронов для инициации даже одного стримера, как показали подробные численные расчеты [Dwyer and Babich, 2011], [Rutjes et. al., 2019]. Поэтому появление лавины убегающих электронов не может быть единственной причиной создания молнии (не может считаться инициацией молнии), но, как показано в главе 7 и [Kostinskiy et al., 2020a], ШАЛ все-таки имеет высокую вероятность играть ключевую роль в инициации молнии и КВР (CID) благодаря другому физическому механизму.

[Rison et al., 2016], по результатам построения по излучению источников VHF-излучения интерферометрических карт внутриоблачных разрядов, обнаружили, по их определению: «разряд короткой продолжительности» («short duration discharge»), который они назвали «прекурсорами» IC-молний, так как такие радиособытия «иногда происходят до того, как IC-молния начнется в том же месте» («sometimes occur seconds before an IC discharge initiates at the same location»). Эти внутриоблачные «разряды короткой продолжительности» никогда не приводили к плазменным каналам длиной в несколько км, то есть, к IC или CG-молниям. И мы также не можем их считать молниями несмотря на то, что они, судя по излучению источников VHF-излучения, представляют из себя значительные плазменные процессы, гораздо большие, чем инициация стримеров, но тем ни менее, не приводящие к классическим молниям.

Другой тип локального кратковременного разряда, о котором мы упоминали выше и который будет подробно рассмотрен на протяжении всей диссертации — это «классический» компактный внутриоблачных разряд — КВР (CID/NBE), изолированный во времени и пространстве [Le Vine, 1980]. Первоначально все КВР считались изолированными в пространстве разрядами (например, [Willett et al., 1989]), но позже было установлено, что 4-12% классических КВР могут инициировать молнии [Lyu et al., 2019], [Bandara et al., 2019]. Классический КВР является самым мощным в природе источником VHF-излучения с токами в десятки кА (до 150-200 кА) и средним переносом заряда около 0.5 Кл. Несмотря на это, появление таких мощных плазменных процессов, как классические КВР, также не может считаться обязательной инициацией молнии. В главе 7 будут приведены аргументы, которые говорят, пользу того, что «прекурсоры» [Rison et al., 2016] и изолированные КВР(CID), вполне вероятно, развиваются с помощью предложенного нами NBE-IE механизма, но они образуются в таких частях облака, где имеются только небольшие объемы с электрическими полями, способными поддерживать движение положительных стримеров. Это обстоятельство может не позволить развиться большим плазменным сетям (сетям UPFs) и большим двунаправленным лидерам, тем самым может не возникнуть важнейшая стадия инициации подавляющего большинства молний ([Mäkelä et al., 2008], [Marshall et al., 2014b]) стадия начальных импульсов пробоя (IBPs). Без IBPs-стадии прекурсоры [Rison et al., 2016] и изолированные КВР (CIDs/NBEs), возможно, не могут развиться в «полноценные» IC или CG-молнии.

В результате, мы видим, что появление первого стримера, лавины убегающих электронов, «прекурсоров» [Rison et al., 2016] или даже КВР (CID/NBE) не может считаться критерием инициации IC или CG-молнии. Далее мы будем считать, что процесс инициации молнии закончен, когда существует плазменный канал длиной в несколько километров, подобный изображенным на Рисунках B.2, B.4-B.7.

**Инициация восходящего с заземленных объектов лидера и инициация**
**двунаправленного лидера протяженными проводящими объектами в**
**электрическом поле грозового облака, как модель инициации молнии**

Целью данной диссертации является последовательное построение механизма инициации молнии в грозовых облаках. Если быть точным, то нужно добавить слова: «без присутствия протяженных проводящих объектов (типа самолетов и ракет) внутри грозового облака (или рядом с ним) и/или присутствия заземленных проводящих объектов длиной более ста метров под грозовым облаком (типа башен, мачт, зданий, ракет с проводом и т.д.)» То есть, без присутствия этих проводящих объектов в грозовом облаке или под ним молния не возникла бы.

Механизм инициации молнии с высотных сооружений впервые экспериментально установлен при наблюдении за восходящими молниями на Эмпайр-стейт-билдинг (Empire State Building, 410 м) в Нью-Йорке. Карл МакИкрон записал с помощью камеры Бойса временные развертки изображений и связанные с ними токи молний и впервые установил, что молнии двигались вверх, то есть были восходящими положительными лидерами [McEachron, 1939], а не нисходящими отрицательными ступенчатыми лидерами. Он подчеркнул, что впервые годом раньше предсказал восходящие лидеры (по результатам исследования длинной искры в лаборатории) Томас Аллибоне [Allibone,1938]. Осциллограммы тока показали, что разряд восходящей молнии длится сотни миллисекунд (вплоть до 1-2 с) и может завершаться обратными ударами, как и у природной молнии, а может и нет (пример осциллограммы тока на Рисунке В.8). Ему удалось измерить скорости восходящего лидера и они были в диапазоне $(0.5-6) \cdot 10^5$ м/с, что близко к измеряемым сейчас значениям. Измеренные пиковые токи были в диапазоне 27-150 кА. Внешний вид типичной восходящей положительной молнии (лидера) длиной не менее 3 км показан на Рисунке В.9. Механизм инициации таких восходящих молний был понятен практически сразу и объяснялся тем, что поляризация высоких сооружений (обычно больших 100 м на равнине) в электрическом поле грозового облака ведет к созданию на вершине башни настолько высокого потенциала (несколько мегавольт), что

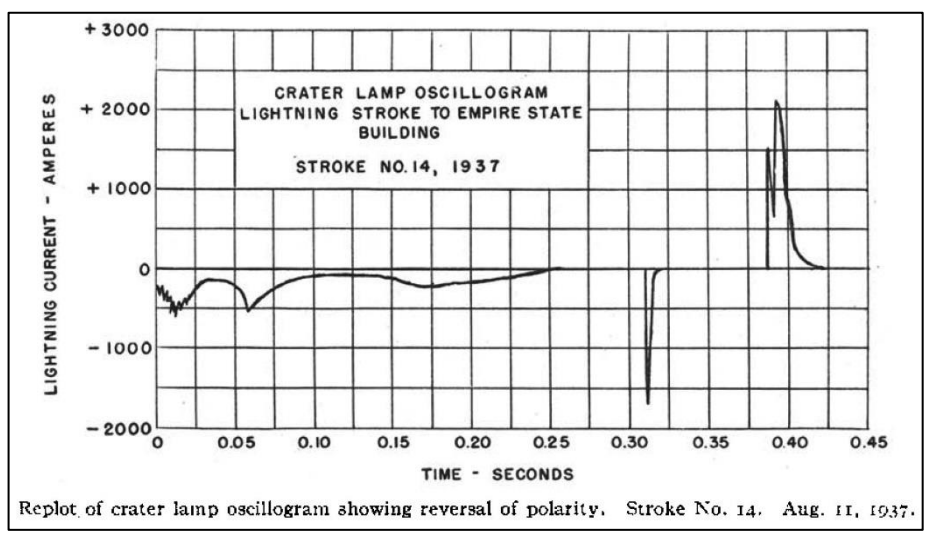

Рисунок В.8. (адаптировано из [McEachron, 1939]). Одна из первых осциллограмм прямого измерения тока восходящей молнии с помощью шунта, снятая 11 августа 1937 г. По оси абсцисс ток в амперах, по оси ординат время в секундах. Продолжительность фазы непрерывного тока превышает 250 мс (0-0.25 с на осциллограмме).

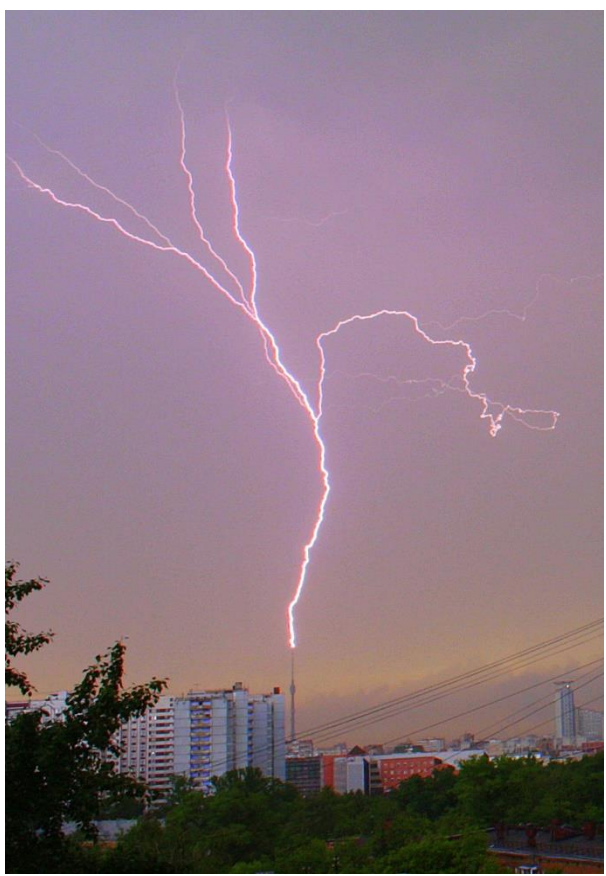

Рисунок В.9. Восходящая с Останкинской башни ветвящаяся положительная молния (положительные лидеры) с длиной ветвей не менее 3 км. Высота башни 540 м, г. Москва. То, что молния является восходящей можно определить по направлению ветвлений, которые направлены вверх. Фотография с выдержкой 3 с.

становилось возможным инициация положительного лидера, как на высоковольтных установках [Allibone,1938], [McEachron, 1939]. Интересно, что МакИкрон, также впервые установил, что положительный лидер двигался вверх ступенями длиной 6-20 м со средним интервалом между ступенями около 50 мкс (подробнее см. главу 6). На Рисунке В.10 мы видим аналогичную типичную картину ступенчатого развития двух восходящих положительных лидеров, которые поднимаются с мачт на горе Сан-Сальвадор (Monte San Salvatore) в Швейцарии [Berger and Vogelsanger, 1966]. Ток восходящего лидера с правой на фотографии мачты длился около 150 мс и не завершился обратным ударом (Рисунок В.10с).

К этому же типу восходящих молний, инициированных из-за поляризации длинного проводника (200-500 м) относятся и триггерные молнии, инициированные ракетой, несущей заземленный провод, описанные выше (Рисунки В.6-В.7, 5.6, 5.7). Это фактически «мгновенно» созданное высотное сооружение, так как для инициации лидера важна общая длина проводника вдоль направления электрического поля и гораздо меньше важна форма. Триггерные восходящие лидеры тоже движутся сотни мс, имеют близкие скорости к измеренным в наблюдениях [McEachron, 1939] и [Berger and Vogelsanger, 1966], но благодаря радиофизическим методам удается проследить траектории молний более, чем на 10 км внутри облаков (например Рисунок В.9, [Hill et al., 2012]). Внешне LMA-карты внутриоблачного движения восходящих положительных лидеров и их ветвлений напоминают привычные фотографии восходящих молний (например, Рисунок В.9).

Тонкая структура стримерной зоны восходящих с высотных сооружений положительных лидеров в настоящее время интенсивно изучается и вызывает споры (см. главу 6), но сам процесс инициации молнии вопросов не вызывает, так как фактически это инициация лидера с длинного заземленного электрода в поле грозового облака (аналогичную схему разрядного промежутка можно реализовать в лаборатории и называется она «перевернутая схема» [Стекольников, 1960, стр.56-57])

Другим, более сложным, но также в принципе качественно понятным механизмом возникновения молнии является механизм инициации молнии самолетами, ракетами (или другими незаземленными длинными проводниками) в электрическом поле грозового облака. Примером такого события может быть инициация молнии взлетающим

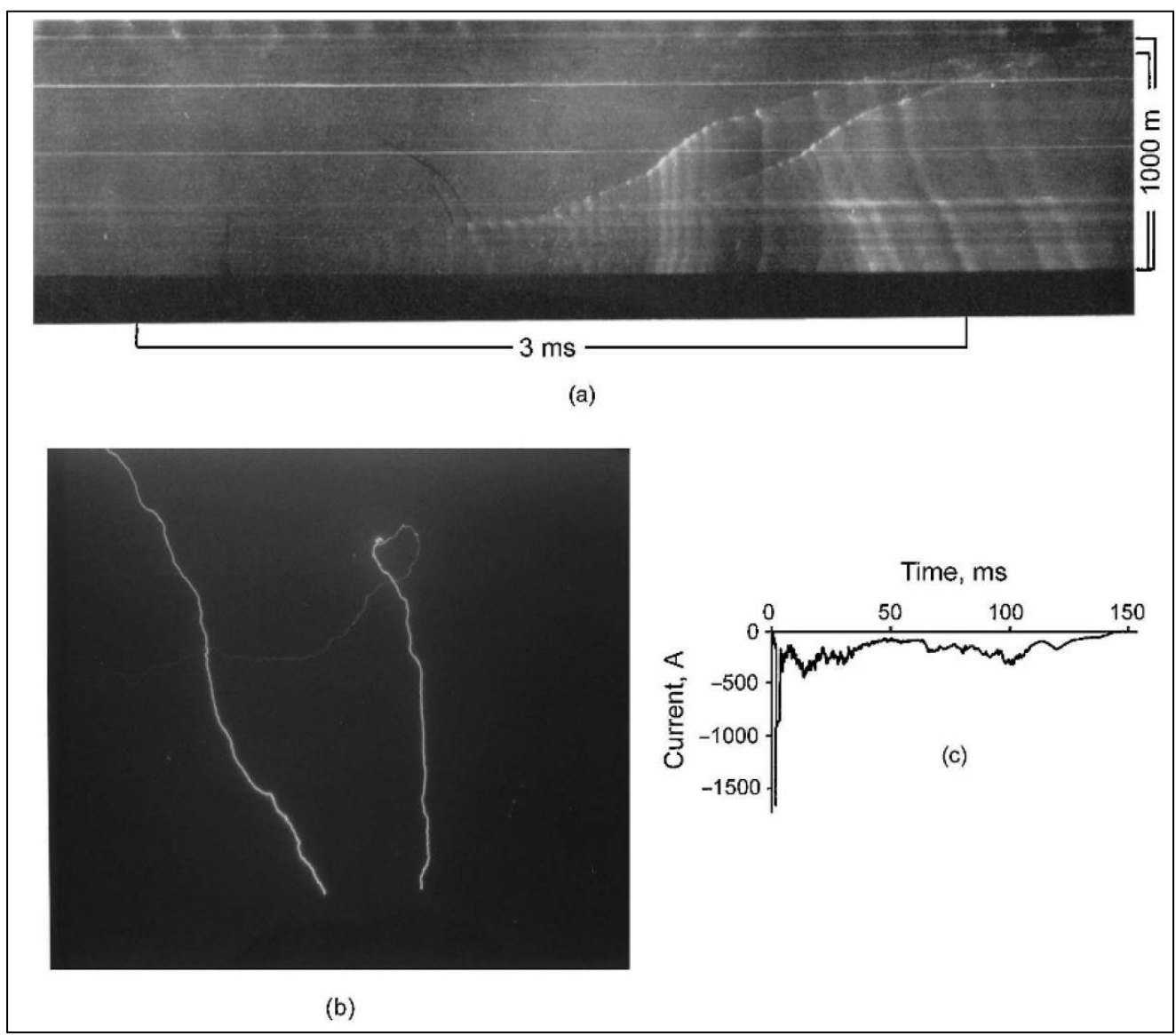

Рисунок В.10. (адаптировано из [Rakov and Uman, 2003], [Berger and Vogelsanger, 1966]) Два восходящих положительных лидера с мачт на горе Монте-Сан-Сальваторе. (а) развертка во времени (слева направо) с помощью стрик-камеры первых 4 мс ступенчатого движения обоих положительных лидеров; (b) интегральная фотография тех же вспышек; (с) осциллограмма тока восходящего лидера с мачты справа на фотографии, который не привел к обратному удару. Длительность непрерывного тока составила около 150 мс.

самолетом (Рисунок 5.1) или ракетой, несущей незаземленный провод (высотно-инициированная триггерная молния, Рисунок 5.2).

На Рисунке В.11 показана упрощенная схема инициации молнии поляризованным самолетом в электрическом поле $E_0$ грозового облака [Laroche at al., 2012] и заряженным от взаимодействия с гидрометеорами. В начале на носу самолета возникают положительные стримерные вспышки и, когда в носовой части самолета потенциал превосходит пороговый, то возникает положительный лидер и начинает распространяться (1). Так как потенциал инициации отрицательного лидера примерно в 1.5-2 раза больше, чем положительного, то отрицательный лидер сразу не возникает. При удлинении положительного лидера на его головке растет потенциал, а на самолете, благодаря поляризации увеличивается локальное поле в хвосте. Когда потенциал хвоста превосходит порог инициации отрицательного лидера, то внутри очередной отрицательной стримерной вспышки стартует отрицательный лидер (2) и начинает свое распространение. Из закона сохранения заряда следует, что суммарный заряд положительного лидера, отрицательного лидера и самолета нейтрален. Таким образом возникает двунаправленная система лидеров (положительный и отрицательный). Этот порядок событий в точности повторяет порядок событий на Рисунке 5.5, который описывает возникновение двунаправленного лидера, инициированного высотно инициированной триггерной молнией. Схема изменений электрического поля и тока во время возникновения и развития молнии, инициированной самолетом, показана на Рисунке В.12. На Рисунке видна довольно сложная структура тока, который состоит из постоянной части с наложенными на нее импульсами. Далее следует длинная по времени часть тока, которая скорее всего соответствует стадии распространения лидеров и называется она в стандарте испытаний самолетов «стадией постоянного тока» (continuing current) как и у молнии, восходящей с высотных сооружений. Длится эта стадия 0.25-1 с, что зафиксировано в стандарте испытаний самолетов.

Несмотря на гораздо более сложный и требующий дальнейшего подробного изучения механизм инициирования молнии летательными аппаратами и высотно инициированными триггерными молниями, принципиально (качественно) он понятен, так как старт лидеров происходит с противоположных «электродов» проводящего объекта благодаря высокому потенциалу, созданному из-за поляризации длинного

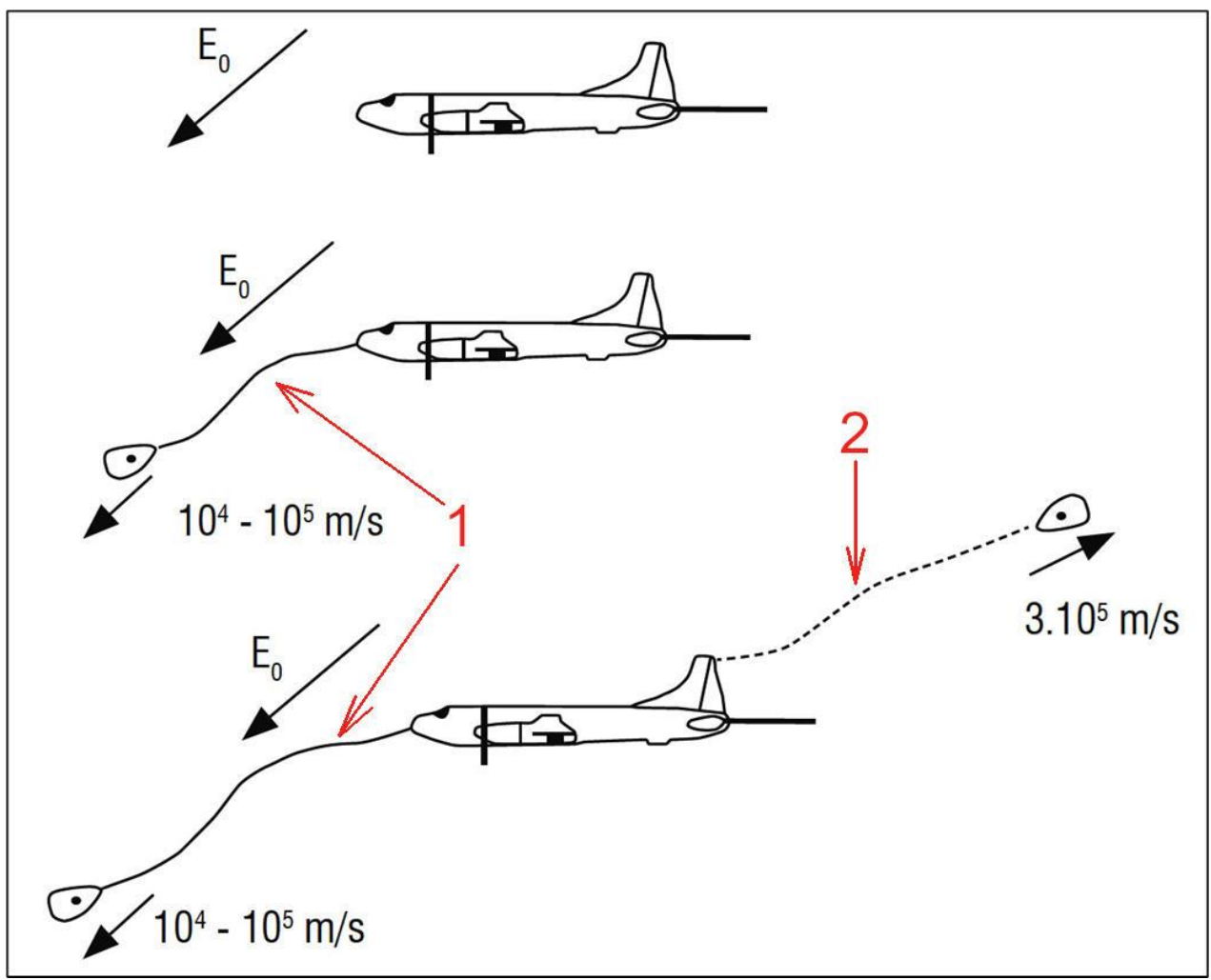

Рисунок В.11. Схема инициации молнии самолетом в электрическом поле грозового облака (адаптировано из [Laroche at al., 2012]). $E_0$ — электрическое поле грозового облака. В начале возникают положительные стримерные вспышки и, когда в носовой части самолета потенциал превосходит пороговый, то возникает положительный лидер (1) и начинает распространяться. Так как потенциал инициации отрицательного лидера примерно в 2 раза больше, чем положительного, то отрицательный лидер сразу не возникает. При удлинении положительного лидера на его головке растет потенциал, а на самолете, благодаря поляризации увеличивается локальное поле в хвосте. Когда потенциал превосходит порог инициации отрицательного лидера, то внутри очередной отрицательной стримерной вспышки с хвоста самолета стартует отрицательный лидер (2) и начинает свое распространение. Суммарный заряд положительного лидера, отрицательного лидера и самолета нейтрален. Цифры около лидеров означают их средние скорости, а конусообразные фигуры на кончиках лидеров обозначают стримерные зоны.

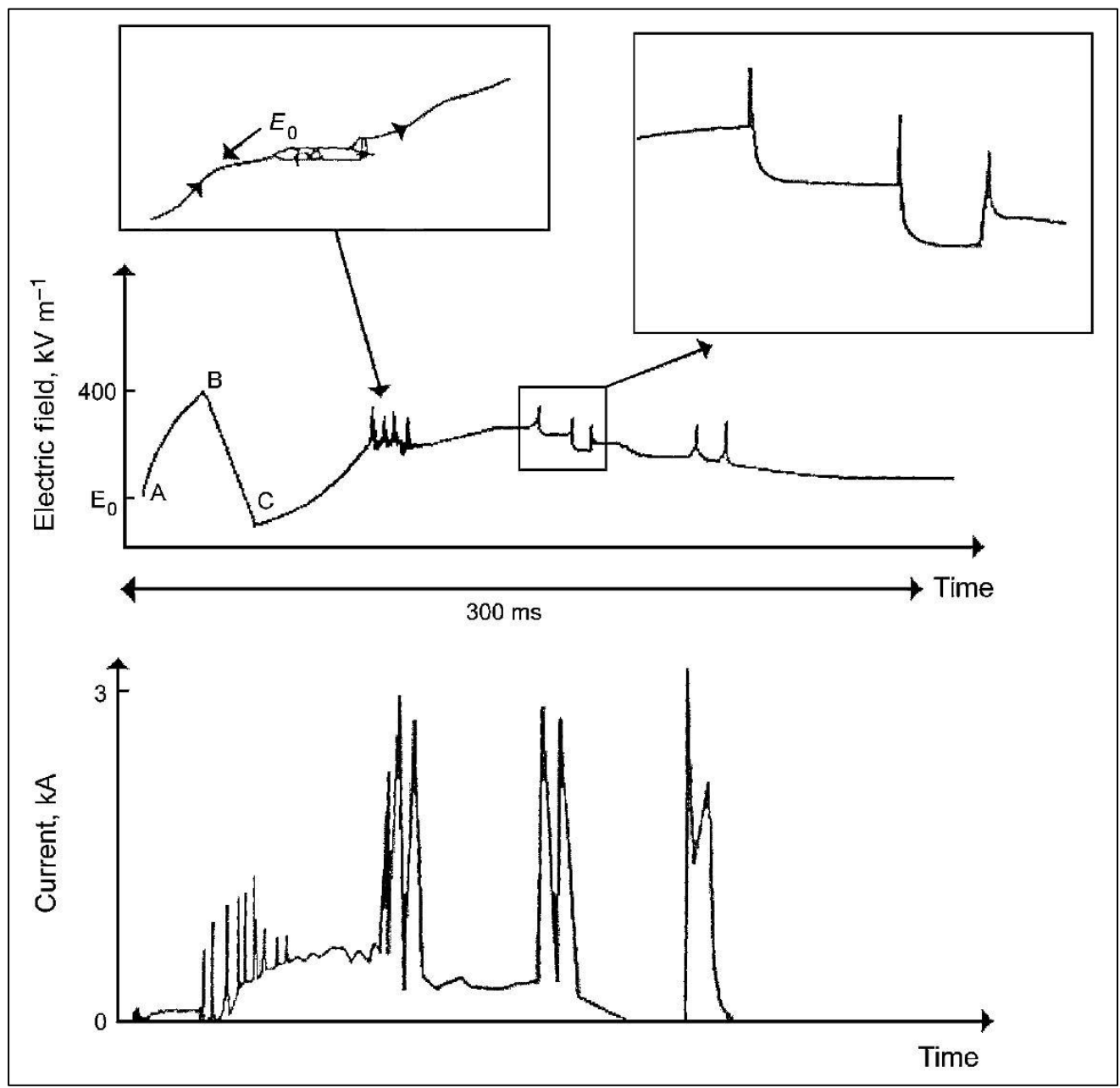

Рисунок В.12. (адаптировано из [Rakov and Uman, 2003, p.358]) Напряженность электрического поля и осциллограммы тока в течение всей продолжительности типичной вспышки молнии, инициированной самолетом. $E_0$ —электрическое поле грозового облака при возникновении молнии. Длительность всей вспышки превышает 300 мс.

проводника во внешнем электрическом поле (аналогично возникновению лидеров с металлического стержня, помещенного на непроводящих нитях внутрь разрядного высоковольтного промежутка, например, [Базелян и Райзер, 2001, стр.114, рис.3.16]).

Таким образом, инициация молнии в электрических полях облаков длинным металлическим объектом (включая внутриоблачные полеты объектов) описывается и подтверждает хорошо известную гипотезу о двунаправленном (bidirectional) лидере Хейнца Каземира [Kasemir, 1960], которая исходила из закона сохранения заряда и представляла молнию, как центральный плазменный канал, по обе стороны которого шли ветвящиеся положительный и отрицательный лидеры (Рисунок 5.3). В частности, и наши результаты по инициации лидеров болтом арбалета в электрическом поле аэрозольного облака (глава 5) и изучение двунаправленных лидеров, инициированных внутри аэрозольного облака (главы 3, 4), также подтверждают гипотезу Каземира.

С середины 1960-х годов и до последнего времени представление о молнии, как о гигантском двунаправленном лидере стало общепризнанной и даже единственной моделью молнии на всех стадиях ее развития. Рисунок В.13 показывает, что молнии, как двунаправленные лидеры схематически изображались совершенно одинаково во всех видах конвективных облаков высотой от 4 до 20 км: небольшом субтропическое «теплом», «типичном» и гигантском грозовом облаке. В то время, когда создавалась эта схема, еще не были развиты методы радиолокации каналов молнии типа LMA или интерферометрических карт. Когда эти методы развились, то они также принципиально не изменили подход к молнии, как двунаправленному лидеру на всех стадиях развития. На типичной LMA-карте внутриоблачной молнии (например, Рисунок В.5) мы видим пусть и сложные, но отрицательный ступенчатый лидер (красным), положительный лидер (синим) и канал между ними (зеленым). Картирование молнии с помощью VHF-интерферометров также позволяло выделить, пусть сложные и запутанные, но в принципе двунаправленные каналы (например, Рисунок В.14 [Stock, 2014, p.109], Рисунок В.39b [Shao et al., 2018]).

Таким образом, сложившееся к 2013-2014 году представление об инициации молнии качественно можно было сформулировать так: внутри облака существует область, где, благодаря заряженным и/или поляризованным гидрометеорам, возникает стример (стримерная вспышка или вспышки) и там через короткое время «рождается»

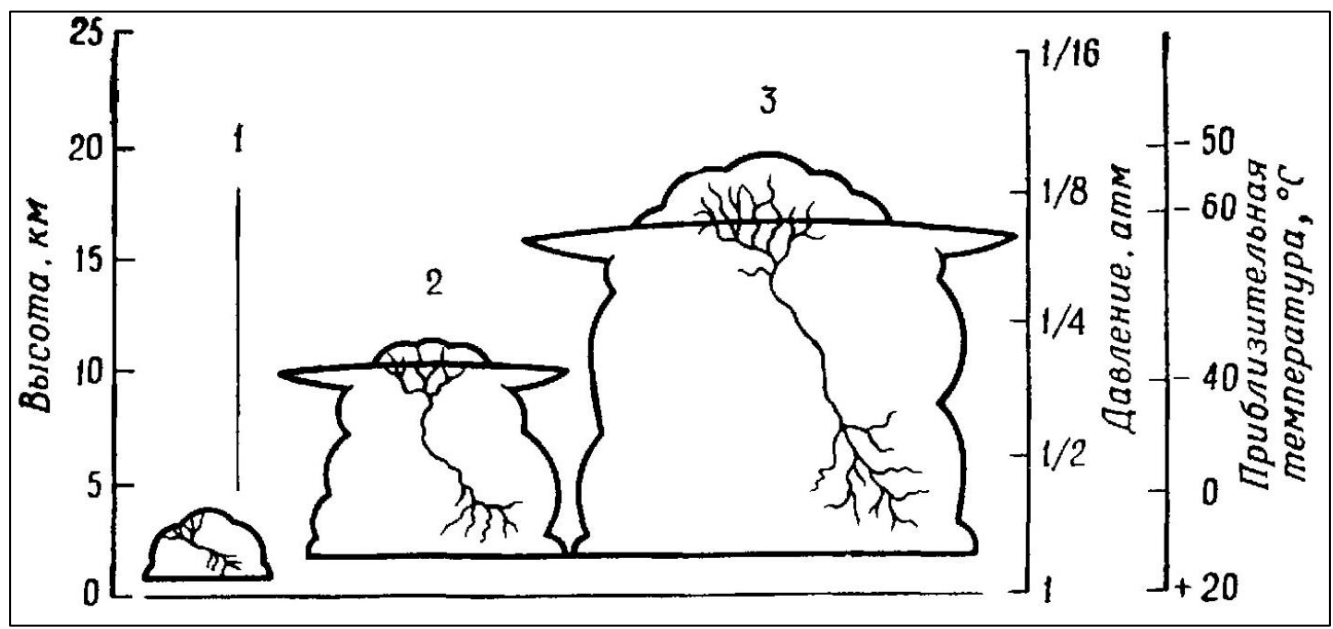

Рисунок В.13 (адаптировано из [Юман, 1972, стр.12]. Молнии, как двунаправленные лидеры в конвективных облаках различных размеров. 1 — субтропическое «теплое» облако; 2 — «типичное» грозовое облако; 3 — «гигантское» грозовое облако. Характерно, что во всех типах облаков молния изображается качественно одинаково, отличаясь только по размеру. В то время, когда рисовалась эта схема, еще не были развиты методы радиолокации каналов молнии (LMA, интерференционное картирование молнии) как, например, на Рисунках В.5, В.29-В.31, В.38-В.43.

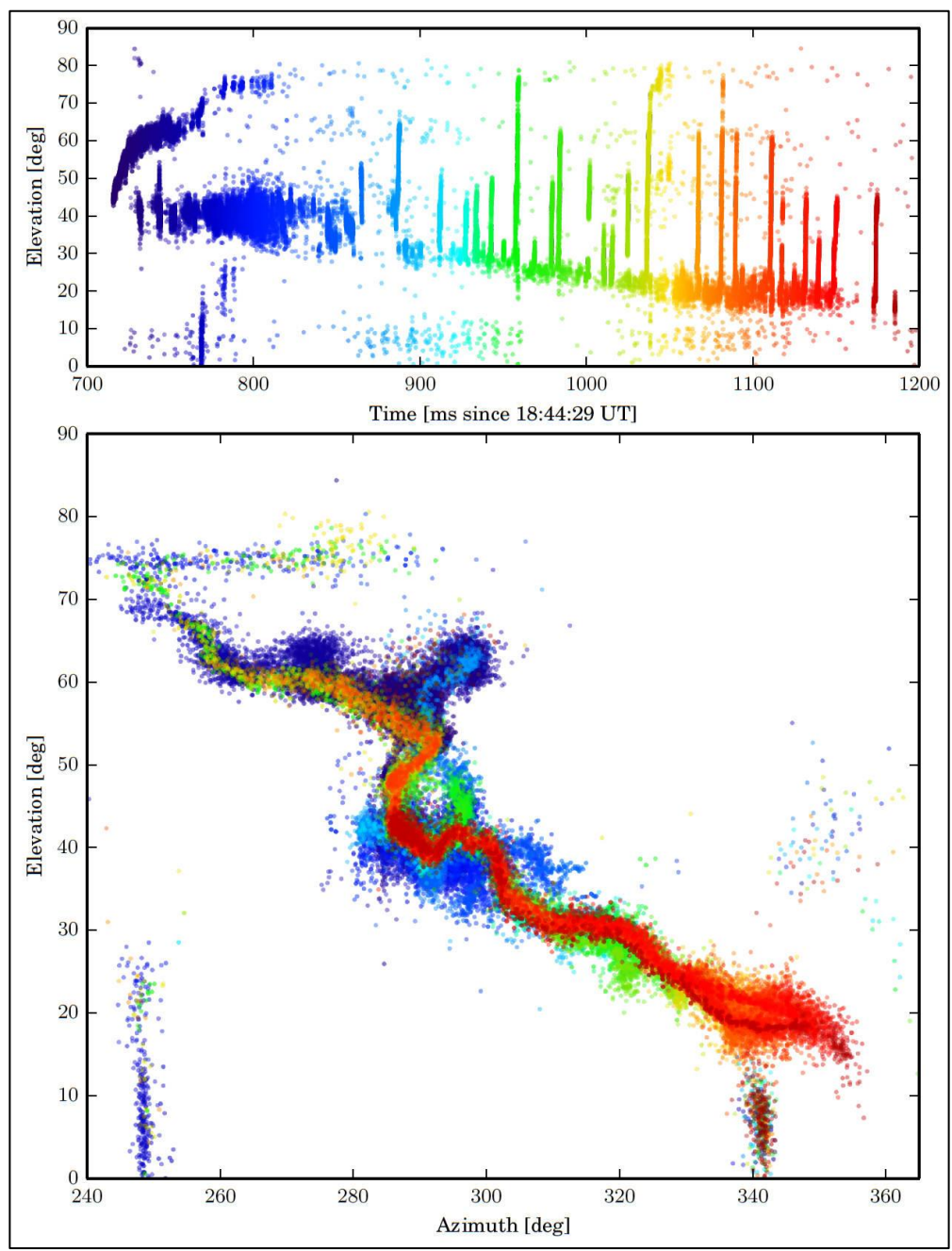

Рисунок В.14 (адаптировано из [Stock, 2014, p.109]). Интерферометрическая карта развития молнии (каждая точка строится на пересечении угловых координат, как на звездном небе, по запаздыванию VHF-сигнала). Временное окно (временное разрешение), в котором выбирается наиболее сильный VHF-сигнал, около 1 мкс, что делает интерферометрические карты гораздо более подробными и позволяет не «пропускать» многие важные VHF-события, а также гораздо лучше следить за быстрыми процессами («реальными» и «кажущимися» движениями VHF-сточников). Время на карте направлено от темно синих цветов к красным и маркировку времени можно контролировать в соответствии с цветом на верхней панели (возвышение-время, Elevation-Time). Вертикальные линии означаю внутриоблачные пробои, когда волна потенциала движется вдоль канала.

небольшой двунаправленный лидер. Его положительный и отрицательный каналы, расширяясь (двигаясь) в течении нескольких сотен миллисекунд аккумулирует в своих чехлах большой заряд (от нескольких кулонов до десятков кулонов). Когда один из лидеров вступает в контакт с наземным сооружением, (благодаря восходящему с земли встречному лидеру), то возникает обратный удар. Если же один из каналов двунаправленного лидера встречает в облаке канал другого двунаправленного лидера, то возникает внутриоблачный разряд (К-процесс, Рисунок В.39b [Shao et al., 2018]). Эта картина событий многократно подтверждалась экспериментально и, казалось бы, исключала другие варианты развития событий. Но оставалось непонятно как все-таки в свободном пространстве, а не на металлическом электроде, стример (стримеры) превращаются в горячий плазменный канал (проблема стримерно-лидерного перехода в облаке). До последнего времени экспериментальные данные и теория стримерно-лидерного перехода касались только превращения стримерной вспышки в лидер на электроде (в стеме стримерной вспышки), а также в головке положительного лидера ([Raizer, 1991], [Bazelyan and Raizer, 1998, 2000], [Bazelyan et al., 2007], [Popov, 2009]). Но в этих случаях на электроде или головке лидера уже существовал большой потенциал (созданный генератором импульсных напряжений), поддерживающий процесс распространения лидера после прохождения ионизационно-перегревной неустойчивости в каждой следующей ступеньке в головке положительного лидера. Далее, было непонятно, что будет с возникшим в грозовом облаке небольшим горячим плазменным образованием и каковы условия его превращения в длинный устойчиво распространяющийся двунаправленный лидер. Дело в том, что в физике длинной искры, — а возникший в облаке начальный лидер вряд ли сразу же будет иметь длину в 30-40 метров как самолет или ракета, — хорошо известно понятие незавершенного разряда (withstand) [Les Renardières Group, 1977, 1981], когда стартовавший с электрода лидер не достигает противоположного электрода. Выше мы уже отмечали, что во время старта триггерной ракеты с заземленным проводом, может произойти несколько десятков попыток инициации устойчиво распространяющегося лидера прежде, чем возникнет положительный лидер, который превратиться в восходящую положительную молнию движущуюся сотни миллисекунд (например, [Lalande et al., 1998]; [Willett et al., 1999]; [Rakov and Uman, 2003 p.274]). То есть, появление небольшого горячего плазменного канала (не то, что стримера) не гарантирует обязательное превращение его в длинный двунаправленный лидер. Таким образом, даже на качественном уровне отсутствовали несколько звеньев в цепочке преобразований плазмы от стримерной вспышки до устойчиво развивающегося двунаправленного лидера (если даже предположить, что процесс инициации молнии можно считать законченным, если возникнет двунаправленный лидер длиной в десятки метров). Возможная подобная цепочка преобразований плазмы рассматривается в разделе 7.6. Кроме того, с начала 1960-х годов существовала проблема раннего появления и чрезвычайно большой интенсивности начальных импульсов пробоя (an initial breakdown pulse — IBPs), что подробно будет рассматриваться в следующем разделе.

Наряду с проблемой IBPs, в последние 7-8 лет появилось все больше экспериментальных данных, которые не укладываются в простую картину инициации молнии, как процесса возникновения небольшого плазменного канала, который разовьется в двунаправленный лидер, и, они настойчиво требует разработки новых подходов к этой проблеме.

**Стадия начального пробоя (initial breakdown — B/IB) и начальные импульсы пробоя (initial breakdown pulse — IBPs)**

Уже после установления в ЮАР в 1930-х годах механизма пробоя многокилометрового «промежутка» облако-земля с помощью отрицательного ступенчатого лидера, сэр Базиль Шонланд с коллегами стали предполагать, что должен существовать какой-то мощный предварительный процесс, создающий отрицательный лидер. [Rakov and Uman, 2003 p.116] описывают эти представления так: «Начальный пробой, часто называемый предварительным пробоем, молнии облака-земля — это внутриоблачный процесс, который инициирует или приводит к инициированию нисходящего ступенчатого лидера». Эти предположения Шонланда и его коллег опирались на следующие эксперименты, проведенные с помощью камеры Бойса. За сотни миллисекунд до появления из облака отрицательного лидера наблюдались внутриоблачные вспышки света. Эти вспышки сопровождались относительно длительными изменениями электрического поля, продолжавшимися более 100 мс, до первого обратного удара. Также они предполагали, что ступенчатый лидер вряд ли живет более нескольких десятков миллисекунд, исходя из измеренной ими скорости движения ступенчатого лидера и представлений о высоте основных электрических зарядов в грозовом облаке, откуда этот отрицательный лидер мог стартовать [Schonland, 1956].

Кларенс и Малан [Clarence and Malan, 1957] предположили, на основе измерений электрического поля с помощью одной антенны, что начальный пробой (обозначенный ими буквой $B$ на Рисунке В.15) представляет собой разряд между центром основного отрицательного заряда и центром небольшого нижнего положительного заряда. Длится $B$(IB)-стадия от 2 до 10 мс. [Clarence and Malan, 1957] считали, что после стадии начального пробоя последует ступенчатый лидер (обозначенный $L$ на Рисунке В.15) либо сразу, либо после так называемого промежуточной стадии (обозначенной $I$), которая может длиться до 400 мс. Промежуточная стадия была интерпретирована ими, как результат зарядки отрицательным зарядом вертикального канала «начальным пробоем» $B$(IB) до тех пор, пока поле в нижней части канала не станет достаточно высоким, чтобы запустить отрицательный ступенчатый лидер, который при контакте с заземленными объектами инициирует обратный удар, обозначенный $R$ на Рисунке В.15. Скорее всего [Clarence and Malan, 1957] не считали канал молнии двунаправленным лидером, но идея, что IB-стадия является подготовительной стадией большого отрицательного лидера, который изображен, например, на Рисунке В.5, оказалась плодотворной. Однако, что это был за процесс им было непонятно даже на качественном уровне.

Прошло около 30 лет, были созданы первые модификации VHF-интерферометров и LMA-систем и возобновились попытки разобраться в начальной стадии пробоя (IB/B-стадии) с использованием нового оборудования. [Proctor et al., 1988] с помощью одной из первых успешных LMA-систем (он, в частности, измерил скорости движения лидеров внутри грозового облака) попытался все объяснить разными режимами движения отрицательного лидера в присутствии различных объемных зарядов облака. Но [Rustan et al., 1980], по результатам LMA-системы, работающей около Космического центра им. Кеннеди, сделали другой вывод: IB/B-стадия значительно отличается от стадии ступенчатого лидера и отрицательным ступенчатым лидером в смысле Шонланда не является. Они сделали этот вывод на основании того, что измеренные ими импульсы начального пробоя характеризуются гораздо более сильными и длительными сигналами

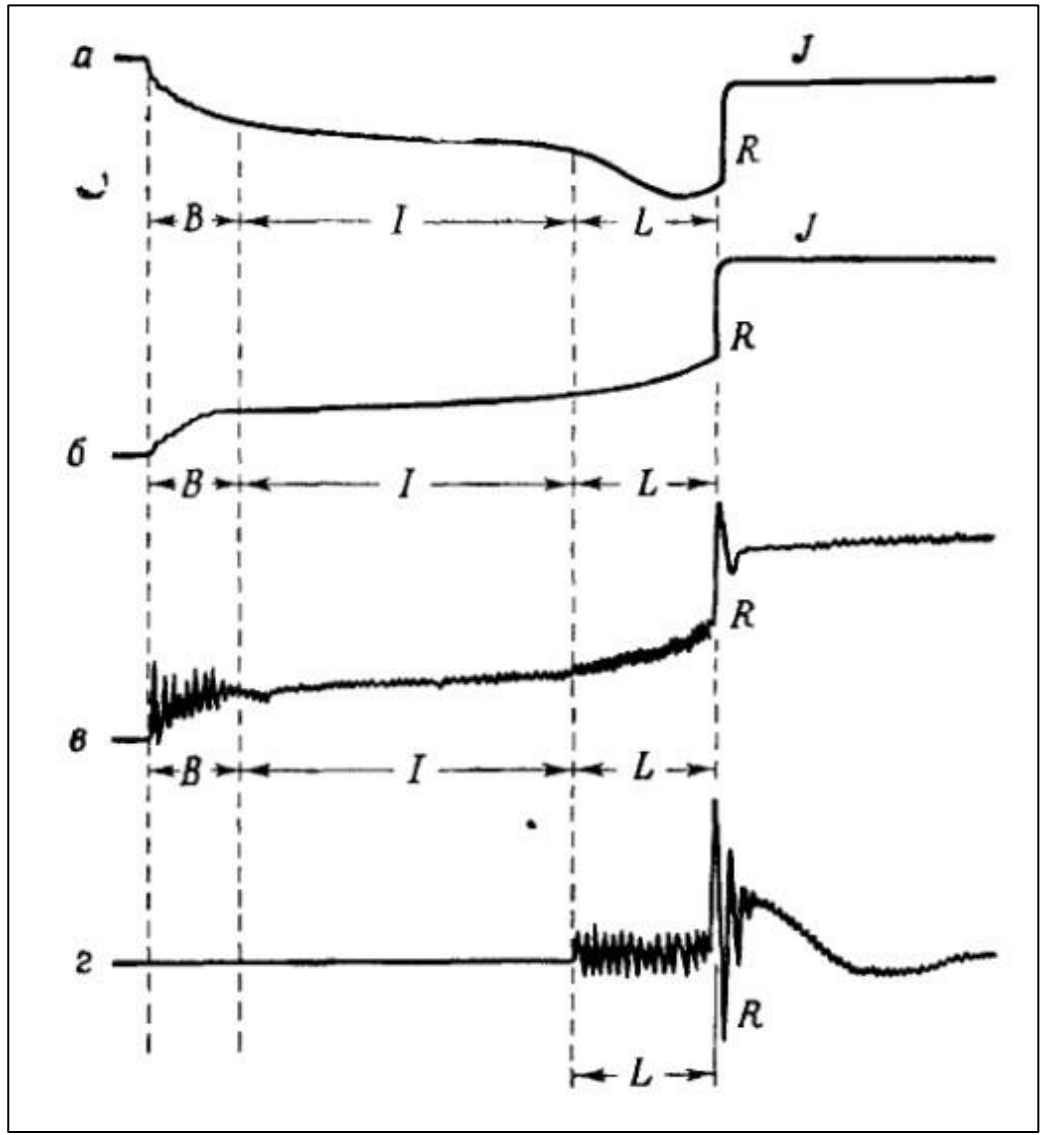

Рисунок B.15 (адаптировано из [Юман, 1972, стр.103] по публикациям [Schonland, 1956], [Clarence and Malan, 1957]). Схема типичных изменений электрического поля молний облако-земля (CG). а — электростатическое поле на расстоянии 2 км, б — электростатическое поле на расстоянии 5 км, в — электростатическое, индукционное и радиационные поля (амплитуда обратного удара $R$ уменьшена), г — радиационные поля на расстоянии 500 км в интервале частот от 200 Гц до 20 кГц. Продолжительность начального пробоя ($B$ или IB) от 2 до 10 мс; продолжительность стадии I от 0 до 400 мс; стадии распространения отрицательного лидера $L$ от 4 до 30 мс.

в VHF-диапазоне (30–50 МГц), чем ступенчатый лидер и при импульсах начального пробоя не происходит значительного изменения электрического поля. [Rhodes and Krehbiel, 1989], используя VHF-интерферометр (274 МГц), также пришли к выводу, что VHF-события (интерпретируемые ими, как каналы) стадии начального пробоя и ступенчатый лидер резко отличаются, причем лидер более рассредоточен (dispersed) и характеризуется «менее четко определенным движением» («less well-defined motion»).

Из стадии начального пробоя (В/IB) было выделено такое важнейшее, как оказалось, явление, — начальные импульсы пробоя (initial breakdown pulse — IBPs). Они непрерывно исследуются, начиная с первых работ [Kitagawa, 1957], [Clarence and Malan, 1957] до сих пор. IBPs являются сериями биполярных импульсов электрического поля относительно большого микросекундного масштаба, как показано на Рисунке В.16. Продолжительность серии, по данным ранних экспериментов, составляла в среднем около 1 мс. Из-за близости по времени IBPs к началу фиксирования антеннами движения ступенчатого лидера их стали считать причиной ступенчатого лидера. Импульсы обычно биполярны. Что крайне важно, и с самого начала вызывало недоумение, так это огромная амплитуда импульсов начального пробоя (IBPs), которая может быть сопоставима или больше, чем амплитудой обратного удара, как хорошо видно на Рисунке В.16а. Это при том, что канал молнии, оканчивающийся отрицательным лидером, имеет длину минимум в несколько километров и понятно, как он может собрать в своем чехле необходимый заряд для производства мощного обратного удара. Но «импульсы начального пробоя» по самому своему названию находятся по времени очень близко к моменту инициации молнии (как уже тогда было понятно, вероятно не более, чем 10 мс). Как может быть собран такой большой заряд за такое короткое время? Когда были проведены более точные измерения, то ситуация оказалась еще более драматичной.

Более поздние измерения (например, [Nag and Rakov, 2008]) с улучшенной аппаратурой подтвердили первоначальные представления о том, что IBPs принципиально отличаются от ступенчатого лидера, хотя также являются импульсным процессом. На Рисунке В.17 видно, что большие IBPs сравнимы с первым обратным ударом и превосходят по амплитуде все остальные обратные удары [Nag and Rakov, 2008]. На этой записи электрического поля нельзя не обратить внимание насколько импульсы

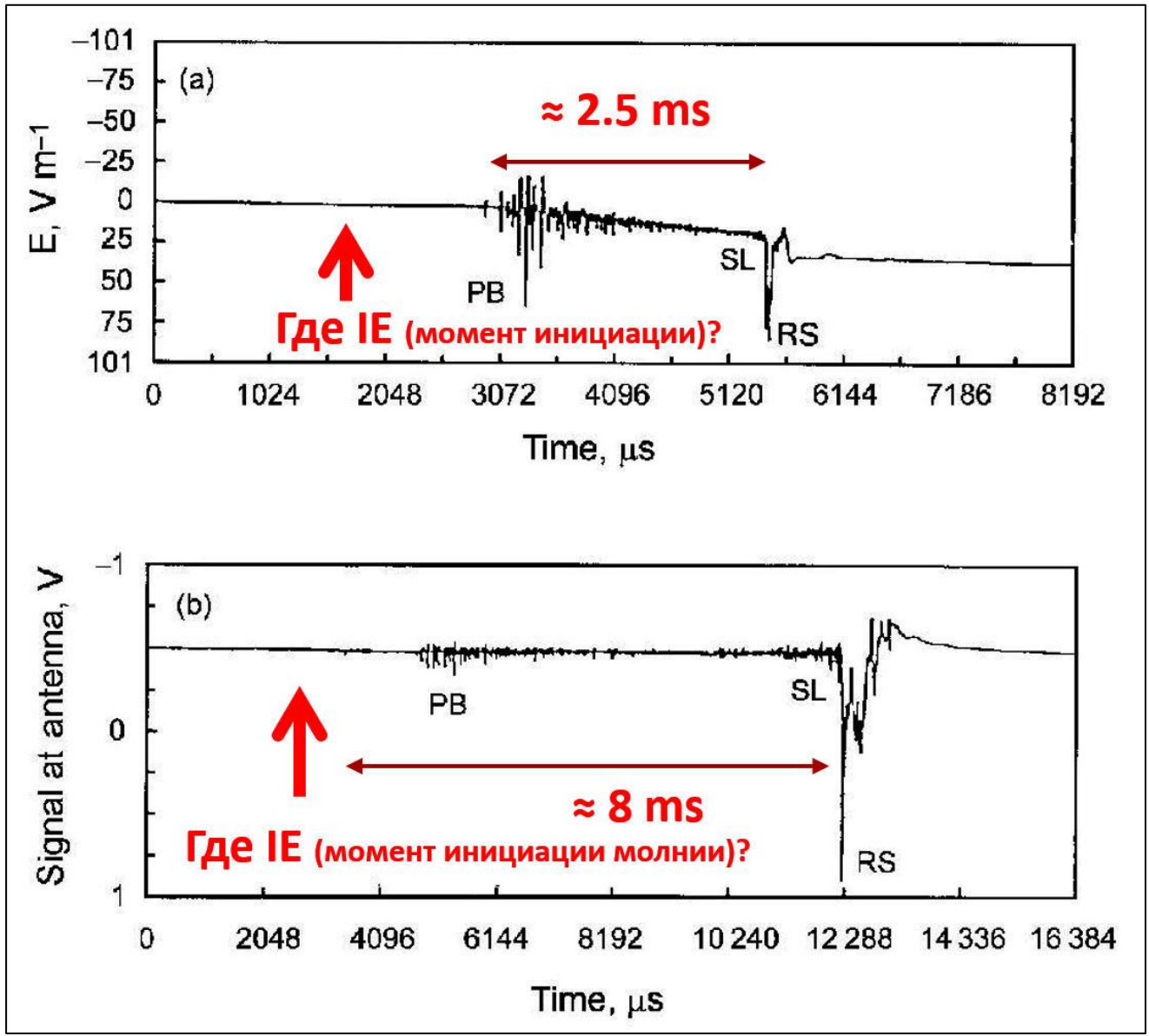

Рисунок В.16 (адаптировано из [Rakov and Uman, 2003 p.119] по данным [Brook, 1992)]). Изменение электрических полей отрицательной молнии «облако-земля»: (а) зимняя молния на расстоянии около 25 км от антенны, (б) летняя молния на неизвестном расстоянии. PB означает стадию импульсов начального пробоя (IBPs), SL — ступенчатый лидер (непосредственно перед разрядом на землю), RS — обратный удар. Общие временные масштабы на панелях *a, b* составляют примерно 8 и 16 мс соответственно. На панели *a* обращает на себя внимание, что IBPs имеют амплитуду близкую к амплитуде обратного удара, когда сформирован плазменный канал длиной в несколько км. На панели *b* IBPs на порядок меньше, чем импульсы обратного удара, но все равно значительно превосходят импульсы ступенчатого лидера.

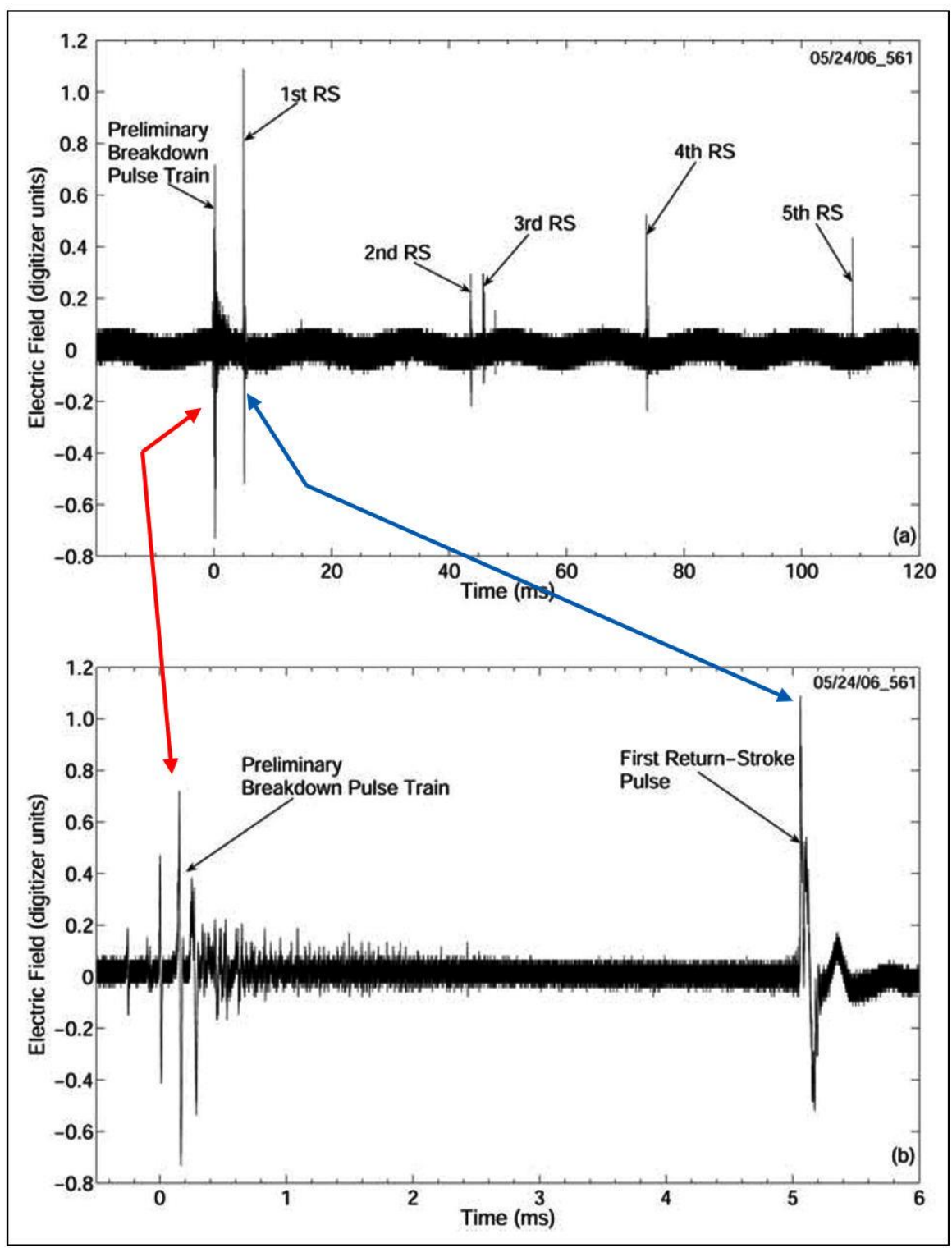

Рисунок В.17 (адаптировано из [Nag and Rakov, 2008]) Изменение электрических полей отрицательной молнии «облако-земля»: (а) запись 120 мс протекания всей молнии, включая пять обратных ударов; (б) запись первых 6 мс (чрезвычайно короткое время развития молнии) от стадии предварительного пробоя (IB-стадии с IBPs) до первого обратного удара. Красные стрелки указывают на серию импульсов начального пробоя (IBPs), RS — обратный удар. Синяя стрелка указывает на первый обратный удар.

отрицательного лидера (лишь иногда немного превышают уровень фона) меньше, чем наибольшие IBPs.

Важнейший шаг в исследовании IBPs был сделан с помощью высокоскоростных видеокамер, которые зафиксировали, что при каждом классическом IBPs фиксируются очень яркие вспышки в видимом диапазоне [Stolzenburg et al., 2013, 2014], [Campos & Saba, 2013]. [Stolzenburg et al., 2013] зафиксировали и подробно описали серии световых вспышек, синхронизованные с несколькими сериями изменения электрического поля, соответствующего IBPs в CG-молниях (Рисунок В.18). Эти вспышки света указывают на быстрое появление (не менее, чем за 20 мкс, что является временем выдержки кадров скоростной камеры) горячих каналов с высокой проводимостью, которые в большинстве случаев исчезают через 40–100 мкс (Рисунок В.18). В этом же событии через 2-5 мс IBPs переходят в отрицательный ступенчатый лидер (порождают ступенчатый лидер). Важно, что переход IBPs в ступенчатый отрицательный лидер записан на той же самой камере с теми же самыми выдержками, что позволяет сделать прямое сравнение светимостей IBPs и ступенчатого лидера, Рисунок В.19 [Stolzenburg et al., 2013]. Эти эксперименты однозначно показывают, что IBPs производятся горячей, сильно проводящей, сильно светящейся плазмой, накопившей большой заряд (наверное, в чехлах своих каналов), суммарно более емких, чем хорошо развитый (длина более 4 км) отрицательный лидер. То есть природа IBPs – это горячая сильно проводящая плазменная система, обладающая емкостью гораздо большей, чем хорошо изученные ступенчатые лидеры молнии. Форма импульса единичного IBP (например, Рисунок В.20, синяя кривая [Karunarathne et al., 2014]) показывает чрезвычайно сильную изрезанность, что может говорить в пользу того, что конкретные пики являются свидетельством контакта плазменных структур (лидеров или сетей), а не излучения одного плазменного канала (линейного осциллятора, зеленая линия на Рисунке В.20).

По последним экспериментальным данным классические IBPs имеют нормированные по дальности (на 100 км) амплитуды поля в среднем около 1 В/м [Smith et al., 2018]; расчетные токи в диапазоне 1–165 кА (пиковые 30-165 кА); переносимый во время каждого импульса большой заряд находится в диапазоне 0.12-1.7 Кл (средний 0.44 ± 0.40 Кл) [Betz et al., 2008], [Karunarathne et al., 2014], [N. Karunarathne et al., 2020]. Но, что может быть самое главное, не менее 90% всех молний имеют стадию начальных

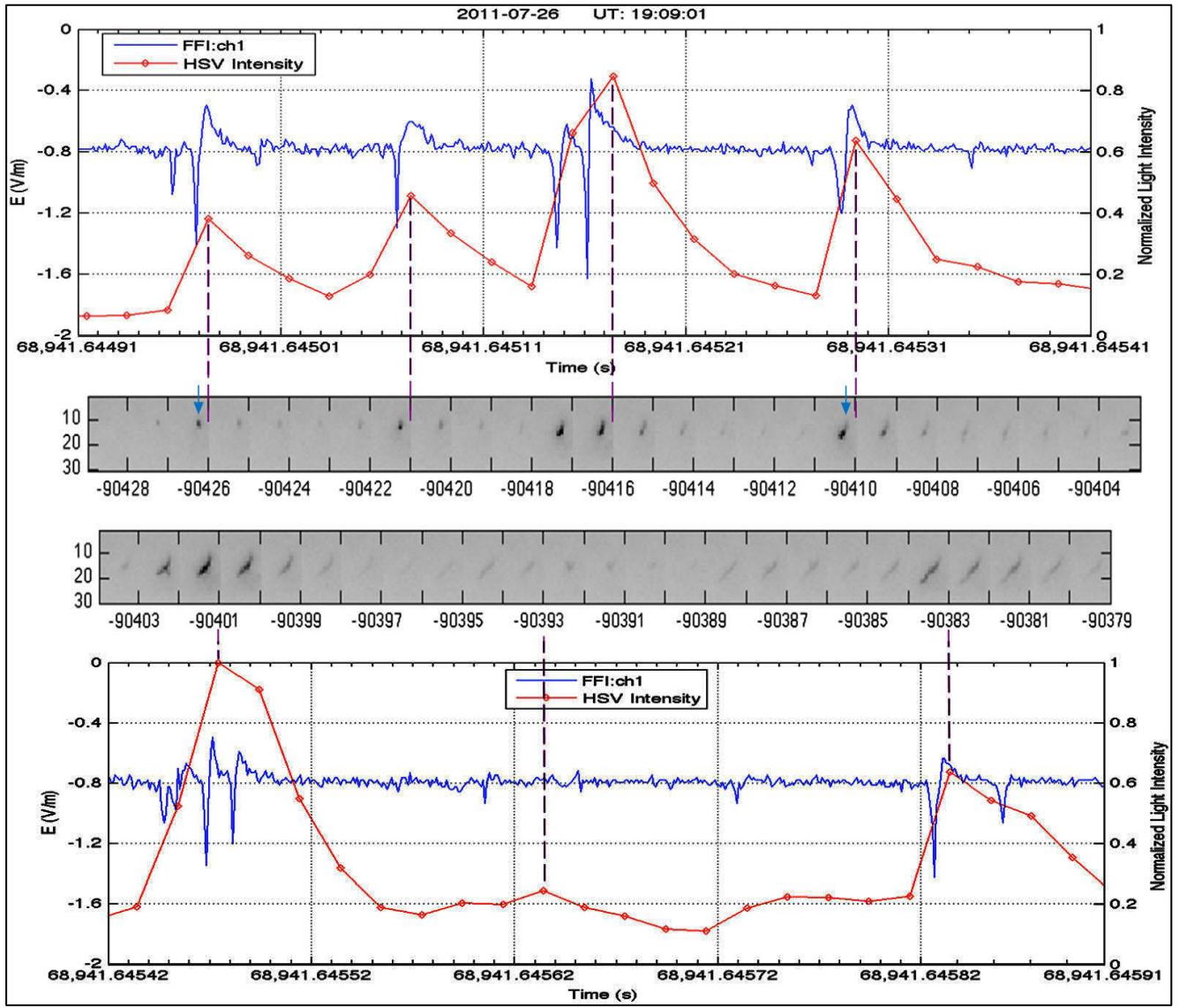

Рисунок В.18 (адаптировано из [Stolzenburg et al., 2013]). Два фрагмента данных, фиксирующих изменение электрического поля IBPs (синие кривые) и соответствующая им (синхронизованная по времени, см. пунктиры) нормализованная в диапазоне (0-1) интенсивность видеосигнала (красные кривые). Выдержка кадров равна 20 мкс, номера кадров идут порядку и пронумерованы автоматически. Изображение инвертировано. Каждый полный кадр обрезается до 20 пикселей в ширину и 31 пикселей в высоту и расположен рядом со следующим кадром. Площадь, показанная на каждом кадре, составляет примерно 482 м в ширину и 747 м в высоту. Первая серия IBPs происходит на высоте 4,6 км. Вертикальные пунктирные линии между панелями указывают несколько соответствующих времен окончания кадра и маркеров на кривой интенсивности свечения.

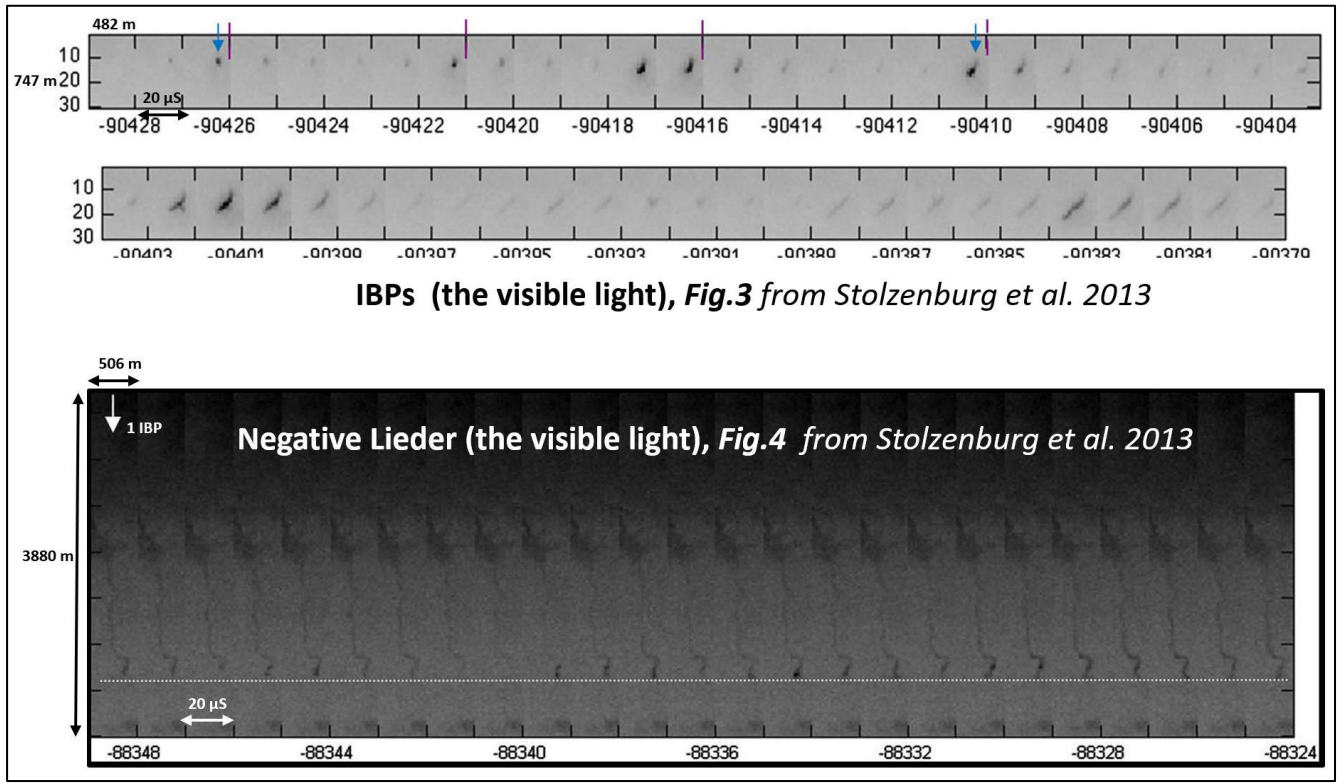

Рисунок В.19 (адаптировано из [Stolzenburg et al., 2013]). Сравнение яркости свечения IBPs (верхняя панель, заимствованная из Рисунка В.18) и ступенчатого отрицательного лидера (нижняя панель). Отчетливо видно, что свечение IBPs, на 1-2 порядка превосходит свечение отрицательного ступенчатого лидера и отрицательный лидер на картинках значительно тоньше.

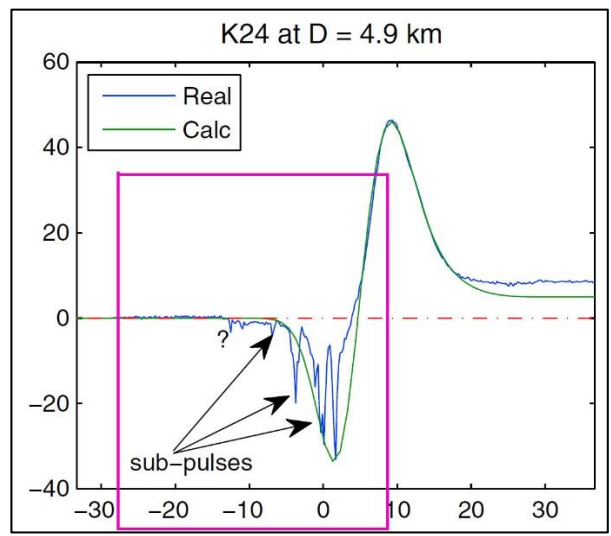

Рисунок В.20. (адаптировано из [Karunarathne et al., 2014]). Форма импульса единичного IBP (синяя кривая) показывает чрезвычайно сильную изрезанность, что может означать, что конкретные пики являются свидетельством контакта плазменных структур (лидеров или сетей). Зеленая кривая – расчет в предположении существования единичного осциллятора, а не контакта каналов (плазменных сетей). По оси абсцисс интенсивность изменения электрического поля в относительных единицах, по оси ординат время (в мкс).

импульсов пробоя (IBPs), которая предшествует появлению отрицательного лидера ([Mäkelä et al., 2008], [Marshall et al., 2014b]), причем для остальных 10% также может присутствовать эта стадия, но экспериментальные данные не были надежными в этих случаях [Mäkelä et al., 2008]. То есть, IBPs являются в подавляющем числе случаев предшественниками и скорее всего причиной появления большого отрицательного лидера, который приведет к CG- и IC-молниям.

Недавно появились свидетельства важной (возможно ключевой) роли IBPs в таком мало изученном явлении, как генерация вспышек рентгена и гамма во время разряда молнии (TGF) [Belz et al., 2020]. Оказалось, что с высокой вероятностью именно начальные импульсы пробоя (IBPs) генерируют TGF. Таким образом еще раз была подтверждена важнейшая потребность дальнейшего изучения IBPs для понимания основных проблем физики молнии.

Качественный механизм инициации IBPs предложен в рамках общего механизма инициации молнии и подробно рассматривается в главе 7.

**Инициирующее молнию событие (IE) и начальное изменение электрического поля (an initial electric-field change — IEC) до появления первого IBP**

Значительные улучшения в измерительном оборудовании, расширение сетей антенн привели к тому, что положение в пространстве и время инициации молнии удалось измерить с точностью до десятков метров и долей мкс. Это привело к еще одному важному открытию. Томас Маршалл и Марибет Стольценбург с коллегами ввели понятие «инициирующее молнию событие (IE)» [Marshall et al., 2014a]. То есть они стали с уверенностью утверждать, что до этого инициирующего события (IE) на их осциллограммах не было зафиксировано существование никаких других протяженных плазменных образований, которые могли бы приводить к измеряемым электромагнитным импульсам в течение сотен миллисекунд (Рисунок В.21а, жирная синяя стрелка). Как обсуждалось ранее, положительный лидер молнии излучает в радиодиапазоне гораздо слабее, чем отрицательный ступенчатый лидер и достаточно длительное время он, может быть не виден на LMA и интерференционных картах молнии, как например, первые 20 мс на Рисунке В.5, а отрицательный лидер двунаправленного канала, например, в этот момент находится далеко от измерительных систем и потому тоже не обнаруживается.

Рисунок В.21 (адаптировано из [Marshall et al., 2014a]) Рисунок показывает IEC-стадию внутриоблачной молнии (IC). (a) показывает все данные предшествующие ионизации в течение 300 мс от каждого прибора и 175 мс развития IC-молнии; (b) первые 4 мс, показывающие IEC и несколько первых импульсов IB; (c) Показаны 2 мс, с подробными данными IEC и первого IBP. На панелях VHF-события, зафиксированные системой картирования LDAR2 нанесены на график времени их прибытия на антенну K02. На панелях "b" и "c" кривая изменений электрического поля $E$ сдвинута по времени к месту возникновения (x, y, z) первого события LDAR2. Красная вертикальная стрелка, помеченная как «инициация» (initiation), указывает время первого события, записанного LDAR2 на каждой панели. Первое событие, записанное LDAR2, произошло на дальностях 0,5 км и 15,8 км соответственно от датчиков K02 и K14.

Однако в случае измерений [Marshall et al., 2014a, 2019], [Rison et al., 2016] и [Lyu et al., 2019] из-за близости инициации молнии к измерительным антеннам такой сценарий можно было уверено исключить [Marshall et al., 2014a], [Rison et al., 2016].

Инициирующее молнию в облаках событие (IE) может быть разной природы. IE может быть слабым weak NBE (КВР/CID), например Рисунок В.21, описанным [Marshall et al., 2014a, 2019], Рисунок 7.12 [Bandara et al., 2019] и [Lyu et al., 2019] или более сильным, классическим NBE (например, Рисунок В.30 [Rison et al., 2016], Рисунок 7.11 [Bandara et al., 2019], а также Рисунок В.44 [Lyu et al., 2019]. Слабые IE имеют VHF-мощность <1 Вт и длительность $\leq$ 1 мкс; если же инициирующее событие является NBE, то они имеют на порядки большую VHF-мощность и длительность 10–30 мкс.

Томас Маршалл с соавторами [Marshall et al., 2014a] также впервые выделил еще одну важную стадию (этап) развития молнии, которую он назвал «начальное изменение электрического поля» (an initial electric-field change — IEC). Как описали этот этап [Marshall et al., 2014a] и [Chapman et al., 2017], IEC представляет собой период (40-9800 мкс), который начинается с IE и заканчивается первым классическим начальным импульсом пробоя (IBP). [Marshall et al., 2019] показали, что во время проведения IEC было много импульсов VHF с длительностью 1–7 мкс и что некоторые совпадающие пары импульсов быстрой антенны (FA) и импульсов VHF, похоже, увеличивают изменение электрического поля IEC, как «усиливающие события» (enhancing events). Пожалуй, наиболее неожиданным и даже поразительным свойством IEC является его чрезвычайная краткость для CG-молний (диапазон 40-1000 мкс, среднее значение 230 мкс), но и для IC-молний IEC также является очень кратким событием по меркам развития двунаправленного лидера (диапазон 0.5-9.8 мс, среднее значение 2.7 мс).

Чтобы наглядно показать поразительность этого результата приведем данные типичных экспериментов по контакту плазменных каналов, инициированных в электрическом поле искусственно заряженного аэрозольного облака, которые сопровождаются событием, аналогичным обратному удару молнии (подробно такие события рассмотрены в главе 3). На Рисунке В.22 показана осциллограмма тока восходящего положительного лидера, который вступает в контакт с отрицательным нисходящим лидером (двунаправленного лидера Каземира). Общее время события составляет 60 мкс, что больше, чем минимальное время IEC для молний облако-земля. За

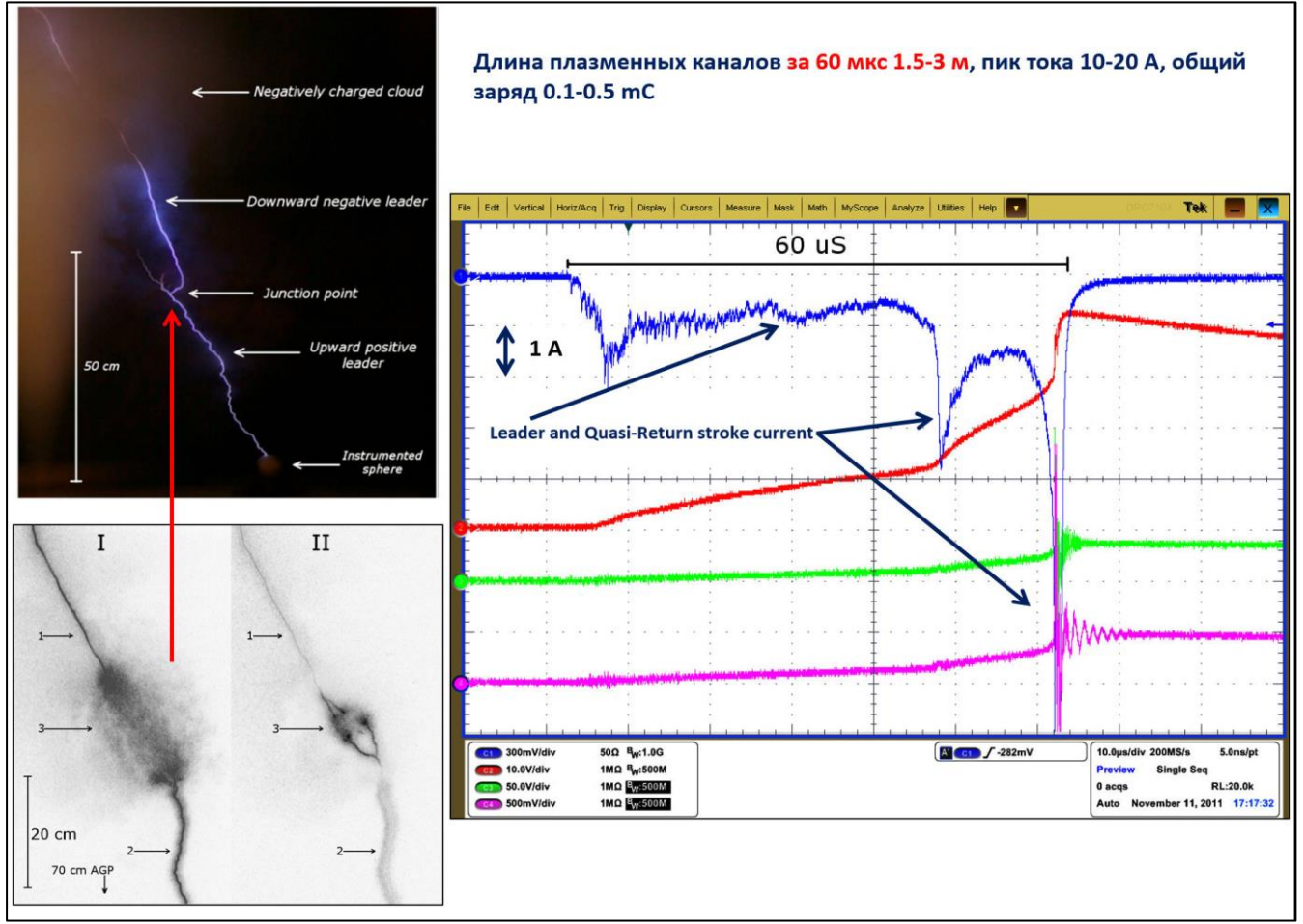

Рисунок В.22. Контакт положительного восходящего лидера и отрицательного нисходящего лидера (отрицательный конец двунаправленного лидера). Осциллограмма тока (синяя линия) показывает, что весь процесс от старта восходящего положительного лидера до образования единого канала занял 60 мкс, закончившись резкими пиками тока квазиобратного удара с током около 7 А. Ток лидера во время его движения находился в диапазоне 0.5-2А. Красная осциллограмма следит за процессом уменьшения заряда аэрозольного облака. Зеленая и фиолетовая осциллограмма показывают сигнал в ФЭУ, которые направлены в другую сторону по отношению движения плазменных каналов в этом событии. Интегральная фотография в верхнем левом углу показывает аналогичное событие встречи двух лидеров. В нижнем левом углу 2 кадра, полученные камерой с усилением изображения 4Picos (Эти два кадра подробно рассмотрены в главе 3, Рисунок 3.6). Все три изображения относятся к разным событиям и выбраны из-за их информативности. Событие, изображенное на осциллограммах, длилось 60 мкс и произвело к переносу заряда 0.1-0.5 мКл, что на три порядка меньше, чем перенос заряда при типичном первом IBP (подготовка, стадия IEC, которого также может длиться 60 мкс), что позволяет выдвинуть гипотезу о параллельном развитии 100-1000 таких каналов на стадии IEC).

это время подобные плазменные каналы накапливают в лучшем случае, судя по осциллограмме тока, заряд 0.1-0.5 мКл, в то время как следующий сразу же за IEC начальный импульс пробоя переносит средний заряд 0.44 Кл (см. раздел «Стадия начального пробоя (B/IB) и начальные импульсы пробоя (IBPs)», стр. 43). То есть, требуется около *тысячи* таких событий, чтобы перенести заряд среднего IBP.

Таким образом, объяснить природу изменений электрических полей и VHF-сигналов IBPs, а также величины токов в десятки кА и зарядов около 0.5 Кл, которые они переносят с помощью представлений о контакте двунаправленных лидеров Каземира очень сложно, так как для накопления таких зарядов в чехлах лидеров необходимо развитие двунаправленных лидеров в течение сотен миллисекунд, а времена протекания начальной стадии развития молнии (IEC) на два порядка меньше. Возможным вариантом решения этого парадокса может быть, действительно, параллельное развитие сотен или тысяч таких небольших каналов, что подробно рассматривается в главе 7, а это в свою очередь потребует одновременной (в пределах нескольких микросекунд) инициации таких небольших каналов в объеме грозового облака (глава 8).

## Компактные внутриоблачные разряды (КВР/CID/NBE) и их возможная роль в инициации молнии

Выше, когда мы обсуждали наиболее важные открытия в физике молнии за последние десятилетия, мы на первое место поставили открытие компактных внутриоблачных разрядов Дэвидом Ле Вайном [Le Vine,1980]. Одиночные биполярные изменения электрического поля продолжительностью (3-30 мкс) в дальней зоне, сопровождающие беспрецедентно сильным VHF-сигналом (на порядок большим, чем у молнии), получили на английском языке название: compact intracloud discharge (CID) [Smith et al., 1999] или narrow bipolar event — NBE [Rison et al., 2016], например, Рисунок B.23. Исключительная важность этого открытия, по нашему мнению, была в том, что CID/NBE по своим электрическим свойствам кардинально отличались от известных и хорошо задокументированных электромагнитных проявлений молнии, но их характеристики были сравнимы с токами и зарядами, переносимыми во время обратных ударов молнии.

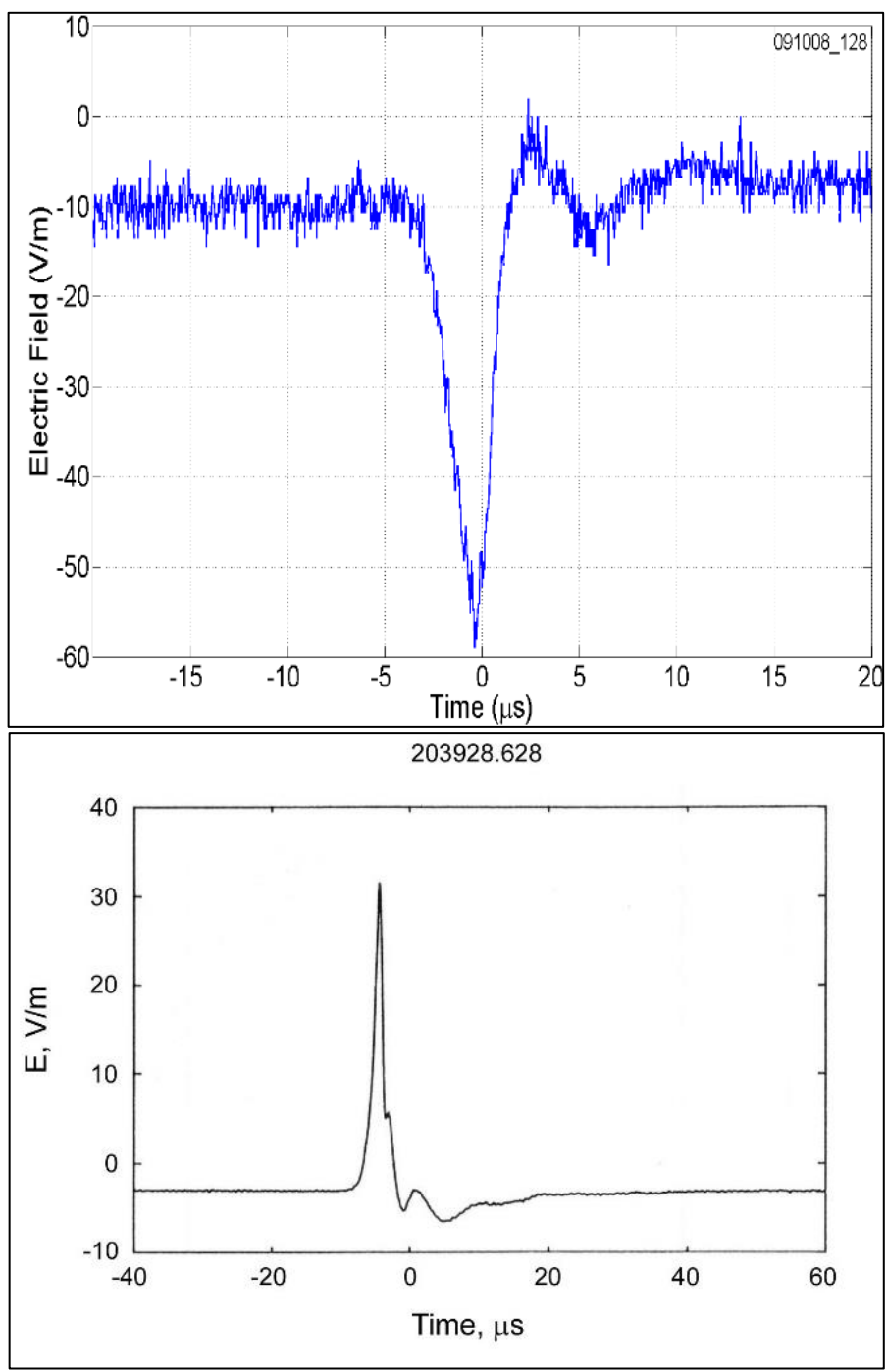

Рисунок В.23. (вверху) широкополосные формы сигналов электрического поля (E) для типичного узкого биполярного события (CID/NBE) с положительной (согласно физическому соглашению о знаке электрического поля) начальной полярностью [Willett et al., 1989], от изолированной грозовой ячейки с центром на расстоянии 45 км от места измерения, (внизу) NBE/CID отрицательной полярности (любезно предоставлено Владимиром Раковым).

Приведем основные свойства положительного КВР (CID/ NBE) [Smith et al., 1999]:

- Время нарастания поля (от 10% до 90% макс. значения) $\approx 2.3 \pm 0.8$ мкс;

- Длительность начальной полуволны электрического поля (по уровню 50% макс. значения) $\approx 4.7 \pm 1.3$ мкс;

- Полная длительность импульса $\approx 25.8 \pm 4.9$ мкс;

- Максимальное значение электрического поля на первой полуволне биполярного импульса (нормировано на 100 км от источника), В/м $\approx 9.5 \pm 3.6$ мкс;

- Максимальное значение электрического поля на второй полуволне биполярного импульса (нормировано на 100 км от источника), В/м $\approx -3.9 \pm 1.6$ мкс;

- Отношение максимальных амплитуд значения электрического поля на первом и второй полуволне биполярного импульса $\approx 2.7$;

- Отношение максимальных амплитуд значения электрического поля CID/ NBE и импульса обратного удара CG-молнии $\approx 0.71$;

- Отношение максимальных амплитуд значения электрического поля CID/ NBE и К-импульса внутриоблачного разряда молнии (IC-молнии) $\approx 0.71$;

Основные свойства VHF-излучения КВР/CID/ NBE:

- Длительность VHF-излучения $\approx 2.8 \pm 0.8$ мкс;

- Максимальное значение поля (нормированное на расстояние 10 км в полосе 1 кГц) $\approx 2.4 \pm 1.1$ мВ/м

- Отношение пиковых амплитуд VHF-излучения КВР и обратного удара CG-молнии $\approx 9.9$

- Отношение пиковых амплитуд VHF-излучения КВР и К-импульса внутриоблачного разряда молнии (IC-молнии) $\approx 29$.

На Рисунке В.24 отчетливо видны отличия КВР от начальных импульсов пробоя (IBPs): уникальная форма сигнала электрического поля и отсутствие периодичности повторений, как у начальных импульсов пробоя (IBPs). Но оба этих явления сходны большими токами и перенесенным зарядом.

На Рисунке В.25 показана схема возможного расположения положительных и отрицательных КВР/CID/NBE внутри грозового облака в соответствие с их знаками (физическое соглашение о направлении электрического поля).

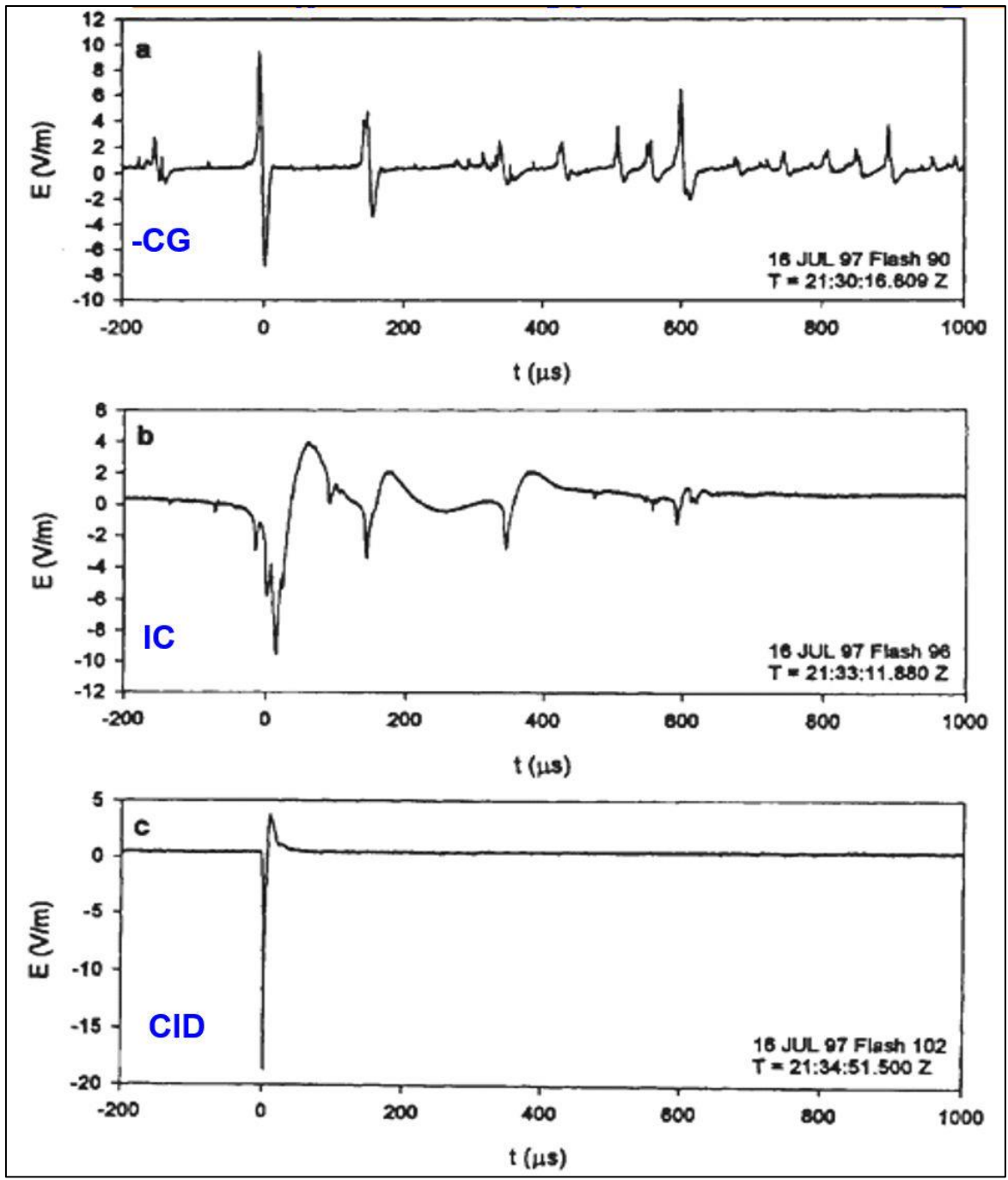

Рисунок В.24 (любезно предоставлен Владимиром Раковым). Примеры осциллограмм электрического поля, характерных для: /а/ IBPs в отрицательных CG-молниях (продолжительность отдельных импульсов 20-40 мкс, межимпульсный интервал 70-130 мкс); /б/ IBPs в IC-молниях (продолжительность отдельных импульсов 50-80 мкс, межимпульсный интервал 600-800 мкс); /в/ КВР/CID/NBE (продолжительность отдельных импульсов 10-30 мкс). Вектор положительного электрического поля направлен вниз, то есть это положительный CID/NBE (соглашение о знаке электрического поля согласно сообществу атмосферного электричества).

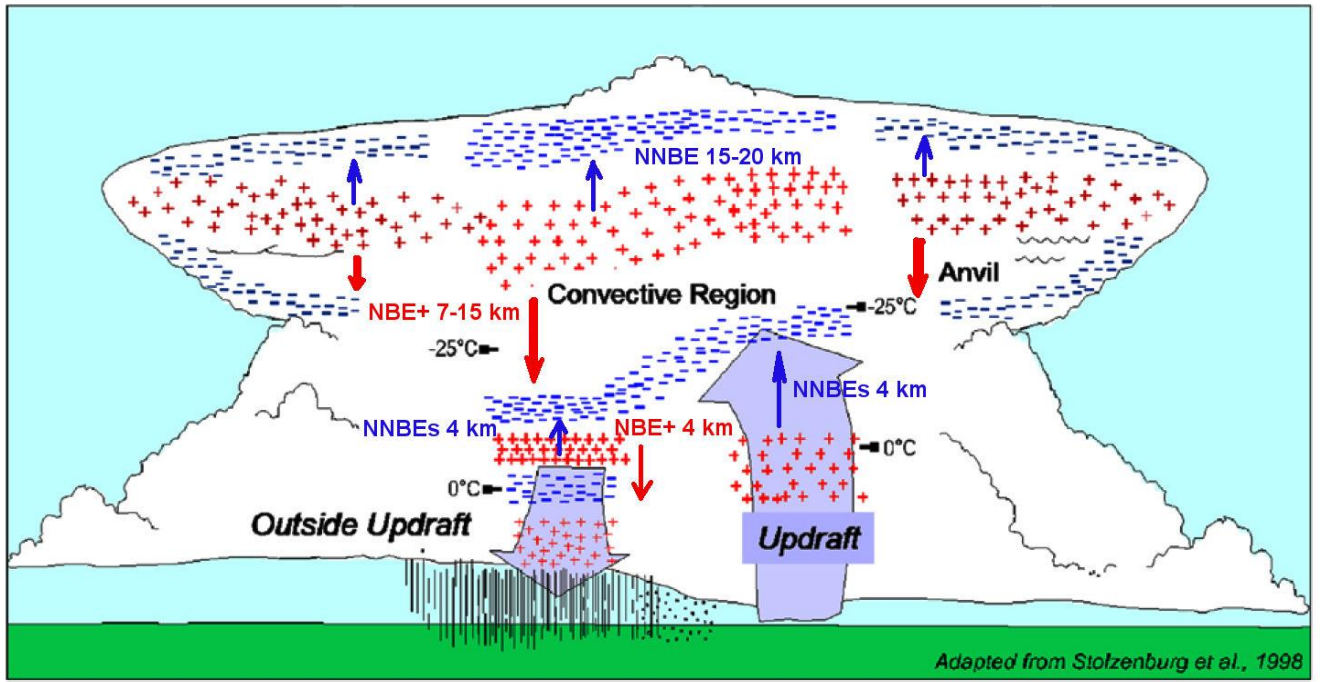

Рисунок В.25. Возможное местоположение КВР/CID/NBE внутри грозового облака. Красные стрелки соответствуют положительным КВР (NBE+), синие стрелки отрицательным КВР (NNBE) (соглашение о знаке электрического поля согласно сообществу атмосферного электричества). Цифры указывают на высоту в км для данного типа КВР. Рисунок сделан на основе схемы распределения зарядов в грозовом облаке, представленной в статье [Stolzenburg et al., 1998].

В настоящее время не очень понятно распределение КВР с высотой. Распределение КВР по высоте, построенное по результатам большого числа измерений (115537 наблюдений), которые фиксировала система обнаружения электромагнитных импульсов Los Alamos Sferics Array (LASA) [Smith et al., 2004] охватывает огромный диапазон высот от 2 до 30 км (Рисунок В.26). Подавляющее большинство событий было зарегистрировано в окрестностях штата Флорида (США). Трудно найти объяснение КВР, которые расположены на высоте выше 20 км (значительно выше тропопаузы), так как туда практически не распространяются даже самые мощные грозовые облака и механизм инициации стримерных вспышек на таких высотах также встречает исключительно большие трудности (глава 8). В случае [Smith et al., 2004] нельзя исключить погрешностей измерений, так как более поздние наблюдения сузили диапазон появления КВР до интервала 6-19 км, Рисунок В.27 (верхняя панель [Leal and Rakov, 2019], нижняя панель [Karunarathna, N. et al., 2015]), но на гораздо меньшей экспериментальной выборке, что не позволяет пока исключить и большие высоты появления КВР. Пока вопрос о предельно высоких и предельно низких высотах появления КВР остается открытым, но все равно 18-19 км это очень большие высоты для успешной инициации стримеров.

Важной и показательной характеристикой КВР являются переносимые заряды и пиковые токи (Рисунок В.28) [Leal and Rakov, 2019)]. Большая часть токов находится в диапазоне от 15 до 65 кА, но также зафиксированы токи от 100 до 160 кА, которые соответствуют очень мощным обратным ударам молнии облако-земля. Заряды находятся в диапазоне 0.5-1 Кл [Rizon et al., 2016], что несколько меньше, чем во время обратных ударов, но КВР не является многокилометровым каналом молнии, движущимся и собирающим заряд сотни миллисекунд, так как общая длительность КВР не превышает, обычно, 30 мкс. Если представлять КВР, как кратковременное появление короткого горячего плазменного канала типа двунаправленного лидера Каземира (100-1000 м), который излучает такой сигнал, то совершенно непонятно, благодаря какому механизму удается такому короткому каналу сконцентрировать внутри своего чехла такой большой заряд за такое короткое время.

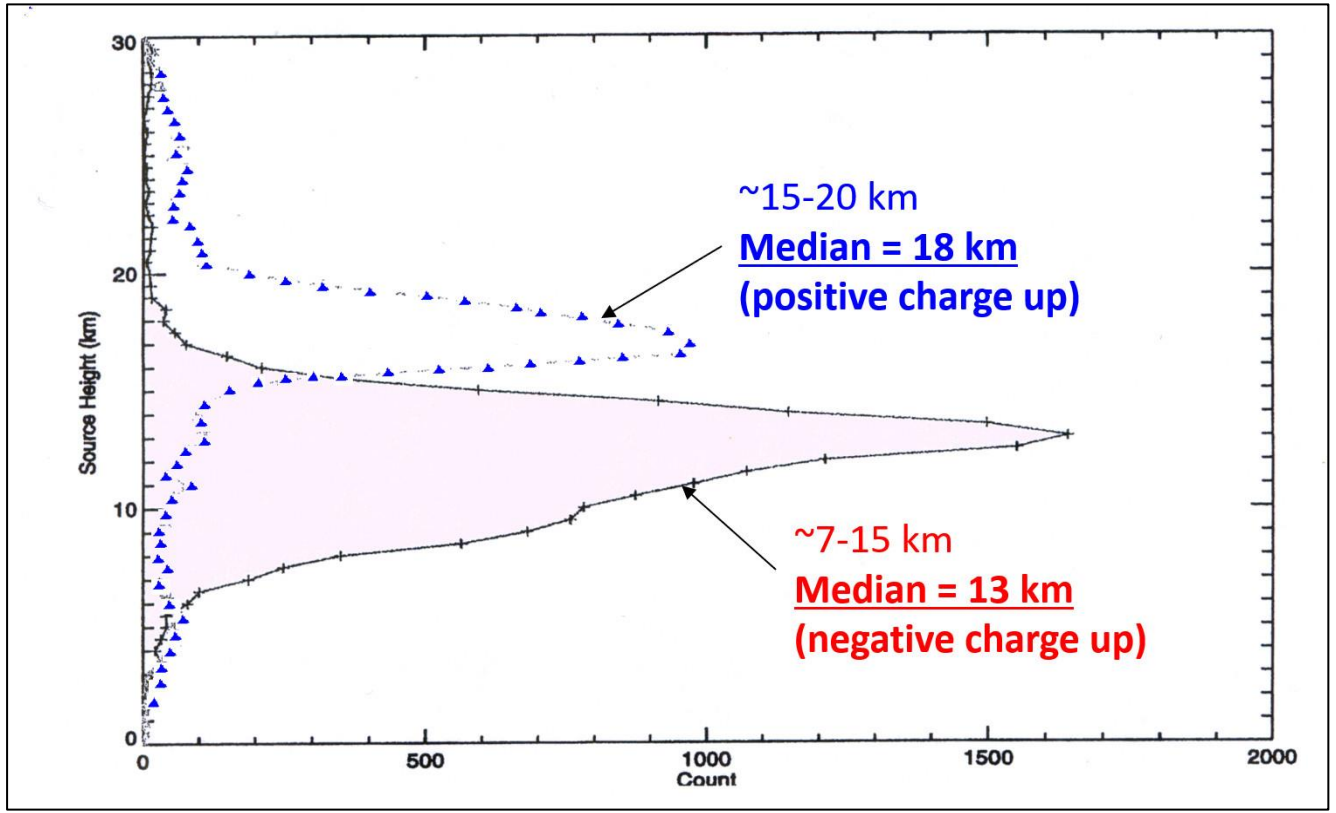

Рисунок В.26. Распределение КВР/CID/NBE по высоте, построенное по результатам работы системы фиксации электромагнитных импульсов Los Alamos Sferics Array (LASA) [Smith et al., 2004]. Сплошная черная кривая показывает положительные NBE (иногда пишут также NBP, подчеркивая, что это кривая импульса), пунктирная синяя кривая показывает отрицательные NBE (NBP). Знаки КВР соответствуют физическому соглашению о знаках. Распределение построено по результатам 115 537 наблюдений и шаг кривой по высоте составляет 0.5 км. Подавляющее большинство событий было зарегистрировано в окрестностях штата Флорида (США). Трудно найти объяснение КВР, которые расположены на высоте выше 20 км (значительно выше тропопаузы), так как туда практически не распространяются даже самые мощные грозовые облака.

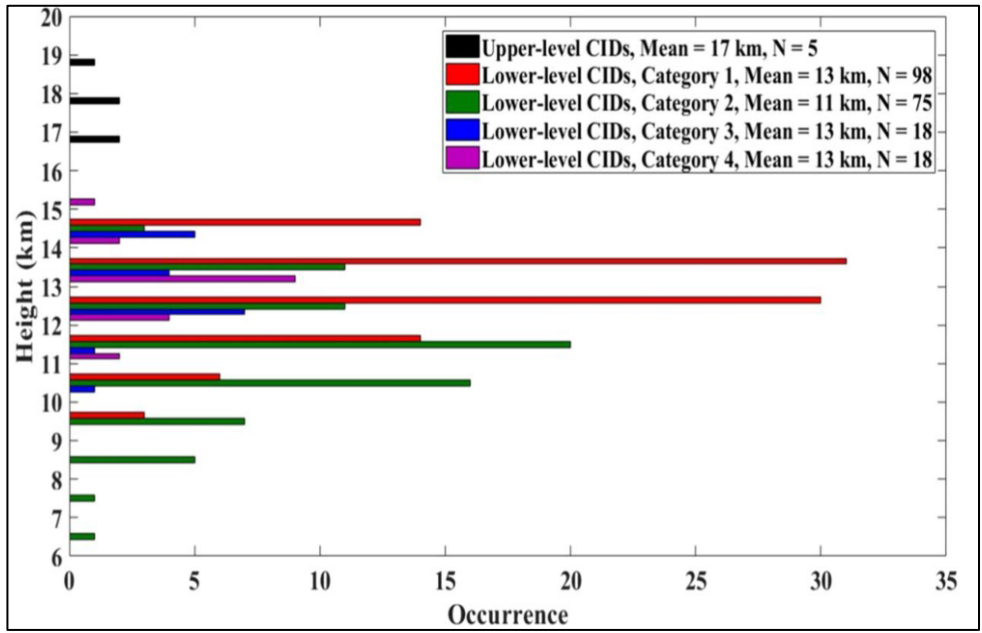

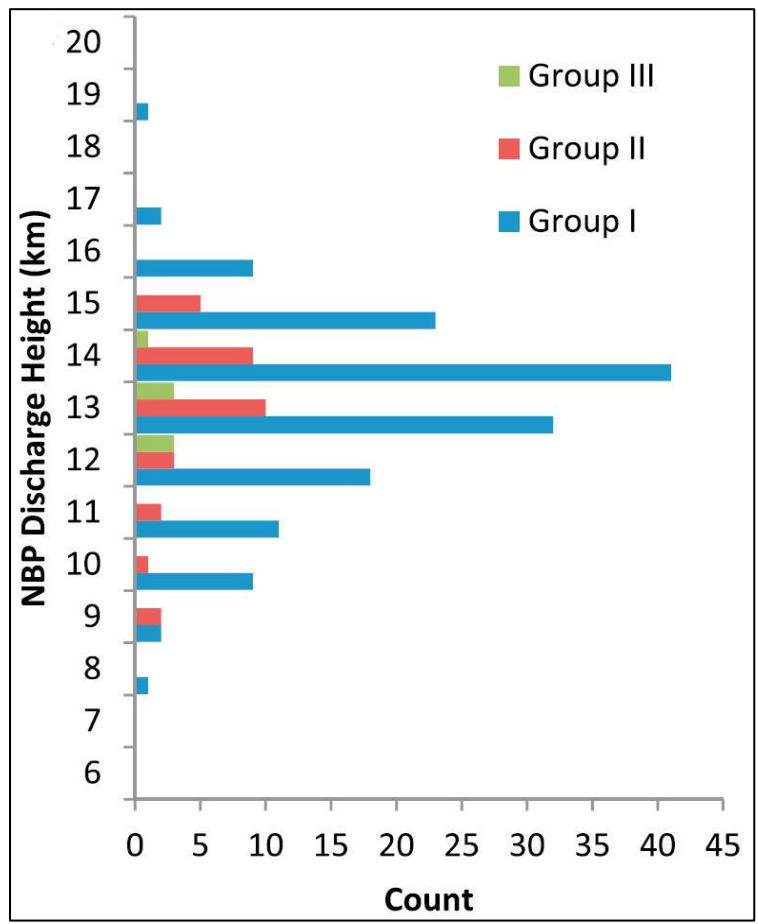

Рисунок В.27. Распределение КВР/CID/NBE по высоте. По оси абсцисс высота, на которой были зафиксированы КВР, по оси ординат число КВР на данное высоте. Верхняя панель построена в работе [Leal and Rakov, 2019)], нижняя панель получена в работе [Karunarathna, N. et al., 2015]. В отличие от Рисунка В.26 здесь интервал обнаружения КВР от 6 до 19 км.

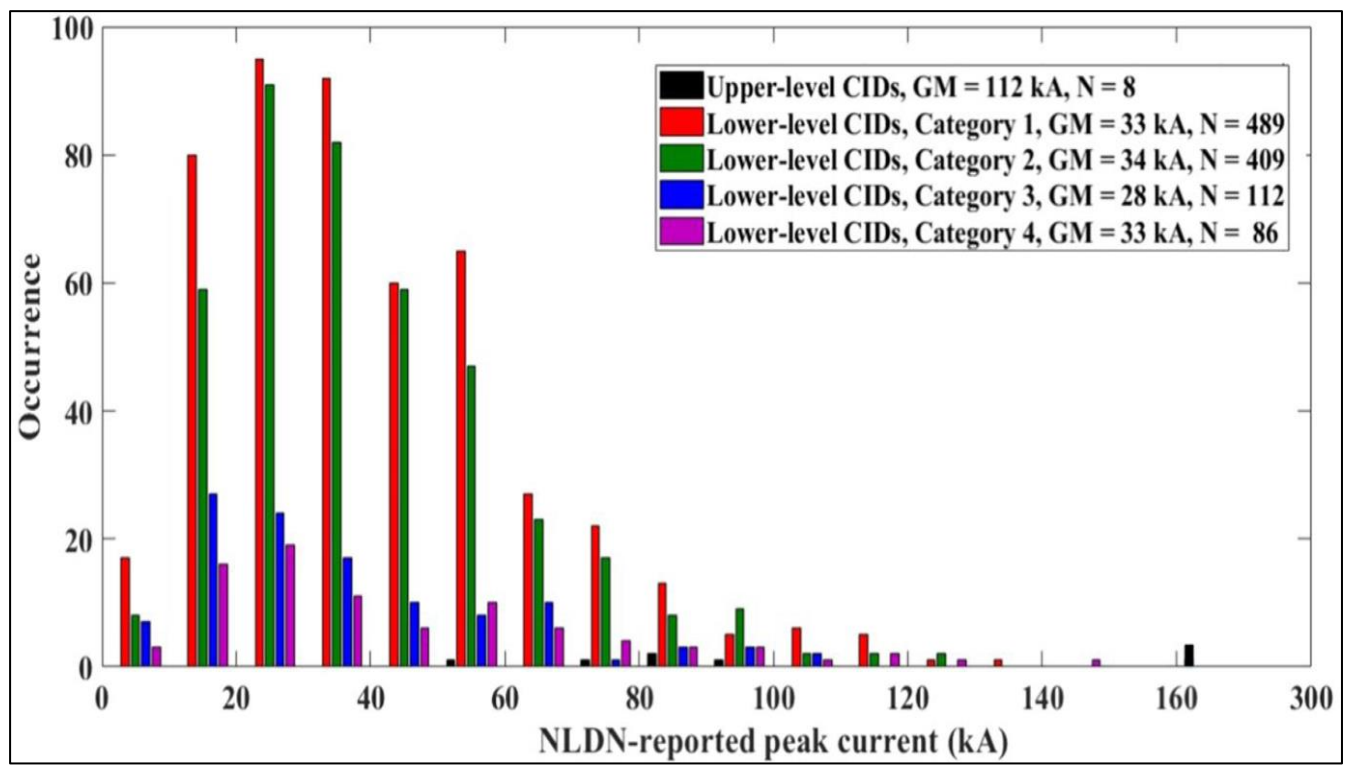

Рисунок В.28. Число КВР (ось абсцисс) в зависимости от величины максимального тока (в кА, ось ординат). [Leal and Rakov, 2019)]. Данное распределение получено Национальной системой определения координат молниевых разрядов США (NLDN). Большая часть токов находится в диапазоне от 15 до 65 кА, но также зафиксированы токи от 100 до 160 кА.

Вопрос связи КВР/CID с инициацией молнии возник сразу же после начала изучения этого явления. Была выявлена, но не доказана связь КВР и молний, как на Рисунке В.29 [Nag, Rakov et al., 2010].

Новый этап в изучении связи КВР и молнии наступил после публикации статьи [Rison et al., 2016]. [Rison et al., 2016], благодаря прецизионным радиоизмерения доказали, что КВР инициирует молнию. Более того, они предположили, что «все или почти все молнии инициируются КВР или их слабыми аналогами». [Rison et al., 2016] использовали быструю антенну (FA) и интерферометр с очень малым окном выборки сигнала (~1 мкс) фиксации наиболее сильного источника VHF, а также программное обеспечение, которое в результате сложных вычислений ставило «точку» на карте в трехмерном пространстве и времени (с помощью данных LMA-системы, так как интерферометрия дает только угловые координаты). Эту «точку» называют центроидом (the centroid). На Рисунке В.30 показана интерферометрическая карта (точки, центроиды) в угловых координатах и изменение электрического поля (красная кривая) во время инициации молнии компактным внутриоблачным разрядом (NBE). Они назвали это событие — NBE2. Максимальный ток этого события NBE2 составлял 57 кА (ток, сравнимый с током сильного обратного удара), а переносимый заряд находился в диапазоне 0.5-0.7 Кл. На панели (а) «возвышение-время», которая отображает центроиды, наложена кривая изменений электрического поля (быстрая антенна FA). Показаны первые 2 мс развития IC-молнии, инициированной КВР (NBE2). На панели 'а' хорошо видно сильное колебание электрического поля КВР в момент инициации молнии, за которым следует постепенное возрастание поля во время начального изменения электрического поля длительностью 1,5 мс (стадия-IEC), а после IEC-стадии следует стадия начальных импульсов пробоя (IBPs). На протяжении IEC-стадии (синие и зеленые источники VHF — центроиды) в среднем расширили область разрядов на 500 м со средней скоростью $3 \cdot 10^5$ м/с, а IBPs-стадия развития молнии (желтые и красные источники VHF) увеличила область разрядов также на 500 м со средней скоростью $1 \cdot 10^6$ м/с. Обращает на себя внимание, что вся IEC-стадия состоит из разрозненных точек и ничем не напоминает кривые движения «каналов» (лидеров) на аналогичных LMA и интерферометрических картах (например, Рисунки В.6, В.7, В.14). Благодаря высокому пространственному и временному разрешению интерферометра и LMA-системы [Rison et al., 2016] удалось исследовать тонкую структуру этого КВР(NBE), который в их работе

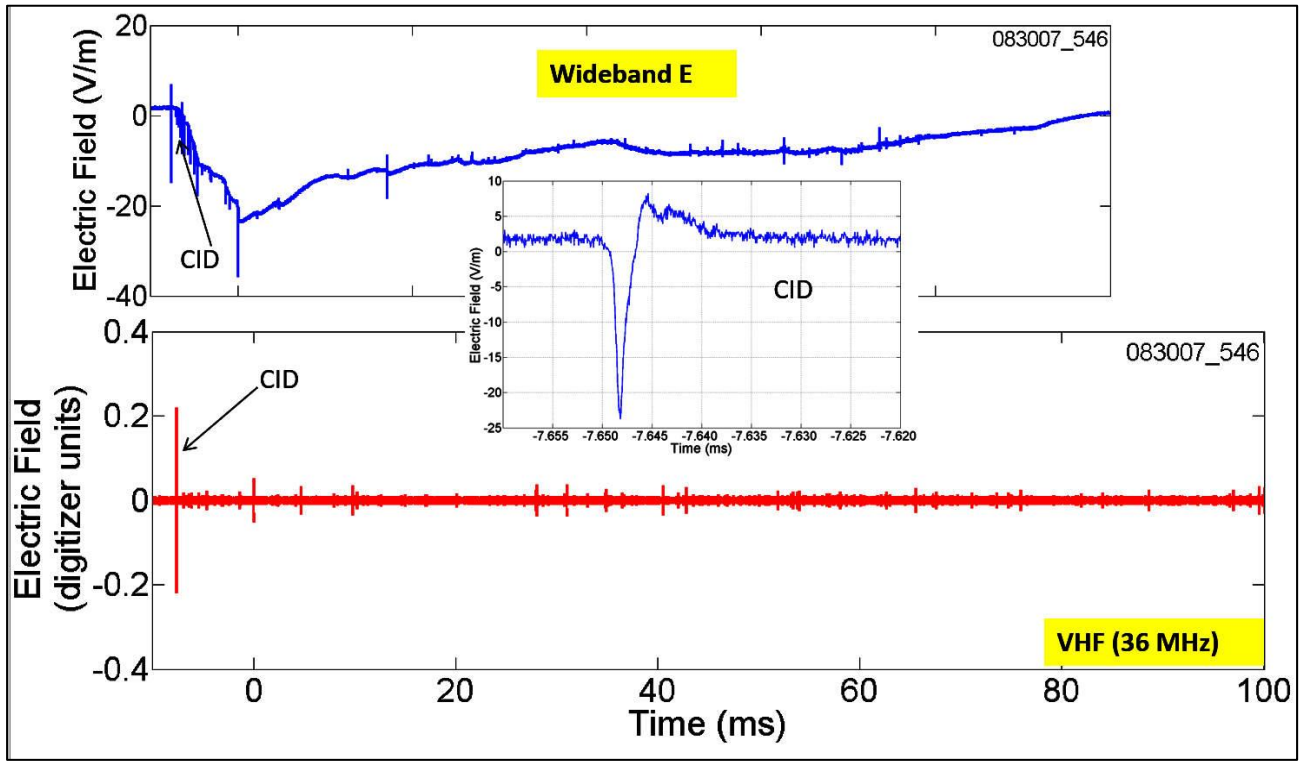

Рисунок В.29 Изменение электрического поля *E* (синие кривые) и VHF-излучения (36 МГц, красная кривая) во время протекания КВР (CID), за которым последовала «нормальная» IC-молния. На вставке показана осциллограмма CID на шкале времени 5 мкс на деление. ([Nag, Rakov et al., 2010], рисунок адаптирован и любезно предоставлен Владимиром Раковым).

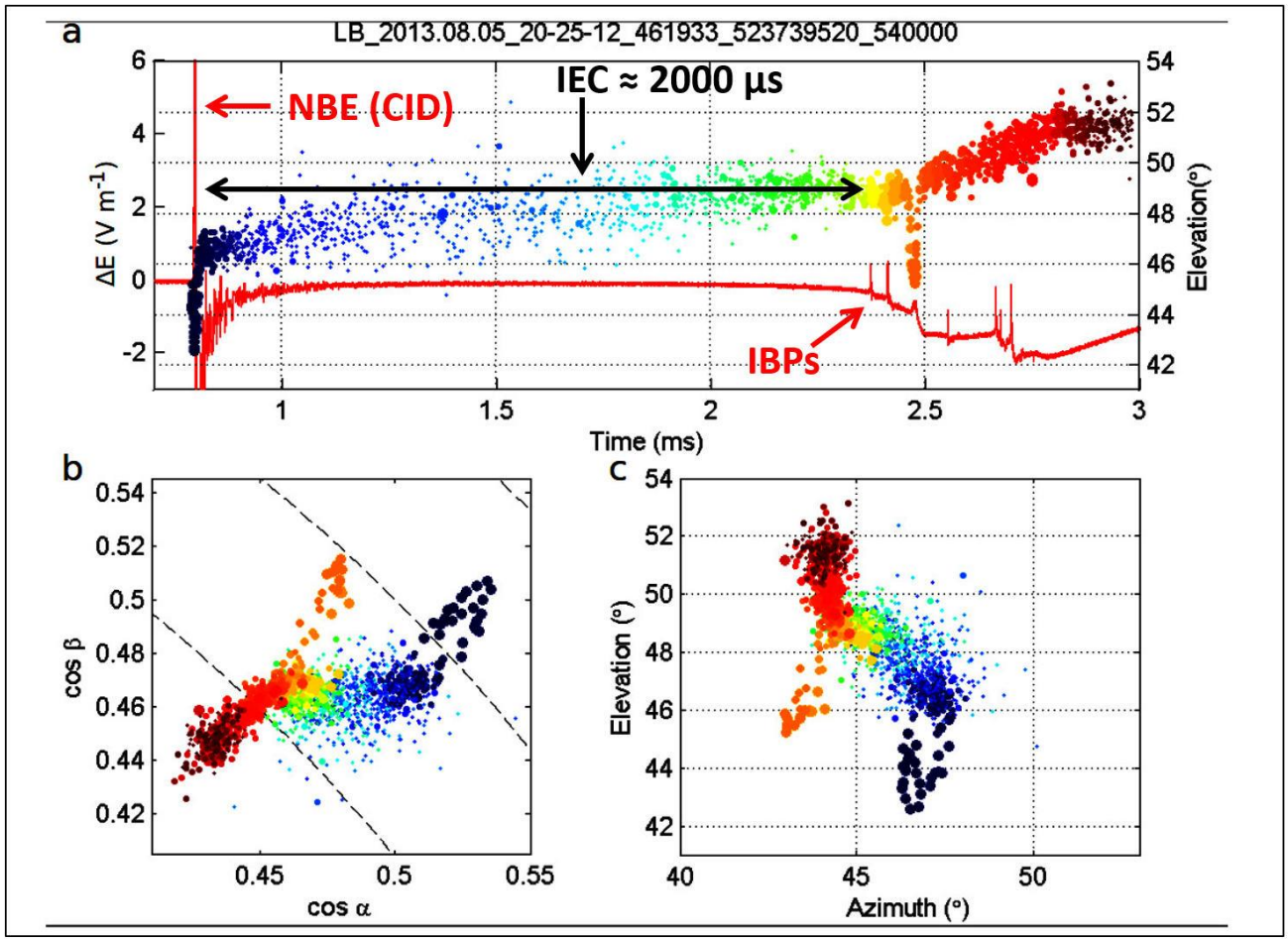

Рисунок В.30 (адаптировано из [Rison et al., 2016], общая LMS-карта разряда на Рисунке В.5). Интерферометрическая карта (точки) в угловых координатах и изменение электрического поля (красная кривая) во время инициации молнии «классическим» компактным внутриоблачным разрядом (КВР/CID/NBE) (в статье [Rison et al., 2016] это событие названо NBE2). (а) на панель «возвышение-время» (точки) наложена кривая измерения электрического поля быстрой антенной (FA) в течение первых 2-х мс развития IC-молнии, инициированной КВР (NBE2). На панели 'а' хорошо видно сильное колебание электрического поля КВР в момент инициации молнии, за которым следует постепенное возрастание поля во время начального изменения электрического поля длительностью 1,5 мс (стадия-IEC), а после IEC-стадии (длительность около 2 мс) следует стадия начальных импульсов пробоя (IBPs). Во время IEC-стадии (синие и зеленые источники VHF) в среднем расширили область разрядов на 500 м со средней скоростью $3 \cdot 10^5$ м/с, а IBPs-стадия развития молнии (желтые и красные источники VHF) увеличила область разрядов также на 500 м со средней скоростью $1 \cdot 10^6$ м/с. Точки (источники VHF) окрашены по времени в соответствие со шкалой времени на панели 'а' и их диаметр соответствует логарифму мощности источников VHF. Множество мелких разбросанных по всему объему (точек) центроидов (интервал времени на панели 'а' примерно 0.6-2.3 мс) может говорить об объемном процессе, протекающем в этой области грозового облака.

маркируется как NBE2. На Рисунке В.31 со шкалой времени 30 мкс показано изменение электрического поля во время КВР с высоким разрешением (красная кривая), которое позволяет увидеть не только сам биполярный импульс, но и колебания, которые длятся еще минимум 25-30 мкс. Центроиды интерферометрических измерений, обозначенные точками с размерами пропорциональными логарифмам мощности в координатах «возвышение-время», с некоторого момента времени показывают «движение» центроидов со скоростью близкой к скорости света $\sim 10^8$ м/с (красная стрелка), а потом центроиды «движутся» вверх со скоростью примерно в два раза меньшей (пространственный диапазон таких движений составлял 500-600 м). Мы пишем слово «движение» в кавычках, так как оно может отображать как реальное движение плазменного канала, например, положительного восходящего лидера, как на Рисунках В.6, В.7, так и «движение» *самого* статистически яркого VHF-события на фронте коллективного движения нескольких (многих) VHF-источников. Дело в том, что, сама методика картирования VHF-сигналов (и LMA-сигналов в еще большей степени) хорошо приспособлена для картирования движения небольшого числа (в идеале одного) плазменных каналов, но принципиально не может хорошо отображать коллективные движения многих источников VHF-излучения, так как в каждом окне, даже при таком коротком окне, как 1 мкс, выбирается один самый сильный сигнал (или несколько с большими оговорками и допущениями). Поэтому, при одновременном движении, например, $10^3$-$10^7$ источников VHF-излучения даже в одном направлении, интерференционная карта все равно представит это движение центроидов (точек) в форме траектории, прочерчивающей приблизительно «канал», как траектория центроидов, показанная красной стрелкой на Рисунке В.31. Несмотря на сложность интерпретации Рисунка В.31, на нем явно виден физический процесс, который распространяется со скоростью, близкой к скорости света, причем, как настаивают [Rison et al., 2016], до момента инициации КВР в этой области грозового облака не существовало горячих плазменных каналов, волна потенциала внутри которых могла бы обеспечить такие большие скорости, подобно скоростям движения потенциала во время обратного удара молнии в наземные сооружения. [Rison et al., 2016] также установили, что интегральный VHF-сигнал нарастает экспоненциально с постоянной времени 0.3 мкс (Рисунок В.32.b). Таким образом, удалось установить, что КВР точно могут инициировать молнии. Однако более обширные наблюдения, которые были проведены

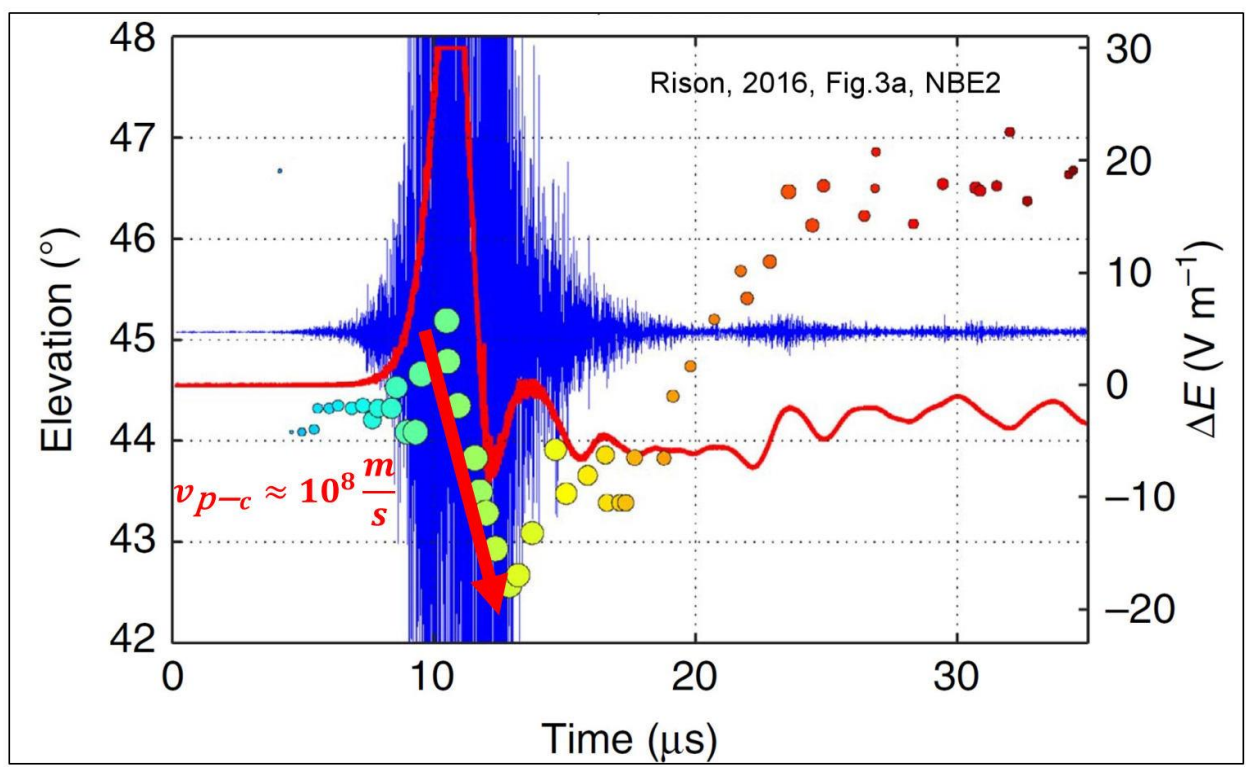

Рисунок В.31 (адаптировано из [Rison et al., 2016], подробное рассмотрение NBE2, который инициирует молнию на Рисунке В.30). Измерение электрического поля быстрой антенной (красная кривая), правая шкала. Центроиды интерферометрических измерений обозначены точками с размерами пропорциональными логарифмам мощности в координатах «возвышение-время». Общая мощность VHF-излучения показана синей кривой, которая сливается в полосу в районе максимума изменения электрического поля. Красной стрелкой показано «движение» центроидов со скоростью близкой к скорости света $\sim 10^8$ м/с.

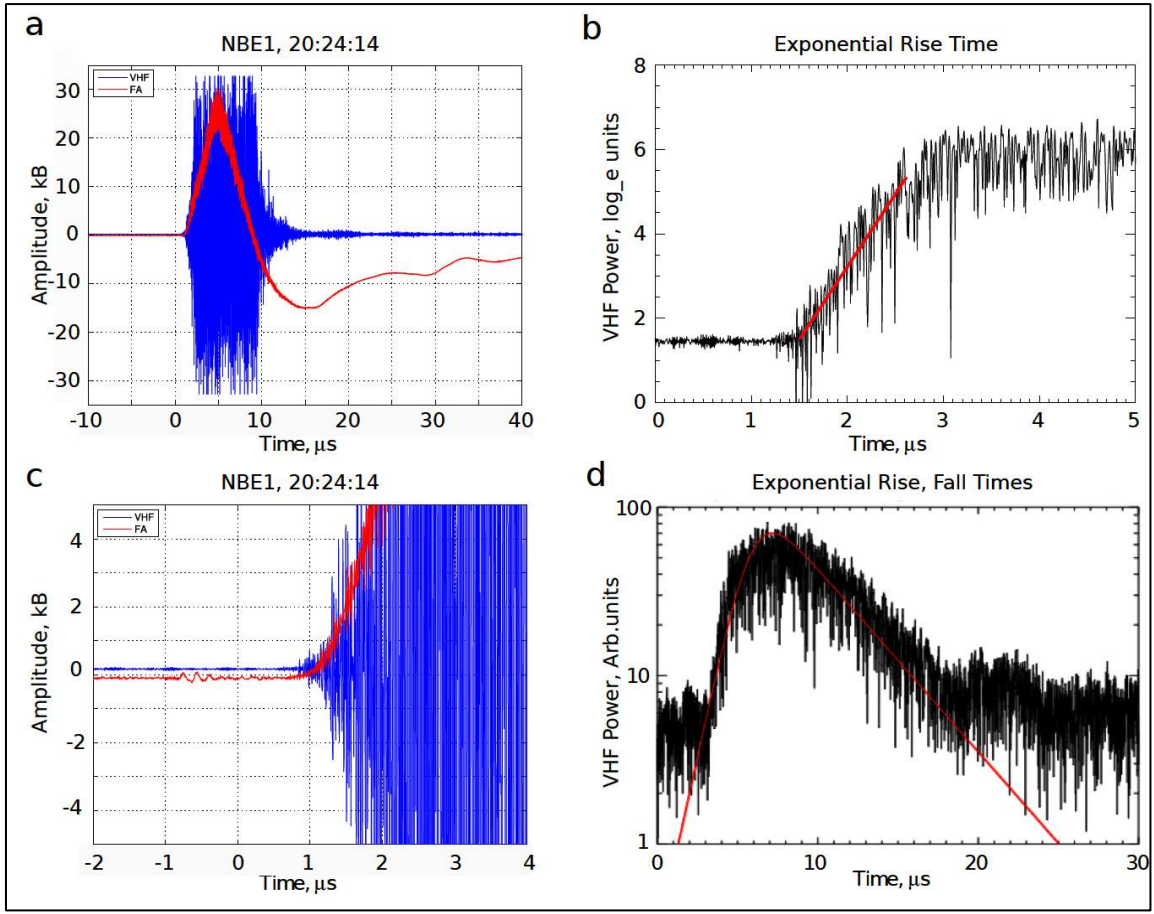

Рисунок В.32 (адаптировано из [Rison et al., 2016]). **(a)** — осциллограмма VHF-антенны (синяя) и электрического поля (FA, красная) для КВР (NBE1 — маркировка другого события); **(b)** — экспоненциальное увеличение мощности VHF в течение первых 1.5 мкс развития КВР перед насыщением сигнала. Красная наклонная прямая соответствует постоянной времени экспоненциального нарастания сигнала 0.3 мкс; **(c)** — график первых нескольких микросекунд КВР (NBE), показывающий i) его быстрое начало, ii) одновременное возникновение излучения, принимаемого быстрой антенной (FA) и VHF-излучения, указывающее на ток, возникающий при пробое, и iii) отсутствие какой-либо измеряемой активности до начала КВР(NBE). И VHF, и сигналы с быстрой антенны электрического поля были на уровне фонового шума на измерительных приборах, который был на 66 дБ ниже амплитуд во время измерений; **(d)** — экспоненциальный рост и падение мощности VHF, полученной из ослабленной формы осциллограммы VHF-сигнала, который также фиксировала быстрая антенна FA, оцифровывающая с той же частотой 180 МГц, что и сигналы интерферометра. Постоянные времени нарастания и спада составляли 1,0 и 4,7 мкс соответственно.

очень активно из-за большой значимости данного наблюдения, показали, что 93-95% молний инициируются слабыми событиями, дающими на несколько порядков более слабый сигнал быстрой антенны и VHF-излучения, чем КВР [Marshall et al., 2014a, 2019], [Bandara et al., 2019] и [Lyu et al., 2019]. И только 5-7% молний (из подробно исследованного массива данных) могут быть инициированы КВР ([Rison et al., 2016], [Lyu et al., 2019]).

[Rison et al., 2016] предложили качественную, а [Attanasio at al., 2019] количественную модель, объясняющую эксперименты [Rison et al., 2016] и тем самым объясняющую, по их мнению, природу КВР (NBE). Они использовали предложенную еще Гриффитсом и Фелпсом [Griffiths & Phelps, 1976] модель инициации молнии, благодаря большой стримерной вспышке, которая сама себя ускоряет и поддерживает электрическим полем на своем фронте. Это своеобразная попытка перенести концепцию газоразрядного стримера, который движется из-за усиления поля в своей головке (во внешнем электрическом поле), на стримерную вспышку, которая может создавать сверхпробойное электрическое поле благодаря многим головкам стримеров (Рисунок В.33). [Rison et al., 2016], [Attanasio at al., 2019] видоизменили модель Гриффитса-Фелпса и предположили, что стримеры могут двигаться с гораздо большими скоростями 5-10·$10^7$ м/с. Применение модели Гриффитса-Фелпса означало, что [Rison et al., 2016] считают КВР коллективным, объемным стримерным процессом (а не развитием линейного двунаправленного лидера), который экспоненциально увеличивает свой заряд на фронте, ток и VHF-сигнал, используя процесс непрерывного ветвления стримеров при сохранении заряда каждой головки постоянным. Предложенная [Attanasio at al., 2019] модель имеет множество слабостей. Во-первых, модель Гриффитса-Фелпса опирается на экспериментальные работы, качество которых в настоящее время нельзя считать удовлетворительным: [Phelps, 1971], [Phelps, 1974], [Phelps & Griffiths, 1976a], [Phelps & Griffiths, 1976b]. 40-45 лет назад исследование длинных стримеров только начиналось и многие предположения, которые легли в основание гипотезы Гриффитса-Фелпса в дальнейшем не подтвердились. Кроме того, Гриффитс и Фелпс не использовали большинство современных экспериментальных методов исследования стримерных вспышек. Гриффитс и Фелпс в этих статьях вообще не измеряли ток и заряд (ключевые для модели параметры). Ни в одной из статей Гриффитса-Фелпса не написано, как они измеряли напряжение. Критическое электрическое поле на промежутке оценивалось с

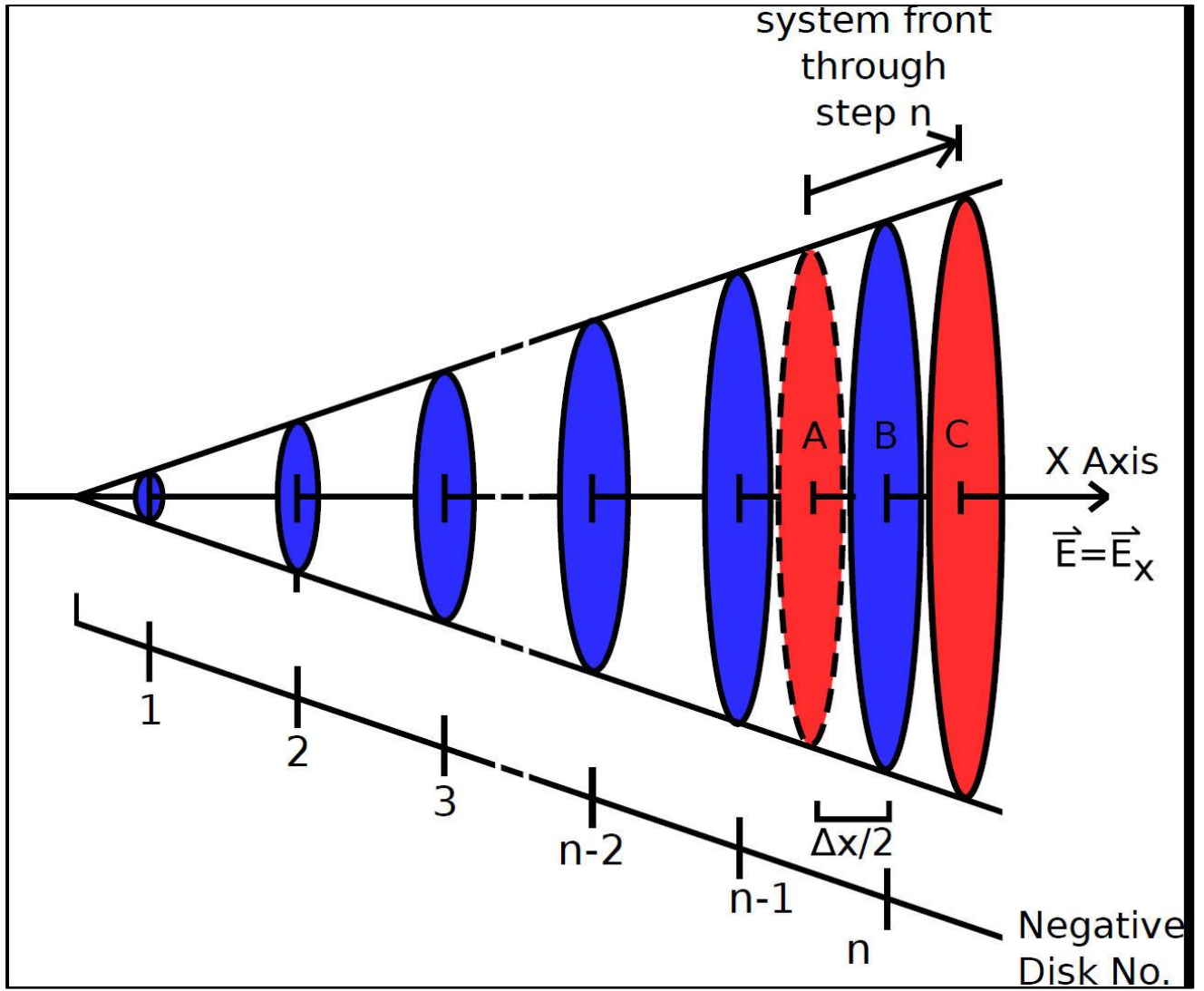

Рисунок В.33 (адаптировано из [Attanasio at al., 2019]). Схематическое изображение модели «быстрого положительного стримерного пробоя» (fast positive breakdown — FPB) в момент n-го шага. Внешнее электрическое поле E в направлении x заставляет положительный заряд на красном пунктирном диске продвигаться от A к C, от задних отрицательных (синих) дисков предыдущих вычислительных шагов, в процессе увеличения его заряда с $q_A$ до $q_C = q_A + \Delta q$, оставляя заряд–$\Delta q$ на B (синий диск).

помощью экспериментов, где основную роль играл человеческий глаз (реально), как измерительный прибор ("The critical field strength was determined by adjusting the dc potential applied between the parallel electrodes upward until the most vigorous streamers just cross the gap. The critical field strength is taken as the quotient of this potential and the gap length. By adjusting the pulse voltage and the vertical position of the needle electrode for each measurement, the onset of streamer crossing was made quite abrupt allowing a reasonably precise estimate of the critical field to be made by observation in total darkness with the *dark-adapted eye*", [Phelps, 1976b]). Стримеры инициировались отдельным источником высокого напряжения (30-200 кВ), который находился внутри заземлённой плоскости на небольшом расстоянии от металлических стенок плоскости. Это расстояние было гораздо меньше, чем расстояние до плоскости под напряжением генератора Ван дер Графа (источник напряжений в экспериментах). При таком напряжении на точечном электроде разряд должен был происходить между точечным источником высокого импульсного напряжения и близкой заземлённой плоскостью. В статьях нет объяснения, как при такой сложной геометрии разрядного промежутка и двух источниках высокого напряжения (импульсном и постоянном) инициируются стримеры, которые пересекают весь промежуток между плоскостями. Такие объяснения необходимы, так как существует разряд даже при электрическом поле между плоскостями равном нулю, [Phelps, 1971, Fig.2a]. Предположения Гриффитса-Фелпса об экспоненциальном размножении зарядов основываются исключительно на непрямых, неколичественных измерениях потоков света из промежутка. При этом они никак не доказывают и не аргументируют почему предполагается линейная связь между зарядом стримерной вспышки и светом из разрядного промежутка [Phelps, 1974]. Это необходимо было сделать, т.к. энерговклад в возбужденные электронные уровни молекул азота $N_2^*(2^+, 1^+, 1^-$ систем) принципиально нелинейно зависит от электрического поля $\varepsilon_{N_2^*} \sim \sigma E^2$. Кроме того, в статье [Phelps, 1974] нигде не обсуждается калибровка линейности оптической системы и ФЭУ в УФ-диапазоне. Неизвестно использовал ли Фелпс хоть какой-нибудь объектив и какова была апертура падения света на поверхность ФЭУ при изменении его положения в разрядном промежутке. Это необходимо было сделать, т.к. ФЭУ 1Р21, используемое в [Phelps, 1974], имеет сложную искривлённую геометрию стеклянного окна, а также резко падающую чувствительность в УФ-диапазоне. Гриффитс и Фелпс ни в одном эксперименте не пользуются электронно-оптическими преобразователями с разверткой и усилением изображения (стрик-камерами). Поэтому, из-за низкой чувствительности фотопленки в УФ-диапазоне, они получали фотографии, на которых изображена сумма около 500 импульсов [Phelps, 1971]. Даже при этом на этих фотографиях нигде не зафиксирован конус распространения стримерной вспышки, [Phelps, 1971, Fig.2], как на Рисунке В.33. Если бы даже конус распространения стримерной вспышки был на этих и других фотографиях по результатам 500 импульсов, то он мог бы быть следствием статистического разброса траекторий стримеров, а не конуса, который возникает в каждой вспышке и используется в модели Гриффитса-Фелпса. Конус, возможно, виден только на фотографиях, где возникает искровой разряд [Phelps, 1971, Fig.4]. На Fig.2f из [Phelps, 1971], скорее всего также был искровой разряд, где отчётливо видна неравномерность свечения и два разных конуса. Если Fig.2 из [Phelps, 1971] не является искровым разрядом, то на ней изображена сильно неоднородная картина разряда. Во всех статьях Гриффитса и Фелпса есть единственное предложение без ссылок на другие статьи, где обосновывается коническое распространение стримерной вспышки на большие расстояния: "The system will expand as it advances, taking a *roughly* conical shape" [Phelps, 1974, p.110]. Заметим также, что в статьях Гриффитса-Фелпса максимальная скорость стримеров даже предположительно не превышает $8 \cdot 10^5$ м/с (нормальные, измеряемые скорости для распространения стримеров при давлениях 0.1-1 атм [Les Renardieres Group, 1972, 1977, 1981]), что сильно противоречит тому, что [Attanasio at al., 2019] распространили представления Гриффитса-Фелпса на скорости вплоть до $10^8$ м/с. Во-вторых, [Bazelyan & Raizer, 1998, pp.227-229; Milikh et al., 2016] убедительно показали, что головки стримеров влияют друг на друга и уменьшают потенциал друг друга, что ведет к оттоку заряда из головок стримеров [Milikh et al., 2016] и уменьшению их скорости и не позволит сформировать широкий плоский однородный слой головок, как на Рисунке В.33. Поэтому модель быстрого положительного пробоя (FPB), как гигантской ветвящейся стримерной вспышки, которая движется со скоростью $10^7$-$10^8$ м/с, трудно считать адекватной моделью, описывающей важные экспериментальные результаты, полученные [Rison et al., 2016].

Другую модель создания КВР предложили Наг и Раков [Nag and Rakov, 2010]. Они предположили, основываясь на замеченных колебаниях электрического поля и его производной, Рисунок В.34, что источником КВР является высокопроводящий плазменный канал длиной 100-1000 м, в котором возникают сильные электрические

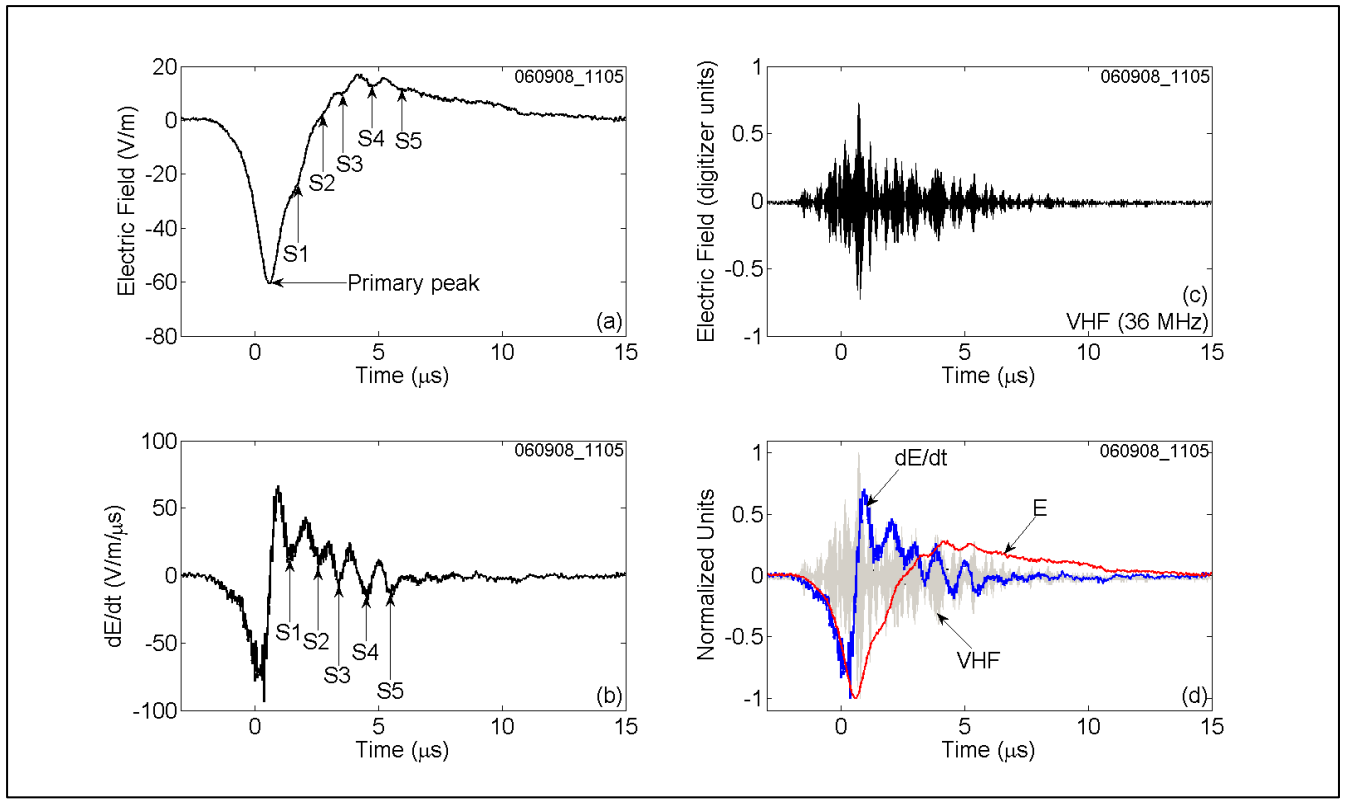

Рисунок В.34 (адаптировано из [Nag and Rakov, 2010]). (a) Вертикальные электрические поля, (b) dE/dt и (c) осциллограммы VHF-излучения КВР(CID), зарегистрированные в Гейнсвилле, штат Флорида. КВР возник на неизвестном расстоянии и продвинул вниз отрицательный заряд. Все три кривые из "a", "b", "c" панелей совмещены на панели (d) для прямого сравнения. S1–S5 — это пять вторичных пиков, проявляющихся в виде ярко выраженных колебаний на панели "b" и в основном в виде «плеч» на "a".

колебания (Рисунок В.35). Электромагнитное излучение этой своеобразной «антенны» и создает, по их мнению, сигнал электрического поля КВР, а излучение стримерных вспышек на концах плазменного канала создает VHF-сигнал КВР. Эта модель не давала ответа на вопрос: откуда взялся канал в грозовом облаке и почему он именно такой формы, поэтому [Nag and Rakov, 2010] назвали свою модель «инженерной». Кроме того, модель не объясняла экспоненциального роста VHF-сигнала на начальном этапе, так как при колебательном (периодическом) процессе в канале стримерные вспышки на концах канала должны также происходить периодически. Но подобная модель может иметь перспективы для исследования начальных импульсов пробоя (IBPs), так как этот процесс, как показали эксперименты, скорее всего является процессом взаимодействия плазменных каналов и/или сетей [Stolzenburg et al., 2013, 2014], [Campos & Saba, 2013].

Иудин и Давыденко предложили модель, основанную на взаимодействии (контакте, сквозной фазе) развивающихся плазменных сетей [Иудин и Давыденко, 2015], [Давыденко и Иудин, 2015]. Они учли неравномерность распределения электрических полей, обусловленную потоковыми неустойчивостями, что значительно увеличило возможности сравнения с экспериментом, но они опирались на предположение, что плазменные каналы (сети) предварительно существуют и КВР является моментом взаимодействия, своеобразной «сквозной фазой» и «обратным ударом» при контакте этих сетей (Рисунок В.36). Как показали эксперименты с помощью современного чувствительного оборудования (например, [Marshall et al., 2014a], [Rison et al., 2016], [Marshall et al., 2019]), КВР происходят в грозовом облаке в момент, когда еще не существуют длинные высокопроводящие каналы (или сети) и поэтому модель, скорее всего не описывает реальный КВР. Но, по нашему мнению, такая развитая модель, при некоторой корректировке, имеет значительные перспективы для описания подготовки (стадия IEC) и протекания начальных импульсов пробоя (IBPs), которые, согласно экспериментам, являются результатом контакта проводящих плазменных сетей или плазменных каналов [Stolzenburg et al., 2013, 2014], [Campos & Saba, 2013], см. также главу 7, Рисунки 7.5, 7.10, 7.13.

Недавно опубликован препринт [Liu et al., 2021] результаты которого определенно говорят в пользу моделей КВР, основанных на первичных стримерных вспышках [Rison et al., 2016], [Kostinskiy et al., 2020a], [Shao et al., 2020], а не на взаимодействии высоко

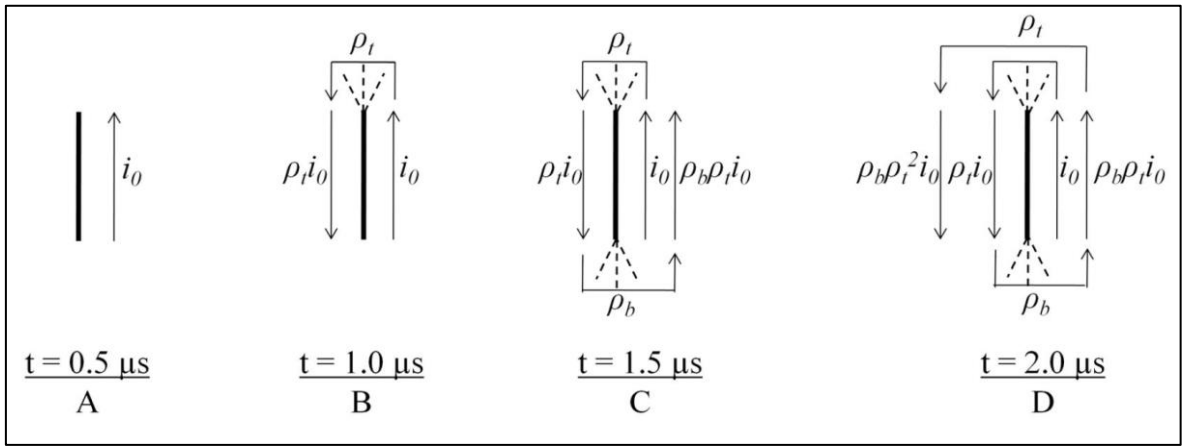

Рисунок В.35 (адаптировано из [Nag and Rakov, 2010]). Схематическое изображение механизма отражающей волны для КВР (CID) для случая длины канала h = 100 м и скорости распространения v = 2 × 10⁸ м/с. Продолжительность волны тока намного больше, чем время прохождения канала. Прямые стрелки представляют собой волны тока в канале CID, а стрелки в форме скобок представляют процесс отражения волн на концах. Ожидается, что отраженные волны тока уменьшат ток на каждом конце, в то время как соответствующее напряжение будет там увеличиваться. В результате на концах канала может произойти разряд, подобный вспышке стримерной короны (показан пунктирными линиями).

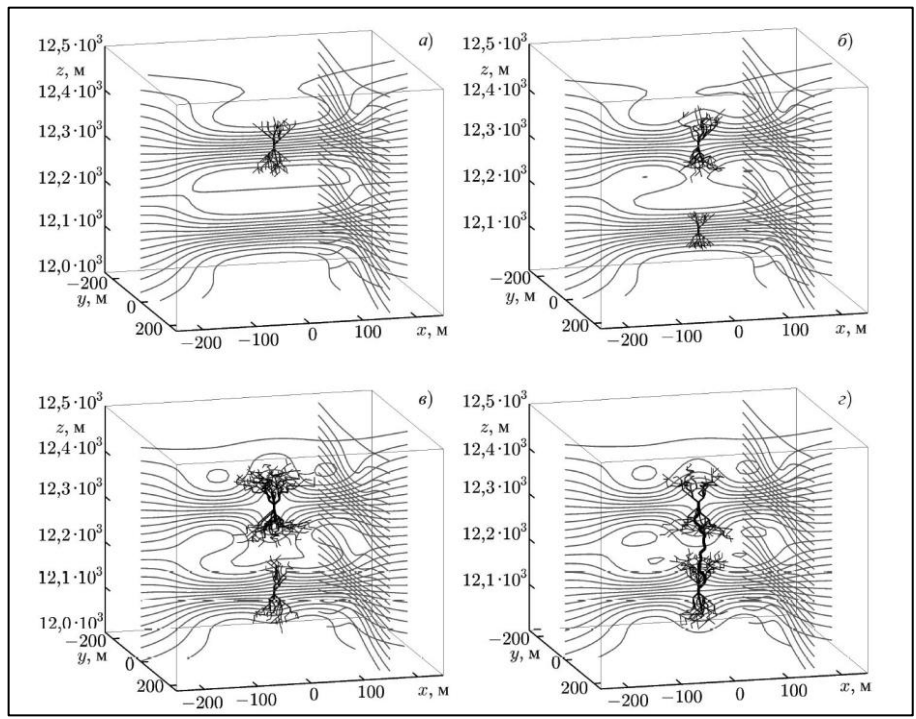

Рисунок В.36 (адаптировано из [Иудин и Давыденко, 2015]). Последовательные стадии развития КВР в среде с пространственно-неоднородным внешним электрическим полем: (а) развитие первого стримерного разряда за семь шагов модельного времени (t = 140 мкс); (б) начало развития второго разряда (t = 5,06 мс); (в) одновременное развитие пары разрядов (t = 7,50 мс); (г) момент электрического контакта разрядов через 556 шагов модельного времени (t = 11,12 мс) с формированием канала с мощным током. Серые линии соответствуют эквипотенциалям в плоскостях y = 0 и x = 200 м.

проводящих горячих плазменных каналов (любых типов «лидеров») или сетей, из чего исходили [Nag and Rakov, 2010], [Иудин и Давыденко, 2015], [Lyu, Cummer et al., 2019]. На Рисунке В.37 показаны линии излучения 337 нм и 777 нм. Если обе линии показывают сильные сигналы, то это, обычно означает, что записаны разряды молнии. Если же линия 337 нм гораздо сильнее, чем линия 777 нм, то с высокой вероятностью удалось записать большой стримерной разряд, что и было зафиксировано в данном случае. Осциллограммы, записывающие электромагнитные сигналы указывают на то, что с высокой вероятностью был записан КВР, одновременно зафиксированный на приборах ASIM на МКС.

Для изучения проблемы инициации молнии в последние три года стала применяться улучшенная интерферометрия с одновременным поляризационным анализом излучения источников VHF [Shao et al., 2018, 2020], что дает дополнительные важные возможности для идентификации свойств и поведения плазменных каналов. На Рисунке В.39 [Shao et al., 2018] отчётливо видны преимущества интерферометрии для анализа мощных линейных процессов типа K-событий, которые являются с высокой вероятностью контактом больших двунаправленных лидеров или плазменных систем, своеобразным аналогом обратного удара при контакте лидера молнии с наземными объектами. [Shao et al., 2018] четко фиксируют благодаря анализу поляризации (белые, перпендикулярные каналу отрезки на Рисунке В.39b), обратный разряд (обратную корону и, возможно, обратные лидеры, которые снабжают током основной канал, пользуясь депонированным ранее объемным зарядом чехлов каналов). С другой стороны для анализа множества параллельных процессов внутри одного временного окна интерферометра, данная методика подходит не очень хорошо и может приводить к неполной картине событий и неверным интерпретациям экспериментальных данных. Этот факт особенно явно себя проявляет при попытке анализировать момент инициации молнии компактным внутриоблачным разрядом (КВР/NBE).

Мы подробно рассмотрим статьи [Shao et al., 2018, 2020], так как гипотеза создания КВР, предложенная в [Shao et al., 2020], является альтернативной гипотезой по отношению к изложенной в главах 7, 8 [Kostinskiy et al., 2020a,b]. [Shao et al., 2020] под непосредственным влиянием препринта [Kostinskiy et al., 2019a] (который был первым вариантом и послужил основой для статьи [Kostinskiy et al., 2020a], а также докладов на двух представительных международных конференциях 2019 года по этой

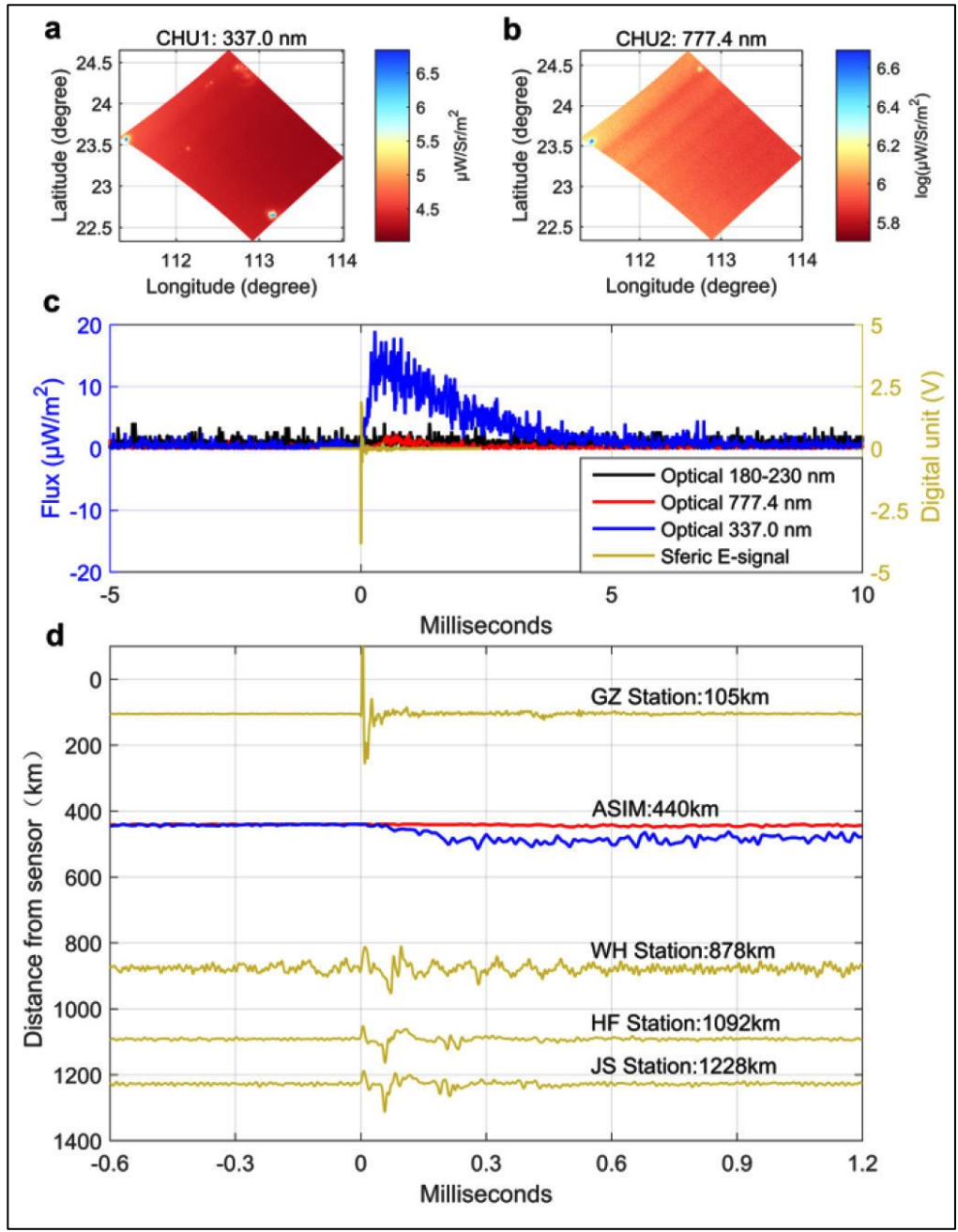

Рисунок В.37 (адаптировано из [Liu et al., 2021]). Оптическая и электрическая фиксация положительного NBE (KBP). (a) изображение, полученное камерой ASIM с фильтром 337 нм; (b) изображение камеры с фильтром 777,4 нм; (c) Сравнение излучения (синий, красный, черный каналы), обнаруженного MMIA, и электромагнитных сигналов (горчичный), зарегистрированных станцией, обозначенной GZ(105 км), t = 0 мс соответствует 07 августа 2019 г., 1306: 02.691042 UTC; (d) Подробная информация об оптическом сигнале и форме электромагнитной волны на четырех станциях. Осциллограммы электромагнитных волн соответствует горчичному цвету (в зависимости от расстояния до события). Сигналы на линии 337 нм и 777,4 нм показаны синими и красными линиями соответственно, полоса 180-230 нм показано черным.

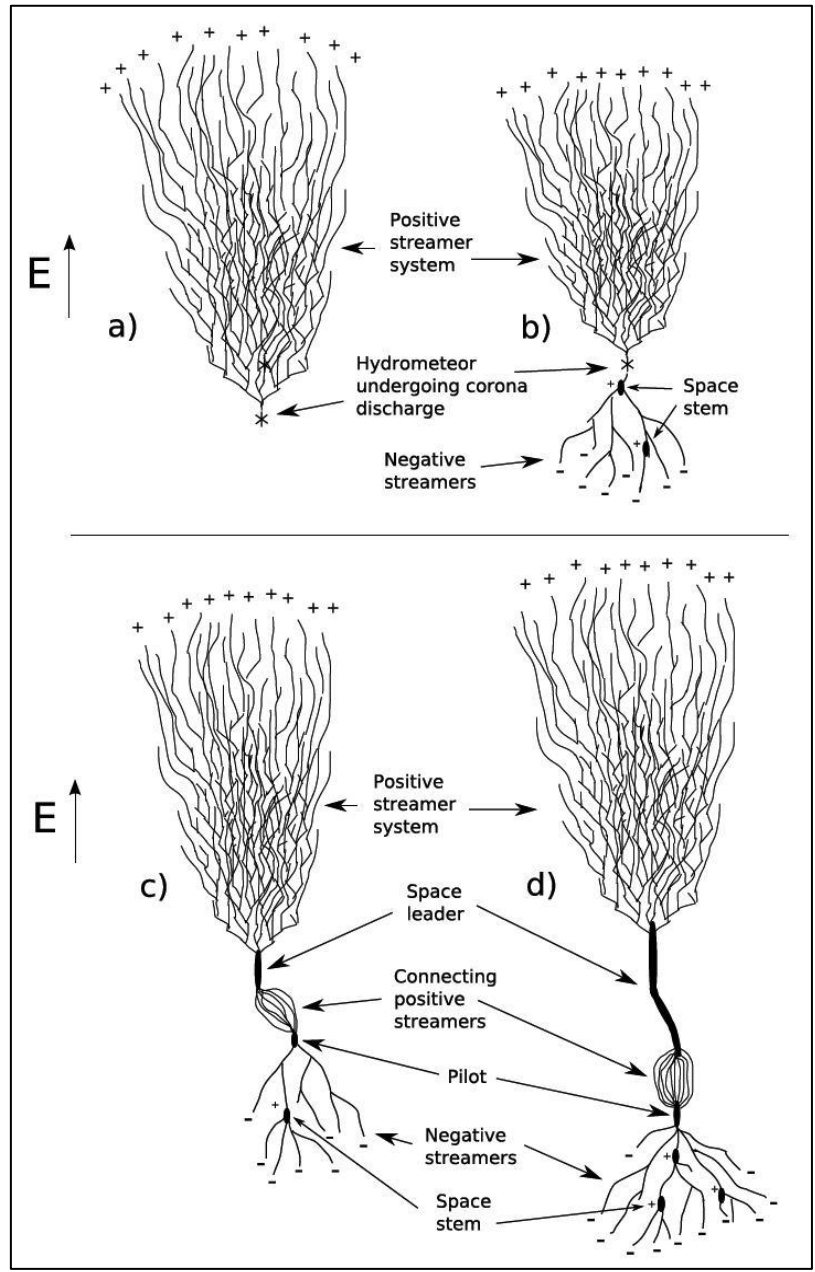

Рисунок В.38 (адаптировано из [Petersen et al., 2008]). Возможная схема инициации лидера молнии по [Petersen et al., 2008]. (а) в области грозового облака с интенсивным электрическим полем корона на гидрометеоре генерирует затравочный (seed) положительный стример. Положительный стример превращается в усиливающую и разветвляющую систему положительных стримеров, дополнительно усиливая электрическое поле вблизи источника стримерной системы (согласно гипотезе [Griffiths and Phelps, 1976] – А.К.); (b) Локальная напряженность электрического поля теперь достаточна для создания как положительных, так и отрицательных стримеров на ближайших гидрометеорах. Положительные стримеры развиваются в усиливающиеся и разветвляющейся системе положительных стримеров ([Griffiths and Phelps, 1976] — А.К.), в то время как отрицательные стримеры развиваются меньше, но создают спейс-стемы; (с) «пилотные разряды» (pilot discharges, по-видимому, авторы имеют ввиду спейс-стем с положительными и отрицательными лидерами на его концах или начальный (только что возникший) спейс-лидер, что является устаревшей концепцией [Schonland, 1956] — А.К.)  инициируются на спейс-стемах с последовательным соединением спейс-стемов, позволяя внутренним стемам нагреваться и развиваться в спейс-лидеры. Стримерная активность продолжается на концах спейс-стем системы; (d) продолжительное нагревание и удлинение спейс-лидеров позволяет им быстро удлиняться и вступать в контакт с более крупными спейс-лидерной системой. Уменьшение градиента потенциала внутри канала спейс-лидера еще больше усиливает электрическое поле на концах спейс-лидера. Возникающее в результате образование стримеров на концах лидеров приводит к созданию новых спейс-стемов на отрицательном конце.

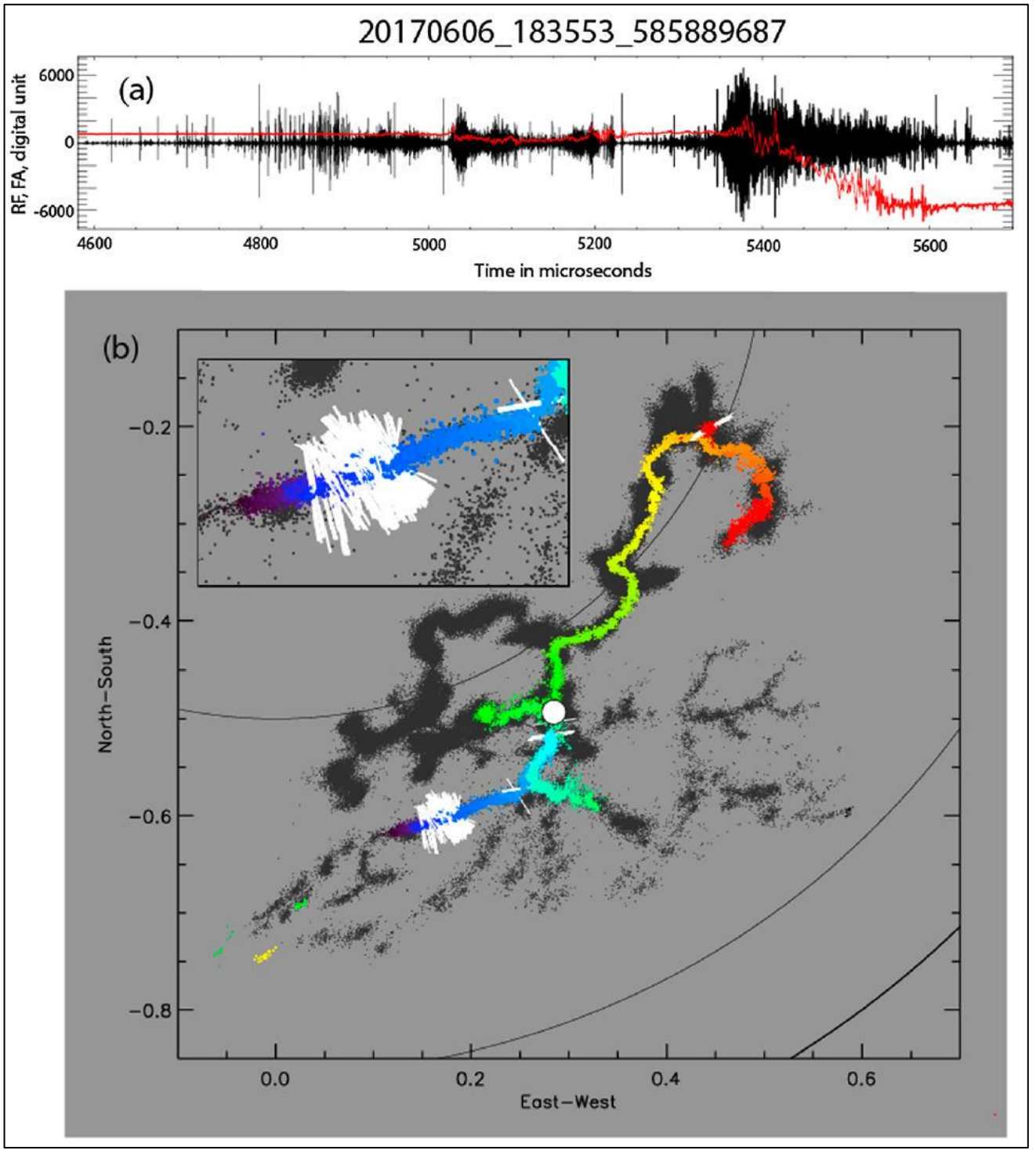

Рисунок В.39 (адаптировано из [Shao et al., 2018]). К-событие при развитии лидера, произошедшее через 229 мс после начала внутриоблачной вспышки. (a) Осциллограммы зависимости от времени сигналов в диапазоне VHF (обозначены — RF) и сигналов, измеряющих электрическое поле быстрой антенной (FA); (b) карта протекания К-события с учетом распространения волны потенциала по длине канала (цветные точки — развитие событий во времени маркировано от фиолетового к красному) и общая структура (черные точки) развития IC-молнии до начала К-события. Белые отрезки показывают ориентацию поляризации, когда обнаруживается, что соответствующие источники поляризованы. Белый кружок в центре обозначает область возникновения молнии.

тематике [Kostinskiy et al., 2019b,c]), не сославшись на препринт и доклады, также предположили, что КВР (как множество объемных стримерных вспышек) инициируется широким атмосферным ливнем космических лучей (ШАЛ). Причем [Shao et al., 2020] использовали ту же метафору для процесса инициации стримерных вспышек («поджиг») — «ignite» («Detailed analysis suggests that lightning-initiating process was *ignited* by a cosmic ray shower» [Shao et al., 2020]), который был использован в препринте [Kostinskiy et al., 2019a]: «The avalanches of runaway electrons act as an initiation wave moving at a speed close to the speed of light, which creates, at different points in the EE-volume, in 1.5–3 µs, a "giant wave" of ordinary positive streamer flashes, which we described above (III.1). It can be said that a flash of a large number of runaway electrons *"ignites"* almost simultaneously many of the $E_{th}$-volumes (see Fig. 6.II(6)), thereby forming the NBE radiation front in the VHF». Как отмечалось выше, [Shao et al., 2020] использовали, быть может, лучшую из имеющихся в настоящее время широкополосных интерферометрических систем, которая позволяла в одном временном окне получать сразу несколько центроидов (точек), а также могла анализировать поляризацию сигналов центроидов (VHF-источников) [Shao et al., 2018, 2020]. Рисунок В.40 иллюстрирует, по мнению [Shao et al., 2020], насколько новая обработка сигналов «интерферометрия с управляемым лучом» (beam steering interferometry) улучшила изображение, уменьшив число шумовых точек. К сожалению, несмотря на довольно подробное описание интерферометрических методик (например, [Shao and Krehbiel, 1996], [Stock et al., 2014], [Shao et al., 2018, 2020]), при анализе полученных конкретных интерференционных карт со «слабыми» событиями (не намного превосходящими уровень шумов) нельзя с уверенностью сказать является ли точка на карте действительным VHF-событием или артефактом обработки сигнала (особенно, если много сигналов излучаются одновременно, что признается самими [Shao et al., 2020, Appendix A]). На временных картинках (Рисунок В.40 b-g), особенно с улучшенной обработкой (d-g) отчетливо видны колебания центроидов во времени, как будто они «движутся» вверх-вниз, дрейфуя в среднем вниз, как указано наклонной черной линией на панели (c). Абсолютные значения размеров на панели (i) определены с большой погрешностью, так как [Shao et al., 2020] не использовали в этих экспериментах LMA-систему, которая могла бы предоставить абсолютные трехмерные координаты. Поэтому размеры брались ими из предположения, что эта IC-молния находится на примерно средней высоте и примерно среднем удалении от измерительной системы.

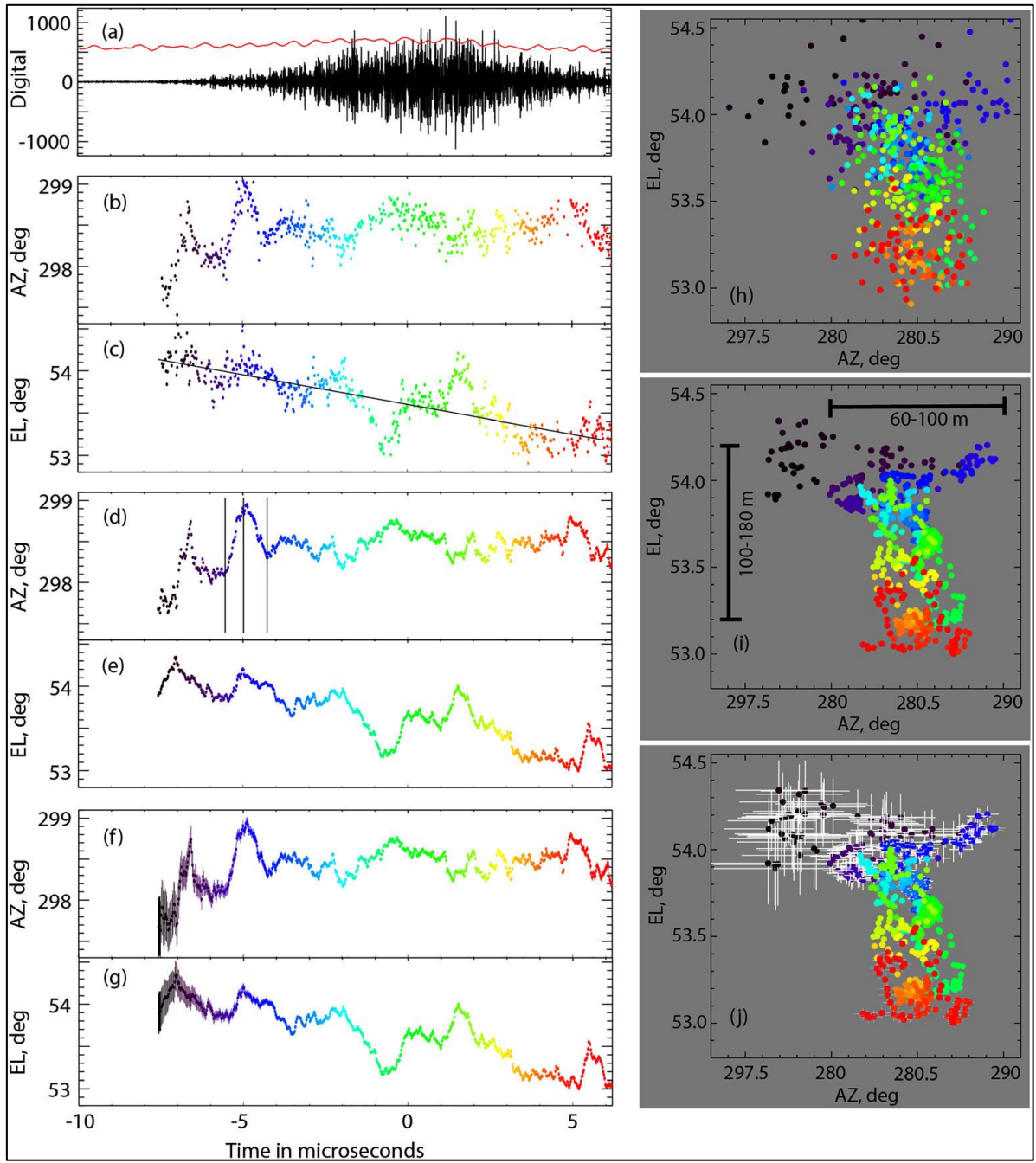

Рисунок В.40 (адаптировано из [Shao et al., 2020] и описывает то же событие, что и на Рисунках В.41, В.43). Первые 14 мкс быстрого положительного пробоя (FPB), который инициирует обычную внутриоблачную молнию (IC-молнию). (a) ЗависимостьVHF-сигнала от времени (черная осциллограмма); (b, c) Зависящие от времени угловые координаты AZ, EL при обычном построении интерференционных карт; (d, e) Зависящая от времени угловые координаты AZ, EL при построении интерференционных карт по методу, предложенному в этой статье [Shao et al., 2020] — интерферометрия с управляемым лучом (beam steering interferometry); (f, g) то же, что (d) и (e), но с указанием погрешностей; (h) График в угловых координатах AZ-EL для «обычной» интерферометрии; (i) График в угловых координатах AZ-EL для интерферометрии с управляемым лучом; (j) То же, что (i), но с указанием погрешностей.

На Рисунке В.41 показано «покадровое» развитие в пространстве события, показанного на Рисунке В.40. Красные центроиды (точки) на каждой панели являются активными в данном временном окне. [Shao et al., 2020] предполагают, что на первых панелях (a-c) наблюдалось движение из левого угла в правый и вниз, как показывают верхние стрелки на панели "о". На панели "c" центроиды напоминают дугу и [Shao et al., 2020] предположили (это исходная экспериментальная информация для главной физической идеи статьи), что это ШАЛ инициирует быстрый положительный пробой (FPB), причем сложной формы, совершающий колебания (Рисунок В.40), и разделенный на части в пространстве (пунктирная линия на панелях «е-о». «Движение» вверх (при колебаниях центроидов) они называют «быстрым отрицательным пробоем» вслед за довольно спорной концепцией отрицательного пробоя, предложенного в статье [Tilles et al., 2019], так как предполагается, что этот пробой происходит благодаря отрицательным стримерам в отсутствие плазменных каналов и положительных стримеров (которым для поддержания движения нужно электрическое поле в два раза меньшее, чем для отрицательных и непонятно почему раньше, до появления отрицательных стримеров, не возникнут положительные стримеры и не разрядят облако). [Shao et al., 2020] хорошо понимают и подробно обсуждают в статье вопрос о том, как интерпретировать с физической точки зрения динамику центроидов на карте Рисунка В.41, полученных методом математической обработки осциллограмм. Они понимают, что данная методика не позволяет построить физические траектории радиоисточников, если внутри временного окна происходит много одновременных событий, а будет строить некоторые средние или даже случайные центроиды суммарных сигналов, или даже артефакты, если структура событий имеет слишком сложную форму [Shao et al., 2020, Appendix A]. И отличить артефакты от реальных сигналов может быть крайне сложно без калибровки на известных плазменных событиях. Это неоднозначность фиксирования некоторых центроидов с одной стороны приводит [Shao et al., 2020] к довольно запутанным объяснениям «траекторий» движений центроидов, чтобы объяснить их неупорядоченность, с другой приводит к утверждению, что «дуга» центроидов на Рисунке В.41, панели "c" свидетельствует именно об инициации FPB благодаря ШАЛ. Из рассуждений авторов нельзя понять, кроме констатации внешнего сходства «дуги» центроидов с «блином» вторичных частиц ШАЛ, почему они считают, что именно в этом месте произошла инициация положительных стримеров (FPB), ведь вторичные

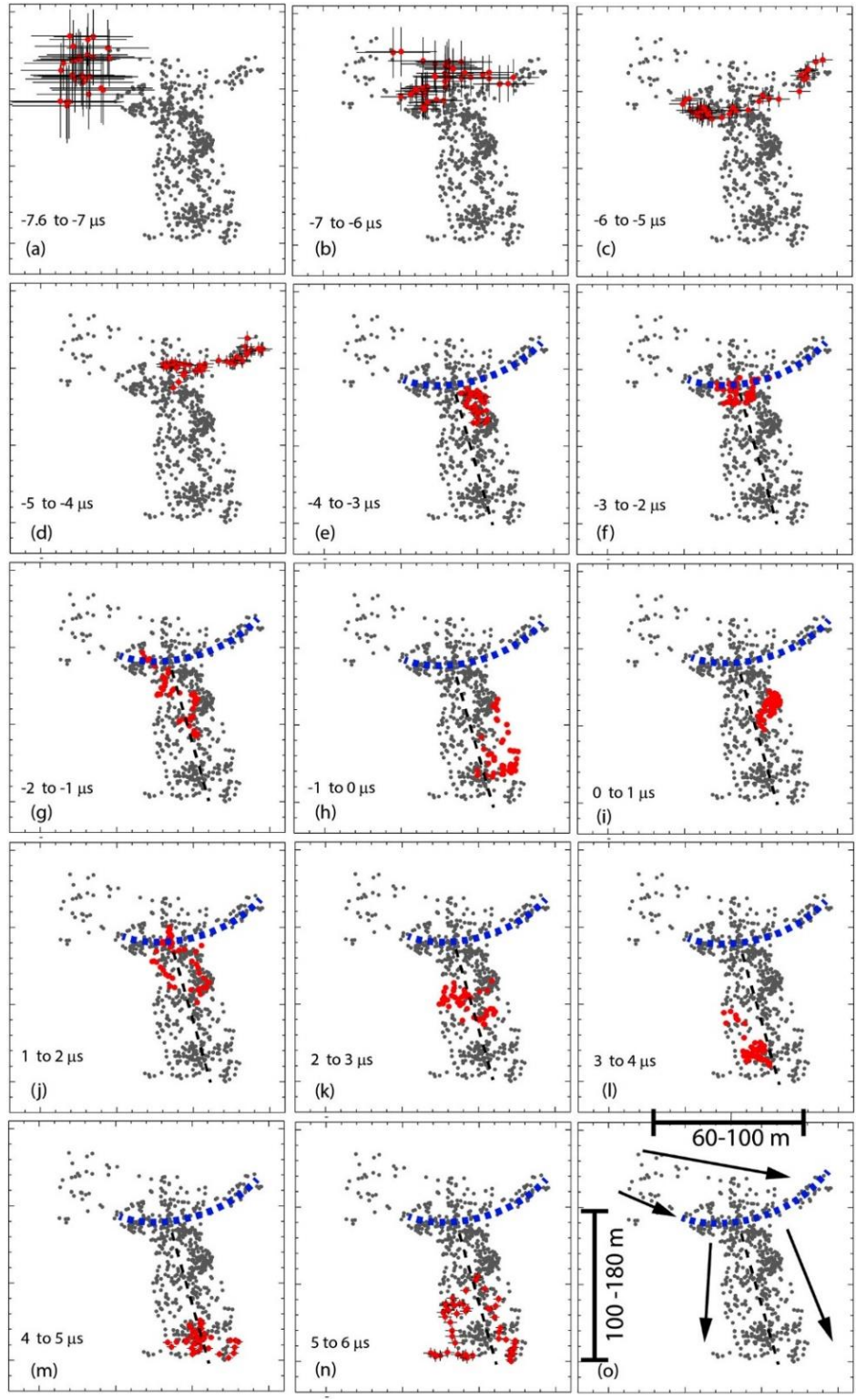

Рисунок В.41 (адаптировано из [Shao et al., 2020] и описывает то же событие, что и на Рисунках В.40, В.43). (а – n) Эволюция во времени («кадр за кадром») источников-VHF, изображенных также на Рисунке В.40. Все панели имеют временное окно 1 мкс. Фоновые серые точки обозначают все события, которые произошли за это время и задают общую структуру распределения центроидов. Красные центроиды на каждой панели — это VHF-источники, активные в диапазоне окна 1 мкс. Перекрестия на центроидах задают размер погрешностей. Синяя пунктирная кривая показывает дугообразную особенность, появившуюся на панели "c". Пунктирная черная линия разделяет источники FPB на левую и правую области; (o) эта панель показывает общее направление развития процесса.

электроны ШАЛ проходят 300 м за одну мкс и весь этот объем пройдут за 500-700 нс. В любом случае, основываясь на этих данных, [Shao et al., 2020] предлагают новую модель инициации быстрого положительного пробоя (FPB), а тем самым и инициации молнии. Это своеобразная модификация модели [Gurevich et al., 1999]. [Gurevich et al., 1999] предполагали, как мы отмечали выше, что относительно большая концентрация вторичных электронов около ствола ШАЛ приведет к появлению стримера. [Shao et al., 2020] предположили, что причиной инициации молнии является усиление электрического поля на фронте ШАЛ в районе ствола (максимума вторичных электронов) из-за поляризации вторичных электронов и ионов в центре «блина» ШАЛ в электрическом поле грозового облака (Рисунок В.42, серый эллипс в центре фронта ШАЛ). [Shao et al., 2020] используют для расчета усиления поля на фронте ШАЛ формулу проводящей металлической сферы, где $\sigma_e$ и $\sigma_0$ являются проводимостями области заполненной вторичными электронами ШАЛ и воздуха соответственно, при этом поле на поверхности такой «сферы» усиливается в три раза:

$$E = E_0 + 2E_0 \frac{\sigma_e - \sigma_0}{\sigma_e + 2\sigma_0}$$  (8),   номер уравнения в [Shao et al., 2020].

Это предположение не является корректным по нескольким причинам и не может применяться к шару, а точнее слабо ионизованному слою, ШАЛ, так как:

{1} этот слой является слоем слабоионизованного газа и не является плазмой, к которой применимо уравнение (8), так как плазмой является ионизованный воздух, в котором выполняется условие Дебая (Raizer, 1991, стр.31). Оценим радиус Дебая для случая «блина» ШАЛ. Численный расчет числа вторичных электронов в ядре ШАЛ с помощью метода Монте-Карло для первичной космической частицы с энергией $10^{16}$ эВ дает приблизительно 3 частицы на 1 $cm^2$ (Figure 4A, [Rutjes et al., 2019]). Каждая частица на пути в 1 см на высоте 8 км производит около 40 тепловых электронов. Таким образом, в каждом кубическом сантиметре в ядре ШАЛ будет находится около 120 электронов в 1 $cm^3$). Тепловые электроны рождаются со средней энергией 24 эВ (Figure 5, [Rutjes et al., 2019]). Радиус Дебая для слоя при этом будет чрезвычайно большим, а это означает, что слой не является плазмой.

$$d[cm] = \left(\frac{kT_e}{4\pi e^2 n_e}\right)^{0.5} = 743 \left(\frac{T_e[eV]}{n_e[cm^{-3}]}\right)^{0.5} \approx 332 \; cm.$$

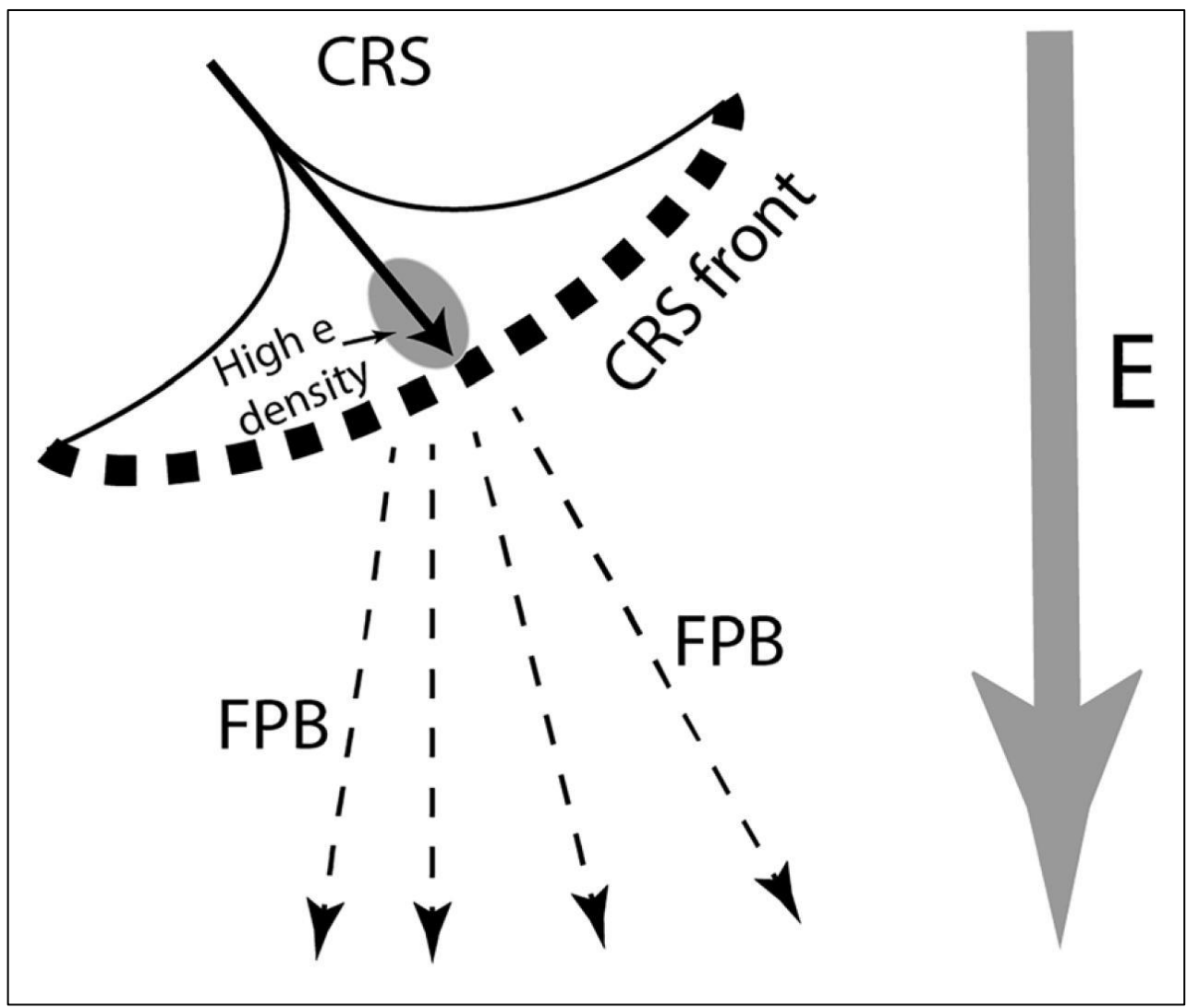

Рисунок В.42 (адаптировано из [Shao et al., 2020]). Иллюстрация гипотетической концепции инициации быстрого положительного пробоя FPB («зажигания» благодаря ШАЛ (CRS)), которая является инициирующим событием (IE) для IC-молнии. Фронт ШАЛ (CRS front) обозначен черными квадратиками, область поляризации электронов показана серым эллипсом, FPB обозначено пунктирными стрелками, электрическое поле E обозначено жирной серой стрелкой.

[Shao et al., 2020] неоправданно считали, что ионизованный газ «блина» ШАЛ ведет себя так же, как плазма канала обратного молнии, но для канала молнии характерна концентрация электронов $10^{17}$-$10^{18}$ см$^{-3}$, что дает радиус Дебая около 1-4 мкм. Именно такие небольшие радиусы Дебая позволяют получить разделение зарядов и обеспечить скорость движения волны потенциала, приближающуюся к скорости света;

{2} для того, чтобы выполнялась формула (8), необходимо также, чтобы плазма успела поляризоваться за время существования свободных электронов ($\sim 10^{-7} s$), то есть, пока электроны не прилипли к молекулам кислорода. Время поляризации определяется максвелловским временем $\tau_M \approx \frac{\varepsilon_0}{\sigma_e}$, ($\varepsilon_0 = 8.854 \cdot 10^{-12} [F \cdot m^{-1}]$, $\sigma_e [S \cdot m^{-1}] = 2.82 \cdot 10^{-2} n_e [cm^{-3}]/(3 \cdot 10^9 p[Torr])$). Для концентрации электронов около 120 см$^{-3}$ и давления на высоте 8 км равном 280 Торр максвелловское время, как легко видеть, будет равно $\tau_M \approx 2\ s$. Если мы даже возьмем неоправданно высокую для этого слоя ШАЛ проводимость (8.6·10$^{-11}$ S/m), которую рассчитали [Shao et al., 2020, p.18], то все равно получается чрезвычайно большое время поляризации ионизованного газа $\tau_M \approx 0.1\ s$, которое отличается в миллион раз от времени существования свободных электронов, что не позволит использовать формулу (8). То есть, для усиления поля на сфере по формуле (8) необходимо, чтобы электроны успели перераспределиться, пока они не прилипли, но для этого время поляризации должно быть в $10^6$ раз меньше. Максвелловское время для канала молнии при концентрации электронов $10^{17}$ см$^{-3}$ будет равно $\tau_M \approx 2.6 \cdot 10^{-15}\ s$. Поэтому аналогия ионизованного слоя вторичных электронов ШАЛ с плазмой канала молнии некорректна на много порядков величин и эффект усиления поля перед фронтом ШАЛ будет чрезвычайно маленьким и никогда не позволит «поджечь» быстрый положительный (стримерный) пробой (FPB) или любой другой разряд;

{3} можно, наконец, оценить сверху какое реально максимальное усиление электрического поля возможно в этом случае, если мы примем все предположения [Shao et al., 2020]. То есть, заменим, как делают [Shao et al., 2020], реальное распределение электронов ШАЛ Нишимуры-Каматы-Грейзена (НКГ) [Kamata & Nishimura, 1958] на шар с радиусом 5 метров, где сосредоточены все низкоэнергетические электроны ионизованного слоя блина ШАЛ ($n_e \cong 6 \cdot 10^{12}$). Ионов в этом шаре будет точно такое же число $n_i \cong 6 \cdot 10^{12}$. Благодаря подвижности электронов за $\tau = 60\ ns$, которые электроны существуют, они сдвинутся на $r = \mu E \tau = v \tau$. Возьмем высокую скорость дрейфа электронов, которая возможна только при пробойных полях $v = 3 \cdot 10^5\ m\ s^{-1}$. Таким образом электроны сдвинуться на $r = v\tau = 1.8 \cdot 10^{-2} m = 18 \; mm$. Усиление электрического поля, которое обеспечит такой дипольный шар вторичных электронов ШАЛ будет равна

$$\Delta E \cong 9 \cdot 10^9 \frac{2 \cdot r \cdot Q}{R^3} = \frac{2 \cdot 9 \cdot 10^9 \cdot 1.8 \cdot 10^{-2} \cdot 1.6 \cdot 10^{-19} \cdot 6 \cdot 10^{12}}{5^3} \approx 2.5 \; V \; m^{-1}.$$

В реальности дополнительное электрическое поле $\Delta E$ будет еще меньше, так как концентрация энергичных электронов (и, следовательно, тепловых электронов) падает плавно согласно НКГ $\propto R^{-2}$ и профиль «блина» ШАЛ является не шаром радиусом 5 метров, а поверхностью, которая имеет кривизну радиусом в десятки метров. Скорость дрейфа электронов в реальных полях облака также будет на порядок ниже. Эти простые оценки показывают, что тепловые электроны «блина» ШАЛ даже в пробойном электрическом поле не смогут увеличить электрическое поле даже на несколько $V \; m^{-1}$. Поэтому гипотеза [Shao et al., 2020] о том, что ШАЛ в электрическом поле грозового облака может инициировать стримерную вспышку, на наш взгляд, является необоснованной, но то, что [Shao et al., 2020], с помощью интерференционных методов обнаруживают множество источников VHF-излучения в объеме во время инициации компактным внутриоблачным разрядом (КВР) IC-молнии, говорит о том, что их эксперименты также не укладываются в простую модель инициации двунаправленного лидера, который потом разовьется в канал длиной в несколько километров. Кроме того, анализ «движения» центроидов и поляризации их излучения во время «движения» (Рисунки B.40b-g, B.43g,h,i [Shao et al., 2020]) можно истолковать как излучение стримерных вспышек и/или плазменных каналов, инициированных преимущественно в направлении локального электрического поля грозового облака.

Таким образом, физику КВР крайне сложно описать с помощью двунаправленного лидера Каземира или взаимодействия плазменных сетей, так как КВР слабо излучает оптический сигнал по сравнению с молнией и начальными импульсами пробоя (IBPs) и непонятна причина создания такого короткого, но мощно излучающего плазменного канала. Поэтому в настоящее время, в попытке объяснить новые экспериментальные данные, появились по крайней мере три новых подхода, основанных на представлении об объемном протекания процесса инициации молнии. Но, первая гипотеза, являющаяся представлением об инициации КВР, как о гигантской стримерной вспышке Гриффитса-Фелпса ([Rison et al., 2016], [Attanasio et al., 2019]) с кило амперными токами, зарядами

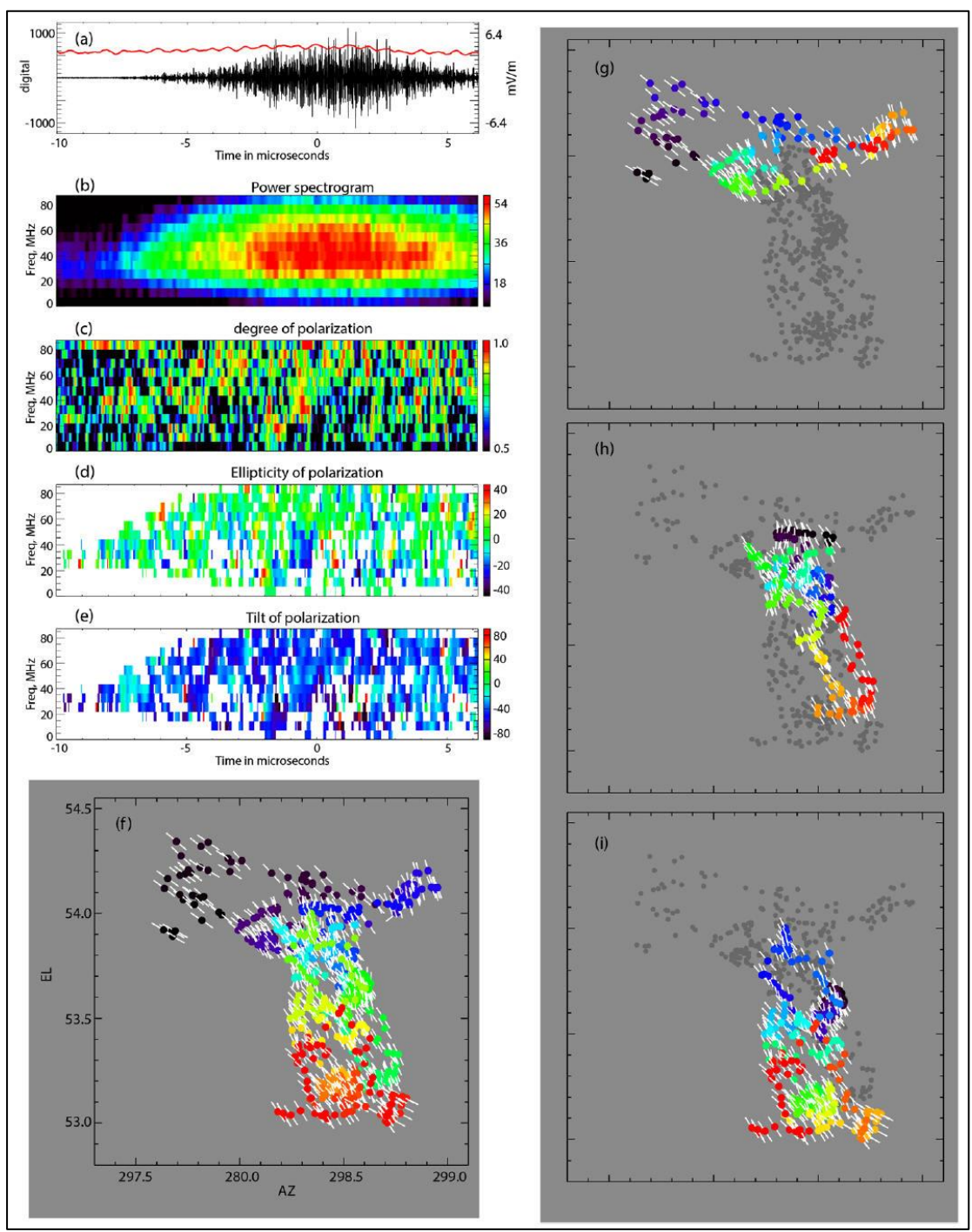

Рисунок В.43 (адаптировано из [Shao et al., 2020] и описывает то же событие, что и на Рисунках В.40, В.41). Наблюдение поляризационной картины излучения VHF-источников (центроидов) во время протекания КВР. (a) Временная форма VHF-сигнала (черная осциллограмма), изменения электрического поля E (красная осциллограмма); (б) энергетическая спектрограмма; (c) степень поляризации: от 0,5 до 1,0; (d) эллиптичность поляризации на открытый угол от -45° до + 45°; (e) ориентации поляризации относительно направления EW, от -90° до +90°; (f) общая карта всего события с ориентацией поляризации (белые отрезки); (g – i) разделение карты "f" на три последовательных интервала времени "g", "h", "i". Развитие во времени на каждой из панелей "f", "g", "h", "i" маркировано цветом от черного к красному. Серые точки показывают уже случившиеся VHF-события (неактивные центроиды).

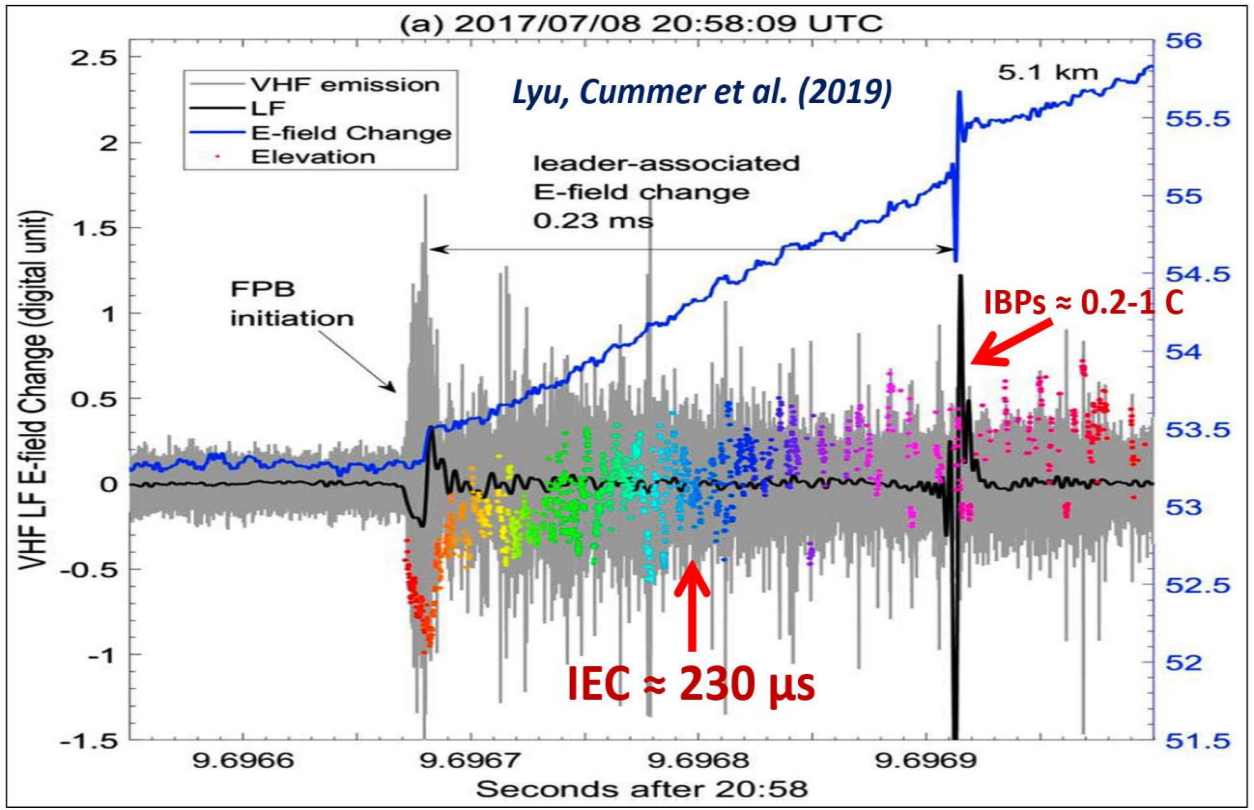

Рисунок В.44 (адаптировано из [Lyu, Cummer et al., 2019]). Пример изменения электрического поля (синяя кривая), связанного с сильным инициирующим событием на VHF-частоте, FPB-initiation (в привычной терминологии — КВР/CID/NBE – А.К.). VHF-излучение (серая кривая), низкочастотное излучение (LF), изменение магнитного поля (черная кривая), осциллограмма изменения электрического поля (синяя кривая) и высота источников VHF-излучения (центроидов, цветные точки) процесса начального пробоя (IB). Рисунок показывает, что «начальный лидер» (IL) IC-молнии начался с относительно более слабого события быстрого положительного пробоя (FPB, жирная синяя стрелка). [Lyu, Cummer et al., 2019] считают, весь показанный на данном рисунке процесс инициации молнии движением предполагаемого «начального лидера» (initial leader — IL) и, по их интерпретации, импульсы начального пробоя (IBPs) являются гигантскими ступенями длиной 200-600 м этого IL. Начинается этот IL с быстрого положительного пробоя (FPB initiation), концепцию которого предложили [Rison et al., 2016], и развили [Attanasio et al., 2019], как гигантскую положительную стримерную вспышку по модели Гриффитса-Фелпса (критика этой концепции представлена в разделе на 56 стр.). Обращает на себя внимание исключительно короткая IEC-стадия развития молнии (около 230 мкс), после которой следует мощный IBP (с зарядом не менее 0.2 Кл), а также рассеянные по всему объему центроиды (точки) VHF-источников, что никак не напоминает карты движения в облаке положительных (Рисунки В6, В.7) или отрицательных лидеров (Рисунок В.4). Недавно опубликован препринт [Liu et al., 2021] результаты которого определенно говорят в пользу моделей КВР (FPB – в терминологии [Lyu, Cummer et al., 2019]), основанных на стримерных вспышках [Rison et al., 2016], [Kostinskiy et al., 2020a], [Shao et al., 2020], а не на взаимодействии высоко проводящих плазменных каналов (любых типов «лидеров») или сетей, из чего исходили [Nag and Rakov, 2010], [Иудин и Давыденко, 2015], [Lyu, Cummer et al., 2019]. Тем не менее, в научном сообществе, изучающем молнию принято любые горячие плазменные образования и процессы в облаке называть лидерами, что на наш взгляд является не очень продуктивным и требует разработки более четкой терминологии при определении плазменных объектов, участвующих в процессе инициации и развития молнии (см. раздел 7.6).

порядка кулона и скоростями около $10^7$-$10^8$ м/с не укладывается в современные представления о физике газового разряда и физике длинной искры. Такой короткий (около 30 мкс) и мощный процесс также трудно объяснить, если не привлекать к его объяснению представления основанные на параллельно инициированных и протекающих плазменных процессах. Другая идея, основанная на инициации объемного стримерного разряда из-за поляризационного усиления электрического поля в слое тепловых электронов ШАЛ [Shao et al., 2020], также не может объяснить данное явление. Поэтому мы считаем полезным развивать подход, изложенный в данной диссертации (главы 7, 8 [Kostinskiy et al., 2020 a, b]), где сделана попытка создать, на основе экспериментальных данных, новую модель инициации молнии, которая принципиально отличается от модели двунаправленного лидера, но использует двунаправленные лидеры, как элементы построения последовательной цепочки преобразований плазмы, ведущих к инициации молнии, как большого ступенчатого лидера.

## Цели работы

1. Изучить возможные формы плазменных образований внутри искусственных, заряженных облаков водного аэрозоля, а также возможные механизмы их инициации.

2. Предложить качественный механизм инициации молнии в грозовых облаках от инициирующего молнию события (IE) через начальную стадию изменения электрического поля (IEC), до первых начальных импульсов пробоя (IBPs), которые переходят в ступенчатый отрицательный лидер (в отсутствие в облаке или рядом с ним протяженных проводящих объектов).

3. Предложить механизм инициации компактных внутриоблачных разрядов (KBP/CID/NBE).

4. Предложить последовательную цепочку переходов плазмы из одного состояния в другое на протяжении всего процесса инициации молнии (от первых газоразрядных лавин, до сквозной фазы взаимодействия плазменных двунаправленных лидеров и/или сетей во время протекания IBPs-стадии развития молнии).

## Задачи работы

1. Обнаружить ключевые недостающие звенья в цепи преобразований (переходов) плазмы из одного состояния в другое на начальной стадии процесса инициации горячих проводящих плазменных каналов внутри искусственных положительно и отрицательно заряженных аэрозольных облаков.

2. Выяснить механизм инициации горячих высокопроводящих плазменных каналов внутри искусственных положительно и отрицательно заряженных аэрозольных облаков.

3. Исследовать механизм квази-сквозной фазы и фазы квази-обратного удара при взаимодействии (контакте) плазменных каналов внутри искусственных положительно и отрицательно заряженных водных аэрозольных облаков.

4. Исследовать структуру двунаправленных лидеров, инициированных внутри искусственных положительно и отрицательно заряженных водных аэрозольных облаков.

5. Исследовать инициацию плазменных образований при движении протяженного проводника около и внутри облака положительно и отрицательно заряженного водного аэрозоля.

6. Исследовать процесс образования ступеней положительного и отрицательного лидера длинной искры применительно к моделированию ступенчатого развития молнии.

7. Построить на основе результатов, полученных в пунктах 1-6, качественный механизм инициации молнии в грозовых облаках от инициирующего молнию события (IE) до первых начальных импульсов пробоя (IBPs), которые переходят в ступенчатый отрицательный лидер (в отсутствие в облаке или рядом с ним протяженных проводящих объектов).

8. Построить механизм инициации компактных внутриоблачных разрядов (KBP/CID/NBE).

## Научная новизна работы

1. Внутри положительно и отрицательно заряженных искусственных водных аэрозольных облаков нами обнаружен новый класс электрических разрядов, которые мы с соавторами назвали «необычные плазменные образования» (unusual plasma formations — UPFs). UPFs представляют из себя сети плазменных каналов, размеров 10-30 см, некоторые из которых нагреты настолько, насколько нагрет восходящий положительный лидер.

2. Найден по крайней мере один механизм инициации UPFs на границе отрицательно заряженного аэрозольного облака. UPFs инициируется внутри восходящей положительной стримерной вспышки до появления других плазменных образований в электрическом поле аэрозольного облака.

3. Впервые представлены два кадра сквозной фазы контакта лидеров, показывающие значительное разветвление лидеров внутри общей стримерной зоны.

4. Впервые показано, что яркость инфракрасного излучения в области контакта восходящего и нисходящих лидеров примерно в 5 раз выше, чем для участков образовавшегося единого канала ниже или выше этой области.

5. Впервые, с помощью ИК-камеры исследована положительная часть (положительный лидер) двунаправленного лидера, находящийся внутри отрицательно заряженного облака водного аэрозоля.

6. Впервые исследован процесс инициации плазменных образований (стримерных вспышек и лидеров) при движении протяженного проводника около и внутри искусственного облака положительно и отрицательно заряженного водного аэрозоля.

7. Впервые во время образования ступеней положительного лидера длинной искры перед ним в зоне стримерной короны обнаружено плазменное образование, сходное со спейс-стемом или спейс-лидером, которые наблюдаются в стримерной зоне отрицательного лидера. Впервые показана морфологическое сходство ступеней положительного и отрицательного лидеров длинной искры.

8. Впервые предложен последовательный качественный механизм инициации молнии в грозовых облаках от инициирующего молнию события (IE) через начальную стадию увеличения электрического поля (IEC), до первых начальных импульсов пробоя (IBPs), которые переходят в ступенчатый отрицательный лидер.

9. Впервые предложен механизм инициации компактных внутриоблачных разрядов (КВР/CID/NBE), который дает непротиворечивое объяснение спектру излучения КВР, характерному для стримерных вспышек, и скорости распространения источников радиоизлучения внутри грозового облака близкой к скорости света. В предложенном механизме ключевую роль играют широкие атмосферные ливни космических лучей (ШАЛ), но механизм их воздействия на инициацию КВР принципиально отличается от предложенных ранее механизмов.

10. Впервые предложена последовательная цепочка переходов плазмы из одного состояния в другое на протяжении всего процесса инициации молнии (от первых газоразрядных лавин, до сквозной фазы взаимодействия плазменных двунаправленных лидеров и/или сетей во время протекания IBPs-стадии развития молнии).

## Теоретическая и практическая значимость работы

Был обнаружен и исследован внутри заряженных искусственных облаков водного аэрозоля новый класс горячих плазменных образований, которые представляют из себя сети взаимодействующих плазменных каналов и двунаправленных лидеров, часто пронизывающих значительную часть аэрозольного облака. Эти результаты открывают новую страницу не только в области физики грозовых облаков, но и в области других заряженных аэрозольных и многофазных сред, таких как выбросы заряженной вулканической тефры, торнадо и пылевые бури, молнии на других планетах солнечной системы.

Проблема физического механизма инициации молнии в облаках не менее ста лет является вызовом физикам и геофизикам. Поэтому построение в данной работе даже качественной теории инициации молнии от первого инициирующего события (IE) через начальную стадию увеличения электрического поля (IEC) до стадии, включающей начальные импульсы пробоя (IBPs) является важным шагом к решению этой проблемы.

Другим важным результатом данной работы является то, что была построена полуколичественная модель компактных внутриоблачных разрядов (КВР/CID/NBE), дающая непротиворечивую интерпретацию их наблюдаемых физических свойств. Данная модель опирается на ключевую роль широких атмосферных ливней космических лучей в инициации молнии и КВР. Было показано, что только ШАЛ, благодаря инициации вторичными электронами и позитронами в небольших объемах грозового облака со сверхпробойными полями множества «обычных» стримерных вспышек, могут обеспечить такие, казалось бы, взаимоисключающие свойства КВР, как излучение свойственное стримерным вспышкам и скорость распространения, свойственную волне потенциала, идущей со скоростью близкой к скорости света по высокопроводящему плазменному каналу, при обратных ударах и стреловидных лидерах.

Практическая значимость работы определяется тем, что продвижение в понимании механизмов инициации молнии и КВР, которые являются мощными электрическими явлениями и генераторами потоков релятивистских частиц, позволит лучше предсказывать опасные явления, вызывающие гибель людей и животных, нарушающие работу линий электропередачи и связи и несущие опасность летательным аппаратам.

Выявленная в работе связь между инициацией молний, КВР и IBPs, и наиболее турбулентными областями облака позволит предложить методики обнаружения наиболее сильных турбулентностей в грозовых облаках, с помощью детектирования электромагнитных сигналов, характерных для IBPs и КВР. Схема предполагаемого эксперимента по обнаружению областей грозового облака, где происходит инициация молний, КВР и происходят IBPs, предположительно наиболее сильно турбулентными областями со встречными многофазными заряженными потоками, изображена на Рисунке В.45. Система антенн (3), работающих по триангуляционному принципу определения абсолютных координат внутриоблачных разрядов с помощью измерения задержек радиосигналов от разных антенн (the time-of-arrival technique — TOA (LMA)), в режиме реального времени с задержкой не хуже 1-3 мс передает координаты управляющей системе активной многолучевой фазированной антенной решётке (АФАР) (4). Лучи АФАР зондируют область грозового облака с поперечником 1-5 км, где благодаря TOA-системе были обнаружены IBPs или КВР и, так как IBPs и, возможно КВР на поздней стадии развития, являются горячими плазменными каналами с концентрацией

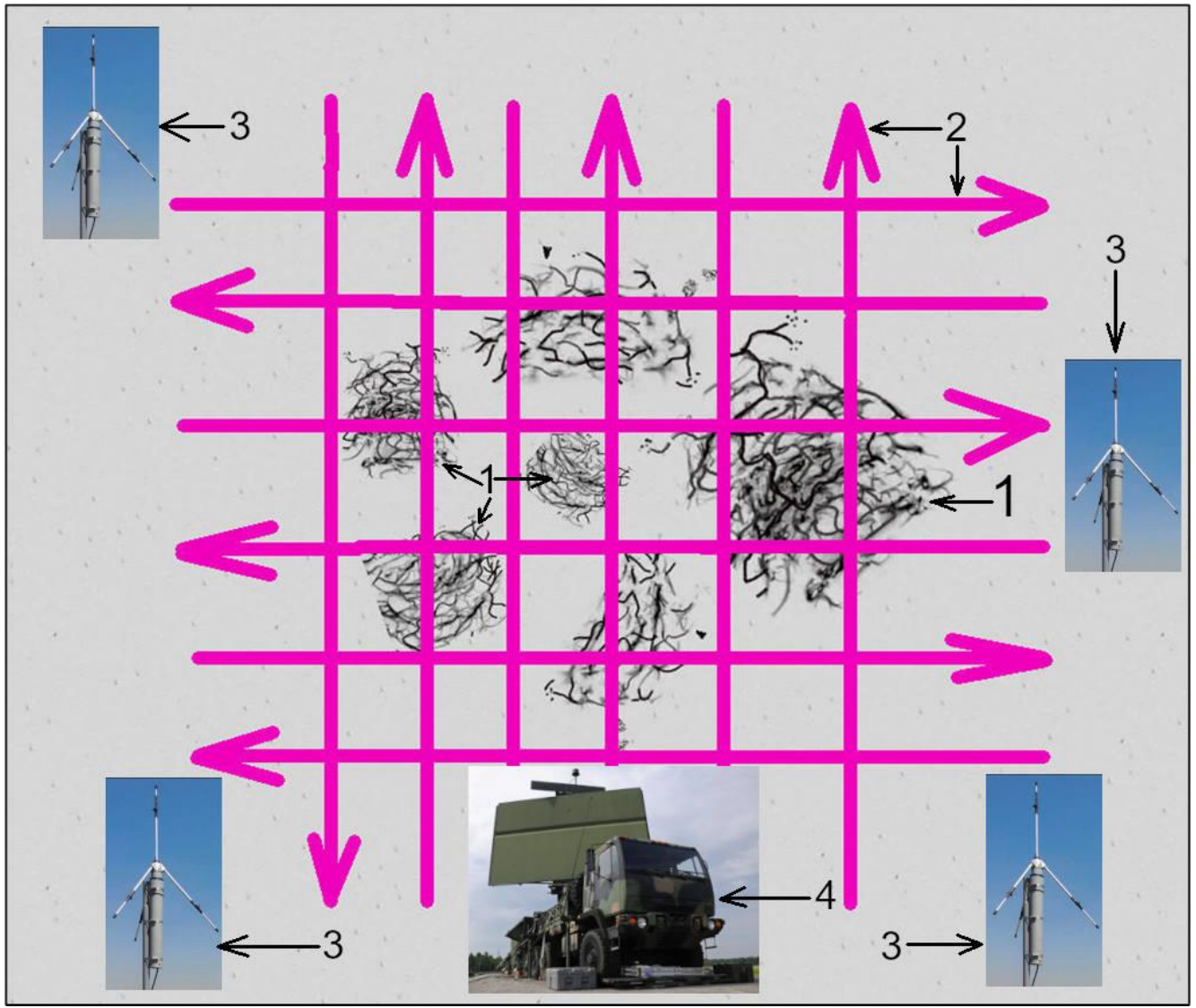

Рисунок В.45. Схема предполагаемого эксперимента по обнаружению областей грозового облака, где происходит инициация молний и КВР (предположительно наиболее сильно турбулентных областей со встречными многофазными заряженными потоками). 1 — области грозового облака, где возможно образуются сети плазменных каналов, которые приводят к IBPs или КВР; 2 — схема лучей активной фазированной антенной решетки (АФАР), которые зондируют область грозового облака, где благодаря TOA (LMA)-системе обнаружены IBPs или КВР; 3 — антенны TOA (LMA)-системы картирования молнии; 4 — АФАР (active electronically scanned array (AESA)), которая получает координаты разрядов от TOA (LMA)-системы и фиксирует отражение от сетей плазменных каналов в области грозового облака размером около 1-5 км, которые создают IBPs или КВР.

электронов не менее, чем $10^{16}$-$10^{17}$ см$^{-3}$, то существует высокая вероятность, что АФАР обнаружит отражающие объекты, которые в поперечнике составляют десятки и сотни метров. Кроме того, можно будет определить каковы отражающие свойства тех областей областей грозового облака, где инициируются молнии и КВР. Фиксирование IBPs на первом этапе представляется гораздо более перспективным, так как подавляющее большинство зафиксированных молний имеет IBPs-стадию (не менее 90% [Mäkelä et al., 2008], [Marshall et al., 2014b]), а «классические» КВР с сильными электромагнитными сигналами фиксируются всего в нескольких процентах случаев и то не на всех широтах. Стадия IBPs отличается настолько сильными изменениями электрического поля, что в средних широтах в 25% молний пиковые значения IBPs превосходили пиковые значения обратных ударов молнии [Mäkelä et al., 2008]. VHF-излучение IBPs также велико и может использоваться для определения их координат, но, конечно, VHF-излучение классических КВР гораздо сильнее.

## Методология и методы исследования

Для достижения целей, определенных для данного исследования и решения задач, поставленных в рамках данной работы, используются как существующие уникальные экспериментальные стенды (ГИН-6 МВ, искусственное, заряженное облако водного аэрозоля[3]) и разработанные методики (осциллографическое измерение токов, электрических полей, потоков оптического излучения), так и впервые применяемые методики для исследования неизвестных типов разрядов внутри аэрозольных облаков, а также длинной искры, создаваемой электрическими полями генераторов импульсных напряжений и заряженных искусственных аэрозольных облаков (ИК-камеры среднего ИК-диапазона, камеры с усилением изображения, скоростные камеры, микроволновая диагностика разрядов[4]). Наиболее важные результаты, такие, как обнаружение внутри аэрозольных облаков нового класса плазменных объектов — «необычных плазменных образований» (UPFs) или разрядов, инициированных протяженными проводящими объектами в электрическом поле аэрозольного облака, фиксировалось не менее двумя независимыми методиками измерений. Поэтому основные полученные экспериментальные результаты обладают высокой степенью достоверности и являются обоснованными, что подтверждено публикациями в ведущих мировых и российских научных журналах по данной тематике исследований (Geophysical Research Letters, Journal of Geophysical Research, Journal of Atmospheric and Solar-Terrestrial Physics, Radiophysics and Quantum Electronics (Известия вузов. Радиофизика)), а также докладами на ведущих мировых конференциях и выступлениями в ведущих научных учреждениях (см. раздел «Апробация результатов работы», стр.102).

---

[3] Работу уникальных установок в этих экспериментах, а также измерения токов, электрических полей обеспечивала группа ультравысоких напряжений Высоковольтного исследовательского центра РФЯЦ – ВНИИ Технической физики имени ак. Е.И. Забабахина под руководством В.С. Сысоева. В работе принимали активное участие М. Г. Андреев, М. У. Булатов, Д. И. Сухаревский, а также Л. М. Макальский.

[4] Микроволновая методика зондирования плазменных объектов разработана, имплементирована и реализована Богатовым Н.А, который также внес большой вклад во все стадии исследований, написания статей и обсуждения ключевых физических проблем.

Методология построения качественного механизма инициации молнии и КВР строилась, как на последовательном учете имеющихся современных экспериментальных данных по физике молнии, КВР и физике широких атмосферных ливней космических лучей, так и на аналитических и численных оценках, использующих результаты исследований по физике плазмы высокого давления и физике длинной искры.

## Объем и структура диссертации

Диссертация состоит из введения, 8 глав, заключения, словаря терминов и списка цитированной литературы. Список цитируемой литературы содержит 296 наименований, список работ по теме диссертации включает 11 статей. Общий объём диссертации составляет 526 страниц и включает 207 рисунков.

## Положения, выносимые на защиту

1. Внутри положительно и отрицательно заряженных искусственных водных аэрозольных облаков существует новый класс электрических разрядов «необычные плазменные образования» (unusual plasma formations — UPFs), которые представляют из себя сети плазменных каналов, размеров 10-30 см, некоторые из которых нагреты настолько, насколько нагрет восходящий положительный лидер.

2. Установлен по крайней мере один механизм возникновения UPFs. UPFs инициируется внутри восходящей положительной стримерной вспышки до появления других плазменных образований в электрическом поле аэрозольного облака.

3. Во время сквозной фазы контакта восходящего и нисходящего лидеров длинной искры, возможно ветвление лидеров внутри общей стримерной зоны.

4. Яркость инфракрасного излучения в области контакта восходящего и нисходящих лидеров длинной искры примерно в 5 раз выше, чем для участков образовавшегося единого канала ниже или выше этой области.

5. При движении протяженного проводника (болта арбалета с проводом и без провода) в окрестностях и внутри искусственного облака положительно и отрицательно заряженного водного аэрозоля возникают стримерные вспышки, инициируются длинные искры (лидеры) и UPFs, и возникают квазиобратные удары в заземленные объекты.

6. Во время образования ступеней (длиной 30-120 см) положительного лидера длинной искры (в электрическом поле высоковольтного генератора импульсных напряжений) перед головкой лидера в зоне стримерной короны может существовать плазменное образование, сходное по морфологии со спейс-стемом или спейс-лидером, которые наблюдаются в стримерной зоне отрицательного лидера. Стримерные зоны, возникающие во время образования ступеней положительного и отрицательного лидеров длинной искры морфологическое сходны.

7. Инициацию молнии в грозовых облаках можно описать, как последовательный качественный механизм от инициирующего молнию события (IE) через начальную стадию увеличения электрического поля (IEC), до первых начальных импульсов пробоя (IBPs), которые переходят в ступенчатый отрицательный лидер.

8. Инициирующим молнию событием (IE) могут быть «классические» компактные внутриоблачные разряды (KBP/CID/NBE) или, в гораздо большем числе случаев, — «слабые» КВР (weak NBE), аналогичные по физическому механизму классическим КВР, но меньшие по масштабу. Механизм инициации компактных внутриоблачных разрядов (КВР/CID/NBE), который дает непротиворечивое объяснение спектру оптического излучения КВР, характерному для стримерных вспышек, и скорости распространения источников радиоизлучения внутри грозового облака близкой к скорости света, можно представить, как инициацию вторичными электронами и позитронами ШАЛ в наиболее турбулентных областях грозового облака в небольших объемах со сверхпробойными полями множества стримерных вспышек, внутри которых возникнут цепочки UPFs, взаимодействующие друг с другом посредством вторичных стримерных корон.

9. Инициацию молнии можно представить, как последовательную цепь переходов плазмы из одного состояния в другое (от первых газоразрядных лавин до стримеров, UPFs, положительных, отрицательных, двунаправленных лидеров, сетей плазменных каналов, которые ведут к сквозным фазам взаимодействия плазменных двунаправленных лидеров и/или сетей во время протекания IBPs-стадии развития молнии).

## Степень достоверности результатов работы

Достоверность результатов по обнаружению и исследованию необычных плазменных образований (UPFs) и длинных искр подтверждена использованием корректных и апробированных независимых методик измерений, а также мнением рецензентов и коллег, верифицировавших результаты этих измерений (см. раздел Апробация результатов работы).

Достоверность предложенного качественного механизма инициации молнии и КВР экспериментально верифицирована:

▪ в части верификации объемного процесса инициации КВР и молнии на стадии инициирующего события (IE):

• [Rison et al., 2016], Рисунки В.30-В.32;

• [Lyu, Cummer et al., 2019], Рисунок В.44;

• [Shao et al., 2020], Рисунки В.40, В.41, В.43;

• [Liu et al., 2021], Рисунок В.37;

▪ в части верификации протекания стадии начального изменения электрического поля (IEC)

● [Rison et al., 2016], Рисунок В.30;

● [Marshall et al., 2019], Рисунок 7.14;

● [Bandara et al., 2019], Рисунок 7.12;

● [Lyu, Cummer et al., 2019], Рисунок В.44;

▪ на стадии начальных импульсов пробоя (IBPs):

● [Marshall et al., 2014a], Рисунок В.21;

● [Rison et al., 2016], Рисунок В.30;

● [Bandara et al., 2019], Рисунок 7.11.

**Апробация результатов работы**

Kostinskiy, A., Marshall, T., and Stolzenburg, M. (2020): The mechanism of the origin the NBE (CID) and the initiating event (IE) of lightning due to the volume phase wave of EAS-RREA synchronous ignition of streamer flashes // EGU General Assembly 2020, 4–8 May 2020, EGU2020-11487, https://doi.org/10.5194/egusphere-egu2020-11487.

Vlasov, A., Fridman, M., and Kostinskiy, A. (2020): Method for the numerical calculation of the mechanism of the origin the NBE (CID) due to the volume phase wave of synchronous ignition of streamer flashes by EAS-RREA // EGU General Assembly 2020, 4‑8 May 2020, EGU2020-22461, https://doi.org/10.5194/egusphere-egu2020-22461, 2020.

Kostinskiy, A. Y., Marshall, T.C., Stolzenburg, M. (2019). The Mechanism of the Origin and Development of Lightning from Initiating Event to Initial Breakdown Pulses // AE13B-3209, AGU Fall Meeting, 9 December 2019.
https://agu.confex.com/agu/fm19/meetingapp.cgi/Paper/551802

Kostinskiy, A. Y., Marshall, T.C., Stolzenburg, M. (2019). The Mechanism of the Origin and Development of Lightning from Initiating Event to Initial Breakdown Pulses // IUGG19-3423, 27th IUGG General Assembly, Montreal, Canada, https://www.czech-in.org/cmPortalV15/CM_W3_Searchable/iugg19/#!abstractdetails/0000757460

Всероссийской конференции по атмосферному электричеству, Санкт-Петербург, 24-28 сент. 2012 г. В 2-х т. Т. 1. СПб: Издательство ГГО, 2012.

Сысоев В., Макальский Л., Андреев М., Булатов М., Сухаревский Д., Иудин Д., Костинский А. Ю. (2012) Физическое моделирование межоблачного разряда в лабораторном эксперименте // Сборник трудов VII Всероссийской конференции по атмосферному электричеству, Санкт-Петербург, 24-28 сент. 2012 г. В 2-х т. Т. 1. СПб.: Издательство ГГО, 2012. С. 244-246.

Сысоев В., Костинский А. Ю., Климашев В., Емельянов А., Иудин Д. (2012) Электрическая структура униполярного облака // Сборник трудов VII Всероссийской конференции по атмосферному электричеству, Санкт-Петербург, 24-28 сент. 2012 г. В 2-х т. Т. 1. СПб.: Издательство ГГО, 2012. С. 238-240.

Andreev M., Bulatov M., Kostinskiy A., Kuhno A., Makalsky L., Mareev E., Syssoev V. (2011) An upward connecting leader at tests of large-scalel ightning-rod models // 7th Asia-Pacific International Conference on Lightning. Chengdu: Chengdu university, 2011.

Также результаты работы регулярно докладывались на семинарах в Институте прикладной физики РАН, Институте общей физики РАН, НИИ Ядерной физики МГУ, ГО «Борок» Института физики Земли РАН, Ереванском физическом институте, Московском институте электроники и математики и др.

## Личный вклад автора

Всего по теме диссертации опубликовано 11 статей (см. раздел «Список публикаций по теме диссертации», стр.525) в ведущих рецензируемых российских и зарубежных научных журналах, из них в 7 автор диссертации является первым автором (в одной — единственным), что показывает его определяющий вклад в эти работы.

В исследованиях, которые представлены в главах 1-5, и 7, автор был основным постановщиком научных задач, отвечал за ключевые оптические измерения с помощью ИК-камер, камер с усилением изображения, скоростных камер, обрабатывал подавляющее большинство экспериментальных данных и обеспечивал их оформление, писал базовые варианты научных статей на основе этих данных, представлял на конференциях и семинарах полученные результаты (см. раздел Апробация результатов работы). Е.А. Мареев и В.А. Раков, Н.А. Богатов активно участвовали в постановке и планировании экспериментов, расширении тематики исследований, верификации полученных данных и написании окончательных английских вариантов статей. Н.А. Богатов планировал, готовил и проводил все эксперименты по микроволновым методикам измерений и многие эксперименты по фотометрированию разрядов с помощью фотоэлектронных умножителей. Работу экспериментальных установок в этих экспериментах, а также измерения токов, электрических полей обеспечивала группа ультравысоких напряжений ВНИЦ Всероссийского энергетического института, г. Истра (ныне Высоковольтный исследовательский центр РФЯЦ – ВНИИ Технической физики имени академика Е.И. Забабахина) под руководством В.С. Сысоева. В работе принимали активное участие М. Г. Андреев, М. У. Булатов, Д. И. Сухаревский, а также Л. М. Макальский. Все они активно участвовали в обсуждении экспериментов и высказали много полезных замечаний и предложений.

В исследованиях, которые представлены в главах 7-8, автор был инициатором построения единого механизма инициации молнии и компактных внутриоблачных разрядов. Автору принадлежит идея рассмотреть КВР, как фазовую волну инициации «обычных» газоразрядных стримерных вспышек, движущуюся со скоростью, близкой к скорости света (благодаря вторичным электронам ШАЛ, экспоненциально усиленным в электрическом поле грозового облака) и определить свойства КВР через свойства множества стримерных вспышек, распределенных по объему. Также автору принадлежит идея и реализация построения последовательной цепочки плазменных превращений с использованием плазменных образований, обнаруженных в экспериментах, описанных в главах 1-5, и 6. Данная цепочка плазменных превращений через короткое время фазы начальных изменений электрического поля (IEC) заканчивалась мощными начальными импульсами пробоя (IBPs), которые также удавалось объяснить трехмерным образованием сетей плазменных каналов, связанных с трехмерной фазой инициации молнии в процессе КВР или слабого КВР. Томасу Маршаллу (Thomas Marshall) и Марибет Стольценбург (Maribeth Stolzenburg) принадлежит четкая постановка задачи построения модели КВР (CID/NBE), которая требовала, чтобы новая модель NBE исключала начальное присутствие горячих высокопроводящих плазменных каналов до момента инициации NBE, то есть NBE не может быть результатом взаимодействия двунаправленных лидеров или плазменных сетей. Также Маршалл и Стольценбург сформулировали все основные ограничения, которые существуют для построения модели NBE (слабое свечение в видимом диапазоне, скорость движения, близкая к скорости света и др.) и верифицировали и согласовали предложенный механизм со всем большим и противоречивым комплексом современных экспериментальных данных по внутриоблачным процессам от IE до IBPs. Они также писали и редактировали окончательный английский вариант статьи.

Автор также предложил математическую модель и методику численного расчета механизма инициации КВР с помощью ШАЛ, вторичные электроны которого экспоненциально усиливаются благодаря механизму убегающих электронов (глава 8). Численную вариант этой математической модели реализовали и оформили графики решений А.А. Власов и М.Л. Фридман.

## Благодарности

И, конечно, автор благодарен своей жене Ольге Орловой, без которой эта диссертация никогда не была бы написана.

# ГЛАВА 1. Открытие нового класса электрических разрядов в облаках искусственно заряженных капель водного аэрозоля и последствия их открытия для инициирования молнии в грозовых облаках

В этой главе приведены эксперименты, где впервые наблюдались «необычные плазменные образования» (UPFs) внутри облаков искусственно заряженного водного аэрозоля с помощью высокоскоростной инфракрасной камеры, работающей в сочетании с высокоскоростной камерой видимого диапазона и измерениями тока и свечения разрядов в оптическом диапазоне [Kostinskiy et al., 2015a]. Параметры инфракрасного излучения каналов плазмы UPFs были близки к параметрам положительных лидеров длинных искр, наблюдаемых в тех же экспериментах, в то время как морфология каналов UPFs отличалась от параметров любых до этого момента известных лидеров, поэтому UPFs можно рассматривать как новый, необычный тип внутриоблачного и облачного разряда. Эти образования, вероятно, оказываются проявлением коллективных процессов создания сложной иерархической сети взаимодействующих каналов на разных стадиях развития (некоторые из них горячие и обнаруживаются в течение миллисекунд). Мы полагаем, что это явление также может происходить в грозовых облаках и может являться одним из звеньев в процессе инициирования молнии внутри грозовых облаков.

Нам удалось обнаружить UPFs в экспериментах, описанных в этом разделе, так как мы использовали (а) отрицательное облако искусственно заряженных капель водного аэрозоля со средним радиусом 0,5 мкм и (б) инфракрасную (ИК) камеру, с матрицей, чувствительной в диапазоне длин волн 2,7–5,5 мкм (длины волн на порядок большие, чем размер капель), что позволило впервые «увидеть», что происходит внутри аэрозольного облака. Облако (размером несколько десятков кубических метров) при отрицательном заряде капель способно инициировать лидеры длинной искры, стартующие с близлежащих заземленной плоскости и металлических объектов на ней ([Верещагин и др., 1988]; [Анцупов и др., 1991]) и, следовательно, может рассматриваться как модель некоторых естественных заряженных аэрозольных систем. Уникальная комбинация параметров аэрозольного облака (относительно небольшой размер капель) и относительно больших длин волн, которые может фиксировать ИК-камера, позволило нам наблюдать новый класс плазменных образований в облаке как при наличии, так и в отсутствие лидерных каналов между облаком и заземленной сферой. В наших экспериментах, обычно, наиболее рельефные изображения этих необычных плазменных образований получаются, когда поблизости образуются длинные лидеры. По этой причине большинство представленных изображений соответствует этому последнему виду явления.

## 1.1. Экспериментальная установка

Эксперименты проводились на установках Высоковольтного исследовательского центра РФЯЦ – ВНИИ Технической физики имени ак. Е.И. Забабахина (http://www.ckp-rf.ru/usu/73578/). Экспериментальная установка, применяемая в экспериментах данного раздела показана на Рисунке 1.1. Заряженное облако (1) создавалось парогенератором (2.1) и источником высокого напряжения (2.2), который подавал напряжение на острие иглы, создающее корону. Острие иглы располагалось в сопле (2.3), через которое проходила паровоздушная струя. Струя имела температуру около 100–120°С и давление 0,2–0,6 МПа. Она двигалась со скоростью около 400–420 м/с углом расширения 28°, образуя благодаря турбулентной струе, заряженное облако, показанное на Рисунке 1.2. Сопло располагалось в центре плоского заземленного металлического экрана (3) диаметром 2 м. В результате быстрого охлаждения пар конденсировался в капли воды со средним радиусом около 0,5 мкм [Верещагин и др., 1988]. Ионы, заряжающие водный аэрозоль, образовывались в коронном разряде между острием и соплом (2.3). На острие подавалось постоянное напряжение 10–20 кВ. Ток, связанный с зарядом, переносимым струей, находился в диапазоне от 60 до 150 мкА. Когда общий заряд, накопленный в облаке, приблизился к 50-60 мкКл или около того, между облаком и находящимися поблизости заземленными объектами спонтанно возникли положительные восходящие лидеры, которые приводили в некоторых случаях к искровым разрядам (под которыми в данном случае мы понимаем взаимодействие восходящих положительных лидеров с

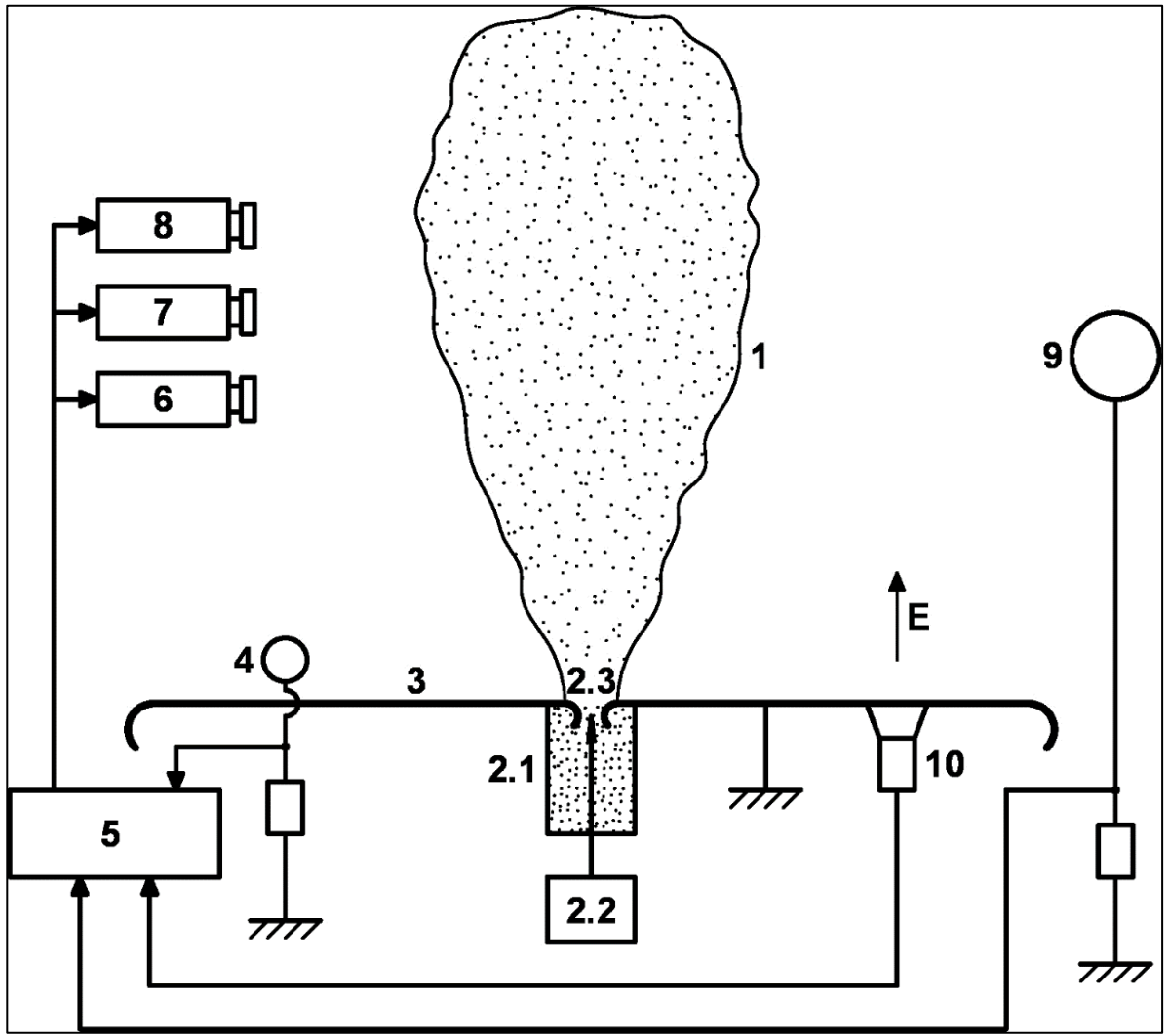

Рисунок 1.1 (адаптировано из [Kostinskiy et al., 2015a]). Экспериментальная установка: 1 — отрицательное искусственно заряженное облако аэрозоля; 2.1 — парогенератор; 2.2 — источник высокого напряжения с коронирующим острием; 2.3 — сопло; 3 — заземленная металлическая плоскость; 4 — 5-сантиметровый заземленный шар с токоизмерительным шунтом; 6 — скоростная 4Picos камера видимого диапазона с усилением изображения; 7 — высокоскоростная инфракрасная камера; 8 — фотоаппарат; 9 — сфера 50 см для отслеживания изменений заряда облака; 10 — измеритель электрического поля (флюксметр).

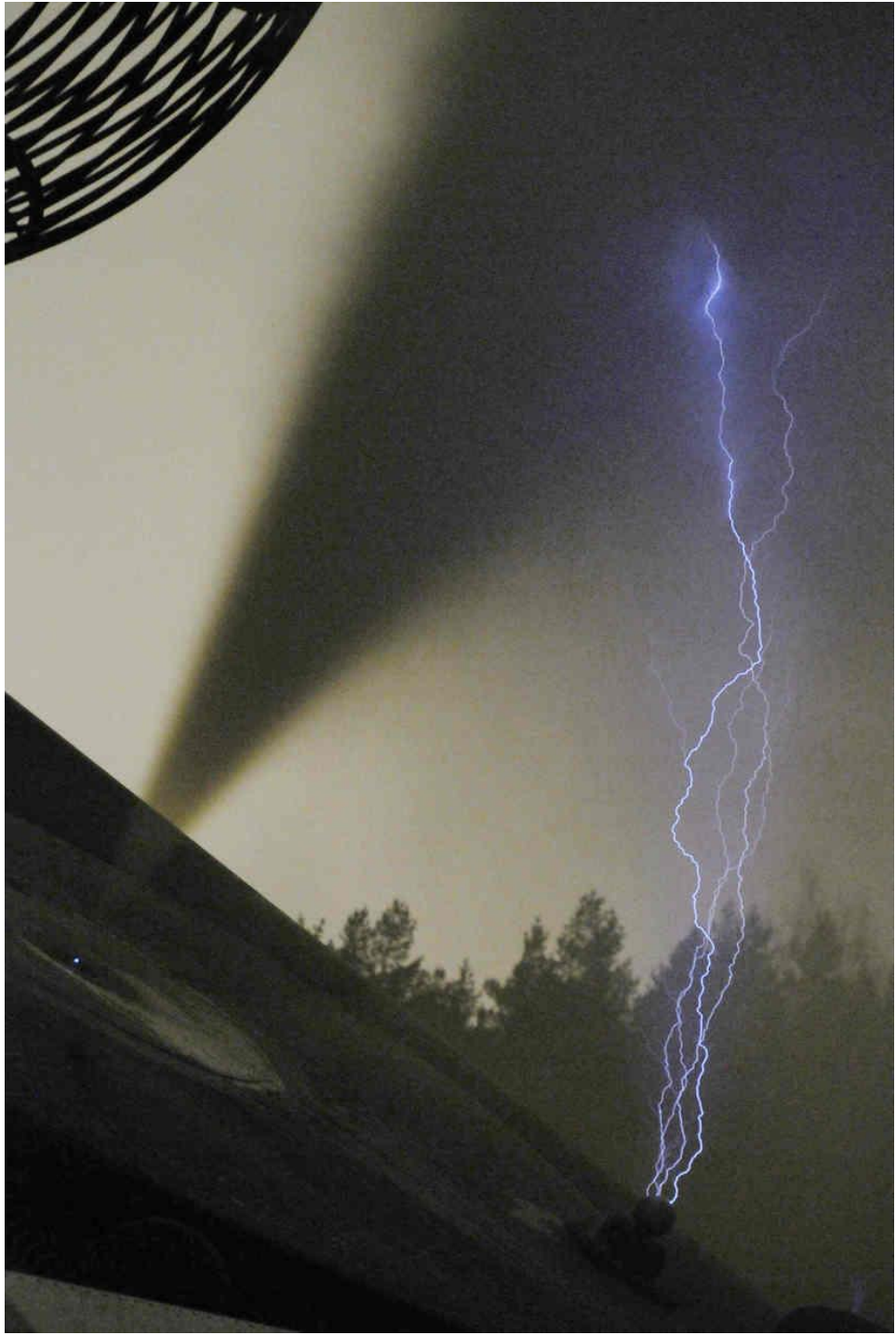

Рисунок 1.2 (адаптировано из [Kostinskiy et al., 2015a]). Снимок, сделанный фотоаппаратом Canon (видимый диапазон; экспозиция — 5 с), отрицательно заряженного облака (наклонная струя, темное образование) и четырех восходящих положительных лидеров длиной около 1,5 м, инициированных заземленной сферой. Внутри облака на фотографии не видны внутриоблачные светящихся образований.

нисходящими отрицательными лидерами, которые были частью двунаправленного лидера, см. глава 3). В случае отрицательно заряженного облака большая часть положительных восходящих лидеров возникла на заземленной металлической сфере (4), как показано на Рисунке 1.2. Сфера имела диаметр 5 см и находилась на расстоянии 0,8 м от центра экрана. Самая верхняя точка сферы (4) находилась на 12 см выше экрана. Важно отметить, что описанное здесь искусственно заряженное облако лишь приблизительно имитирует реальное грозовое облако, которое имеет другие размеры, температуру, содержит кристаллы льда и другие гидрометеоры с широким диапазоном размеров, форм, фазовых состояний и т.д. Токи в лидерах и искрах, возникающих на сфере (4), измерялись омическим шунтом сопротивлением 1 Ом, сигнал с которого передавался на оцифровывающий осциллограф Tektronix (5) с полосой пропускания 1 ГГц. Когда ток превышал заданное значение, запускался осциллограф, который, в свою очередь, генерировал импульс, который использовался для запуска высокоскоростной камеры 4Picos с усилением изображения, работающей в видимом диапазоне (6) и высокоскоростной инфракрасная камера FLIR 7700M (7). ИК-камера FLIR 7700M работала со скоростью 115 кадров в секунду (длительность кадра 8,7 мс, время экспозиции 6,7 мс) с разрешением 640×512 пикселей. Скоростная камера, работающая в видимом диапазоне (4Picos), выдавала 2 кадра по 1360×1024 пикселей каждый. Общая картина разряда регистрировалась фотоаппаратом Canon (8). Все камеры были установлены на расстоянии около 3 м от оси струи, образующей облако. Изображение, представленное на Рисунке 1.2, было получено с помощью камеры Cannon (выдержка 5 с). Изображения, представленные на Рисунках 1.3, 1.5-1.9 были получены с использованием FLIR 7700M (экспозиция 6,7 мс). Изображения большего и меньшего размера, показанные на Рисунке 1.11, были получены с использованием ИК-камеры FLIR 7700M (выдержка 7,7 мс) и 4Picos (выдержка 1 мкс). Для отслеживания динамики зарядки (и разрядки) облака использовалась медная сфера (9) диаметром 50 см, заземленная через сопротивление 100 МОм и расположенная в 6 м от облака. Сигналы с сопротивления 100 МОм, свидетельствующие об изменении электрического потенциала, наведенного на сфере диаметром 50 см облачным зарядом, регистрировались осциллографом (5). Электрическое поле на поверхности заземленной плоскости измерялось флюксметром (10).

## 1.2. Результаты экспериментов

### 1.2.1. Двухкадровая запись камерой с усилением изображения и инфракрасная запись восходящих положительных лидеров и необычных плазменных образований (UPFs) внутри облака

На Рисунках 1.3a и 1.3b показаны два последовательных кадра, снятых ИК-камерой в верхней части облака (нижняя граница кадра находилась примерно на 70 см выше заземленной плоскости). Все процессы разрядов, наблюдаемые на этих кадрах, происходили внутри облака и, следовательно, не отображались на камерах и фотоаппаратах в видимом диапазоне (наблюдались только вспышки рассеянного света). Каждый кадр имел экспозицию 6,7 мс и мертвое время 2 мс, так что два изображения могли быть разделены во времени на 2–15,4 мс. Отсюда следует, что большинство разрядных процессов, зарегистрированных на двух кадрах, были видны в ИК-диапазоне не менее 2 мс. Процессы, наблюдаемые на ИК-изображениях, включают: (1) верхнюю часть в облаке восходящего положительного лидера от заземленной сферы, нижняя часть которой, развивающаяся в чистом воздухе, находилась вне поля зрения ИК-камеры, (2) большая стримерная зона, пересекающая каждый кадр от нижнего правого до верхнего левого угла (предположительно, положительная стримерная корона от первых стримерных вспышек и от восходящего положительного лидерного канала, включая его ветви, которые находятся вне поля зрения камеры), и (3) необычные плазменные образования (UPFs), которым посвящена данная глава. По результатам измерения тока, который шел через заряженную сферу и шунт, стримерная вспышка, в результате которой образовался лидер имела максимум тока около 10 А, восходящий положительный лидер имел пик тока около 5 А и во время распространения лидера ток лидера опускался до примерно 0.2 А. Все разрядные события длились около 35 мкс и с облаком провзаимодействовал общий положительный заряд первой стримерной вспышки и стримерной короны лидера размером около 15 мкКл заряда (Рисунок. 1.4).

На Рисунке 1.3 хорошо видно, что структура и форма UPFs (3) принципиально отличаются от восходящего положительного лидера (1) и стримерных вспышек (2). Наиболее яркие части UPFs намного (на порядок) ярче стримеров. Кроме того, на Рисунке 1.3a хорошо видно, что интенсивность ИК-излучения от UPFs (3), близка по

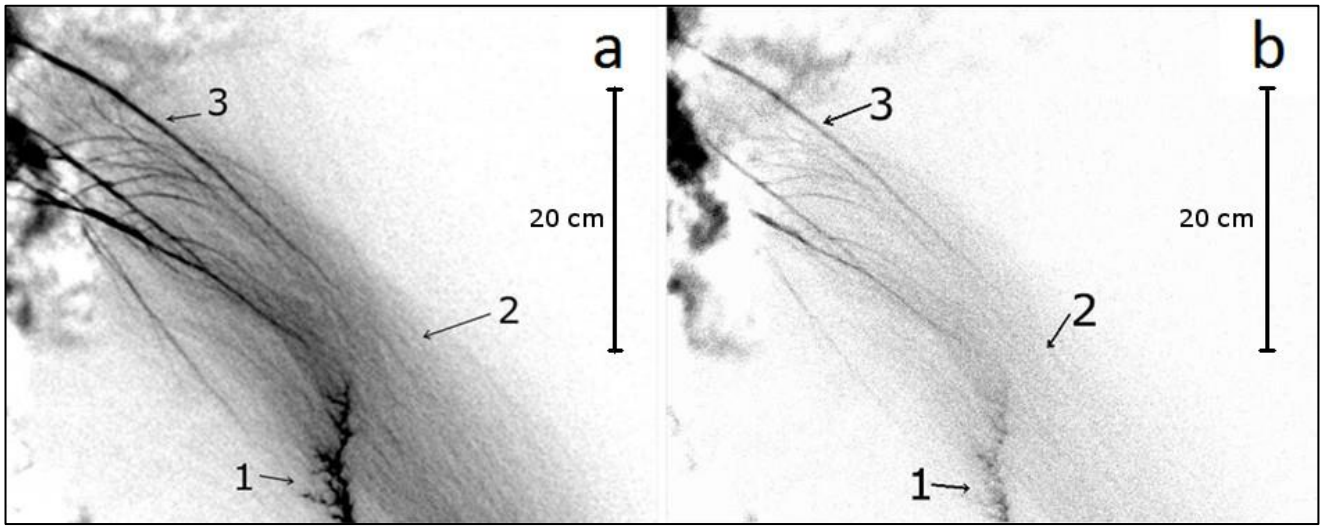

Рисунок 1.3 (адаптировано из [Kostinskiy et al., 2015a]). Два последовательных инфракрасных изображения (инвертированы), полученных с выдержкой 6,7 мс и разделенных интервалом 2 мс, которые показывают различные разрядные процессы внутри облака. Во время этого события в видимом диапазоне наблюдались только вспышки рассеянного света, а не отдельные каналы. 1 — верхняя часть восходящего положительного лидера (его нижняя часть, фиксируемая в видимом диапазоне, находится вне поля зрения ИК-камеры), 2 — стримерная зона первых стримерных вспышек, 3 — необычные плазменные образования (UPFs).

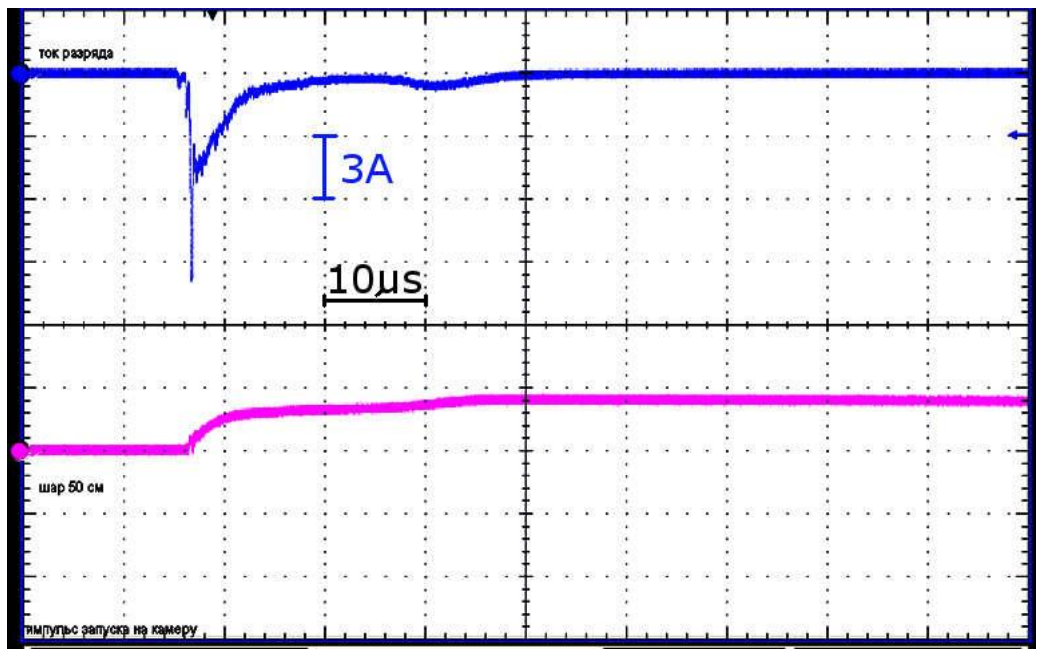

Рисунок. 1.4 (адаптировано из [Kostinskiy et al., 2015a]). Осциллограмма события, зафиксированного на Рисунке 1.3. Синий луч — ток восходящего положительного лидера, идущий через шунт (одно вертикальное большое деление соответствует току — 3 А), горизонтальное большое деление соответствует интервалу времени — 10 мкс). Фиолетовый луч показывает динамику заряда облака (прибор не калиброван и служит для качественного понимания процессов ухода заряда с облака).

интенсивности от горячего восходящего положительного канала лидера (Рисунок 1.5а2), но его морфология (сложная сеть разных по интенсивности каналов, пронизывающих относительно большую область облаков) не похожа на морфологию положительного лидера (основной канал с боковыми ветвлениями, указывающими направление распространения лидера, со стримерными зонами, начинающимися на головках лидерных каналов). На Рисунке 1.5b2, который фиксирует интенсивность ИК-излучения каналов через несколько мс после разряда, видно, что UPFs излучают даже немного сильнее, чем распадающийся канал восходящего лидера.

ИК-изображения UPF, показанные на Рисунках 1.3а и 1.3b, типичны в присутствии восходящего положительного лидера от заземленной сферы, входящего в отрицательно заряженное облако. На сегодняшний день зарегистрировано более 100 таких событий.

## 1.2.2. UPF расположенный внутри облака ниже верхней части канала восходящего положительного лидера

На Рисунке 1.6 показан пример UPF длиной 3–4 см, который ориентирован вдоль канала положительного лидера и расположен значительно ниже его верхней части. Оба полностью находятся внутри аэрозольного облака. От верхнего конца UPF вверх к отрицательному облаку распространяются положительные стримеры. С нижним, отрицательным концом UPF взаимодействуют положительные стримеры стримерной зоны нижних ветвей восходящего положительного лидера. Анализ интенсивностей ИК-излучения UPF и положительного восходящего лидера на Рисунке 1.7 показывает, что нагрев UPF сравним с нагревом положительного лидера. UPF, показанный на Рисунках 1.6-1.7, качественно похож на спейс-лидер, участвующий в формировании ступеней отрицательного лидера в длинных искрах и молниях, хотя они зарождались и развивались в совершенно ином контексте, например: [Les Renardières Group, 1981]; [Gamerota et al., 2014]; [Petersen and Beasley, 2013].

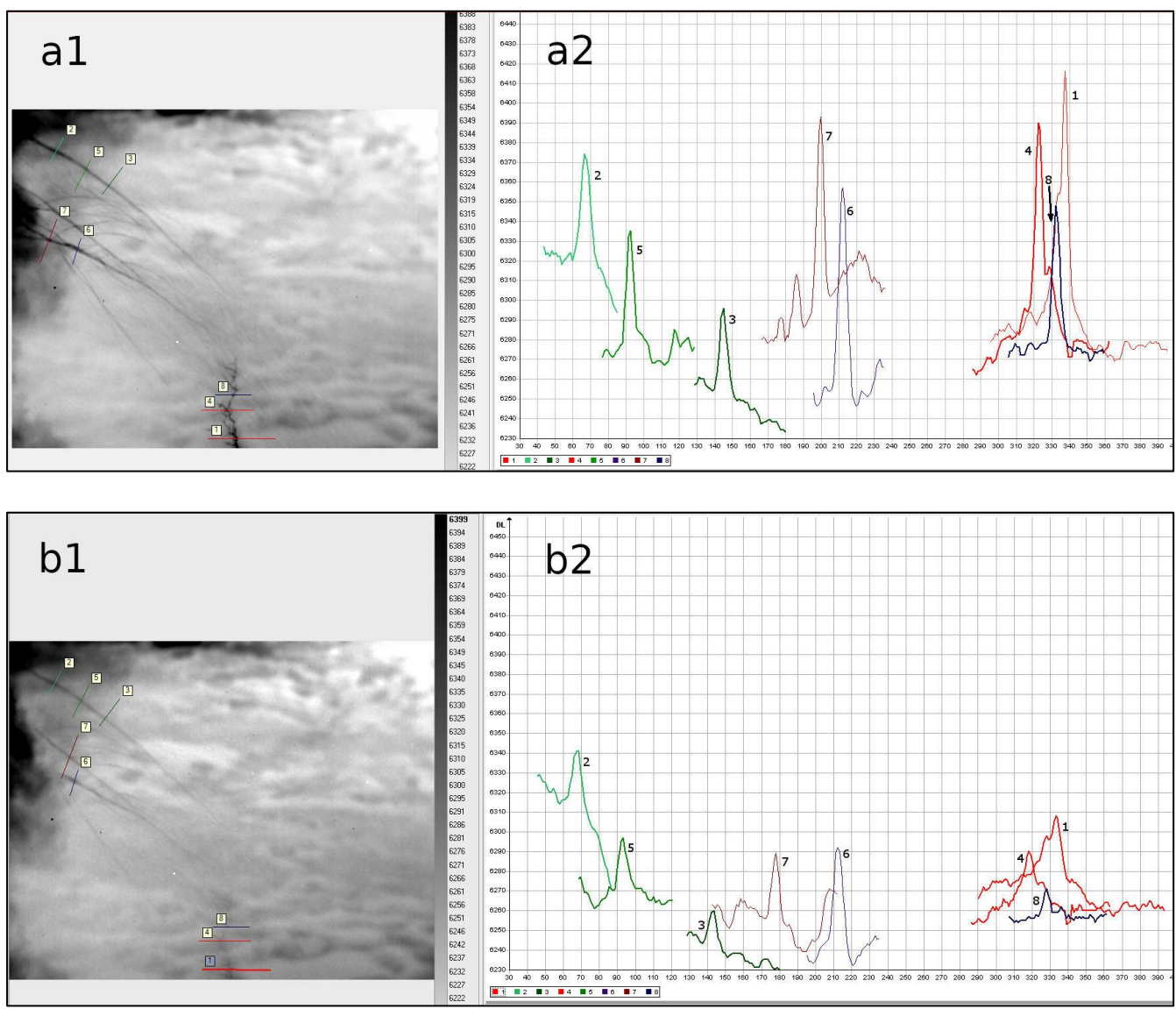

Рисунок 1.5 (адаптировано из [Kostinskiy et al., 2015a]). Два последовательных инфракрасных изображения (a1 — то же изображение, что и на Рисунке 1.3a, a b1 — то же изображение, что и на Рисунке 1.3b), полученных с выдержкой 6,7 мс и разделенных интервалом 2 мс, которые показывают различные разрядные процессы внутри облака. Анализ интенсивности ИК-излучения вдоль линий на изображениях b1 и b2 позволяет сравнить нагрев UPFs и восходящего положительного лидера. Вертикальная шкала на b2 является линейной и цифры приведены в относительных единицах, а горизонтальная шкала соответствует номеру пикселя.

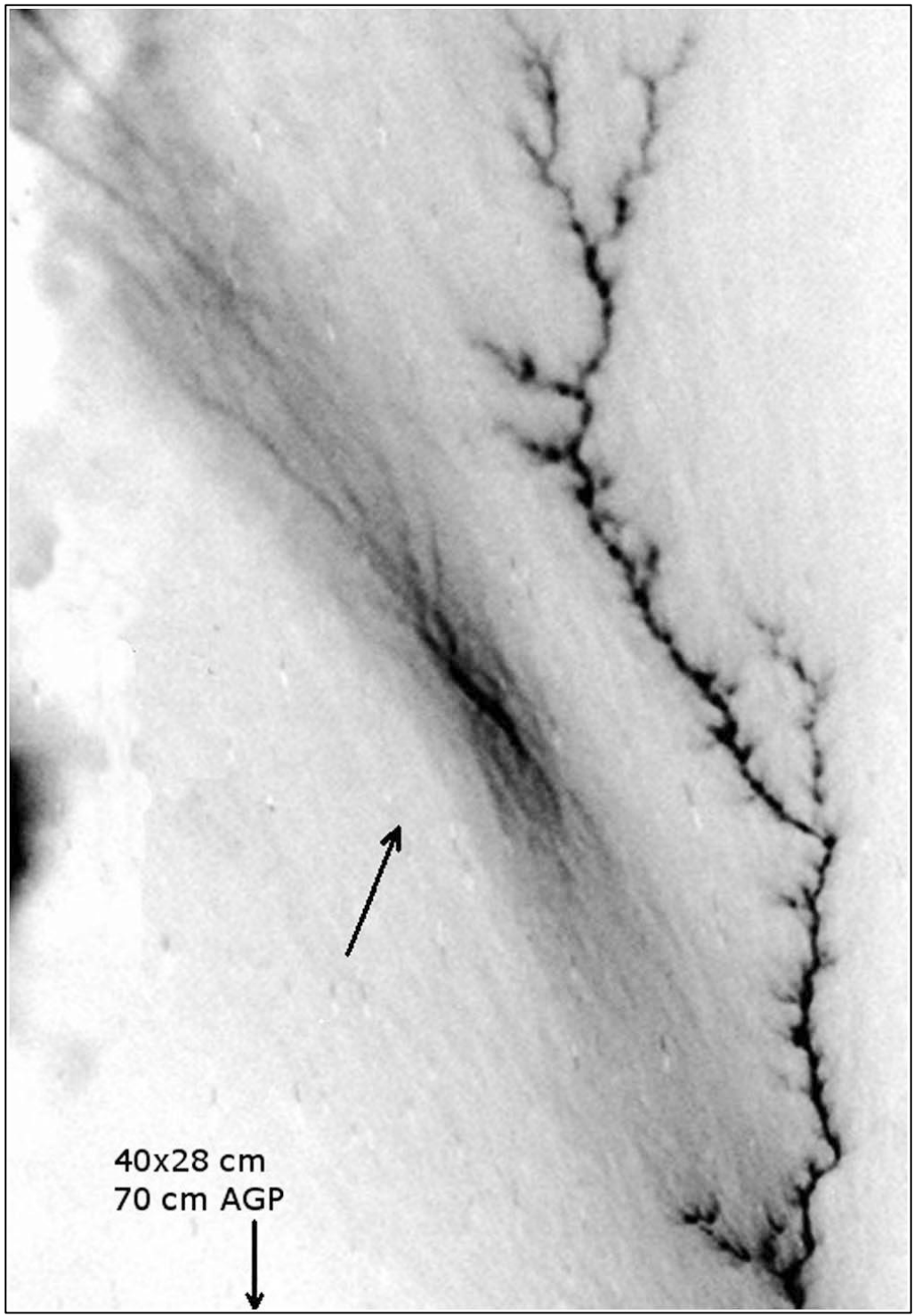

Рисунок 1.6 (адаптировано из [Kostinskiy et al., 2015a]). Инфракрасное изображение (инвертированное), полученное с выдержкой 6,7 мс, которое показывает верхнюю часть восходящего положительного (правого) лидера и (слева) UPF, оба внутри облака; наклонная стрелка указывает на самую яркую часть UPF. Похоже, что эти два процесса представляют собой разные разрядные процессы, которые взаимодействуют через свои стримерные зоны в нижней части изображения. AGP означает — над заземленной плоскостью.

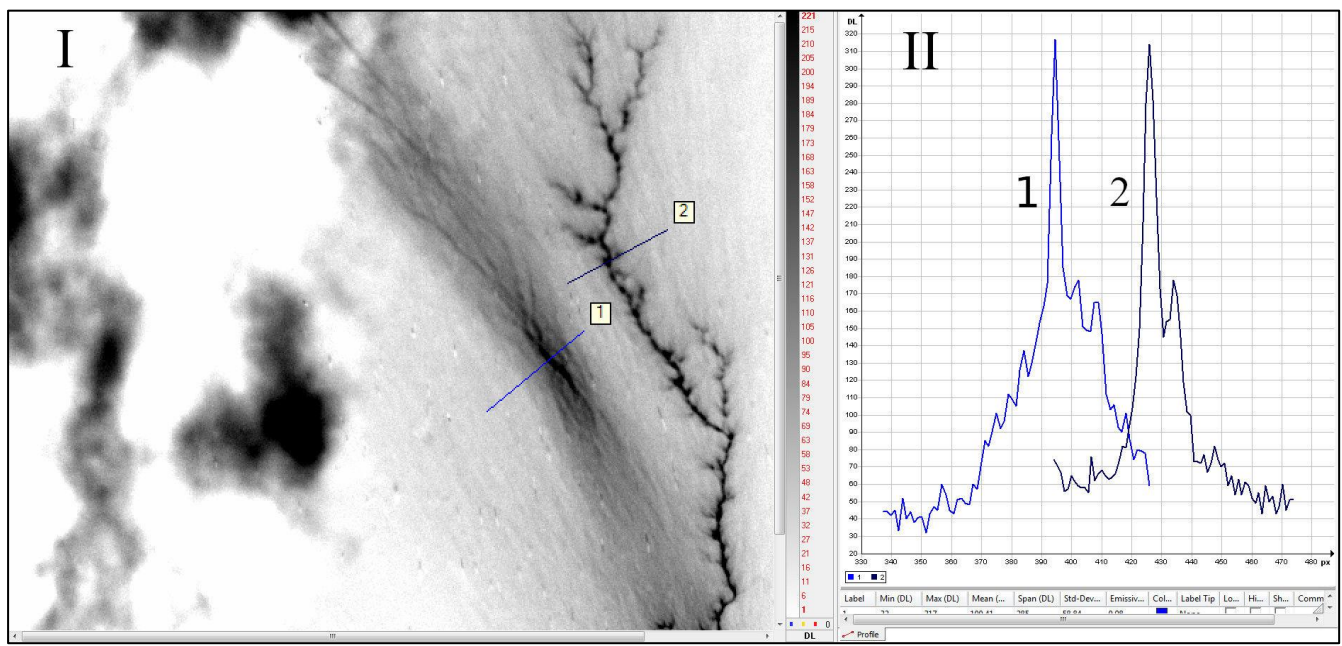

Рисунок 1.7 (адаптировано из [Kostinskiy et al., 2015a]). Часть инфракрасного изображения 2.1.1.7, полученного с выдержкой 6,7 мс, которое показывает верхнюю часть восходящего положительного (правого) лидера и (слева ниже) UPF, оба внутри облака. Эти два плазменных образования, скорее всего, представляют собой разные разрядные процессы, которые взаимодействуют в нижней части изображения. Анализ интенсивности ИК-излучения вдоль линий на изображениях I и II позволяет сравнить нагрев UPFs и восходящего положительного лидера и показывает, что они сравнимы. Вертикальная шкала на Рисунке II является линейной и цифры на ней приведены в относительных единицах, а горизонтальная шкала на Рисунке II соответствует номеру пикселя.

### 1.2.3. ИК-зондирование облака и его окрестностей в поисках местоположения инициации UPFs

Наблюдения показали, что UPFs могут иметь разные формы и возникать в разных контекстах, практически в любом месте облака и в непосредственной близости от него, вместе (в рамках выдержки камеры) с лидерными каналами или без них (пример UPF без лидерных каналов показан на Рисунке 1.8). Поэтому было последовательно исследовано все пространство аэрозольного облака, благодаря изменению поля зрения ИК-камеры, начиная с заземленной сферы (с которой часто инициируется восходящий положительный лидер), двигаясь ступенчато на 50 см вверх и 20 см влево (в сторону более

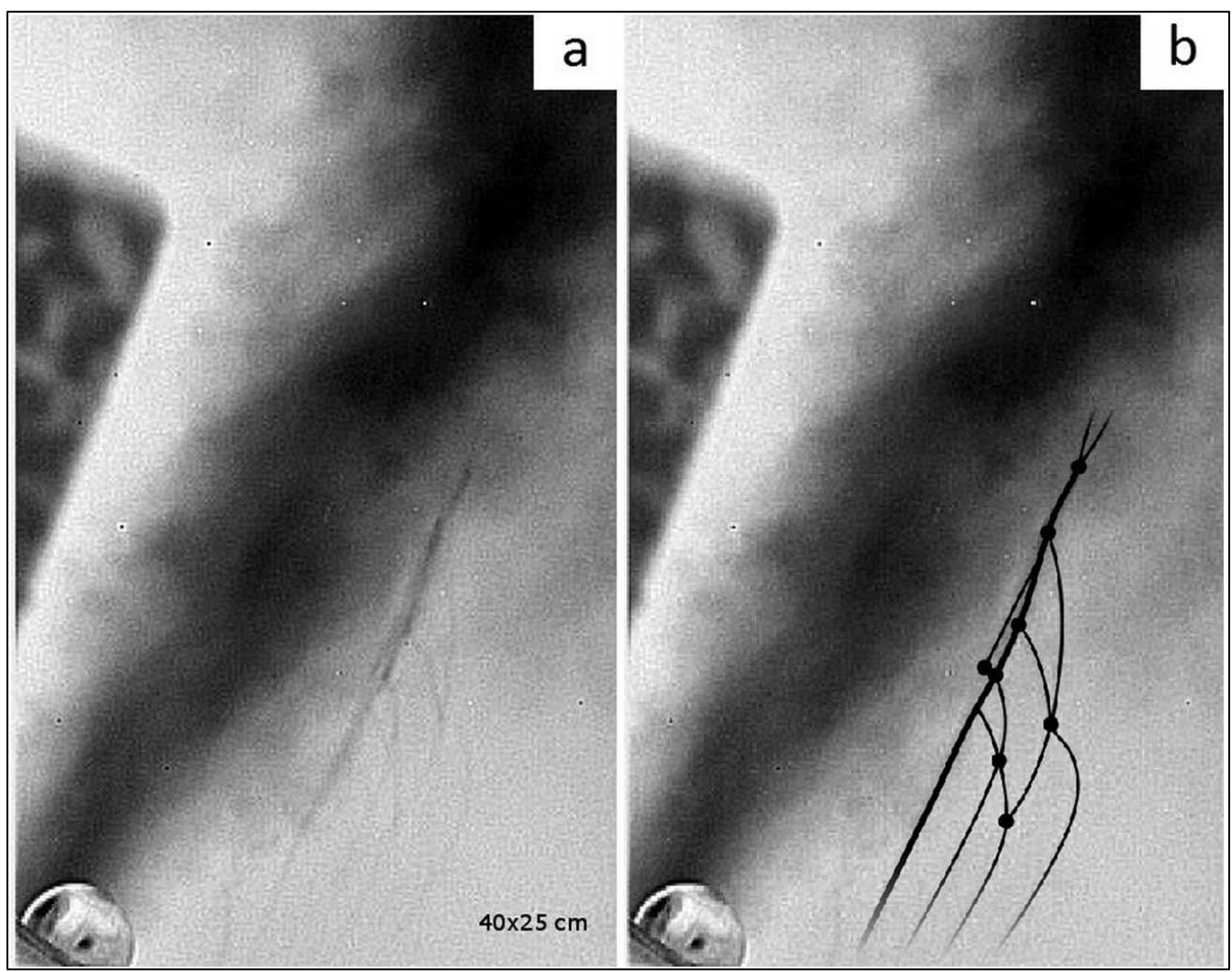

Рисунок 1.8 (адаптировано из [Kostinskiy et al., 2015a]). (a) ИК-изображение UPF, записанное при отсутствии восходящего положительного лидера. (б) То же самое изображение, что и на Рисунке «а», но дополнительно показывающий скетч UPF, наложенный на изображение для улучшения визуализации UPF. Неясно, достигают ли нисходящие, расширяющиеся ветви UPF, заземленной плоскости или нет. Обращает на себя внимание ячеистая структура сети каналов UPF, которая не наблюдается в лидерах.

плотной центральной части облака). Результаты этого «сканирования» показаны на Рисунках 1.9a-d. На Рисунке 1.9а можно увидеть только направленный вверх положительный лидер, стартующий от заземленной сферы, выходящий за верхний край поля зрения камеры. UPFs не видно, хотя, нельзя исключить, что они присутствовали на больших высотах. Рисунок 1.9b показывает верхнюю часть восходящего положительного лидера (внизу справа) и UPFs (вверху слева). Эти два процесса, по-видимому, представляют собой разные по происхождению разряды, которые взаимодействуют благодаря стримерным зонам. Как и на Рисунках 1.5 и 1.7, ИК-яркость наиболее интенсивных элементов UPFs сопоставима с яркостью лидерного канала. Нижнюю часть

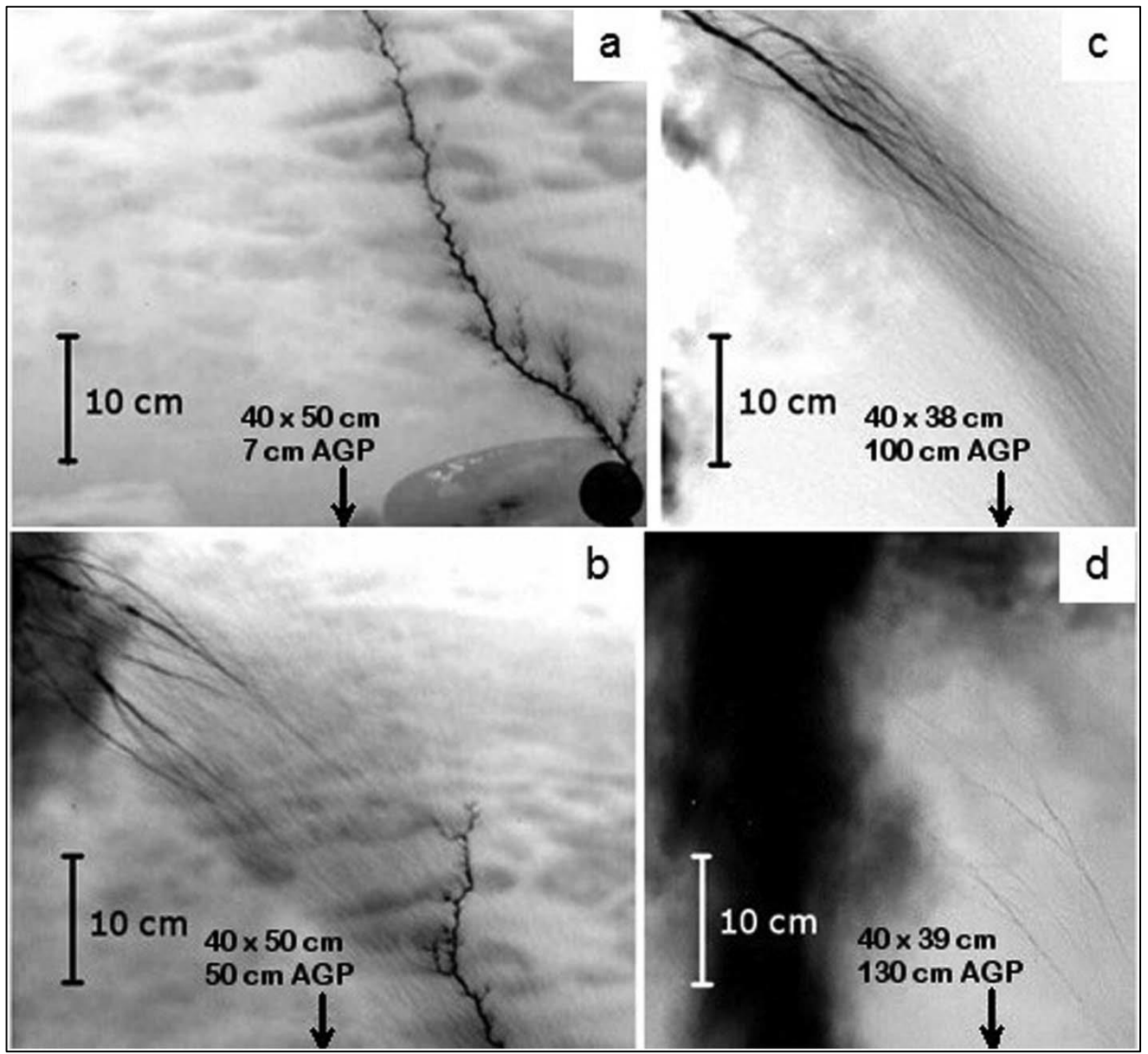

Рисунок 1.9 (адаптировано из [Kostinskiy et al., 2015a]). Инфракрасные изображения (инвертированные), полученные с экспозицией 6,7 мс, на которых видны процессы на различной высоте над заземленной плоскостью (AGP) и на разных горизонтальных расстояниях от оси облака: a) восходящий положительный лидер от заземленной сферы. UPFs не видно; b) верхняя часть восходящего лидера (внизу справа) и UPFs (вверху слева), оба внутри облака. По сравнению с Рисунком «a», поле зрения ИК-камеры было сдвинуто вверх и влево к оси облака; c) нижняя часть UPFs (внутри облака). Восходящий положительный лидер находится за пределами поля зрения ИК-камеры, которая была перемещена (относительно Рисунка «b») дальше вверх и влево; (d) То же, что и на Рисунке «c», но для верхней части UPF вблизи центральной части облака (см. темное образование слева). Обращает на себя внимание, что относительно слабые каналы UPFs разветвляются по направлению к оси облака, но не пересекают ось. Это направление разветвления противоположно тому, что показано на Рисунке «c», что говорит о том, что UPFs могут расширяться в разных направлениях.

UPFs можно увидеть на Рисунке 1.9c (обращает на себя внимание направление ветвления UPFs к восходящему положительному лидеру) и Рисунке 1.9d (ветвление UPFs направлено к самой плотной центральной части облака) соответственно. Рисунок 1.9d соответствует самому верхнему и крайнему левому полю зрения камеры в «сканирующем» эксперименте.

### 1.2.4. UPF, одновременно наблюдаемые, как в ИК, так и в видимом диапазонах

Поскольку обнаруженные в ИК-области необычные плазменные образования (UPFs) никогда ранее не наблюдались, желательно было обнаружить их и в видимом диапазоне, когда они выходят за видимые границы аэрозольного облака.  Как отмечалось выше, наблюдение UPFs в видимом диапазоне затруднено, потому что размер капель аэрозольного облака, обычно около 0,5 мкм, близок к длинам волн видимого света, так что капли эффективно рассеивают свет. Изображения, обычно, получаются размытыми, как на Рисунке 1.10 (вверху), если вообще получаются. Используемые изображения в видимом диапазоне встречаются редко. На Рисунке 1.11 мы показываем пример одновременного изображения в инфракрасном и редком видимом диапазоне одного и того же UPFs, который был инициирован близко к видимому краю облака. Видимое изображение соответствует фрагменту инфракрасного, но основные черты ИК-изображения четко различимы и в видимом диапазоне. Вероятно, что получение изображений в видимом диапазоне стало возможным в этом случае благодаря близости UPFs к краю облака, где оптическая плотность облака была относительно низкой. Время экспозиции ИК-камеры составляло 7,7 мс против 1 мкс для камеры видимого диапазона. Обращает на себя внимание, что ИК-камера обеспечивает значительно больший уровень детализации, хотя некоторые дополнительные детали изображения могут быть связаны не только с ее частотным диапазоном, но и с гораздо более длительным временем экспозиции. Основываясь на выдержке камеры видимого диапазона в 1 мкс и времени записи, UPFs, видимый на левой (меньшей) панели рисунка Рисунке 1.11, сформировался уже в пределах 1,4 мкс после первой вспышки короны с заземленной сферы (это

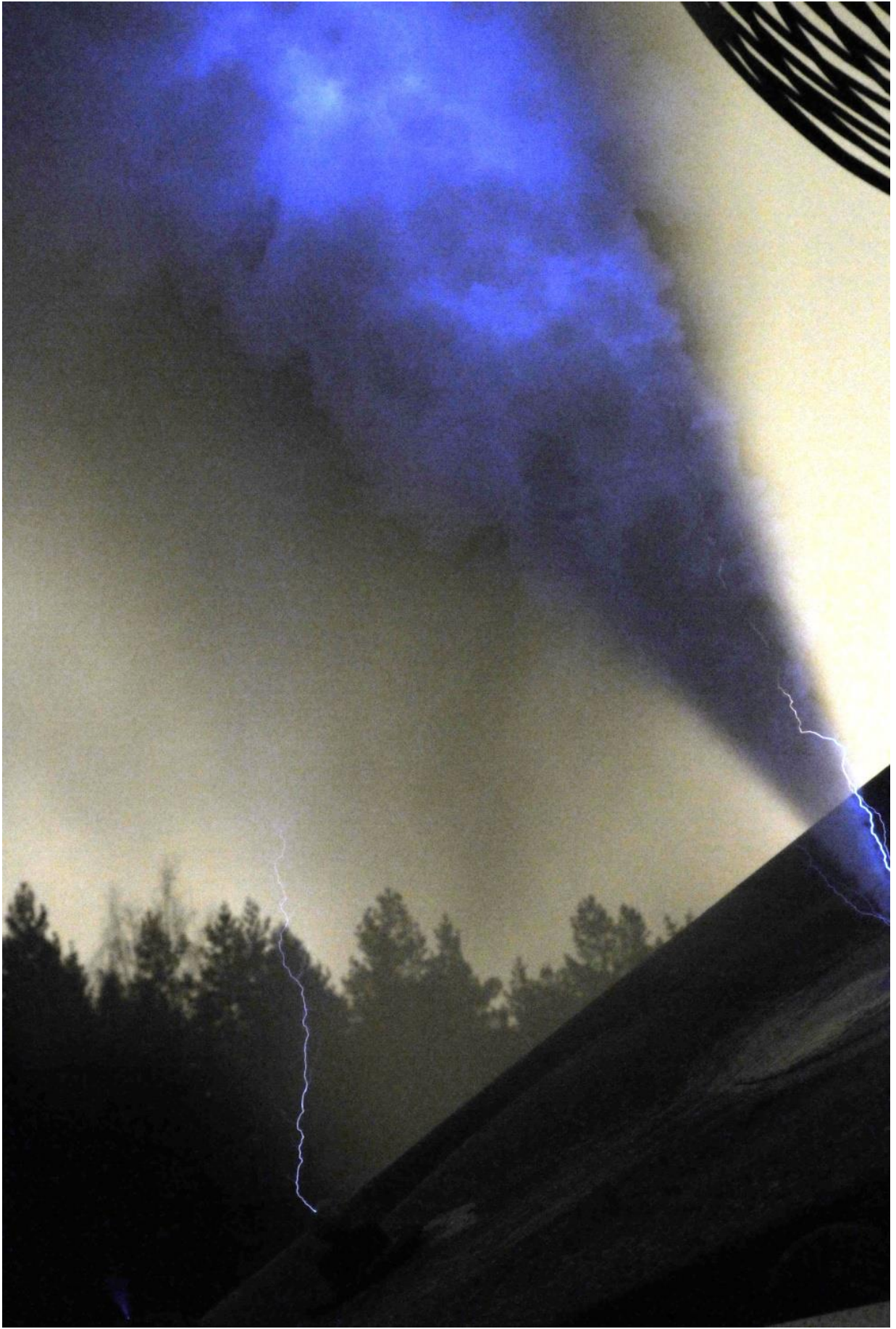

Рисунок 1.10. Фотография разрядов внутри облака заряженного аэрозоля (вверху облака).

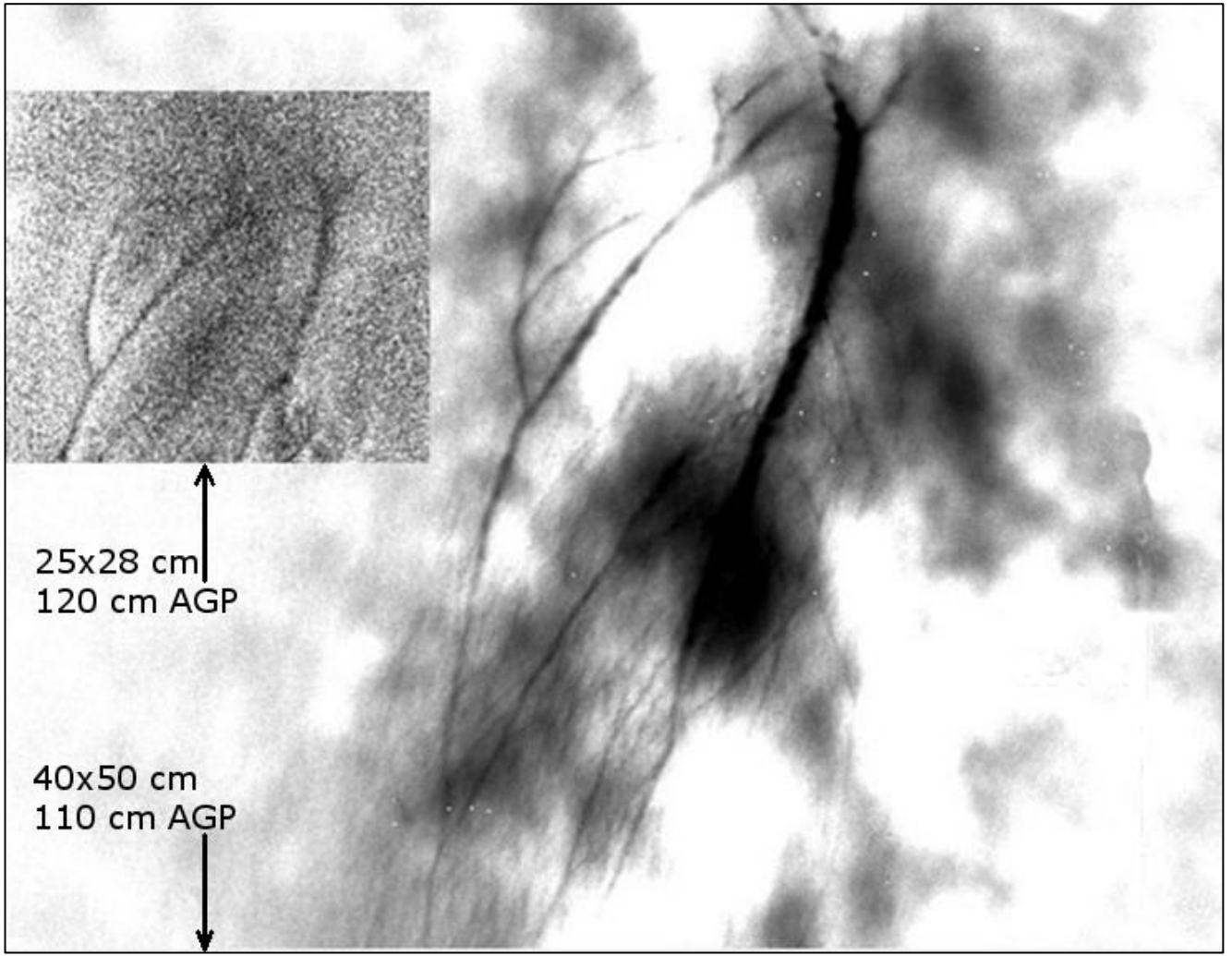

Рисунок 1.11 (адаптировано из [Kostinskiy et al., 2015a]). Одновременные изображения UPFs на краю облака в инфракрасном (справа) и видимом диапазоне (слева). Последнее (меньшее) изображение соответствует фрагменту первого, но основные черты ИК-изображения четко различимы в видимом диапазоне. Время экспозиции ИК-камеры составляло 7,7 мс против 1 мкс для камеры видимого диапазона. Обращает на себя внимание гораздо больший уровень детализации, обеспечиваемый ИК-камерой. AGP означает — «выше заземленной плоскости». Изображения инвертированы.

предположение будет подтверждено и уточнено более детальными экспериментами в главе 2).

Другим примером одновременного наблюдения UPFs в инфракрасном и видимом диапазоне является подробно описанное в главе 2 событие, изображенное на Рисунках 2.2. (видимое), 2.6 (ИК-изображение с анализом ИК-излучения из каналов лидера и UPFs).

Обратим внимание на то, что UPFs могут быть соизмеримы по размерам с положительными восходящими лидерами, и связаны с ними длинными положительными стримерами, как хорошо видно на Рисунке 1.12. UPFs могут быть даже длиннее, чем положительные восходящие лидеры, имея сложную иерархическую форму, как хорошо видно на Рисунке 1.13.

Также важно еще раз подчеркнуть принципиально объемный и сетевой характер возникновения и развития сразу нескольких UPFs со сложной пространственной структурой, который отличает их от развития положительных и отрицательных лидеров. Этот объемный характер хорошо виден, как на выше приведенных рисунках этого раздела, так и на Рисунках 1.14, 1.15.

## 1.2.5. Несколько UPFs, взаимодействующих между собой и с положительными лидерами в рамках одного события

Довольно частым явлением является появление в рамках одного события двух UPFs, которые взаимодействуют друг с другом и восходящим положительным лидером. Инициация нескольких плазменных образований, которые могут образовывать взаимодействующие цепочки плазменных объектов, внутри положительной стримерной вспышки является важным звеном в процессе превращений плазмы в длинные плазменные каналы  На Рисунке 1.16 в рамках одного события показаны два UPFs (указаны красными стрелками), которые взаимодействуют, благодаря положительным стримерам со стримерной короной восходящего положительного лидера (нижний UPFs) и друг с другом, образуя цепочку взаимосвязанных плазменных образований различной природы. Благодаря взаимодействию стримерных корон этих плазменных образований

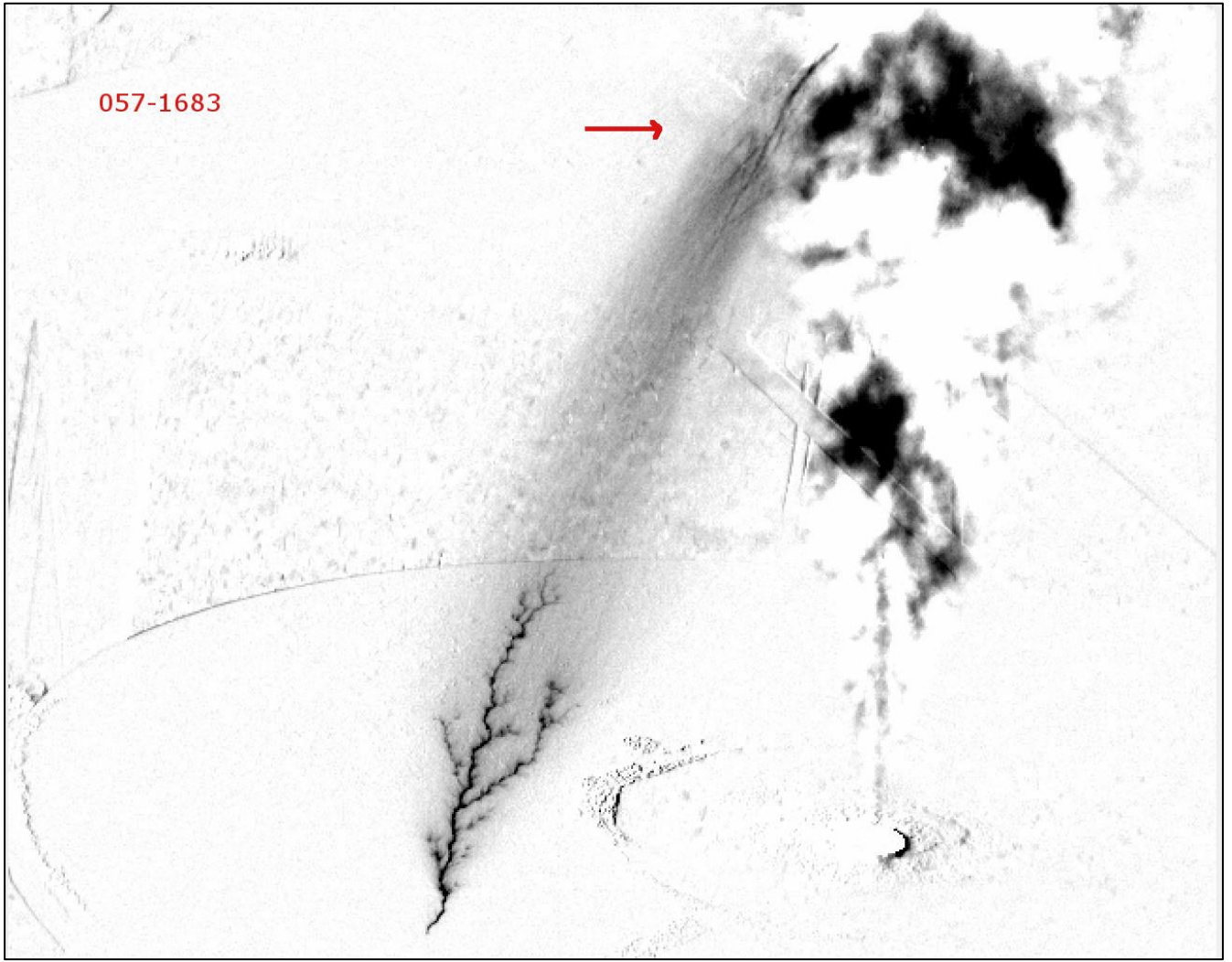

Рисунок 1.12 (адаптировано из [Andreev et al., 2014]). UPFs (размером около 30 см, указаны стрелкой) соизмеримы по размерам с положительным восходящим лидером (размер около 40 см). Между UPFs и лидером расстояние около 45 см. Изображение инвертировано.

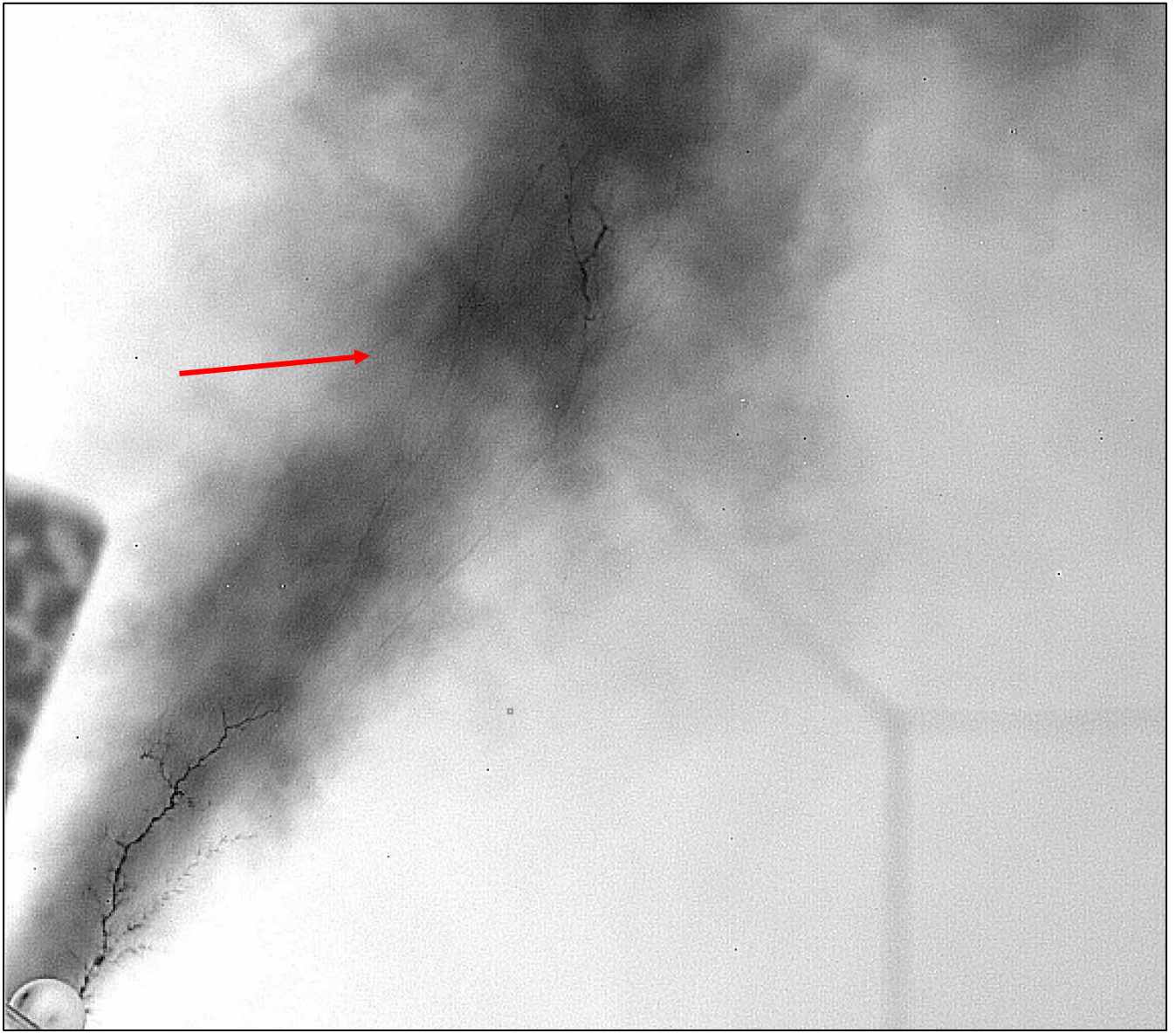

Рисунок 1.13. UPFs (указаны стрелкой) превышают по размерам положительный восходящий лидер. UPFs имеют сложную разветвленную форму. Изображение инвертировано.

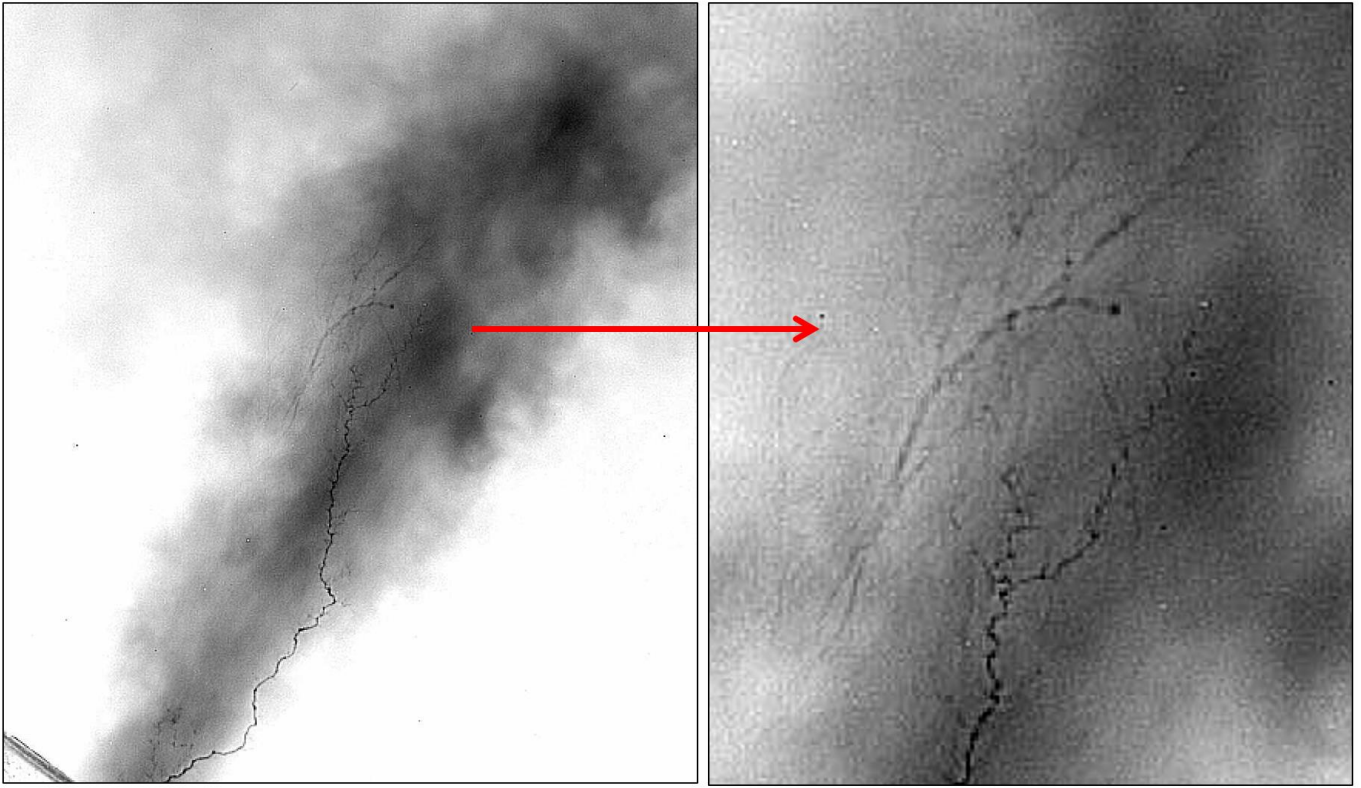

Рисунок 1.14 (адаптировано из [Andreev et al., 2014]). UPFs (вверху над положительным восходящим лидером и на правом увеличенном фрагменте, который указан стрелкой) показывают принципиально объемный характер возникновения и развития сразу нескольких UPFs со сложной пространственной структурой. Изображение инвертировано.

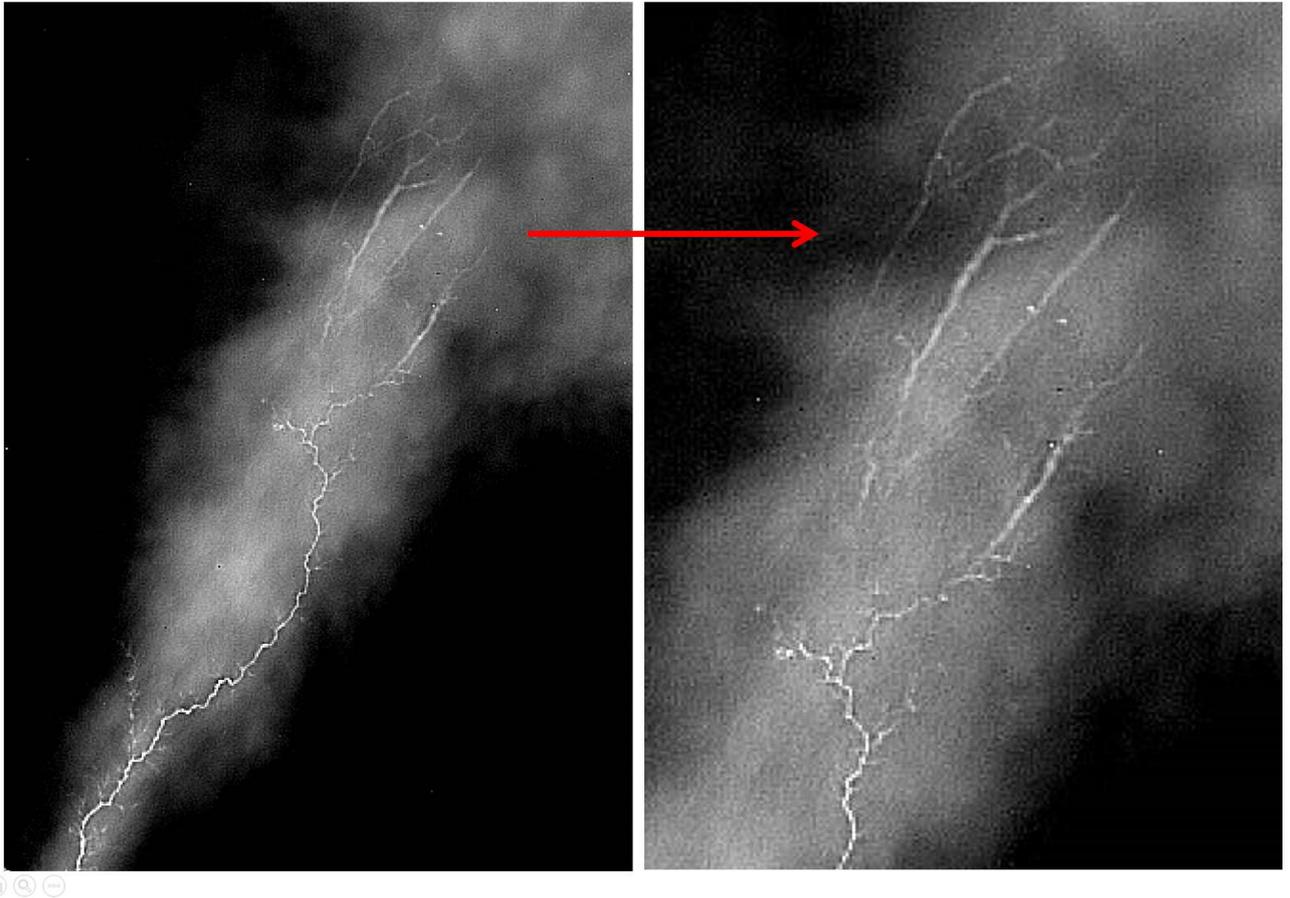

Рисунок 1.15 (адаптировано из [Andreev et al., 2014]). UPFs (вверху над положительным восходящим лидером и на правом увеличенном фрагменте, который указан стрелкой) показывают принципиально объемный характер возникновения и развития нескольких UPFs со сложной пространственной структурой. Верхние концы UPFs ветвятся в сторону отрицательного облака

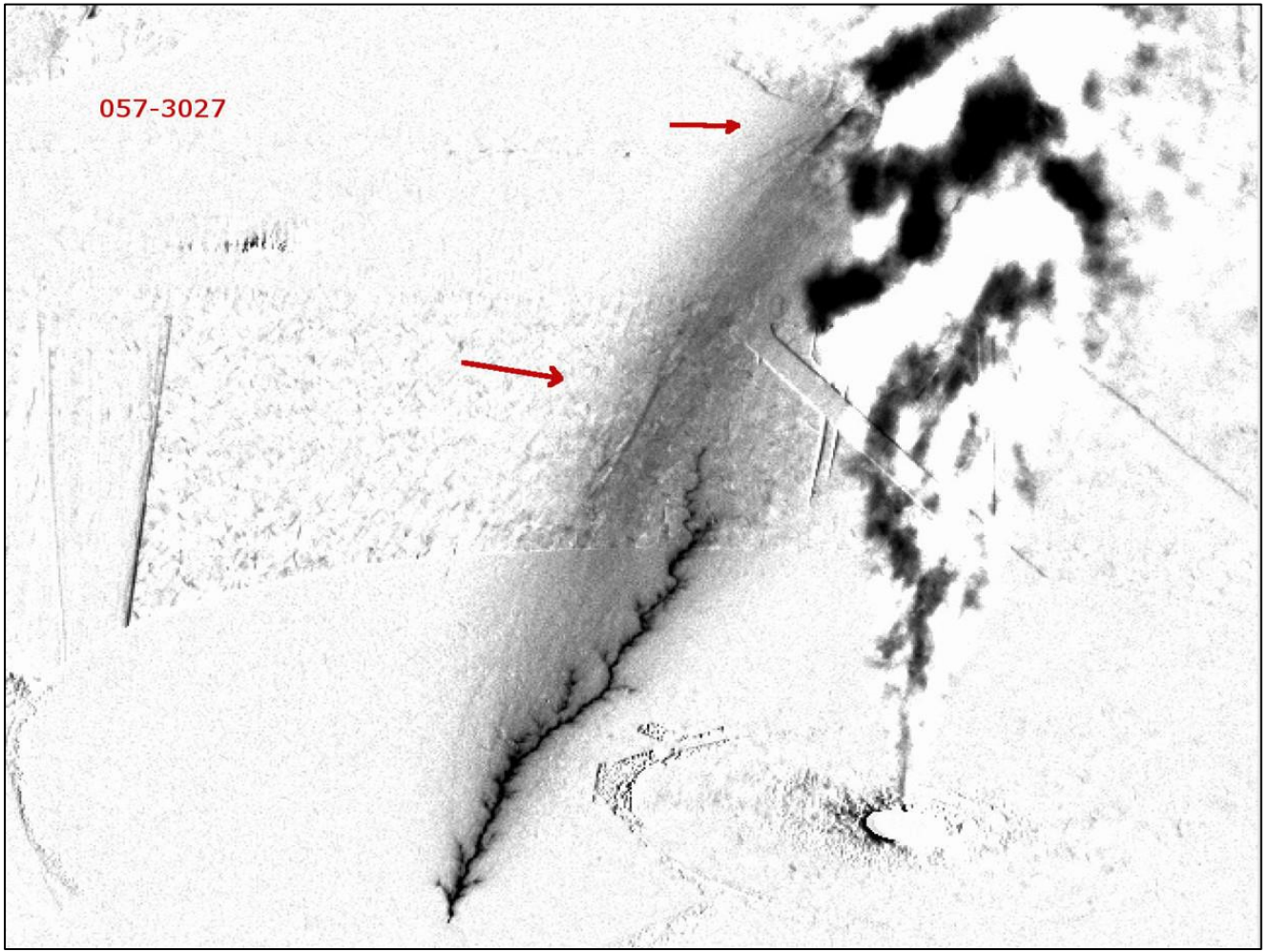

Рисунок 1.16 (адаптировано из [Andreev et al., 2014]). Два UPFs (указаны стрелками), которые взаимодействуют, благодаря положительным стримерам со стримерной короной восходящего положительного лидера и друг с другом, образуя цепочку взаимосвязанных плазменных образований различной природы. Характерно, что при этом общая стримерная корона является достаточно яркой. UPFs находятся на одной линии, идущей от заземленной плоскости, на расстоянии около 25 см друг от друга (в плоскости рисунка). Расстояние от нижнего UPFs до заземленной плоскости около 50 см (в плоскости рисунка). Изображение инвертировано.

их время жизни может увеличиваться. Вероятно поэтому, стримерная корона, соединяющая эти плазменные образования является достаточно яркой. UPFs находятся на одной линии, идущей от заземленной плоскости, на расстоянии около 25 см друг от друга (в плоскости рисунка), что, скорее всего связано с тем, что они инициировались внутри одной стримерной вспышки (как будет подробно показано в главе 2). Расстояние от нижнего UPFs до заземленной плоскости около 50 см. На Рисунках 1.17-1.18 мы также видим аналогичные события, где показаны два UPFs (указаны стрелками), которые взаимодействуют, благодаря положительным стримерам со стримерной короной восходящего положительного лидера. Они также образуют цепочку взаимосвязанных плазменных образований. UPFs также находятся примерно на одной линии, идущей от заземленной плоскости. Расстояние между UPFs на Рисунке 1.17 около 33 см друг от друга, а расстояние от нижнего UPFs до заземленной плоскости около 50 см. На Рисунке 1.18 расстояние между UPFs меньше и равно 17 см, а расстояние от нижнего UPFs до заземленной плоскости около 54 см.

На Рисунке 1.19 при взаимодействии двух UPFs (указаны стрелками), общая стримерная корона между UPFs является гораздо более яркой, чем между нижним UPFs и нижней частью положительного лидера. Возможно, UPFs и восходящий положительный лидер находятся не в одной плоскости в трехмерном пространстве. Особенно хорошо видна яркость стримерной короны между UPFs на фрагменте изображения 1.19.а (указан синей стрелкой). Расстояние между UPFs около 18 см, а расстояние от нижнего UPFs до заземленной плоскости около 58 см.

На Рисунке 1.20 хорошо видно, что стримерная корона между UPFs (стрелка на 2-м кадре) была настолько интенсивной, что она зафиксирована и на втором кадре вместе с положительным лидером, хотя между кадрами прошло не менее 2 мс.

## 1.3. Обсуждение результатов, полученных в главе 1

Чтобы проверить, действительно ли наблюдаемая яркость ИК-изображений UPFs свидетельствует об их высокой температуре, мы оценили различные источники ИК-излучения электрических разрядов в воздухе, включая тормозное излучение и

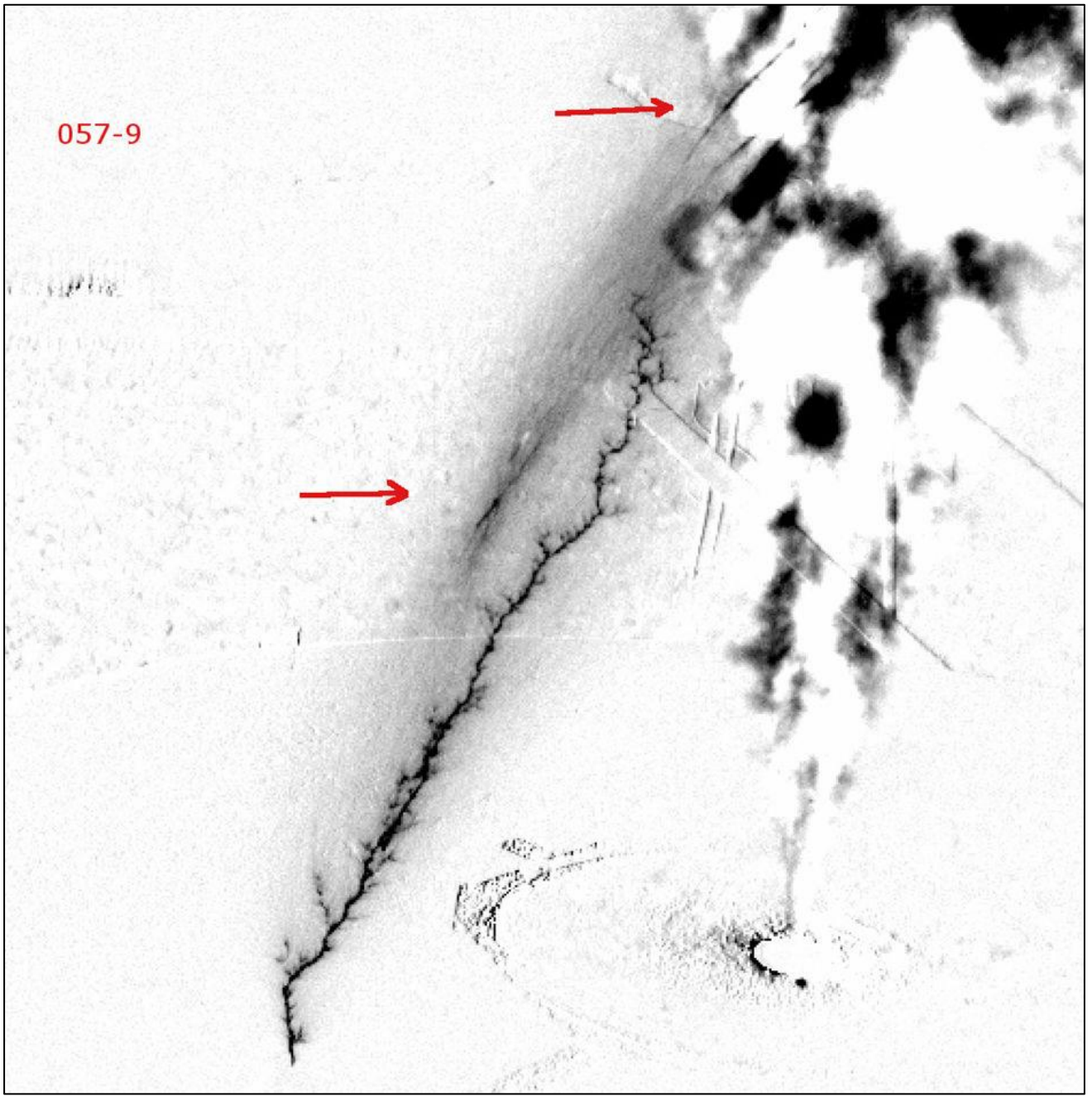

Рисунок 1.17 (адаптировано из [Andreev et al., 2014]). Два UPFs (указаны стрелками), которые взаимодействуют, благодаря положительным стримерам со стримерной короной восходящего положительного лидера и друг с другом, образуя цепочку взаимосвязанных плазменных образований различной природы. При этом общая стримерная корона между UPFs является более яркой, чем между нижним UPFs и нижней частью положительного лидера. Расстояние между UPFs около 33 см друг от друга, а расстояние от нижнего UPFs до заземленной плоскости около 50 см. Изображение инвертировано.

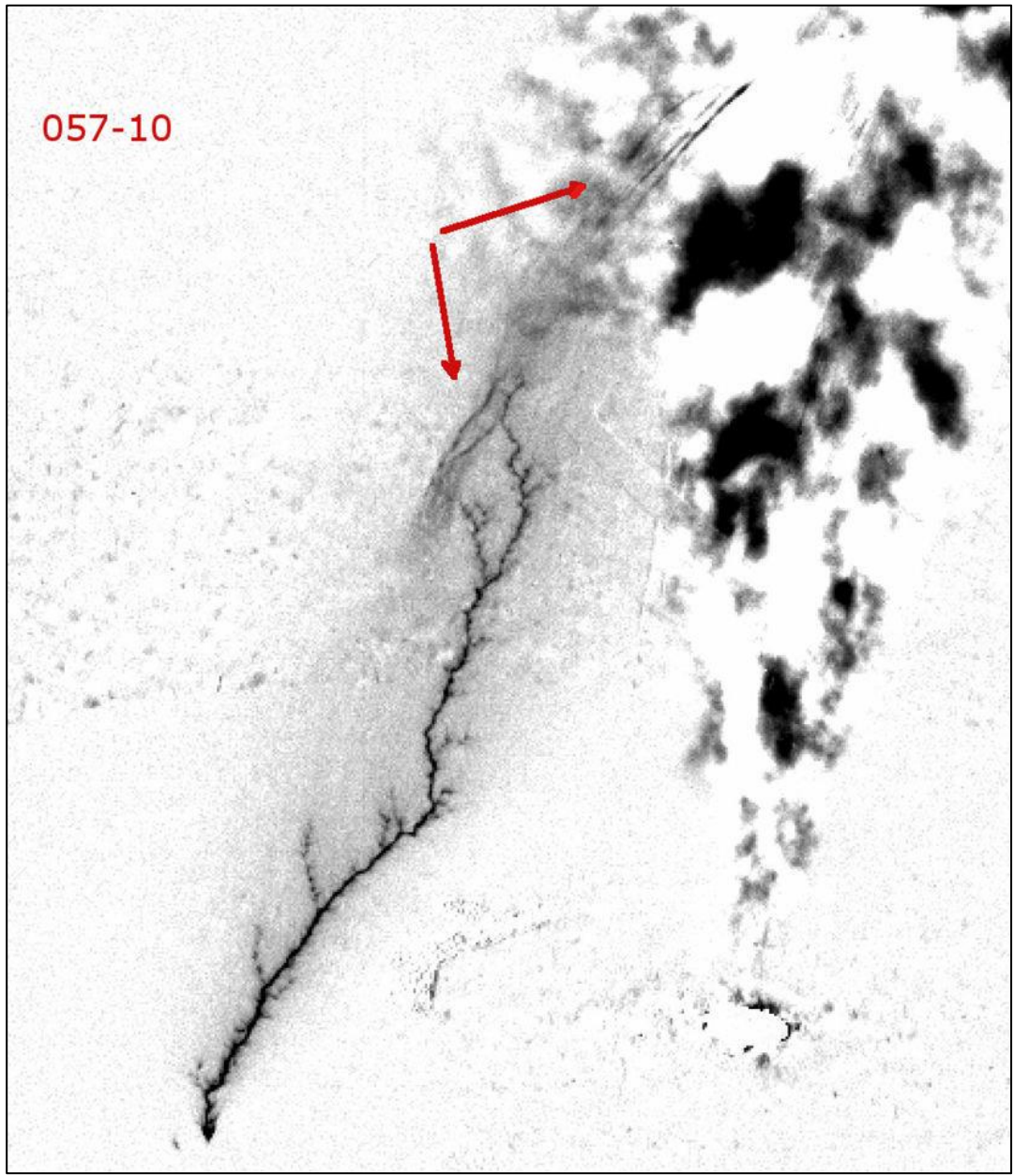

Рисунок 1.18 (адаптировано из [Andreev et al., 2014]). Два UPFs (указаны стрелками), которые взаимодействуют, благодаря положительным стримерам со стримерной короной восходящего положительного лидера и друг с другом, образуя цепочку взаимосвязанных плазменных образований различной природы. При этом общая стримерная корона между UPFs является более яркой, чем между нижним UPFs и нижней частью положительного лидера. Возможно, UPFs и восходящий положительный лидер находятся не в одной плоскости в трехмерном пространстве. Расстояние между UPFs равно 17 см, а расстояние от нижнего UPFs до заземленной плоскости около 54 см. Изображение инвертировано.

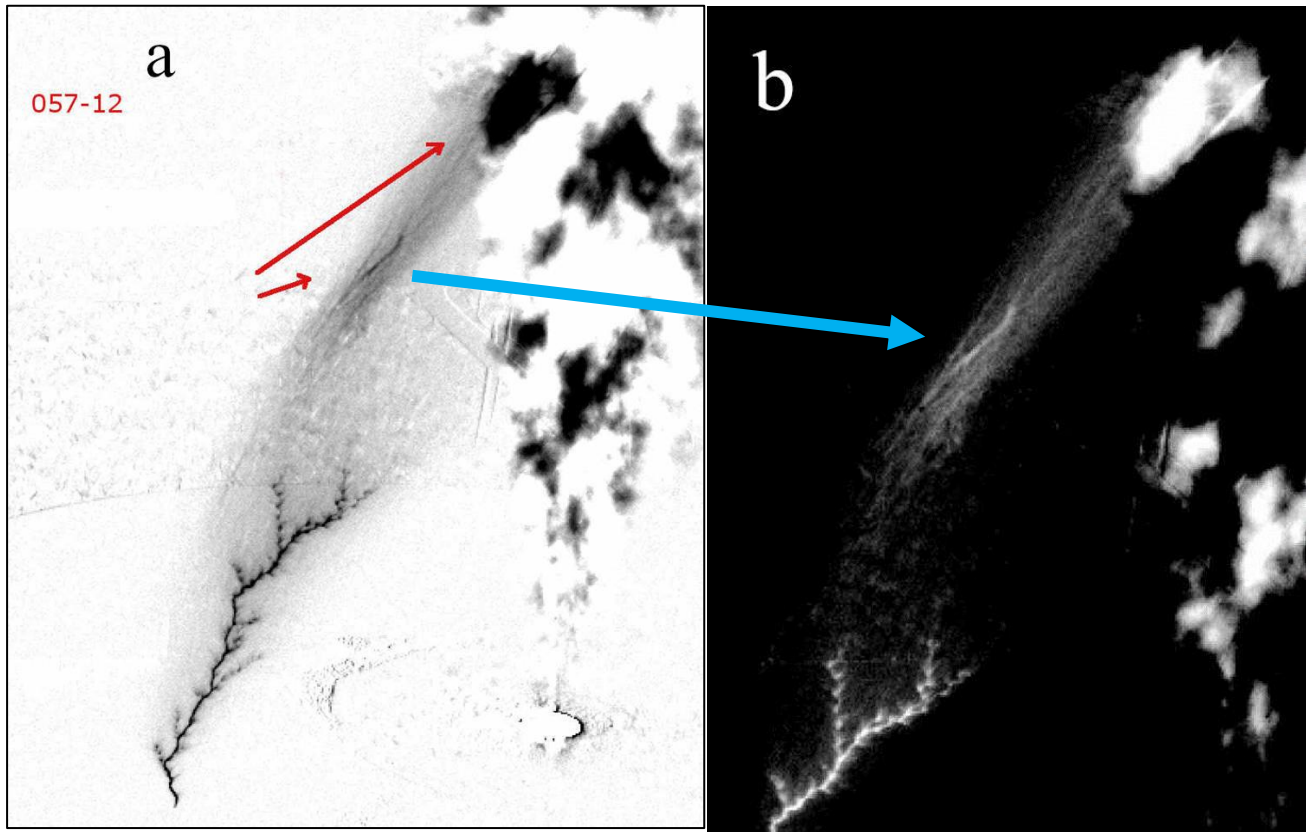

Рисунок 1.19 (адаптировано из [Andreev et al., 2014]). Два UPFs (указаны красными стрелками), которые взаимодействуют, благодаря положительным стримерам со стримерной короной восходящего положительного лидера и друг с другом, образуя цепочку взаимосвязанных плазменных образований различной природы. При этом общая стримерная корона между UPFs является более яркой, чем между нижним UPFs и нижней частью положительного лидера. Возможно, UPFs и восходящий положительный лидер находятся не в одной плоскости в трехмерном пространстве. a) Изображение «a» инвертировано; b) фрагмент изображения «a» (указан синей стрелкой), где более отчетливо видна стримерная корона между двумя UPFs.

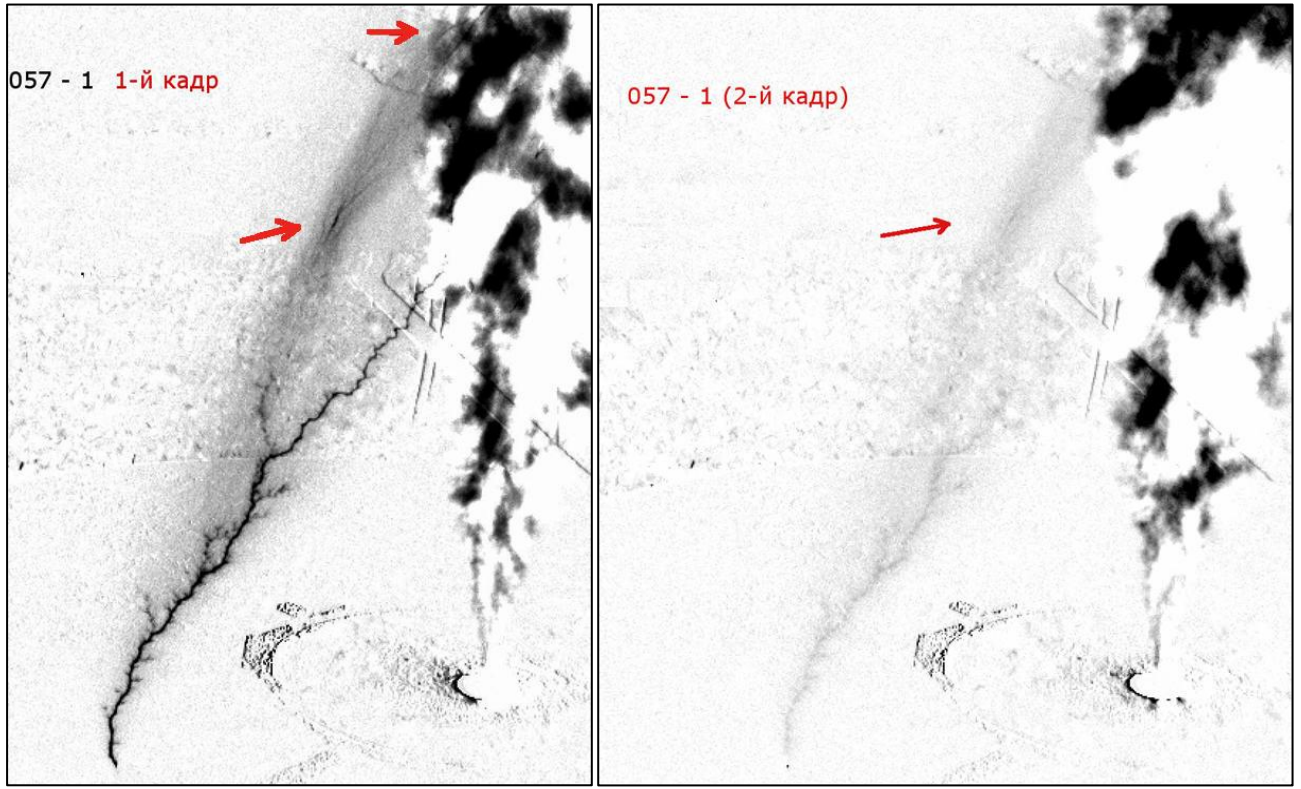

Рисунок 1.20 (адаптировано из [Andreev et al., 2014]). 1-й кадр: Два UPFs (указаны стрелками на первом кадре), которые взаимодействуют, благодаря положительным стримерам со стримерной короной восходящего положительного лидера и друг с другом, образуя цепочку взаимосвязанных плазменных образований различной природы. 2-й кадр: изображение того же процесса, записанное через минимум 2 мс после 1-го кадра. Стримерная корона между UPFs (красная стрелка) была настолько интенсивной, что она зафиксирована и на втором кадре вместе с положительным лидером.

рекомбинационное излучение (в обоих случаях непрерывные спектры), а также излучение за счет электронных переходов между высшими электронными уровнями молекул и излучение колебательно-возбужденных молекул (в обоих случаях дискретные спектры). Предварительные оценки показывают, что в условиях неравновесного разряда, когда нагрев и возбуждение газа незначительны, основной вклад в излучение в диапазоне чувствительности нашей ИК-камеры ($\lambda$ = 2,7–5,5 мкм) вносят молекулы $CO_2$, возбужденные прямым электронным ударом в несколько нижних уровней антисимметричной колебательной моды. Поскольку излучение, возникающее при переходе с первого колебательного уровня в основное состояние, сильно поглощается окружающим воздухом, ИК-камера должна регистрировать только излучение, связанное с переходом с более высокого колебательного уровня на соседний более низкий уровень.

В отсутствие ионизации электронным ударом (когда разряд переходит в стадию диссипации) возбуждение колебательных уровней $CO_2$ происходит за счет эффективного колебательного обмена с колебательно-возбужденными молекулами $N_2$. Релаксация ИК-излучения на этой стадии связана с тушением колебательно-возбужденного $N_2$, что в основном осуществляется за счет столкновений с молекулами $H_2O$ (водяным паром). Характерное время тушения составляет около 10 мс, что сравнимо со временем экспозиции кадра ИК-камеры и, вероятно, намного больше типичной продолжительности активной (ионизация электронным ударом) стадии разряда. В результате послесвечение на стадии диссипации разряда дает основной вклад в ИК-изображение разряда.

Интенсивность ИК-излучения на стадии неравновесного послесвечения определяется степенью колебательного возбуждения $N_2$ и конечной температурой газа, которую он приобретает после колебательно-поступательной (VT) релаксации, поскольку основная часть полной электрической энергии разряда в сильно подпороговых полях идет на колебательное возбуждение $N_2$. После VT-релаксации устанавливается частичное термодинамическое равновесие (что означает, что температура газа, колебательная температура и температура электронов равны друг другу) и, следовательно, интенсивность ИК-излучения существенно определяется степенью нагрева газа, которая аналогична ситуации при неравновесном свечении воздуха, рассмотренной выше.

Основываясь на нашем анализе, мы пришли к выводу, что яркость ИК-изображений UPFs разумно представляет собой конечную температуру газа, которая достигается в результате процесса разряда (с учетом их размера, поскольку яркость оптически тонких объектов пропорциональны их пространственной протяженности вдоль луча зрения).

В настоящее время проводится более детальный анализ кинетики колебаний воздуха в течение всего процесса разряда (включая послесвечение), необходимый для определения количественной зависимости между интенсивностью ИК-излучения и конечной температурой газа.

UPFs часто включают в себя несколько более или менее параллельных каналов относительно высокой яркости, которые постоянно связаны между собой большим количеством более слабых каналов или ответвлений (см., например, Рисунки 1.3, 1.8, 1.9, 1.11). В результате общая структура UPFs выглядит как сеть каналов нерегулярно изменяющейся яркости, которые пронизывают относительно большую облачную область, в отличие от лидеров, у которых есть основной канал с ветвями, указывающими направление его распространения, более или менее регулярно меняющейся яркости со стримерными зонами на его концах. Таким образом, морфология UPFs явно отличается от морфологии лидеров. Кроме того, вторичные каналы часто появляются и заканчиваются в разных точках одного и того же магистрального канала или исходят из общей точки в пространстве и заканчиваются на соседнем канале, образуя своего рода петли или разрывы в общей структуре канала (см., в частности, сотовую структуру относительно простой сети каналов UPFs на рисунках 1.8, а также на рисунках 1.9.b и c). Последнее поведение никогда не наблюдалось у лидеров. На основании вышеизложенного мы заключаем, что UPFs — уникальное явление, резко отличающееся от лидера. Поэтому нами этим плазменным образованиям было дано собственное название. Кроме того, ИК-яркость стримеров в искровых разрядах существенно отличается от яркости лидера, и наблюдается довольно резкая граница между ярким лидерным каналом и зоной слабосветящихся стримеров. Напротив, UPFs обычно демонстрируют более постепенное изменение яркости вдоль канала (сравните, например, Рисунки 1.9.a (лидер) и 1.9.c (UPFs)).

Как интерпретировать наблюдаемую морфологию UPFs? Известно (например, [Базелян и Райзер, 1997]), что основной причиной контракции каналов как в стримерах, так и в лидерах является сильная зависимость скорости ионизации электронным ударом от так называемой приведенной напряженности электрического поля ($E/N$), которая равна соотношение реальной напряженности электрического поля $E$ и плотности молекул воздуха $N$. Эта зависимость обеспечивает положительную обратную связь, приводящую к быстрому уменьшению кривизны ионизированной области (радиуса вершины канала) с ростом поля при переходе от лавины к стримеру и к быстрому уменьшению плотности воздуха с повышением температуры (из-за так называемой ионизационно-перегревной неустойчивости) в случае перехода стример-лидер. Различные части UPFs развиваются в областях с разной плотностью объемного заряда, что может привести к неоднородному нагреву плазменного канала и срабатыванию механизма ионизационно-перегревной неустойчивости только на определенных участках канала (процесс может быть прерван в случае недостаточной величины местного заряда). Описанный механизм, учитывающий неоднородное распределение заряда в облаке, может объяснить широкий спектр яркости большинства наблюдаемых UPFs и изменчивость их структуры. Природа UPFs типа спейс-лидера (см. Рисунок 1.7), которые появляются недалеко от заземленной плоскости, менее ясна и требует дальнейших исследований.

## 1.4. Результаты, полученные в более ранних работах, в которых, возможно, также фиксировались UPFs

[Temnikov, 2012a], возможно, также ранее зафиксировал UPF. К сожалению, из-за плохого пространственного разрешения, это нельзя утверждать с уверенностью. На Рисунке 1.21, кадр (5) можно видеть ярко светящийся объект внутри промежутка между заземлённой плоскостью и облаком. Возможно, данный плазменный объект является UPFs (подобный зафиксированному на Рисунке 1.7). [Temnikov, 2012a] называет его «плазменным стемом, который первоначально возник в зазоре». По-видимому, из этого UPFs («стема») разовьётся двунаправленный лидер, который соединится с восходящим положительным лидером, что приведет к квазиобратному удару на кадре (7).

## 1.5. Выводы главы 1

- Впервые внутри заряженных аэрозольных облаков обнаружены «необычные плазменные образования» (UPFs), которые представляют из себя

иерархические сети плазменных каналов, ранее не обнаруженные в других экспериментах

- Морфология UPFs принципиально отличается от морфологии всех ранее открытых разрядных каналов в квазипостоянных электрических полях в частности, от отрицательных и положительных лидеров

- UPFs могут рождаться парами и взаимодействовать друг с другом и восходящим положительным лидером, что повышает время их жизни

- UPFs могут «пронизывать» трехмерной сетью значительные объемы аэрозольного облака

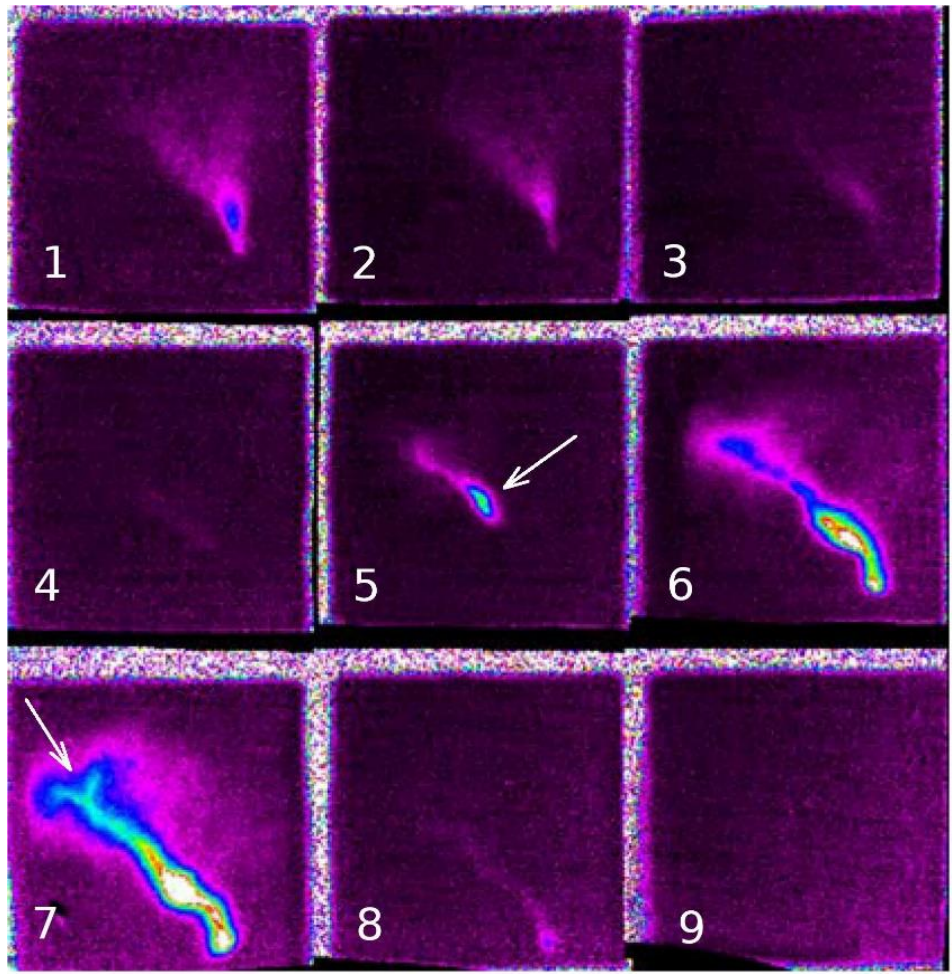

Рисунок 1.21. (рис. 18 из [Temnikov, 2012a]). Возможно, на этом рисунке зафиксирован на кадре (5, стрелка) UPFs (подобный зафиксированному на Рисунке 1.6). [Temnikov, 2012a] называет его «плазменным стемом», который первоначально возник в зазоре. По-видимому, из этого UPFs («плазменного стема») разовьётся двунаправленный лидер, который приведет к квазиобратному удару на кадре (7). Размер кадров 70х70 см, длительность кадра 3,0 мкс, интервал между кадрами 0,1 мкс.

**ГЛАВА 2. Инициация необычных плазменных образований (UPFs) внутри положительной стримерной вспышки, поддерживаемой электрическим полем отрицательно заряженного облака водного аэрозоля**

В главе 1 и статьях [Kostinskiy et al., 2015a; Kostinskiy et al., 2015b] было описано, как внутри отрицательного искусственно заряженного аэрозольного облака, благодаря ИК-камерам диапазона 2.5-5.5 мкм, был обнаружен класс новых «необычных плазменных образований» (в главе 4 описаны UPFs, обнаруженные внутри положительного аэрозольного облака [Kostinskiy et al., 2015b]). Данные плазменные образования получили название — «unusual plasma formations» (UPFs) [Kostinskiy et al.,2015a]. Причина и механизм образования UPFs в этих работах не были установлены. В данной главе описывается результаты установления одной из причин образования UPFs. Нами было зафиксировано, что UPFs возникают внутри объёма первой положительной стримерной вспышки, стартующей с заземленного электрода, с начальным током в диапазоне 1-1.5 А. UPFs появляются заведомо до того, как стримерная корона или канал восходящего с заземленного электрода положительного лидера смогут повлиять на эти процессы, происходящие в непосредственной близости от края или внутри аэрозольного облака. Сравнение интенсивностей ИК-излучения канала «обычного» положительного восходящего лидера, восходящего с заземленного электрода, и ИК-излучения UPFs позволяет предположить, что, по крайней мере, некоторые UPFs являются такими же горячими плазменными каналами, как и восходящий положительный лидер. Время возникновения UPFs оказалось не более, чем 1-1.5 мкс. В данных экспериментах расстояние, которая проходила стримерная вспышка от электрода до оси облака, составляло 1-1.5 м. Первые 30-40 см стримерная вспышка была расходящейся под углом 5-7°, а затем ее диаметр оставался приблизительно постоянным в диапазоне 5-10 см. Средняя скорость движения фронта стримерной вспышки была в диапазоне $5\text{-}7 \cdot 10^5$ м/с. К моменту вхождения стримерной вспышки в облако, при длине вспышки 1-1.5 м, ее заряд был в диапазоне 0.2-0.3 мкКл. На большей части своей длины, начиная от заземленного электрода, стримерная вспышка была квази-однородна, то есть не имела внутри видимых неоднородностей с масштабом, меньшим поперечного размера вспышки. Вблизи границы облака внутри стримерной вспышки могли появляться нагретые плазменные каналы (UPFs), которые протягивались внутрь облака.

Надо отметить, что на ИК изображениях UPFs имели близкую или даже большую яркость, чем положительные восходящие лидеры [Kostinskiy et al., 2015a; Kostinskiy et al., 2015b]. Это дало основание предполагать, что UPFs являются плазменными каналами, нагретыми не менее, чем восходящие положительные лидеры. В экспериментах с искусственно заряженным отрицательным аэрозольным облаком положительные стримерные вспышки и восходящие положительные лидеры стартовали с заземлённой плоскости и распространялись по направлению к облаку. Полученные изображения UPFs [Kostinskiy et al., 2015a; Kostinskiy et al., 2015b] были интегральными изображениями, поскольку длительность экспозиции ИК-камеры, 2-3 мс, значительно превышала характерное время развития внутриоблачных разрядов. Поэтому увидеть динамику зарождения и развития UPFs и сделать какие-либо заключения о механизме их появления, основываясь на экспериментах, было невозможно.

Задачей данной главы было установление процессов, которые приводят к возникновению UPFs. Для исследования причин возникновения UPFs экспериментальная установка, аналогичная описанной в [Kostinskiy et al., 2015a], была дополнена микроволновой диагностикой, что, вместе с другими приборами, позволило получить экспериментальные данные, которые указывают на возможный механизм возникновения UPFs, как горячих плазменных каналов, образованных внутри длинной положительной стримерной вспышки.

Длинными стримерами мы в данной статье называем стримеры, потерявшие гальваническую (токовую) связь с точкой их инициации (возникновения). Характерную длину проводящего канала стримеров можно оценить исходя из скорости движения их головки $v_{str}$ и времени прилипания электронов $\tau_a$ в канале за головкой стримера.

$$L_{str} \approx v_{str}\tau_a = 2 - 10 \cdot 10^7 \frac{\text{см}}{\text{с}} \cdot 10^{-7}\text{с} \approx 2 - 10 \text{ см}, \qquad (2.1)$$

[Bazelyan & Raizer, 1998)], [Kossyi et al., 1992].

## 2.1. Экспериментальная установка

Эксперименты проводились на установках Высоковольтного исследовательского центра (г. Истра) РФЯЦ – ВНИИ Технической физики имени ак. Е.И. Забабахина (http://www.ckp-rf.ru/usu/73578/). Экспериментальная установка, аналогичная описываемой в главе 1, применяемая в данных экспериментах показана на Рисунке 2.1. Облако заряженного водного аэрозоля (1) было создано парогенератором (2.1) и высоковольтным источником (2.2), связанным с иглой, создающей плазменную корону между иглой и заземленной плоскостью (3). Игла была расположена в сопле (2.3), через которое вылетал поток пара. Пар в сопле имел температуру около 120–140°C и давление в пределах 0.2–0.6 МПа. Поток двигался с начальной скоростью около 340-400 м/с с апертурой угла 28°, формируя затопленную турбулентную струю. Сопло с иглой было расположено в центре заземлённой плоскости (3) диаметром 2 м. В результате быстрого охлаждения, пар конденсировался в водяные капли со средним радиусом около 0.5 мкм. Ионы коронного разряда между кончиком иглы и соплом (2.3) заряжали водяные капли. Коронный разряд возникал от источника постоянного напряжения 10–20 кВ, который был подключен к игле. Ток, переносимый заряженной струей аэрозоля, находился в диапазоне от 100 до 160 мкА. Когда общий заряд, аккумулированный в облаке, достигал величины ~50 мкКл, стримерные вспышки и/или лидеры спонтанно возникали между заземлёнными объектами и облаком. В случае отрицательно заряженного облака, большинство стримерных вспышек и восходящих положительных лидеров развивалось от заземлённой металлической сферы (4). Металлическая сфера диаметром 5 см (4) находилась на расстоянии 0.85 м от центра заземлённой плоскости (3), так, чтобы положительные стримерные вспышки и положительные лидеры, восходящие с металлической сферы, распространялись перпендикулярно направлению диагностического микроволнового пучка. Верхняя точка сферы находилась на расстоянии 12 см выше плоскости.

Токи стримерной вспышки или восходящего положительного лидера, появляющихся на поверхности сферы (4), измерялись малоиндуктивным омическим шунтом с сопротивлением 1 Ом, сигнал от которого приходил на цифровой осциллограф (5). Как только ток превышал определенное установленное значение, осциллограф (5)

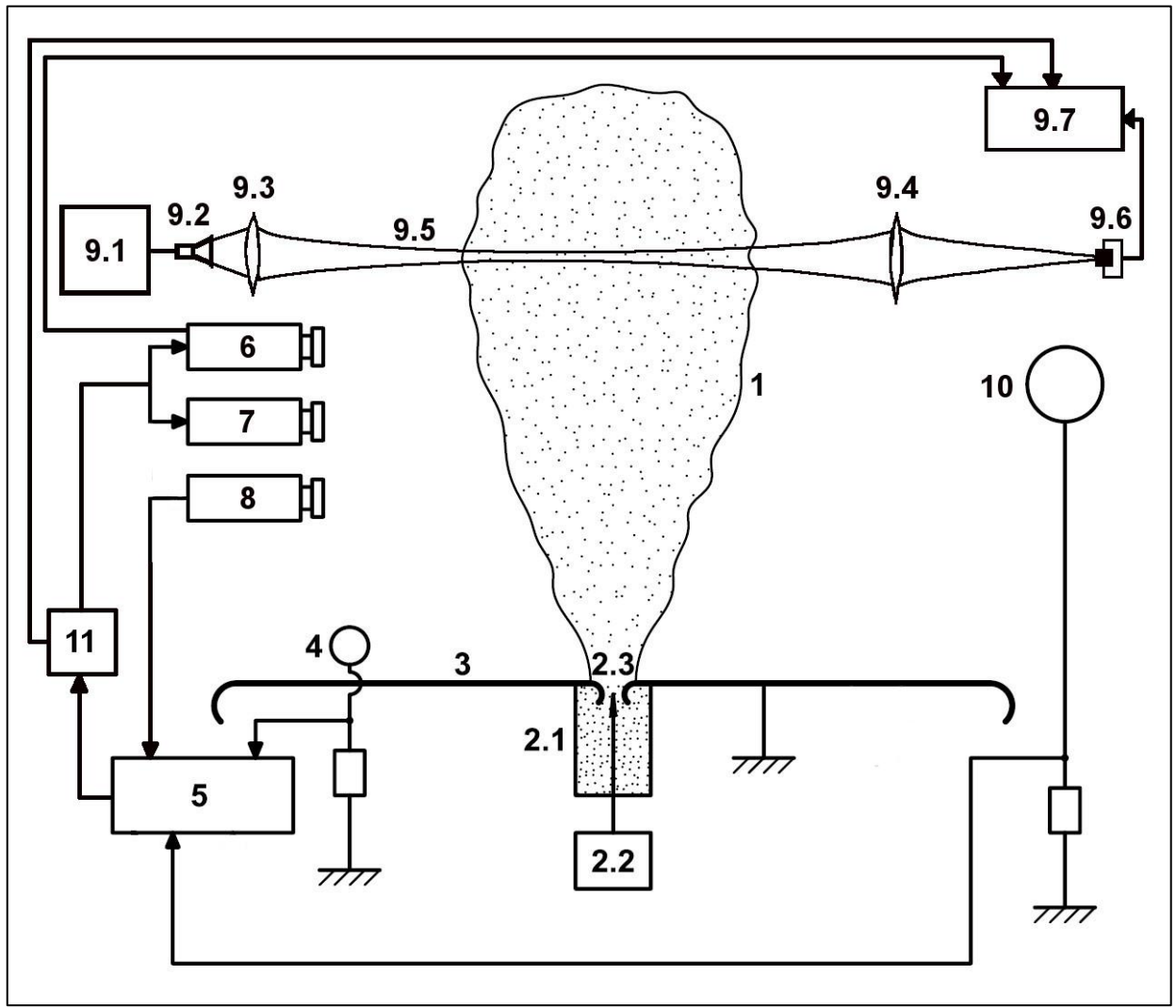

Рисунок 2.1 (адаптировано из [Kostinskiy et al., 2021]). Экспериментальная установка: 1 — облако заряженного водного аэрозорля, 2.1 — генератор пара, 2.2 — высоковольтный источник, подающий напряжение для создания коронного разряда в сопле, 2.3 — сопло, 3 — заземленная металлическая плоскость, 4 — 5 см шарик, соединенный с измерительным шунтом, 5 — осциллограф, 6 — высокоскоростная камера с усилением изображения 4Picos, 7 — высокоскоростная ИК-камера FLIR-7700, 8 — ФЭУ, 9.1 — микроволновой генератор G4-91, 9.2 — рупорная антенна, 9.3-9.4 — диэлектрические линзы, 9.5 — СВЧ пучок, 9.6 — открытый конец приемного волновода, СВЧ усилитель и СВЧ диод, 9.7 — осциллограф, 10 — 50 см сфера для мониторинга изменений заряда облака, 11 — генератор импульсов записывал ток, сигнал с фотоэлектронного умножителя — ФЭУ (8) и сигнал с металлического шара диаметром 50 см (10), а также подавал триггерный сигнал на генератор импульсов (11). Генератор импульсов (11) формировал триггерный TTL сигнал для запуска высокоскоростной камеры видимого диапазона (300-800 нм) с усилением изображения 4Picos (6), инфракрасной скоростной камеры FLIR SC7700M (7), а также второго осциллографа (9.7), который записывал сигнал с микроволнового диода (9.6), фиксировавшего микроволновое излучение (9.5), прошедшее через облако (1). Инфракрасная камера FLIR SC7700M ($\lambda\approx2.5$-5.5 мкм) работала со скоростью 412 кадров в секунду (время экспозиции составляло 2.4 мс), размер изображения на матрице был 320 × 256 пикселей. Высокоскоростная камера видимого диапазона 4Picos фиксировала изображение на матрицу 1360×1024 пикселей с экспозицией от 50 нс до 10 мкс. Камеры были установлены на расстоянии 8.5 м от сопла, формирующего аэрозольное облако (2.3) в направлении оси распространения микроволнового пучка. ИК-камера была снабжена германиевым объективом с фокусным расстоянием 50 mm и апертурой f/2. Для камеры видимого изображения использовался стеклянный объектив с фокусным расстоянием 50 мм и апертурой f/0.95. Угол зрения (диаграмма направленности) ФЭУ составлял $\sim10^{o}$, размер области зрения ФЭУ в месте положения облака составлял $\sim1$ м$^2$.

Источником микроволнового излучения был генератор G4-91 (9.1). Выходная мощность генератора была 5 мВт, частота излучения 35 ГГц (длина волны 8.5 мм). Генератор работал в непрерывном режиме, относительный уровень флуктуации выходной мощности составлял $\sim10^{-3}$. Сходящийся микроволновой пучок с гауссовым профилем формировался рупорной антенной (9.2) и диэлектрическими линзами (9.3, 9.4). Перетяжка микроволнового пучка (9.5) находилась на оси аэрозольного потока. Угол между осью микроволнового пучка и осью аэрозольного потока составлял 85-87$^0$. Диаметр микроволнового пучка в области перетяжки был $\sim10$ см (в исследуемой области пучок был почти цилиндрическим, а в районе видимого края аэрозольного облака он был всего на 3% шире, чем в его центре). Расстояние от оси микроволнового пучка до заземлённой плоскости было $\sim1$ м. Поляризация СВЧ-излучения была линейной (вертикальной). Микроволновое излучение, прошедшее через облако, фокусировалось диэлектрической линзой (9.4) в открытый конец приемного волновода. После этого излучение усиливалось с помощью микроволнового усилителя 20 дБ и детектировалось микроволновым диодом (9.6). Выходной уровень сигнала с микроволнового диода регистрировался осциллографом (9.7). Относительное ослабление проходящего через облако микроволнового излучения определялось отношением значения выходного уровня сигнала с диода к невозмущенному уровню (в отсутствие облака). Основным источником шума, определяющим чувствительность микроволновой диагностики в целом, была нестабильность выходной мощности микроволнового генератора, которая составляла $\sim 10^{-3}$; это же значение определяло минимальный уровень поглощения зондирующего микроволнового излучения, который мы могли зарегистрировать. Незаряженное облако и заряженное облако в промежутках между внутриоблачными разрядами не вносило заметного ослабления в зондирующее СВЧ излучение. Фотоэлектронный умножитель был направлен в верхнюю половину облака на высоту около 0.8-1 м над плоскостью. Постоянная времени ФЭУ составляла несколько миллисекунд и его данные были важны только на фронте оптических событий.

## 2.2. Экспериментальные результаты

На Рисунке. 2.2 изображены первая стримерная вспышка (Рисунок 2.2.I) и последующее развитие положительного восходящего лидера со стримерной короной (Рисунок 2.2.II). Эти два последовательных кадра, получены камерой с усилением изображения 4Picos. Выдержка первого кадра (Рисунок 2.2.I) составляла 2 мкс, второго кадра (Рисунок 2.2.II) составляла 10 мкс, промежуток времени между кадрами был равен 1 мкс. На первом кадре Рисунок 2.2.I изображена первая стримерная вспышка (1), которая возникла на заземленной сфере (2) и распространялась до самого видимого края облака (3). После этого стримеры вспышки вошли в облако и пересекли под почти прямым углом область прохождения микроволнового пучка (5). Стримерные головки прошли расстояние по дуге около $S_{st} \approx 1.2$ м от заземленной сферы (2) до области прохождения микроволнового пучка примерно за $\tau_{st} \approx 1.7$ мкс. Факт движения стримеров внутри аэрозольного облака фиксируется по пику поглощения микроволнового излучения на Рисунке 2.3(4). Полуширина по полувысоте пика микроволнового поглощения была около 135 нс и пик (Рисунок 2.3(4)) находился внутри времени экспозиции первого кадра 4Picos (Рисунок 2.3(6)) примерно за 0.85 мкс до окончания экспозиции. Таким образом, средняя скорость стримеров первой вспышки стримерной короны в плоскости

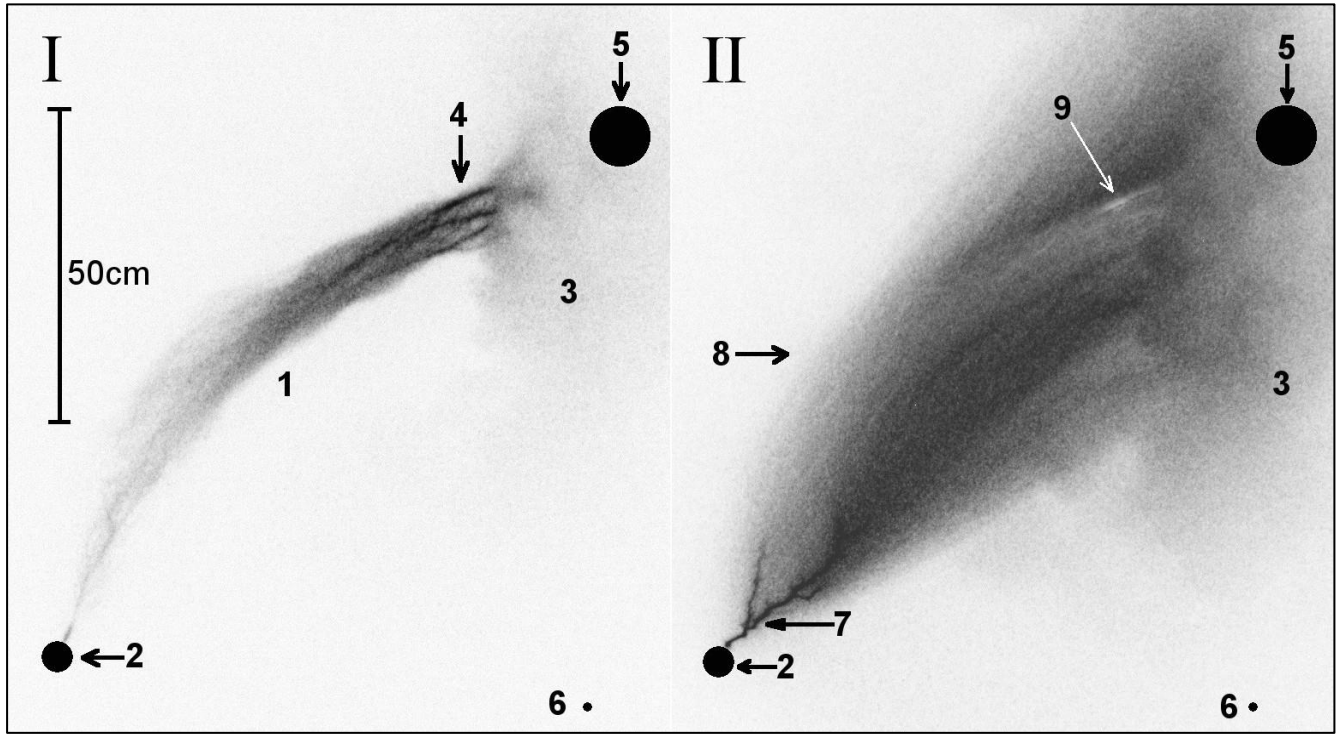

Рисунок 2.2 (адаптировано из [Kostinskiy et al., 2021]) (событие 2015-12-04_03). I — первый кадр камеры 4Picos с выдержкой 2 µs; II — второй кадр камеры 4Picos с выдержкой 10 µs; промежуток времени между кадрами 1 µs, оба кадра инвертированы.1 — первая вспышка положительной стримерной короны; 2 — заземлённый металлический шарик на плоскости; 3 — заряженное аэрозольное облако; 4 — UPFs; 5 — область прохождения микроволнового пучка; 6 — центр заземлённой плоскости, где расположено сопло; 7 — восходящий положительный лидер; 8 — стримерная корона восходящего положительного лидера; 9 — светлая неоднородная полоса на фоне изображении стримерной короны положительного восходящего лидера (8) является артефактом.

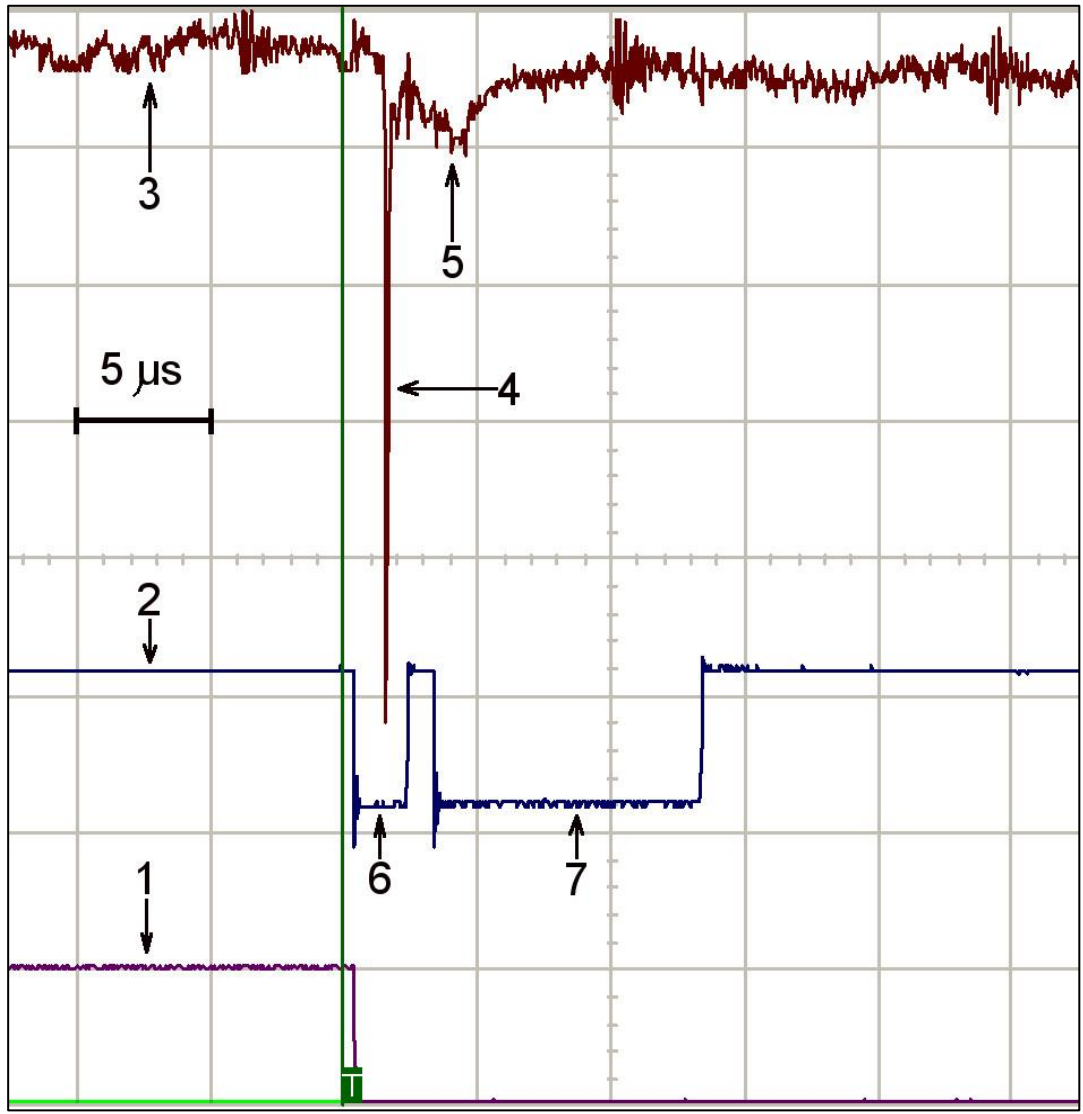

Рисунок 2.3 (адаптировано из [Kostinskiy et al., 2021]) (событие 2015-12-04_03). 1 — осциллограмма синхроимпульса, который пришел от первого осциллографа (Рисунок 2.1(5)) на второй осциллограф (Рисунок 2.1(9.7)); 2 — осциллограмма экспозиции кадров камеры 4Picos; 3 — осциллограмма сигнала с микроволнового диода (Рисунок 2.1(9.6)), фиксирующего микроволновое излучение, которое прошло сквозь аэрозольное облако; 4 — пик поглощения микроволнового излучения стримерами первой положительной вспышки; 5 — поглощение микроволнового излучения стримерами восходящего положительного лидера; 6 —время экспозиции первого кадра 4Picos; 7 — время экспозиции второго кадра 4Picos изображения была около $v_{st} \approx S_{st}/\tau_{st} \approx 7 \cdot 10^5$ м/с. Сразу после прохождения стримерной короны образовались каналы UPFs (Рисунок 2.2.I(4)), которые частично находятся за пределами оптически непрозрачного облака и поэтому хорошо видны на снимке скоростной камеры 4Picos. То, что UPFs не образовались раньше, чем произошла первая вспышка стримерной короны, а были её следствием, показывает осциллограмма, полученная с РМТ (Рисунок 2.4(2)). ФЭУ начинает фиксировать свет через $\approx 1.7$ мкс после первой вспышки стримерной короны. Начало сигнала ФЭУ находилось в пределах экспозиции (Рисунок 2.4(7)) первого кадра 4Picos (Рисунок 2.2.I). Световое излучение начинается до начала второй и третьей стримерной вспышки (Рисунок 2.5(7,8)), которые стали причиной рождения положительного восходящего лидера (Рисунок 2.2.II(7)). Во время записи этого события, у ФЭУ было большое собственное время, поэтому ФЭУ корректно отслеживает только момент начала свечения в облаке. Заземленный шар, с которого стартовала стримерная вспышка, и пространственная область $\sim 0.5$ м вблизи него находились за пределами поля зрения ФЭУ и свет разрядов из этой области ФЭУ не регистрировал. Интенсивность света разряда на электроде, рассеянного облаком, была ниже порога чувствительности ФЭУ в этом эксперименте.

На кадре на Рисунке 2.2.I, кроме стримерной вспышки, отчетливо видны три UPFs (4), которые аналогичны тем, что были зафиксированы в предыдущих экспериментах [Kostinskiy et al., 2015a]. Так как UPFs находятся от точки старта стримерной вспышки на расстоянии около 1 м, а средняя скорость распространения стримеров около $7 \cdot 10^5$ м/с, то процесс перехода стримеров в UPFs начался примерно через 1.4 мкс после старта стримерной вспышки. Судя по кадру на Рисунке 2.2.I, UPFs сформировались до окончания экспозиции этого кадра. С момента старта первой стримерной вспышки до окончания экспозиции первого кадра прошло около 2.5 мкс, следовательно время формирования UPFs не более, чем 1.1 мкс. На втором кадре, полученном скоростной камерой 4Picos (Рисунок 2.2.II), продолжительностью выдержки 10 мкс, записанном через 1 мкс после окончания экспозиции первого кадра, хорошо виден восходящий положительный лидер (7) и его стримерная корона (8), сформировавшиеся за 10 мкс. Длина лидера в плоскости второго кадра (Рисунок 2.2.II) от точки старта до самой удалённой точки около 29±2 см.

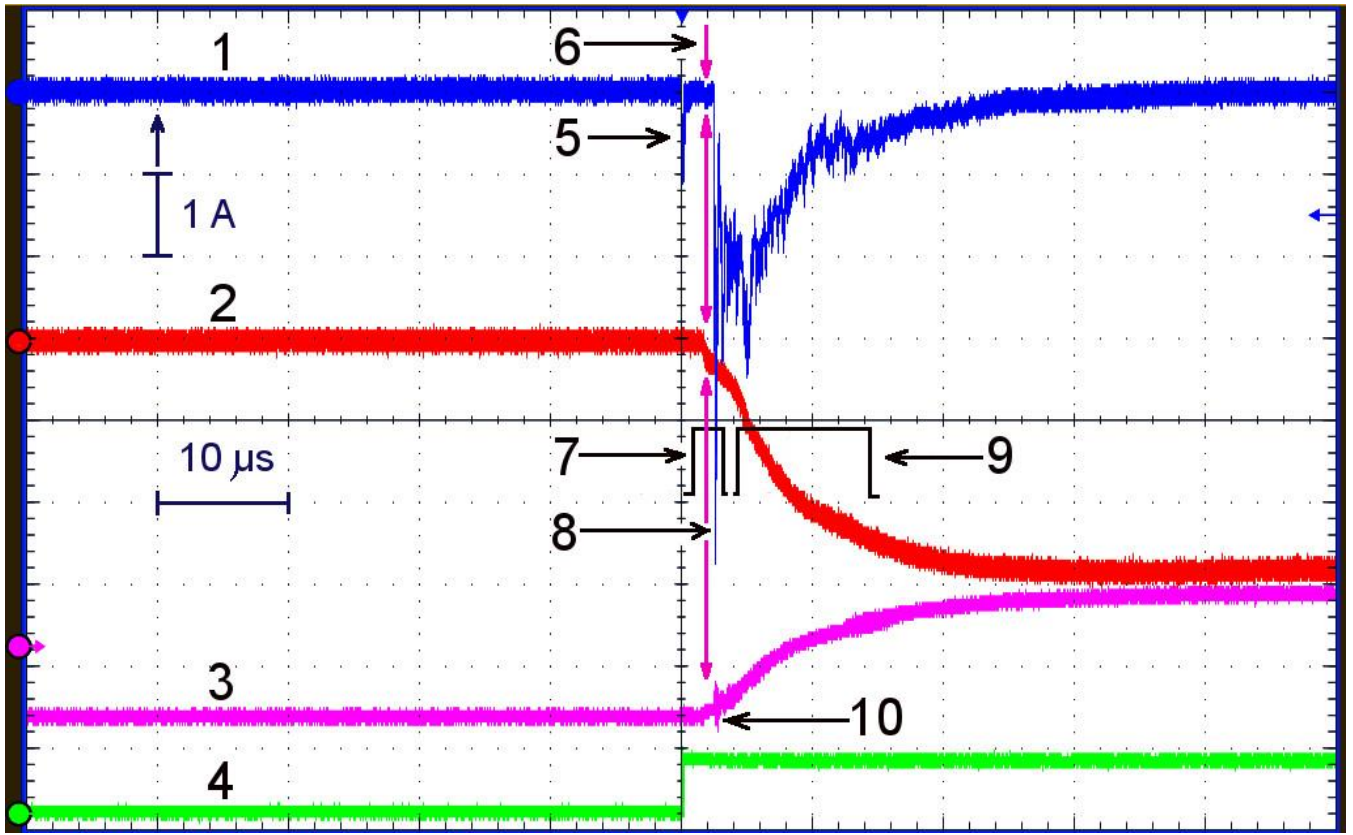

Рисунок 2.4 (адаптировано из [Kostinskiy et al., 2021]) (событие 2015-12-04_03). 1 — осциллограмма тока, идущего через заземлённую сферу (Рисунок 2.1(4)) на плоскости на первый осциллограф (Рисунок 2.1(5)); 2 — осциллограмма сигнала с ФЭУ (Рисунок 2.1(8)); 3 — сигнал, идущий с шара диаметром 50 см (Рисунок 2.1(10)), который контролирует зарядку аэрозольного облака; 4 — синхроимпульс, который посылает первый осциллограф (Рисунок 2.1(5)) на генератор импульсов (Рисунок 2.1(11)); 5 — ток первой вспышки положительной короны; 6 — момент максимального поглощения микроволнового излучений (показан на этой осциллограмме фиолетовыми стрелками); 7 — время экспозиции первого кадра 4Picos (нарисован на этой осциллограмме); 8 — ток второй стримерной вспышки; 9 —время экспозиции второго кадра 4Picos (нарисован на этой осциллограмме), 10 — наводка на сферу диаметром 50 см.

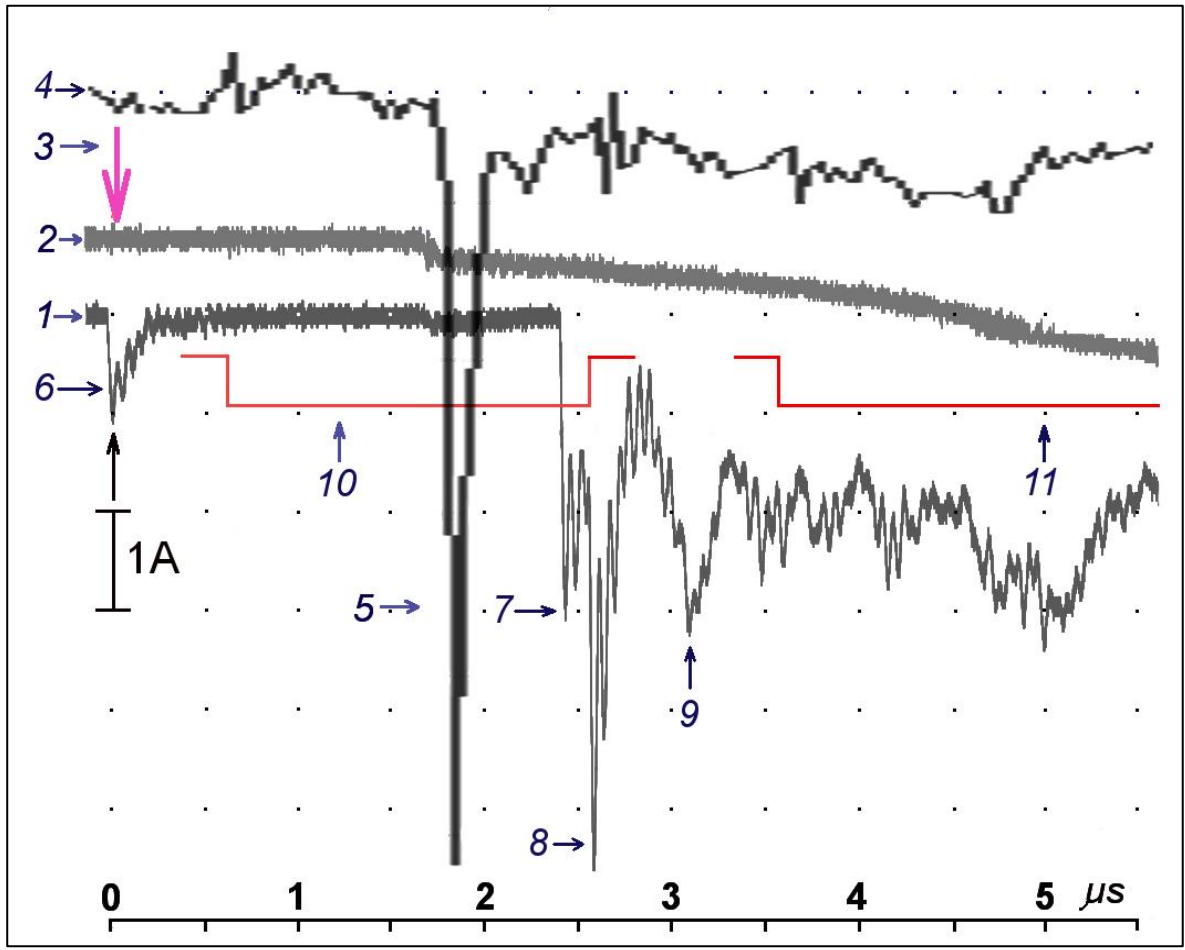

Рисунок 2.5 (адаптировано из [Kostinskiy et al., 2021]) (событие 2015-12-04_03). 1 — осциллограмма тока, идущего через заземлённую сферу (Рисунок 2.1(4)) на первый осциллограф (Рисунок 2.1(5)); 2 — осциллограмма сигнала с ФЭУ (Рисунок 2.1(8)); 3 — (указано стрелкой) момент генерации синхроимпульса первым осциллографом (Рисунок 2.1(5)), который посылается на второй осциллограф (Рисунок 2.1(9.7)); 4 — осциллограмма сигнала с микроволнового диода (Рисунок 2.1(9.6)), фиксирующего микроволновое излучение, которое прошло сквозь аэрозольное облако; 5 — пик поглощения микроволнового излучения стримерами первой положительной вспышки; 6 — величина тока первой вспышки положительной стримерной короны; 7 — величина тока второй вспышки стримерной короны; 8 — величина пика третьей вспышки стримерной короны; 9 — четвертый пик тока; 10 — время экспозиции первого кадра 4Picos (нарисовано на этой осциллограмме); 11 — время экспозиции второго кадра 4Picos (нарисовано на этой осциллограмме). Колебания на осциллограмме тока (1) с периодом около 50 нс, особенно заметные на стримерных вспышках 6, 7, 8 являются собственными колебаниями измерительного контура и являются артефактами.

Последовательность событий, изображенных на первом и втором кадре 4Picos (Рисунок 2.2), можно также проследить благодаря осциллограммам тока через заземлённую сферу (Рисунок 2.4(1), Рисунок 2.5(1)), а также благодаря интенсивности видимого излучения, регистрируемого ФЭУ, (Рисунок 2.4 (2), Рисунок 2.5(2)) и электрического потенциала (Рисунок 2.4(3)), наведенного зарядом облака на шаре диаметром 50 см (Рисунок 2.1(10)). Ток первой вспышки стримерной короны Рисунок 2.4(5), Рисунок 2.5(6) был относительно небольшим, около 1.1 А (передний фронт тока составил 30±5 нс, полуширина пика по полувысоте (FWHM) — 90±10 нс, задний фронт — 147±10 нс).

На Рисунок 2.5 изображены осциллограммы того же события, что и на Рисунок 2.4, но с расширенным масштабом времени. Поэтому на Рисунок 2.4 можно с погрешностью не хуже 0.1 мкс определить, что пик сильного поглощения микроволнового излучения (5) произошёл через 1.75 мкс после первой стримерной вспышки Рисунок 2.5(6), внутри которой позже сформировались UPFs (Рисунок 2.2.I(4)), и за 0.56 мкс до начала второй Рисунок 2.5(7), и 0.71 мкс третьей стримерной вспышки Рисунок 2.5(8), после которых вблизи поверхности заземленной сферы был инициирован восходящий положительный лидер. Небольшое изменение заряда облака фиксируется по изменению потенциала электростатической антенны (шар диаметром 50 см, Рисунок 2.4(3)) в момент, когда стримеры первой вспышки (Рисунок 2.4(5), Рисунок 2.5(6)) достигли облака (момент времени близкий к максимальному поглощению микроволнового излучения (Рисунок 2.4(6)) указан фиолетовыми стрелками на Рисунок 2.4. Это относительное изменение потенциала облака приблизительно соответствует соотношению заряда облака, около 60 мкКл, и заряда стримерной вспышки, около 0.12 мкКл, найденного интегрированием тока на Рисунок 2.5(6). Средний заряд облака определялся по зарядному току высоковольтного источника (Рисунок 2.1(2.2)), в тот момент, когда заряд облака переставал изменяться и ток зарядки компенсировался током дрейфа заряженных капель на землю, [Antsupov et al., 1991]. Положительный лидер Рисунок 2.2.II(6) возник после второй или третьей стримерной вспышки (Рисунок 2.5(7,8)), которые произошли через 2.42 и 2.58 мкс после первой стримерной вспышки. У этих двух стримерных вспышек, также как и у первой вспышки, фронты возрастания тока имеют типичную для длинных стримерных вспышек в существенно подпороговом поле длительность 30±5 нс. У второй вспышки (Рисунок 2.5(7)) трудно определить задний фронт и полуширину импульса тока по полувысоте, т.к.

происходит наложение третьего импульса на второй импульс тока. Максимум тока второй вспышки превосходит максимум первой в три раза и достигает 3.14 А. У третьего импульса (Рисунок 2.5(7)) можно оценить и задний фронт, длительность которого составляла около 180±20 нс. Ток третьего импульса в максимуме достиг 5.8 А и вызвал наводку (Рисунок 2.4(10)), которая видна на осциллограмме изменения потенциала шара диаметром 50 см (Рисунок 2.4(3)).

Четвертый пик тока величиной 3.3 А (Рисунок 2.5(9)) также представляет интерес, так как он заметно отличается от значений тока первых трёх стримерных вспышек характеристиками переднего фронта. Передний фронт тока (Рисунок 2.5(9)) составил 195±10 нс (что в 6 раз больше, чем у первых трёх пиков тока стримерной короны), полуширина пика по полувысоте — 180±10 нс, задний фронт — 210±10 нс. Возможно, этот фрагмент осциллограммы тока с медленным фронтом связан со стримерно-лидерным переходом, в результате которой в стеме стримерной вспышки рождается лидер [Bazelyan & Raizer, 1998].

Заряд, который стримеры первых вспышек и стримеры Рисунок 2.2.II(8) восходящего положительного лидера Рисунок 2.2.II(7) перенесли в район аэрозольного облака в течении всего времени разряда Рисунок 2.2.II(7) можно оценить по осциллограмме тока Рисунок 2.4(1) и этот заряд будет около 15 мкКл, что составляет около четверти всего отрицательного заряда облака. Этот заряд в течение 20 мкс значительно изменил положительный потенциал электростатической антенны (Рисунок 2.4(3)). Изображение восходящего лидера в течение экспозиции второго кадра 4Picos (10 мкс), хорошо видно на Рисунок 2.2.II(7). В первые несколько микросекунд, когда лидер развивался наиболее быстро и ток лидера был 1.5-2А, часть стримеров, видимо, попало в зону прохождения микроволнового пучка (Рисунок 2.2.II (5)), о чём говорит небольшое поглощение микроволнового излучения Рисунок 2.3(5). Любопытно, что более, чем в 3 и 5 раз более сильные токи второй и третьей стримерной вспышки по сравнению с током первой вспышки, не дали заметного микроволнового поглощения, что может говорить о том, что большая часть стримеров прошли мимо зоны, которую микроволновой пучок пересекает.

На Рисунок 2.6.I показано интегральное по времени ИК-изображение данного события (размер картинки 320х256 пикселей, частота следования кадров — 412 Гц,

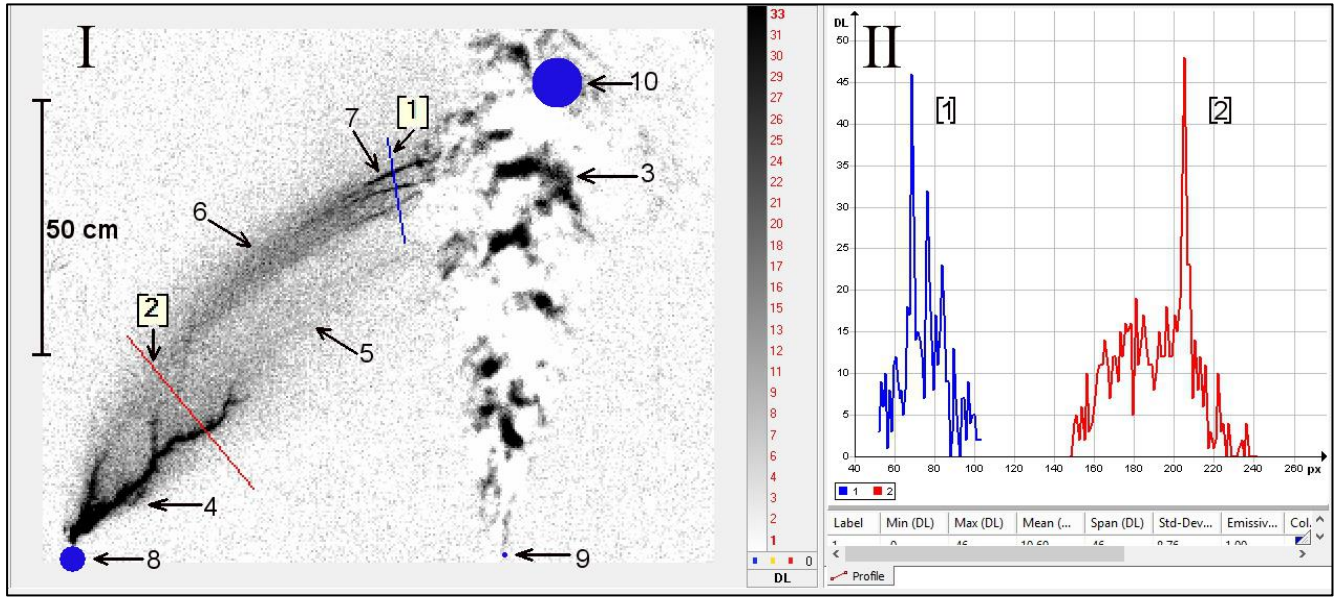

Рисунок 2.6. (адаптировано из [Kostinskiy et al., 2021]), (событие 2015-12-04_03). Изображение события в ИК-диапазоне. I — кадр ИК-камеры FLIR SC7700M (диапазон длин волн — 2.5-5.5 мкм, размер картинки 320х256 пикселей, размер пикселя 14х15 мкм, глубина изображения — 14 бит, частота следования кадров — 412 Гц, выдержка кадра — 2.4 мс, фокусное расстояние объектива — 50 мм, f/2). ИК-изображение получено вычитанием из данного кадра предыдущего кадра и инвертировано; II — Яркость ИК-излучения вдоль синей линии пикселей ИК изображения, включающего UPFs — (1) и вдоль красной линий пикселей ИК-изображения, включающего канал лидера в месте, где нет ветвлений — (2); 3 — заряженное аэрозольное облако; 4 — восходящий положительный лидер; 5 — стримерная корона положительного лидера; 6 — стримерная корона первой стримерной вспышки; 7 — UPFs; 8 — заземлённый металлическая сфера на плоскости (нарисована в масштабе); 9 — центр заземлённой плоскости, где расположено сопло (нарисован); 10 — область прохождения микроволнового пучка (нарисована в масштабе).

выдержка кадра — 2.4 мс). Для большего контраста ИК-изображение получено вычитанием из данного кадра (Рисунок 2.6.I) предыдущего кадра и инвертировано. При длительности экспозиции ИК-кадра 2.4 мс, в него попали практически все элементы разрядного процесса. От ветвящегося восходящего положительного лидера (4) поднимается вверх к облаку стримерная корона (5). Выше стримерной короны лидера видна первая стримерная вспышка (6), внутри которой мы можем отчётливо наблюдать три UPFs (7), такие же по очертаниям, как и на изображении в видимом диапазоне (Рисунок 2.2.I(4)), но с худшим пространственным разрешением.

Как и в предыдущих наших экспериментах [Kostinskiy et al., 2015a; Kostinskiy et al., 2015b], яркость ИК-излучения, а, значит, и температура воздуха наиболее горячих UPFs достигает температуры в горячем восходящем положительном лидере, о чём говорит на Рисунок 2.6.II сравнение уровня ИК-излучения вдоль линий пикселей, Рисунок 2.6.II(1), и канала восходящего лидера в месте, где нет ветвлений, Рисунок 2.6.II(2). Длина лидера без учёта ветвления на ИК-кадре в плоскости изображения от точки старта до самой удалённой точки около $42 \pm 1$ см. Это означает, что, за всё следующее за вторым кадром камеры 4Picos (Рисунок 2.2.II) время движения, лидер продвинулся ещё примерно на 10 см. ИК-изображение Рисунок 2.6.I имеет меньшее пространственное разрешение и яркость по сравнению с ИК-изображениями, полученными в предыдущих экспериментах [Kostinskiy et al., 2015a; Kostinskiy et al., 2015b], т.к. в данной серии экспериментов они сделаны с в 2.5 раза большего расстояния, в изображениях в 4 раза меньше пикселей и в 3-4 раза меньшая выдержка кадров при том же самом объективе. Несмотря на это, все основные элементы разряда удаётся наблюдать и на этом ИК-изображении.

Важной характеристикой длинной стримерной вспышки является ее геометрическая форма. Мы считаем, что напряженность электрического поля в облаке и его окрестности находится в диапазоне значений 500-1000 кВ/(м·атм), так как значение напряженности электрического поля $E \approx 500$ кВ/м является минимальным для поддержания распространения положительных стримерных вспышек в воздухе при атмосферном давлении [Bazelyan & Raizer, 1998)] и стримеры движутся во всем пространстве от заземленной плоскости до оси аэрозольного облака. С другой стороны, значение электрического поля не превышает 1000 кВ/(м·атм), т.к. из отрицательного аэрозольного облака не распространяются вниз к заземленной плоскости отрицательные стримерные вспышки (E ≈ 1000 кВ/(м·атм) — минимальное значение напряженности электрического поля, которое может поддержать движение длинных отрицательных стримеров при атмосферном давлении [Bazelyan & Raizer, 1998)]). На Рисунок 2.7.I представлена интегральная фотография нескольких последовательных положительных стримерных вспышек (5), которые в течение 1 секунды стартовали от заземленной плоскости (3) из точки (7) к отрицательно заряженному аэрозольному облаку (1) (фотоаппарат Nicon D3S, f/4.8, f =65 мм, выдержка — 1 с). Общая длина вспышек (5) от заземленной плоскости до видимого края облака находится в диапазоне 1.3-1.4 м. Максимальная ширина конуса всех вспышек на фотографии (Рисунок 2.7.I) около 12 см. Угол конуса всех вспышек на фотографии (Рисунок 2.7.I) не превосходит 5°. Точность оценки ширины и угла конуса вспышек с помощью интегральной фотографии очень невысока.

Для корректных измерений необходимо измерять параметры единичной стримерной вспышки. На Рисунке 2.7.II изображен верхний фрагмент типичной индивидуальной стримерной вспышки, соответствующий области пространства (6) на Рисунке 2.7.I. Изображение получено камерой с усилением изображений 4Picos, выдержка составляет 100 нс. На Рисунке 2.7.II хорошо видно, что ширина индивидуальной положительной стримерной вспышки находится в диапазоне 5-6 см и стримерная вспышка на всей длине ~30 см (Рисунок 2.7.II) не расширяется, то есть ее ширина остается приблизительно одинаковой. Аналогичное поведение длинных стримерных вспышек, зафиксированное камерой 4Picos, изображено на Рисунок 2.8. Изображение верхнего фрагмента индивидуальной положительной стримерной вспышки, соответствующего области 6 на Рисунке 2.7.I, получено с выдержками — 3 мкс и — 100 нс. Исходя из этих экспериментов и этой геометрии разряда, форма длинной стримерной вспышки имеет следующую динамику. После инициации на заземленном электроде, стримерная вспышка расширяется под углом 5-7°, двигаясь в таком режиме первые 30-40 см. После этого диаметр индивидуальной стримерной вспышки стабилизируется и составляет 5-10 см.

Другое событие (14_2015-12-04) изображено на Рисунках 2.9-2.11. В этом случае за 40 мкс до стримерной вспышки и через 160 мкс после стримерной вспышки СВЧ диагностика и измерение тока на заземленном шарике не регистрируют каких-либо

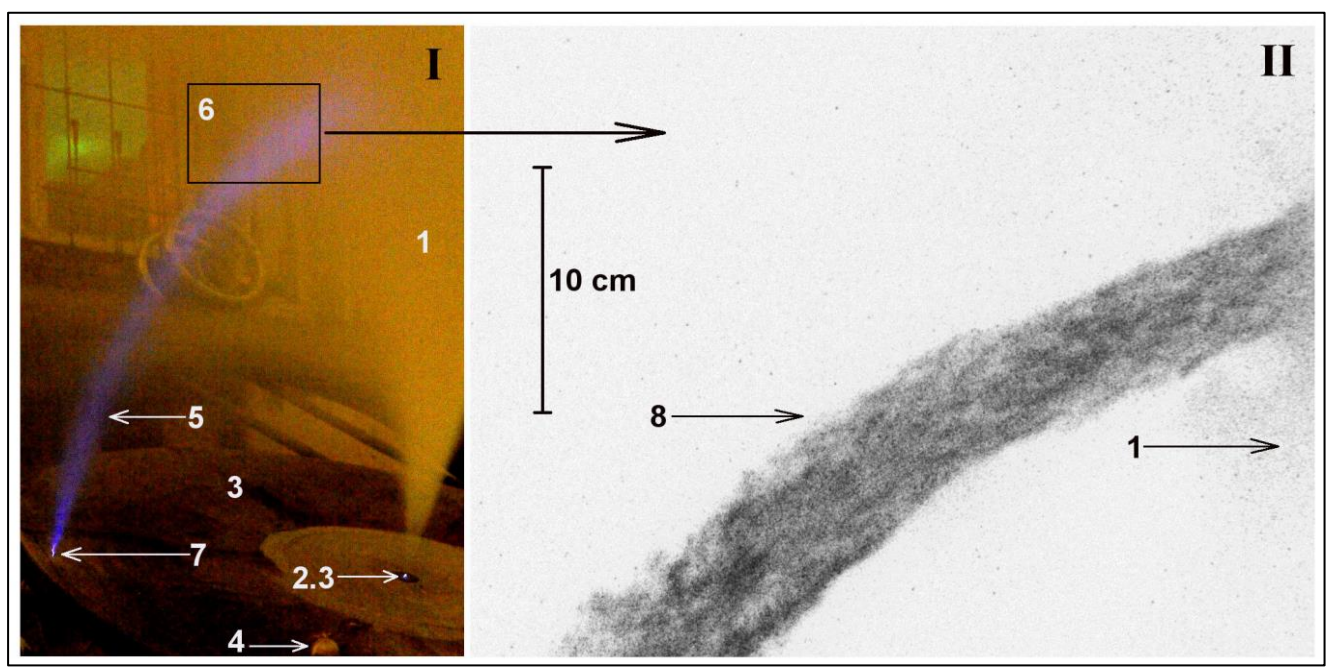

Рисунок 2.7 (адаптировано из [Kostinskiy et al., 2021]). I — фотография нескольких последовательных положительных стримерных вспышек, которые движутся от заземлённой плоскости к отрицательно заряженному аэрозольному облаку (фотоаппарат Nicon D3S, f/4.8, фокусное расстояние —65 мм, выдержка — 1 с); II — изображение верхнего фрагмента одной положительной стримерной вспышки, полученная камерой с усилением изображений 4Picos, выдержка - 100 нс (фрагмент соответствует в пространстве области 6); 1 — отрицательно заряженное облако аэрозоля водяных капель, 2.3 — сопло, 3 — заземленная металлическая плоскость (расстояние от сопла до края плоскости равно 1 м), 4 — 5 см шарик, соединенный с измерительным шунтом, 5 — изображение нескольких последовательных стримерных вспышек, 6 — область положительной стримерной вспышки соответствующая изображению II, 7 – стем положительной стримерной вспышки; 8 - верхний фрагмент положительной стримерной вспышки (нижний край фотографии находится на высоте около 0.8 м над заземленной плоскостью, ширина стримерной вспышки около 7 см).

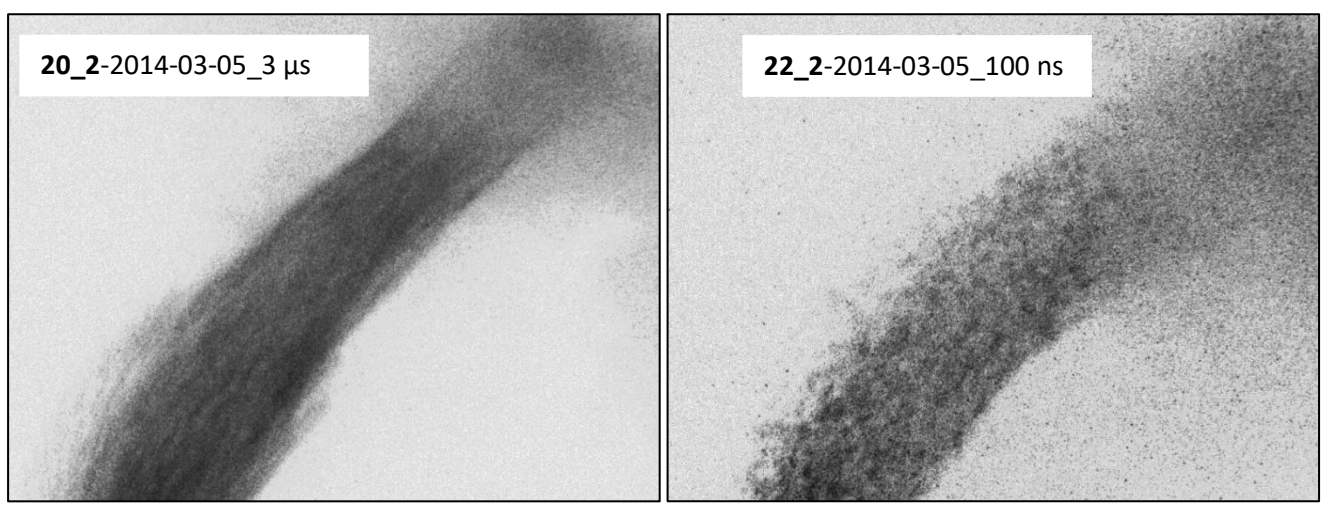

Рисунок 2.8 (адаптировано из [Kostinskiy et al., 2021]). Изображение верхнего фрагмента двух положительных стримерных вспышек, соответствующих в пространстве области 6 на Рисунок 2.7.I. Изображения получены камерой с усилением изображений 4Picos, выдержка события 20_2 - 3 мкс, выдержка события 22_2 - 100 нс.

разрядных событий (Рисунок 2.9(5,6), Рисунок 2.10(1), Рисунок 2.11(6,7)). В частности, в результате стримерной вспышки не произошел стримерно-лидерный переход и не был инициирован восходящий положительный лидер, как в событии 03_2015-12-04 изображённом на Рисунок 2.2.II. На Рисунке 2.10 изображена часть стримерной вспышки (1), которая возникла на заземленной сфере (2) и распространялась до видимого края облака (3). После этого стримеры вспышки вошли в облако и пересекли почти перпендикулярно область прохождения микроволнового пучка (5). В этом случае стримерные головки долетели от заземленного шарика (2) до области прохождения микроволнового пучка (5) примерно за $\tau_{st} \approx 2.4$ мкс (точность измерений ±50 нс), пройдя расстояние по дуге около $S_{st} \approx 1.2$ м. Факт движения стримеров внутри аэрозольного облака фиксируется по пику поглощения микроволнового излучения на Рисунке 2.11(8). Длительность микроволнового поглощения (FWHM) была немного больше, чем на Рисунке 2.5 и равнялась 160±20 нс. Пик поглощения находился на самом краю экспозиции кадра 4Picos (Рисунок 2.11(8,9)). Средняя скорость стримеров в этом случае в плоскости изображения была несколько меньше, чем на Рисунке 2.2 и равнялась $v_{st} \approx S_{st}/\tau_{st} \approx 5 \cdot 10^5$ м/с. На Рисунке 2.10 недалеко от края облака образовались хорошо различимые в видимом диапазоне UPFs (Рисунок 2.10(4)). От времени старта стримерной вспышки до начала экспозиции первого кадра 4Picos (Рисунок 2.10) прошло около 600 нс, то есть, начало стримерной вспышки не попало в кадр. Поэтому, на Рисунке 2.10 нижняя часть стримерной вспышки, которую стримеры прошли в первые 600 нс, не видна и изображение данного события нельзя использовать для идентификации формы стримерной вспышки. Изображение единичной стримерной вспышки на Рисунке 2.2.I отличается от изображения на Рисунке 2.10 потому, что в кадр (Рисунок 2.2.I) попали также вторая и третья стримерные вспышки (Рисунок 2.4(8)).

Максимум тока стримерной вспышки (Рисунок 2.11(6)) был около 1.5 А (передний фронт тока составил 35±5 нс, длительность пика тока по полувысоте (FWHM) — 100±10 нс, задний фронт — 190±10 нс). На осциллограмме потенциала электростатической антенны (Рисунок 2.9(7)) в момент, когда стримеры вспышки достигли облака, можно заметить небольшое изменение заряда облака. Общий заряд аэрозольного облака был около 60 мкКл, а общий заряд стартующей стримерной вспышки, судя по осциллограмме тока Рисунок 2.11(6), был около 0.3 мкКл и за время движения заряда к облаку он также существенно не увеличился. Так как лидер не образовался, то и дальнейшего уменьшения

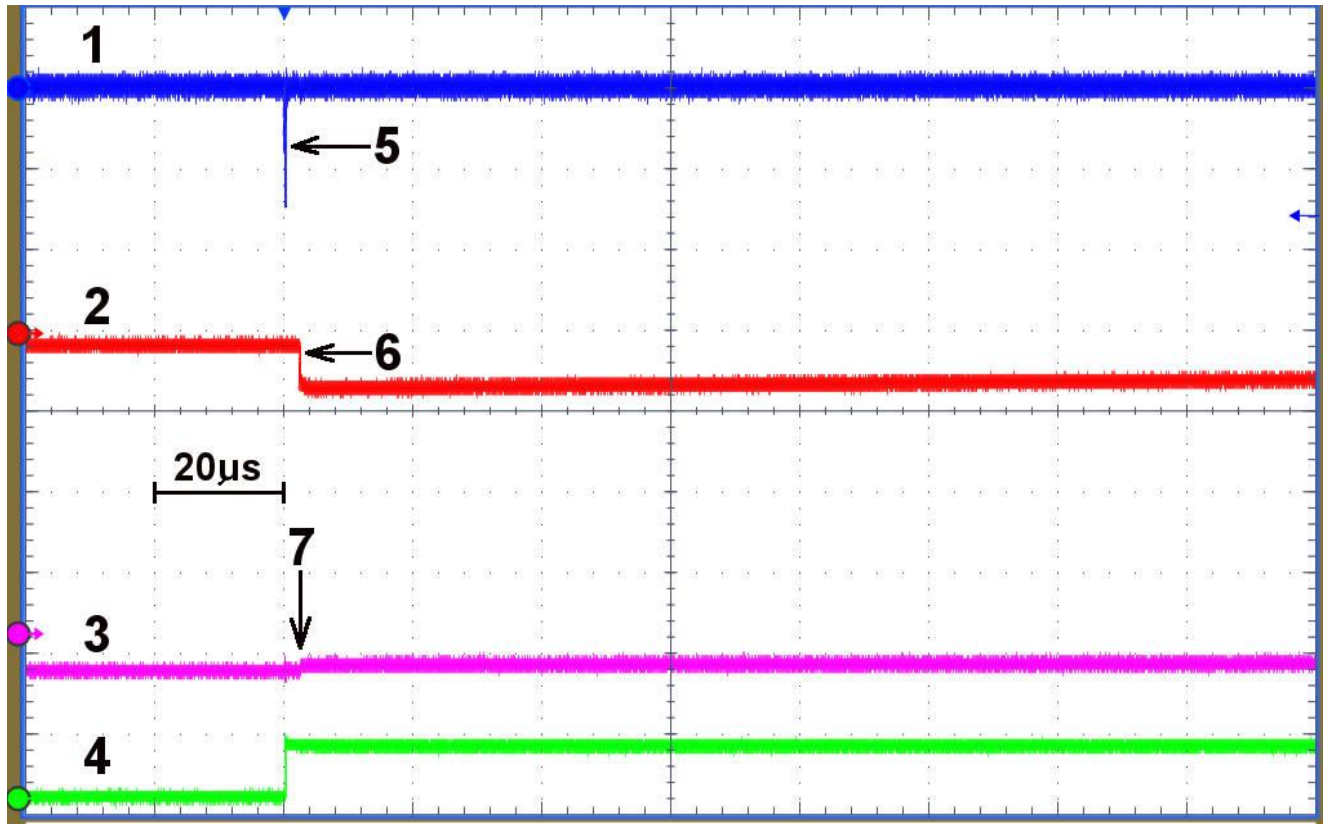

Рисунок 2.9 (адаптировано из [Kostinskiy et al., 2021]), (событие 14_2015-12-04, то же, что и на Рисунке 2.10). 1 — осциллограмма тока, идущего через заземлённую сферу (Рисунок 2.1(4)) на плоскости на первый осциллограф (Рисунок 2.1(5)); 2 — осциллограмма сигнала с ФЭУ (Рисунок 2.1(8)); 3 — сигнал, идущий со сферы диаметром 50 см (Рисунок 2.1(10)); 4 — синхронизирующий импульс, который посылает первый осциллограф (Рисунок 2.1(5)) на генератор импульсов (Рисунок 2.1(11)); 5— ток положительной стримерной вспышки; 6 — сигнал с ФЭУ, фиксирующий прохождение стримерной вспышки; 7 — слабое изменение заряда аэрозольного облака после контакта со стримерной вспышкой.

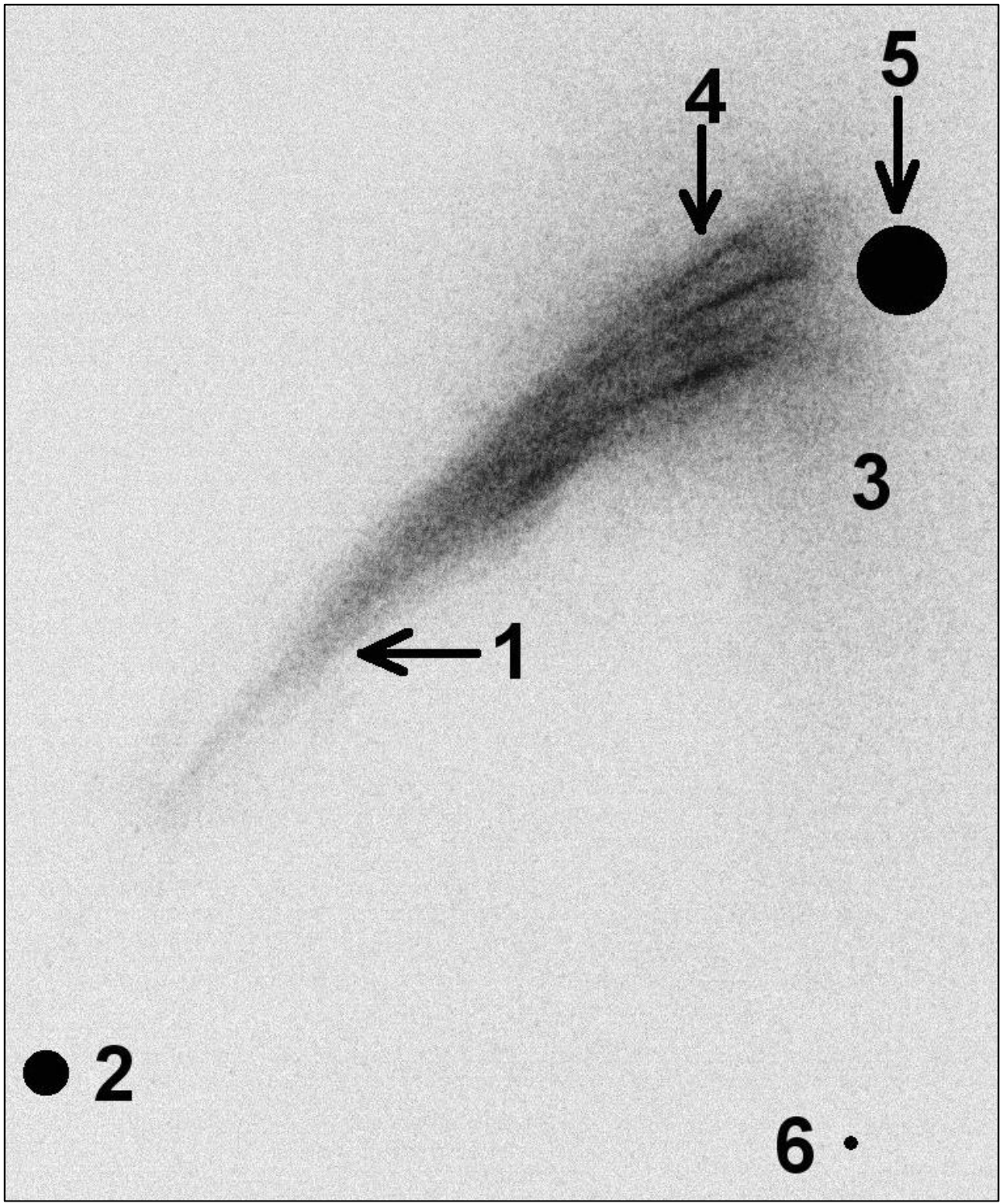

Рисунок 2.10 (адаптировано из [Kostinskiy et al., 2021]), (событие 14_2015-12-04, то же что и на Рисунок 2.9). Кадр получен камерой 4Picos с выдержкой 2 мкс, кадр инвертирован. 1 — вспышка положительной стримерной короны; 2 — заземлённая металлическая сфера на плоскости (рисунок); 3 — видимая граница отрицательно заряженного аэрозольное облака; 4 — UPFs; 5 — область прохождения микроволнового пучка; 6 — центр заземлённой плоскости, где расположено сопло (картинка).

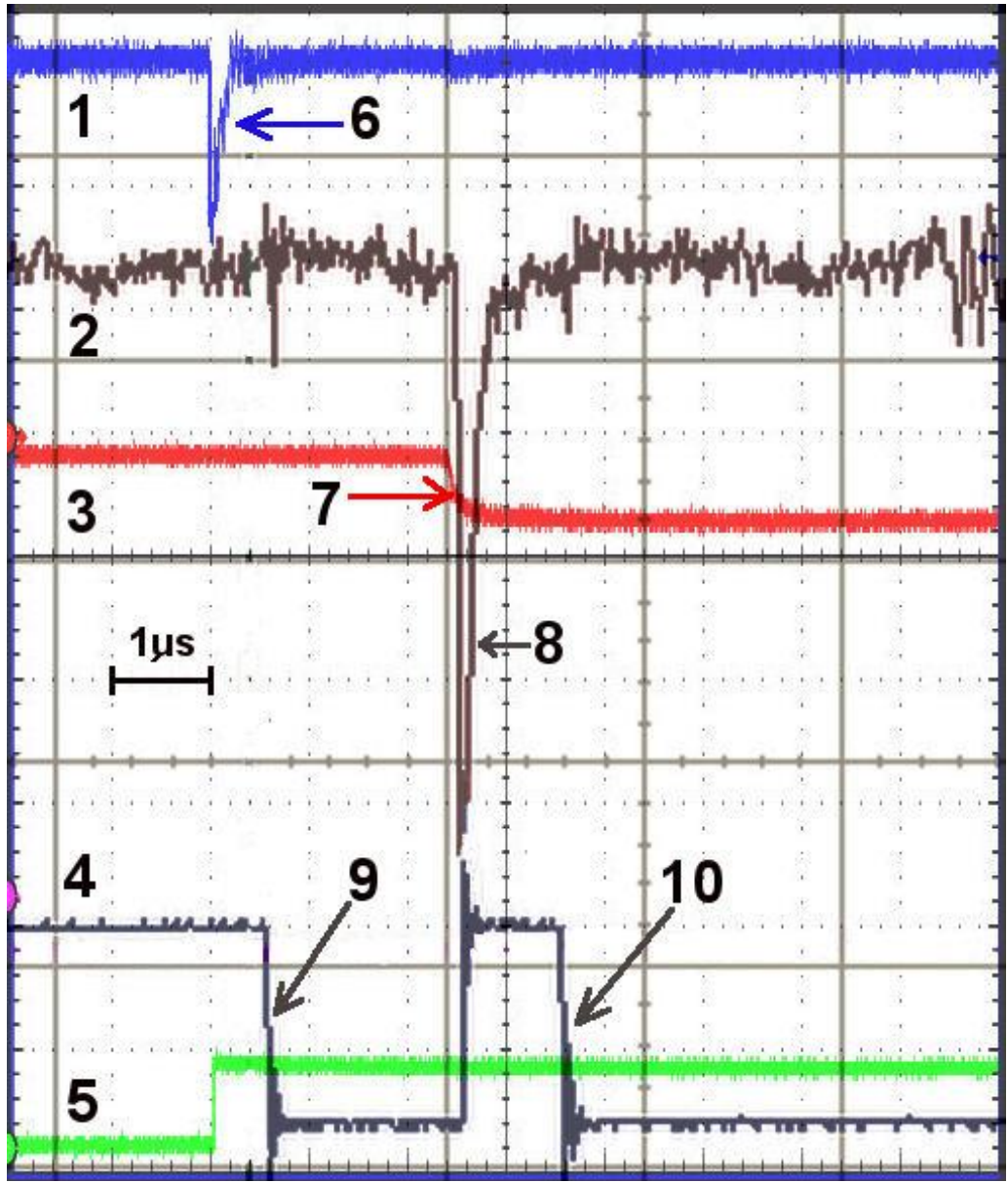

Рисунок 2.11 (адаптировано из [Kostinskiy et al., 2021]), (событие 14_2015-12-04, то же что и на Рисунок 2.9 и Рисунок 2.10). 1 — осциллограмма тока, идущего через заземлённый шарик; 2 — осциллограмма сигнала с микроволнового диода, фиксирующего микроволновое излучение, которое прошло сквозь аэрозольное облако; 3 — осциллограмма сигнала с ФЭУ; 4 — осциллограмма времени экспозиции камеры 4Picos; 5 — осциллограмма синхронизующего импульса, который подаёт на генератор первый осциллограф; 6 —ток первой стримерной вспышки; 7 — сигнал с ФЭУ во время прохождения стримерной вспышки; 8 — пик поглощения и рассеяния микроволнового излучения стримерами первой положительной вспышки; 9 —время экспозиции первого кадра 4Picos; 10 — часть времени экспозиции второго кадра камеры 4Picos.

заряда облака не произошло, как на осциллограмме Рисунок 2.4(3). Таким образом, и в данном случае UPFs (Рисунок 2.10(4)) образовались внутри траектории стримерной вспышки, так как других плазменных событий не было зафиксировано. UPFs (Рисунок 2.10(4)) выглядят не такими яркими, как на Рисунке 2.2.I(4), так как они образовались в конце времени экспозиции кадра (Рисунок 2.11(9)), на что указывает сигнал ФЭУ (Рисунок 2.11(7)) и сигнал поглощения микроволнового излучения (Рисунок 2.11(8)).

## 2.3. Сравнение с результатами, полученные в более ранних работах, в которых, возможно, также фиксировались физические проявления UPFs

[Анцупов и др., 1990] с помощью фотоэлектронной камеры с развёрткой и усилением изображения (ФЭР-14) до начала развития лидера обнаружили плазменные образования, которые возникали после первой стримерной вспышки (Рисунок 2.12) и последующих стримерных вспышек (см. также Рисунок 3.12 и анализ этого явления в связи с инициацией двунаправленного лидера внутри облака в главе 3). Эти плазменные образования [Анцупов и др., 1990] назвали «ядром» и так описали их свойства: «…наблюдалось возникновение на высоте 0,3-0,6 м ярко светящегося протяженного образования длиной 0.1-0.2 м которое условно назовем «светящееся ядро». Эти «ядра» возникали с интервалом в несколько десятков микросекунд. При определенной интенсивности процесса со стержня (на заземленной плоскости – А.К.) в направлении ядра начинает развиваться лидер, имеющий характерную скорость 1.5-2.5 см/мкс. Более яркая светящаяся головка оставляет заметный след на фотохронограмме».

Так как восходящий положительный лидер образуется позже появления «ядра» (в такой же последовательности, как описано в этом разделе возникновение UPFs) то, скорее всего «ядро» является свечением стримерной вспышки в верхней части вспышки (около границы облака), где электрическое поле выше, чем около плоскости и поэтому стременная вспышка светится ярче. Скорее всего, UPFs рождаются в результате перехода этой стримерной вспышки в горячие плазменные каналы.

В пользу того, что «ядро» является стримерной вспышкой говорит изображение «обычной» стримерной вспышки, поднимающейся с заземленного шарика, на

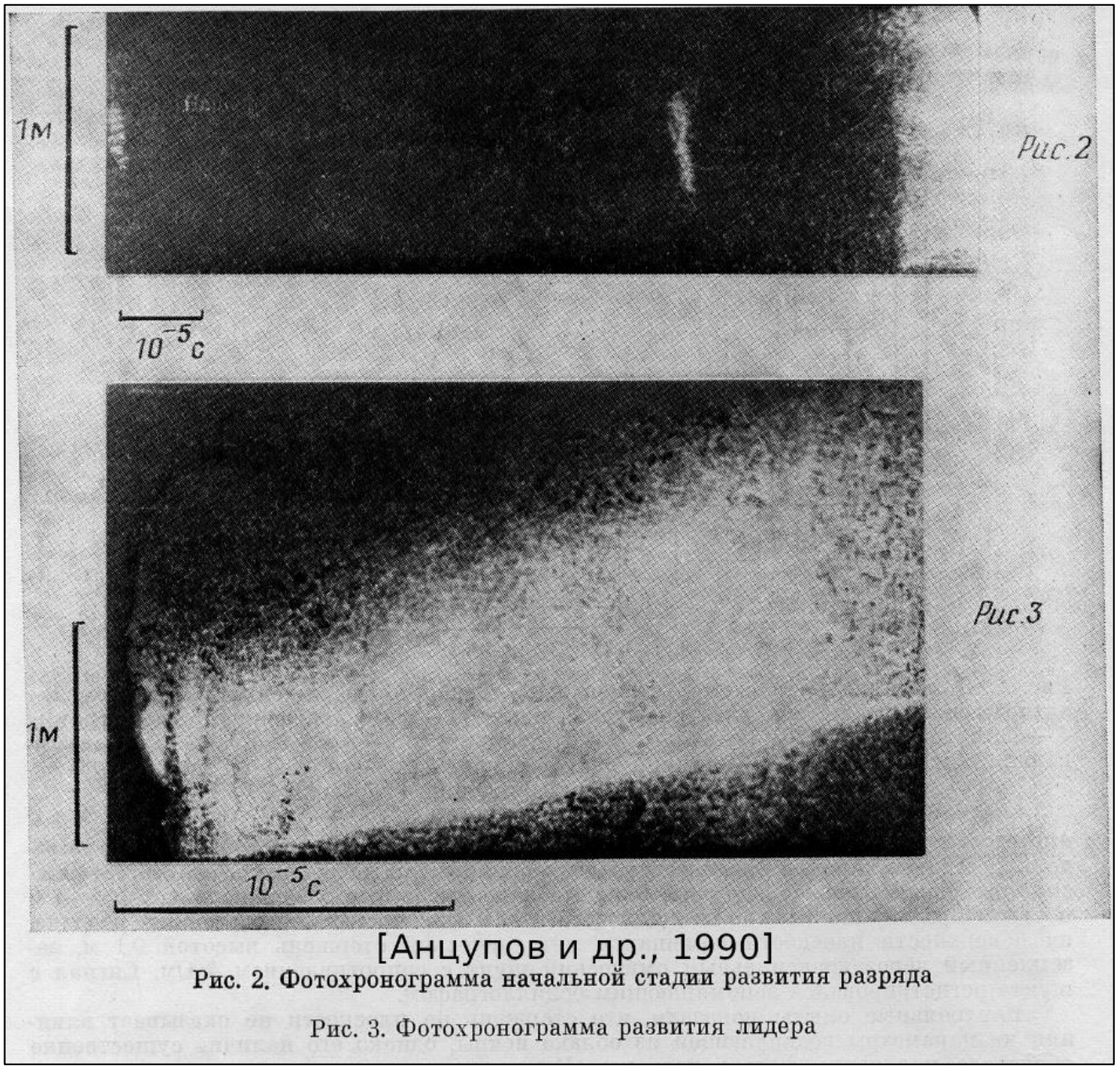

Рис.2

Рис.3

[Анцупов и др., 1990]
Рис. 2. Фотохронограмма начальной стадии развития разряда

Рис. 3. Фотохронограмма развития лидера

Рисунок 2.12 (адаптировано из [Анцупов и др., 1990]). Фиксация стримерных вспышек и восходящего положительного лидера фотоэлектронной камеры с развёрткой и усилением изображения (ФЭР-14). До начала развития лидера в правом нижнем углу (*Рис.2*) видны 2 протяженные оптические вспышки, которые возникали после или во время первой стримерной вспышки и последующих стримерных вспышек (см. также Рисунок 3.12). Эти светящиеся образования [Анцупов и др., 1990] назвали «ядром». «Ядра» возникали с интервалом в несколько десятков микросекунд и скорее всего являлись изображениями стримерных вспышек около края облака, где электрическое поле было более сильным. При определенной интенсивности процесса зарядки облака со стержня (на заземленной плоскости – А.К.) в направлении «ядра» начинает развиваться лидер (*Рис.3*), имеющий характерную скорость 1.5-2.5 см/мкс. Более яркая светящаяся головка оставляет заметный след на фотохронограмме. ФЭР-14 имел фотокатод, чувствительный в УФ области и хорошо фиксировал стримерную корону лидера. Поэтому хорошо различима не только стримерная вспышка («ядро»), но и стримерная корона лидера в процессе его движения

Рисунке_2.13 (событие 14-2_2014-12-17), которая очень похожа на «ядро» на верхнем изображении Рисунка 2.12 из [Анцупов и др., 1990]), но сделано это изображение с гораздо лучшим пространственным разрешением камеры 4Picos. На осциллограмме тока через шунт этой стримерной вспышки, ток стримерной вспышки соответствует току «обычной» стримерной вспышки (Рисунок 2.14). Лидер стартует с заземленной сферы через 180 мкс. На Рисунке 2.15 мы также можем видеть стримерную вспышку, похожую на «ядро» на нижнем изображении Рисунка 2.12 из [Анцупов и др., 1990]. На осциллограмме тока (желтый луч, Рисунок 2.16) видно, что стримерная вспышка появилась примерно за 7 мкс до старта лидера с заземленной сферы. Осциллограмма сигнала с ФЭУ (фиолетовый луч) фиксирует свечение стримерной вспышки в течение примерно 200 нс. Время экспозиции кадров фиксируется пиками голубого луча (колебания на голубом луче происходят из-за наводки и поэтому экспозиция не выглядит П-образно).

Также в пользу того, что стримерная вспышка, похожая на «ядро», предшествует рождению UPFs и скорее всего является его причиной, говорит Рисунок 2.17. Судя по осциллограмме тока (Рисунок 2.18), лидер возникнет на заземленной сфере через 7 мкс. В пользу рождения UPFs в стримерной вспышке говорит веретонообразная форма стримерной короны восходящего лидера (нижний рисунок), которая следует по силовым линиям электрического поля. Синяя стрелка указывает возможное положение UPFs на нижнем кадре. Красной стрелкой указано место, где, по-видимому, образовался еще один проводящий UPF, поляризованная плазма которого изменила силовые линии электрического поля, которые визуализируются благодаря траекториям стримеров. UPFs, поляризованные в электрическом поле, искажают первоначальную картину силовых линий облака. Выдержка кадров 200 нс, между кадрами 1.9 мкс.

## 2.4. Обсуждение полученных результатов главы 2

Первой стримерной вспышкой мы называем вспышку стримерной короны с заземлённой сферы, в электрическом поле отрицательно заряженного аэрозольного облака, до появления которой в изучаемом объёме не было зафиксировано никаких плазменных процессов в течение, как минимум, 40 мкс. В экспериментах, результаты которых представлены на Рисунках 2.3-2.6 удалось установить, что UPFs инициируются после

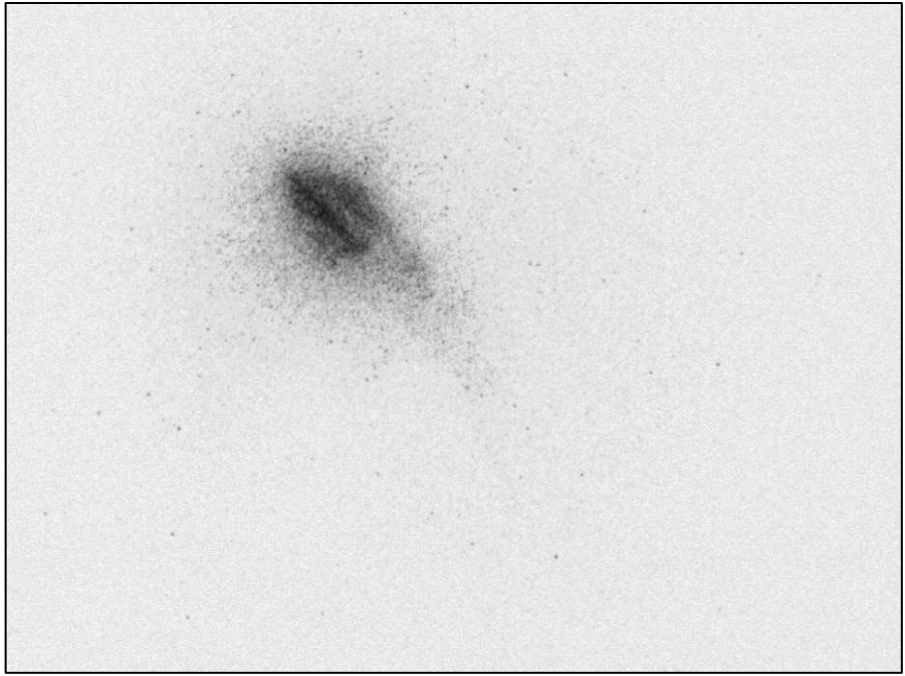

Рисунок 2.13 (событие 14-2_2014-12-17). Стримерная вспышка, которая похожа на «ядро» из работы [Анцупов и др., 1990], см. Рисунок 2.12 (верхний рисунок). Выдержка кадра камеры 4Picos равна 100 нс.

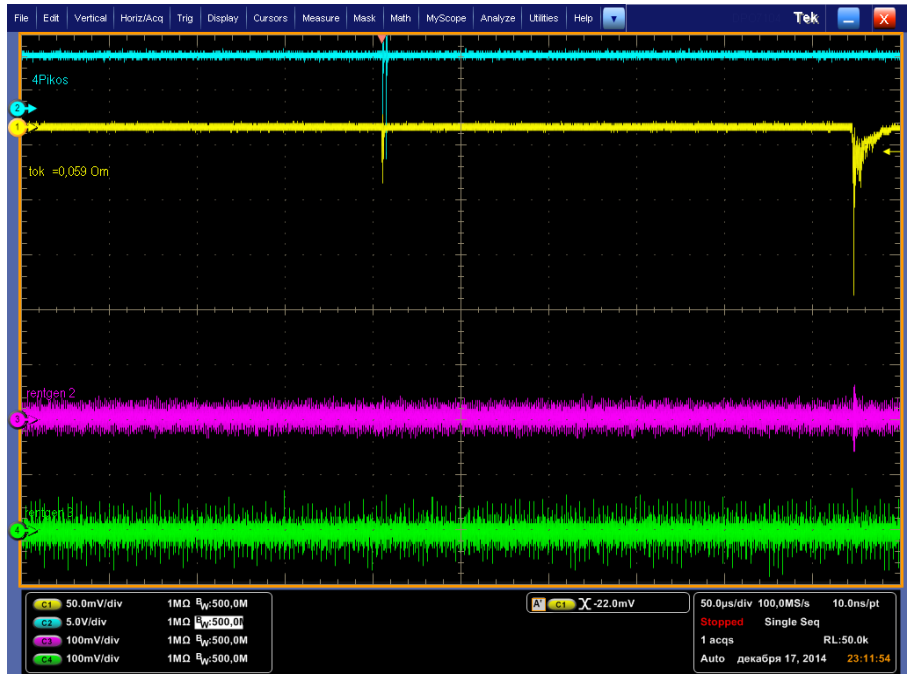

Рисунок 2.14 (событие 14-2_2014-12-17). Осциллограмма тока (желтый луч) стримерной вспышки (Рисунок 2.13) с развитием лидера через 180 мкс. Время экспозиции кадров фиксируется пиками голубого луча. Большое деление по горизонтали соответствует 50 мкс, по вертикали соответствует 1А

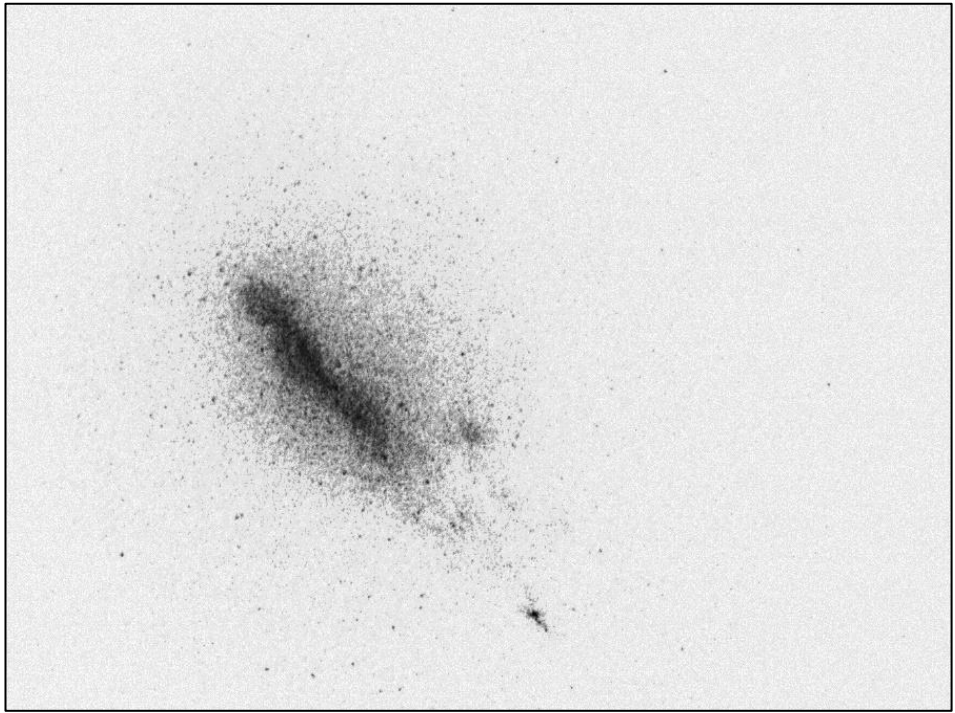

Рисунок 2.15 (событие 43-2_2014-12-17). Стримерная вспышка, которая похожа на «ядро» из работы [Анцупов и др., 1990], см. Рисунок 2.12 (нижний рисунок). Выдержка кадра равна 100 нс.

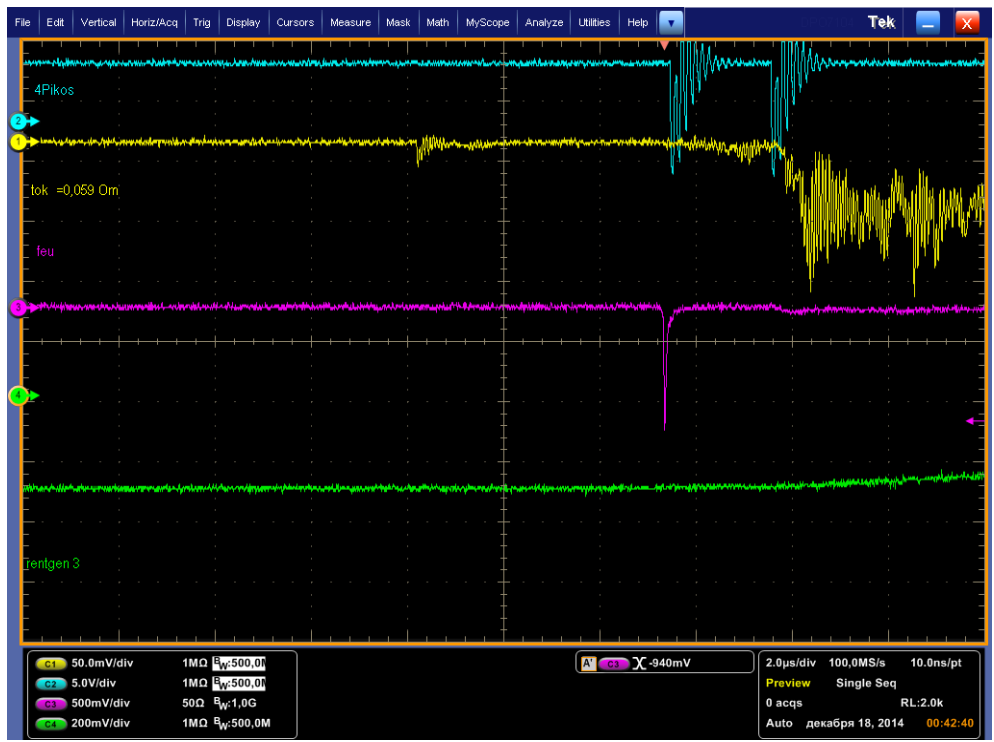

Рисунок 2.16 (событие 43-2_2014-12-17). Осциллограмма тока (желтый луч) стримерной вспышки (Рисунок 2.15) с развитием лидера через 3 мкс. Осциллограмма свечения стримерной вспышки (фиолетовый луч). Время экспозиции кадров фиксируется пиками голубого луча (колебания на голубом луче происходят из-за наводки).

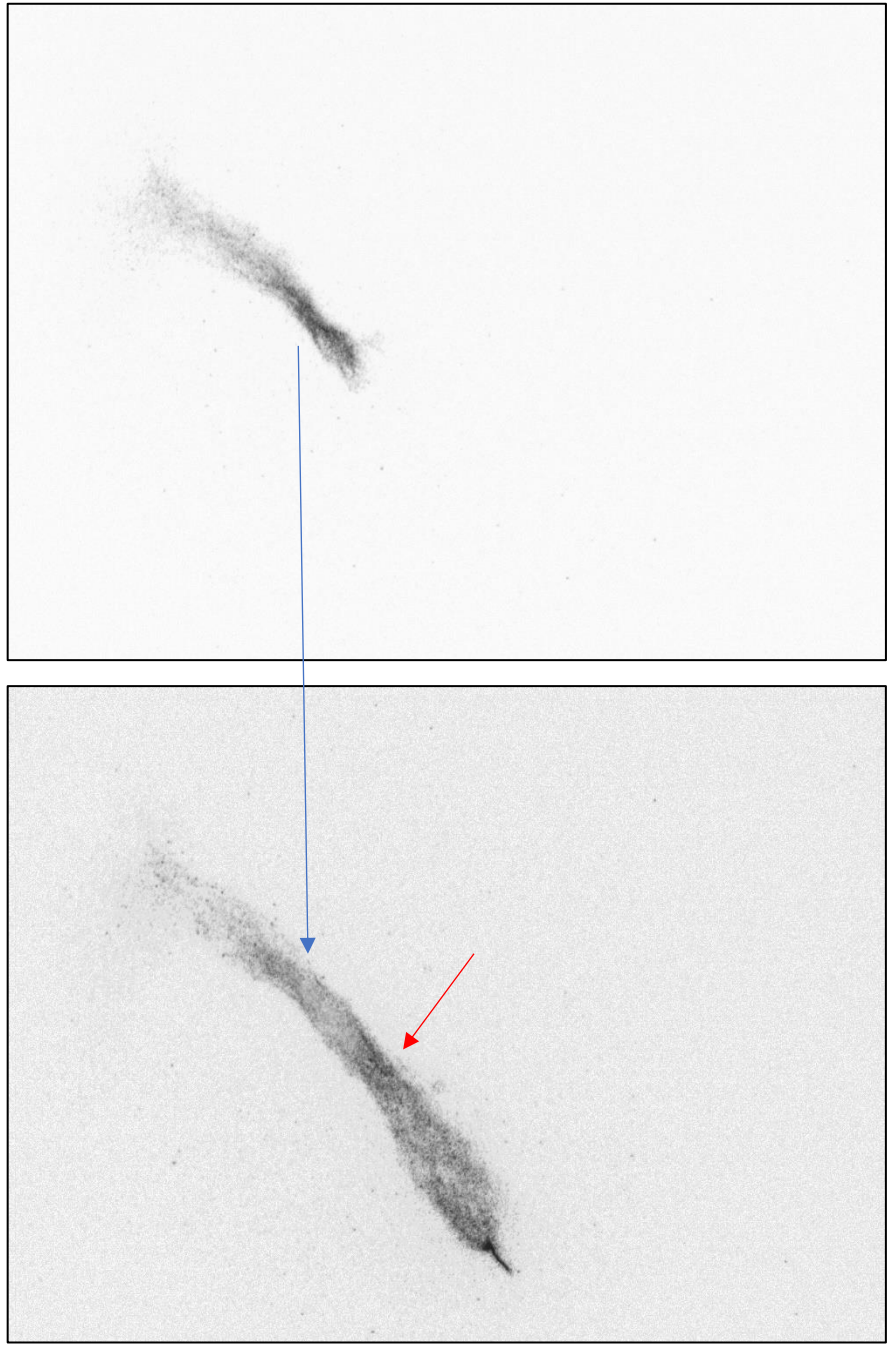

Рисунок 2.17 (событие 46_2014-12-17). Стримерная вспышка, которая похожа на «ядро» из работы [Анцупов и др., 1990], предшествует рождению UPFs (верхний рисунок). Судя по осциллограмме тока (Рисунок 2.18), лидер на заземленной сфере возникнет на заземленной сфере через 7 мкс. В пользу рождения UPFs в стримерной вспышке также говорит веретонообразная форма стримерной короны восходящего лидера (нижний рисунок), которая следует по силовым линиям электрического поля. Синяя стрелка указывает возможное положение UPFs на нижнем кадре. Красной стрелкой указано место, где, по-видимому, образовался еще один UPF. UPFs, поляризованные в электрическом поле, искажают первоначальную картину силовых линий облака, делая ее веретонообразной (или волнообразной). Выдержка кадров 200 нс, между кадрами 1.9 мкс.

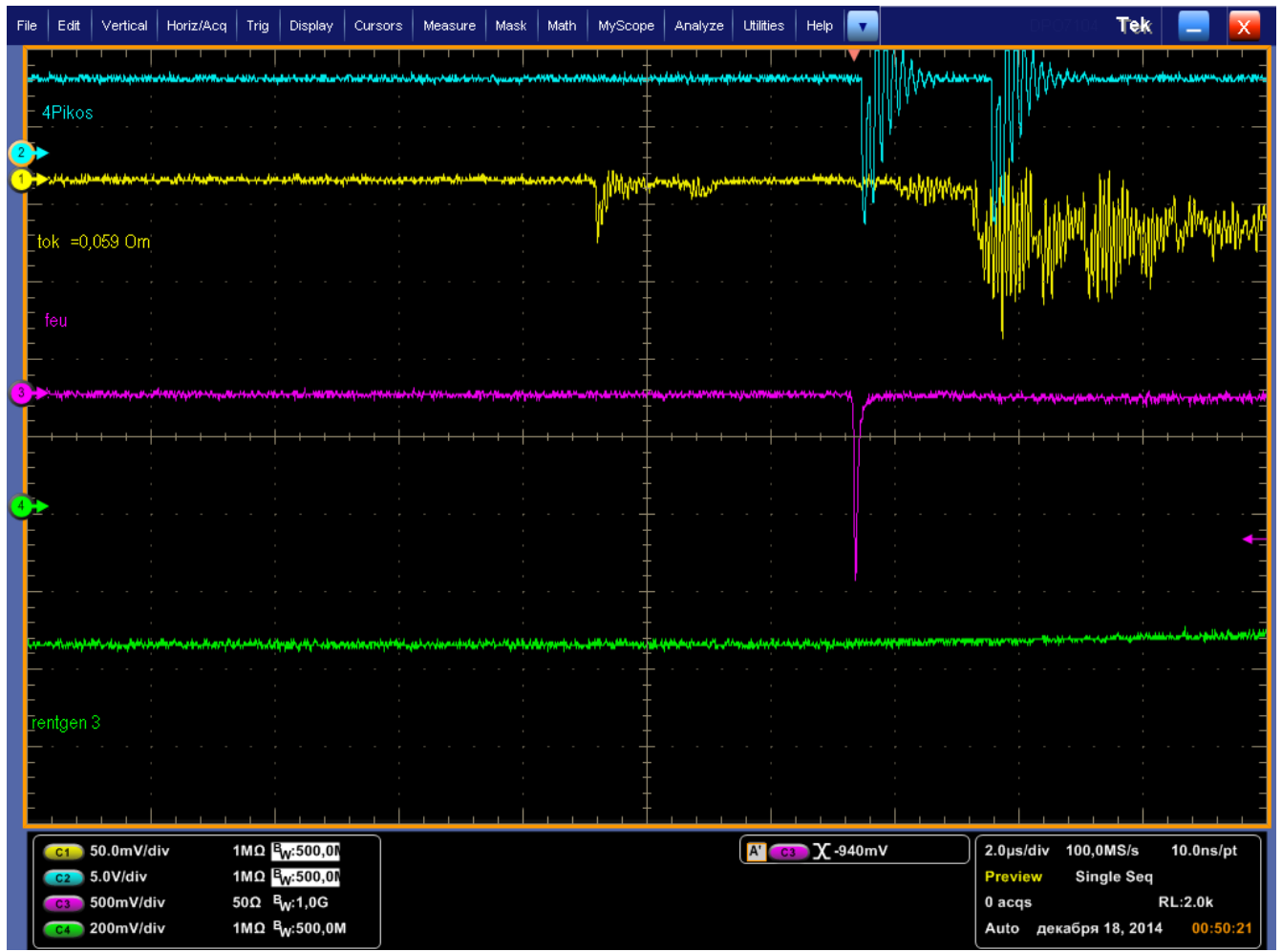

Рисунок 2.18 (событие 46_2014-12-17). Осциллограмма тока (желтый луч) стримерной вспышки и последующего восходящего лидера через 7 мкс (Рисунок 2.17). Осциллограмма свечения стримерной вспышки в течение примерно 200 нс (фиолетовый луч). Время экспозиции кадров фиксируется пиками голубого луча (колебания на голубом луче происходят из-за наводки). Видно, что стримерная вспышка попадает в экспозицию первого кадра, а лидер попадает в экспозицию второго кадра.

прохождения первой стримерной вспышки в подпороговых полях с напряженностью электрического поля, находящегося в диапазоне 500-1000 кВ/(м·атм). UPFs инициируются за время не большее, чем 1.1 мкс после прохождения первой стримерной вспышки. Эти времена соответствуют характерному времени развития стримерно-лидерного перехода в воздухе при атмосферном давлении [Bazelyan et al., 2007, Popov et al., 2009]. При этом стримеры первой вспышки до начала образования UPFs проходят расстояние в пределах 0.88-1.05 м в плоскости изображения. Судя по ИК-изображениям (Рисунок 2.6), UPFs являются горячими и проводящими настолько же, насколько нагрет и имеет проводимость положительный восходящий лидер. До начала первой стримерной вспышки в течение 40 мкс не обнаружено никаких событий (Рисунок 2.4, Рисунок 2.9), которые могли бы быть причиной образования UPFs. Первые стримерные вспышки, которые приводят к появлениям UPFs имеют следующие параметры: относительно небольшой ток 1-1.5 А, передний фронт около 30 нс, задний фронт около 130-150 нс. Длительность пика тока по полувысоте в диапазоне 90-120 нс.

Таким образом, нам удалось установить один из механизмов образования UPFs. Важно отметить, что рождаются UPFs не поодиночке, а сразу несколько каналов, образуя небольшую сеть (Рисунок 2.2, Рисунок 2.6). Не исключено, что появление одного UPF стимулирует появление других, т.к. все они расположены близко друг к другу.

Метод микроволновой диагностики, применяемый в данной работе, хорошо зарекомендовал себя, как способ идентификации различных плазменных процессов внутри заряженного аэрозольного облака.

Термин UPF был введен в работе [Kostinskiy et al., 2015a], глава1, для обозначения необычных разрядов внутри облака заряженных капель. В данной работе мы наблюдали разрядные каналы UPF, которые в части своей длины находились за пределами регистрируемой в видимом свете границы облака. Вероятно, необычные свойства UPFs, отличающие их от обычных стримеров, обусловлены наличием в искусственном заряженном облаке значительного пространственного электрического заряда, удерживаемого на каплях облака. Капли искусственного облака, непрерывно внедряемые в центр облака струей пара, разлетаются на периферию облака, постепенно испаряясь и уменьшаясь в диаметре. Поэтому наблюдаемое в видимом свете искусственное заряженное облако должно быть окружено облаком мелких капель низкой объемной плотности, которые несут те же заряды, что и капли в центре облака. Вероятно, пространственный электрический заряд присутствует не только внутри видимого облака, но и за его границами, поэтому там тоже могут иметь место UPFs, которые мы и наблюдали в данных экспериментах.

## 2.5. Выводы главы 2

- В данной работе удалось установить одну из причин образования необычных плазменных формирований (UPFs), наблюдаемых ранее в электрическом поле искусственного облака заряженных водяных капель [Kostinskiy et al., 2015a; Kostinskiy et al., 2015b]. Было зафиксировано, что UPFs появляются внутри или вблизи края облака за время не большее чем 1.4 мкс, в объеме первой положительной стримерной вспышки, стартующей с заземленного электрода, с начальным током на электроде в диапазоне 1-1.5 А.

- Сравнение ИК-излучения идущего из канала положительного восходящего лидера и из UPFs позволяет предположить, что, по крайней мере, некоторые UPFs являются такими же горячими и проводящими плазменными образованиями, как и восходящий положительный лидер.

- Заряд, выносимый стримерной вспышкой с электрода, составлял 0.1-0.3 мкКл. Средняя скорость стримерной вспышки была меньше, чем $10^6$ м/с. Двигаясь первые 30-40 см под коническими углами 5-7°, далее стримерная вспышка перестает расширяться и стабилизирует свой диаметр на уровне 5-10 см.

**ГЛАВА 3. Наблюдение взаимодействия (контакта) положительных и отрицательных лидеров в электрических разрядах метрового масштаба, генерируемых электрическими полями облаков отрицательно заряженного водного аэрозоля**

В данной главе представлены подробные наблюдения взаимодействия между положительными и отрицательными лидерами в электрических разрядах метрового масштаба, генерируемых электрическими полями облака отрицательно заряженного водного аэрозоля, и обсуждаются их возможные последствия для процесса контакта молнии с наземными объектами [Kostinskiy et al., 2016]. В экспериментах использовались оптические изображения, полученные с помощью трех различных высокоскоростных камер (видимого диапазона, сдвинутого в УФ, с усилением изображения; скоростная камера видимого диапазона с CMOS-матрицей; и инфракрасного диапазона), и соответствующие им записи тока. Впервые представлены два снимка сквозной фазы лидеров, показывающие значительное разветвление лидера внутри общей стримерной зоны. Параметры фронта и длительности тока подобны параметрам тока главной стадии длинной искры. Скорости положительных и отрицательных лидеров внутри общей стримерной зоны для двух событий оказались схожими. Более высокие скорости лидера обычно связаны с более высокими токами лидера. Контакт происходил между отрицательным лидером (двунаправленного лидера), нисходящим из отрицательно заряженного облака водного аэрозоля и положительным лидером, который поднимается с заземлённой плоскости (заземленной сферы). Противоположная положительная сторона двунаправленного лидера (по отношению к каналу нисходящего отрицательного лидера) представляет собой восходящий ветвящийся положительны лидер, развивающийся внутрь отрицательного облака. Внутри отрицательного облака обнаружено множество плазменных каналов. Получены дополнительные данные о характеристиках контакта лидеров (сквозной и главной фазы), благодаря использованию скоростных видео и инфракрасных камер. В случае соединения лидеров «головка к головке» яркость инфракрасного излучения в области сквозной фазы (вероятно, пропорциональная температуре газа и, следовательно, подводу энергии) обычно была примерно в 5 раз выше, чем для участков канала ниже или выше этой области. В 16% случаев нисходящий отрицательный лидер (двунаправленного лидера) соединялся с восходящим положительным лидером ниже его головки (контактировал с боковой поверхностью положительного лидера), при этом контакт осуществлялся через сегмент канала, который казался перпендикулярным одному или обоим каналам лидеров.

### 3.1. Введение в главу 3

Известно, например, [Berger, 1977, p. 178]; [Golde, 1977, p. 555]; [Анцупов и др., 1990], что феноменология длинных лабораторных искр сходна с феноменологией молнии. По этой причине многие термины, впервые введенные для физики молнии (например, лидер и обратный удар), были приняты впоследствии для длинных искр и наоборот термины, введенные для длинных искр (например, спейс-стем, спейс-лидер) теперь используются для описания процессов в стримерной зоне ладеров молнии. Некоторые основные механизмы развития каналов молнии (например, тонкая структура образования ступеней отрицательных лидеров, включая очень сложное взаимодействие спейс-стемов и спейс-лидеров) были впервые обнаружены в длинных искрах ([Stekolnikov and Shkilyov,1963], [Gorin et al., 1976]) и только недавно наблюдались похожие структуры в стримерных зонах молний [Biagi et al., 2010]; [Gamerota et al., 2014]. По понятной причине лабораторные наблюдения за длинными искрами можно проводить на гораздо более близких расстояниях и для гораздо большего количества событий. Термины "attachment process" и «сквозная фаза» ("breakthrough phase"), применяются как к молнии (природное явление), так и к лабораторным длинным искрам (искусственное явление). Та же терминология используется для триггерных молний, инициируемых ракетой с проводом, которые, хоть и искусственно инициированы, представляют собой полномасштабный разряд молнии, источником энергии которого является естественное грозовое облако (см., например, [Rakov and Uman, 2003, Ch. 4 and 7] и ссылки в них).

Искусственно заряженные облака водного аэрозоля, генерирующие «безэлектродные» электрические искры (например, [Верещагин и др., 1988]; [Antsupov et al., 1991]; [Temnikov et al., 2007]; [Syssoev et al., 2014]) недавно использовались для изучения множества разрядных процессов, которые могут произойти при развитии молнии (или предполагается, что могут произойти), но с трудом поддаются фиксации имеющимся методами наблюдений. Схема экспериментов во всех этих работах была подобно той, что описана в главе 1.

Впервые гипотеза о существовании явления подобного главной стадии при разряде молнии и длинной искры, инициированного электрическим полем облака отрицательно заряженного аэрозоля была высказана в работе [Анцупов и др., 1990]. Характеристики фронта нарастания тока этого разряда и его длительность с учётом масштаба явления, оказались подобны характеристикам главной стадии в длинной искре, инициируемой высоковольтными генераторами импульсного напряжения (ГИН) и аналогичны по физической природе обратному удару молнии. Это встреча (контакт) горячих, высокопроводящих плазменных каналов. Далее в этой диссертации под словами «главная стадия» или «обратный удар» мы будем понимать разряд в облаке заряженного аэрозоля, в котором встречаются положительный и отрицательный лидер, в результате чего появляется резкий пик тока (с фронтом десятки наносекунд) и нейтрализуется накопленный ранее в каналах и их чехлах заряд. [Анцупов и др., 1990] делали вывод, что главная стадия возникает в результате контакта между отрицательным лидером, нисходящим из искусственного облака отрицательно заряженного водного аэрозоля, и положительным лидером, который инициируется и поднимается с токоприёмника (стержня, сферы) на заземлённой плоскости в электрическом поле этого облака. Интегральная фотография процесса представлена на Рисунке 3.1, где отчётливо виден извилистый канал положительного восходящего лидера, который движется с заземленной сферы вверх к облаку. Навстречу ему под углом к его траектории движется отрицательный лидер, нисходящий из облака. В месте, где расстояние между двумя каналами приближается к минимальным, стримерные короны лидеров начинают усиленно взаимодействовать. При подобной траектории движения лидеров в пространстве их взаимодействие заканчивается почти перпендикулярной перемычкой между ними (плазменным мостиком). На временной развёртке события с помощью фоторегистратора ФЭР-14 Рисунок 3.2. (адаптировано из [Анцупов и др., 1990]) видна траектория головки положительного лидера (1), восходящего с плоскости в течение примерно 28 мкс (сам канал лидера находится ниже головки и не виден все это время). За это время лидер проходит расстояние не менее 65 см и после этого, в сквозной фазе, канал положительного лидера (1') становится виден благодаря резкому усилению свечения, также как и канал отрицательного лидера (3). Между положительным и отрицательным

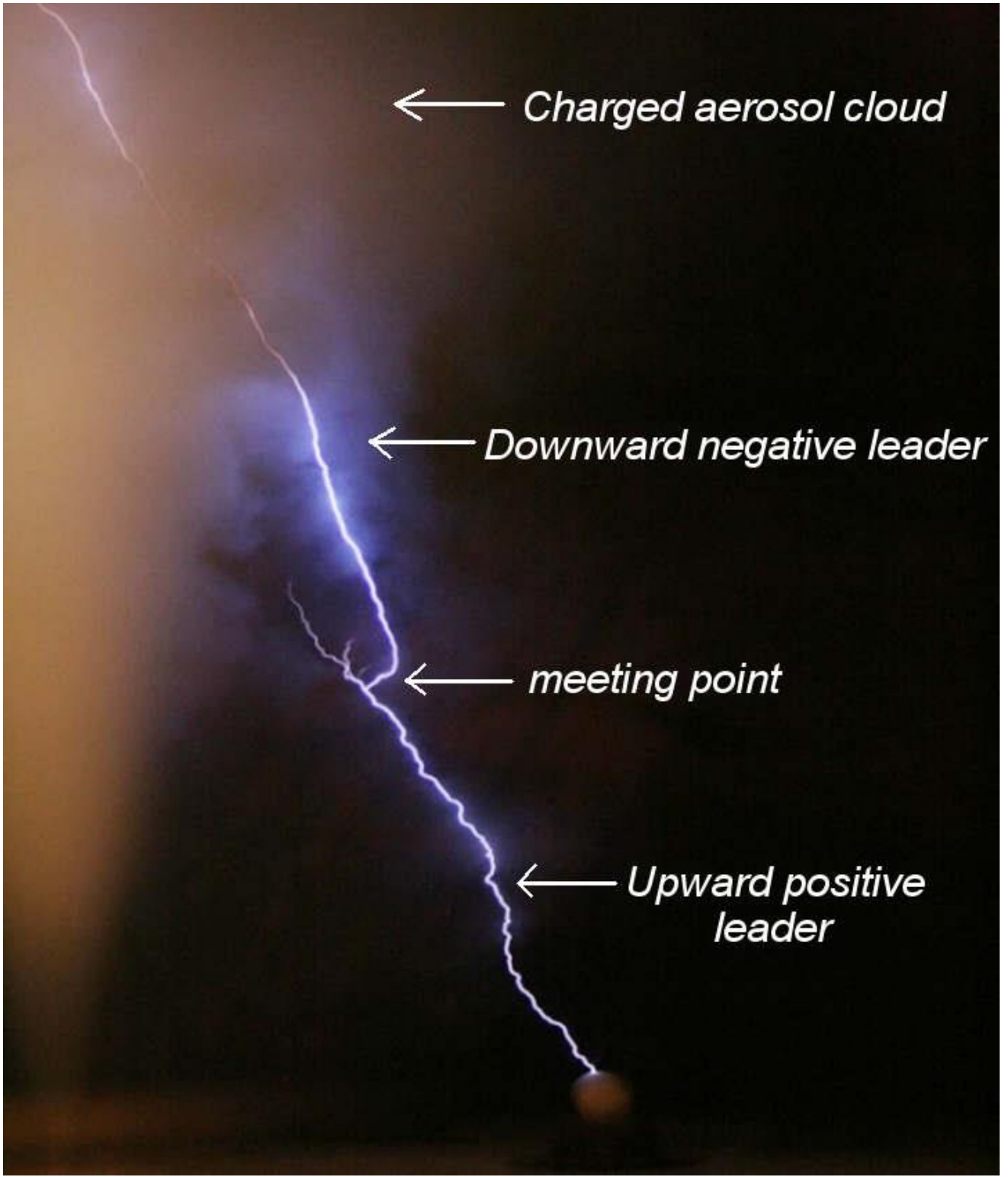

Рисунок 3.1 (адаптировано из [Kostinskiy et al., 2016]). Интегральная фотография встречи нисходящего отрицательного лидера и восходящего положительного лидера с последующим обратным ударом (главной стадией).

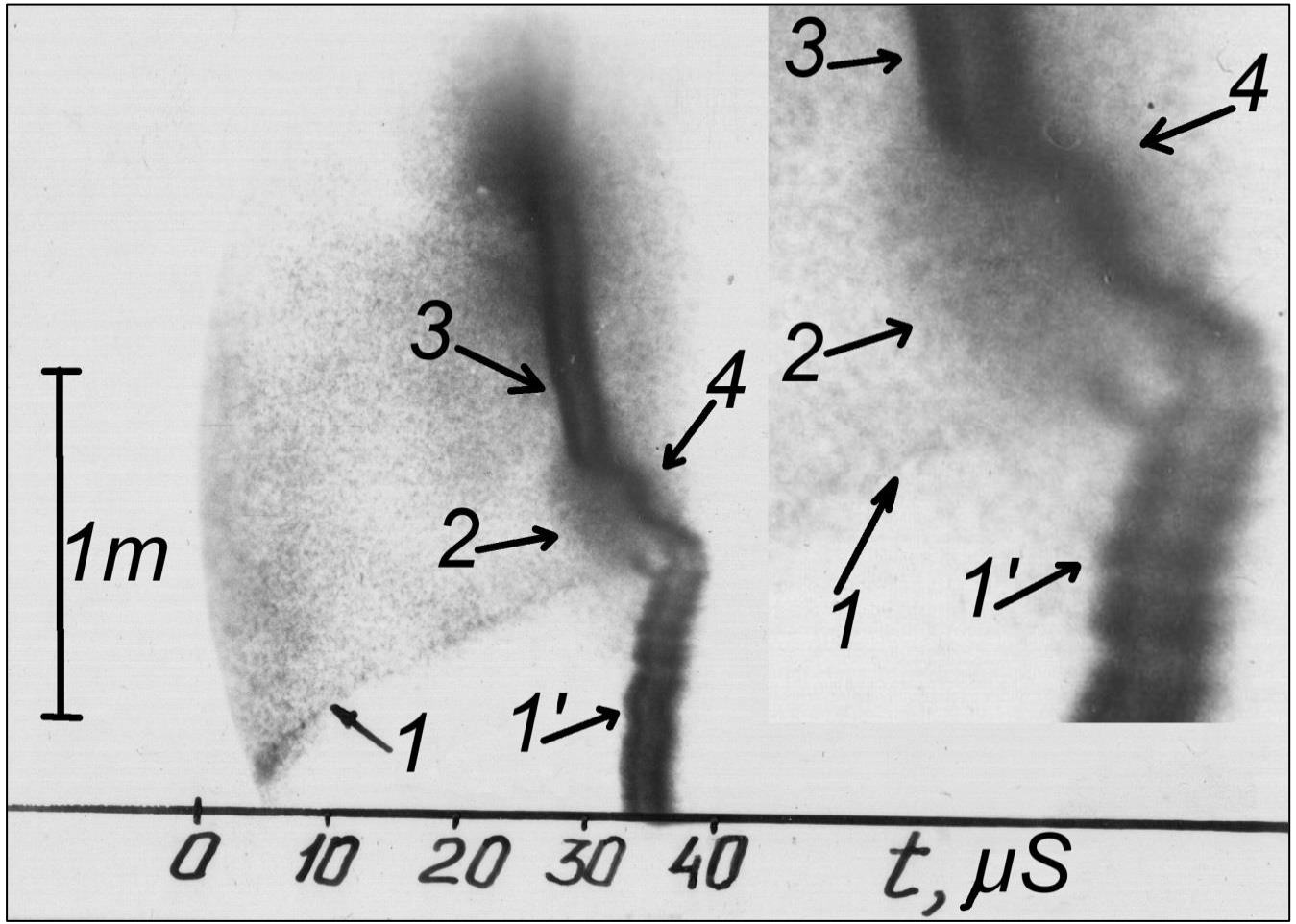

Рисунок 3.2 (адаптировано из работы [Анцупов и др., 1990], оригинал фотографии любезно предоставлен Сысоевым В.С.). Встреча двух лидеров. Развёртка во времени. *1* – Головка положительного лидера; *1'* – канал положительного лидера во время сквозной фазы; *2* – свечение короны во время сквозной фазы, *3* – канал отрицательного лидера во время сквозной фазы; *4* – канал разряда во время главной стадии. Интересно, что здесь отчетливо видно, что яркость каналов может упасть внутри сквозной фазы (ослабление свечения, а не непрерывный процесс увеличения яркости), как позже будет видно на падающих осциллограммах тока внутри сквозной фазы (Рисунок 3.5).

лидером видна соединяющая их положительная корона (2), что однозначно говорит о начале сквозной фазы разряда. Примерно через 1-2 микросекунды наступает главная стадия длинной искры, во время которой лидеры соединяются в единый канал (4). В момент начала сквозной фазы расстояние между положительным и отрицательным лидерами было около 30÷35 см. Таким образом, можно сказать, что уже авторы работы [Анцупов и др., 1990] имели серьёзные основания считать, что зафиксирован факт встречи нисходящего из отрицательного облака отрицательного лидера и восходящего с плоскости положительного лидера. Позднее [Temnikov et al., 2007] на аналогичной установке, реализующей облако отрицательно заряженного аэрозоля, с помощью скоростной камеры с усилением изображения в режиме покадровой съёмки также зафиксировали события, подобные описанной [Анцупов и др., 1990] встрече двух лидеров. К сожалению, невысокое пространственное разрешение кадров не позволяло более подробно изучить это явление. В частности, по развёрткам [Анцупов и др., 1990] и покадровым фотографиям [Temnikov et al., 2007] не удавалось чётко зафиксировать явление, подобное статическим фотографиям Рисунок 3.1. — существование особо выделенной перемычки (мостика) между лидерами, которое однозначно указывает, что произошла встреча двух лидеров и это не точка ветвления восходящего положительного лидера. Кроме того, оставались неизученными два ключевых вопроса подобного процесса. Первый, что происходит на противоположном (исходя из гипотезы Каземира [Kasemir, 1960]) положительном конце внутриоблачного канала (двунаправленного лидера), существует ли там положительный лидер и какова его морфология? Второй вопрос: как и в какой момент происходит рождение внутриоблачного канала в аэрозольном облаке? (Второй вопрос рассматривается в главе 2)

Эти исследования получили дальнейшее развитие в 2013-2018 годах с использованием нового оборудования. В частности, [Kostinskiy et al., 2015a] (см. главу 1), используя высокоскоростную инфракрасную камеру, наблюдали необычные плазменные образования (UPFs) в искусственных отрицательно заряженных облаках водного аэрозоля. Предполагаемые параметры плазмы UPFs были близки к параметрам лидеров, наблюдаемых в тех же экспериментах, в то время как морфология каналов очень сильно отличалась от таковой у лидеров. Эти плазменные образования (UPFs) скорее всего являются проявлениями коллективных процессов (после прохождения через объем облака длинной положительной стримерной вспышки), создающих, ранее не исследованную, сложную иерархическую сеть взаимодействующих каналов (см. подробнее главы 1, 2). [Kostinskiy et al., 2015a] предположили, что это явление также должно происходить в грозовых облаках и представляет собой недостающее звено во все еще плохо изученном механизме инициирования молнии (см. главу 7). [Kostinskiy et al., 2015a] получили детальные инфракрасные изображения двунаправленных лидеров, образованных облаком положительно заряженного водного аэрозоля (см. главу 4). Двунаправленный лидер в этом случае состоял из идущей вниз положительной части канала (лидера) и идущей вверх отрицательной части канала (лидера), причем эти два канала соединены одноканальной средней частью (скорее всего UPFs, из которого впоследствии в обе стороны развились лидеры). Восходящий лидер был связан с разветвленной объемной сетью не так сильно светящихся каналов, пронизывающих верхнюю часть облака (см. глава 4). [Kostinskiy et al., 2015b] продемонстрировали возможность инициирования электрических разрядов арбалетным болтом, движущимся в электрическом поле облака отрицательно заряженного водного аэрозоля (см. главу 5).

В этой главе мы представляем подробные наблюдения сквозной фазы положительных и отрицательных лидеров в электрических разрядах метрового масштаба, генерируемых облаками отрицательно заряженного водного аэрозоля, и обсуждаем их возможные последствия для процесса контакта нисходящей молнии с заземленными объектами. Процесс контакта нисходящей молнии с заземленными объектами, который можно рассматривать как переход от лидерной стадии к обратному удару, является одним из недостаточно понимаемых процессов развития молнии, в первую очередь из-за его короткой продолжительности и частого появления в том же кадре высокоскоростного видео, что и обратный удар. Обычно предполагается, что процесс контакта нисходящей молнии с заземленными объектами при отрицательных ударах включает положительный восходящий соединяющий лидер (positive upward connecting leader (UCL)) и сквозную фазу [Rakov and Uman, 2003, Ch. 4]. К настоящему времени в длинных искрах наблюдалась стримерная корона сквозной фазы вместе с положительным восходящим лидером [Горин и Шкилев, 1974], [Les Renardières Group, 1977], [Lebedev et al., 2007]; [Shcherbakov et al., 2007] и в триггерной молнии [Biagi et al., 2009]; [Hill et al., 2016], но пока не наблюдалась для природной нисходящей молнии.

В любой молнии и длинной искре сквозная фаза начинается, когда стримерные зоны положительных и отрицательных лидеров с относительно низкой проводимостью соприкасаются. После этого образуется общая стримерная зона, которую составляют положительные стримеры, идущие от головки положительного лидера к головке отрицательного лидера. Последующее движение двух плазменных каналов с высокой проводимостью навстречу друг другу происходит внутри общей стримерной зоны. Общая стримерная зона может быть сформирована при разряде молнии, направленной вниз от облака к земле (около земли), при восходящей молнии (вверху) и во время процесса формирования ступени отрицательного лидера (между положительным концом спейс-лидера и головкой отрицательного лидера). Как только два плазменных канала соприкасаются друг с другом, начинается обратный удар (также называемый главной, основной стадией), процесс, подобный обратному удару молнии. Стримерная зона положительного лидера поддерживается средним электрическим полем при атмосферном давлении не менее 450-500 кВ/м (4.5-5 кВ/см), независимо от типа электрического разряда [Bazelyan and Raizer, 1998]. Именно это поле внутри стримерной зоны, а не электрический потенциал или длина разрядного промежутка, во многом определяет физические процессы внутри общей стримерной зоны. Поэтому мы считаем, что понимание характеристик и динамики сквозной фазы длинных искр может быть полезно для улучшения нашего понимания сквозной фазы молнии и других плазменных явлений в грозовых облаках.

## 3.2. Экспериментальная установка

Эксперименты выполнялись на уникальных установках Высоковольтного исследовательского центра РФЯЦ – ВНИИ Технической физики имени ак. Е.И. Забабахина. В данной работе используется подобная описанным в главах 1 и 2 схема экспериментального стенда генерации заряженного аэрозольного облака отрицательной полярности, дополненная другими измерительными приборами (Рисунок 3.3). Выходное сопло (2.3) генератора заряженного аэрозоля располагалось в центре плоского металлического экрана диаметром 2 м с закруглёнными краями (3). Генератор заряженного аэрозоля состоял из генератора пара (2.1) и зарядного устройства (2.2). Паровоздушная струя из паропровода при температуре около 100-120 $^0C$ под давлением

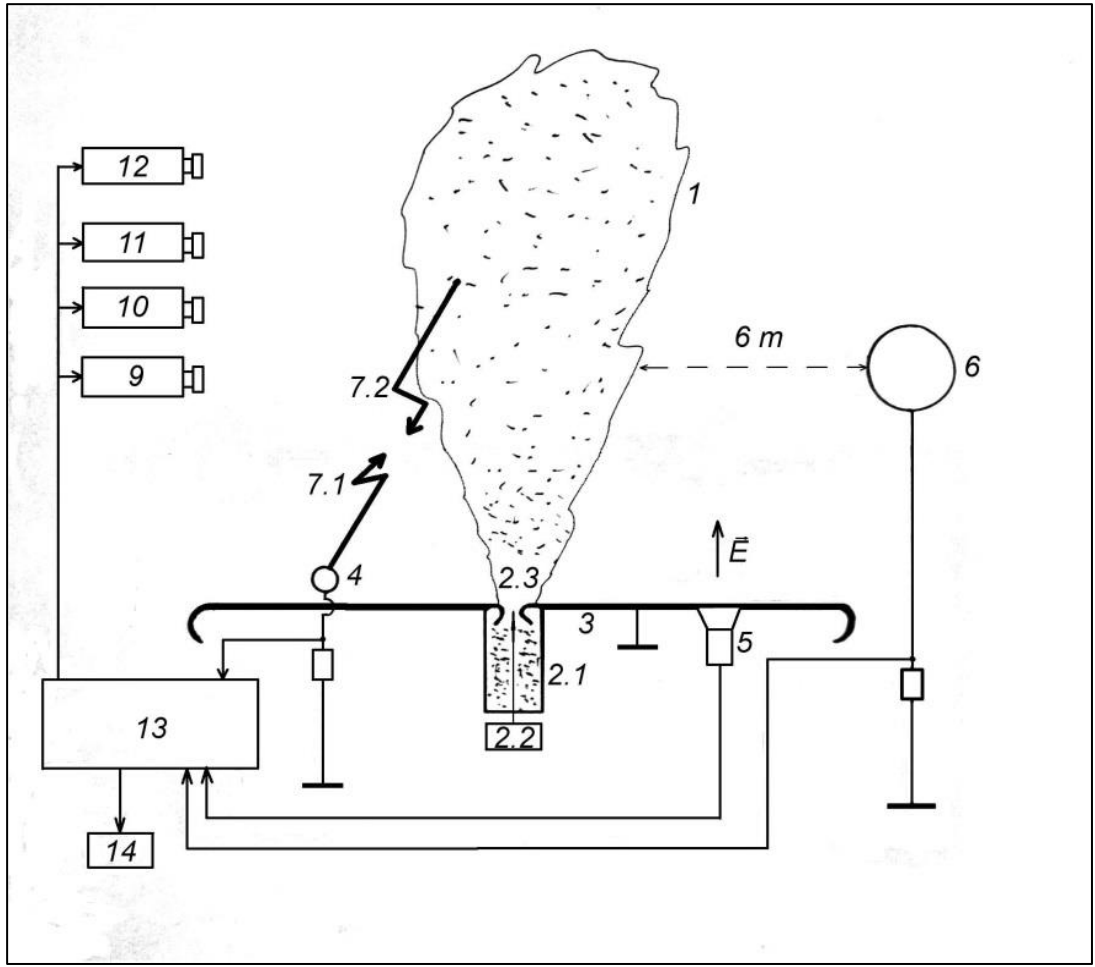

Рисунок 3.3 (адаптировано из [Kostinskiy et al., 2016]). Схема эксперимента. 1— облако заряженного аэрозоля; 2.1 — генератор паровоздушной струи; 2.2 — источник высокого напряжения, подаваемого на иглу в сопле; 2.3 — сопло, через которое выносится паровоздушная струя с коронирующей иглой посредине; 3 — заземленная металлическая плоскость; 4 — приёмный электрод (металлический шарик диаметром 5 см); 5 — измеритель электрического поля, медленный (флюксметр); 6 — зонд электрического поля, быстрый (металлический шар диаметром 50 см); 7.1 — восходящий положительный лидер; 7.2 — отрицательный нисходящий лидер; 9 — FASTCAM SA4 скоростная видеокамера; 10 — FLIR 7700М, скоростная ИК-камера; 11 — ФЭУ (фотоэлектронный умножитель); 12 — 4Picos, скоростная камера с ЭОП; 13 — блок синхронизации и регистрации сигналов (осциллограф Tektronix DPO 71004, генератор сигналов Tektronix AFG 3252); 14 — блок хранения данных (компьютер).

0.2÷0.6 МПа вылетала со скоростью, близкой к скорости звука в данной смеси (около 400÷450 м/с) из сопла (2.3) с углом раскрыва 28°, образуя адиабатически расширяющуюся затопленную струю. При этом в атмосфере создавалось облако заряженного аэрозоля (1). В результате быстрого охлаждения пар конденсировался в капли средним размером 0.3÷1 мкм (в зависимости от влажности окружающего воздуха). Ионы, заряжающие аэрозоль, образовывались в коронном разряде, между тонкой заострённой иглой, расположенной в сопле (2.3) и соплом. На иглу от высоковольтного источника (2.2) подавалось постоянное напряжение 10÷20 кВ отрицательной полярности. Ток выноса заряда паровоздушной струей находился в пределах 60÷150 мкА. При накоплении в аэрозольном облаке общего заряда до значений около 60 мкКл самопроизвольно возникали различные разряды. Для измерения тока, использовался шунт с сопротивлением 1 Ом, сигнал с которого подавался на цифровой осциллограф Tektronix DPO с полосой 500 МГц (13). Шунт был подсоединен к приемному электроду в виде металлического шарика (4) диаметром 5 см, верхняя точка которого возвышалась над плоским экраном на расстоянии 12 см. Шарик находился от центра экрана на расстоянии 0,8÷0,9 м. При превышении тока в шунте заданного значения, запускался осциллограф, который в свою очередь, выдавал импульс запуска на скоростные камеры видимого диапазона 4Picos (12) и FASTCAM SA4 (9) и инфракрасную камеру FLIR7700M (10). Цветная камера FASTCAM SA4 работала во время измерений в режиме непрерывной записи «по кругу» со скоростью 50 000 ÷ 500 000 кадров в секунду (fps) с разрешением 320 × 192 пикселей и останавливалась на том кадре, во время записи которого на неё приходил импульс синхронизации с осциллографа. Черно-белая камера 4Picos с электронно-оптическим усилением изображения позволяла фиксировать два кадра высокого пространственного разрешения (1360x1024 пикселей) с выдержкой каждого кадра от 0.2 нс до 80 с и минимальным временем между кадрами 500 нс (оптическое усиление было в 5000 раз). Пространственное разрешение (размер пикселя) составляло 0,63 мм для 4Picos, 1,2 мм для FLIR и 1,46 мм для FASTCAM. Камеры срабатывали, когда ток через 5-сантиметровую сферу превышал заданное пороговое значение. Для контроля динамики общего заряда аэрозольного облака, использовался изолированный медный шар (6) диаметром 50 см, соединённый через сопротивление 100 МОм с осциллографом (13). Это позволяло фиксировать динамику накопления заряда в процессе зарядки облака и быстрые процессы ухода заряда из него. Шар находится на расстоянии 6 метров от аэрозольного облака. Динамика свечения регистрировалась с помощью ФЭУ видимого диапазона (11). Электрическое поле на поверхности заземлённой плоскости измерялось флюксметром (5). Кроме того, изменение интенсивности света регистрировалось с помощью фотоумножителя (ФЭУ), работающего в видимом диапазоне и направленного в том же направлении, что и высокоскоростные камеры. Общие снимки разрядов были получены с помощью цифрового фотоаппарата (Canon EOS 5D Mark III). Камеры были установлены на расстоянии около 3 м от оси струи, создающей облака.

Во время данных экспериментов аэрозольное облако заряжалось до 50÷100 мкКл. По измерениям с помощью флюксметра, электрическое поле, создаваемое аэрозольным облаком на поверхности заземленного экрана на расстоянии около 0,8 м от оси струи, имело значение 4÷5 кВ/см и слабо росло по направлению к облаку. При этом с поверхности шарика самопроизвольно возникали восходящие к аэрозольному облаку положительные стримерные вспышки и восходящие лидеры. Температура воздуха была в районе 10-15 $^0$С, а относительная влажность составляла 90-95%.

Все разряды, рассмотренные в этой статье, демонстрируют связь между нисходящим отрицательным лидером (который является нижней частью двунаправленного лидера, инициированного в электрическом поле облака) и восходящим положительным лидером, инициированным с заземленной плоскости. В этом отношении они похожи на нисходящие отрицательные удары молнии. В обоих случаях соединение приводит к яркому освещению всего канала и быстро нарастающему импульсу тока, измеренному на земле, что считается проявлением процесса обратного удара в случае молнии. Мы будем применять термин «обратный удар» или «квазиобратный удар», к аналогичному процессу в разрядах, генерируемых искусственно заряженными облаками. Обращает на себя внимание, что максимумы тока для обратных ударов в искусственно заряженном облаке составляют от 5 до 50 А, а длительность импульса тока составляет порядка микросекунды по сравнению с десятками кило Ампер и сотнями микросекунд, соответственно, для молнии. Кроме того, типичный перенос заряда при первом ударе отрицательной молнии составляет около 2-5 Кл, в то время как общий заряд искусственно заряженного облака в этих экспериментах был менее 100 мкКл (в несколько раз меньше, чем заряд, внедряемый одной ступенью лидера молнии). Кроме того, в отличие от нисходящей отрицательной молнии, возможно, в разрядах в поле аэрозольного облака восходящий положительный лидер был инициирован до начала нисходящего отрицательного лидера в облаке. Важно отметить, что существенные различия в величинах заряда и формы волны тока обратного удара, перечисленные выше, вряд ли существенно повлияют (по крайней мере, качественно) на процессы в стримерной зоне, которым в основном и посвящен данный раздел. Действительно, разрядные процессы в стримерной зоне лидера определяются электрическим полем, создаваемым зарядами головки лидера, зарядами на коротком участке лидерного канала (включая чехол — его коронирующую оболочку) сразу за головкой и зарядами стримеров, образующих стримерную зону; они слабо зависят от крупномасштабного внешнего электрического поля, создаваемого зарядами в облаке. Во время сквозной фазы напряженность электрического поля внутри общей стримерной зоны увеличивается и влияние внешних электрических полей там становится еще менее значимым.

### 3.3. Экспериментальные результаты

### 3.3.1. Сквозная фаза взаимодействия лидеров, зафиксированная камерой с усилением изображения

Здесь будут представлены по два кадра 4Picos с соответствующими записями силы тока и видимого излучения из разряда. Неопределенность по времени составляла 20–40 нс.

Первые два кадра показана на Рисунке 3.4. Время экспозиции каждого кадра составляло 100 нс, а временной интервал между кадрами составлял 2 мкс. На кадре (I) изображена стримерная зона нисходящего отрицательного лидера (1) и восходящего положительного лидера — UPL (2), которые находятся в контакте, образуя общую стримерную зону (3). Обращает на себя внимание, что UPL, общая длина которого составляет около 1 м, является разветвленным, а нисходящий отрицательный лидер — нет. Длина общей стримерной зоны около 17 см. Стримеры в основном излучают в УФ-диапазоне (300–430 нм). Диапазон чувствительности фотокатода 4Picos со стеклянным объективом (315–850 нм) покрывает значительную часть УФ-диапазона излучения стримеров. На кадре (II) видна более поздняя стадия процесса, похожего на обратный

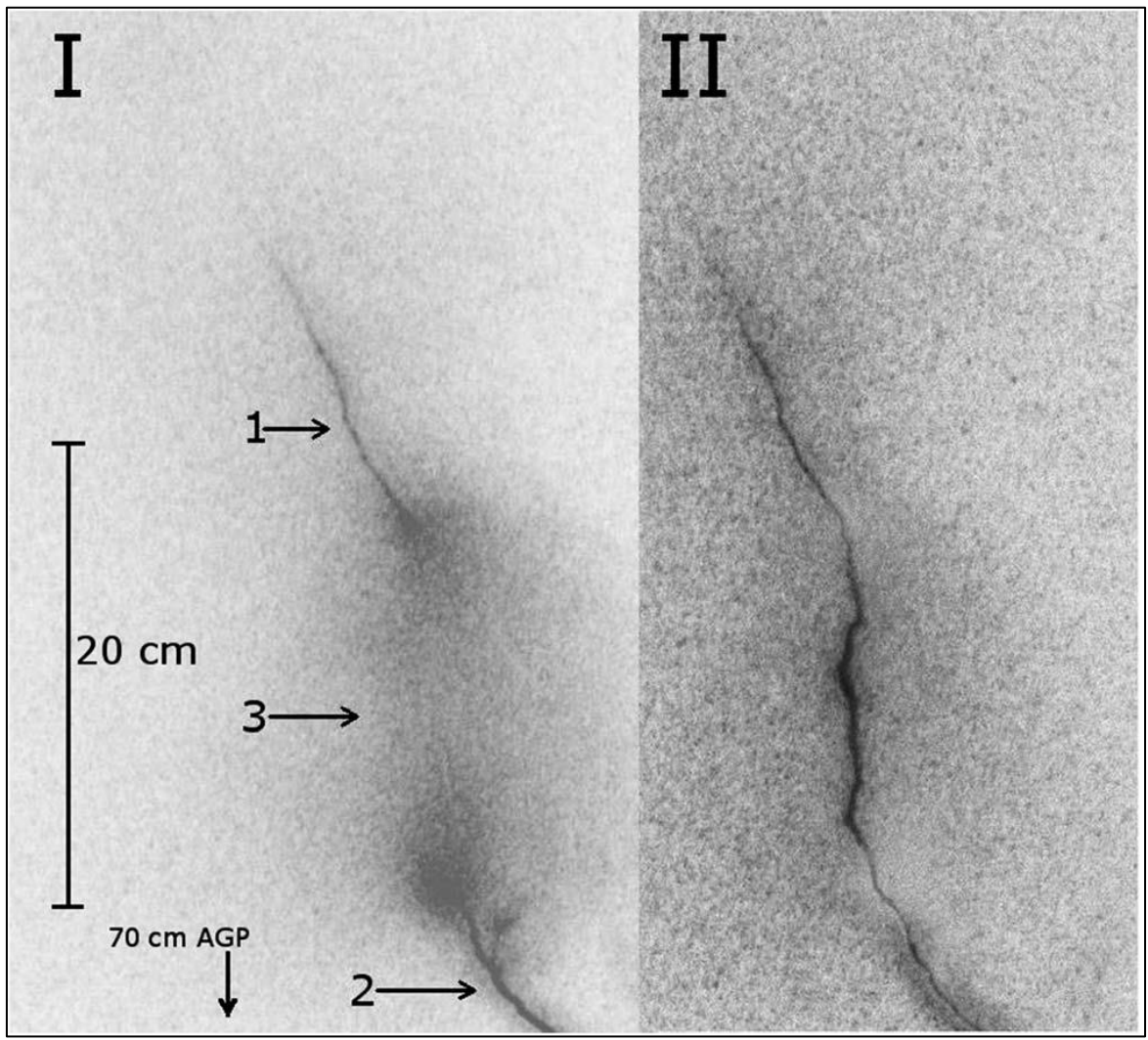

Рисунок 3.4 (адаптировано из [Kostinskiy et al., 2016]). Два кадра камеры 4Picos, показывающие сквозную фазу (I) и более позднюю стадию процесса, похожего на возвратный удар (II), отрицательного разряда на землю, создаваемого облаком искусственно заряженного водного аэрозоля. Время экспозиции каждого кадра составляет 100 нс, а временной интервал между кадрами составляет 2 мкс (см. (4) на Рисунок 3.5). Нисходящий отрицательный лидер — 1, восходящий положительный лидер —2 и общая стримерная зона — 3. Изображение (II) было значительно слабым, чем изображение (I) (из-за значительно более слабого тока, соответствующего изображению (II) (см. значения тока (1) в моменты съемки (4) на Рисунке 3.5b) и было усилено контрастом больше, чем изображение (I), чтобы улучшить его визуализацию. AGP ("above the grounded plane") означает «над заземленной плоскостью». Другие параметры события см. на Рисунке 3.5.

удар. Пятна относительно низкой яркости в кадре (II) в местах, где были яркие головки лидеров в кадре (I) являются артефактами камеры. Причина артефакта в том, что времени между кадрами недостаточно для полного восстановления люминофора, используемого в усилителе изображения камеры. Соответствующие записи тока и интенсивности света, а также сигнала запуска камеры и времени экспозиции двух кадров, изображенных на Рисунке 3.4, показаны в двух разных временных масштабах на Рисунках 3.5a и 3.5b. Осциллограммы величины тока и интенсивности света имеют два пика: первый, меньший, соответствует сквозной фазе, а второй, больший, — процессу, подобному обратному удару (импульсы тока ограничены шкалой измерения и превышают 8 А). Временной интервал между двумя пиками составляет около 1,4 мкс. Ток в сквозной фазе уменьшается после пика, превышающего 8 А, и составляет 4,9 А (в минимуме колебаний) непосредственно перед началом процесса, подобного обратному удару. (Здесь и далее по тексту раздела все токи имеют абсолютные значения, а начало обратного удара определяется по резкому увеличению тока). Спад тока после его первого пика (сквозная фаза) демонстрирует колебания с размахом амплитуды в диапазоне от 0,6 до 1,9 А. В среднем ток снижается с 7,8 А в конце первого кадра до 5,7 А в начале обратного удара. Интересно, что на Рисунке 3.2 также отчетливо видно, что яркость каналов падает внутри сквозной фазы (ослабление свечения, а не непрерывный процесс), аналогично поведению тока на Рисунке 3.5.

На Рисунке 3.6 показаны два кадра камеры 4Picos, на которых показаны два снимка сквозной фазы отрицательного разряда на землю, создаваемого облаком искусственно заряженного водного аэрозоля. На сегодняшний день это единственный двухкадровый снимок сквозной фазы соединения противоположно заряженных лидеров длинной искры или молнии. Время экспозиции для кадра (I) составляло 100 нс, для кадра (II) — 50 нс. Временной интервал между кадрами составлял 2 мкс. Обозначены нисходящий отрицательный лидер (1), восходящий положительный лидер (2) и общая стримерная зона (3). Длина общей стримерной зоны (расстояние между концами лидеров) составляет около 20 см в кадре (I) и около 4,5 см в кадре (II). Обращает на себя внимание, что только восходящий положительный лидер на Рисунке 3.6(I) является разветвленным, в то время как на Рисунке 3.6 (II) как восходящие положительные, так и нисходящие отрицательные лидеры демонстрируют явное ветвление. Удивительно, но фактически, внутри общей стримерной зоны можно выделить две стримерные зоны сквозной фазы на кадре (II) (одна

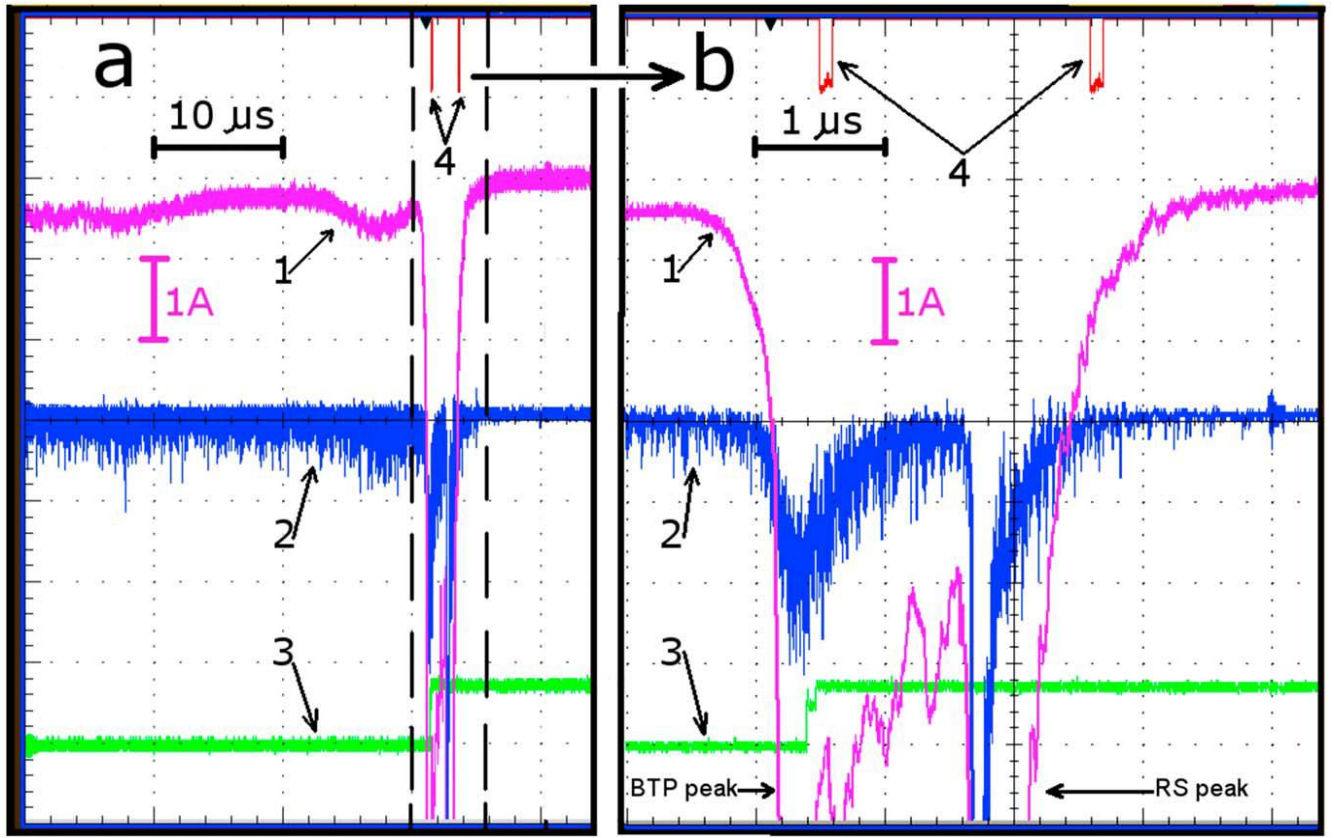

Рисунок 3.5 (адаптировано из [Kostinskiy et al., 2016]). Осциллограммы тока — 1 и интенсивности света — 2, соответствующие двум кадрам 4Picos, показанным на Рисунке 3.4. Также показаны сигнал запуска камеры — 3 и время экспозиции двух кадров — 4. Все записи показаны на двух временных шкалах: (а) — 10 мкс на деление и (б) — 1 мкс на деление. Вертикальные пунктирные линии на Рисунке «а» показывают общий временной интервал, показанный на Рисунке «b». BTP — обозначает сквозную фазу и RS обозначает — «процесс, похожий на обратный удар».

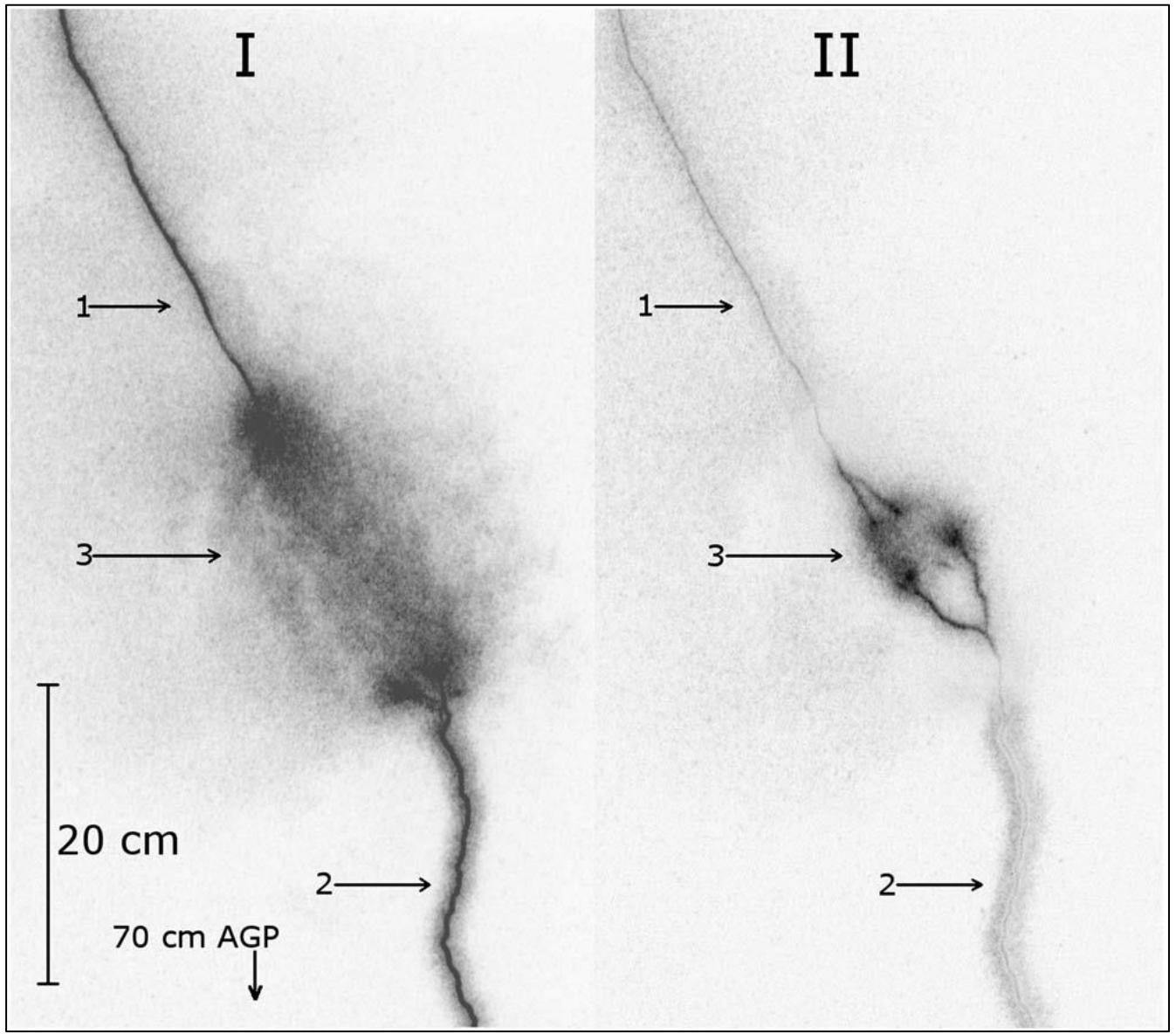

Рисунок 3.6 (адаптировано из [Kostinskiy et al., 2016]). На двух кадрах камеры 4Picos показана сквозная фаза отрицательного разряда на землю, создаваемого облаком искусственно заряженных капель воды. Время экспозиции для кадра (I) составляет 100 нс, для кадра (II) — 50 нс. Временной интервал между кадрами составляет 2 мкс. Обозначены нисходящий отрицательный лидер — 1, восходящий положительный лидер — 2 и общая стримерная зона — 3. AGP ("above the grounded plane") означает «над заземленной плоскостью».

общая стримерная зона, видимая на кадре (I), преобразовалась в две стримерные зоны на кадре (II)) и вполне вероятно, что могут образоваться две точки соединения каналов, ведущие к петле или разделению канала на два в процессе обратного удара (хотя для данного события нет изображения единого канала после обратного удара), особенность, которая иногда наблюдается как в лабораторных искрах, так и в молнии. Соответствующие осциллограммы тока и интенсивности света, а также сигнала запуска камеры и времени экспозиции двух кадров (см. Рисунок 3.6) показаны в двух разных временных масштабах на рисунках 3.7a и 3.7b. Как и на Рисунке 3.5, осциллограммы силы тока и света показывают два пика, первый соответствует сквозной фазе, а второй — обратному удару.

Однако в этом случае пик интенсивности света в процессе, подобном обратному удару, меньше, чем в сквозной фазе (возможно из-за исчерпания заряда в канале двунаправленного лидера). Временной интервал между двумя пиками составляет около 2,4 мкс. Импульсы тока сквозной фазы и обратного удара ограничиваются 8 А шкалы осциллографа. Ток в сквозной фазе во время первого кадра и в течение 0,3 мкс после этого кадра превышает 8 А. Затем он показывает уменьшение с колебаниями, аналогично, но менее выражены, чем на рисунке 3.5b, и за 0,15 мкс до начала второго кадра достигается минимум 3,2 А. Во время второго кадра ток составляет 3,4 А, а затем постепенно увеличивается до 4 А в начале обратного удара.

## 3.3.2. Яркость области контакта каналов относительно верхней и нижней частей объединенного канала (соединение головка-головка)

На Рисунке 3.8 (слева) представлен пример инфракрасного (ИК) изображения (одного из нескольких сотен), показывающего область контакта, расположенного полностью внутри облака, между нисходящим отрицательным и восходящим положительным лидером и профили ИК-яркости (Рисунок 3.8, справа), соответствующие ИК-интенсивностям восходящего положительного лидера (1), нисходящего отрицательного лидера (2) и области соединения лидеров (3). Положительные и отрицательные лидеры были идентифицированы на основе разной степени извилистости их каналов: положительные лидерные каналы в наших экспериментах всегда были более

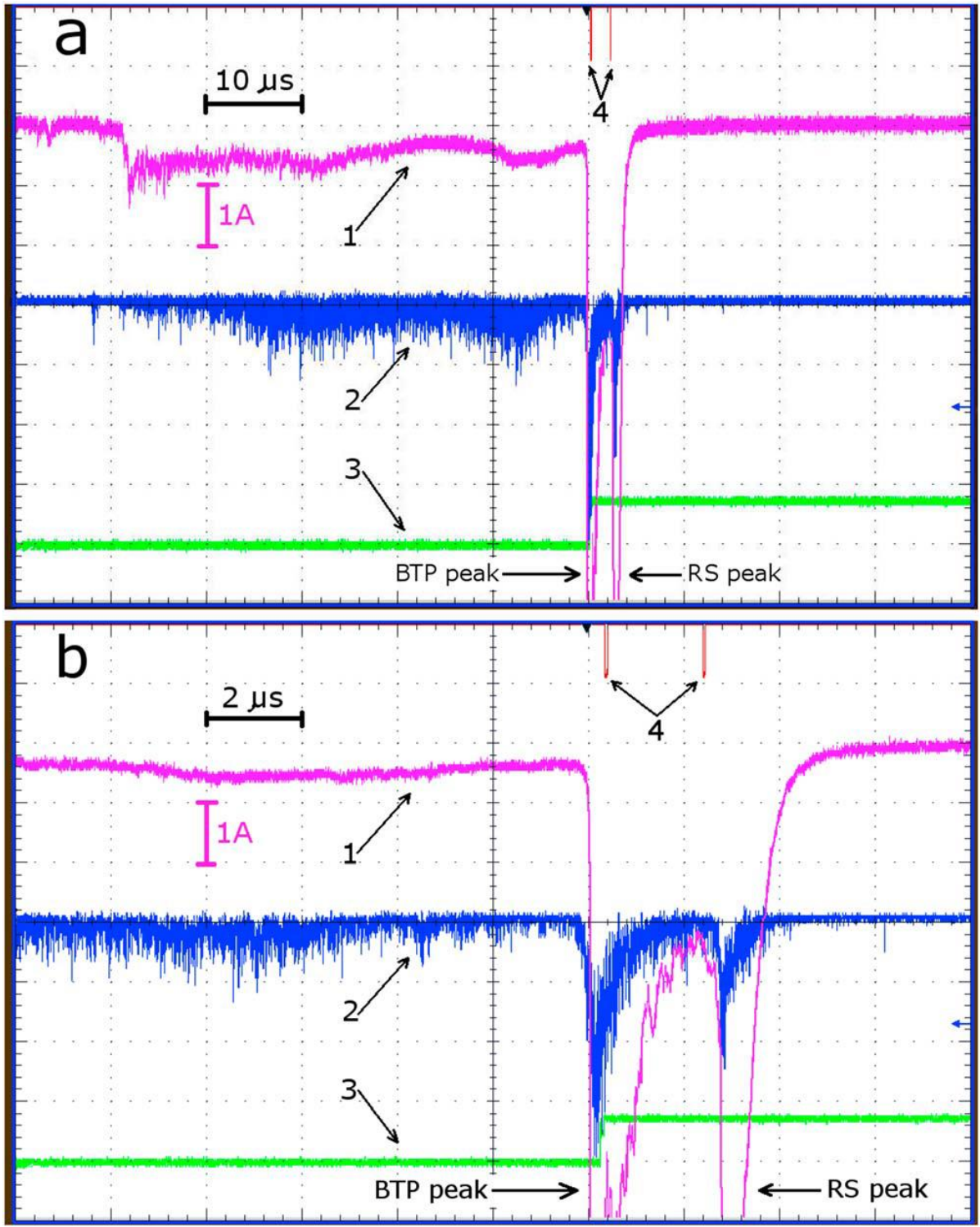

Рисунок 3.7 (адаптировано из [Kostinskiy et al., 2016]). Осциллограммы тока — 1 и интенсивности света — 2, соответствующие двум кадрам 4Picos, показанным на Рисунке 3.6. Также показаны сигнал запуска камеры — 3 и время экспозиции двух кадров — 4. Все записи показаны на двух временных шкалах: (a) — 10 мкс на деление и (b) — 2 мкс на деление. BTP и RS обозначают сквозную фазу и обратный удар, соответственно.

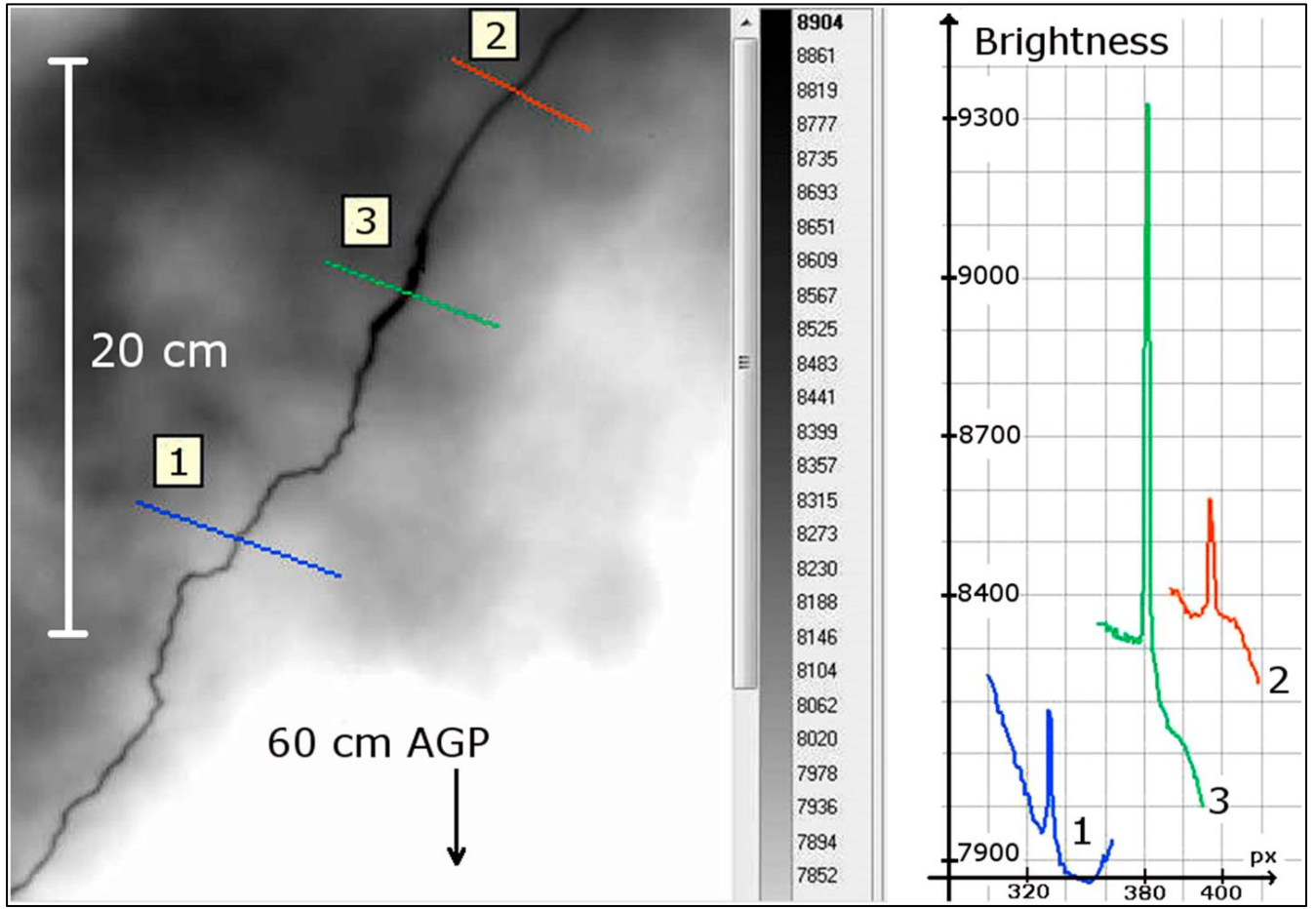

Рисунок 3.8 (адаптировано из [Kostinskiy et al., 2016]). (слева) Инфракрасное изображение, показывающее область соединения (находится полностью внутри облака) между нисходящим отрицательным и восходящим положительными лидерами (соединение головка-головка); (справа) профили ИК-яркости, соответствующие восходящему положительному лидеру — 1, нисходящему отрицательному лидеру — 2 и область соединения — 3. По горизонтальной оси отложено расстояние (в пикселях) вдоль наклонных линий, обозначенных 1, 2 и 3 на рисунке слева, а по вертикальной оси отложена ИК-яркость в относительные единицы. AGP означает — «над заземленной плоскостью».

извилистыми, чем отрицательные. Это заключение основано на нашем анализе нескольких сотен изображений, для которых мы знали полярность восходящего лидера по измеренному току и видели процесс контакта в видимом диапазоне. В 100% случаев положительный канал лидера был явно более извилистым, чем отрицательный. При движении канала молнии извилистость отрицательного лидера, связана с его ступенчатым движением — процессом, которого скорее всего не было в наблюдаемых нами искрах метрового масштаба. Это объяснение является наиболее вероятным в наших экспериментах, где наблюдалась низкая извилистость отрицательных лидеров. Отрицательный лидер двигался без ступеней вероятно потому, что его ток непрерывно обеспечивали восходящие положительные стримеры, которые возникали на головке встречного восходящего положительного лидера либо стримеры поднимались с заземленной плоскости, так как электрическое поле, вероятно превышало 5 кВ/см (поле поддержания положительных стримеров) на всей длине от заземленной плоскости до головки отрицательного лидера (см. главу 2). Обращает на себя внимание, что ИК-яркость области контакта (вероятно, пропорциональная нагреву газа; см. [Kostinskiy et. al., 2015a]) в 5 раз выше, чем яркость, соответствующая положительному или отрицательному лидеру выше и ниже области контакта.

### 3.3.3. Связь нисходящего отрицательного лидера с боковой поверхностью восходящего положительного лидера

На рисунке 3.9 показаны два последовательных кадра цветной скоростной камеры FASTCAM SA4, работающей со скоростью 50000 кадров в секунду с разрешением 320×192 пикселей. Время экспозиции каждого кадра составляло 20 мкс при практически нулевом мертвом времени между кадрами. На кадре 1 виден только положительный лидер, восходящий с заземленной сферы. На кадре 2 восходящий положительный лидер немного длиннее и ярче и находится в контакте с нисходящим отрицательным лидером. Положительные и отрицательные лидеры были идентифицированы на основе анализа тока и относительного уровня их извилистости, как обсуждалось выше. Изображение на кадре 2 фиксирует обратный удар. Соответствующая этому событию осциллограмма тока показана на рисунке 3.10. Начальный пик тока соответствует вспышке восходящих

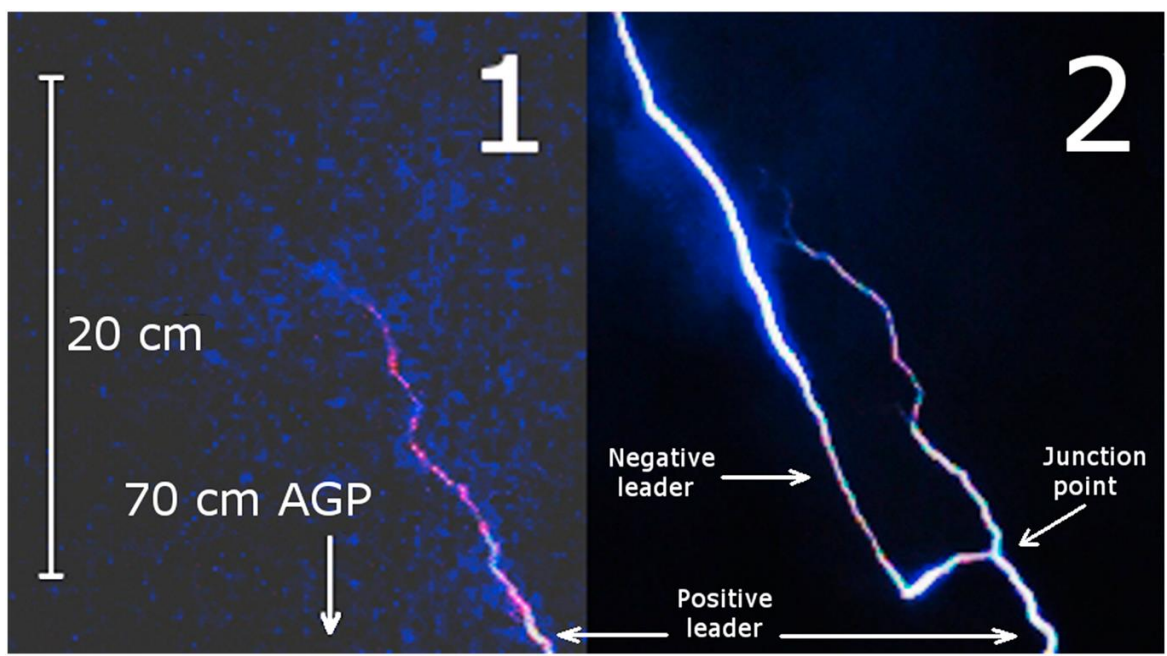

Рисунок.3.9 (адаптировано из [Kostinskiy et al., 2016]). Два последовательных кадра с выдержкой 20 мкс камеры FASTCAM SA4, показывающие контакт нисходящего отрицательного лидера с восходящим положительным лидером ниже его головки. Примерно 16% наших разрядов демонстрируют такое поведение. Положительные и отрицательные лидеры были идентифицированы на основании измерений знака тока и экспериментально определенной зависимости извилистости лидерного канала от полярности (для нескольких сотен исследованных случаев положительный лидерный канал всегда был заметно более извилистым, чем отрицательный). AGP означает — «над заземленной плоскостью». Соответствующая данному событию осциллограмма тока показана ниже на Рисунке 3.10.

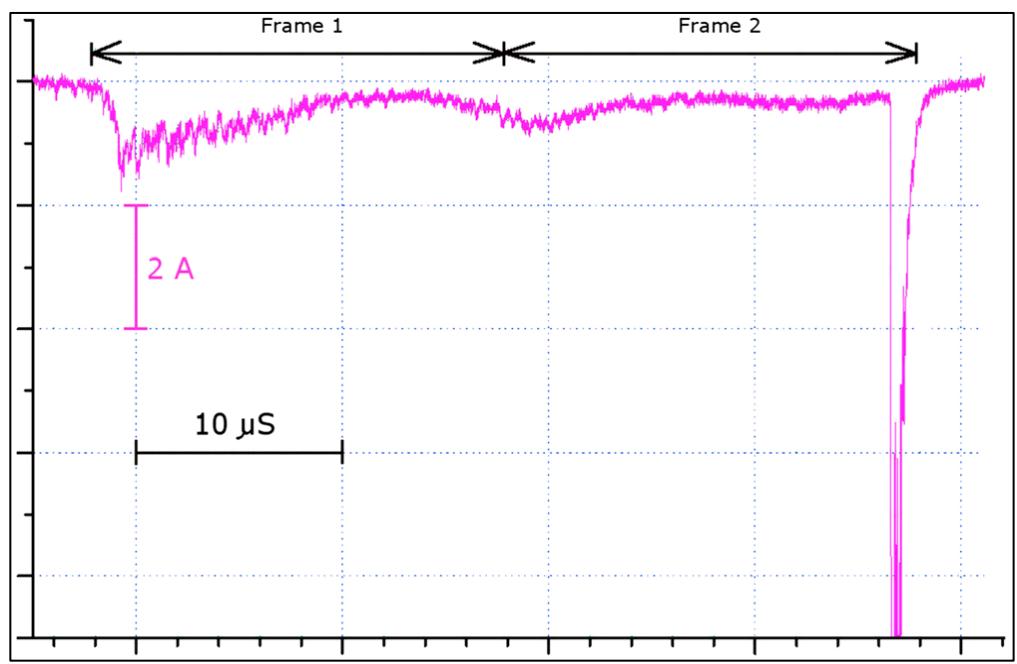

Рисунок 3.10 (адаптировано из [Kostinskiy et al., 2016]). Осциллограмма тока, соответствующая двум кадрам FASTCAM SA4, показанным на рисунке 3.9.

положительных стримеров (пиковый ток около 1,8 А, ширина пика по полувысоте равна 50–100 нс), за которым следует восходящий положительный лидер, развивающийся в течение примерно 37 мкс с уменьшением тока примерно до 0,2–0,4 А. Резкий импульс тока, соответствующий процессу, подобному обратному удару, имеет пик более 9 А, время нарастания фронта 80–100 нс и полуширина пика по полувысоте около 400 нс при общей длительности 1,5–1,7 мкс. Как отмечалось выше, кадр 1 на Рисунке 3.9 соответствует восходящему положительному лидеру, когда величина тока была относительно небольшой, а кадр 2 включает в себя процесс, похожий на обратный удар молнии. Отрицательный лидер на Рисунке 3.9(2) соединен с положительным лидером значительно ниже его верхнего конца (с боковой поверхностью положительного лидера) и что соединение осуществляется через сегмент канала длиной 4,1 см, который кажется более или менее перпендикулярным, как к положительному, так и к отрицательному каналам лидеров. Мы неоднократно наблюдали такую геометрию на фотографиях отрицательных разрядов на землю, создаваемых искусственно заряженными облаками. Пример такой фотографии показан на Рисунке 3.11.

### 3.3.4. Возможная причина возникновения внутриоблачного двунаправленного лидера

Возникает вопрос, за счёт каких процессов возникает двунаправленный лидер внутри аэрозольного облака? В настоящее время на данный вопрос можно ответить предположительно, опираясь на материал, изложенный в главе 2 и на временные развертки, снятые ранее стрик-камерой (электронно-оптическим преобразователем или фотоэлектронным регистратором – ФЭР-14) [Анцупов и др., 1990].

На Рисунке 3.12 [адаптировано из Анцупов и др., 1990] изображена развертка во времени процесса появления в объёме и встречи двух лидеров, полученная ФЭР-14. Обращает на себя внимание важный факт, что за несколько микросекунд до старта лидеров на Рисунке 3.12 фиксируется короткая (0.5÷0.8 мкс) конусообразная стримерная вспышка на высоте 0.5÷1 м над заземлённой плоскостью (1) (подробное обсуждение этого явления проводится в разделе 2.3). [Анцупов и др., 1990] в тексте статьи называют ее «ядро».  После этого положительный восходящий лидер (2) начинает двигаться навстречу

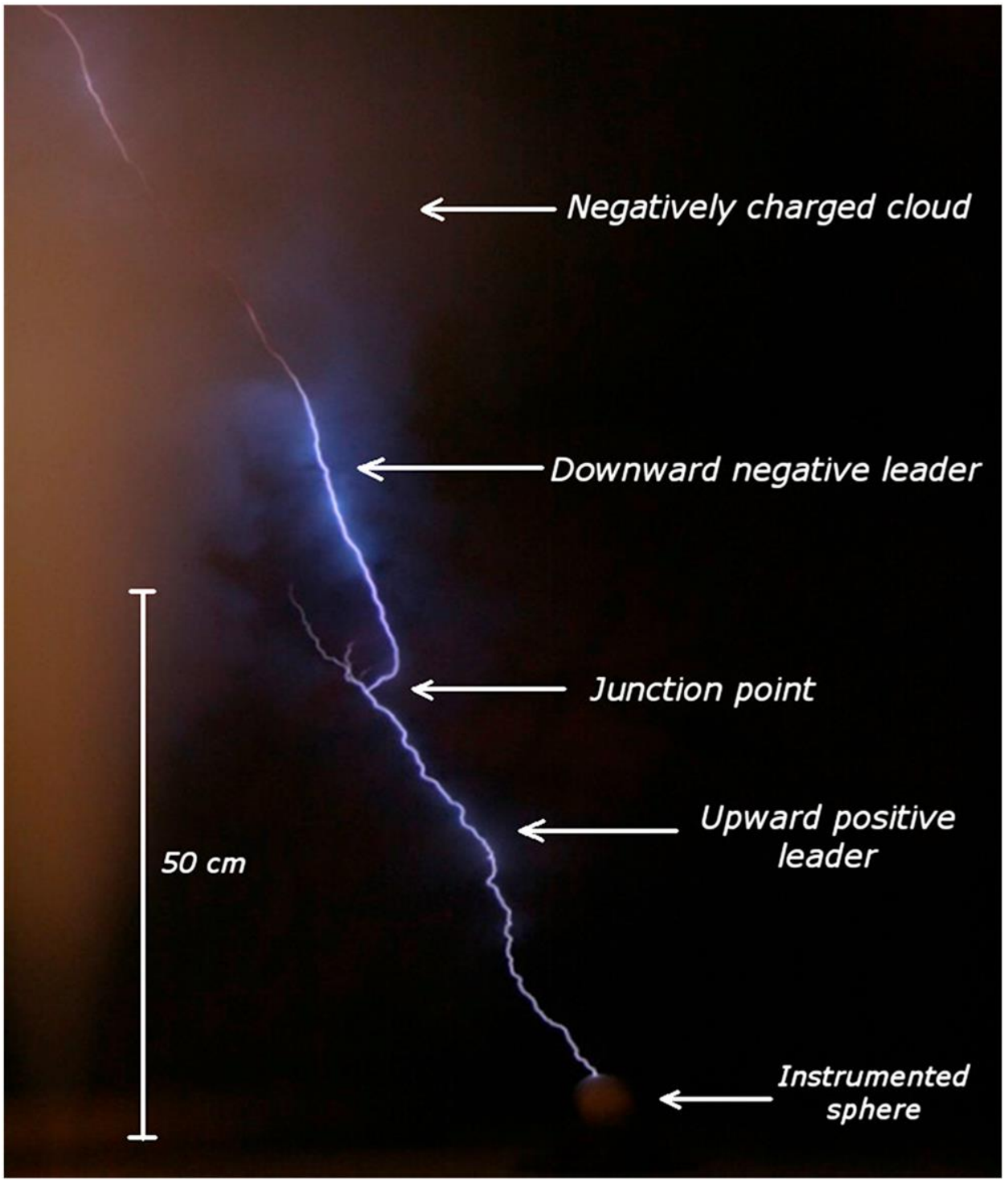

Рисунок 3.11 (адаптировано из [Kostinskiy et al., 2016]). На фотографии показан нисходящий отрицательный лидер, контактирующий с восходящим положительным лидером ниже своего верхнего конца. Полярность лидера определялась на основании осциллограммы тока и экспериментально определенной зависимости извилистости лидерного канала от полярности (для нескольких сотен исследованных случаев положительный лидерный канал всегда был заметно более извилистым, чем отрицательный). Время выдержки 2 с.

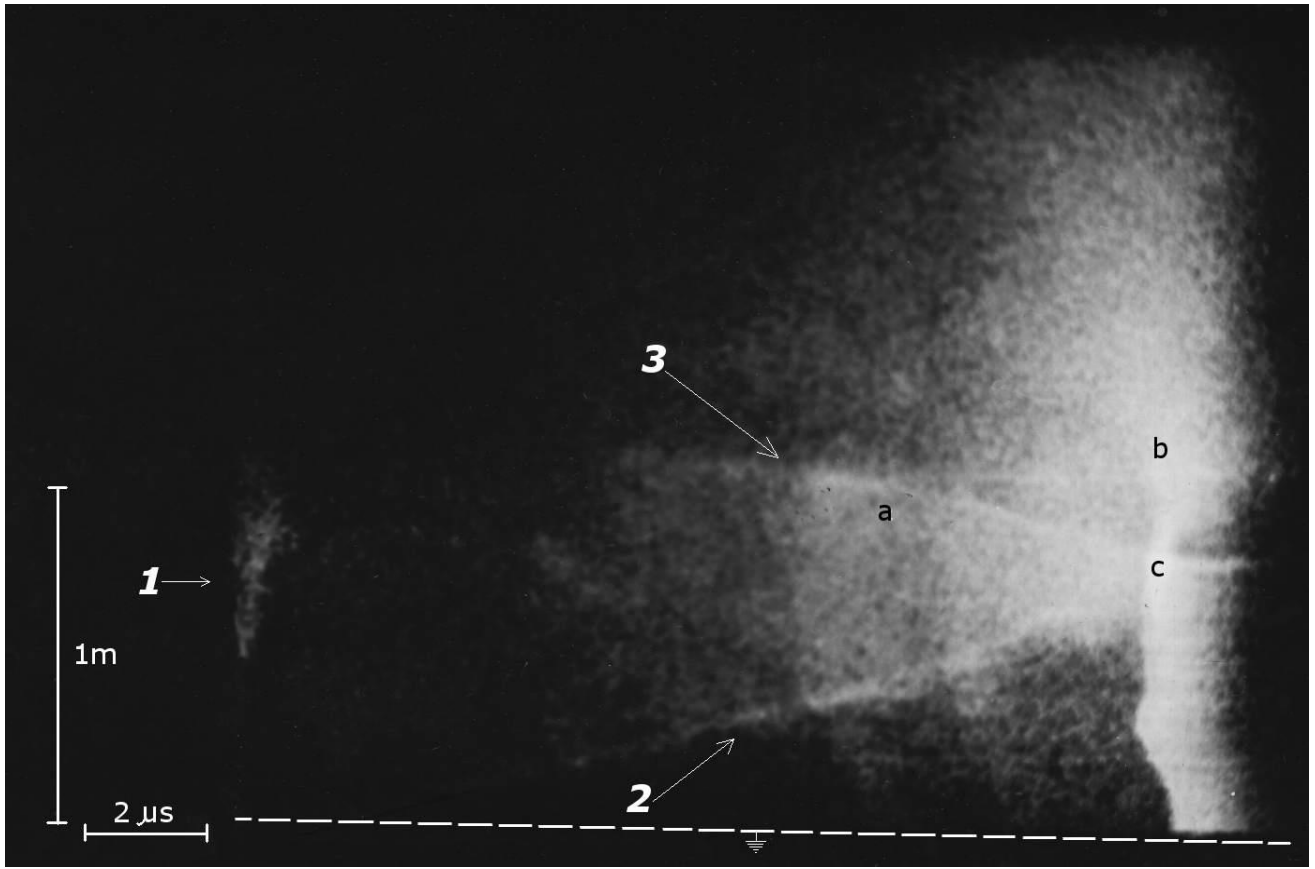

Рисунок 3.12 (адаптировано из [Анцупов и др., 1990], оригинал фотографии любезно предоставлен В.С. Сысоевым). Развертка во времени процесса инициации стримерной вспышки около облака и последующей инициации положительного восходящего и отрицательного нисходящих лидеров, полученная стрик-камерой. Движение лидеров друг к другу ведет к их контакту и явлению, похожему на обратный удар. 1 – Конусообразная стримерная вспышка на высоте 0.5÷1 м над заземлённой плоскостью. 2 – положительный восходящий лидер, 3 – плазменное образование (возможно UPFs), из которого стартует лидер, нисходящая часть которого *ac* — отрицательный лидер. *abc* – характерный треугольник двунаправленного лидера, который при развёртках движения лидера отрицательной искры соответствует спейс-лидеру [Анцупов и др., 1990].

внутриоблачному плазменному образованию (1), которое через примерно 5 мкс после появления на высоте 1 метр в точке *a* превращается в двунаправленный лидер (3), нисходящая часть которого *ac* — отрицательный лидер, а восходящая часть *ab* — положительный лидер. Динамика развития двунаправленного лидера похожа на спейс-лидер отрицательной длинной искры ([Stekolnikov and Shkilyov (1963)], [Gorin et al., 1976]). На начальной стадии движения, судя по наклону траекторий движения головок лидеров, их скорости изменяются слабо. После возникновения более яркого свечения в пространстве между лидерами (усиление сквозной фазы) их скорость начала увеличиваться вплоть до момента активной сквозной фазы, который характеризуется яркой вспышкой излучения (в этот момент ФЭР-14 закрывает электронно-оптический преобразователь, чтобы не повредить прибор во время яркой вспышки при квазиобратном ударе). Также на изображении, полученном с ФЭР-14 хорошо виден характерный треугольник двунаправленного лидера *abc*, который при временных развёртках движения лидера отрицательной искры соответствует спейс-лидеру. Надо отметить, что на других изображениях подобных событий перед стартом лидеров может быть зафиксировано несколько стримерных вспышек, похожих на вспышку (1), Рисунок 2.12 [Анцупов и др., 1990], и только потом стартуют лидеры. Интервал времени между этими предварительными стримерными вспышками может находиться в широком диапазоне от 0,3 до 80 мкс. С высокой вероятностью эти предварительные свечения в объёме облака начинаются со вспышки стримерной короны с заземлённой плоскости. Интегральная фотография такого события приведена на Рисунке 3.13, где хорошо видна и конусность данного свечения, и подобие геометрических размеров изображению на Рисунке 3.12 (расстояние от сопла в центре, откуда инжектируется аэрозольная струя, до края заземленной плоскости составляет 1 м). Так как аэрозольное облако создаёт достаточно равномерное электрическое поле с напряжённостью от 5 до 7÷10 кВ/см, то положительные стримеры, возникающие на заземлённой плоскости (и, в частности, на токоприёмнике — сфере) достигают облака и могут стать причиной последующего возникновения горячих плазменных образований в результате развития ионизационно-перегревной неустойчивости [Bazelyan and Raizer, 1998], которые позже стимулируют появление внутриоблачных лидеров. На Рисунках 3.14.I и 3.14.II, камерой 4Picos зафиксирована вспышка положительной стримерной короны, вслед за которой последовало образование лидеров и обратный удар при встрече этих лидеров.

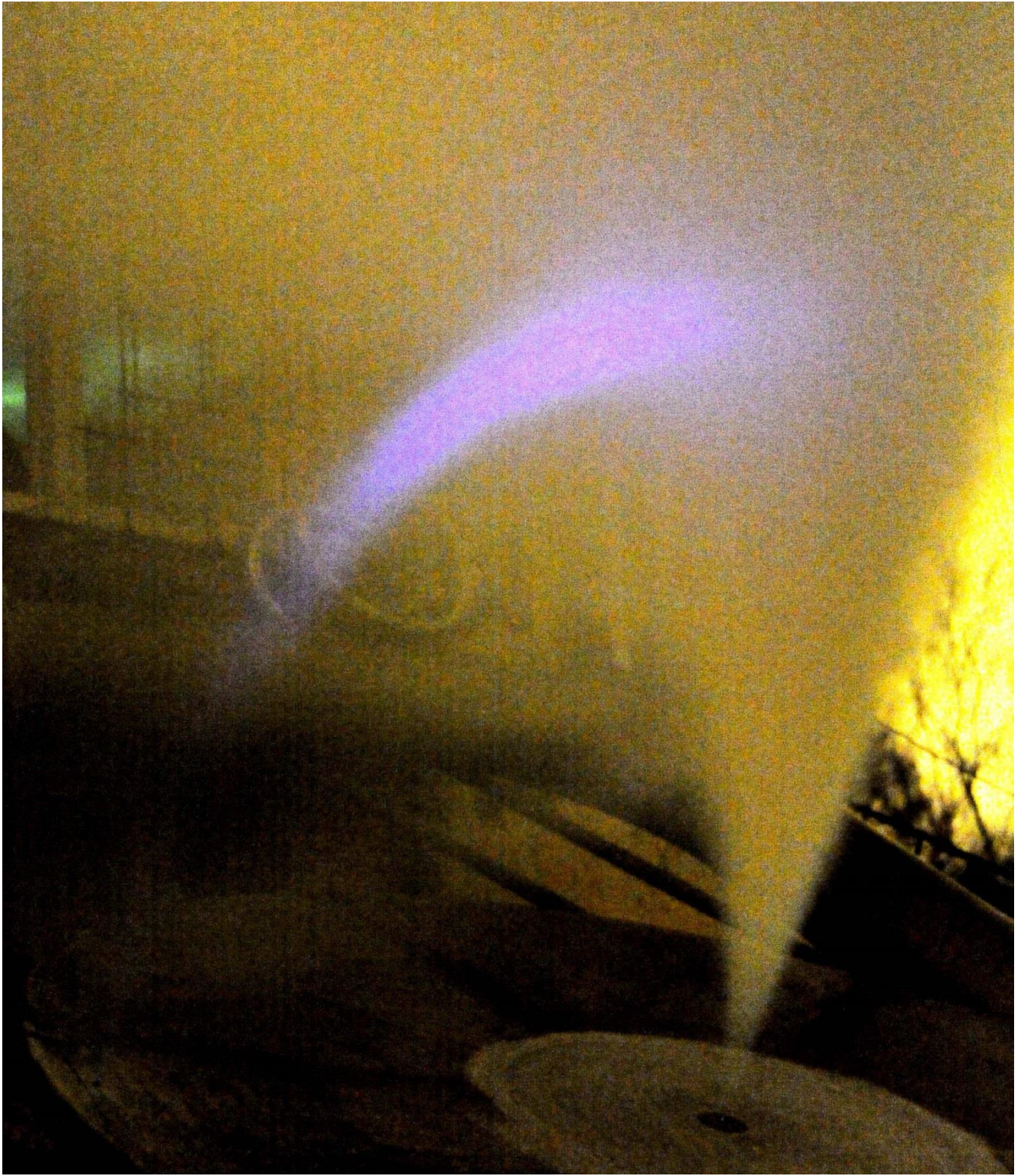

Рисунок 3.13. Фотография положительной стримерной вспышки, восходящей с плоскости к отрицательно заряженному облаку водного аэрозоля.

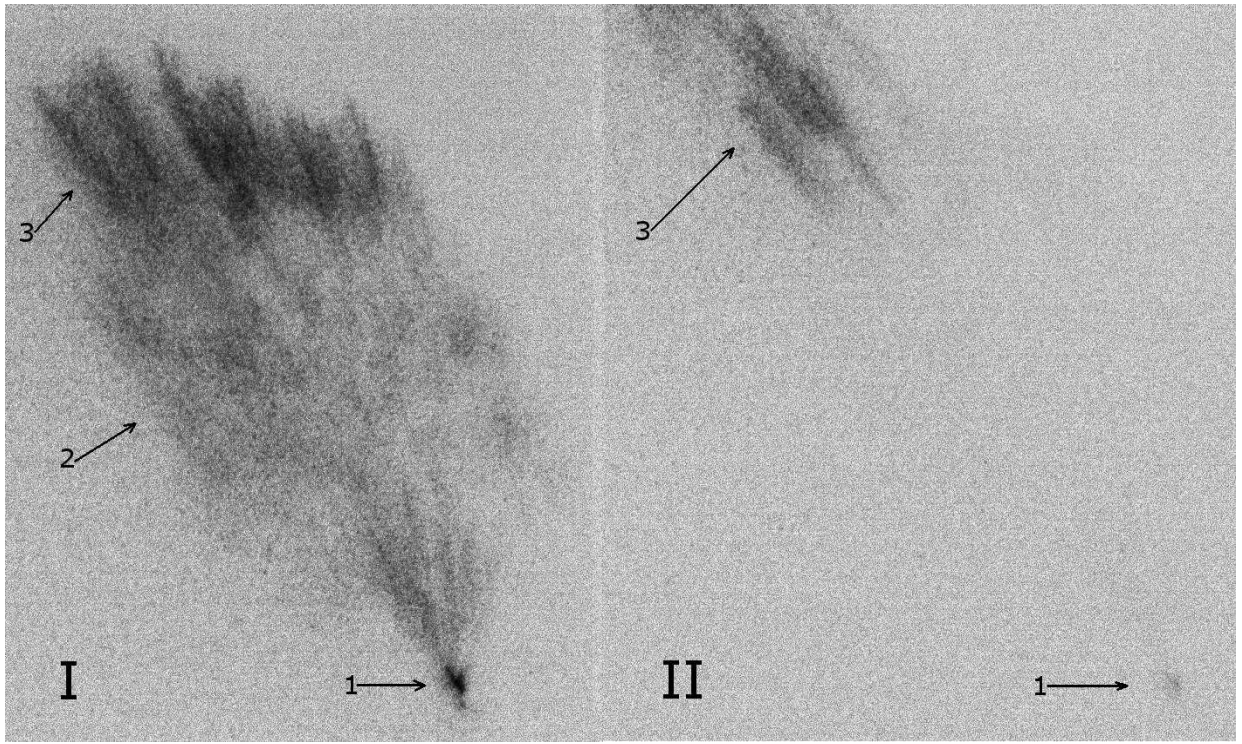

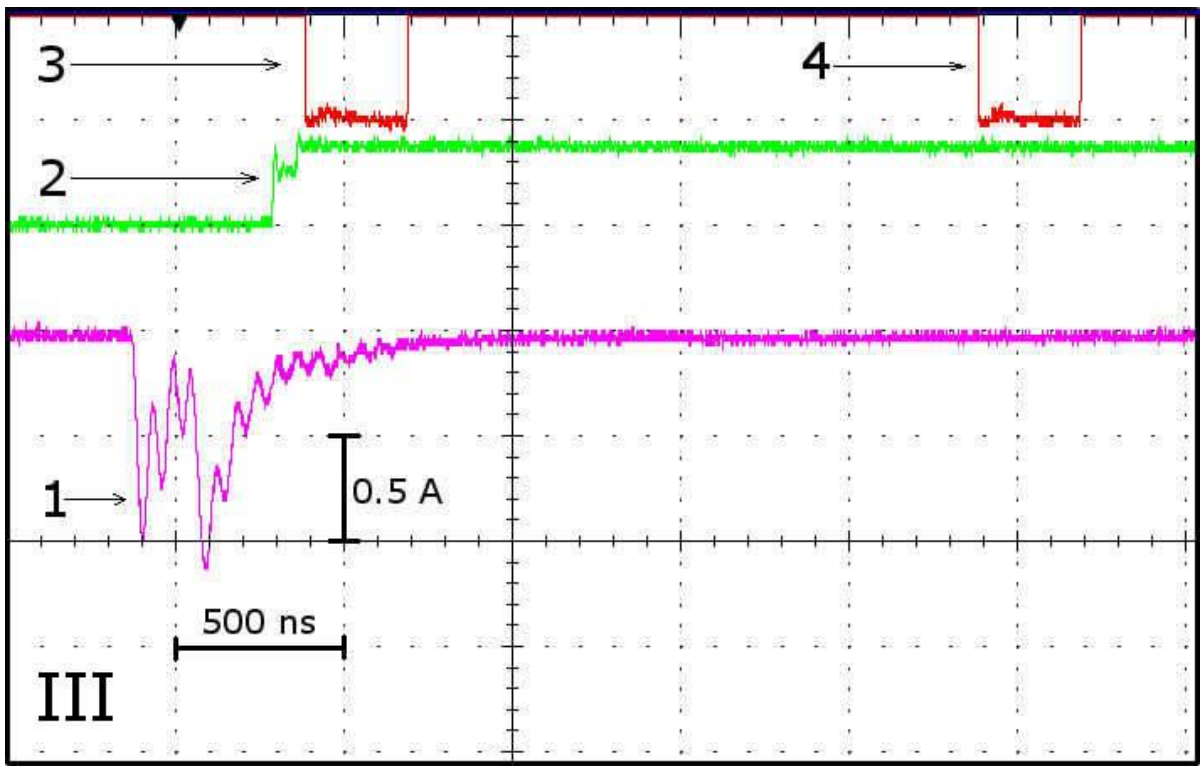

Рисунок 3.14. На I и II камерой 4Picos зафиксирована вспышка положительной стримерной короны, вслед за которой последовало образование лидеров и обратный удар при встрече этих лидеров. Выдержка кадров 300 нс, между кадрами I и II прошло 1,7 мкс, Высота кадров I, II составляет 68-70 см. III – осциллограммы процессов, зафиксированных I и II. 1 – осциллограмма тока, текущего через шунт, и положительный лидер до момента сквозной фазы, которая характеризуется резким возрастанием тока. 2 – сигнал с осциллографа на запуск камеры 4Picos, 3 – осциллограмма, фиксирующая моменты открытия и закрытия затвора камеры 4Picos.

На Рисунке 3.14.I можно выделить стем (стебель) стримерной короны (1) на заземленной сфере, который на панели 2.1.3.14.II через 1.7 мкс практически погас, стримерную корону (2) и более яркие плазменные образования (3), некоторые из них скорее всего также стримеры, а некоторые, возможно, через несколько микросекунд эволюционируют в UPFs, горячие плазменные внутриоблачные образования, впервые описанные в [Kostinskiy et al., 2015a]. Характерно, что на втором кадре (3.14.II), который следует за первым через 1.7 мкс, мы видим свечение (3), не связанное с токоприемником (ток (1) на соответствующей событию на осциллограмме 3.14.III во время выдержки второго кадра (4) равен нулю с точностью до чувствительности осциллографа в этом эксперименте — 0.05 А). Данное свечение сложно связать с восходящей с плоскости стримерной короной и не исключено, что зафиксирован момент формирования горячего плазменного образования (UPFs) в облаке. Судя по осциллограмме тока (1) на Рисунке 3.14.III, стримерная вспышка состояла из двух импульсов тока величиной около 1 А в максимуме и длительностью тока по полувысоте $100 \div 120$ нс. Общая продолжительность вспышки $0.7 \div 0.8$ мкс, что также близко к параметрам свечения плазменного образования (1), зарегистрированного на Рисунке 3.12. На осциллограмме (1) Рисунок 3.14.III не показан ток лидера с последующим обратным ударом, т.к. эти события произошли на несколько микросекунд позже стримерной вспышки.

Таким образом, анализируя результаты эксперимента можно предположить, что вспышка стримерной короны с заземлённой плоскости, достигающая облака, является начальным звеном, в цепи событий, приводящих в результате к появлению объёмных плазменным образованиям, встречным лидерам, сквозной фазе и обратному удару. Как происходит переход стримерной вспышки, аналогичной зафиксированной на Рисунках 3.12-3.14, в развивающиеся горячие плазменные образования (UPFs) подробно описано в главе 2.

### 3.3.5. Верхняя, положительная, часть двунаправленного внутриоблачного лидера

Нисходящий отрицательный лидер при встрече с восходящим с плоскости положительным лидером порождают процесс, похожий на обратный удар (главную стадию) в молнии и длинной искре, но, крайне интересен вопрос, как выглядит скрытая в аэрозольном облаке противоположная, положительно заряженная часть внутриоблачного канала (лидера), которую ранее никто не наблюдал? Тонкую структуру данного канала впервые удалось установить в данной работе с помощью ИК-камеры. Все ИК-изображения, приведенные на Рисунках 3.15-3.17, соответствуют событиям с осциллограммами, которые фиксируют квазиобратный удар при встрече лидеров в аэрозольном облаке, описанные выше (Рисунки 3.7- Рисунок 3.11).

На Рисунке 3.15.I и 3.15.II (увеличенный фрагмент) мы видим центральную (1) и верхнюю (2) часть внутриоблачного канала во время разряда, который закончился квазиобратным ударом (нижняя часть этого канала — нисходящий отрицательный лидер). Верхняя часть данного канала очень похожа на сильно ветвящийся положительный лидер, стартующий с заземленной плоскости, и резко отличается от центральной части (1), на которой при данной контрастности изображения не фиксируются ветвления.

На Рисунке 3.16.I и 3.16.II (увеличенный фрагмент) мы также видим центральную (1) и верхнюю многократно ветвящуюся часть положительного канала (2). Также можно отметить не поддающиеся однозначной интерпретации плазменные образования (3), которые похожи на очень интенсивное ветвление канала положительного лидера или объёмные плазменные образования – UPFs. На Рисунке 3.15 и Рисунке 3.16 ветвящееся часть разрядного канала, похожая на положительный лидер, заметно меньше, чем центральная, длинная часть канала.

На рис. Рисунке 3.17.I (изображение получено путём вычитания предыдущего кадра) и Рисунке 3.17.II (необработанный нижний фрагмент Рисунка 3.17.I) наоборот, практически весь кадр занимает длинный положительный лидер (1) с различимым нижним ветвлением посередине (2), точкой верхнего ветвления (3) двух длинных ветвей (4) и плазменным образованием (5), которое скорее всего относится к UPFs, а не ветвлениям положительного лидера.

На Рисунке 3.18 хорошо видна различная природа разных частей канала. Мало извилистая отрицательная часть канала (1), которая продолжается примерно до точки (4), где начинает быть различима извилистая структура восходящего предположительно положительного лидера (2) с более сильными ветвлениями (3) и гораздо менее извилистая структура нисходящего отрицательного лидера.

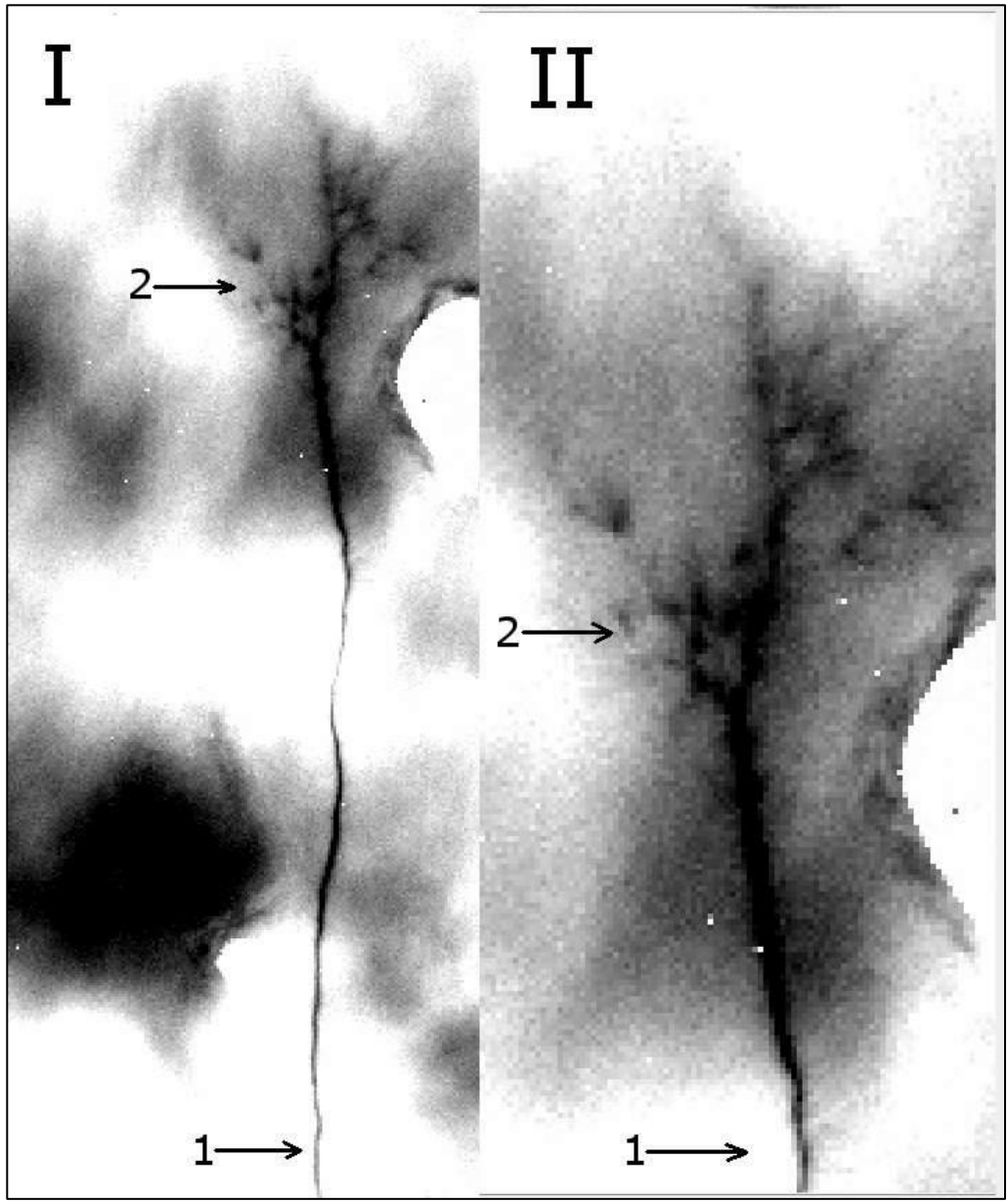

Рисунок 3.15 (событие 266-2014-09-01). Верхняя, положительная часть (положительный лидер) двунаправленного внутриоблачного разряда (лидера). Центральная – (1) и верхняя (положительная) – (2) части внутриоблачного канала, который закончился обратным ударом (нижняя, не видная на изображении часть этого канала — нисходящий отрицательный лидер). Верхняя часть (2) данного канала очень похожа на сильно ветвящийся положительный лидер и резко отличается от центральной части (1), на которой при данной контрастности изображения не фиксируются ветвления.  II – увеличенный фрагмент изображения – I.

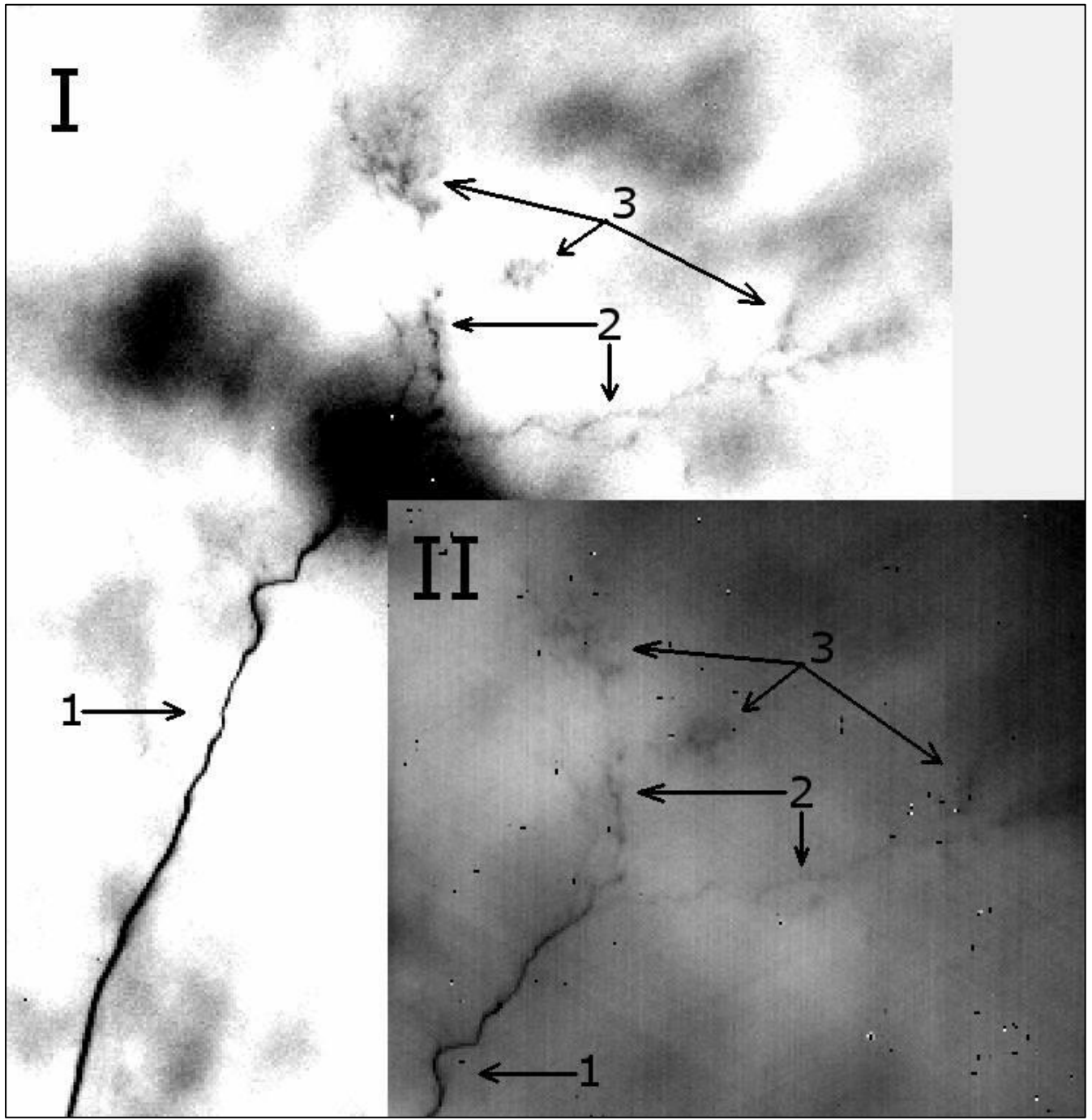

Рисунок 3.16 (событие 286-2014-09-01). Верхняя часть двунаправленного внутриоблачного разряда (лидера). Центральная – (1) и верхняя (положительная) – (2) части внутриоблачного канала, который закончился обратным ударом (нижняя, не видная на изображении часть этого канала — нисходящий отрицательный лидер). 3 – трудно интерпретируемые плазменные образования. Верхняя часть 2 данного канала очень похожа на сильно ветвящийся положительный лидер и резко отличается от центральной части (1), на которой при данной контрастности изображения не фиксируются ветвления. II – увеличенный фрагмент изображения – I.

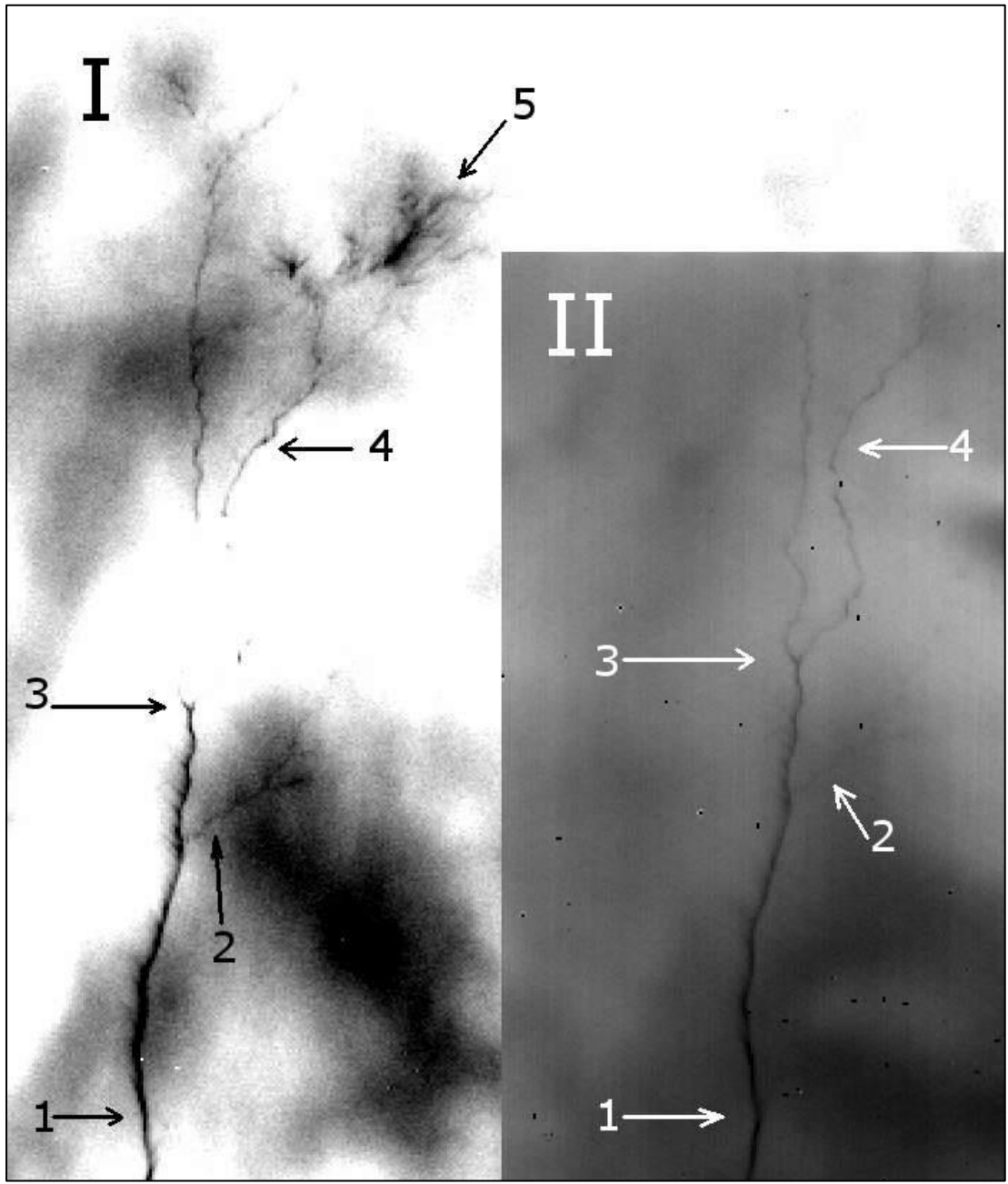

Рисунок 3.17 (событие 271_2014-09-01). Верхняя часть двунаправленного внутриоблачного разряда (лидера). I — изображение получено путём вычитания предыдущего кадра и II — необработанный нижний фрагмент I. Практически весь кадр занимает длинный положительный лидер (1) с различимой веткой посередине (2), точкой ветвления (3) двух длинных ветвей (4) и плазменным образованием (5), которое по внешнему виду можно отнести к UPFs, а не сложным ветвлениям положительного лидера.

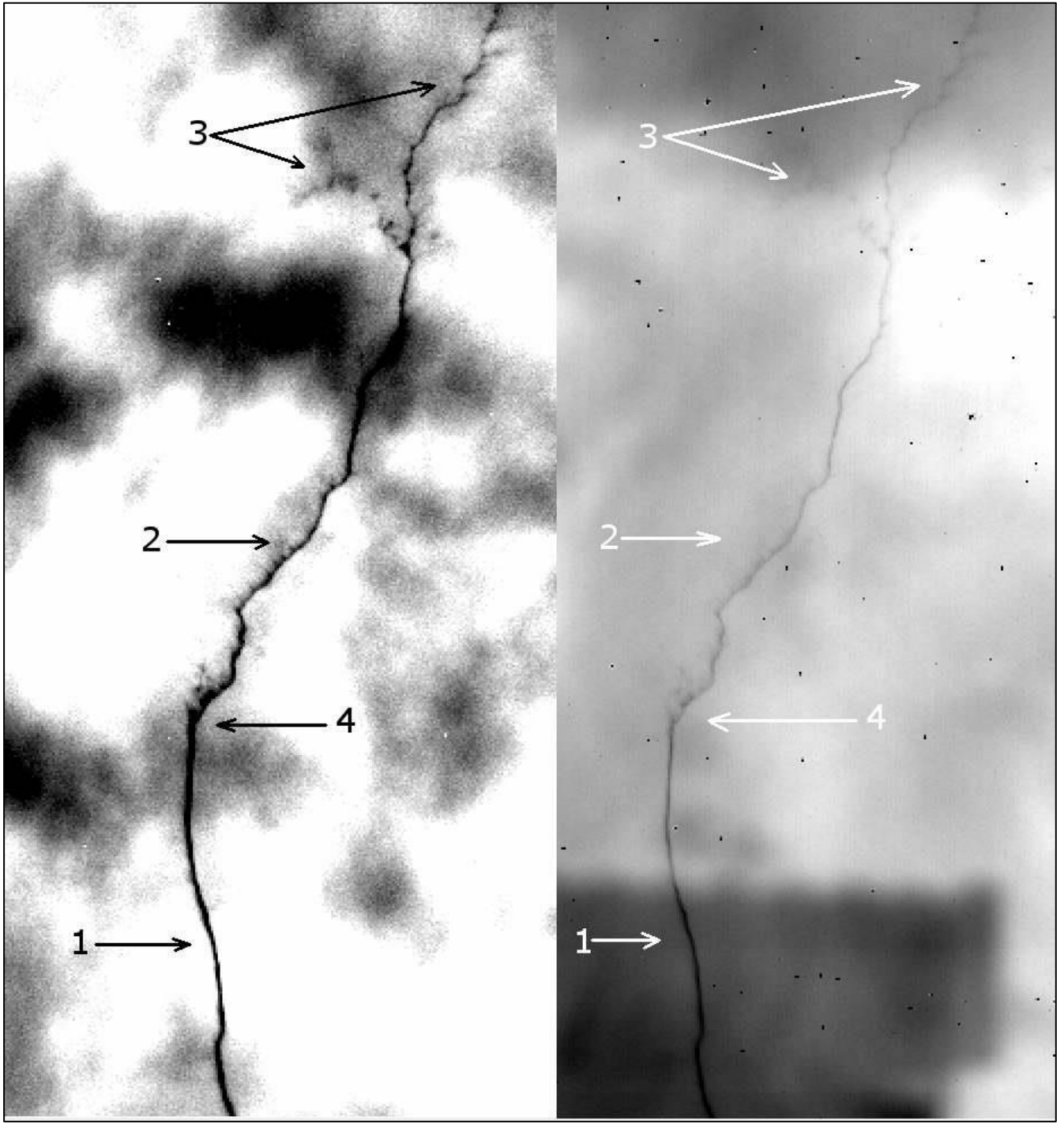

Рисунок 3.18 (событие 349_2014-09-01). Средняя и верхняя часть двунаправленного внутриоблачного разряда. Мало извилистая часть канала (1), которая продолжается до точки (4), где отчетливо различима извилистая структура восходящего предположительно положительного лидера (2) с сильными ветвлениями (3).

**3.4. Сравнение с результатами, полученными в более ранних работах, где также исследовался контакт двух лидеров, инициированных в электрическом поле облака искусственно заряженного аэрозоля**

Перед тем, как анализировать более ранние публикации, касающиеся облака искусственно заряженного аэрозоля ([Верещагин и др., 2003] [Temnikov et al. 2007], [Temnikov 2012a], [Temnikov et al. 2012b]) необходимо сделать два важных замечания.

Первое, довольно длительное время, например, в работах [Верещагин и др., 2003], [Temnikov 2012b] предполагалось, что осциллограммы тока без квазиобратного удара являются осциллограммами тока стримерной короны, которая длится от 1.14 до 15 мкс [Верещагин и др., 2003, Таблица 1]. Пример таких осциллограмм приведен на Рисунках 3.18, 3.19 (судя по форме и всем приведенным характеристикам, скорее всего это одна и та же осциллограмма, где ток продолжается 8 мкс). Это предположение в корне расходится со всем корпусом экспериментальных данных, полученным, как при исследовании стримерных вспышек длинной искры ([Les Renardières Group, 1972], ([Les Renardières Group, 1974], ([Les Renardières Group, 1977]), так и исследовании стримерных вспышек в электрическом поле искусственно заряженного аэрозоля (например, осциллограммы на Рисунках 1.4, 2.4 (5), 2.5(6), 2.11 (6), 3.14.III). Во всех случаях фронт тока стримерной вспышки имеет продолжительность 25-35 нс, а сама вспышка имеет продолжительность 150-200 нс, а в случае двойной вспышки не более 300-400 нс. Поэтому на осциллограммах Рисунков 3.18-3.19 (из [Верещагин и др., 2003], [Temnikov 2012b]) только первый и может быть второй пик тока соответствуют стримерной вспышке, а ток всей остальной осциллограммы обеспечивает восходящий положительный лидер, который может продолжаться в этих экспериментах от нескольких микросекунд до десятков микросекунд без появления квазиобратного удара. Кроме того, некорректная оценка времени продолжительности стримерных вспышек в [Верещагин и др., 2003], [Temnikov 2012b] говорит о том, что методика покадровой высокоскоростной съемки с помощью камеры К011 (например, [Temnikov 2012b]) имеет невысокую чувствительность, раз на Рисунке 3.19 (слева, [Figure 6, Temnikov 2012b]) авторам не

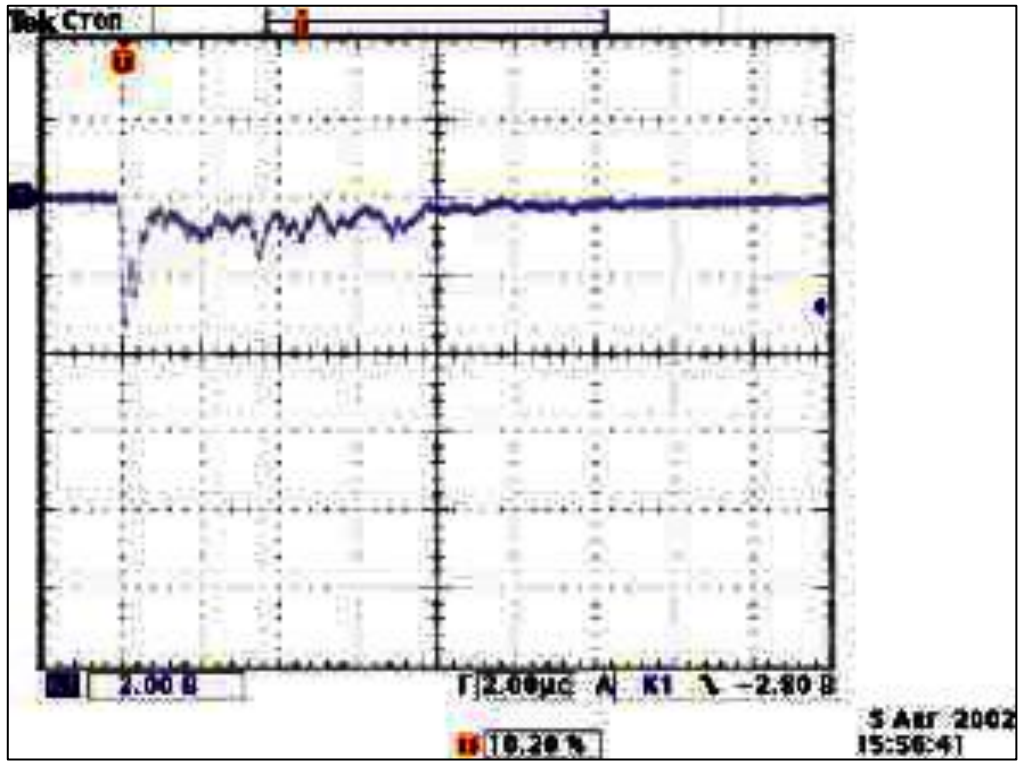

Рисунок 3.18. (адаптировано из рис.5 из [Верещагин и др., 2003] с подписью: «Осциллограмма тока стримерной короны, протекающего через стебель, *без перехода в лидерный разряд»*.)

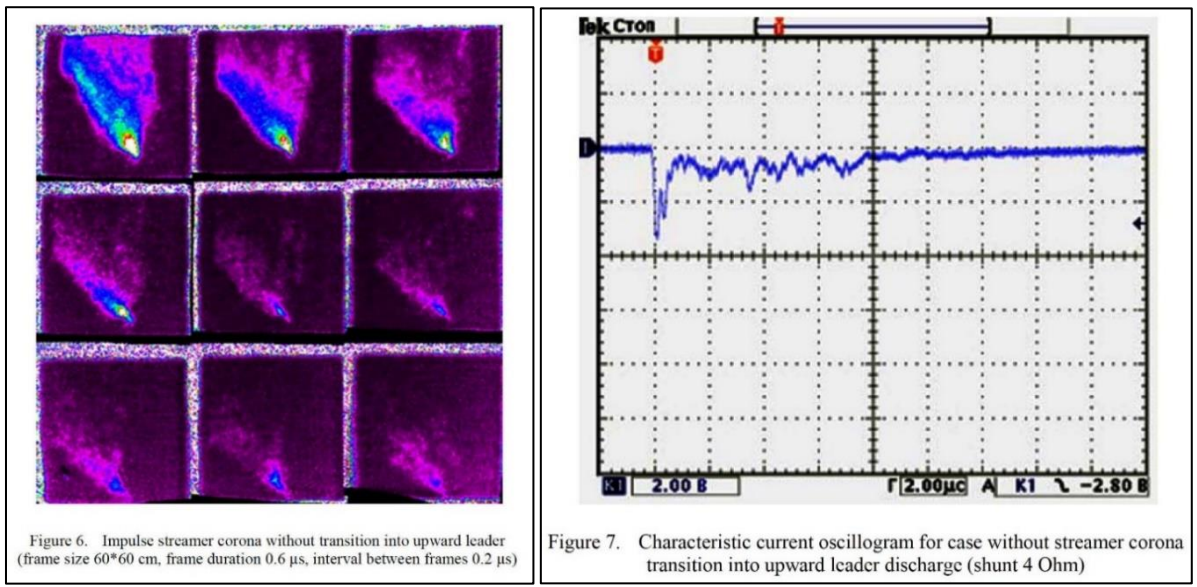

Figure 6. Impulse streamer corona without transition into upward leader (frame size 60*60 cm, frame duration 0.6 μs, interval between frames 0.2 μs)

Figure 7. Characteristic current oscillogram for case without streamer corona transition into upward leader discharge (shunt 4 Ohm)

Рисунок 3.19 (адаптировано из Figure 6, 7 из [Temnikov et al., 2012b])

удается надежно зафиксировать восходящий положительный лидер и они считают, что весь промежуток времени продолжается стримерная вспышка (или серия стримерных вспышек с чрезвычайно короткими интервалами между вспышками). Недооценка низкой оптической чувствительности и низкого пространственного разрешения камеры К011 будет и в других случаях вести к некоторым некорректным интерпретациям оптических изображений. Несмотря на эти ограничения, авторами были получены интересные и важные результаты (см. ниже).

Второе, авторы этих работ (см., например, [Верещагин и др., 2003], [Temnikov 2012b]) считают, что электрическое поле на границе облака заряженного аэрозоля может превышать 20 кВ/см (что следует из более ранних численных расчетов [Vereshchagin et al., 1995], [Орлов и Темников,1996]]). Такие высокие электрические поля вряд ли реализуемы на границе отрицательного облака, так как, в случае столь высоких полей в эксперименте можно было бы наблюдать отрицательные стримерные вспышки, которые распространяются от отрицательного облака к заземленной плоскости, что наблюдается в положительно заряженном облаке, где для поддержания движения положительных стримеров требуется поле 4.5-5 кВ/см. Так как во всем корпусе экспериментов с отрицательно заряженными аэрозольными облаками, которые проводились почти 40 лет различными научными группами, никогда не наблюдались отрицательные стримерные вспышки, идущие от отрицательного облака к заземленной плоскости, то это означает, что электрическое поле в разрядном объеме никогда не превышает 10-12 кВ/см (поле поддержания отрицательных стримеров) на расстояниях более нескольких сантиметров (исключить локальные высокие электрические поля на размерах менее 1-2 см мы не можем).

На Рисунке 3.20 [адаптировано из Fig. 9, Temnikov et al. (2007)], зафиксирована сквозная фаза и квазиобратный удар при контакте восходящего с заземленной плоскости положительного лидера с нисходящим из облака заряженного аэрозоля отрицательным лидером (выдержка кадров 0.2 мкс, межкадровая пауза 2 мкс, высота кадра 70 см). Поведение лидеров во время сквозной фазы Рисунок 3.20 [адаптировано из Fig. 9, Temnikov et al. (2007)] аналогично нашим измерениям, например, зафиксированным на Рисунке 3.6. Осциллограмма, сопровождающая Рисунок 3.20 [адаптировано из Fig. 10, Temnikov et al. (2007)] имеет два пика, подобно нашей осциллограмме на Рисунке 3.7,

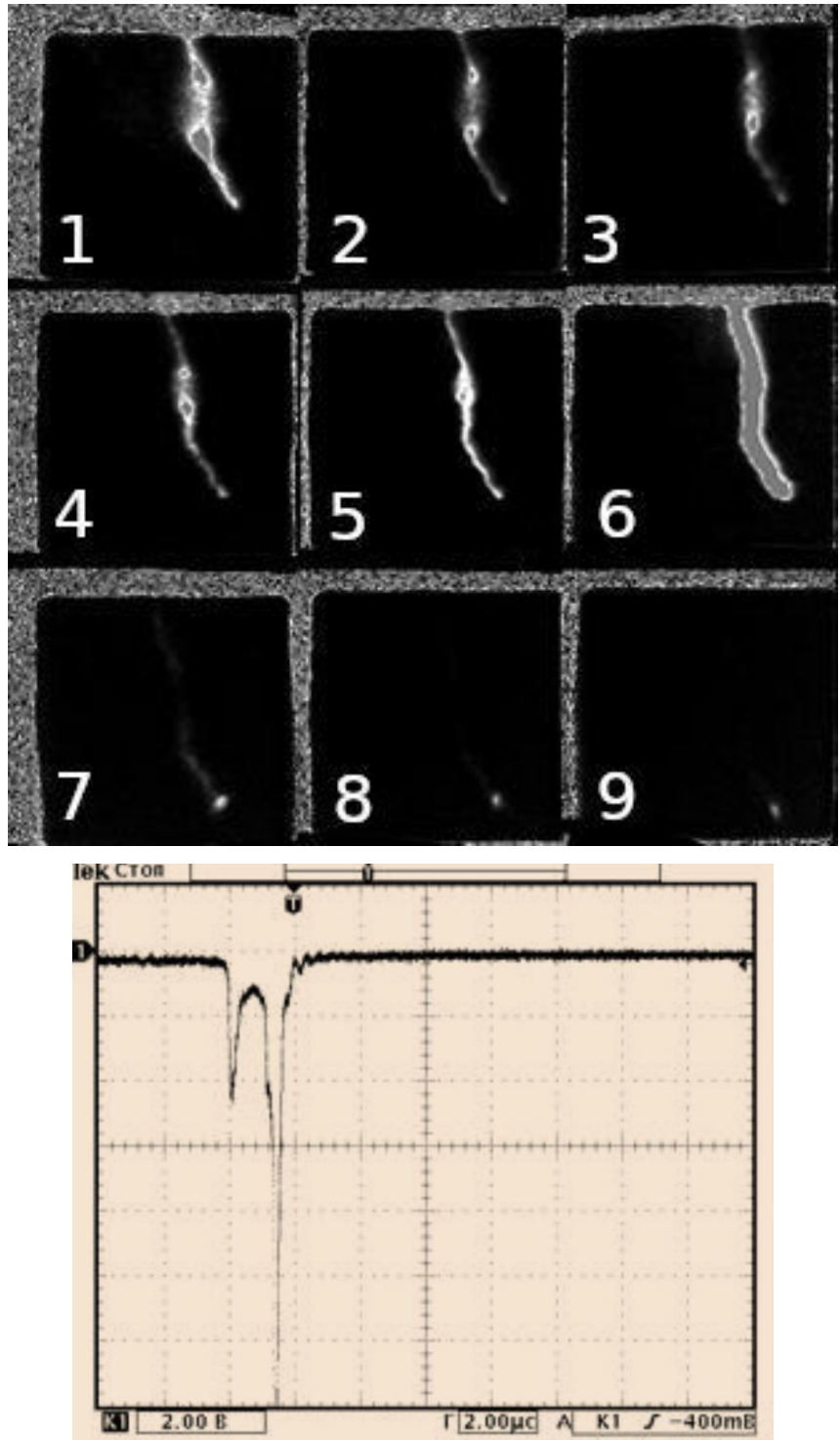

Рисунок 3.20 (адаптировано из Fig. 9,10 [Temnikov et al., 2007], подписи под рисунками Образование финальной стадии разряда третьего типа, когда восходящий лидер взаимодействует с нисходящим лидером из облака заряженного аэрозоля (выдержка кадров 0.2 мкс, межкадровая пауза 2 мкс). Высота кадра 70. Типичная осциллограмма тока финальной стадии (вероятно, не этого конкретного события, изображенного на покадровой съемке) третьего типа разряда (сопротивление измерительного шунта R = 1.39 Ohm) которая фиксирует ток события, зафиксированного на Рисунке 3.6). [Temnikov et al. (2007)] называют осциллограмму на их Fig. 10 «типичной осциллограммой тока финальной стадии», а не осциллограммой тока финальной стадии данного события, но мы предполагаем, что [Temnikov et al., 2007] считают, что и для этого конкретного события осциллограмма имеет схожую форму.

В работе [Temnikov, 2012а], благодаря скоростной покадровой съемке и осциллограмме тока, был, по-видимому, впервые зафиксирован контакт трех каналов. На Рисунке 3.21 (адаптированные Fig.6-Fig.7 из [Temnikov, 2012а]) хорошо видно, как на кадре (3) к восходящему положительному лидеру присоединяется нисходящий отрицательный лидер, а позже на кадрах (8,9) можно видеть резкое возрастание тока, которое, по мнению [Temnikov, 2012а] соответствует резким пикам тока на осциллограмме, которые свидетельствуют о двух квазиобратных ударах.

[Temnikov, 2012а, Таблица VI] также установили влияние второго заряженного аэрозольного облака (дополнительно к работающему первому облаку) на параметры разряда, что является важным продолжением исследований. По их данным амплитуда тока разрядов увеличилась в 1.3-1.4 раза, число контактов лидеров возросло в 1.7-2.8 раза, переносимый заряд вырос в 1.25-1.4 раза, а продолжительность финальной стадии практически не изменилась.

TABLE VI.    RELATION OF CHARACTERISTICS (AVERAGE) OF FINAL STAGE OF DISCHARGE BETWEEN CHARGED AEROSOL CLOUD AND GROUND WITH SECOND UPPER CHARGED AEROSOL CELL AND WITHOUT IT $C_{two}/C_{one}$

| Parameters of final stage, $C_i$ | $C_{two}/C_{one}$ |
|---|---|
| Current amplitude | 1.3-1.4 |
| Maximal current steepness | 1.7-2.8 |
| Charge | 1.25-1.4 |
| Duration | 0.95-1.1 |

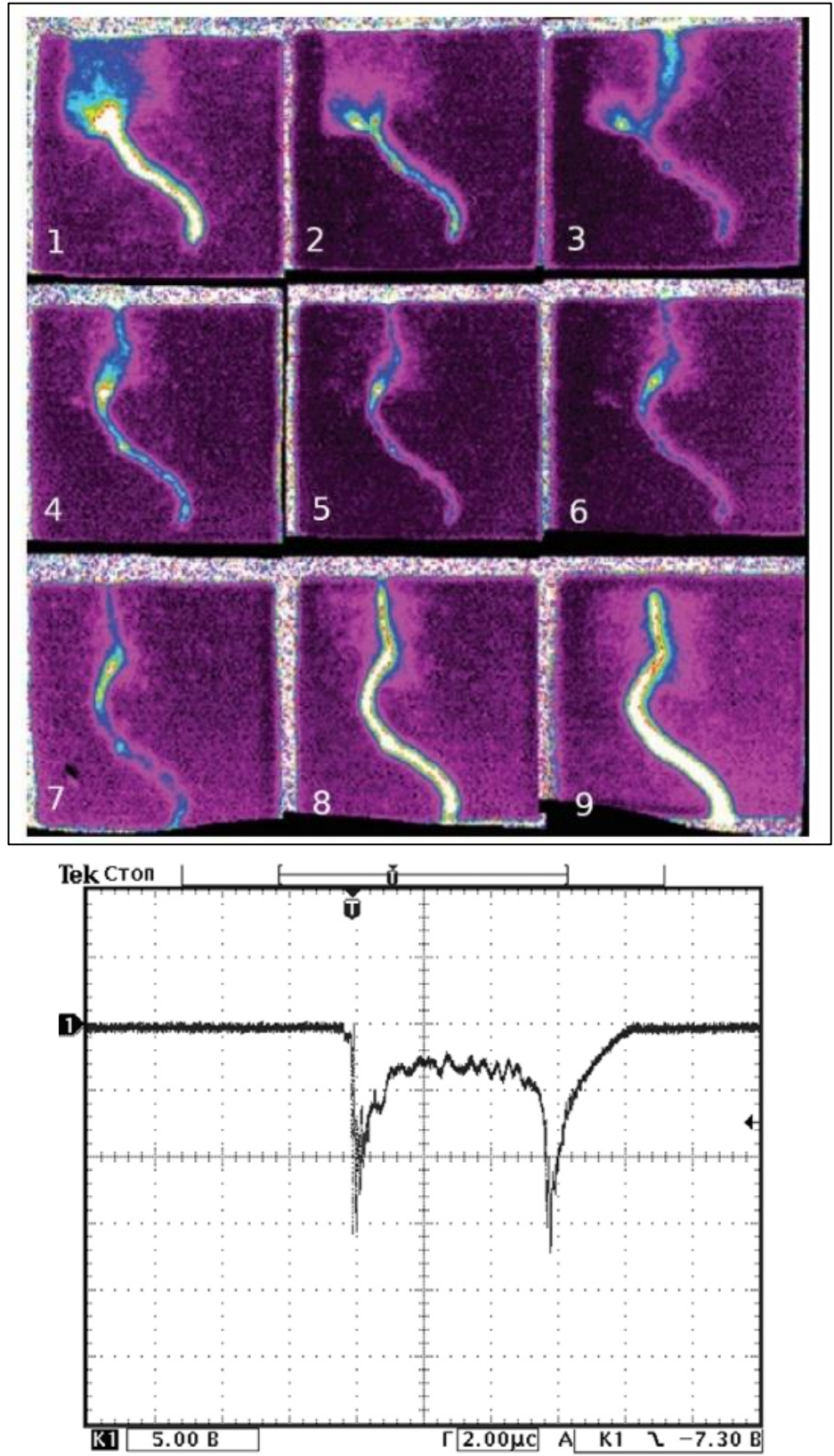

Рисунок 3.21 (адаптировано Fig.6-Fig.7 из [Temnikov, 2012a]). Подпись под Fig.6: «Подключение незаряженного участка искусственного облака заряженного аэрозоля и формирование повторного разряда (размер кадра 70х70 см, длительность кадров 0.6 мкс, паузы между кадрами 0.2 мкс)». Подпись под Fig.7: «Осциллограмма тока финальной стадии разряда из искусственного облака заряженного аэрозоля отрицательной полярности при формировании повторного разряда (сопротивление шунта 0.95 Ом)

**3.5. Обсуждение результатов экспериментов главы 3**

Как отмечалось в разделе 3.1, процесс взаимодействия молнии с заземленными объектами, особенно его сквозная фаза, является одним из наименее изученных процессов молнии. Существует ограниченное количество опубликованных оптических изображений общей стримерной зоны восходящего и нисходящего лидера для молний или лабораторных искр, и на сегодняшний день в литературе нет двухкадровых изображений сквозной фазы. В связи с этим два снимка, сделанные во время сквозной фазы и показанные на рисунке 3.6, дают новую информацию о поведении лидеров, развивающихся в общей стримерной зоне. Сквозная фаза началась, когда стримерные зоны с относительно низкой проводимостью восходящих и нисходящих лидеров соприкоснулись, образуя общую стримерную зону. Мы предполагаем, что последующее расширение двух плазменных каналов с относительно высокой проводимостью по направлению друг к другу внутри общей стримерной зоны может привести либо к одному каналу (см. Рисунки 3.4 и 3.8), либо к нескольким каналам (что приводит к разделению канала, как показано на рисунке 4.28 из [Rakov and Uman, 2003] для естественной молнии; на рисунке 15 из [Howard et al., 2010] для триггерных молний; и на рисунке 3b из [Shcherbakov et al., 2007] для длинных лабораторных искр). В случае нескольких контактов (соединений) каналов возможны два сценария: (1) несколько соединений формируются последовательно (вероятно, потому что сопротивление одного соединения слишком велико) и (2) множественные соединения являются результатом ветвлений, как положительных, так и отрицательных каналов. Этот последний сценарий кажется более вероятным на Рисунке 3.6, хотя изображение результирующего канала обратного удара, отсутствует. Соответствующие токи непосредственно перед началом обратного удара, составляют около 5,7 А и 4 А. Непонятно, почему ток в сквозной фазе значительно уменьшается после своего максимума. Возможно, контакт стримеров не приводит к прямому преобразованию общей стримерной зоны в горячий канал, что способствовало бы отводу облачного заряда за один импульс. На эту конкуренцию, возможно, влияет объемный заряд стримеров, который уменьшает электрическое поле вблизи канала горячего лидера, из которого берут начало стримеры, и тем самым ограничивает ток. Наблюдаемое явление, возможно, предполагает, что последовательность пробоев может означать перекрытие общей стримерной зоны. Действительно, если после первоначального контакта полное сопротивление области контакта и результирующее падение напряжения остаются достаточно высокими, дополнительный пробой в этой области может создать дополнительный контакт, параллельный первоначальному. В случае молнии есть некоторые свидетельства того, что сквозная фаза не является единым и непрерывным движением (final jamp), а является серией «прыжков», которые могут быть результатом последовательных ионизационно-перегревных неустойчивостей. Очевидно, это связано с множественными контактами, о чем свидетельствуют измерения dE/dt на нескольких измерительных пунктах для природных и триггерных молний, о которых сообщили [Howard et al., 2010]. Они исследовали так называемый импульс быстрого перехода (fast-transition (FT)), а также множественные импульсы, возникающие во время предшествующего медленного фронта (slow front (SF)), длительность которых для первых ударов обычно составляет несколько микросекунд и начало которых означает начало сквозной фазы молнии. Предполагалось, что импульс FT и импульсы SF имеют одну и ту же природу и связаны с множеством контактов, последовательно устанавливаемых между нисходящими отрицательными и восходящими положительными лидерами во время сквозной фазы. [Howard et al., 2010] представили на их рисунке №15 оптическое изображение (видеокадр), показывающее разделенный канал инициированной молнии с двумя первичными контактами, один из которых показывает меньшее разделение (два подсоединения), так что общее количество отображаемых контактов составляло три. Они связали эти три контакта с тремя основными импульсами SF/FT, замеченными в их записях dE/dt (см. их рисунок №14), что предполагает, что эти соединения были установлены в разное время (последовательно) в течение 2,1 мкс длительности сквозной фазы. Наше событие, показанное на Рисунке 3.6, может быть такого же типа, как описано [Howard et al., 2010], хотя у нас нет изображения обратного удара, чтобы считать так с абсолютной определенностью. Интересно, что событие триггерной молнии, изученное [Howard et al., 2010] показало резкое увеличение тока основного канала на 20 кА, связанное с импульсом SF. Еще один резкий скачок тока произошел позже, во время импульса FT. Общий пик тока составил 45 кА (необычно высокий). Недавно [Hill et al., 2016] связали каждый из изученных ими импульсов SF/FT в ударах триггерной молнии с быстрым, в килоамперном масштабе, увеличением тока основного канала с последующим уменьшением скорости нарастания тока (хотя и не уменьшением тока, как и в наших событиях, показанных на Рисунках 3.5-3.7). Средняя арифметическая длительность сквозной фазы в исследовании [Hill et al., 2016] составляла 1,77 мкс, а средний ток непосредственно перед началом сквозной фазы составлял 16,7 А. Начало сквозной фазы было связано с увеличением тока в основании канала до типичного значения во многие сотни ампер.

Во время развития молнии, как только плазменные каналы нисходящего и восходящего лидеров входят в контакт друг с другом внутри общей стримерной зоны, начинается стадия обратного удара. Она начинается с двух волн обратного тока, инициированных точкой соединения двух лидеров. Одна волна тока движется вверх, к облаку, а другая - вниз, к земле [Wang et al., 1999, 2013, 2014a]; [Jerauld et al., 2007]. Ожидается, что нисходящая волна будет короткой и после отражения от земли, вероятно, догонит фронт восходящей волны (например, [Rakov, 2013]). В конечном итоге образуется единственная восходящая волна обратного удара. Также могут быть волны тока, связанные с множественными контактами во время сквозной фазы, описанной выше. [Hill et al., 2016] сделали вывод из своих коррелированных записей dE/dt и dI/dt, что каждый импульс SF/FT был связан с токовой волной, распространяющейся от точки контакта (области контакта каналов) до земли. По их оценкам, для импульсов SF скорости этих волн составляли от $4,3 \times 10^7$ до $1,6 \times 10^8$ м/с, а для импульсов FT — от $1,2 \times 10^8$ до $1,6 \times 10^8$ м/с.

Наши наблюдения за контактами "головка-головка" показывают, что в инфракрасном диапазоне участок плазменного канала, который остается после существования общей стримерной зоны, значительно ярче — примерно в 5 раз, чем участки канала выше и ниже него (см. Рисунок 3.8). [Kostinskiy et al., 2015a] пришли к выводу, что яркость инфракрасных изображений, полученных с помощью камеры FLIR, можно рассматривать как индикатор конечной температуры газа в отображаемом плазменном канале. Это означает, что канал в окрестности области соединения горячее и/или толще, чем участки канала выше или ниже области соединения. Похоже, что наибольший вклад энергии в канал молнии у земли связан со сквозной фазой. Интересно, что о повышенной яркости вблизи области контакта никогда не сообщалось на основе интегрированных во времени изображений молний в видимом диапазоне, хотя [Wang et al., 2014b] обнаружили, что максимумы светового излучения наиболее велики вблизи области контакта.

[Lu et al., 2013], используя высокоскоростную фотографию, сообщил о процессе контакта молнии, включающем контакт нисходящего отрицательного лидера с боковой поверхностью противоположно заряженного восходящего соединяющего лидера (UCL). UCL был около 400 м в длину и развивался от вершины здания сопоставимой высоты. Он был таким же ярким или ярче, чем нисходящий отрицательный лидер. Они интерпретировали это явление (соединение ниже верхней части восходящего лидера) как то, что UCL управляется (поддерживается) общим электрическим полем нескольких ветвей нисходящего лидера, а не какой-то одной ветвью, которая в конечном итоге устанавливает соединение. Связи между нисходящим отрицательным лидером и восходящим положительным лидером, показанные на наших Рисунках 3.9 и 3.11, очень похожи на те, которые наблюдались при нисходящей природной молнии [Lu et al., 2013]. В следующем исследовании [Lu et al., 2013] сообщили о двух основных типах контакта между нисходящими отрицательными и восходящими положительными лидерами естественной молнии: соединение кончик к кончику (головка к головке) и соединение отрицательного лидера с боковой поверхностью восходящего положительного лидера. Из 24 событий, наблюдаемых с помощью высокоскоростных видеокамер, 10 (42%) продемонстрировали первый тип подключения, 12 (50%) - второй тип и 2 (8%) — комбинацию первого и второго типов. Они никогда не наблюдали соединения восходящего положительного лидера с боковой поверхностью нисходящего отрицательного лидера. Чтобы сравнить возникновение различных типов лидерных соединений в наших искрах метрового масштаба с результатами [Lu et al., 2016] для нисходящей молнии, мы изучили дополнительный набор данных, полученных 13 мая 2016 года. Всего 268 разрядов показали контакт нисходящего отрицательного и восходящего положительного лидеров. Из этих 268 в 78,7% случаев лидеры контактировали головка к головке, а в 15,7% случаев отрицательный лидер контактировал с положительным лидером ниже его вершины. Кроме того, в 5,6% случаев лидеры развивались параллельно друг другу и не могли сформировать отчетливо контактирующий общий плазменный канал. Подобно наблюдениям [Lu et al., 2016], мы никогда не видели контакта восходящего положительного лидера с боковой поверхностью нисходящего отрицательного лидера, скорее всего потому, что котакт гораздо легче формируется короной, идущей от положительного лидера, так как ей требуется в два раза меньшее электрическое поле для поддержания стримеров.

## 3.6. Выводы главы 3

1. Представлены наблюдения с помощью высокоскоростных видеокамер, а также синхронизированные записи тока при контакте (сквозной фазе) между положительными и отрицательными лидерами в электрических разрядах метрового масштаба, генерируемых облаками отрицательно заряженного водного аэрозоля, и обсуждаются их возможные последствия для понимания процесса сквозной фазы молнии.

2. Впервые представлены два снимка сквозной фазы контакта лидеров, показывающие значительное разветвление лидеров внутри общей стримерной зоны.

3. В случае соединения лидеров типа головка-головка яркость инфракрасного излучения в области контакта (вероятно, пропорциональная температуре газа и, следовательно, подводу энергии) была примерно в 5 раз выше, чем для участков канала ниже или выше этой области.

4. В 16% случаев нисходящий отрицательный лидер соединялся с восходящим положительным лидером ниже его верхней части (контактировал с боковой поверхностью положительного лидера), причем соединение осуществлялось через сегмент канала, который казался перпендикулярным одному или обеим каналам лидера.

# ГЛАВА 4. Плазменные образования, включая двунаправленные лидеры, инициированные в электрическом поле *положительно* заряженного водного аэрозоля, обнаруженные внутри облака, благодаря ИК-камерам в диапазоне 3-6 мкм

Данная глава посвящена плазменным явлениям внутри аэрозольного облака *положительно* заряженных водяных капель. Предыдущие три главы касались отрицательно заряженного облака.

В данной главе в начале будет сделан обзор более ранних исследований положительно заряженного облака и различных форм плазмы, которые возникают в электрическом поле этого облака (стримерные вспышки и лидеры, идущие от положительно заряженного облака к заземленной плоскости).

Далее будут представлены, полученные в рамках исследований [Kostinskiy et al., 2015b], подробные инфракрасные (3-6 мкм) изображения плазменных образований (включая двунаправленные (bidirectional) лидеры, предсказанные для молнии Каземиром [Kasemir, 1960]), инициированных облаком положительно заряженного водного аэрозоля (средний радиус капель около 0,5 мкм). Во многих случаях зафиксированный результирующий двунаправленный лидер состоял из идущей вниз положительной части и восходящей отрицательной части, причем эти две части (обе разветвленные, хотя и по-разному) были соединены одноканальной средней частью. Нисходящая часть включала извилистый положительный лидерный канал (аналогичный его восходящему извилистому аналогу, наблюдаемому при отрицательной полярности облака), который часто сопровождался гораздо менее извилистыми, но часто столь же яркими, идущими вниз плазменными образованиями неизвестной природы. Также наблюдалась положительная стримерная зона, идущая от облака к заземленной плоскости.

В этих экспериментах, как и в экспериментах с отрицательным облаком (глава 1) были впервые обнаружены в положительном облаке сильно излучающие в ИК-диапазоне плазменные образования различной морфологии, которые ранее были названы необычными плазменными образованиями (unusual plasma formations — UPFs) [Kostinskiy et al., 2015a]. Некоторые UPFs, обнаруженные в электрическом поле положительно заряженного облака не наблюдались внутри отрицательно заряженного облака.

Либо положительный лидерный канал, либо UPFs могут контактировать с заземленной плоскостью. Восходящая часть связана с большой разветвленной сетью каналов, часто выходящих веером из верхней части обычно более яркого основного канала и, по-видимому, пронизывающих всю верхнюю часть облака. Некоторые из этих слабых каналов могут быть длинными и яркими отрицательными стримерами, в то время как другие могут быть похожи на UPFs. ИК-светимость вдоль самой яркой части двунаправленного лидерного канала часто неоднородна. Некоторые изменения яркости канала локализованы и предполагают участие процессов взаимодействия нескольких каналов (типа, известного взаимодействия спейс-лидеров) в нескольких местах вдоль канала.

## 4.1. Введение в главу 4

Разряды, инициированные положительно заряженным аэрозольным облаком, изучались на экспериментальных установках, аналогичных изучению разрядов, инициированных отрицательным облаком. Важным отличием было то, что разряды, инициированные положительным облаком «не чувствовали» заземленную сферу (не стартовали с нее) и в данном случае не удавалось измерить динамику и величину тока разрядов.

Длинные искры, инициированные отрицательно заряженным облаком аэрозоля, гораздо лучше изучены, чем разряды, инициированные положительно заряженным облаком. Это связано с тем, что в электрическом поле отрицательно заряженного облака легче инициируются восходящие с токоприёмника (заземленной сферы или стержня) на плоскости положительно заряженные стримерные и лидерные разряды, т.к. пороги их возбуждения значительно меньше, чем у отрицательных [Базелян и Райзер, 1997]. Возбуждение положительных разрядов, восходящих с токоприёмника, позволяет напрямую измерять текущий через канал разряда ток, который является важнейшей информативной характеристикой плазменных процессов. Также по уровню тока на осциллографе удобно синхронизировать запуск других измерительных приборов, что позволяет одновременно сопоставить различные физические процессы в разряде.

Положительно заряженное аэрозольное облако, в отличие от отрицательного, практически «не чувствует» возвышенность на плоскости и возникающие разряды замыкаются по всей поверхности заземлённой плоскости. Это происходит из-за высокого порога распространения отрицательных стримеров, вспышки которых становятся источником зарождения лидеров [Les Renardieres Group, 1981]. Вследствие этого, прямое измерение тока в таких разрядах затруднительно.

В первых же исследованиях было установлено, что разряды, инициированные положительным облаком, сильно отличаются от разрядов, инициированных отрицательным облаком [Верещагин и др. 1988]: «Формы разряда при положительном объемном заряде более разнообразны. Существует канальная стадия в виде кратковременной искры, при которой разряд развивается вдоль оси струи (Рисунок 4.1(3)). Наряду с этим наблюдается форма, аналогии которой в разрядах между электродами нет. Эта форма представляет собой изогнутый столб, относительно слабо светящийся фиолетовым светом (Рисунок 4.2 (2 нижний разряд)). Он существует в течение нескольких секунд. Диаметр столба меньше, а яркость освещения больше в области объемного заряда облака, к экрану (заземленной плоскости – А.К.) столб расширяется, а интенсивность свечения снижается. Возникает такой объемный заряд из области на границе струи примерно на высоте 1.5 м от плоскости и заканчивается на расстоянии 0.5 м от оси струи. Со временем объемный разряд исчезает или чаще переходит в канальную стадию (Рисунок 4.1(3,4)). Более тщательный анализ объемного разряда показывает, что свечение столба не является вполне однородным: можно выделить более ярко светящиеся нити и ветви. В некоторых случаях разряд имеет незавершенный вид, при этом свечение соответствует форме стримерной короны (Рисунок 4.1(4)). <…> Разряд при положительной полярности в большей мере зависит от влажности окружающего воздуха, чем при отрицательной полярности. При низкой влажности разряды распространяются исключительно вдоль оси струи. Длина электрических разрядов и их частота существенно зависят от напряжения на игле. При повышении напряжения выше напряжения, соответствующего возникновению разрядов, частота разрядов увеличивается, а их длина уменьшается».

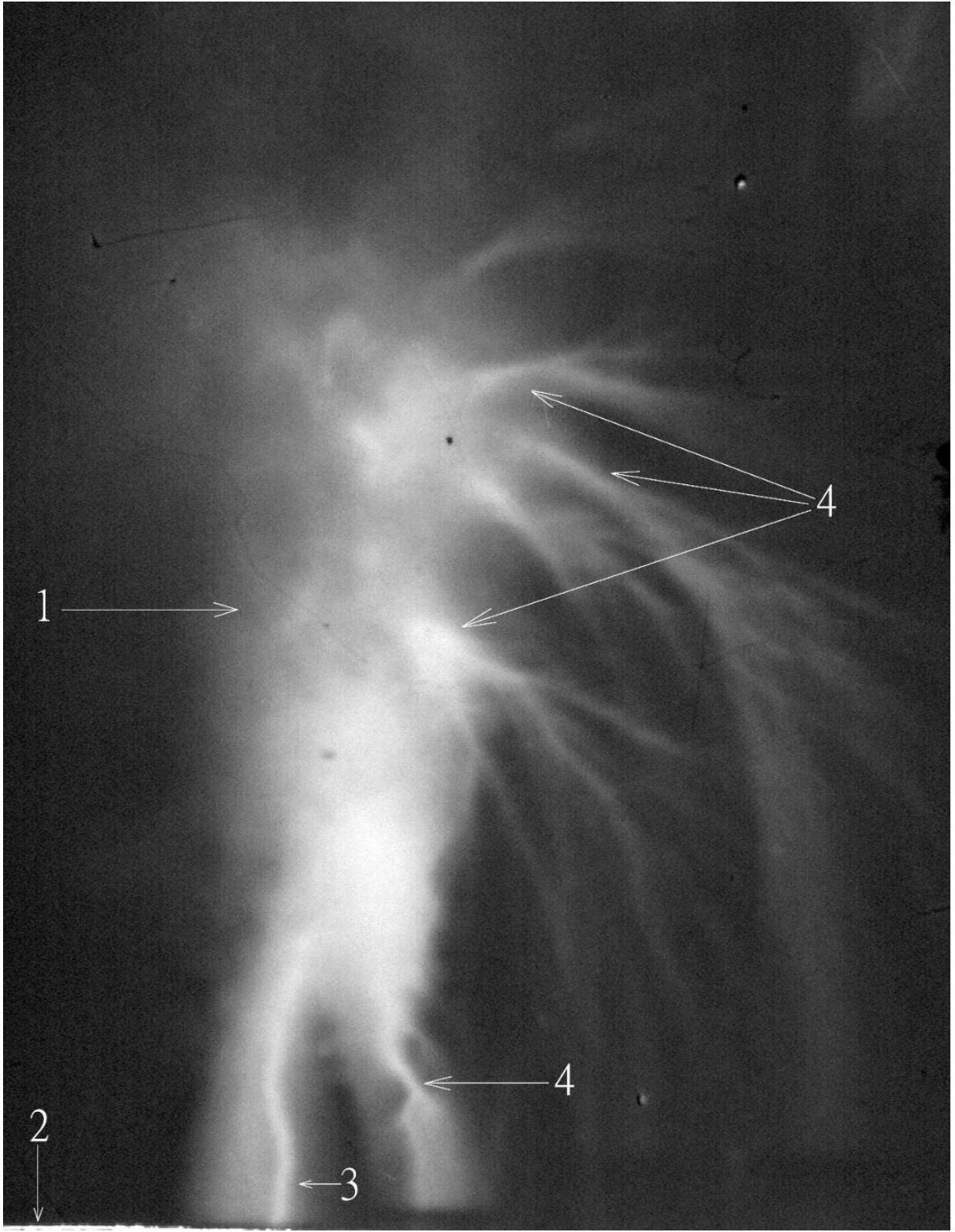

Рисунок. 4.1 (адаптировано из [Верещагин и др. 1988], оригинал фотографии любезно предоставлен В.С. Сысоевым). Интегральная, фотография разрядов с выдержкой 3 секунды. 1 – положительно заряженное аэрозольное облако; 2 – заземленная плоскость; 3 – канальная стадия разряда в виде кратковременной искры, 4 – незавершенные лидеры.

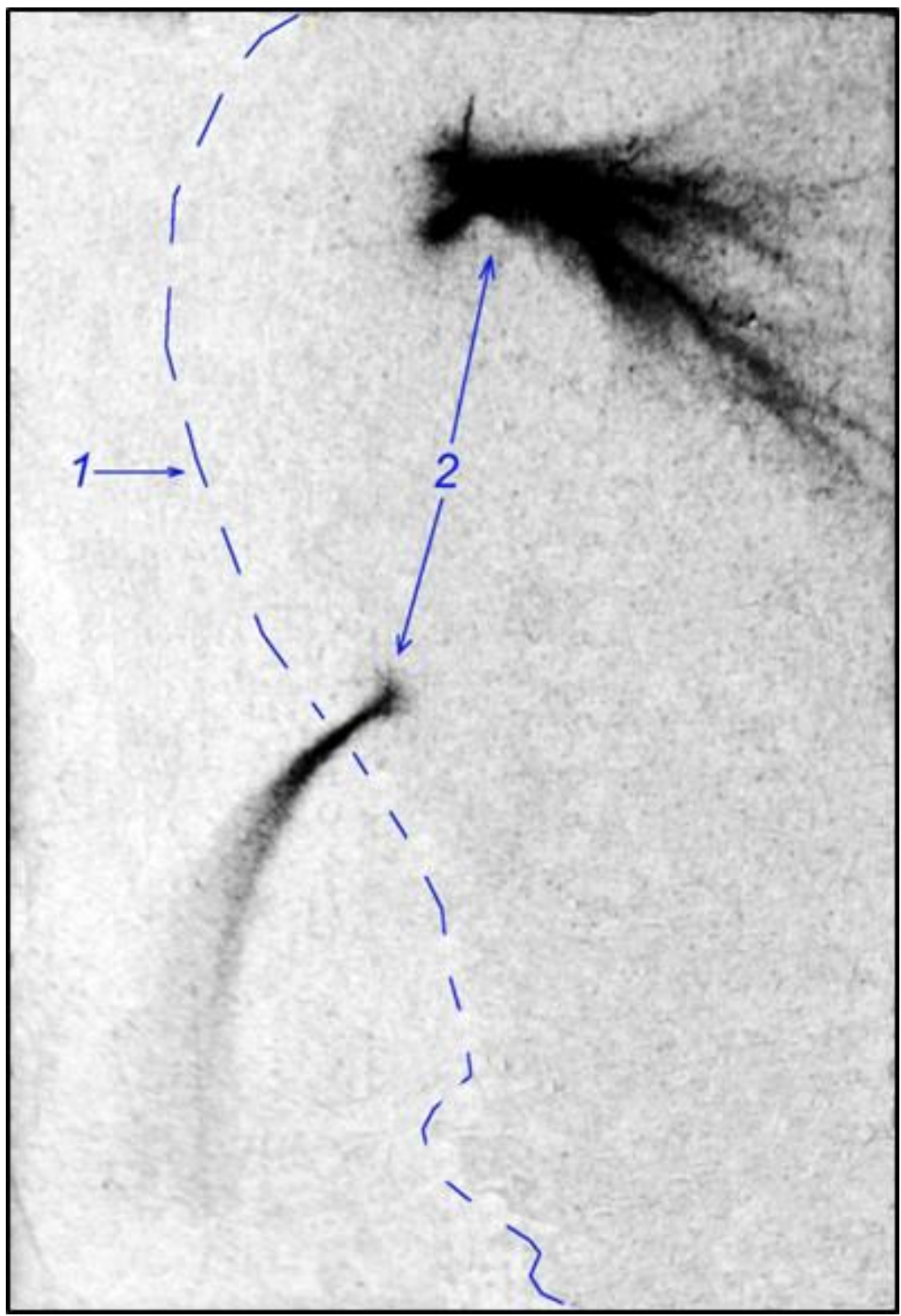

Рисунок. 4.2 (адаптировано из [Верещагин и др. 1988], оригинал фотографии любезно предоставлен В.С. Сысоевым). Интегральная, инвертированная фотография разрядов с выдержкой 3 секунды. 1 – видимый контур положительно заряженного аэрозольного облака; 2 – разряды, возникшие в аэрозольном облаке: стримерная вспышка (внизу), незавершенные лидеры (вверху).

Таким образом, [Верещагин и др. 1988] выделяют три основных типа разряда: стримерный разряд, идущий от положительно заряженного облака к заземлённой плоскости; лидерный, незавершенный разряд, идущий от положительно заряженного облака к заземлённой плоскости; завершенный разряд в канальной стадии, когда горячий плазменный канал, скорее всего касается заземленной плоскости (этот разряд описан не подробно). Обратим внимание на главное отличие разрядов в поле отрицательного и положительного облака. В положительном облаке разряды инициируются на границе (или внутри) аэрозольного облака и распространяются по направлению к заземленной плоскости. Происходит это, скорее всего потому, что поле поддержания положительных стримеров (4.5-5 кВ/см) минимум в 2 раза меньше, чем поле поддержания отрицательных стримеров (10-12 кВ/см) [Bazelyan and Raizer, 1998] и поэтому инициация положительных стримеров даже на границе облака гораздо более вероятна, чем на заземленной плоскости. Соответственно и положительные лидеры возникают в районе видимых границ заряженного облака, двигаясь вниз к заземленной плоскости. В случае отрицательно заряженного аэрозольного облака складывается обратная ситуация. Положительные стримерные вспышки инициируются на заземленной плоскости и движутся вверх в сторону отрицательно заряженного аэрозольного облака. В ту же сторону (вверх к отрицательному аэрозольному облаку) движутся и восходящие положительные лидеры.

[Анцупов и др., 1990] подтвердили результаты первых исследований и показали, что в случае положительного облака нет инициации разрядов с заземленных стержней даже длиной 50 см (при отрицательной полярности облака заземленный объект высотой 10 см хорошо инициирует разряды). [Анцупов и др., 1990] также обнаружили, что: «При увеличении тока выноса ГЗА (генератора заряженного аэрозоля – А.К.) более 60 мкА возникал разряд в виде конусообразной трубки, изогнутой по силовым линиям электрического поля, отличающийся характерным фиолетовым свечением (Рисунок 4.2/2/ нижний разряд). Длина трубки достигала 2-3 м, а диаметр у основания 0.5-1 м. Такой объемный заряд возникает из ярко светящегося ядра размером в несколько сантиметров из области на границе струи и завершается на плоскости в 0.5 м от оси струи. Свечение разряда не является однородным: яркость значительно больше в области объемного заряда, причем можно выделить более ярко светящиеся нити и ветви, а в основании трубки преобладает однородное диффузное свечение. Особенностью этого разряда является возможность его длительного существования (десятки секунд), что говорит о наличии динамического равновесия системы, когда подпитка аэрозольного облака током выноса ГЗА уравновешивается суммарными потерями заряда за счет осаждения заряженных аэрозольных частиц на заземленные предметы и протеканием тока разряда трубки. В силу естественных флуктуаций облака, изменения газодинамических параметров ГЗА трубка может погаснуть, либо перейти в стадию искрового (лидерного) разряда (Рисунок 4.1(3)). При увеличении тока выноса ГЗА трубка трансформировалась в разветвленный канальный разряд, напоминающий по форме разряд молнии (Рисунок 4.1(4)). Из-за трудности локализации канала разряда в требуемой области пространства не удалось получить фотохронограммы разряда в начальной стадии. Из осциллограмм токов главной стадии разряда, когда головка положительного лидера, развивающегося из облака, уже достигла заземленной плоскости, следует, что амплитуда тока изменяется в пределах 1-4 А, и длительность его протекания составляет 2-4 мкс». Таким образом, в этих исследованиях были обнаружены ранее открытые формы разрядов и подтверждено, что в случае положительно заряженного облака разряды инициируются преимущественно на границе или внутри аэрозольного облака. Трудность измерения токов позволяет говорить только о порядке величины тока длинных искр и примерном времени разряда.

Перед дальнейшими исследованиями стояла проблема более подробного изучения природы разрядов и обнаружения их частей, которые находятся внутри облаков.

Как отмечалось в [Kostinskiy et al., 2015a], выбор данной экспериментальной установки для изучения разрядов внутри заряженного аэрозольного облака любой полярности в данном исследовании не случаен. Заряженное аэрозольное облако принципиально отличается от любой высоковольтной электродной системы тем, что электрический заряд в нём распределен между множеством капель, электрически не связанных друг с другом. Оно стоит в ряду природных заряженных аэрозольных систем, таких как грозовое облако, торнадо, вулканическая тефра и т. д. Для осуществления разряда в таких системах необходимо, чтобы в процессе разряда часть заряда была «снята» с аэрозоля (или нейтрализована разрядом другого знака) каким-то быстрым коллективным процессом. В данном исследовании, как и в исследовании отрицательно заряженного облака [Kostinskiy et al., 2015a], используется водный аэрозоль со средним размером капель около 0.5 мкм, что даёт возможность «видеть» внутри облака с помощью современных инфракрасных камер (ИК-камер) с матрицами, чувствительными в диапазоне длин волн 3-14 мкм (средний ИК-диапазон). Сочетание параметров аэрозоля и параметров диагностического излучения позволило обнаружить внутри положительно заряженного аэрозольного облака, также как и отрицательного, несколько классов новых внутриоблачных плазменных образований, названных в [Kostinskiy et al., 2015a] – UPFs. Многие UPFs, обнаруженные в положительном аэрозольном облаке, мало похожи по морфологии на UPFs, инициированные разрядами в отрицательном облаке. Некоторые из UPFs в положительно заряженном облаке настолько длинны, что позволяют развиться с их концов двунаправленным разрядам (лидерам), которые были предложены Каземиром [Kasemir, 1960] в качестве физической модели внутриоблачных разрядов и молнии. Однако, картина, зафиксированная в положительно заряженном аэрозольном облаке, значительно сложнее предложенной Каземиром и включает в себя также до этого не обнаруженную и теоретически не предсказанную сложную внутриоблачную сеть плазменных каналов.

## 4.2. Инфракрасные изображения плазменных образований, включая двунаправленные лидеры, инициированных электрическим полем облака заряженного водяного аэрозоля

Считается, что около 90% разрядов молний, ударяющих в самолеты, инициируются самими самолетами. В инициировании, по-видимому, участвует двунаправленный лидер, положительная и отрицательная части которого развиваются с противоположных сторон самолета (см. подробно главу 5). Кроме того, двунаправленный лидер развивается из верхнего и нижнего концов незаземленной тонкой проволоки в так называемых высотно инициированных молниях [например, Lalande et al., 1998]. Также считается, что разряды молний в облаках и ступенчатые нисходящие лидеры молнии облако-земля являются по существу двунаправленными лидерами (например, [Kazemir, 1950]; [Mazur and Ruhnke, 1993]. Недавно было замечено, что так называемые «возвратные лидеры» (recoil leaders), возникающие в распадающихся частях (ветвях) ранее активных каналов молнии, демонстрируют (но не всегда) двунаправленное расширение [например, Warner et al., 2012]. Наконец, инициация молнии в облаке обязательно, (не исчерпываясь им), должна включать в себя процессы с участием двунаправленных лидеров, например, [Rakov, 2006]. Таким образом, двунаправленный лидер является одним из фундаментальных процессов в физике молнии, понимание которого важно для его адекватного моделирования и в ряде других областей, таких как молниезащита летательных аппаратов и интерпретация изображений канала молнии, полученных с помощью систем локации молнии [Rison et al., 2016], [Lyu et al., 2019]. На сегодняшний день детали физики развития двунаправленного лидера молнии до конца неясны. Положительная часть двунаправленного лидера обычно долго «молчит» в VHF-диапазоне и, следовательно, не отображается, поскольку ее излучение значительно слабее, чем излучение отрицательной части (отрицательного ступенчатого лидера). Кроме того, VHF -интерферометры обычно не могут разрешить источники, излучающие одновременно, особенно, если их много [Rison et al., 2016], [Lyu et al., 2019] и выдают некоторое усредненное значение. Оптические наблюдения обычно проводятся на относительно больших расстояниях, поэтому детали процессов распространения лидеров не видны или видны очень нечетко. Кроме того, большую часть времени двунаправленные лидеры полностью или частично скрыты внутри облака, что делает невозможным их оптическое отображение.

В этой главе мы представляем инфракрасные изображения двунаправленных лидеров, созданных искусственными облаками заряженных капель воды, полученные на очень близких расстояниях (3 ÷ 6 м) от облака. Длины волн (2,7–5,5 мкм) регистрируемого ИК-излучения были значительно больше, чем размер облачных водяных капель (типичный радиус 0,5 мкм), что дало уникальную возможность визуализировать плазменные образования внутри облака, которые не наблюдались до этого в видимом диапазоне. Стоит отметить, что такие плазменные образования, как правило, невозможно обнаружить (из-за рассеяния) даже в инфракрасном диапазоне в естественных облаках с типичными размерами гидрометеоров в 5-30 мкм и более. На ИК-изображениях показаны каналы горячих лидеров, стримерные зоны и необычные плазменные образования (UPFs) как в нижней, так и в верхней частях двунаправленного лидера. Мы не могли полностью отобразить нижнюю и верхнюю части двунаправленного лидера для одного и того же события (требовались разные поля обзора и расстояния), но представленные изображения нижней или верхней части представляют большое количество событий, инициированных в электрическом поле положительно заряженного аэрозольного облака.

## 4.2.1. Экспериментальная установка

Экспериментальная установка была аналогична описанной в главах 1-3. Она показано на Рисунке 4.3. Облако заряженных водяных капель (1) было создано парогенератором (2.1) и источником высокого напряжения (2.2), соединенным с острием, создающим корону. Последнее располагалось в сопле (2.3), через которое проходила паровоздушная струя. Струя имела температуру около 100–120 °C и давление 0,2–0,6 МПа. Он двигался со скоростью около 400–420 м/с с углом раскрытия 28°, образуя облако. Сопло располагалось в центре плоского металлического экрана (3) диаметром 2 м. В результате быстрого охлаждения пар конденсировался в капли с типичным радиусом около 0,5 мкм. Ионы, заряжающие капли, образовывались в коронном разряде между острием и соплом (2.3). На острие подавалось постоянное напряжение 10–20 кВ. Ток, связанный с зарядом, переносимым струей, находился в диапазоне от 60 до 150 мкА. Когда общий заряд, накопленный в облаке, приближался к 60 мкКл, между облаком и находящимися поблизости заземленными объектами спонтанно возникли разряды. В случае отрицательно заряженного облака большая часть разрядов возникла из заземленной металлической сферы (4), а в случае положительно заряженного облака разряды обычно были в виде стримерных вспышек, направленных вниз от кромки облака и двунаправленного лидеров. Сфера имела диаметр 5 см и находилась на расстоянии 0,8 м от центра экрана. Его самая верхняя точка находилась на 12 см выше экрана.

В случае отрицательно заряженного облака токи лидеров, возникающих на заземленной сфере (4), измерялись резистивным шунтом сопротивлением 1 Ом, сигнал с которого передавался на цифровой осциллограф (13). Когда ток превышал заданное значение, запускался осциллограф, который, в свою очередь, генерировал импульс, который использовался для запуска высокоскоростной кадрирующей камеры 4Picos с усилением изображения, работающей в видимом диапазоне (9) и инфракрасной высокоскоростной камеры FLIR SC7700M (10). В настоящих экспериментах с положительно заряженными облаками (в отличие от случая с отрицательно заряженным облаком) разряды не инициировались на сфере (4), снабженный токоизмерительным

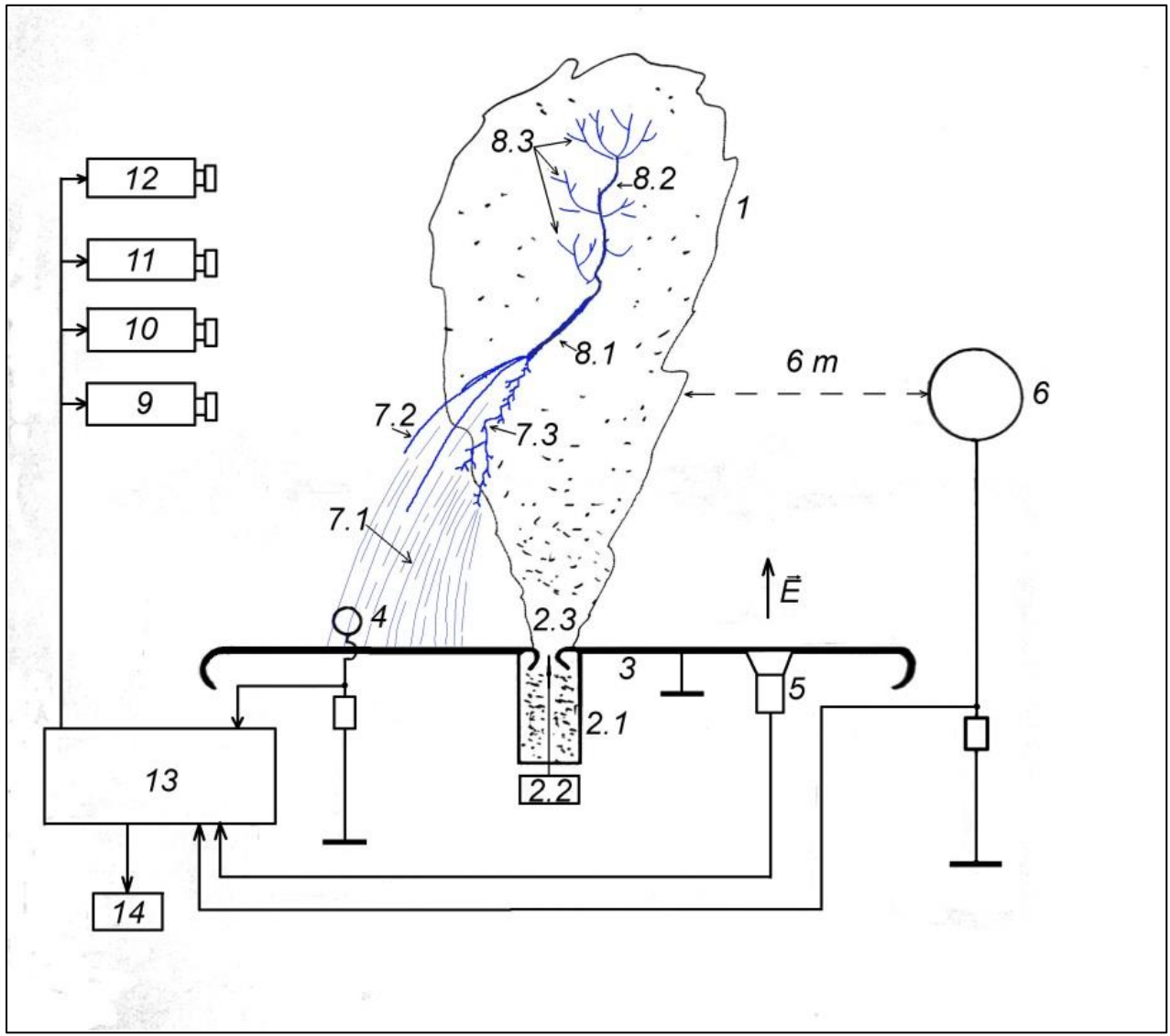

Рисунок 4.3 (адаптировано из [Kostinskiy et al., 2015b]). Схема эксперимента. 1— облако отрицательно заряженного аэрозоля; 2.1 — генератор паровоздушной струи; 2.2 — источник высокого напряжения, подаваемого на иглу в сопле; 2.3 — сопло, через которое выносится паровоздушная струя с коронирующей иглой посредине; 3 — заземленная металлическая плоскость; 4 — приёмный электрод (металлический шарик диаметром 5 см), в данном эксперименте не использовался; 5 — измеритель электрического поля, медленный (флюксметр); 6 — зонд электрического поля, быстрый (металлический шар диаметром 50 см); 7.1 — стримерная корона нисходящего положительного лидера и сталкера, ветвящегося вниз; 7.2 — сталкер, ветвящейся вниз по силовым линиям; 7.3 – нисходящий положительный лидер; 8.1 – сталкер, с концов которого стартуют положительный и отрицательный лидеры; 8.2 – отрицательный восходящий лидер; 8.3 – объёмная сеть сталкеров, взаимодействующих с отрицательным восходящим лидером; 9 — 4Picos скоростная видеокамера; 10 — FLIR 7700M, скоростная ИК-камера; 11 — ФЭУ (фотоэлектронный умножитель); 12 — фотоаппарат Canon EOS 5D Mark III; 13 — блок синхронизации и регистрации сигналов (осциллограф Tektronix DPO 71004, генератор сигналов Tektronix AFG 3252); 14 — блок хранения данных (компьютер).

шунтом. В результате невозможно было измерить ток и использовать его для запуска камер и других приборов, показанных на Рисунке 4.3. В случае положительно заряженного облака высокоскоростные камеры запускались световыми сигналами, фиксируемыми фотоумножителями (11).

Инфракрасная (ИК) камера FLIR SC7700M (10) работала со скоростью 115 кадров в секунду (длительность между кадрами 8,7 мс и время экспозиции 6,7 мс) с разрешением 640 × 512 пикселей. Скоростная камера, работающая в видимом диапазоне, выдавала 2 кадра размером 1360 × 1024 пикселей каждый. Общий снимок разряда фиксировался с помощью цифрового фотоаппарата Canon EOS 5D Mark III (12). Камеры устанавливались на расстоянии 3 м или 6 м от облака. Все изображения, представленные в этой статье, были получены с помощью ИК-камеры (экспозиция кадров была 6,7 мс). Использовался германиевый объектив с фокусным расстоянием 50 мм. Диафрагма была установлена на значении f/2.

## 4.2.2. Экспериментальные результаты

### 4.2.2.1. Части двунаправленного лидера и плазменных образований, движущиеся вниз

В случае положительно заряженного облака направленная вниз часть двунаправленного лидера должна быть положительным лидером, поскольку электрическое поле между положительно заряженным облаком и заземленной плоскостью под ним направлено вниз (к плоскости). В случае отрицательно заряженного облака положительный лидер может быть легко идентифицирован, так как он инициируется заземленной сферой (Рисунок 4.3 (4)) и его ток можно напрямую измерить. Пример ИК-изображения восходящего положительного лидера, развивающегося от сферы к отрицательному облаку, показан на Рисунке 4.4, где хорошо видны небольшие ступеньки положительного лидера размером 0.5-1 см.

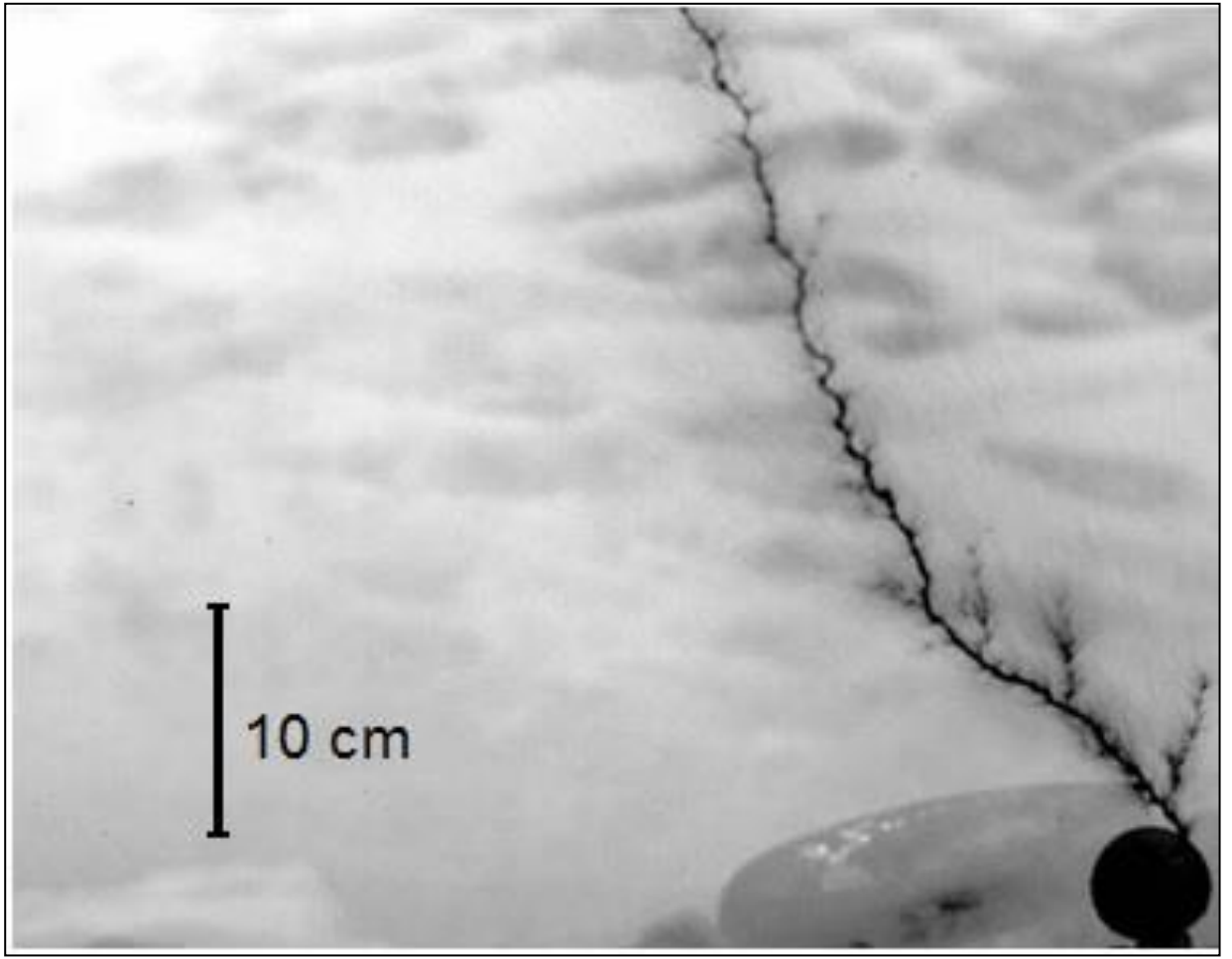

Рисунок 4.4 (адаптировано из [Kostinskiy et al., 2015b]). Инфракрасное изображение (инвертированное), полученное ИК-камерой с выдержкой 6,7 мс, показывает восходящий положительный лидер от заземлённой сферы в электрическом поле отрицательно заряженного облака.

Отличительными чертами восходящих положительных лидеров в экспериментах с искусственно заряженными облаками являются выраженная извилистость и характерное ветвление их каналов, как видно на Рисунке 4.4, где ветвление направлено вверх, что указывает на направление развития лидера. Мы будем использовать эти особенности развития лидеров для выявления положительных лидеров в системе двунаправленного лидера. Направление ветвления положительного лидера в поле между положительно заряженным облаком и заземленной плоскостью под ним, как ожидается, будет направлено вниз. Наши наблюдения согласуются с этим ожиданием.

Восходящие положительные лидеры, инициированные в поле отрицательного облака, характеризуются пиковым током в несколько ампер (до нескольких десятков ампер), продолжительностью тока в несколько десятков микросекунд и типичной передачей заряда 20-30 мкКл, если за ними не следует процесс, подобный обратному удару молнии (квазиобратный удар). Примеры осциллограмм тока восходящего положительного лидера показаны на Рисунке 4.5 (для события без квазиобратного удара) и на Рисунке 4.6 (для события с квазиобратным ударом). Ток квазиобратного удара может достигать десятков ампер, когда относительная влажность приближается к 100%.

На Рисунке 4.7 показана нижняя часть двунаправленного лидера, инициированного в положительно заряженном облаке (его центральная, более плотная часть видна в виде вертикально-струйной структуры на панели I). На Рисунке 4.7 (одинаковая нумерация на всех трех панелях) обозначены средняя часть двунаправленного лидера (1), восходящий отрицательный лидер (2), необычное плазменное образование, UPF, (3), нисходящий положительный лидер (4) и положительная стримерная корона (5). Все элементы двунаправленного лидера расположены внутри облака, за исключением нижней части положительной стримерной короны, которая контактирует с заземленной плоскостью. Панели II - III представляют собой увеличенные версии панели I, а панель III была получена путем вычитания из кадра I предыдущего кадра, который не содержал изображений разряда (для повышения контраста). Нисходящий положительный лидер (4) на панели III очень похож (за исключением направления движения) на его восходящий аналог на Рисунке 4.4. UPF (3) не сильно отличается от положительной стримерной короны (5), но отличие его каналов в сравнении с более однородной светимостью, создаваемой большим количеством положительных стримеров отчетливо видна.

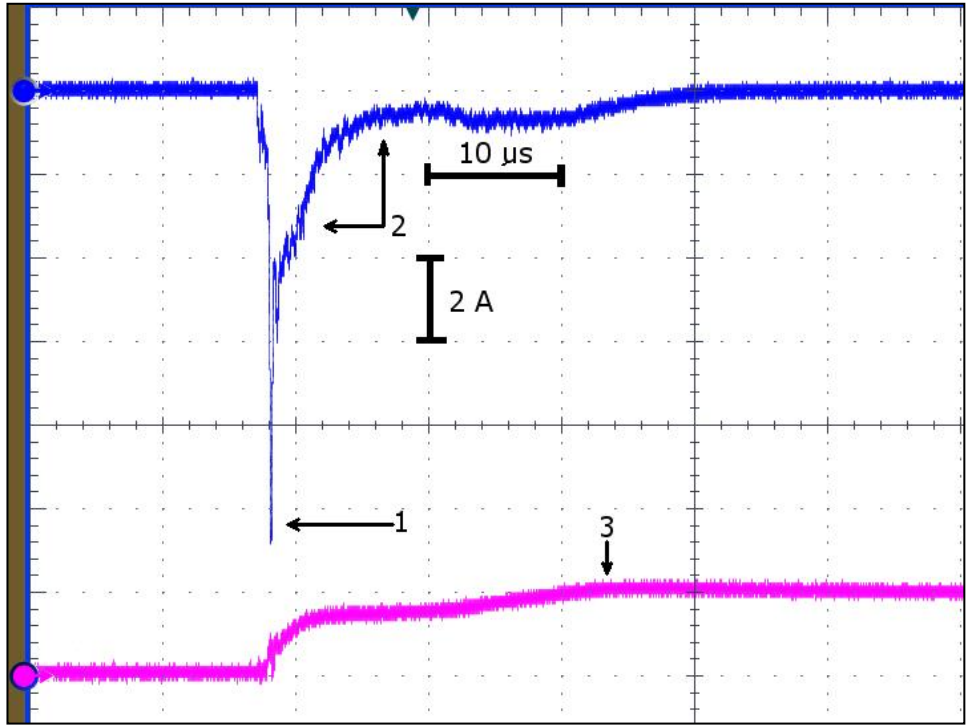

Рисунок 4.5. Осциллограмма тока (2) и сигнал от 50-сантиметровой сферы ((6) Рисунок. 4.3), используемой для мониторинга изменений заряда облака (3) для события, показанного на Рисунке 4.4 (облако заряжено отрицательно). Обозначены: начальный импульс стримерной короны (1), ток восходящего положительного лидера (2) и наведенный потенциал (3) на сфере, указывающий на снижение (частичную нейтрализацию) отрицательного заряда в облаке. Процесса, похожего на обратный удар, не происходило.

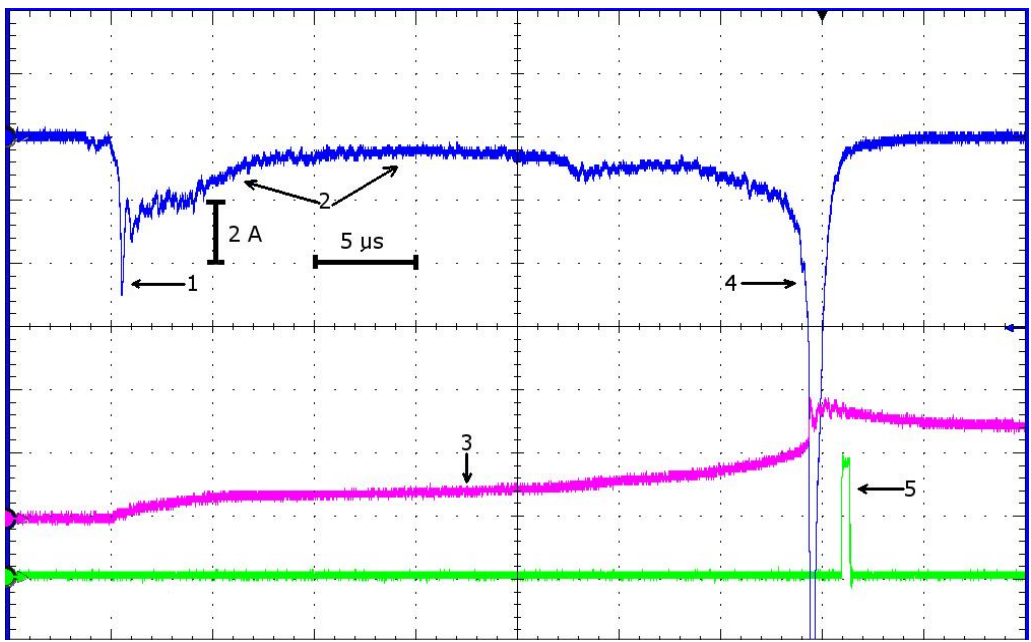

Рисунок 4.6. Осциллограмма тока (2), сигнал от 50-сантиметровой сферы ((6) Рисунок. 4.3), используемой для мониторинга изменений заряда облака (3) и импульс запуска для высокоскоростной камеры (5). Облако было заряжено отрицательно. Обозначены: начальный импульс стримерной короны (1), ток восходящего положительного лидера (2), и наведенный потенциал (3) на сфере, указывающий на снижение (частичную нейтрализацию) отрицательного заряда в облаке, ток квазиобратного удара (4) и импульс, запускающий камеру (5).

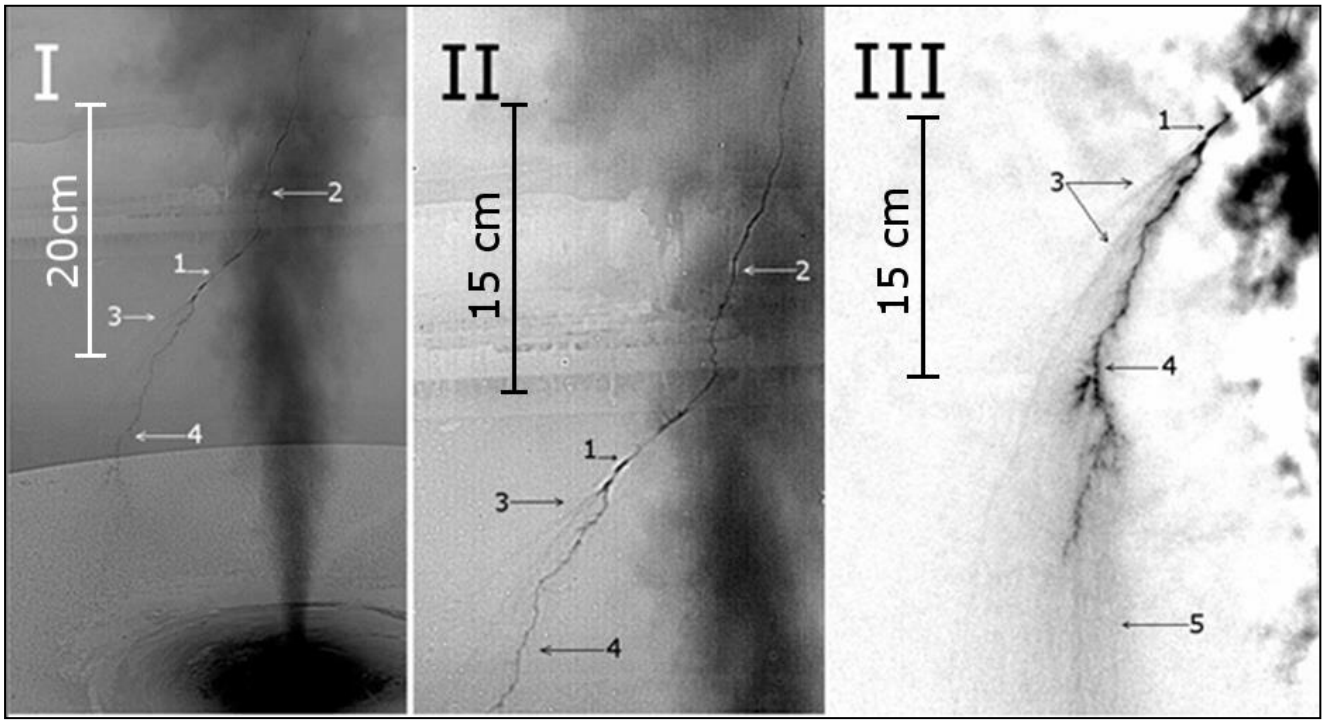

Рисунок 4.7 (адаптировано из [Kostinskiy et al., 2015b]), (событие 125-1276-1). Инфракрасные изображения (инвертированные), полученные при экспозиции 6,7 мс, показывают нижнюю часть двунаправленного лидера, инициированного в положительно заряженном облаке. Обозначены средняя часть двунаправленного лидера (1), восходящий отрицательный лидер (2), необычное плазменное образование, UPF, (3), нисходящий положительный лидер (4) и положительная стримерная корона (5). Панели II и III представляют собой увеличенные версии панели I, а панель III была получена путем вычитания предыдущего кадра, который не содержал изображений разряда.

Отметим, что средняя часть, где, по-видимому, был инициирован двунаправленный лидер, трудно отличимая от участков верхней части двунаправленного лидера. Поэтому определение средней части является довольно субъективным. Мы обозначили его приблизительное (ожидаемое) положение на ИК-изображениях, предполагая, что оно должно быть менее извилистым, чем нижняя (положительная) или верхняя (отрицательная) части. Граница между средней и нижней частями в целом была более заметной, чем между средней и верхней частями.

На Рисунке 4.8 показана средняя часть двунаправленного лидера (1), которая расположена примерно на 40 см выше заземленной плоскости, UPF (2), положительный нисходящий лидер (3) и отрицательный восходящий лидер (4). UPF (2) выглядит как раздвоенный канал (гораздо менее извилистый, чем канал положительного лидера (3)), некоторые части которого имеют интенсивность ИК-излучения, сравнимую или превышающую интенсивность инфракрасного излучения положительного нисходящего лидера (3). Вполне вероятно, что у этих двух плазменных образований также сравнимые температуры. Интересно, что UPF, скорее всего, контактирует с заземленной плоскостью, а положительный лидер — нет (возможно контактирует стримерная корона). Вставка в правом нижнем углу Рисунка 4.8 получена вычитанием предыдущего кадра.

Еще один пример нижней части двунаправленного лидера показан на Рисунке 4.9. Он похож на событие, показанное на Рисунке 4.7, за исключением того, что положительный нисходящий лидер (1) значительно короче, чем UPF (2), которое составляет около 1 м в длину и соприкасается с заземленной плоскостью.

На Рисунке 4.10 зафиксировано событие похожее на Рисунок 4.7, но имеющее важные особенности. На Рисунке 4.10 и увеличенных фрагментах рис. Рисунке 4.10.II, Рисунке 4.10.III можно выделить следующие элементы: (1) — центральный плазменный канал (возможно, UPF), его яркость более, чем в два раза превосходит яркость всех остальных частей канала, включая лидеры. Вверх из центрального UPF (1) в положительное облако уходит отрицательный лидер (2); около центрального UPF (1) можно видеть еще несколько каналов, похожих на UPF переменной яркости и структуры (3). Заметим, что в отрицательном облаке также часто фиксировалось несколько рядом лежащих UPFs разной яркости. Также по морфологии развития можно зафиксировать с

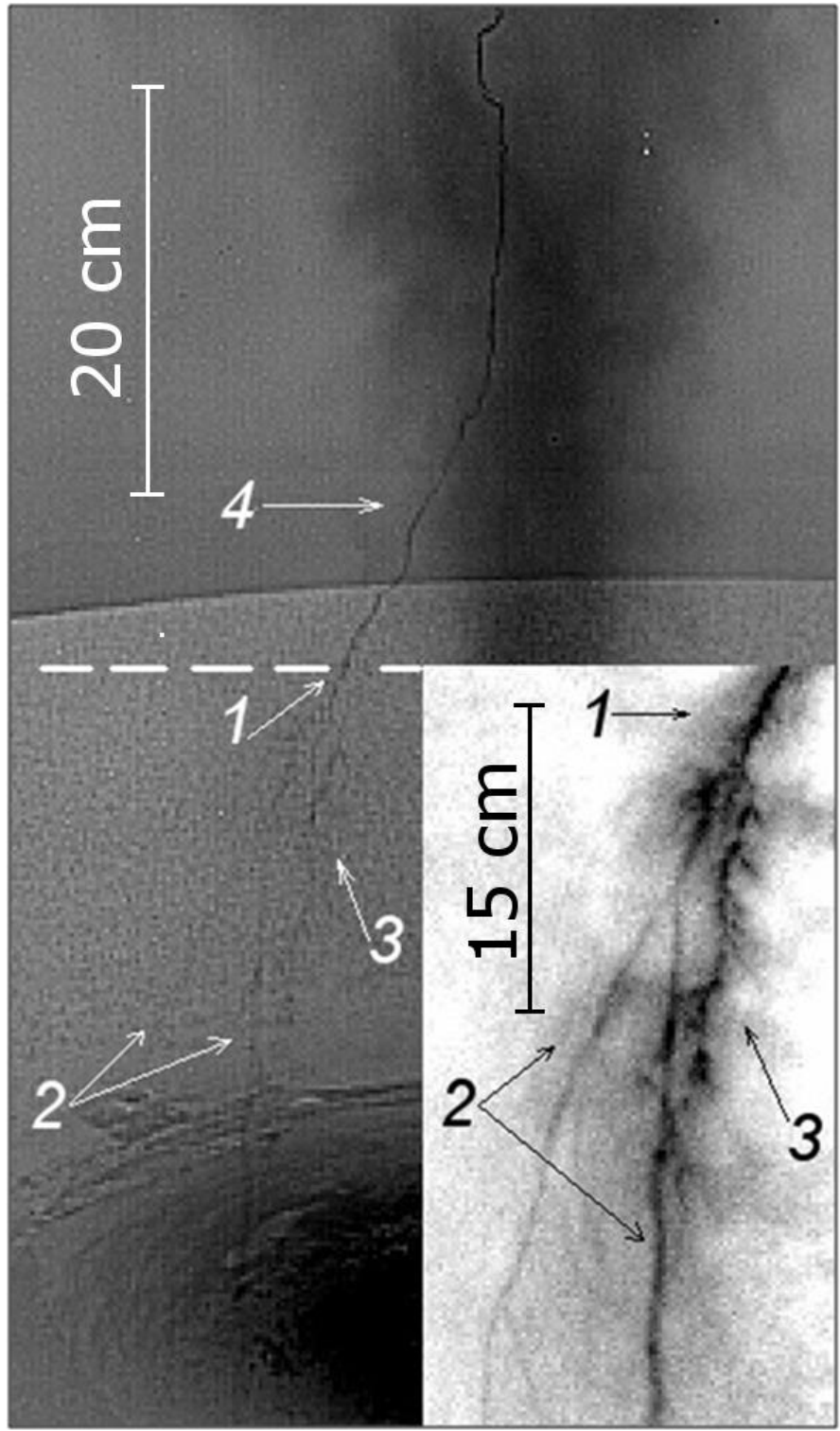

Рисунок 4.8 (адаптировано из [Kostinskiy et al., 2015b]). Инфракрасные изображения, полученные с выдержкой 6,7 мс, показывают нижнюю часть двунаправленного лидера, инициированного в положительно заряженном облаке (событие 125-941). Обозначены: средняя часть двунаправленного лидера (1), необычное плазменное образование, UPF, (2), которое контактирует с заземленной плоскостью, положительный нисходящий лидер (3) и отрицательный восходящий лидер (4). Вставка в правом нижнем углу получается вычитанием предыдущего кадра и является увеличенным фрагментом изображения.

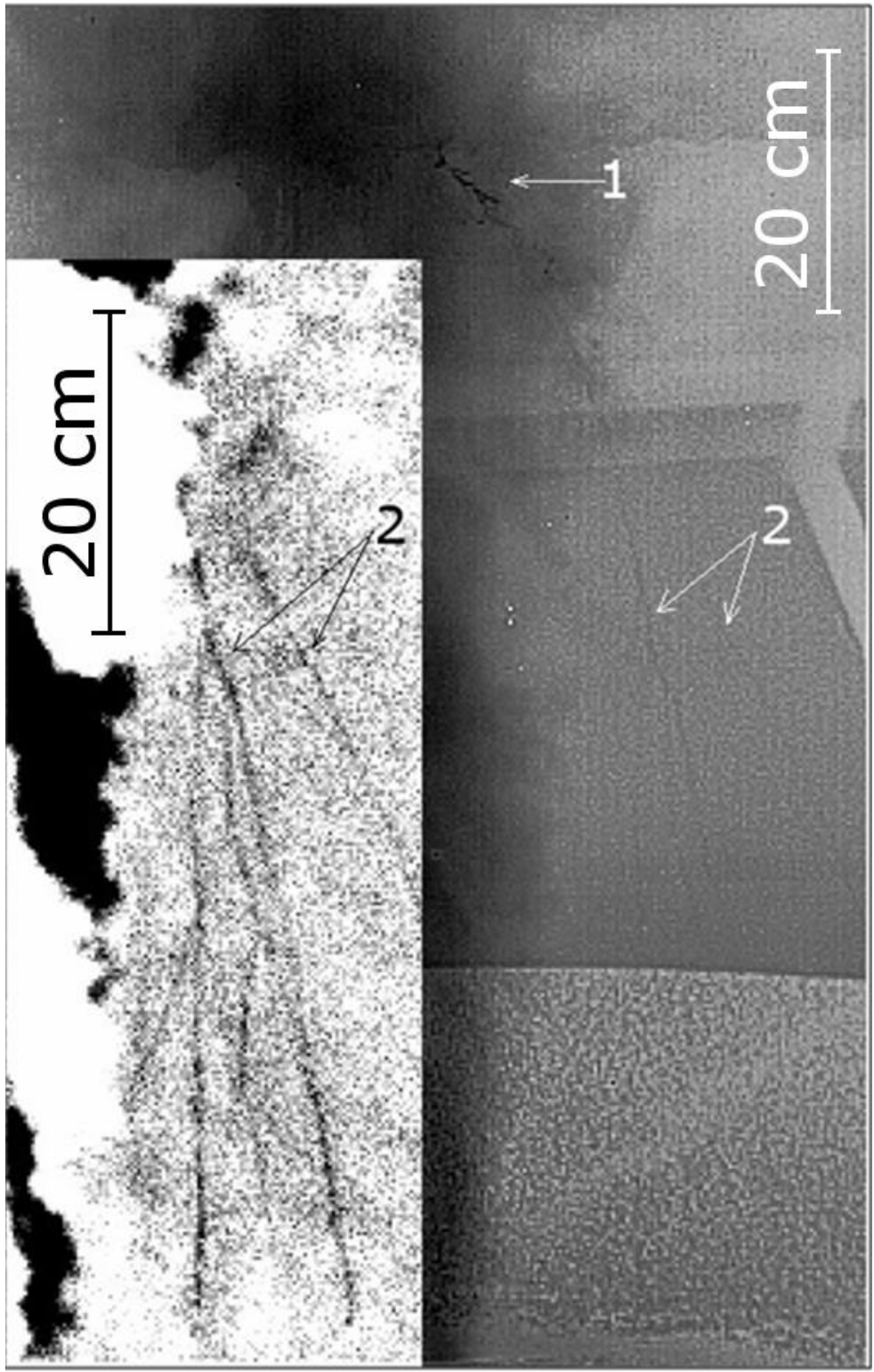

Рисунок 4.9 (адаптировано из [Kostinskiy et al., 2015b]). Инфракрасное изображение, полученное с выдержкой 6,7 мс, показывает нижнюю часть двунаправленного лидера, инициированного положительным заряженным облаком (событие 125-1746). Обозначены: короткий нисходящий положительный лидер (1) и необычное плазменное образование, UPF, (2), которое контактирует с заземленной плоскостью. Вставка в левом нижнем углу получается вычитанием предыдущего кадра.

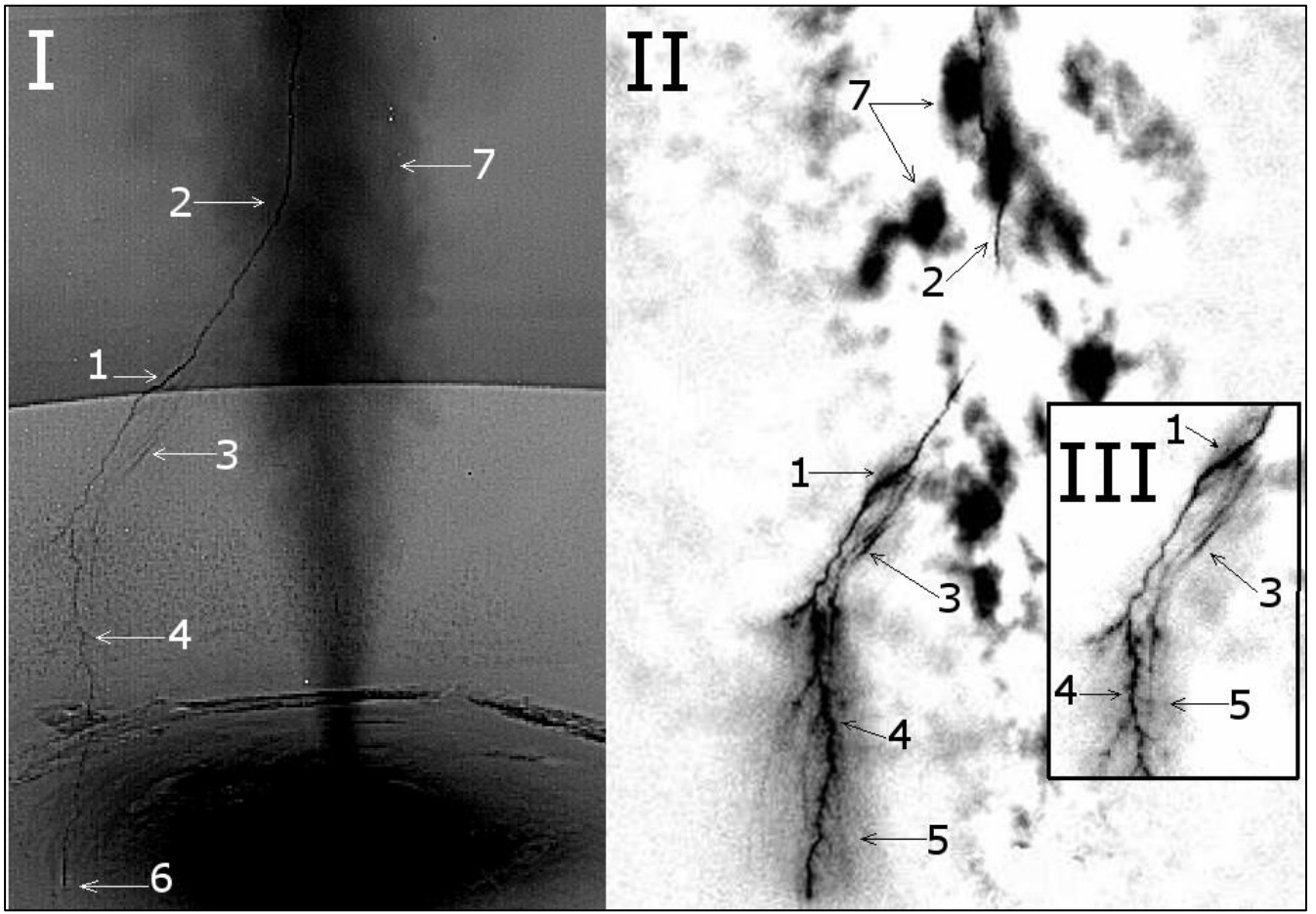

Рисунок 4.10 (адаптировано из [Kostinskiy et al., 2015b]), (событие 125-2584). 1 – центральный плазменный канал (UPF), отрицательный восходящий лидер – 2; UPF переменной яркости и структуры – 3, нисходящий положительный лидер – 4, положительная стримерная корона – 5, точка контакта положительного лидера с плоскостью – 6, центральная часть аэрозольной струи – 7. Большая часть разряда находится внутри облака и не фиксируется в видимом диапазоне. Режим записи: частота кадров — 115 Гц, выдержка — 6.7 мс, число пикселей в исходном кадре — 640x512 (на данном рисунке изображена только часть кадра с разрядом).

высокой вероятностью извилистый, ветвящийся положительный лидер (4), который в данном разряде, в отличие от события на Рисунке 4.7, достигает заземлённой плоскости в точке — (6). На обработанных вычитанием из данного кадра предыдущего частях (Рисунок 4.10.II-4.10.III) отчётливо видна нисходящая положительная корона (5), положительного лидера (4). Плотность аэрозольной струи неоднородна и ИК-камера фиксирует наиболее плотную часть струи близкую к оси, перпендикулярной плоскости (7). Большая часть разряда также находится внутри облака и не фиксируется в видимом диапазоне.

На Рисунке 4.11, как и на Рисунке 4.9(1) впервые зафиксирована тонкая структура инициации положительных лидеров вместе с другими типами внутриоблачных разрядов. На Рисунке 4.11 можно выделить почти прямой по форме центральный плазменный канал (UPF) по краям более яркий, в середине менее (1) (возможно данный UPF поляризован, как и на Рисунке 4.7). Вверх из верхней части центрального UPF (1) в положительное облако уходит, вероятно (изображение размыто), восходящий отрицательный лидер (4). Из нижней части центрального UPF (1) выходят нисходящий UPF (2) и хорошо идентифицируемый извилистый положительный лидер (3). На Рисунке 4.9(1) также хорошо виден короткий извилистый нисходящий положительный лидер (1) и длинные, величиной около метра нисходящие положительные UPF (2), доходящие до заземлённой плоскости.

Данные экспериментальные результаты с использованием ИК-камеры соответствуют ранее полученным фотографиями разрядов, поддерживаемых электрическим полем положительно заряженного облака [Верещагин и др. 1988]. Мы можем видеть нисходящие от края облака положительные лидеры движущиеся к заземленной плоскости, которые остаются незавершенными или касаются плоскости (Рисунки 4.12(1), 4.1(3)), нисходящие стримерные короны (Рисунки 4.12(2), 4.2(2-нижний)), расположенные в верхней части облака относительно короткие незавершенные лидеры (Рисунки 4.1(4), 4.2(2-верхний), 4.12(5)). Однако эти фотографии видимого диапазона не позволяют изучать тонкую структуру разрядов в месте инициации лидеров и не позволяют фиксировать UPFs.

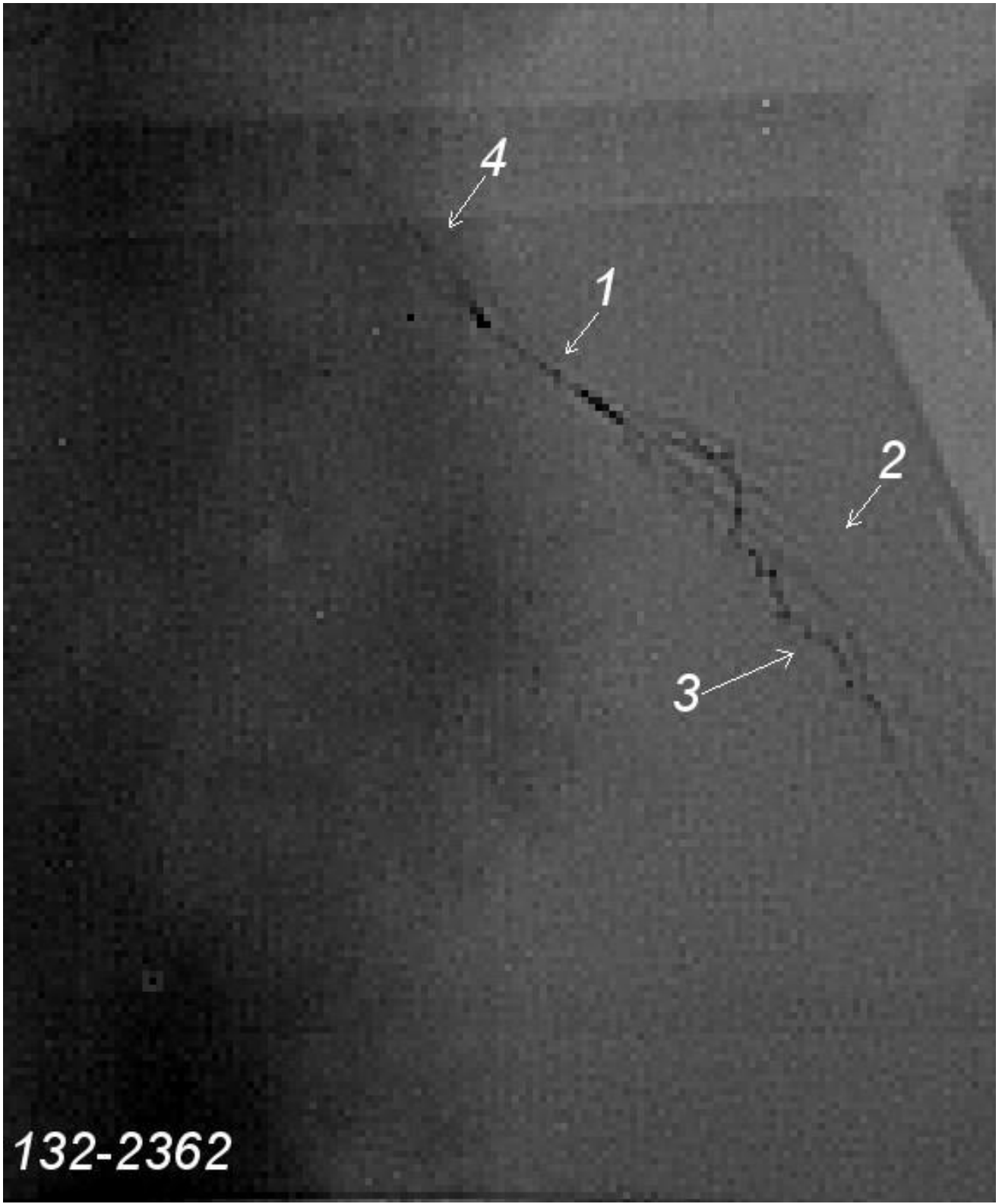

Рисунок 4.11 (событие 132-2362). Рождение двунаправленной структуры внутри облака. 1 – «центральный» UPF, длина ≈ 5÷6 см, находится на высоте ≈ 60 см над заземлённой плоскостью ;  2 – нисходящее продолжение «центрального» UPF, возможно с нисходящей положительной стримерной короной; 3 – нисходящий положительный лидер, возникающий из «центрального» UPF; 4 – возможно восходящий отрицательный лидер (изображение сильно размыто). Режим записи: частота кадров — 115 Гц, выдержка — 6.7 мс, число пикселей в исходном кадре — 640x512 (на данном рисунке изображена только часть кадра с разрядом).

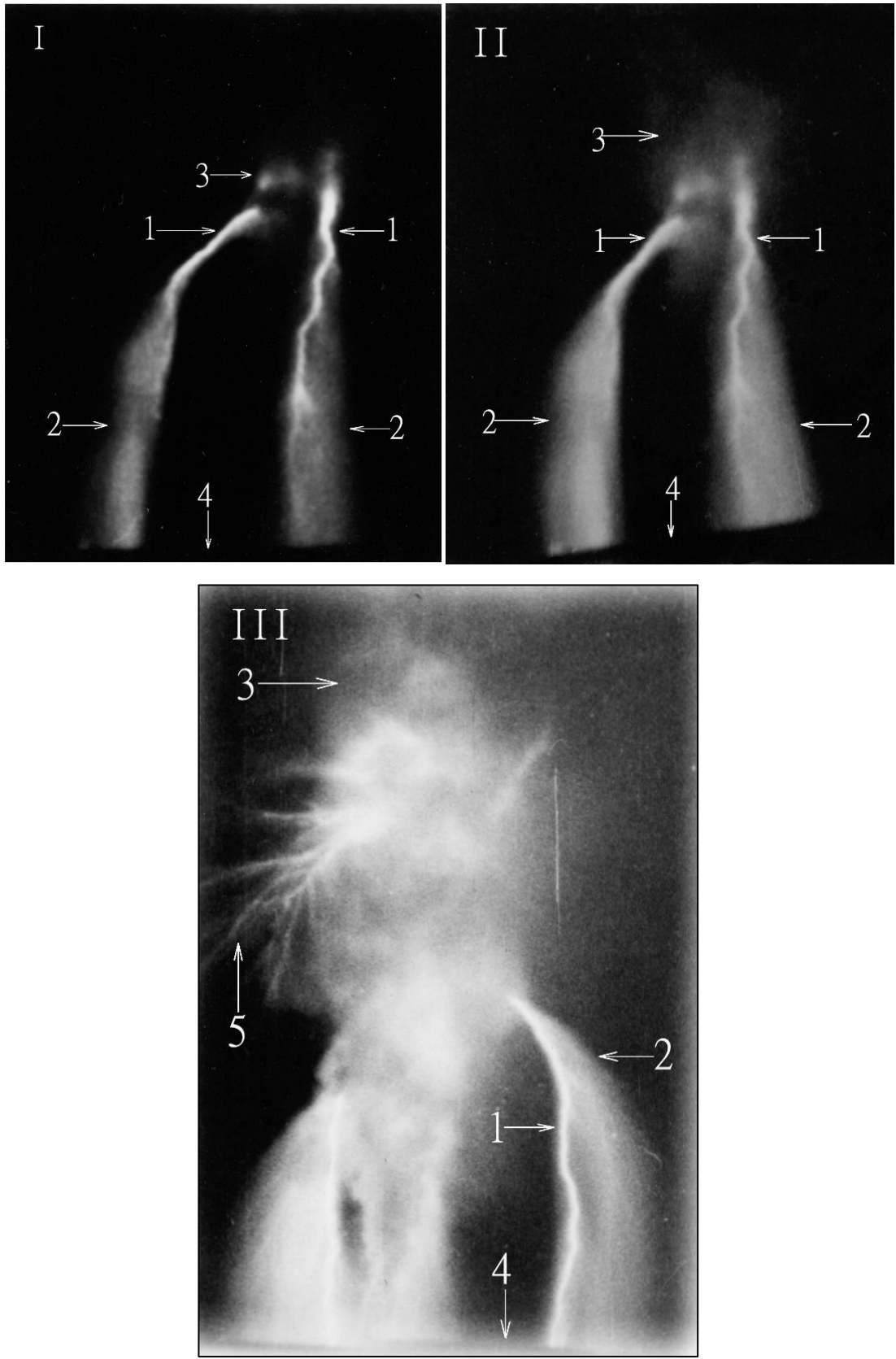

Рисунок 4.12 (адаптировано из [Верещагин и др. 1988], оригиналы фотографий любезно предоставлен В.С. Сысоевым). Интегральные фотографии разрядов с выдержкой 3 секунды. 3 – положительно заряженное аэрозольное облако; 4 – заземленная плоскость; 1 – положительный нисходящий лидер, 2 – стримерные короны, 5 – незавершенные лидеры.

Изучая все приведённые Рисунки 4.7-4.11, снятые с расстояния 6 м, можно в первом приближении реконструировать структуру двунаправленного лидера и других внутриоблачных разрядов в положительном аэрозольном облаке. Она отличается от простой структуры двунаправленного лидера [Kasemir, 1969], [Edens et al., 2012], хотя в ней и можно выделить нисходящий положительный лидер и восходящий отрицательный. Кроме них, обычно в центре, находится гораздо более нагретое, скорее всего поляризованное плазменное образование, из которого, возможно, и возникают оба лидера (мы называли его выше «центральным UPF»). Кроме того, из центрального UPF вместе с положительным лидером возникают нисходящие положительные каналы, искривлённые, возможно, по силовым линиям электрического поля, подобно положительной стримерной короне, но данные каналы на порядок более яркие в ИК-диапазоне, чем стримеры (их мы называем положительными нисходящими UPFs), причём их яркость часто не уступает или даже превышает яркость горячих положительных лидеров, что не позволяет отнести их к положительной стримерной короне.

## 4.2.2.2. Части двунаправленного лидера, и других плазменных образований, движущиеся вверх

ИК-изображения, показанные на Рисунках 4.7-4.11, были получены с расстояния 6 м с вертикальным полем зрения высотой 1.2 м над заземленной плоскостью. Чтобы лучше фиксировать верхнюю внутриоблачную часть двунаправленного лидера, расстояние от ИК-камеры до облака было уменьшено до 3 м.

На Рисунке 4.13 показана верхняя часть двунаправленного лидера, включая нижнюю часть плазменного канала с относительно высокой (1) а другую часть с относительно низкой (2) интенсивностью ИК-излучения, соединенной с сетью плазменных каналов (3) в верхнем левом углу кадра. Все каналы расположены глубоко внутри аэрозольного облака. Похоже, что разветвленная сеть плазменных каналов пронизывает значительный объем облака и не является только продолжением начальных каналов или стримерной короны лидеров. ИК-яркость некоторых частей сети сравнима с

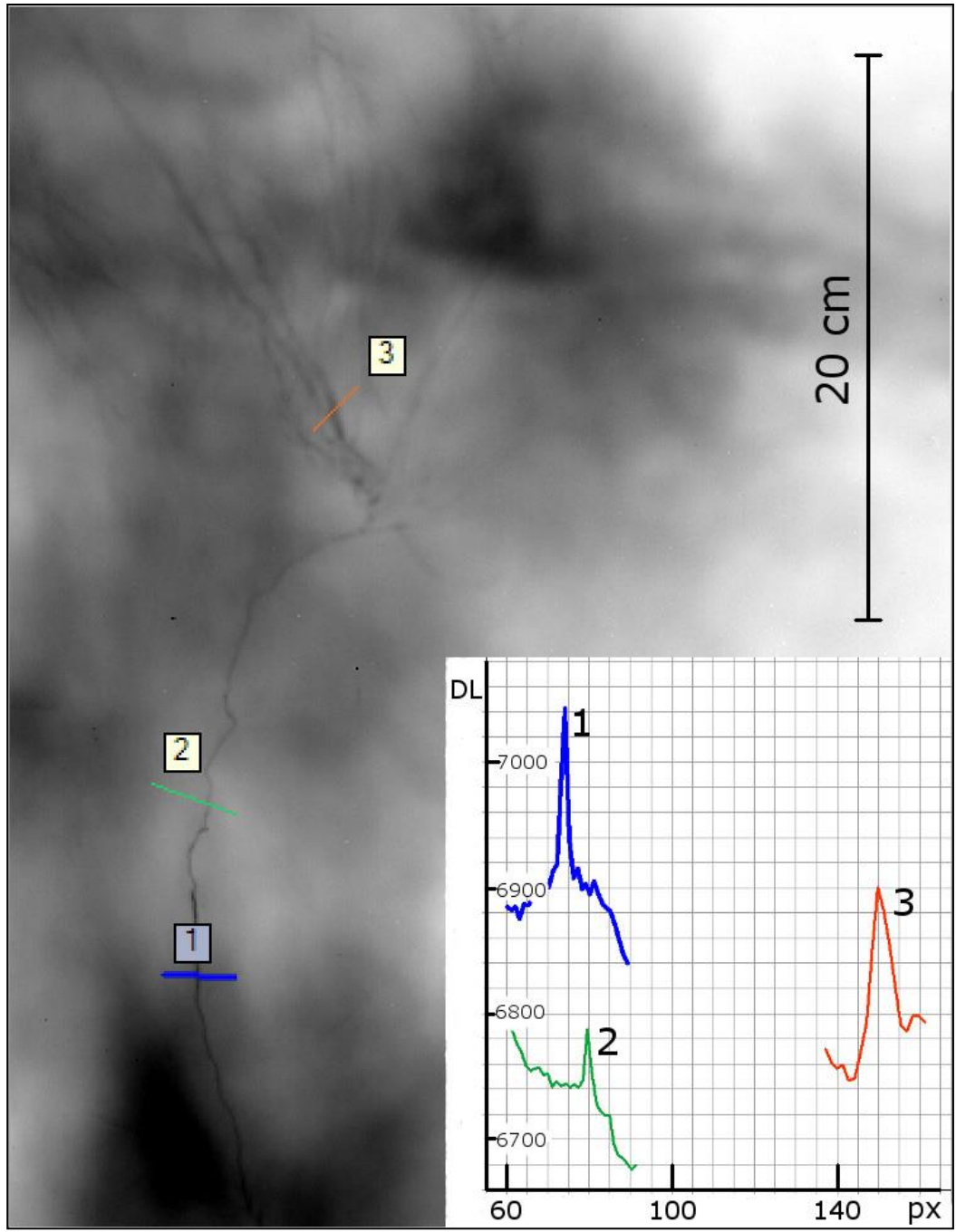

Рисунок 4.13 (адаптировано из [Kostinskiy et al., 2015b]), (событие 051-3013). Инфракрасное изображение (инвертированное), полученное с выдержкой 6,7 мс, которое показывает верхнюю часть двунаправленного лидера, инициированного полем положительно заряженного облака (событие 051-3013). В левом верхнем углу кадра можно увидеть плазменный канал (2 – восходящий отрицательный лидер) с относительно высокой (внизу) и относительно низкой (вверху) интенсивностью ИК-излучения, соединенный с сетью плазменных каналов. На вставке в правом нижнем углу показаны приблизительные профили ИК-яркости для участков канала, обозначенных 1, 2 и 3 (горизонтальная ось - номер столбца пикселей в матрице камеры, а вертикальная ось - яркость в относительных единицах).

яркостью канала восходящего отрицательного лидера, что не позволяет интерпретировать разветвленную сеть, как стримерную зону отрицательного лидера. Яркость некоторых ветвей сети (3) превышает яркость канала восходящего отрицательного лидера (2). Морфология сети (3) резко отличается от морфологии отрицательной стримерной вспышки длинной искры. Возможно, поэтому некоторые каналы в верхней части кадра не относятся к стримерной короне восходящего отрицательного лидера. Это обстоятельство и сложная морфология всей сети (включая значительные вариации яркости, свидетельствующие о неравномерном нагреве) позволяют предположить, что наблюдаемые плазменные структуры могут иметь ту же природу, что и UPFs, описанные в главах 1-2. Последнее утверждение в первую очередь основано на том факте, что то, что мы называем UPFs, значительно отличается по морфологии от лидеров или стримеров. Действительно, в сравнении с нижней положительной частью двунаправленного лидера, UPFs гораздо менее извилисты и разветвляются по-другому в сравнении с положительными лидерами, но при этом UPFs значительно ярче, чем стримерная зона положительного лидера. В верхней части UPFs образуют сложную сеть каналов, которая скорее всего не является лидером, хотя каналы сети выглядят гораздо ярче, чем отрицательные стримеры. Мы не можем исключить возможность того, что нижние и верхние UPFs имеют разную природу, хотя их отличия друг от друга могут быть связаны с разной полярностью и разным положением относительно более плотной части облака и заземленной плоскости. Интересно, что положительный лидер имеет одинаковый внешний вид независимо от того, распространяется ли он вниз или вверх, и инициируется ли он от заземленного объекта или составляет нижнюю часть двунаправленного лидера.

Другой пример верхней части двунаправленного лидера показан на Рисунке 4.14. Он аналогичен событию, показанному на Рисунке 4.13, за исключением того, что восходящий отрицательный лидерный канал более однороден и имеет яркий кончик (головку), от которого (которой) веером расходятся более слабо светящиеся каналы. В нижней части канала имеется очень яркий элемент (черная стрелка), который может свидетельствовать о контакте плазменных каналов (ниже этот вопрос будет рассмотрен более подробно). Обращаем внимание на явно независимые от этого канала и более яркие элементы сложной по форме сети каналов в верхнем левом углу кадра (указаны белыми стрелками). Некоторые слабо светящиеся каналы трудно различить на приводимом

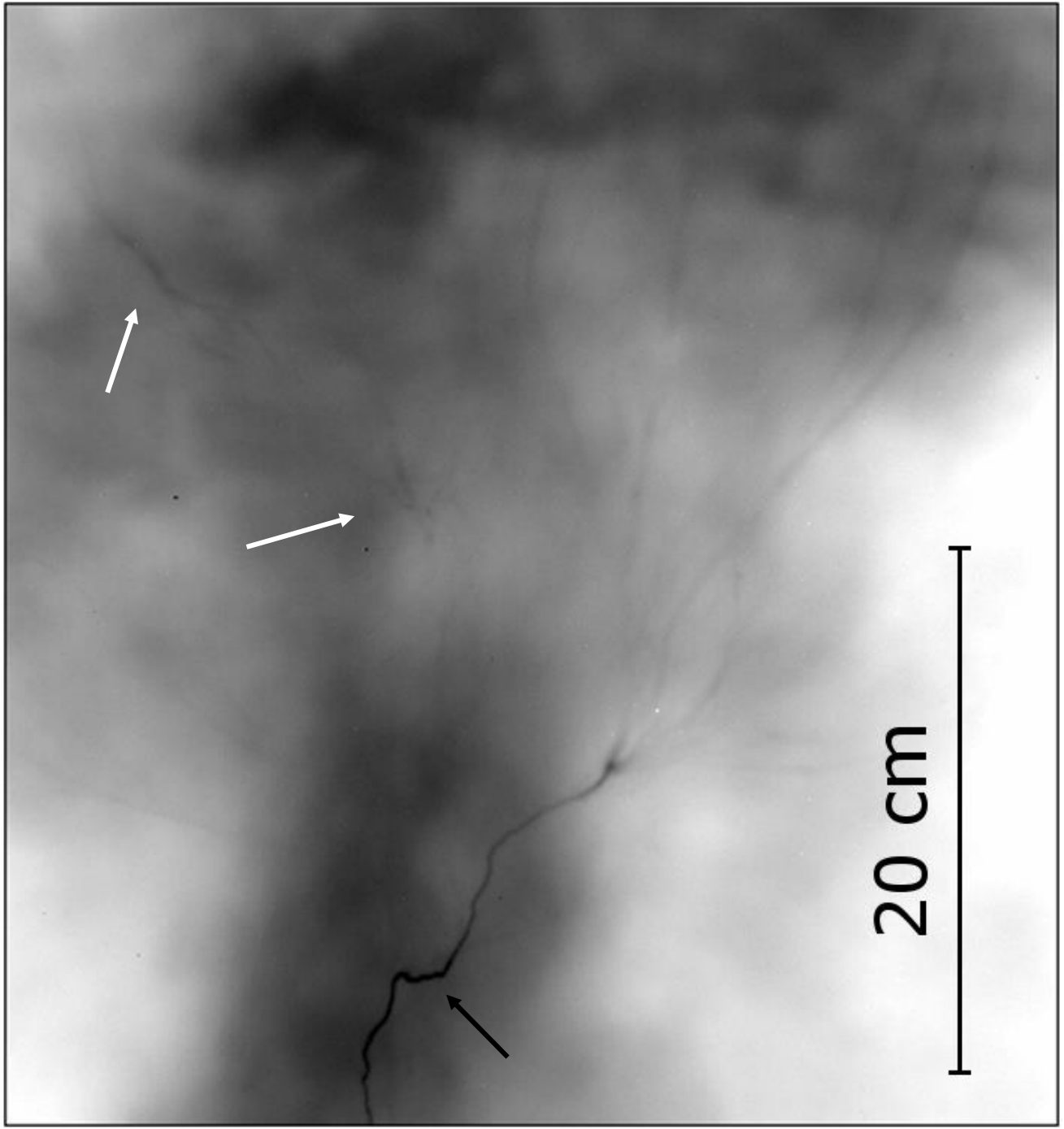

Рисунок 4.14 (адаптировано из [Kostinskiy et al., 2015b]), (событие 053-854). Инфракрасное изображение (инвертированное), полученное с выдержкой 6,7 мс, показывает верхнюю часть двунаправленного лидера, инициированного положительным заряженным облаком. В правой верхней части кадра виден плазменный канал (восходящий отрицательный лидер) переменной ИК-интенсивности, связанный с сетью плазменных каналов. Обращают на себя внимание явно независимые и более яркие элементы сети слабых каналов в верхней левой части кадра (указаны белыми стрелками). В нижней части канала имеется очень яркий элемент (черная стрелка), который может свидетельствовать о контакте плазменных каналов.

рисунке, но они явно обнаруживаются при контрастной обработке кадра. К сожалению, используемый ранее метод вычитания кадров, который был полезен при изучении нижней части двунаправленного лидера, не всегда работал для его верхней части, расположенной глубже в облаке, из-за движения частей облака внутри времени экспозиции, что снижало контрастность изображения.

На Рисунке 4.15 мы видим событие, аналогичное событию на Рисунке 4.14. От верхнего конца (головки) отрицательного лидера расходятся много каналов, а также можно различить множество каналов, пронизывающих почти весь объем облака, которые, скорее всего, не связаны с отрицательным каналом. Анализ инфракрасного излучения вдоль основного канала показывает сильный локальный нагрев канала в области (2). ИК-излучение в точке (2) примерно в три раза больше, чем в точках (1) и (3), которые лежат ниже и выше точки (2). Этот сильный локальный нагрев может свидетельствовать о контакте разных плазменных каналов в области (2), что мы неоднократно наблюдали при сквозной фазе с последующим квазиобратным ударом при контакте восходящих положительных лидеров и нисходящих отрицательных лидеров в электрическом поле отрицательно заряженного облака (см. главу 3, и, например, Рисунок 3.8). Это может говорить о том, что фиксируемые на ИК-изображениях каналы имеют сложную составную структуру и некоторые их части или они целиком могут формироваться благодаря слиянию нескольких каналов.

На Рисунке 4.16, как и на Рисунке 4.15 мы видим в середине плазменного канала в несколько раз более ярко светящуюся в ИК-диапазоне область (1). В этой более нагретой области существует еще одно важное свидетельство возможного контакта двух плазменных каналов — это более ярко светящаяся область, по-видимому, ранее существовавшей стримерной зоны взаимодействия каналов в сквозной фазе (2). Событие, запечатленное на Рисунке 4.16 также говорит в пользу возможной составной структуры каналов, которые станут в результате эволюции отрицательными и положительными частями двунаправленного лидера. Это также говорит в пользу существования одной или нескольких сквозных фаз и квазиобратных ударов при формировании разрядов в электрическом поле положительно заряженного аэрозольного облака, как наблюдалось ранее в главе 3 для контакта каналов в электрическом поле отрицательного облака (Рисунки 3.6, 3.7).

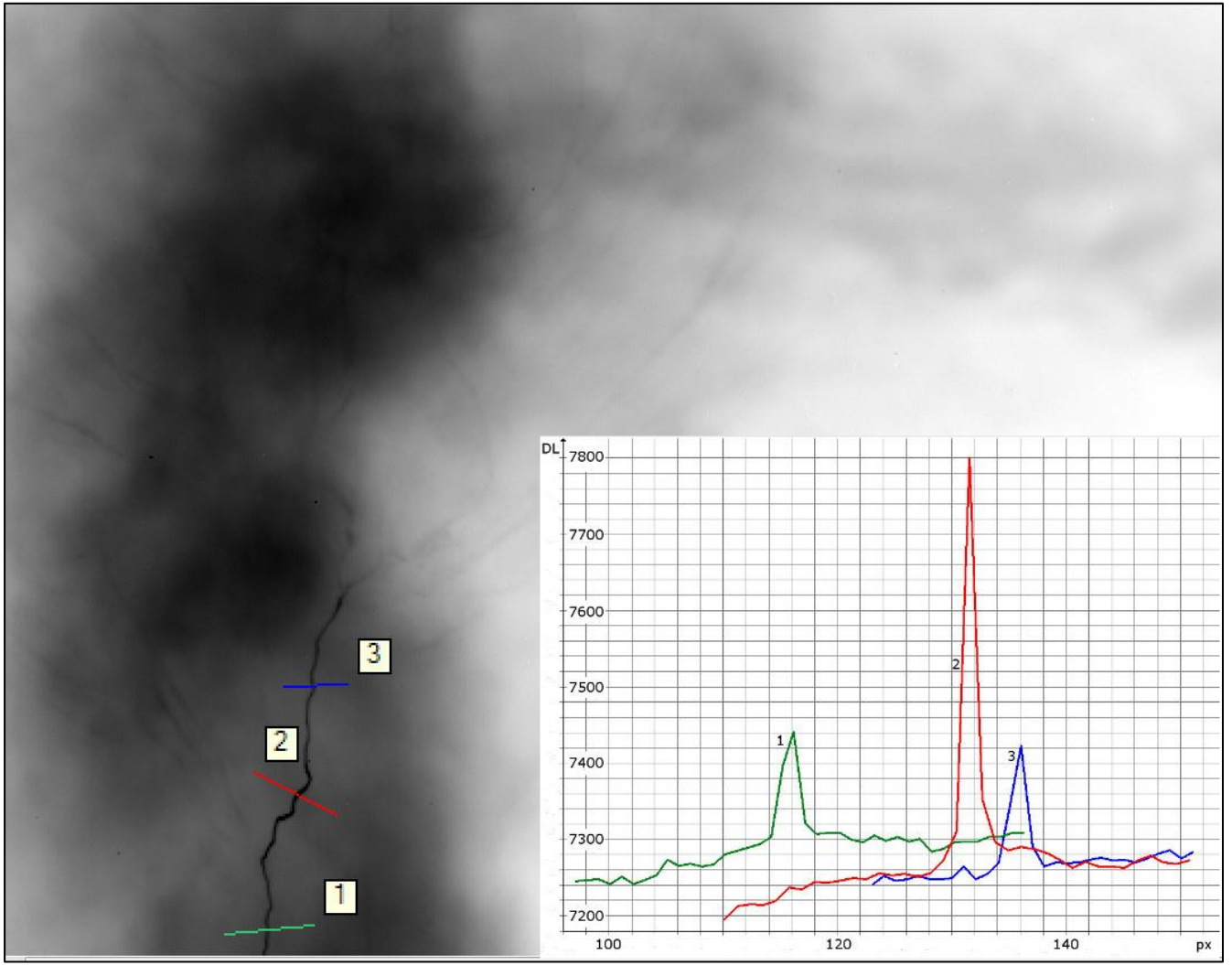

Рисунок 4.15 (событие 051-356). Верхняя часть разряда. Вдоль цветных линий 1,2,3, пересекающих плазменный канал, измеряется яркость излучения в относительных единицах (ось абсцисс). По оси ординат откладываются номера пикселей. Все события происходят глубоко внутри аэрозольного облака. Режим записи: частота кадров — 115 Гц, выдержка — 6.7 мс, число пикселей в исходном кадре — 640х512. Инфракрасное изображение инвертировано.

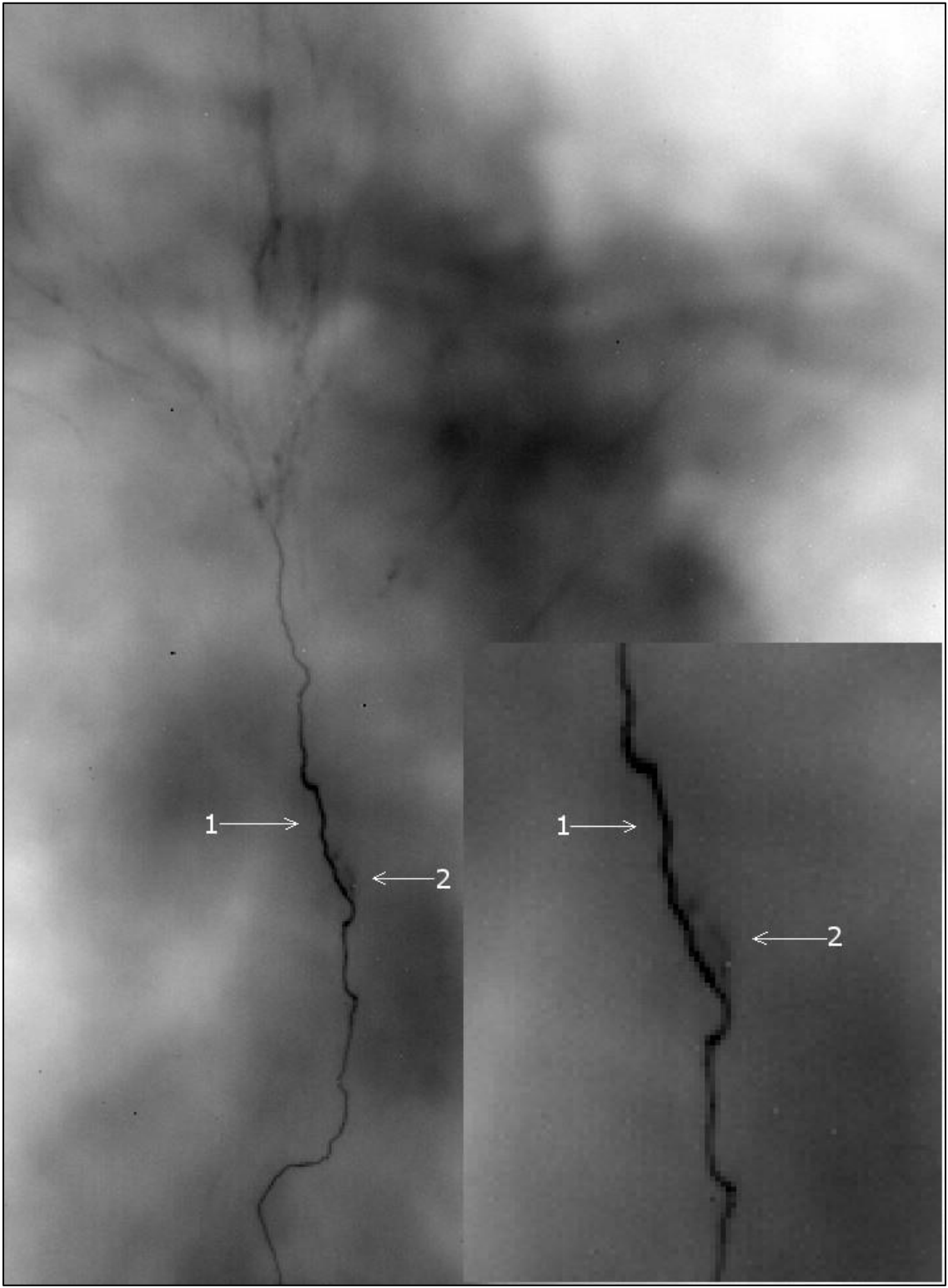

Рисунок 4.16 (событие 052-3016). Ярко светящаяся в ИК-диапазоне область в середине канала – 1. Область – 2, возможно является местом контакта плазменных структур, что видно по взаимодействующей стримерной короне сквозной фазы. Вверху кадра характерный веер каналов разной яркости. Режим записи: частота кадров — 115 Гц, выдержка — 6.7 мс, число пикселей в исходном кадре — 640x512 (на данном рисунке изображена только часть кадра с разрядом).

ИК-светимость вдоль основного канала (включая среднюю часть и нижнюю часть, восходящего лидера) двунаправленного канала может иметь не одну, а несколько областей со значительно повышенной яркостью. Наиболее яркий длинный канал на Рисунке 4.17 имеет, по крайней мере, три коротких (относительно общей длины канала) сегмента с гораздо большей интенсивностью ИК-излучения, чем области канала выше и ниже сегмента. ИК-яркие сегменты отмечены на Рисунке 4.17 белыми стрелками. Изображение на Рисунке 4.17 также говорит в пользу возможной составной структуры результирующего двунаправленного лидера. То есть, результирующий двунаправленный лидер, имеющий положительный и отрицательный лидер на своих концах, может формироваться в результате слияния и последующей поляризации нескольких плазменных каналов, а не расти от одного сегмента, как например, растет двунаправленный лидер от поверхности самолета [Lalande et al., 1998], что отличается от простой схемы двунаправленного лидера Каземира. То есть, более длинный двунаправленный лидер может не развиваться из одной точки, а быть каналом, составленным из ранее независимых, инициированных отдельно друг от друга каналов, которые позже объединились (слились) в один общий канал, который после объединения и конечной поляризации также будет выглядеть как двунаправленный лидер.

На увеличенном фрагменте Рисунка 4.17 (Рисунок 4.18) также отчетливо видно, что некоторые сложные внутриоблачные плазменные образования не связаны с основным каналом и имеют морфологическое строение, которое очень сложно описать в устоявшихся терминах стримеров и лидеров.

### 4.2.2.3. Структура внутриоблачных разрядов у основания облака, вблизи заземлённой плоскости и квазиобратные удары в электрическом поле положительно заряженного облака

Как указывалось выше, [Верещагин и др. 1988] выделили три основных типа разряда в поле положительно заряженного облака: стримерный разряд, идущий от облака к заземлённой плоскости (Рисунок 4.2(2 нижний)); лидерный, незавершенный разряд, идущий от положительно заряженного облака к заземлённой плоскости (Рисунки 4.1(4), 4.12(1)); завершенный разряд в канальной стадии, когда горячий плазменный канал, скорее всего касается заземленной плоскости (этот разряд в публикациях описан не

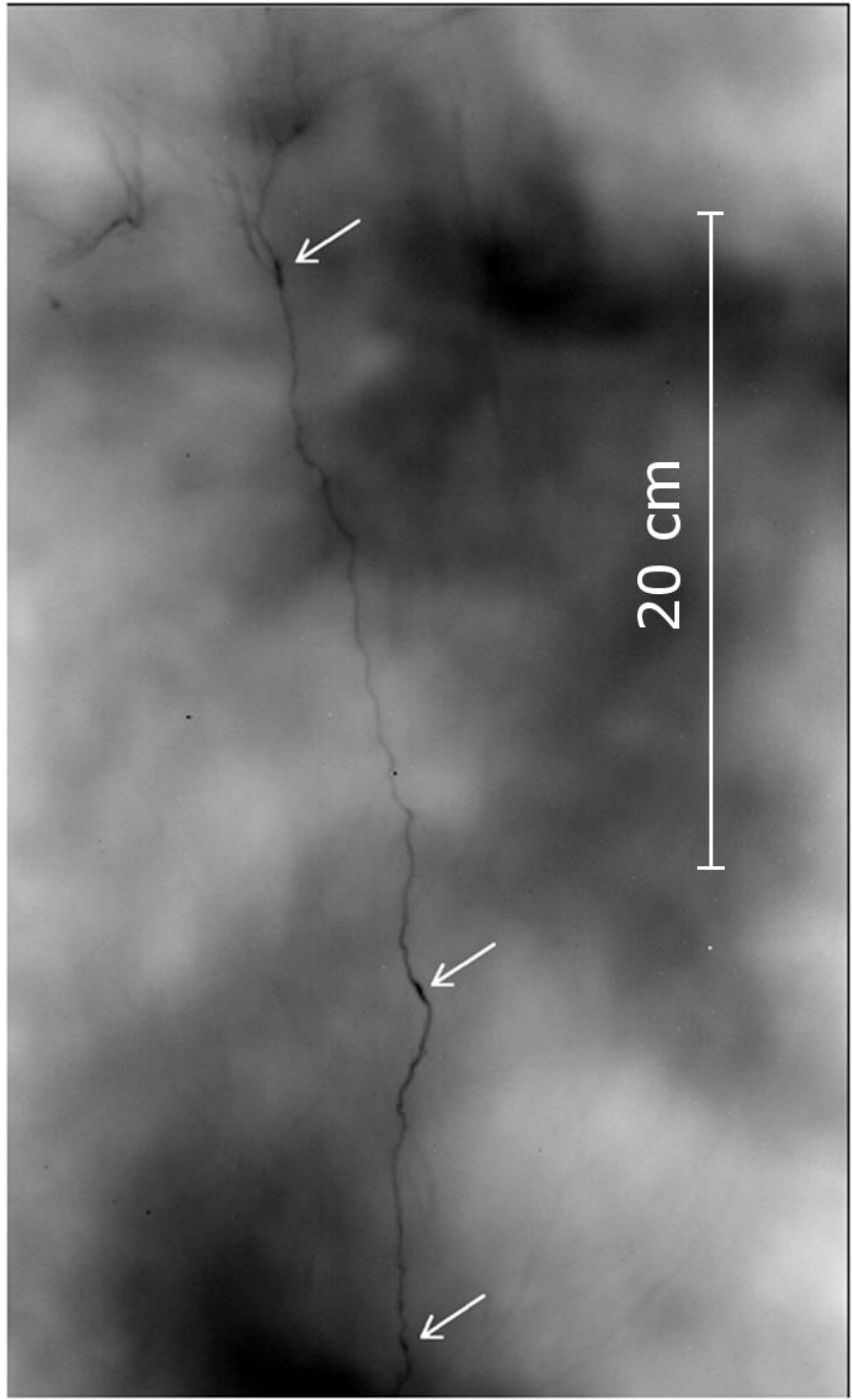

Рисунок 4.17 (адаптировано из [Kostinskiy et al., 2015b]), (событие 051-2836). Инфракрасное изображение (инвертированное), полученное с выдержкой 6,7 мс, показывает верхнюю часть двунаправленного лидера, инициированного положительно заряженным облаком. Три коротких участка с гораздо большей интенсивностью ИК-излучения вдоль основного канала отмечены белыми стрелками. В левом верхнем углу кадра виден одиночный плазменный канал (восходящий отрицательный лидер) переменной ИК-интенсивности, связанный с сетью плазменных каналов. Наблюдается также множество плазменных каналов по всему объему, которые могут быть и не связаны с основным каналом. Слева от основного канала есть независимое, относительно яркое плазменное образование (см. увеличенное изображение на Рисунке 4.18)

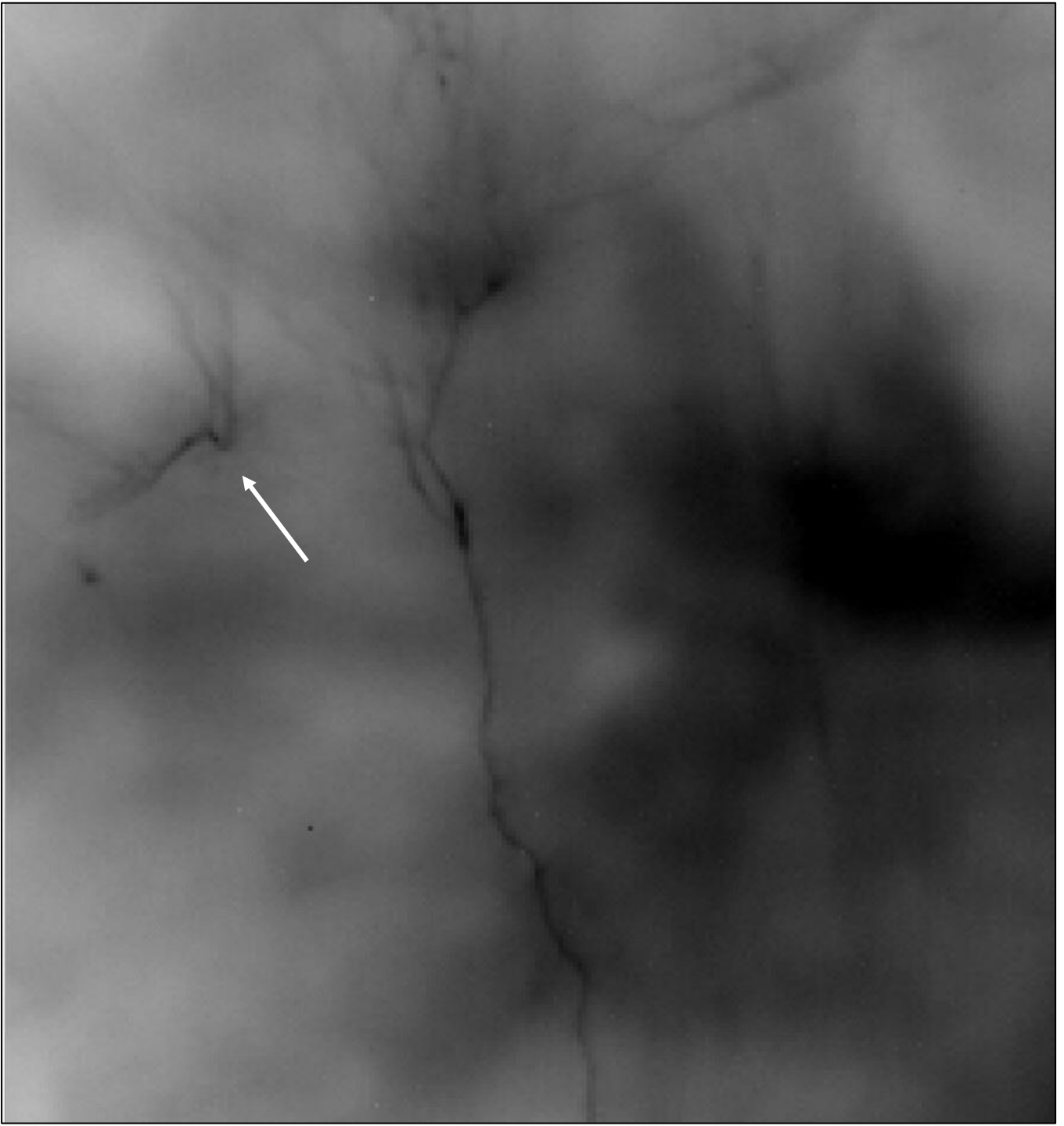

Рисунок 4.18 (адаптировано из [Kostinskiy et al., 2015b]), (увеличенный фрагмент Рисунка 4.17). Верхняя часть разряда. На увеличенном фрагменте хорошо видно, что некоторые сложные плазменные образования не связаны с основным каналом и имеют морфологическое строение, которое очень сложно описать в терминах стримеров и лидеров. Режим записи: частота кадров — 115 Гц, выдержка — 6.7 мс, число пикселей в исходном кадре — 640x512.

подробно и надежные фотографии этого явления, к сожалению, не приведены). «Завершенный тип разряда», по-видимому, должен включать в себя не только контакт с плоскостью, но и контакт двух или нескольких плазменных каналов, что приводит к квазиобратному удару и яркому свечению.

На Рисунке 4.19 хорошо видна структура канала, состоящего, как минимум из двух сегментов. Мы можем выделить верхнюю яркую часть канала (1), которая превосходит в среднем в 4 ÷ 5 раз ИК-яркость нижней части канала (3), а также изгиб канала (2), который похож на место встречи положительного и отрицательного лидеров в отрицательном облаке Рисунок 3.11. Также на Рисунке 4.19 изображен веер каналов (4), который виден не очень чётко из-за низкого контраста на фоне плотного аэрозоля. Неравномерность нагрева канала указывает, что, по крайней мере, верхняя часть канала (1) не является восходящим с заземлённой плоскости разрядом, иначе нагрев в нижней части канала был бы больше, чем в верхней, так как через него ток тек бы гораздо большее время. Также надо отметить, что разряды, формирующиеся около заземленной поверхности гораздо короче, чем двунаправленные лидеры, которые мы рассматривали в главе 3.

На Рисунке 4.20 можно видеть картину, аналогичную, зафиксированной на Рисунке 4.19, но еще более сложную. Средняя, очень яркая часть канала (1), превосходит в ИК-диапазоне в среднем в 5÷7 раз яркость нижней части канала (3), а также она в 2 раза превосходит верхнюю часть канала (2). В свою очередь излучение в ИК-диапазоне верхней части канала (2) превосходит в 2-3 раза излучение нижней части (3). Разветвленный веер более тонких каналов (4) поднимающихся с верхнего конца лидера хорошо виден на увеличенном фрагменте 4.20.II. Характерно, что на Рисунке 4.20 окончание верхней части разряда (5) не имеет головки, а имеет «острое» окончание (обычно головкой считают круглое или эллипсоидное окончание канала), в отличие, например, от Рисунка 4.14. И на Рисунке 4.20 резкая неоднородность нагрева значительных участков канала по длине не может быть объяснена только особенностями траектории, т.к. слишком велики различия в нагреве частей (1) и (3). Это также позволяет предположить, что, по крайней мере, средняя часть канала (1) не восходящий с плоскости разряд (он возникает, скорее всего, внутри облака), иначе нагрев в нижней части канала (3) не отличался бы настолько от средней части (1).

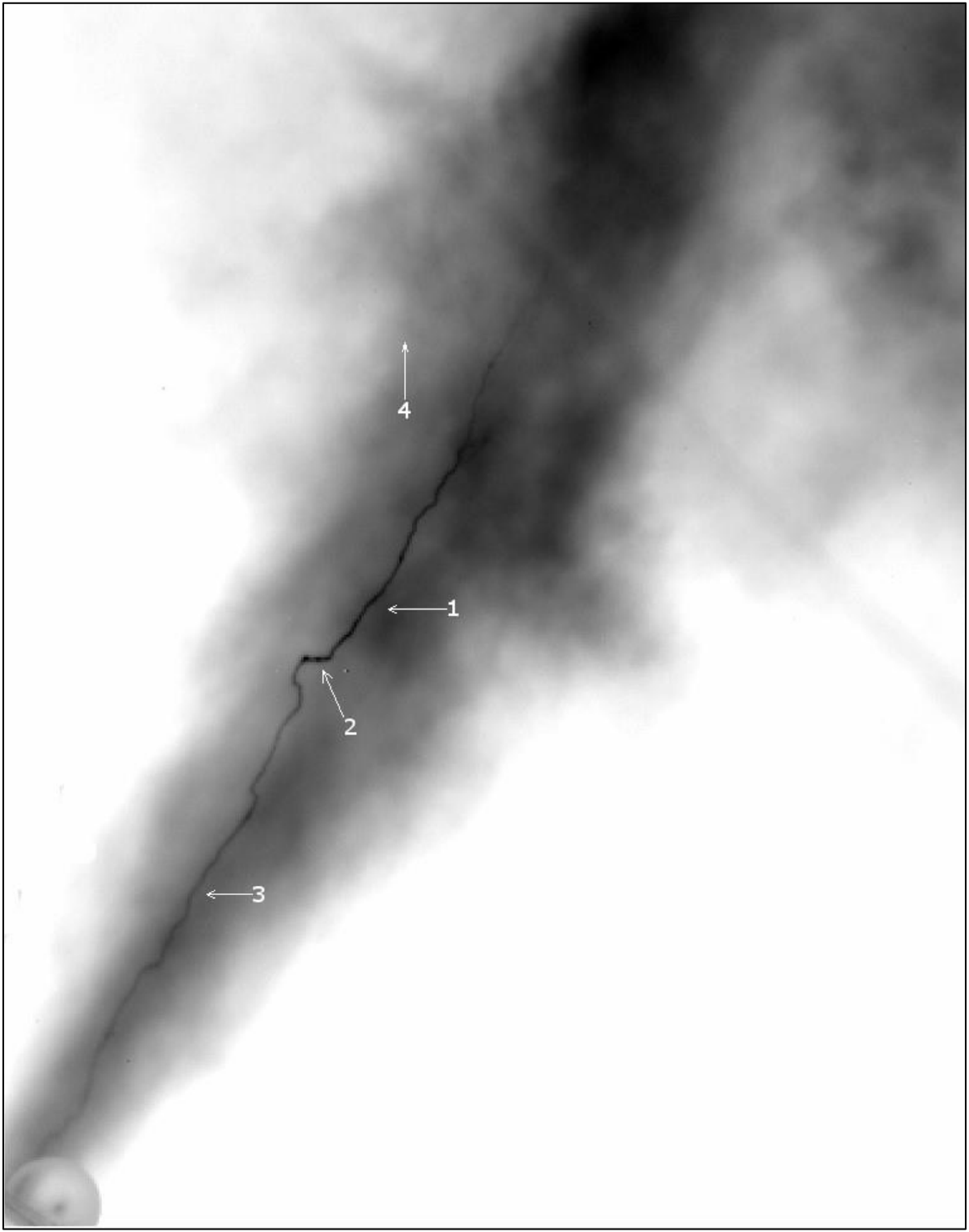

Рисунок 4.19 (событие 031-2968). Разряд вблизи заземлённой плоскости. Верхняя яркая часть канала – 1 (превосходит в среднем в 4-5 раз яркость нижней части канала – 3, перемычка – 2, веер расходящихся от головки отрицательного лидера каналов – 4. Диаметр сферы в левом нижнем углу – 5 см. Режим записи: частота кадров — 111 Гц, выдержка — 8.7 мс, число пикселей в исходном кадре — 640х512 (на данном рисунке изображена только часть кадра с разрядом).

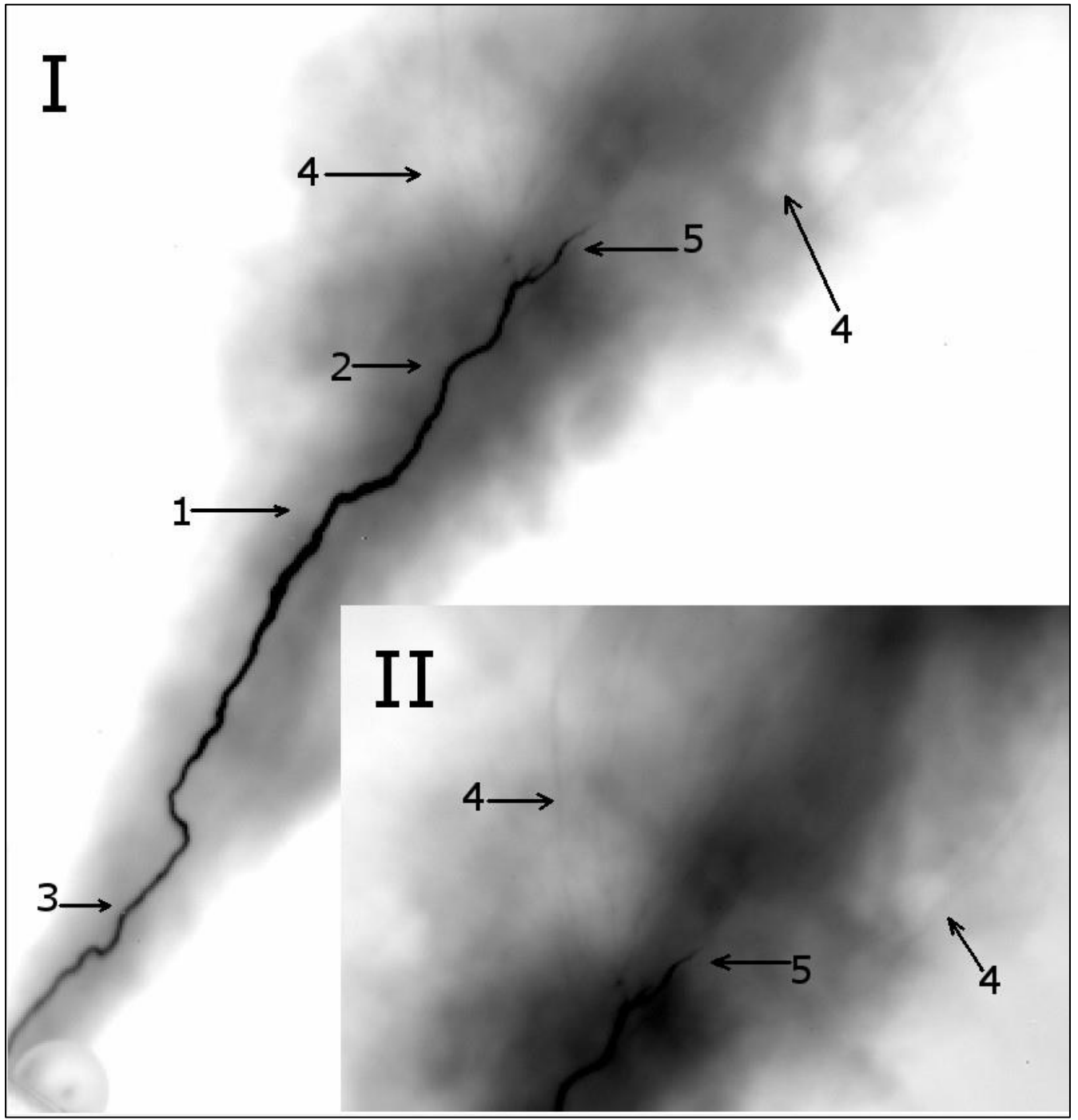

Рисунок 4.20 (событие 028-2281). Разряд вблизи заземлённой плоскости. Верхняя яркая часть канала – 1 (превосходит в среднем в 5-7 раз яркость нижней части канала – 3, и в 2-3 раза верхнюю часть канала – 2, веер каналов, распространяющихся вверх – 4, окончание канала не имеет «головки» – 5. Диаметр сферы в левом нижнем углу – 5 см. Режим записи: частота кадров — 111 Гц, выдержка — 8.7 мс, число пикселей в исходном кадре — 640x512.

На Рисунке 4.21 фиксируется ещё более сложная структура конечного канала. По уровню яркости в ИК-излучении можно выделить три структурных элемента разрядного канала. Здесь наиболее яркой является верхняя часть канала, Рисунок 4.21.I(6-5-8-4), II (10), IV(10),  в центре менее яркая область Рисунок 4.21.I (9-7), II (11), IV(11)) и внизу наиболее слабо светящаяся часть канала, Рисунок 4.21.I(1-3-2), II(12), IV(12). Части канала II, IV (10 и 11) в середине заметно ярче, чем по краям. На рис. Рисунок 4.21.IV можно заметить, что в местах стыка частей канала (13) и (14) яркость быстро изменяется примерно в два раза. Выше верхнего острого конца канала на рис. Рисунок 4.21.I различима сеть более тонких каналов. Головки в верхней части канала также нет. На Рисунке 4.21 резкая неоднородность свечения в ИК-области участков канала по длине также не может быть объяснена только особенностями распространения разряда, т.к. в местах стыков каналов (13) и (14) нет заметных изгибов канала.  Нижняя часть канала (12), как и на предыдущих Рисунках 4.19-4.20 значительно уступает в яркости верхним частям канала (10) и (11), что может говорить о том, что квазиобратный удар (с наибольшей плотностью энерговыделения) в поле положительного облака возникает между каналами, а не при контакте канала и плоскости.

На Рисунке 4.22 также можно выделить три структурных элемента разрядного канала. Здесь также, как и на Рисунке 4.21, наиболее яркой является верхняя часть канала (I ( 5-6-7-8), II (12)),  в центре менее яркая область (I(2-3-4), II(11)) и внизу наиболее слабо светящаяся часть канала (I(1), II(10)). Отчётливо яркую область можно выделить только внутри наиболее яркой части канала (12), причём, самая яркая часть канала (5) может относиться либо к перемычке (месту контакта каналов), либо к резкому изменению направления распространения канала. Выше верхнего конца канала (8) фиксируется разветвленная сеть более тонких каналов (9). Головки в верхней части разряда нет.

Рисунок 4.23 отличается от Рисунков 4.21-4.22, так как на Рисунке 4.23 хорошо видны два структурных элемента (7) и (8) разрядного канала, у которых центральные части гораздо ярче в ИК-диапазоне, чем области выше и ниже этих сегментов. Максимум структурного элемента II(7) находится в точке I(2), а максимум структурного элемента II(8) находится в районе точек I (4, 5). Характерно, что даже при такой малой общей длине канала (около 25 см) и близости его к заземлённой плоскости, основание канала I(1) светится почти в три раза слабее, чем область выше него I(2). Над разрядом видна широкая разветвлённая сеть каналов (9), которые заметно длиннее, чем

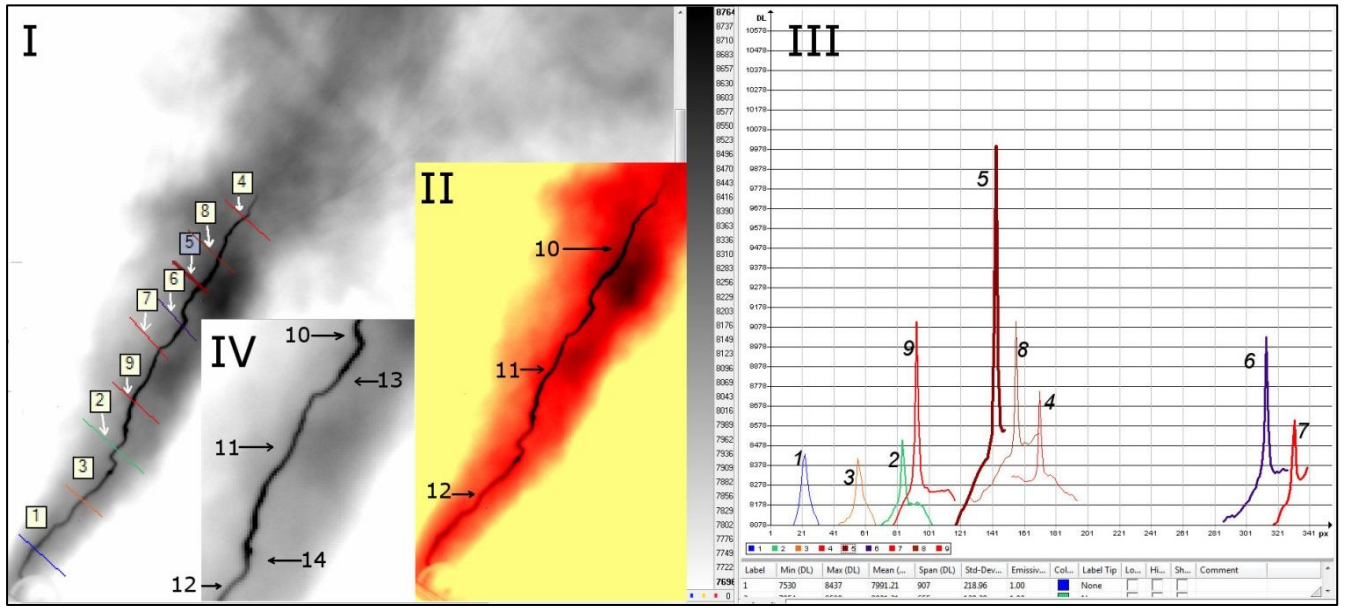

Рисунок 4.21 (событие 026-1324). Разряд вблизи заземлённой плоскости. 1 ÷ 9 – линии, вдоль которых измеряется яркость излучения в ИК-области (в относительных единицах). 10, 11, 12 – области канала с различной яркостью. 13, 14 – области, где яркость меняется скачком. Режим записи: частота кадров — 111 Гц, выдержка — 8.7 мс, число пикселей в исходном кадре — 640x512.

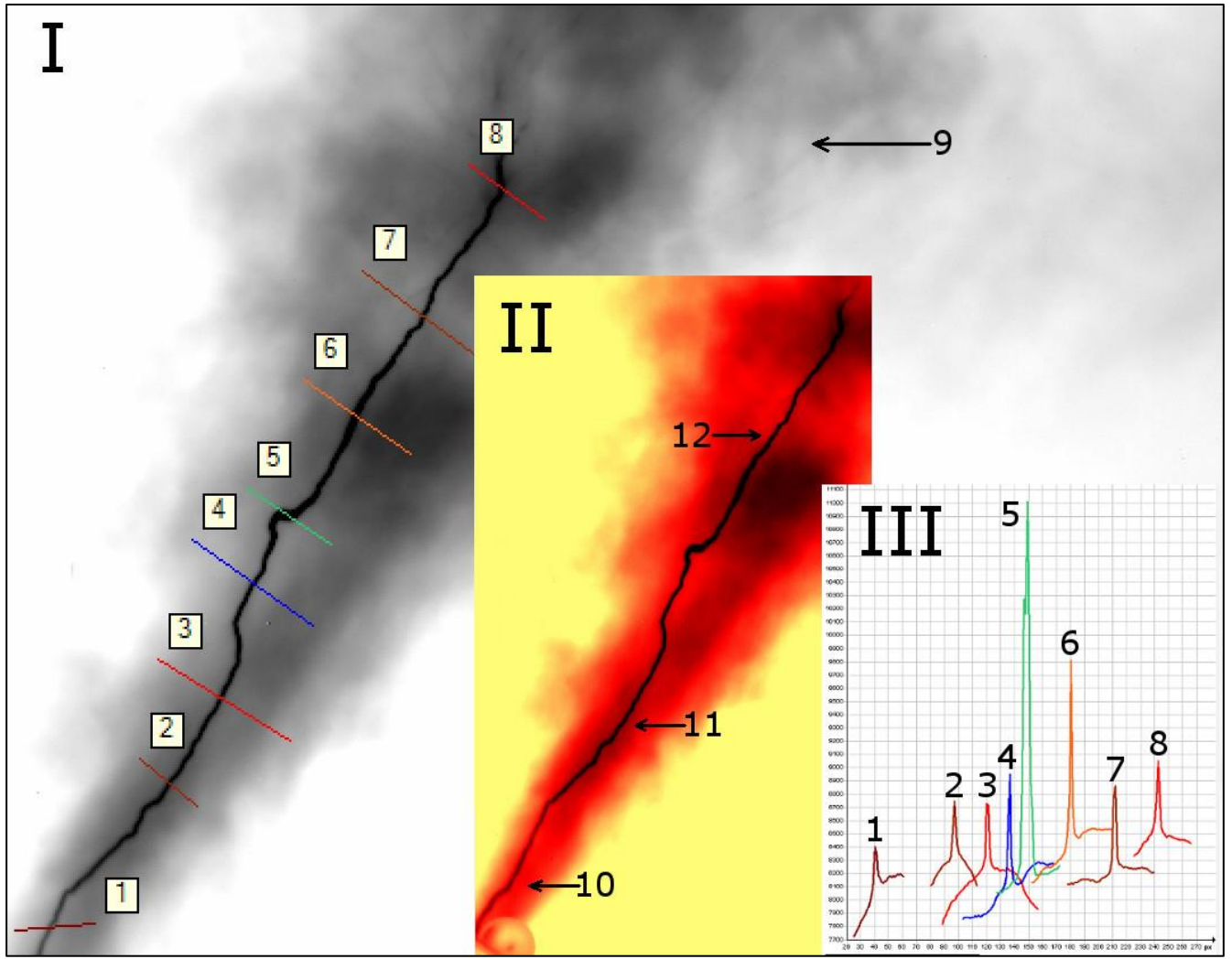

Рисунок 4.22 (событие 027-443). Разряд вблизи заземлённой плоскости. 1 ÷ 8 – линии, вдоль которых измеряется яркость излучения в ИК-области (в относительных единицах). 9 – веер каналов, в верхней части канала; 10, 11, 12 – области канала с различной яркостью. Диаметр сферы в левом нижнем углу – 5 см. Режим записи: частота кадров — 111 Гц, выдержка — 8.7 мс, число пикселей в исходном кадре — 640x512 (на данном рисунке изображена только часть кадра с разрядом).

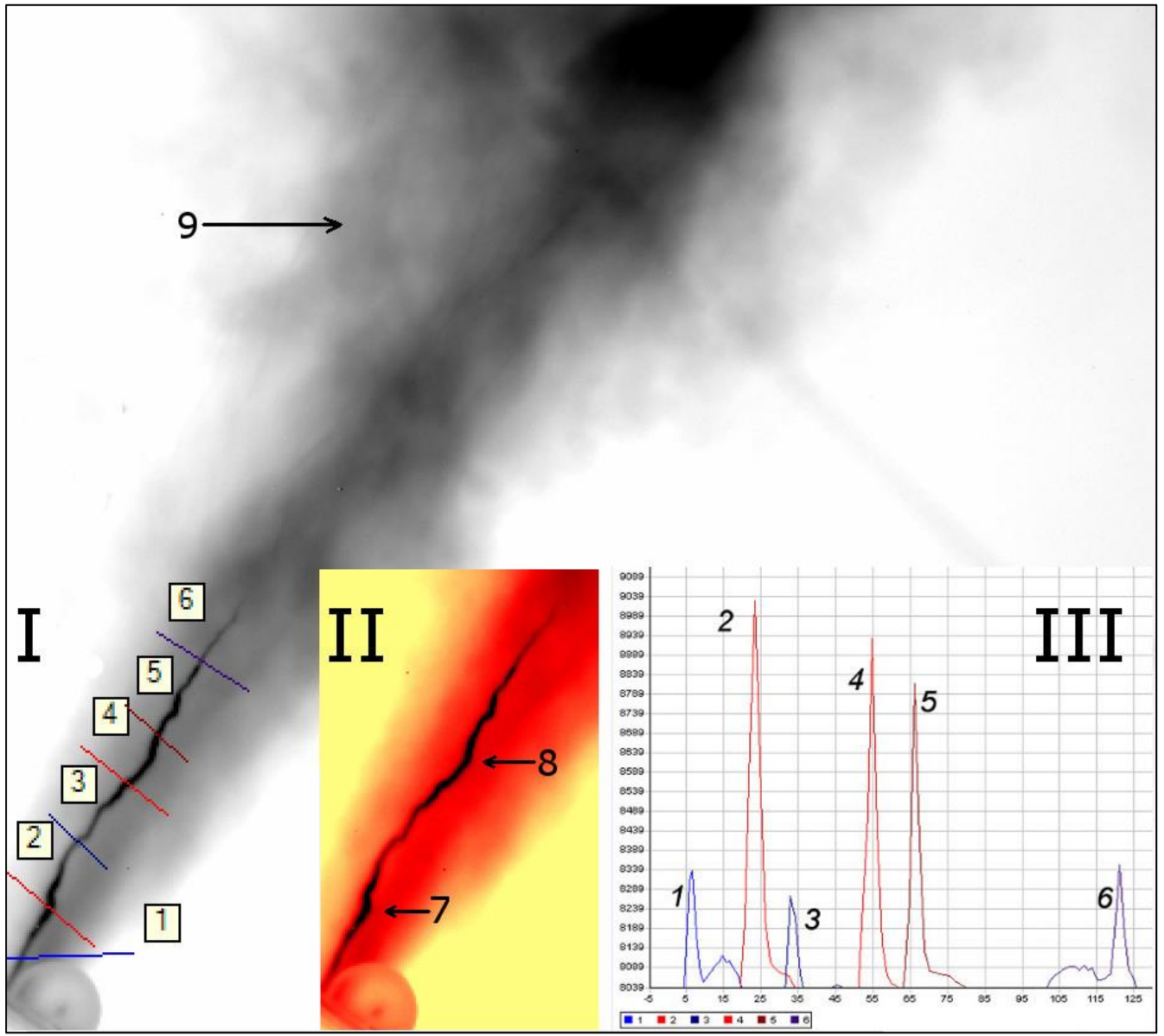

Рисунок 4.23 (событие 027-3165). Разряд вблизи заземлённой плоскости. 1 ÷ 6 – линии, вдоль которых измеряется яркость излучения в ИК-области (в относительных единицах). 7,8 – области канала с максимальной яркостью, 9 – веер каналов, в верхней части канала. Диаметр сферы в левом нижнем углу — 5 см. Режим записи: частота кадров — 111 Гц, выдержка — 8.7 мс, число пикселей в исходном кадре — 640x512 (на данном рисунке изображена только часть кадра с разрядом).

сам основной разряд. Весьма вероятно, что в данном случае на Рисунке 4.23 мы фиксируем взаимодействие двух плазменных структур, образовавшихся в объёме недалеко от заземлённой плоскости.

### 4.2.2.4. Наблюдения разрядов, инициированных положительно заряженным облаком в видимом диапазоне

До последнего времени, исследования разрядов в поле положительного облака в видимом диапазоне производилось в основном с помощью интегральных фотографий (например, Рисунки 4.1, 4.2, 4.12 из [Верещагин и др. 1988]). Также были сделаны попытки получить фотохронограммы с разверткой изображения, однако проблемы с синхронизацией фоторегистратора с процессами развития разряда привели к тому, что не были опубликованы полученные фотохронограммы разрядов в поле положительного облака [Верещагин и др., 1988], [Анцупов и др., 1990].

Мы попытались получить изображения разрядов в электрическом поле положительного облака с помощью камеры 4Picos. Синхронизация с разрядом производилась благодаря импульсу света от разрядов, фиксируемых фотоэлектронным умножителем (ФЭУ).

На Рисунке 4.24 зафиксирована стримерная вспышка, которая направлена от положительного облака к заземленной плоскости. Выдержка кадра равна 2 мкс. Это поведение соответствует описанию и фотографиям стримерных вспышек, зафиксированных [Верещагин и др., 1988], [Анцупов и др., 1990], например, фотографии стримерной вспышки на Рисунке 4.2.

На Рисунке 4.25 также зафиксирована стримерная вспышка, которая направлена от положительного облака к заземленной плоскости. Выдержка кадра равна 2 мкс. Это стримерная вспышка также аналогична, как поведению стримерных вспышек на фотографиях зафиксированных [Верещагин и др., 1988], [Анцупов и др., 1990], так и нашим измерениям в ИК-диапазоне, например, Рисунок 4.7.III. Важным отличием является то, что на Рисунках 4.25, 4.26 свечение стримерной вспышки около заземленной плоскости очень слабо. Создается впечатление, что большая часть стримеров стримерной

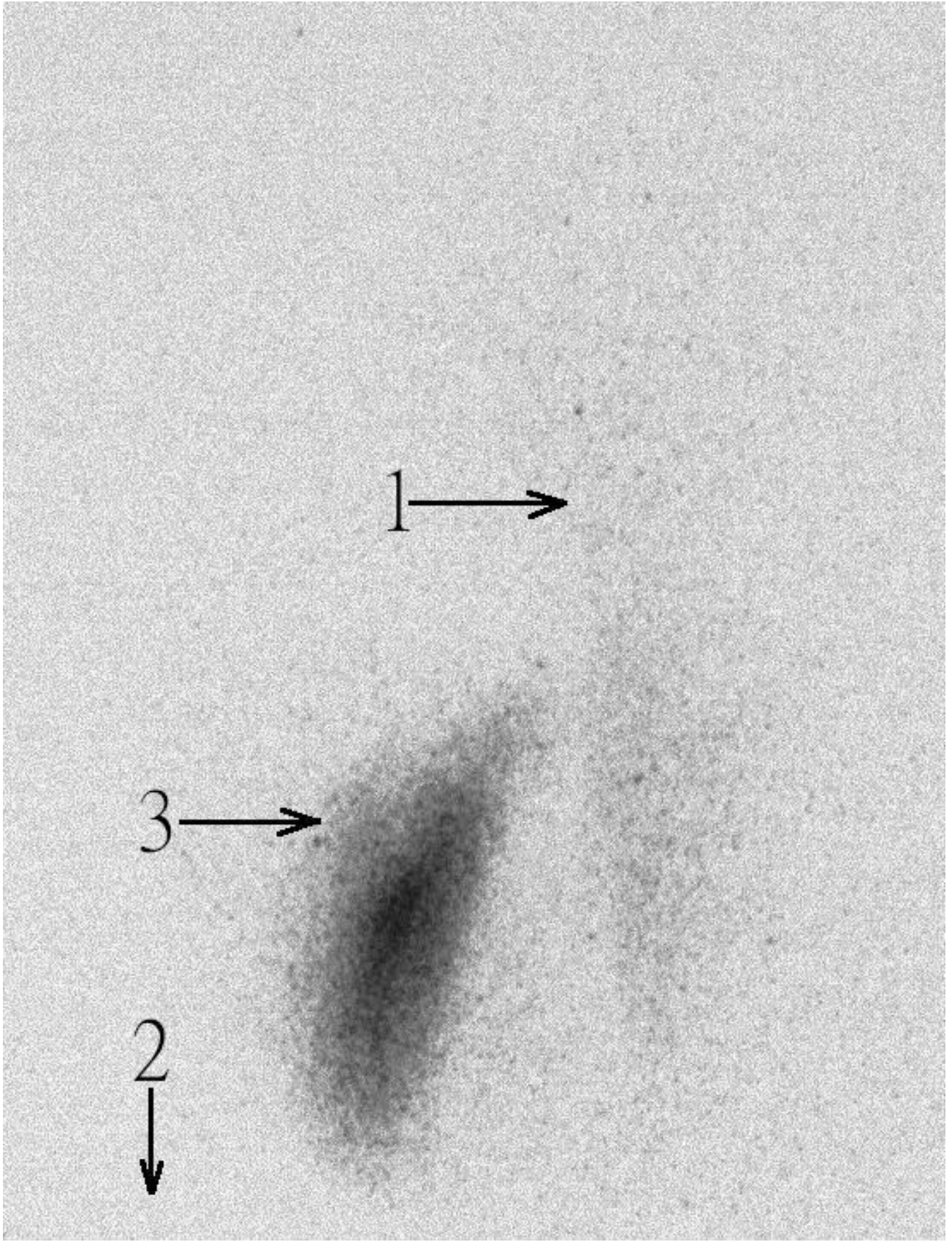

Рисунок 4.24 (событие 01_2015-10-25). Стримерная вспышка (3), идущая от облака (1) к заземленной плоскости (2). Выдержка кадра — 2 мкс. Изображение инвертировано.

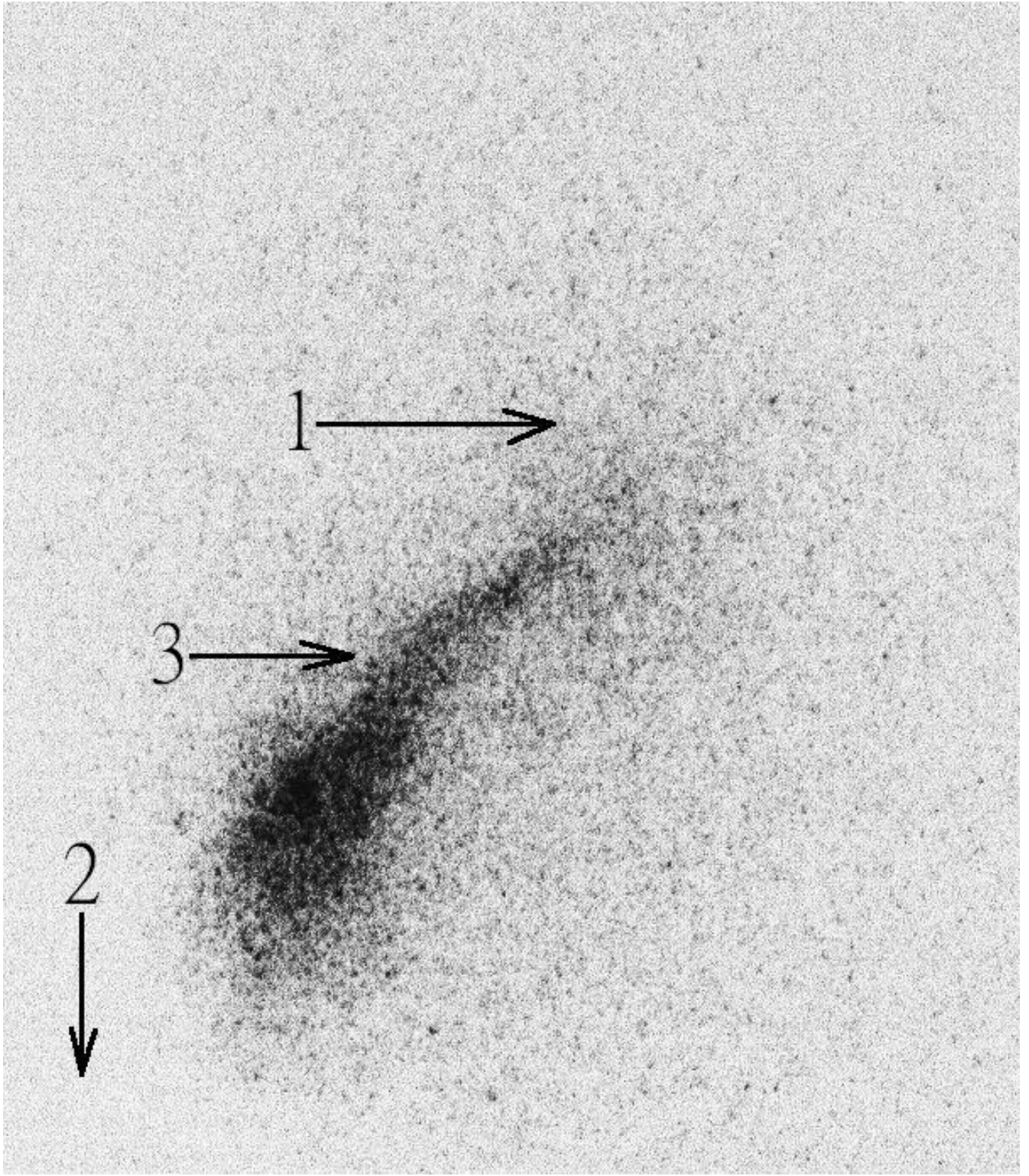

Рисунок 4.25 (событие 02-1_2015-10-25). Стримерная вспышка (3), идущая от облака (1) к заземленной плоскости (2). Выдержка кадра — 2 мкс. Изображение инвертировано.

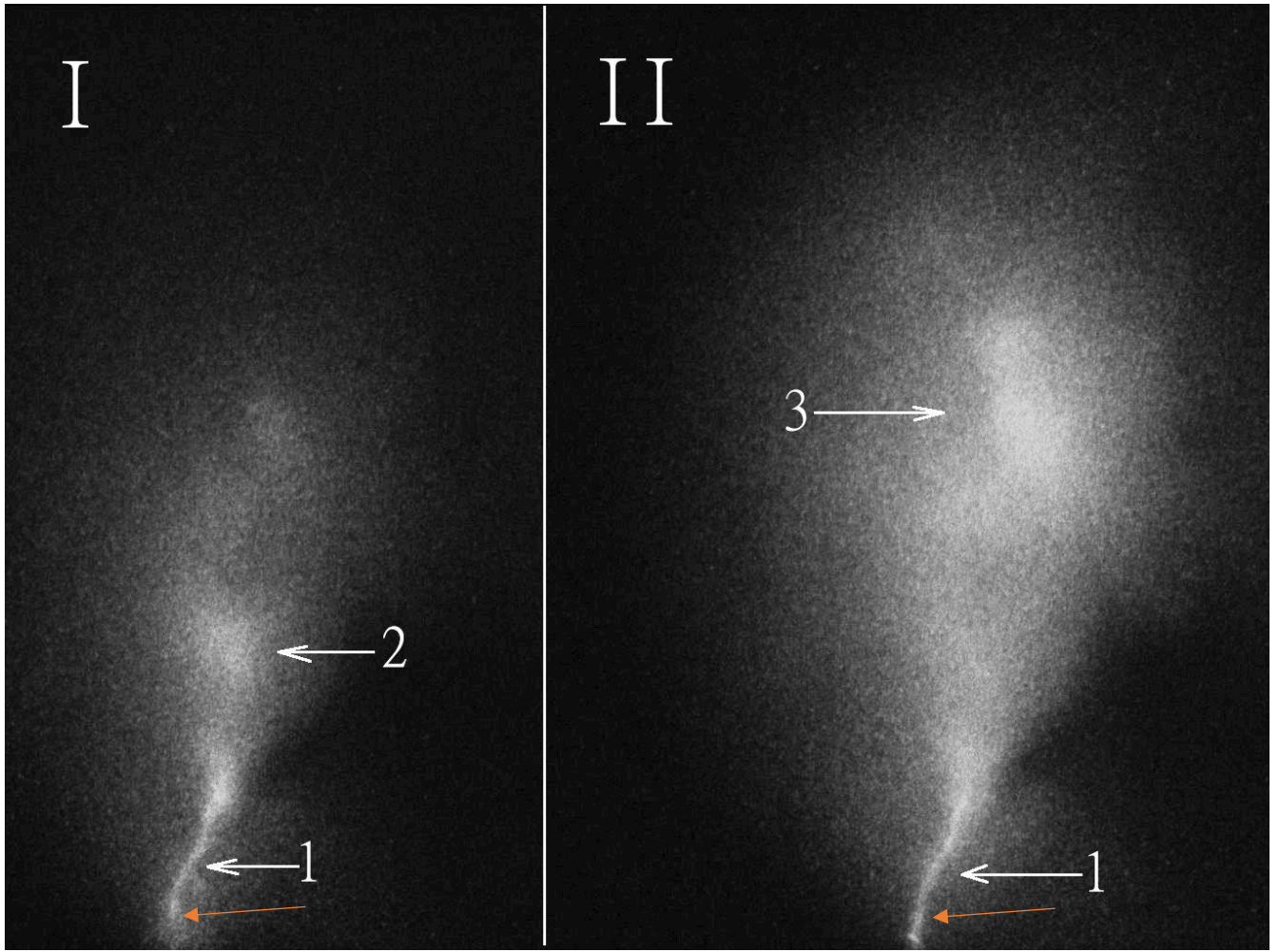

Рисунок 4.26 (событие 06_2015-10-25). Плазменный канал (1), внутриоблачная вспышка (2), внутриоблачная вспышка (3). Выдержка кадров — 2 мкс, между кадрами — 1 мкс. Коричневая стрелка показывает одно и тоже место на кадре.

вспышки может даже не доходить до поверхности. Выдержка кадра равна 2 мкс достаточно большая, чтобы стримерная вспышка полностью сформировалась и достигла поверхности, так как скорости стримерной вспышки обычно близка к 100 см/мкс ($10^6$ м/с), а расстояние от облака до плоскости не превышает 50-70 см.

На Рисунке 4.26 мы видим развитие канала (1) около оси струи. Изображение канала и плазменных образований внутри облака размыты из-за рассеяния света на каплях. Скорее всего по более яркому свечению видно, что внутри облака также развиваются плазменные образования (2 и 3), но они дают только общую засветку облака. Любопытно, что, видимо, канал (1) распространяется вниз, так как его нижний конец заметно продвинулся вниз на кадре II (коричневые стрелки). Также, исходя из анализа свечения облака можно сделать предварительный вывод, что скорее всего канал (или каналы) распространяются внутри положительного облака, так как его засветка за 3 мкс значительно увеличилась и распространилась вверх. Однако, также видно насколько более информативны ИК-кадры, которые дают гораздо больше возможностей исследовать тонкую структуру внутриоблачных разрядов.

На Рисунке 4.27.I мы видим засветку от квазиобратного удара, когда из-за рассеянного света на видимом изображении невозможно выделить ни один элемент разряда, так как разряд происходит около оси аэрозольной струи и во многом проходит внутри облака. На Рисунке 4.27.II плазма распадается и можно видеть размытое изображение плазменного канала (1). Поэтому разумно предположить, что квазиобратный удар также является встречей двух плазменных каналов, которые можно было видеть на ИК-изображениях.

Таким образом, изображения в видимом свете подтверждают существование ранее обнаруженных типов разрядов, с помощью фотографий и ИК-изображений, но с их помощью нам пока не удалось получить какую-нибудь значимую новую информацию.

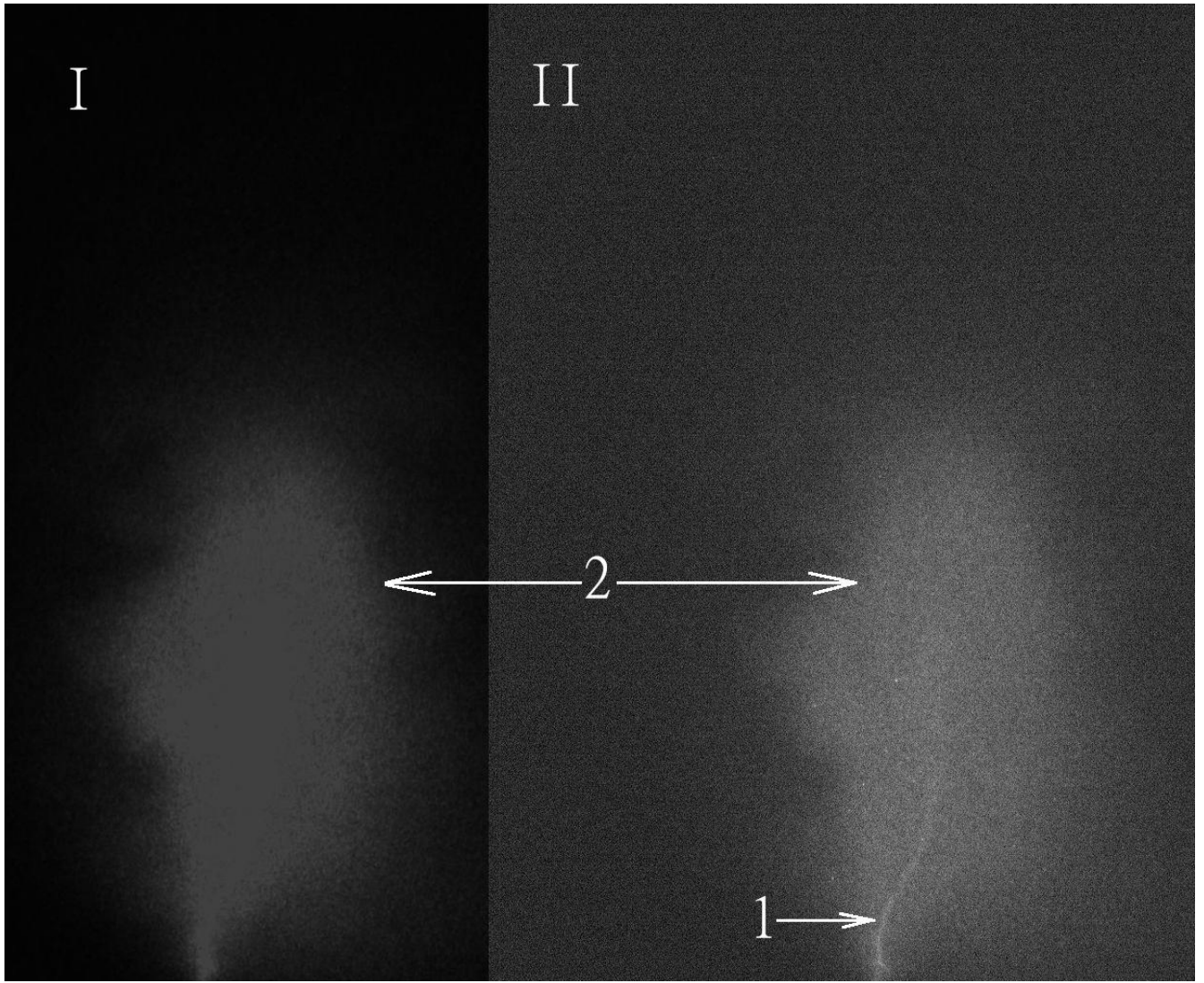

Рисунок 4.27 (событие 20_2015-10-25). Плазменный канал (1), внутриоблачная вспышка, засвечивающая большую часть облака (2). Выдержка кадров — 1 мкс, между кадрами — 2 мкс. Изображение на Рисунке II имеет гораздо более сильное повышение контрастности изображения, чтобы его можно было увидеть на фоне изображения I.

## 4.3. Обсуждение результатов, полученных в главе 4

В данном разделе описаны разнообразные плазменные образования, обнаруженные внутри положительно заряженного аэрозольного облака, включая новые, до этого момента никогда не наблюдаемые. Впервые об этом классе разрядов внутри заряженного аэрозоля было сообщено в статье [Kostinskiy et al., 2015b]. Некоторые из обнаруженных плазменных образований похожи на двунаправленные (bidirectional) лидеры, предсказанные Каземиром [Kasemir, 1960] и результаты их наблюдений подробно обсуждаются ниже. Другие плазменные образования образуют сложные сетевые структуры, которые нельзя отнести к известным по опыту изучения длинной искры и молнии плазменным образованиям, таким как стримерные вспышки и лидеры (положительные и отрицательные). Их мы, как и в главах 1, 2, мы будем называть необычными плазменными формированиями (UPFs). Также, как и в отрицательном аэрозольном облаке, в положительно запряженном облаке UPFs не являются редким явлением, а, наоборот, сети плазменных каналов внутри положительно заряженного облака гораздо гуще и обширнее. При использовании ИК-методики наблюдений, они обнаруживаются практически при каждом внутриоблачном аэрозольном разряде положительной полярности. Их можно разделить в первом приближении на два класса: «центральные UPFs», из которых в обе стороны (преимущественно вверх и вниз) уходят каналы, напоминающие положительные и отрицательные лидеры, а также «стримероподобные» UPFs, которые занимают своей сетью значительные объёмы пространства. Однако существуют и другие формы плазменных образований, морфологию которых описать сложно в принятых терминах, например, такие, как на Рисунке 4.10(3) или на Рисунке 4.18(белая стрелка).

Плазменное образование с центральным UPF в середине (например, Рисунок 4.7(1)), некоторые элементы которого мы будем из-за морфологического сходства описывать в терминах двунаправленного лидера, отличается от классической модели двунаправленного лидера, предложенного Каземиром [Kasemir, 1960] и исследованного в связи с изучением молниевых разрядов, инициируемых летательными аппаратами [Lalande et al, 1998]. Действительно, в этом комплексном плазменном образовании можно выделить каналы, похожие на нисходящий положительный лидер и восходящий отрицательный, вписывающиеся в модель двунаправленного лидера, но кроме них, из того же центрального UPF в том же месте, где стартует положительный нисходящий лидер, практически всегда обнаруживаются нисходящие положительные каналы, искривлённые по силовым линиям электрического поля, подобные положительной стримерной короне (см., например, Рисунки 4.7 (3), 4.8(2), 4.9(2). Но данные каналы всегда являются гораздо более яркими в ИК-диапазоне, чем положительные стримеры, причём, что важно, часто их яркость не уступает или даже превышает яркость горячих положительных лидеров (см., например, Рисунок 4.8(2)), что не позволяет отнести их ни к положительной стримерной короне, ни к положительным лидерным каналам (положительный лидер, Рисунок 4.8(3), радикально отличается по форме от этих дугообразных каналов, Рисунок 4.8(2), что хорошо видно, когда они движутся в одном поле зрения). Поэтому мы также предлагаем считать эти дугообразные плазменные каналы необычными плазменными образованиями (UPFs), плазма которых близка по параметрам к плазме лидеров.

С другой стороны, если предположить, что центральный UPF и/или дугообразный UPF создают первичный плазменный канал, из которого разовьются положительный нисходящий и отрицательный восходящий лидер, то представление о двунаправленном лидере может оказаться плодотворным для анализа развития плазменных каналов. Если описывать плазменное образование в терминах классического двунаправленного лидера, то двунаправленный лидер состоит из идущей вниз положительной части, идущей вверх отрицательной части, эти две части (обе разветвленные, хотя и по-разному) соединены средней частью неоднородной интенсивности (центральный и/или дугообразные UPFs).

Один из плазменных сегментов мы можем определить с высокой вероятностью из-за высокого морфологического сходства с восходящим в электрическом поле отрицательного облака положительным лидером. Нисходящая часть плазменного образования включала извилистый положительный лидерный канал (например, Рисунки 4.7(4), 4.8(3), 4.10(5)), аналогичный его восходящему аналогу, наблюдаемому при отрицательной полярности облака (Рисунок 3.15, 3.18). Выше мы отметили, что практически всегда, нисходящий положительный лидер сопровождался гораздо менее извилистыми, но часто столь же или более яркими, идущими вниз плазменными образованиями, которые ранее не наблюдались при изучении длинной искры и молнии. Мы предположили, что эти необычные каналы имеют ту же природу, что и необычные плазменные образования (UPFs), которые мы обнаружили внутри отрицательно заряженного облака (глава 1). Также наблюдалась слабая положительная стримерная зона, направленная вниз. Любопытно, что нисходящий положительный лидерный канал, либо необычные плазменные образования (UPFs) могут соприкасаться с заземленной плоскостью, не вызывая там выраженного квазиобратного удара. Но, как отмечалось в [Верещагин и др, 1988], [Анцупов и др., 1990], квазиобратные удары фиксировались в разрядах, инициированных положительно заряженным аэрозольным облаком и нам также удалось зафиксировать подобные квазиобратные удары в разрядах около оси аэрозольного потока, как в ИК-диапазоне, так и в видимом диапазоне (см. обсуждение ниже).

## 4.4. Выводы главы 4

Благодаря использованным новым методикам ИК-измерений впервые удалось исследовать верхнюю, внутриоблачную, часть плазменных образований, инициированных электрическим полем положительно заряженного облака, Рисунки 4.13-4.18 [Kostinskiy et al., 2015b]. Кроме основного восходящего отрицательного плазменного канала, каждому разряду сопутствует разветвлённая сеть более тонких плазменных каналов (и сетей каналов) разной формы, яркости и длины, принизывающая почти весь наблюдаемый в ИК-диапазоне объём облака. Эти каналы могут развиваться параллельно отрицательному лидерному каналу или почти перпендикулярно ему. Некоторые из слабых каналов могут быть интерпретированы как необычно длинные и яркие отрицательные стримеры, в то время как другие (особенно показывающие сложную морфологию и те, которые явно не связаны с восходящим отрицательным лидером) могут быть похожи на UPFs.

Природа UPFs в положительном облаке в настоящее время только выясняется, но уже сейчас мы можем сказать, что:

(1) UPFs обнаруживаются в облаках заряженного водного аэрозоля положительной и отрицательной полярности,

(2) UPFs могут принимать разные формы, резко отличающиеся от известных форм стримеров и лидеров длинной искры и молнии,

(3) по крайней мере некоторые их части настолько горячи, насколько горячи лидерные каналы, и

(4) их морфология заметно отличается от стримерных вспышек и лидеров. Вероятно, направленные вниз дугообразные UPFs и нисходящие положительные лидеры произошли (или, по крайней мере, инициировались) в разное время в течение 6.7-мс экспозиции кадра, хотя они, в принципе, могут конкурировать с положительными лидерами в процессе контакта с заземленной плоскостью.

ИК-светимость вдоль самой яркой части двунаправленного лидерного, в основном состоящей из канала отрицательного лидера часто неоднородна. На Рисунках 4.15-4.17 вдоль траектории основного канала наблюдались сегменты гораздо более высокой яркости (превышение в 2-7 раз) по сравнению с лежащими рядом сегментами каналов. Эти сегменты могут говорить о взаимодействии двух и более плазменных каналов в процессе внутриоблачных квазиобратных ударов. Если это предположегние верно, то результирующий двунаправленный лидер может быть результатом взаимодействия нескольких плазменных элементов (возможно, аналогов спейс-лидеров) до момента формирования объединенного классического двунаправленного лидера. То есть, результирующий двунаправленный лидер, имеющий положительный и отрицательный лидер на своих концах, может формироваться в результате слияния и последующей поляризации нескольких плазменных каналов (или более сложных образований), а не расти от одного начального сегмента, как например, растет двунаправленный лидер от поверхности самолета [Lalande et al., 1998].

Основная цель этого исследования состояла в том, чтобы впервые описать все плазменные образования внутри положительного облака, включая части двунаправленных лидеров (и их тонкую структуру), инициированные без каких-либо проводников. Раньше при исследовании молнии с помощью VHF-методик надежно и почти непрерывно во времени фиксировалось развитие только одного сегмента – отрицательной части двунаправленного лидера (ступенчатого отрицательного лидера), например, [Rison et al., 2016], а оптические методы позволяли наблюдать только одну часть (любой полярности), выходящую из облака [Stolzenburg et al., 2013, 2014]. Кроме того, мы неожиданно обнаружили необычные плазменные образования, некоторые из которых, казалось, конкурировали с нисходящим положительным лидером за установление контакта с заземленной плоскостью, в то время как другие образовывали сложные сети каналов, очевидно, пронизывающих всю верхнюю часть облака. Очевидно, что для полной интерпретации наблюдаемых свойств двунаправленных лидеров необходимы дальнейшие исследования.

Нам также удалось проверить предположение [Верещагин и др. 1988], о том, что электрическое поле облака положительно заряженного аэрозоля может инициировать квазиобратные удары с высоким уровнем оптического сигнала. Рисунки 4.19-4.22 с помощью ИК-измерений фиксируют у каналов с общей длиной 25-80 см несколько хорошо различаемых по яркости излучения отдельных сегментов. Это говорит о том, что в этих местах было сильное энерговыделение, которое может соответствовать контактам плазменных каналов, подобное зафиксированным квазиобратным ударам в поле отрицательно заряженного аэрозольного облака. Существенным отличием таких каналов, созданных в поле положительного облака, является то, что мы не обнаруживаем сильного энерговыделения при контакте этих каналов с заземленной плоскостью. Наблюдения за разрядами в видимом диапазоне также подтвердило наличие квазиобратных ударов у каналов, находящихся около заземленной плоскости вблизи оси аэрозольной струи (например, Рисунок 4.27). При квазиобратном ударе была засвечена рассеянным на каплях светом практически все аэрозольное облако Рисунок (4.27.I), а через 3 мкс можно было обнаружить распадающийся плазменный канал, формирование которого и вызвало такую мощную вспышку света.

Полученные в данном разделе результаты могут быть полезны также для понимания природных внутриоблачных разрядов в грозовых облаках. Они позволяют предположить, что структура внутриоблачных разрядов в грозовом облаке может быть гораздо сложнее той, что обнаружена в многочисленных экспериментах в радиодиапазоне [Rison et al., 2016], [Hare et al., 2019], [Shao et al., 2020] где, по-видимому, фиксировались только яркие плазменные образования (лидеры) и, возможно, некоторые самые мощные более тонкие внутриоблачные каналы, которые взаимодействуют с основным лидерным каналом [Hare et al., 2019].

**ГЛАВА 5. Моделирование в лабораторных экспериментах аналогов высотно-инициированных триггерных молний (altitude-triggered lightning) и «классических» триггерных молний в электрическом поле отрицательно и положительно заряженного облака заряженного водного аэрозоля**

В данной главе представлены результаты [Kostinskiy et al., 2015c], где впервые продемонстрирована возможность моделирования в лабораторных условиях высотно-инициированных триггерных молний (altitude-triggered lightning) и триггерных молний с помощью инициации болтом арбалета (моделирующего летательный аппарат) разрядов, в электрическом поле облака отрицательно заряженного водного аэрозоля. Зафиксировано более сотни подобных событий. В рамках каждого события одновременно измерены: ток, текущий из облака заряженного аэрозоля через болт по плазменному каналу; динамика восходящих положительных лидеров со стримерной короной с последующим квазиобратным ударом и распадом плазмы, а также получены интегральные фотографии событий. Обнаружена высокая степень подобия разряда, инициированного болтом в электрическом поле искусственного облака заряженного аэрозоля и высотно-инициированной молнии в природных условиях. Также исследованы некоторые особенности инициирования положительных лидеров триггерных молний. В частности, получены параметры токов прекурсоров (стримерных вспышек) восходящих положительных лидеров.

В данном разделе также представлены результаты впервые реализованной инициации необычных плазменных образований длинными заземленными проводящими предметами. Благодаря использованным новым методикам ИК-измерений удалось впервые зафиксировать необычные плазменные образования (UPFs), инициированные длинными проводящими предметами (болтом арбалета с заземленным металлическим проводом) или восходящими с них лидерами, внутри отрицательного и положительного облака заряженного аэрозоля, причем пиковые токи, фиксируемые благодаря проводу, прикрепленному к болту арбалета, оказались в районе 30-40 А. Эти инициированные UPFs также кардинально отличаются по морфологии каналов (сетей) от стримерных вспышек и отрицательных и положительных лидеров длинной искры, как и ранее обнаруженные UPFs (см. главы 1-4). UPFs, обнаруженные в отрицательно заряженном облаке образуют мощные разветвленные иерархические сети каналов, похожие на зафиксированные внутри облака без заземленного болта, но большие по числу каналов и более нагретые, судя по ИК-изображениям и общим текущим токам. UPFs, обнаруженные в положительно заряженном облаке также образуют мощные разветвленные иерархические сети каналов, направленные в сторону оси облака, но они инициируются непосредственно с заземленного болта (а не благодаря лидеру) и не похожи на UPFs, возникающие самопроизвольно (без болта) внутри положительно заряженного облака аэрозоля (глава 4).

## 5.1. Введение в главу 5

Разряды молнии, инициированные летательными аппаратами, находящимися в электрическом поле грозового облака (Рисунок 5.1), представляют серьезную опасность для безопасности летательных аппаратов и размещенного на них оборудования [Rakov and Uman, 2003].

Изучение таких разрядов также имеет большой самостоятельный интерес для изучения физики молнии и грозы, как один из способов создания длинных двунаправленных (bidirectional) лидеров (длиной сотни метров, например, Рисунок 5.2), которые предложил в качестве модели молнии Каземир [Kasemir, 1960] (Рисунок 5.3) и, которые по современным представлениям должны играть ключевую роль во время инициации и развития молниевых каналов [Rakov and Uman, 2003] и формирования начальных импульсов пробоя (IBPs) [Kostinskiy et al., 2020a]. Подобный разряд инициируется ракетой (например, Рисунок 5.4), несущей за собой незаземленный провод. Последовательность событий во время высотно-инициированной триггерной молнии, следующая (Рисунок 5.5): ракета, несущая металлический провод, поднимается за 2-3 секунды на высоту 400-600 м, в результате чего провод поляризуется в поле грозового

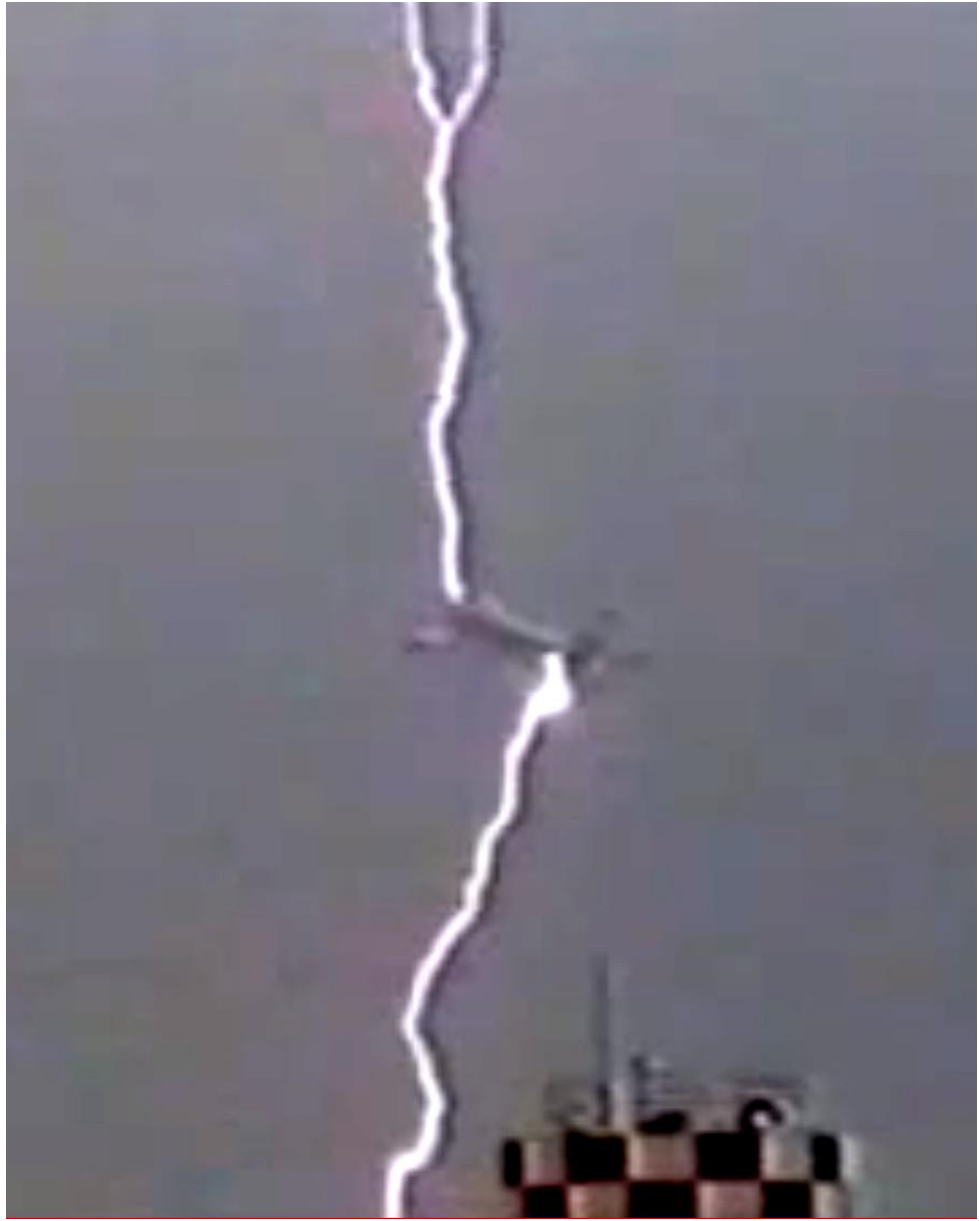

Рисунок 5.1. Молнии, инициированные самолётом на взлёте (скорость 65-100 м/с). С носа самолета движется восходящий положительный лидер, черной стрелкой отмечено возможное место контакта восходящего лидера и хвоста самолета (Любезно предоставлено Z.I. Kawasaki).

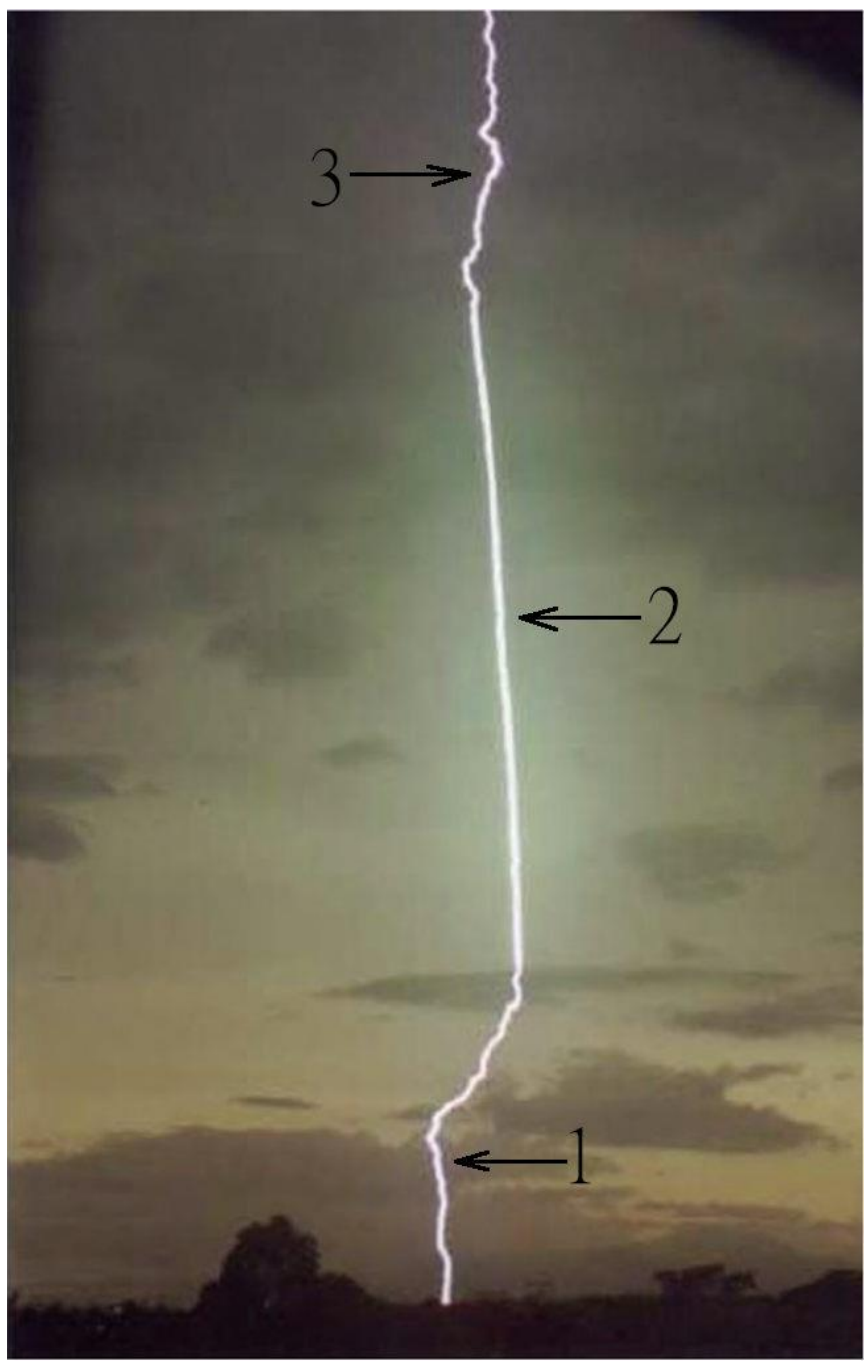

Рисунок 5.2. Интегральная фотография высотно-инициированной триггерной молнии (altitude-triggered lightning), полученная в Бразилии 23 ноября 2000 г. (любезно предоставлена др. Осмаром Пинто младшим, INPE). 1 – нижний конец двунаправленного лидера, образованного в результате встречи восходящего положительного лидера и нисходящего отрицательного лидера; 2 – плазма, испаренного медного провода; 3 – восходящий положительный лидер

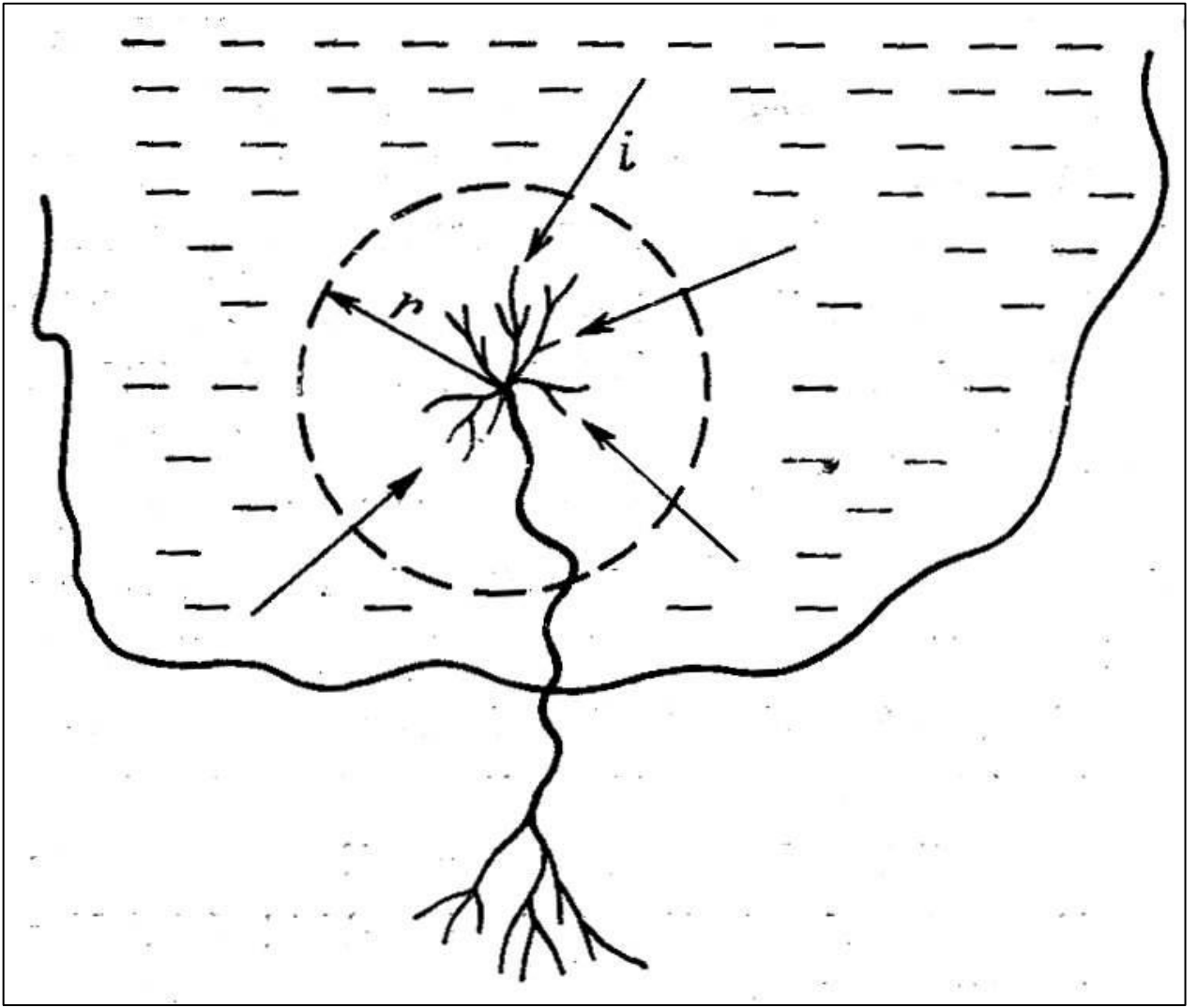

Рисунок 5.3. Схематичное изображение молнии, как двунаправленного (bidirectional) лидера, согласно гипотезе Каземира [Kasemir, 1960].

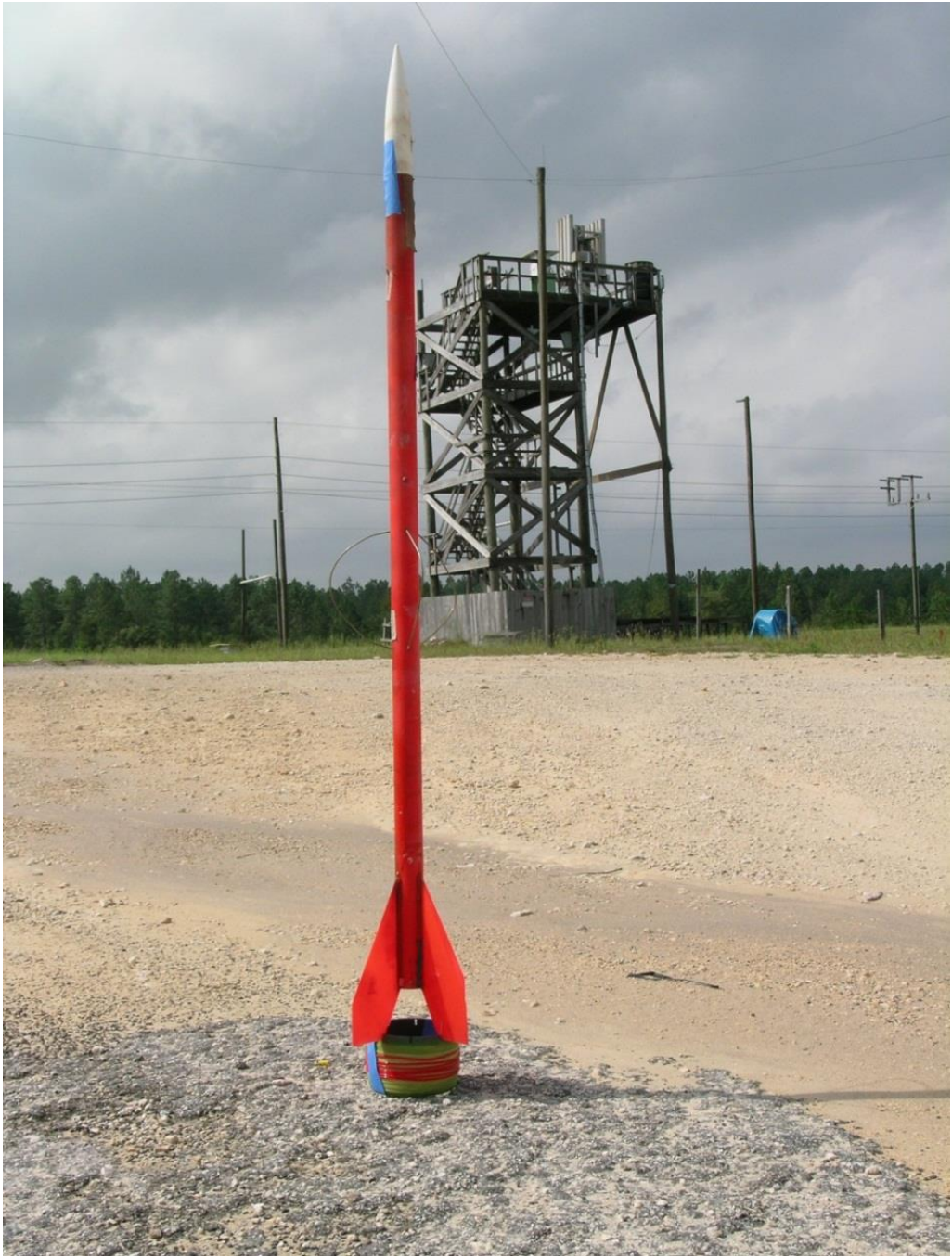

Рисунок 5.4 (любезно предоставлен Владимиром Раковым). Ракета длиной 1 м, несущая провод (намотан на шпульке в хвостовой части) и инициирующая триггерные и высотно-инициированные триггерные молнии. Фотография любезно представлена Владимиром Раковым.

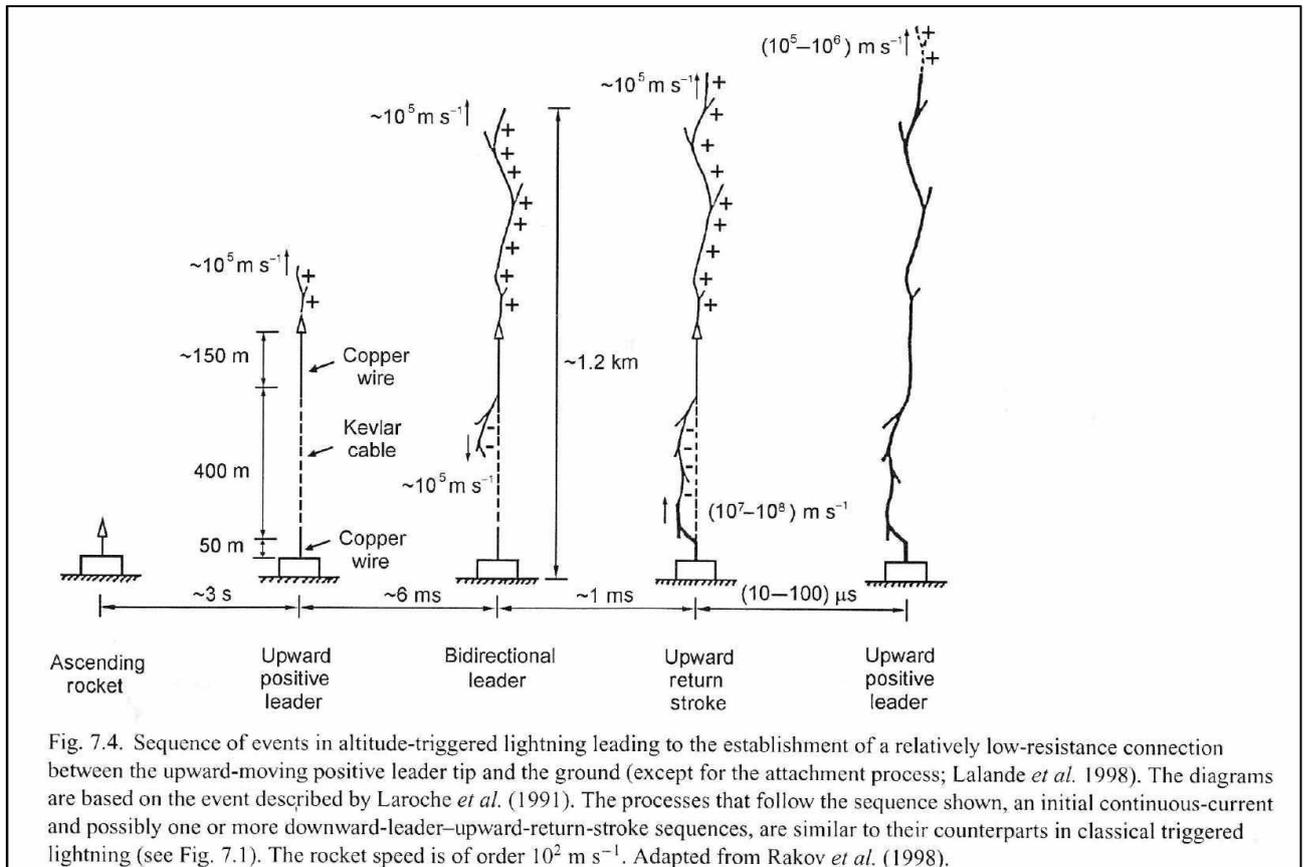

Fig. 7.4. Sequence of events in altitude-triggered lightning leading to the establishment of a relatively low-resistance connection between the upward-moving positive leader tip and the ground (except for the attachment process; Lalande *et al.* 1998). The diagrams are based on the event described by Laroche *et al.* (1991). The processes that follow the sequence shown, an initial continuous-current and possibly one or more downward-leader–upward-return-stroke sequences, are similar to their counterparts in classical triggered lightning (see Fig. 7.1). The rocket speed is of order $10^2$ m s$^{-1}$. Adapted from Rakov *et al.* (1998).

Рисунок 5.5 [Rakov and Uman, 2003], (любезно предоставлен Владимиром Раковым): ракета, несущая металлический провод, поднимается 2-3 секунды на высоту 400-600 м, в результате чего провод поляризуется в поле грозового облака и стартует восходящий положительный лидер; в среднем через 5-6 мс с нижнего конца провода стартует отрицательный нисходящий лидер; через примерно 1 мс навстречу нисходящему отрицательному лидеры стартует с земли восходящий положительный лидер и они образуют сквозную фазу (которая продолжается 10-100 мкс); после контакта нисходящего отрицательного лидера и восходящего положительного лидера следует обратный удар [Lalande et al., 1998], [Laroche et al., 1991].

облака, на его верхнем конце формируется потенциал, достаточный для инициации положительного лидера и после этого стартует восходящий положительный лидер; в среднем через 5-6 мс после старта положительного лидера с нижнего конца провода стартует ступенчатый нисходящий отрицательный лидер; через примерно 1 мс навстречу нисходящему отрицательному лидеры стартует с заземленных сооружений восходящий положительный лидер и они образуют сквозную фазу; после контакта нисходящего отрицательного лидера и восходящего положительного лидера следует обратный удар, который продолжается 10-100 мкс [Lalande et al., 1998], [Laroche et al., 1991].

«Обычная» триггерная молния имеет более простую последовательность развития разрядов (Рисунки 5.6, 5.7): ракета, несущая металлический провод, поднимается за 1-2 секунды на высоту 300-500 м, в результате чего между верхним концом заземленного провода (или кончиком ракеты) с нулевым потенциалом и точкой воздуха перед кончиком ракеты формируется потенциал, достаточный для инициации положительного лидера и после этого стартует восходящий положительный лидер, который движется сотни мс; после этого следует пауза в десятки мс; после паузы вниз стартует нисходящий стреловидный отрицательный лидер; когда нисходящий лидер достигает поверхности, следует обратный удар, после которого может быть еще несколько ударов [Rakov and Uman, 2003]. Мы несколько упростили картину развития триггерной молнии, так как не рассмотрели испарение проволоки при превышении током значения примерно 300 А и связанный с ним резкое уменьшение и возрастание в течение нескольких сотен мкс электрического тока до 1 кА (the initial current variation — ICV) [Rakov and Uman, 2003, p.274], так как в наших экспериментах испарение провода не происходит.

Однако, описанные выше исследования данного явления в природных условиях затруднены из-за небольшого числа событий, которые можно зафиксировать в грозовой сезон, и дороговизны натурных экспериментов. Поэтому, важное значение приобретают лабораторные методы исследования, моделирующие данные явления с хорошей степенью подобия и значительно облегчающие изучение основных параметров процесса.

Метод моделирования процессов развития молниевого разряда в лабораторных условиях с помощью генераторов Маркса (для решения задач теории и практики молниезащиты) достаточно широко используется исследователями, например, [Les Renardières Group, 1981]. Такие исследования позволяют получать ценную информацию

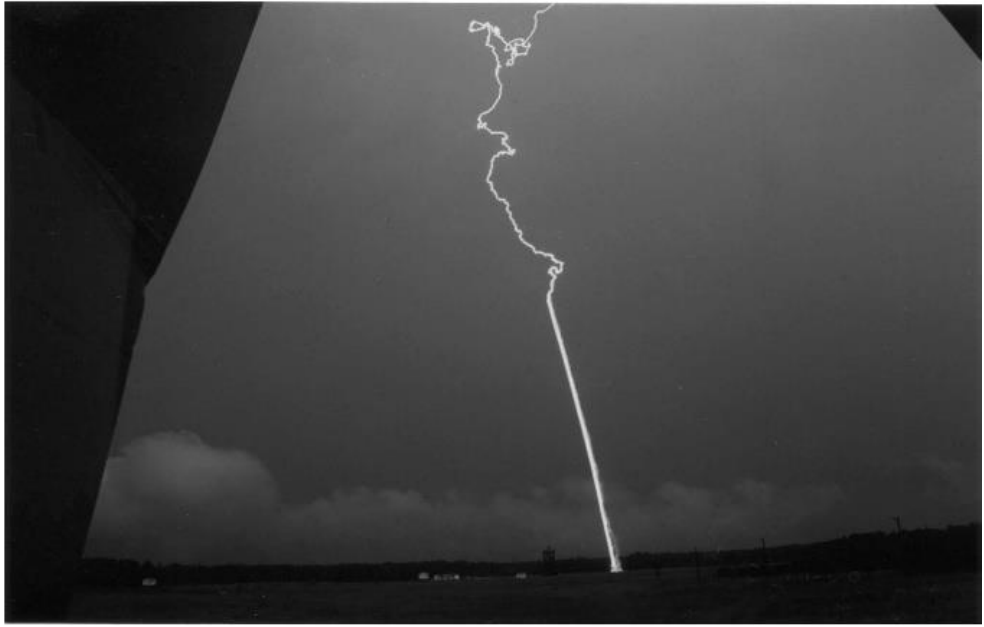

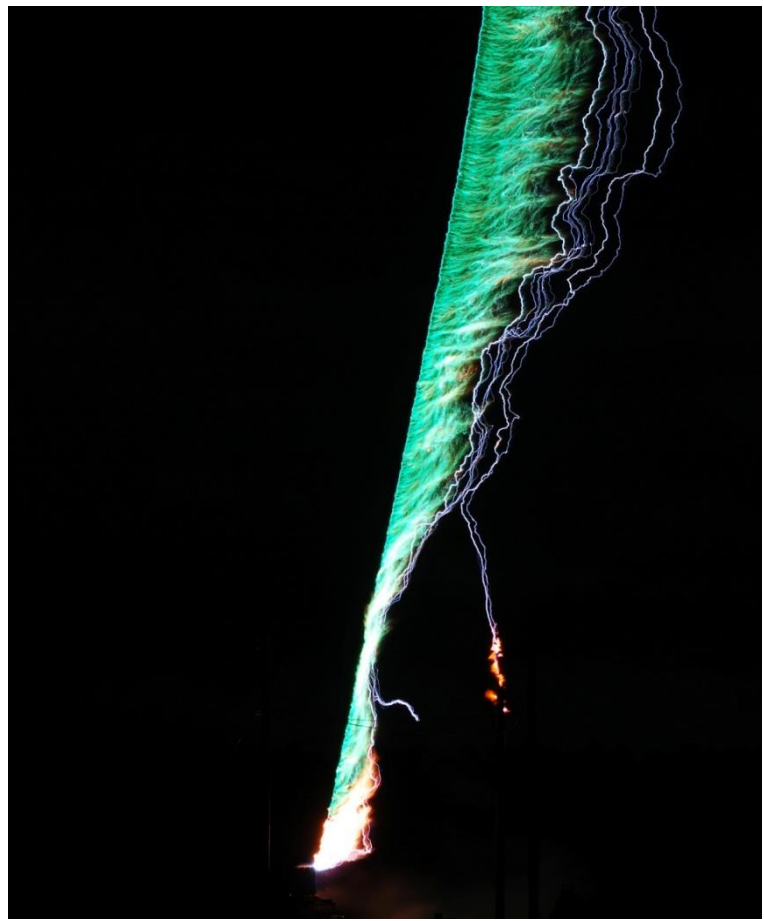

Рисунок 5.6 [Rakov and Uman, 2003], (любезно предоставлен Владимиром Раковым): Верхняя фотография является интегральной фотографией «обычной» триггерной молнии (triggered lightning), полученная в США (Camp Blanding, Florida). От поверхности земли начинается прямой участок плазмы испаренного медного провода с верхнего конца, которого (ракета) стартует извилистый, восходящий положительный лидер. На нижнем кадре видна заземленная часть провода и хорошо различаются 8 последовательных ударов молнии, первый из которых испаряет медь (зеленое свечение).

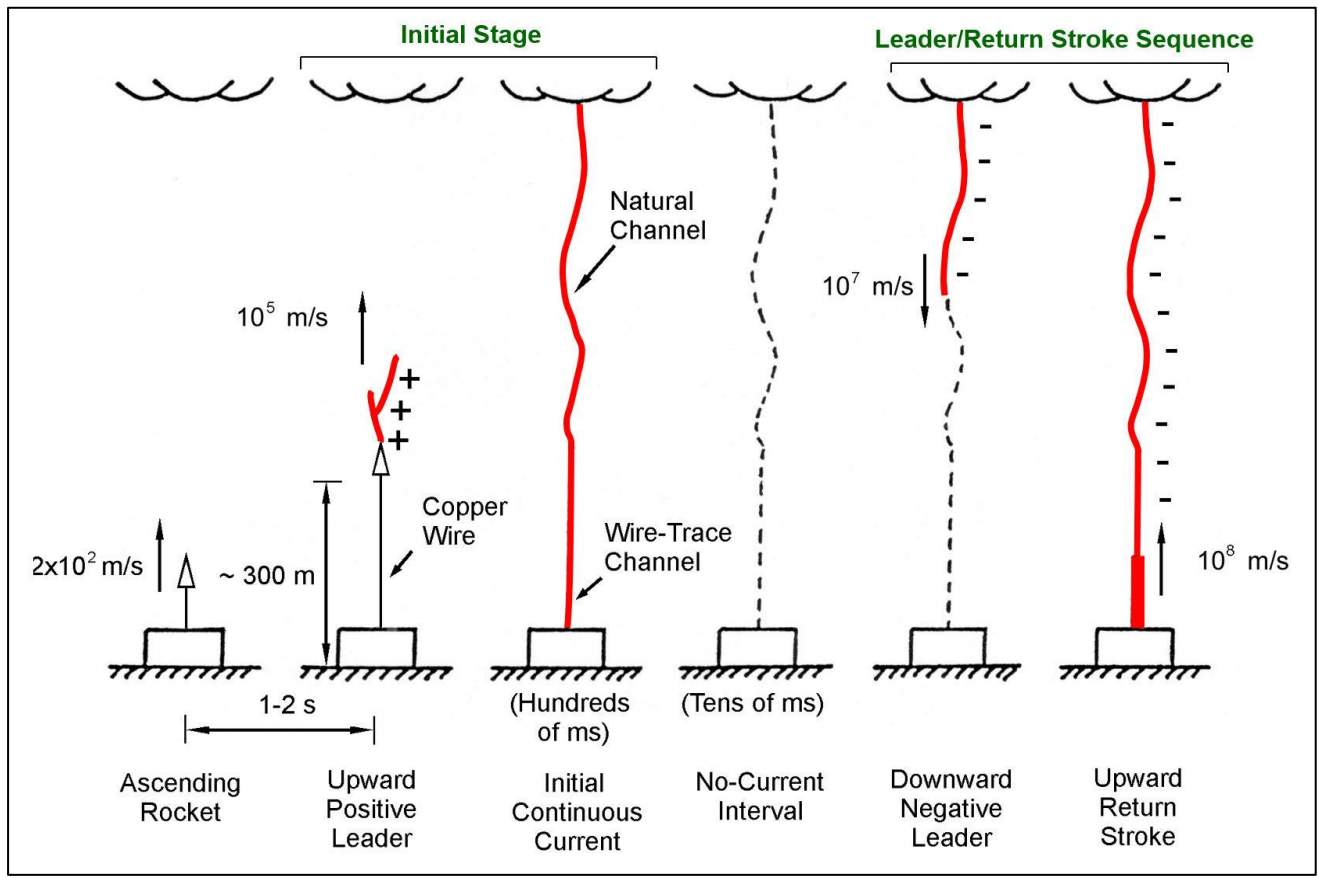

Рисунок 5.7 (любезно предоставлен Владимиром Раковым): ракета, несущая металлический провод, поднимается 1-2 секунды на высоту 300-500 м, в результате чего провод поляризуется в поле грозового облака и стартует восходящий положительный лидер, который движется сотни мс; после этого следует пауза в десятки мс; после паузы вниз стартует нисходящий стреловидный отрицательный лидер; когда нисходящий лидер достигает поверхности, следует обратный удар, после которого может быть еще несколько ударов [Rakov and Uman, 2003].

о физических процессах при развитии разряда длинной искры и применять полученные данные для понимания физики молнии и молниезащиты. При этом остаётся вопрос: насколько процессы в длинной лабораторной искре подобны процессам в молнии, где нет конденсаторов и металлических электродов.

Обычно в лабораторных экспериментах используют следующую схему испытаний летательных аппаратов на молниезащищенность. Неподвижную модель летательного аппарата подвешивают на изоляторах между электродами генератора импульсных высоких напряжений и изучают наиболее подверженные поражению области летательного аппарата. Несмотря на частое применение, в данной схеме моделирования отсутствует движение летательного аппарата, а также возникает значительное отличие условий пробоя в импульсном электрическом поле электродной системы лабораторной установки по сравнению с пробоем в квазипостоянном электрическом поле грозового облака, где заряд расположен на гидрометеорах и разряд чаще всего инициируется самим летательным аппаратом [Rakov and Uman, 2003], [Lalande et al., 1998].

В данном разделе приведены исследования, где электрические разряды инициировались арбалетным болтом, движущимся в электрическом поле облака отрицательно заряженных капель воды (с прикрепленным заземленным и незаземленным проводом или без провода). Были инициированы и исследованы сотни разрядов. В этих событиях высокоскоростная видеокамера записала изображения восходящих положительных лидеров, формирующихся как на поверхности заземленной сферы, так и на поверхности летящего болта, с последующим квазиобратным ударом. Соответствующие токи были измерены и обработаны. Были также получены интегральные фотографии событий. Результаты могут помочь улучшить наше понимание такого ключевого для инициации и развития молнии плазменного образования, как двунаправленный лидер, улучшить понимание инициирования молнии летательными аппаратами и вертикальным проводником, быстро поднятым под грозовым облаком для того, чтобы инициировать молнию с помощью заземленной ракеты с проводом (триггерные молнии).

Кроме этого, в данном разделе будут рассмотрены результаты впервые реализованной инициации необычных плазменных образований длинными заземленными проводящими предметами. Благодаря использованным новым методикам

ИК-измерений удалось впервые зафиксировать необычные плазменные образования (UPFs), инициированные длинными проводящими предметами (болтом арбалета с заземленным металлическим проводом) или восходящими с них лидерами, внутри отрицательного и положительного облака заряжённого аэрозоля.

## 5.2. Экспериментальная установка

Длинные искры протяженностью до нескольких метров, стартующие с заземленной плоскости или возвышающегося над заземлённой плоскостью проводящего предмета, в поле искусственно созданного облака заряженного аэрозоля были получены достаточно давно [Верещагин и др., 1988], [Анцупов и др., 1990]. В данном исследовании использовалась схема экспериментального стенда генерации заряженного аэрозольного облака, подобная описанной в главах 1-4, но имеющая существенные различия в постановке эксперимента и используемом измерительном оборудовании (Рисунок 5.8). Выходное сопло (5) генератора заряженного аэрозоля располагалось в центре плоского металлического экрана диаметром 2 м с закруглёнными краями (6). Генератор заряженного аэрозоля состоял из двух основных частей: генератора пара (2) и зарядного устройства (3). Паровоздушная струя из паропровода (7) при температуре около $150^0$С под давлением 0,2-0,6 МПа вылетала со скоростью, близкой к скорости звука данной смеси (около 400-450 м/с) из сопла (5) с углом раскрыва $28^0$, образуя адиабатически расширяющуюся затопленную струю. При этом в атмосфере создавалось облако заряженного аэрозоля (1). В результате быстрого охлаждения, пар конденсировался в капли размером около 0,5 мкм [Верещагин и др., 1988], [Анцупов и др., 1990]. Ионы, заряжающие аэрозоль, образовывались в коронном разряде, между тонкой заострённой иглой (4), расположенной в сопле (5) и соплом. На иглу от высоковольтного источника (3) подавалось постоянное напряжение 10÷20 кВ отрицательной полярности. Ток выноса заряда затопленной струей находился в пределах 60÷100 мкА. При накоплении в аэрозольном облаке общего заряда до значений около 60-70 мкКл самопроизвольно на плоскости возникали длинные искровые разряды.

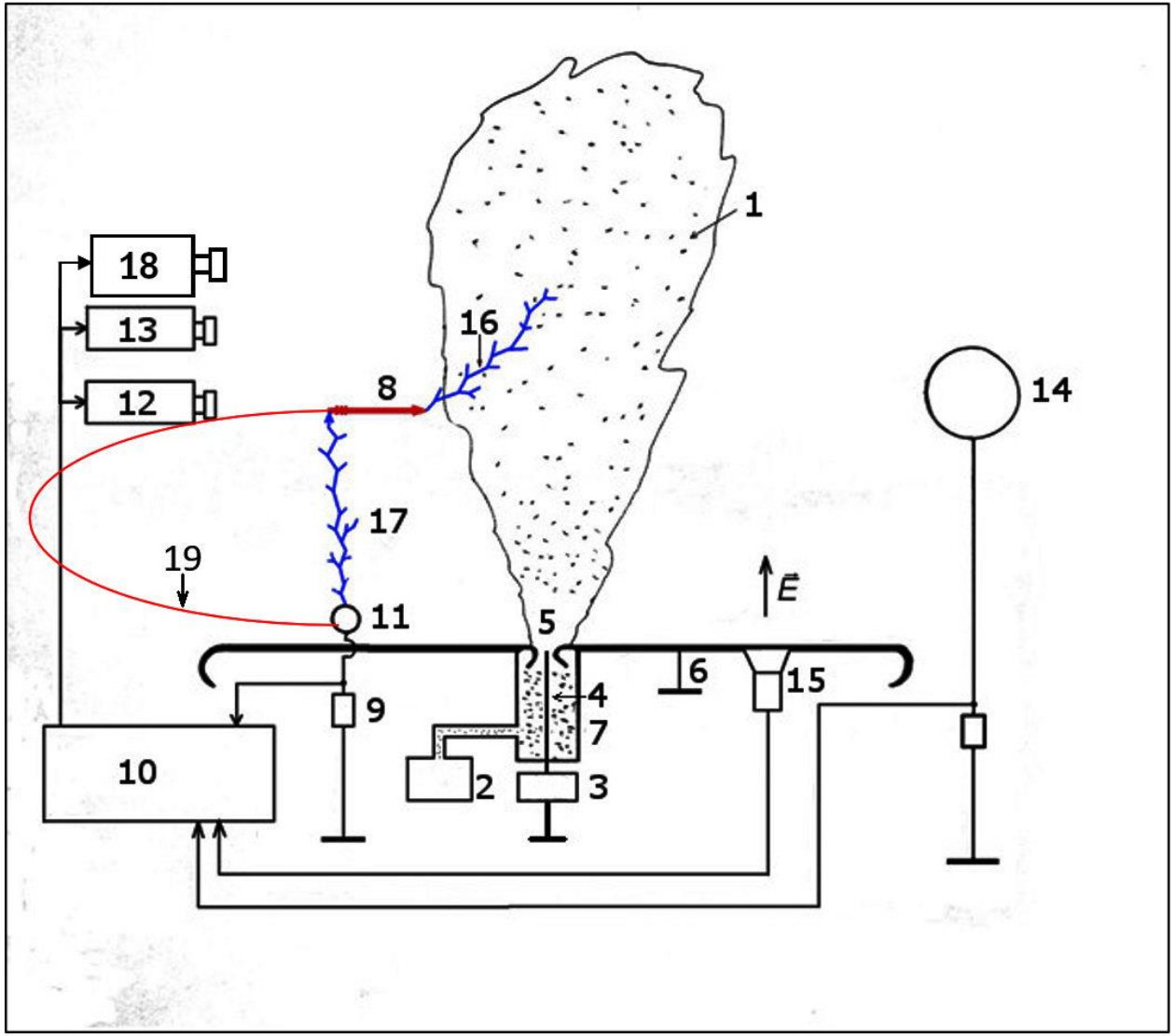

Рисунок 5.8 (адаптировано из [Kostinskiy et al., 2015c]). Схема эксперимента. 1 – облако заряженного аэрозоля; 2 – генератор пара; 3 – источник высокого напряжения; 4 – игла, которая создает коронный разряд; 5 – сопло; 6 – заземленная плоскость; 7 – паропровод; 8 – проводящий болт арбалета (для высотно-инициированных разрядов используется без провода (19), а для триггерных разрядов с проводом (19); 9 – измерительный шунт; 10 – осциллограф; 11 – металлическая сфера; 12 – скоростная видеокамера FASTCAM SA4; 13 – фотоаппарат; 14 – медная сфера диаметром 50 см для мониторинга электрического поля облака; 15 – флюксметр; 16 – плазменный канал восходящего лидера с наконечника болта; 17 – восходящий положительный лидер от заземленной сферы. 18 – ИК-камера FLIR 7700, 19 – заземленный провод, прикрепленный к арбалетный болту (используется для инициации триггерных разрядов, а для высотно-инициированных разрядов не используется).

В облако заряженного аэрозоля (1) с расстояния 10÷12 метров от оси струи, на высоте 0,5÷1,2 м от плоского заземлённого металлического экрана (6) выпускали из арбалета проводящие болты (8) длиной 0.58 м, диаметром 8.8 мм, весом 0.03 кг со скоростью 75 м/с. Схема запуска болта представлена на Рисунке 5.9.

Некоторые болты в целях моделирования триггерных молний имели прикрепленный к ним медный провод (19) диаметром 0,1 мм. Для измерения тока, проходящего через летящий болт, использовался шунт (9) с сопротивлением 1 Ом, сигнал с которого подавался на цифровой осциллографом Tektronix DPO с полосой 500 МГц (10). Шунт был подсоединен к приемному электроду в виде металлической сферы диаметром 5 см, верхняя точка которой возвышалась над заземленной плоскостью на расстоянии 10 см. Шарик находился от центра плоскости на расстоянии 0,8 м. При превышении током в шунте заданного значения, запускался осциллограф, который в свою очередь, выдавал импульс остановки записи кадров на скоростную камеру видимого диапазона FASTCAM SA4 (12). Цветная камера FASTCAM SA4 работала во время измерений в режиме непрерывной записи «по кругу» со скоростью 225 000 кадров в секунду (fps) и останавливалась на том кадре, во время записи которого на неё приходил импульс синхронизации с осциллографа. При этой скорости записи каждый кадр фиксировал изображение на часть матрицы камеры размером 128х64 пикселя (128 – по вертикали, 64 по горизонтали). Каждый кадр имел выдержку 4,44 мкс и следовал за предыдущим встык, без временного пробела. Для исследования инициации плазменных образований внутри аэрозольного облака использовалась ИК-камера FLIR 7700 (18). Для контроля динамики общего заряда, находящегося на аэрозольном облаке, использовался изолированный медный шар (14) диаметром 50 см, соединённый через сопротивление 100 МОм с осциллографом. Это позволяло фиксировать динамику накопления заряда в процессе зарядки облака и быстрые процессы ухода заряда из него. Общая картина разряда фиксировалась цветным цифровым фотоаппаратом (13). Скорость полёта болта была измерена с помощью скоростной видеокамеры.

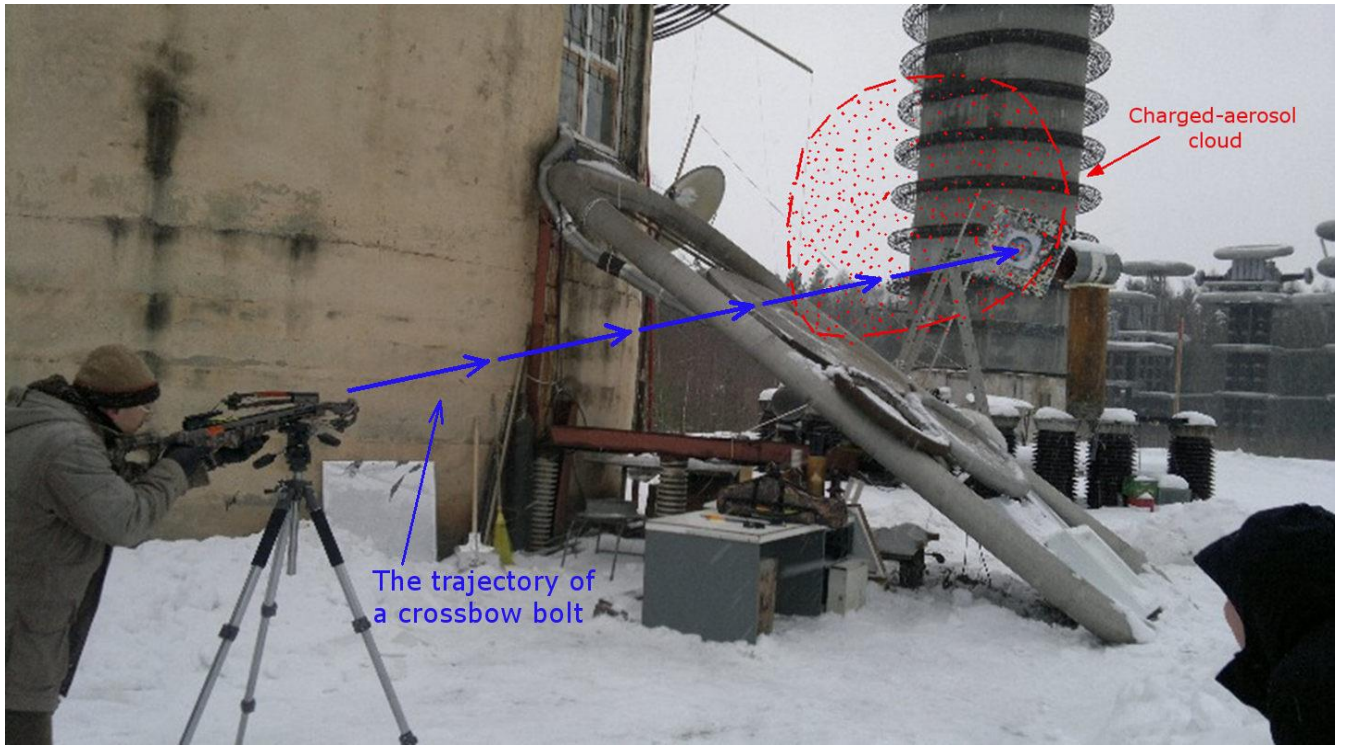

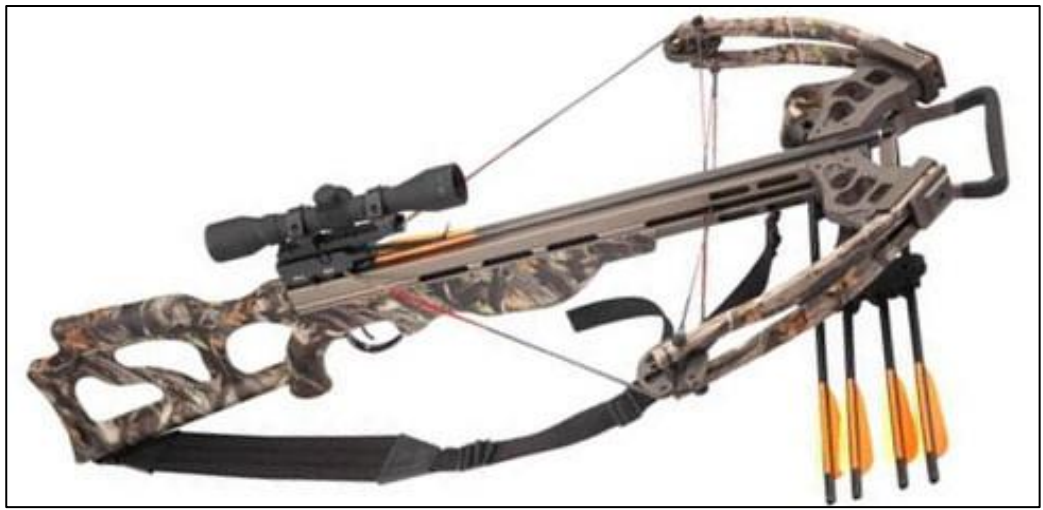

Рисунок 5.9. Схема запуска болта из блочного арбалета «Архонт» (нижнее фото). Траектория болта обозначена синими стрелками. Область аэрозольного облака обозначена красным.

## 5.3 Результаты экспериментов

### 5.3.1. Разряд, который может являться аналогом высотно-инициированной триггерной молнии (ATL)

Во время данных экспериментов отрицательно заряженное аэрозольное облако заряжалось до 60÷80 мкКл, при которых с заземлённой плоскости самопроизвольно инициировались длинные искровые разряды. Электрическое поле, создаваемое аэрозольным облаком на поверхности заземленной плоскости на расстоянии около 0,8 м от оси струи, имело значение 4÷5 кВ/см и слабо росло по направлению к облаку. При этом с поверхности сферы и плоскости самопроизвольно возникали положительные стримерные вспышки и искровые разряды. Электрическое поле на видимой границе облака не превосходило 10-11 кВ/см. В пользу этого говорит отсутствие вспышек отрицательной стримерной короны с границ аэрозольного облака в сторону заземлённой плоскости, так как движение длинных отрицательных стримеров поддерживается полем 10-11 кВ/см (в положительно заряженном облаке такие вспышки положительных стримеров надежно зафиксированы, например, Рисунки 4.2, 4.12).

Болт, выпущенный из арбалета, летел сквозь аэрозольное облако со скоростью близкой к 75 м/с почти параллельно плоскости на высоте 0,5÷1 м над сферой с измерительным омическим шунтом. Скорость болта измерялась с помощью скоростной камеры FASTCAM SA4. В среднем в 20% случаев запуска болта удавалось инициировать разряд аэрозольное облако-болт-земля, а примерно в 10% случаев инициировался разряд облако-болт-сфера, что позволяло измерять ток инициированного разряда. В большинстве остальных случаев удавалось зафиксировать серию стримерных вспышек с наконечника заземленного через провод болта при приближении болта к облаку на расстояние меньшее метра от оси струи, при этом стримерная вспышка не приводила к длинной искре. Верхняя точка сферы, возвышающегося на 10 см над плоскостью, в данной конфигурации электрических полей играет роль «высотного сооружения» в природных условиях под грозовым облаком. В процессе эксперимента одновременно измерялись ток со сферы, синхронный сигнал с шара диаметром 50 см (отображающий динамику изменения заряда облака), велась видеозапись развития разряда и получались интегральные фотографии. Угловой обзор камеры FASTCAM SA4 при записи на скорости 225000 кадров в секунду не позволял одновременно снимать движение лидеров с обеих концов болта. Поэтому скоростная видеокамера синхронно с осциллограммой тока со сферы фиксировала в одном эксперименте движение возникающих лидеров либо с наконечника болта к облаку, либо со сферы к оперению болта (с последующими в обоих случаях квазиобратными ударами и распадом плазмы). Синхронизация осциллографа и видеокамеры позволяла по одновременному наблюдению за током и кадрами видеозаписи восстановить последовательность событий с обоих концов болта с точностью не хуже 1-2 мкс. Одновременно с помощью фотоаппарата получали интегральную фотографию всего разряда.

На интегральных фотографиях разряда (5.10a и 5.10b) видны все характерные элементы высотно-инициированной триггерной молнии (altitude-triggered lightning), зафиксированной в природных условиях [Lalande et al., 1998], [Laroche et al., 1991]. Плазменный канал, между измерительной сферой (или заземленной плоскостью) перекрывает промежуток от сферы до конца болта, (или до провода, если он прикреплён к болту). Далее ток замыкается по проводящему болту. Ближе к облаку на фотографии виден плазменный канал (положительный лидер), который движется с наконечника болта в аэрозольное облако по восходящей траектории под возрастающим углом к болту $20 \div 60^0$. Болт в данных экспериментах (Рисунок 5.10) пролетал над сферой на расстоянии $0,5 \div 0,6$ м.

Приведём наиболее вероятную последовательность событий, исходя из данных, полученных в эксперименте. Вероятно, первоначальным событием в разряде облако-болт-земля является стримерная вспышка и старт положительного лидера с наконечника болта в сторону аэрозольного облака. В пользу этого предположения говорит то, что в тех случаях, когда не удавалось инициировать длинный искровой разряд облако-болт-земля, в большинстве событий наблюдалась вспышка с наконечника болта, направленная к облаку. После первоначальной вспышки, с наконечника болта формируется положительный лидер. Движение лидера, восходящего к облаку, фиксируется скоростной видеокамерой, кадры которой синхронизированы с осциллограммой тока, который течёт в это время со сферы через шунт на осциллограф (Рисунок 5.11). Практически одновременно (в интервале не более 1,5 мкс после старта лидера с наконечника болта) со

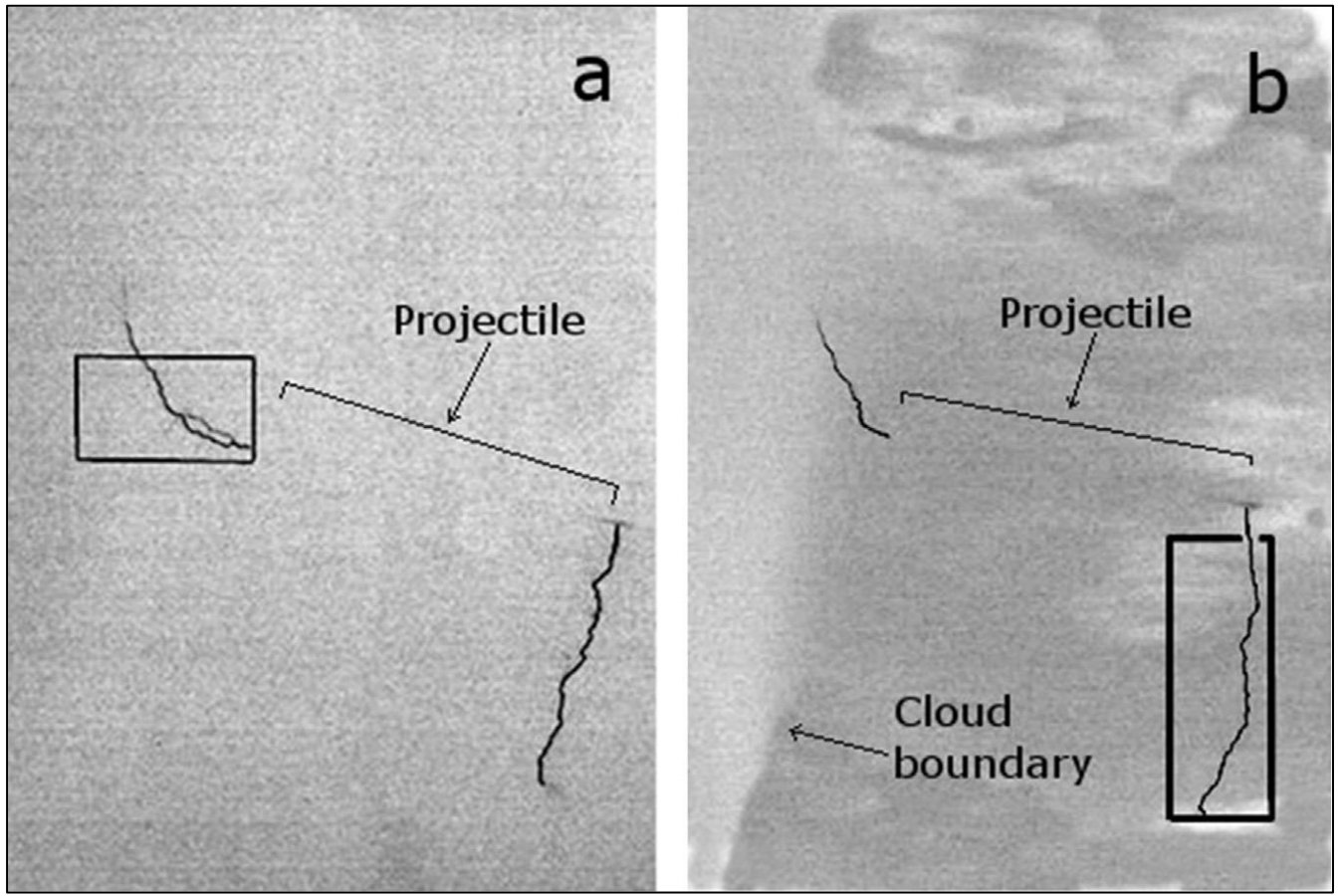

Рисунок 5.10 (адаптировано из [Kostinskiy et al., 2015c]). Примеры интегрированных фото (инвертированы) инициированных болтом разрядов. Два отдельных разряда показаны на панелях (a) и (b) в прямоугольных областях, показывающих поле зрения высокоскоростной видеокамеры FASTCAM SA4, отображающей лидеры, стартующие от наконечника болта и от заземленной сферы, соответственно. Проводящий болт (projectile) на фотографиях не виден, но его положение обозначено квадратной скобкой. Длина болта равна 0,58 м. Граница облака заряженного аэрозоля видна и отмечена на (b) стрелкой. (Данной фотографии соответствует скоростная съемка и осциллограмма на Рисунке 5.11).

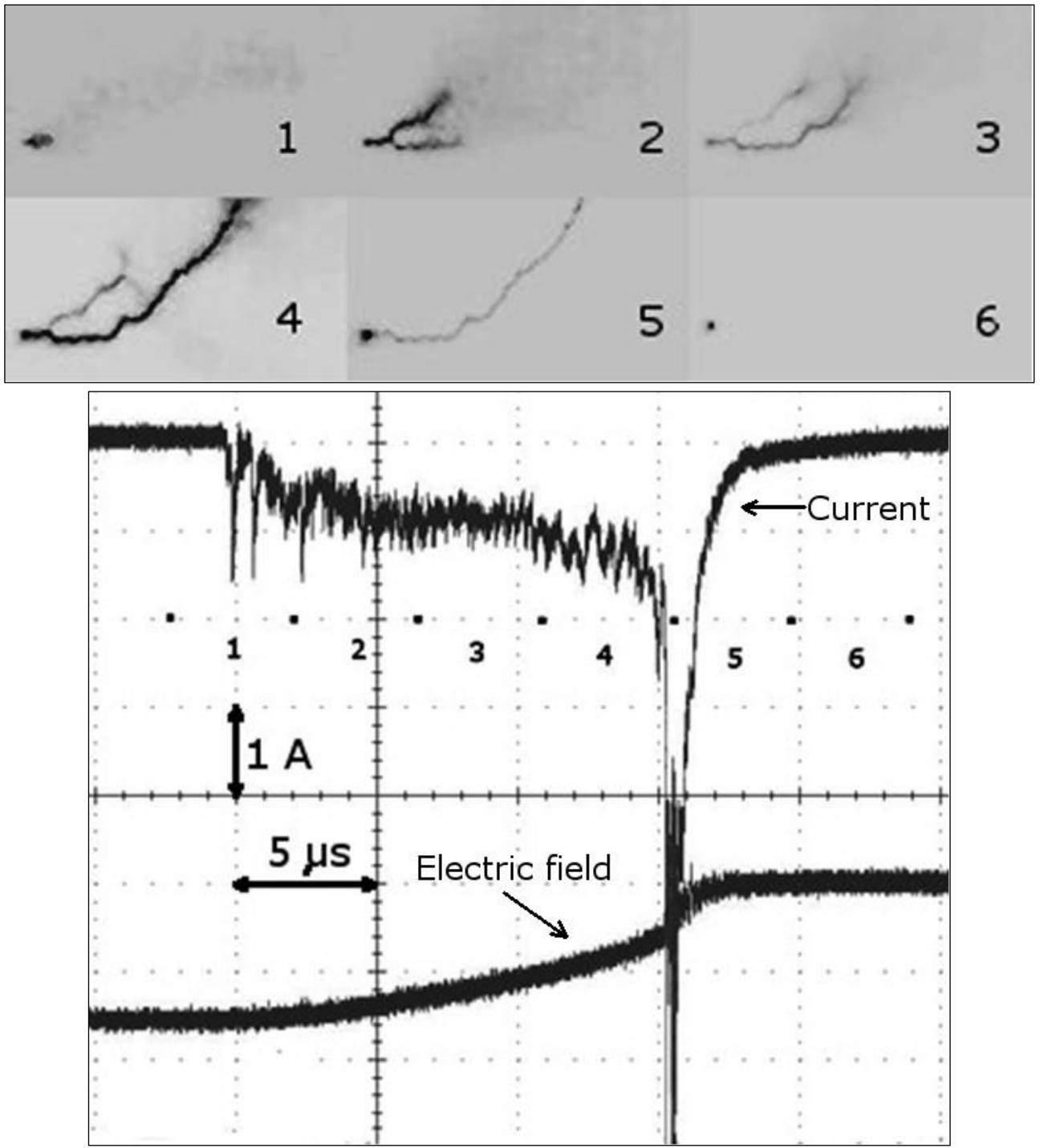

Рисунок 5.11 (адаптировано из [Kostinskiy et al., 2015c]). Шесть кадров камеры FASTCAM SA4 выдержкой 4,44 мкс (вверху), показывающие процессы развития лидера между болтом и отрицательным облаком и соответствующий событию ток через шунт (внизу). Время экспозиции каждого из шести видеокадров указано на текущей записи цифрой. Первые три кадра соответствуют развитию положительного лидера от болта. Четвертый кадр соответствует контакту канала восходящего лидера и болта, когда наблюдается резкий максимум тока. Также показана динамика изменения заряда облака. Событие соответствует интегральной фотографии на Рисунке 5.10а (левый прямоугольник, показывающий примерное поле зрения видеокамеры). Фотография была сделана под углом 150 градусов относительно направления съемки высокоскоростной видеокамерой.

сферы в сторону хвоста болта стартует другой восходящий положительный лидер (в некоторых случаях также стартует несколько лидеров с плоскости около сферы). Распространение это лидера фиксируется в другом эксперименте последовательностью кадров видеокамеры, синхронизованной с осциллограммой тока со сферы (Рисунок 5.12). Рисунок 5.11 и Рисунок 5.12 относятся к разным экспериментам, но последовательность событий надёжно устанавливается и в том, и в другом случае, благодаря синхронизации видеокамеры с током разряда. При имеющемся временном разрешении видеокамеры можно сделать заключение, что лидер с наконечника стрелы и лидер со сферы движутся большую часть времени одновременно с близкими скоростями до момента квазиобратного удара (контакта восходящего со сферы лидера с болтом). После замыкания плазменным каналом промежутка сфера-хвост болта следует квазиобратный удар, которому соответствует резкое увеличение тока за время примерно 150 нс при полуширине времени квазиобратного удара по полувысоте тока около 400-500 нс. Квазиобратному удару соответствует и резкое увеличение свечения разряда. В случае инициированного пролетающим болтом разряда облако-болт-земля во всех экспериментах наблюдался квазиобратный удар. Этим инициированный разряд отличается от самопроизвольно возникающего на плоскости положительного лидера, который приводит к обратному удару не более, чем в 5-15% разрядов (в зависимости от относительной влажности воздуха).

Характер распространения и параметры тока положительного лидера, восходящего со сферы (в присутствии болта) и стартующего самопроизвольно (без присутствия болта) значительно отличаются от того же лидера, инициированного пролетающим болтом. Осциллограммы тока со сферы (Рисунок 5.12) и синхронные с ней кадры, записанные видеокамерой, показывают, что при пролёте болта над сферой и инициации положительного лидера с наконечника болта, со сферы стартует положительный лидер и движется почти перпендикулярно вверх около 12 мкс до перекрытия промежутка между сферой и хвостом болта. После первого максимума тока на осциллограмме (Рисунок 5.12, кадр 1), фиксирующего момент инициации в стримерной вспышке восходящего лидера со сферы, ток несколько падает, а потом ток лидера возрастает вплоть до момента перекрытия лидером промежутка сфера-хвост болта, после чего следует фаза разряда, похожая на обратный удар в молнии и длинной искре (квазиобратный удар). При этом резко возрастает ток и свечение разряда (Рисунок 5.12, кадр 4). На первых трёх кадрах

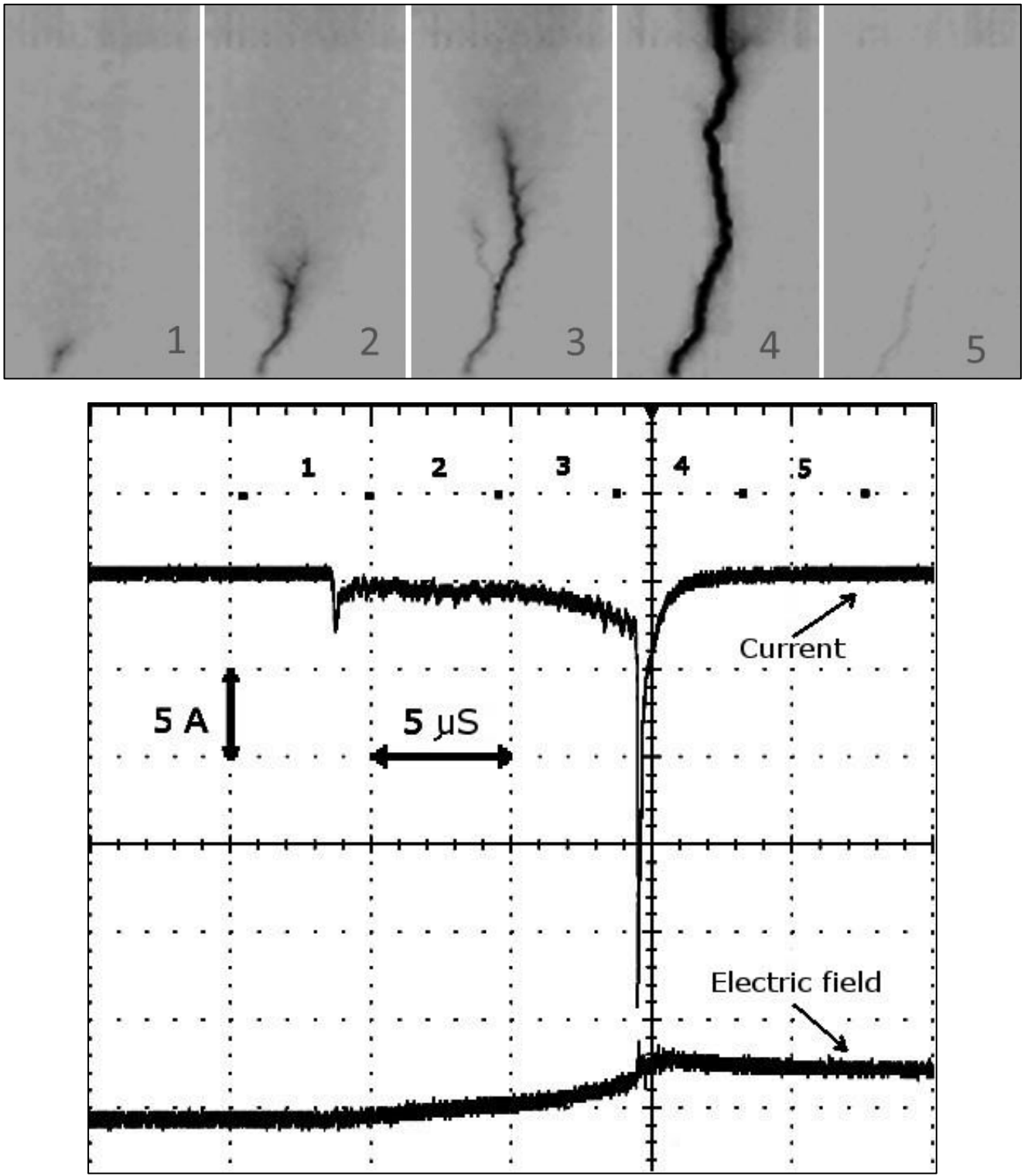

Рисунок 5.12 (адаптировано из [Kostinskiy et al., 2015с]). Пять видеокадров 4,44 мкс (вверху), показывающих развитие разрядных процессов между землей и снарядом и соответствующий ток на землю (внизу). Время экспозиции пяти видеокадров указано на текущей записи. Процессы разряда между снарядом и облаком для этого события не отображались. Первые три кадра соответствуют положительному лидеру с земли. Отрицательного лидера от снаряда не было. Также показана динамика изменения заряда облака. Событие соответствует объединенной фотографии на рис. 2б, на которой показано примерное поле зрения видеокамеры.

(Рисунок 5.12, кадры 1-3) кроме растущего канала положительного лидера с отмирающими боковыми ветвями видна длинная стримерная корона, которая, по крайней мере, со второго кадра (Рисунок 5.12) перекрывает весь промежуток от шарика до конца болта. Так как матрицы скоростных видеокамер имеют максимумы чувствительности в видимом диапазоне, то они плохо фиксируют УФ-излучение стримерной короны даже на расстояниях в несколько метров, хотя все-таки стримерную корону можно различить при сильной обработке полученных изображений. Всё это время заряд аэрозольного облака уменьшается (что видно по нижнему лучу на осциллограмме Рисунка 5.12, который фиксирует общее уменьшение поля от аэрозольного облака). Особенно быстро заряд уменьшается в момент квазиобратного удара. В результате этого процесса в аэрозольном облаке нейтрализуется значительная часть общего внедренного в объём заряда (до $20 \div 30\%$). Для восстановления электрического поля облака до прежних значений требуется время около $0,5 \div 1$ с. После квазиобратного удара происходит распад плазмы, что видно на кадре 5 (Рисунок 5.12) и следующем за ним (не приведены на Рисунке 5.12).

Оставшаяся от верхнего положительного лидера светящаяся, видимо, горячая точка на хвосте болта видна на 6 кадре (Рисунок 5.11) и ещё более десяти кадров, следующих после окончания свечения плазмы. Ни в одном эксперименте не удалось зафиксировать нисходящий отрицательный лидер с хвоста болта навстречу восходящему положительному лидеру со сферы. Однако, основываясь только на приведённых экспериментах нельзя исключить существование нисходящего отрицательного лидера, т.к. скорость записи видеокамеры недостаточно быстрая для исследуемых процессов (4,44 мкс на кадр).

В том случае, когда болт пролетал над сферой и заземленной плоскостью на высоте около метра (заметно выше, чем в первом случае (Рисунки 5.11 и 5.12), характер распространения восходящего со сферы лидера изменялся. Это хорошо видно на осциллограмме тока (Рисунок 5.14, жёлтая линия). Интегральная фотография этого события (9911_05-01-2013) изображена на Рисунке 5.13. Первая фаза постоянного роста тока лидера, длящаяся около 20 мкс, происходит аналогично случаю низкого пролёта болта над сферой (осциллограммы тока на Рисунках 5.11 и 5.12), но потом ток лидера выходит на почти стационарное значение (полку), достигая величины 1,5 А. При этом продолжается фаза почти постоянного тока (1,5 А) при распространении лидера (около

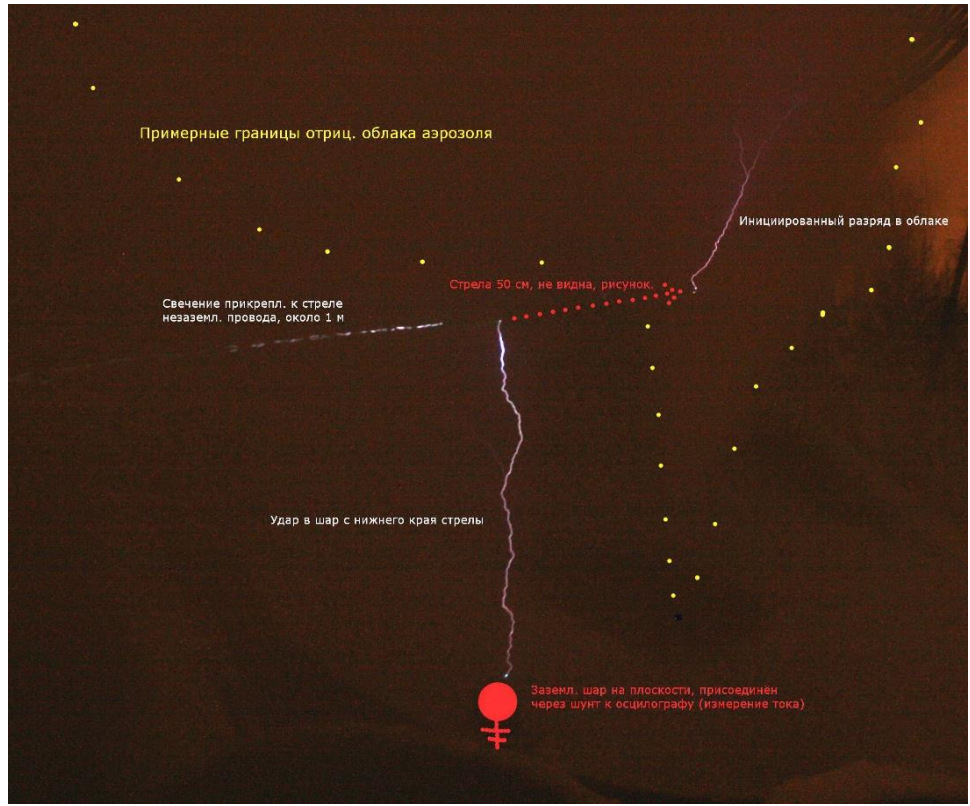

Рисунок 5.13 (событие 9911_05-01-2013). Интегральная фотография разряда, который является аналогом высотно-инициированной триггерной молнии. Разряд характеризуется высотой над сферой около метра. Осциллограммы тока и динамики заряда облака во время разряда приводятся на Рисунке 5.14.

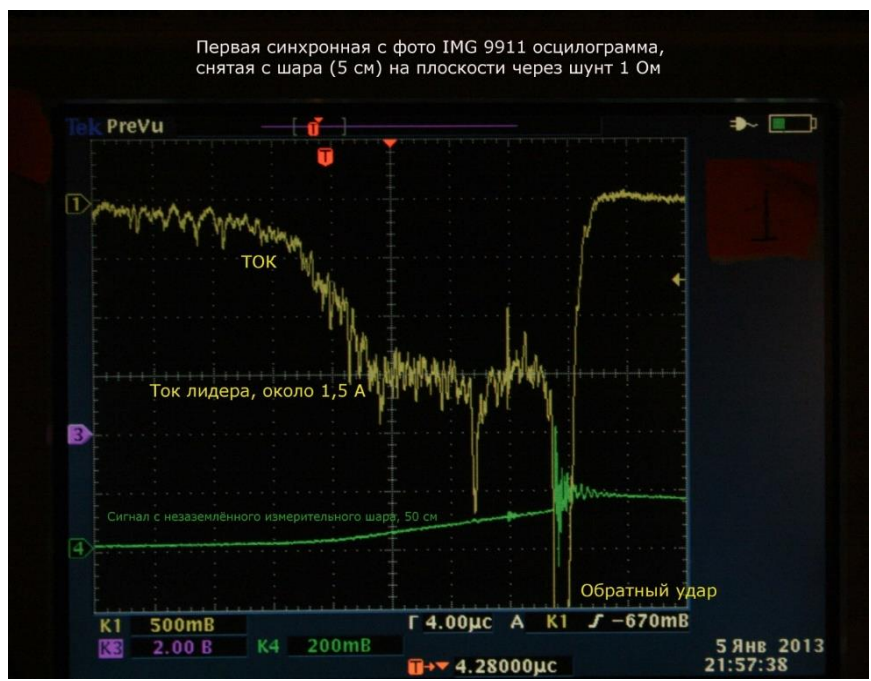

Рисунок 5.14 (осциллограммы события 9911_05-01-2013, интегральная фотография которого изображена на Рисунке 5.13). Желтый луч показывает динамику тока через заземленную сферу и шунт (одно большое деление соответствует току 0.5 А, одно большое горизонтальное деление соответствует отрезку времени 4 мкс). Зеленый луч качественно показывает динамику заряда аэрозольного облака.

12 мкс) до начала квазиобратного удара. Также на этой постоянной фазе тока можно выделить четкий максимум тока (до начала квазиобратного удара) величиной около 1 А и длительностью около 500 нс. Зелёный луч осциллографа (Рисунок 5.14) показывает, что и в этом случае всё время движения лидера уменьшается заряд аэрозольного облака, скачком уменьшаясь в момент квазиобратного удара, который длился около 1.5 мкс.

### 5.3.2. Разряд, который может являться аналогом классической («обычной») триггерной молнии

В обзоре этой главы были описаны основные свойства триггерных молний, инициированных в электрическом поле грозового облака и даны их фотографии и схема последовательности событий после инициации триггерных молний (Рисунки 5.6, 5.7) [Rakov and Uman, 2003]. В соответствие с этой схемой нами проводился и эксперимент по моделированию триггерных молний триггерными разрядами, инициированными в электрическом поле заряженного аэрозоля. Схема эксперимента для изучения триггерных разрядов была дополнена (Рисунок 5.8).  К болту арбалета (8) крепился медный провод диаметром 0.1 мм (19) и болт с медным проводом летел к заряженному положительно или отрицательно аэрозольному облаку. С другой стороны медного провода находилась заземленная сфера, через которую ток фиксировался осциллографом (10). При превышении заданного уровня тока, осциллограф (10) подавал синхронизирующий сигнал на ИК-камеру FLIR 7700 (18). ИК-камера использовалась для того, чтобы фиксировать инициацию заземленным болтом плазменных образований внутри облака. В этой схеме эксперимента медный провод имел форму петли (19), что затрудняло расчет потенциала между проводом и окружающим воздухом.

Типичная интегральная фотография триггерного разряда при отрицательной полярности облака представлена на Рисунок 5.15. На фотографии видна белая точка инициации положительного восходящего лидера на кончике болта (стрелы). Так как эксперименты проводились ночью, то болт и прикрепленный к нему провод не видны на фотографии.  Восходящий, извилистый положительный лидер имеет длину около 1.5 метра, поднимаясь вверх к аэрозольному облаку.

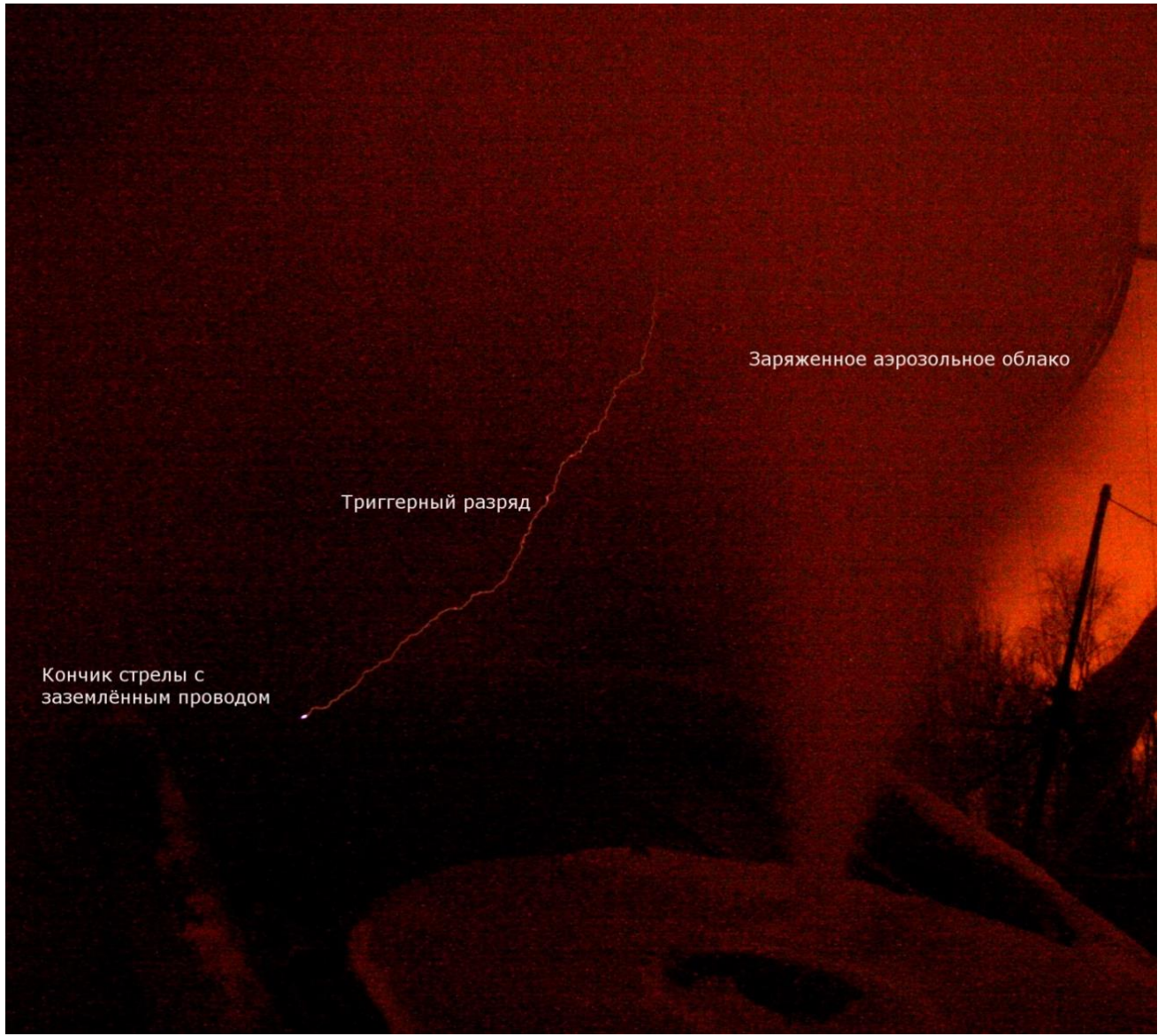

Заряженное аэрозольное облако

Триггерный разряд

Кончик стрелы с
заземлённым проводом

**Рисунок 5.15.** На интегральной цветной фотографии видна белая точка инициации положительного восходящего лидера с кончика болта (стрелы), летящего слева-направо. Так как эксперименты проводились ночью, то болт и прикрепленный к нему провод не видны. Восходящий положительный лидер имеет длину около 1.5 метра, поднимаясь вверх к отрицательному аэрозольному облаку.

**5.3.2.1. Прекурсоры положительного восходящего лидера, инициированного заземленным болтом в электрическом поле отрицательно заряженного аэрозольного облака**

В триггерной молнии, инициированной электрическим полем грозового облака, по мере движения ракеты и увеличения разности потенциалов между ее кончиком и воздухом, начинают фиксироваться осцилляции тока величиной 20-60 А и продолжительностью 10-25 мкс (Рисунок 5.16) [Rakov and Uman, 2003]. Их назвали «прекурсорами» (precursor pulses) положительных лидеров [Rakov and Uman, 2003, стр. 275]. Мы также зафиксировали подобное явление в наших экспериментах, причем некоторые его параметры оказались очень близки к параметрам прекурсоров грозовых триггерных молний.

Зафиксированные в наших экспериментах колебания тока «прекурсоров» лидеров (precursor pulses), при полете заземленного болта к аэрозольному облаку (Рисунки 5.17, 5.18) имели ту же амплитуду, что и у реальных триггерных молний – 10-30 А несмотря на то, что пространственный масштаб реальной триггерной молнии на 2-3 порядка больше, чем наши триггерные разряды. Но время колебаний триггерных разрядов было около 1 мкс (Рисунки 5.17, 5.18), что на порядок меньше, чем у прекурсоров триггерных молний (Рисунок 5.16). По нашему мнению, с высокой вероятностью прекурсоры являются токами стримерных вспышек, которые порождают колебания в проводе и измерительной цепи. Сходство величин токов природных триггерных молний и триггерных разрядов говорит о том, что ключевую роль играет не размер системы, а электрическое поле, в котором распространяется ракета или болт. Меньшее время колебаний для триггерных разрядов может определяться длиной менее 10 м короткого провода и емкостью тонкого болта по сравнению с емкостью ракеты и длиной провода в сотни метров.

Во время движения заземленного арбалетного болта к аэрозольному облаку (около 100 мс) прекурсоры возникают многократно (десятки раз до старта устойчиво движущегося положительного лидера), как и в реальной триггерной молнии [Rakov and Uman, 2003], причем их токи, как отмечалось выше, достаточно велики (10-40 А). Так как

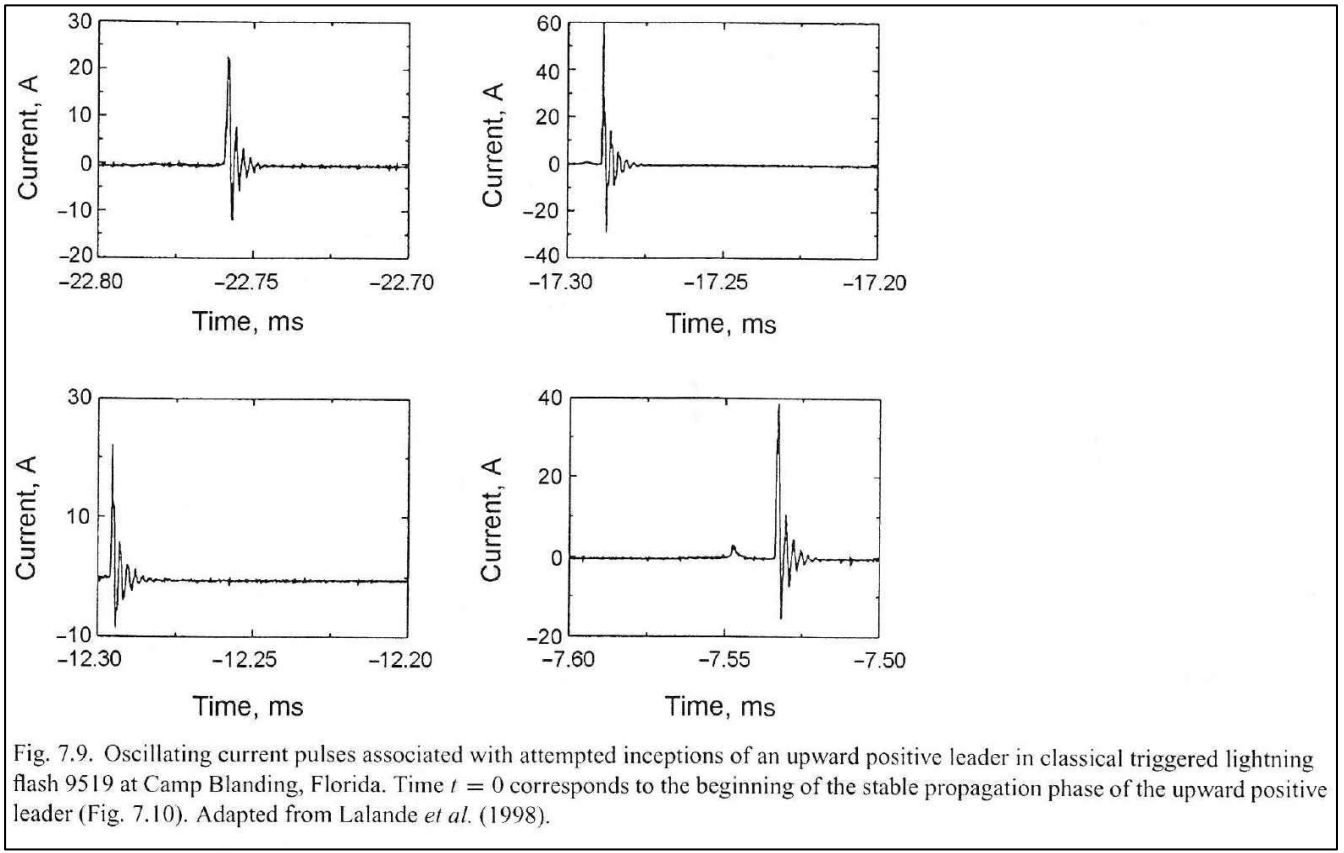

Fig. 7.9. Oscillating current pulses associated with attempted inceptions of an upward positive leader in classical triggered lightning flash 9519 at Camp Blanding, Florida. Time $t = 0$ corresponds to the beginning of the stable propagation phase of the upward positive leader (Fig. 7.10). Adapted from Lalande *et al.* (1998).

Рисунок 5.16 (любезно предоставлен Владимиром Раковым). «Прекурсоры» лидеров (precursor pulses, [Rakov and Uman, 2003], стр. 275), которые являются импульсными осцилляциями тока, ассоциированными с попыткой инициации положительного лидера в классической триггерной молнии [Rakov and Uman, 2003]. Скорее всего они являются стримерными вспышками.

Рисунок 5.17. «Прекурсоры» лидеров (precursor pulses), которые были зафиксированы при полете заземленного болта триггерных разрядов. Верхний луч – измерение тока с шунтом 1 Ом. Нижний луч – шар 50 см, который используется для контроля зарядки облака. Большое горизонтальное деление равно 500 нс, а большое вертикальное деление равно 5 А.

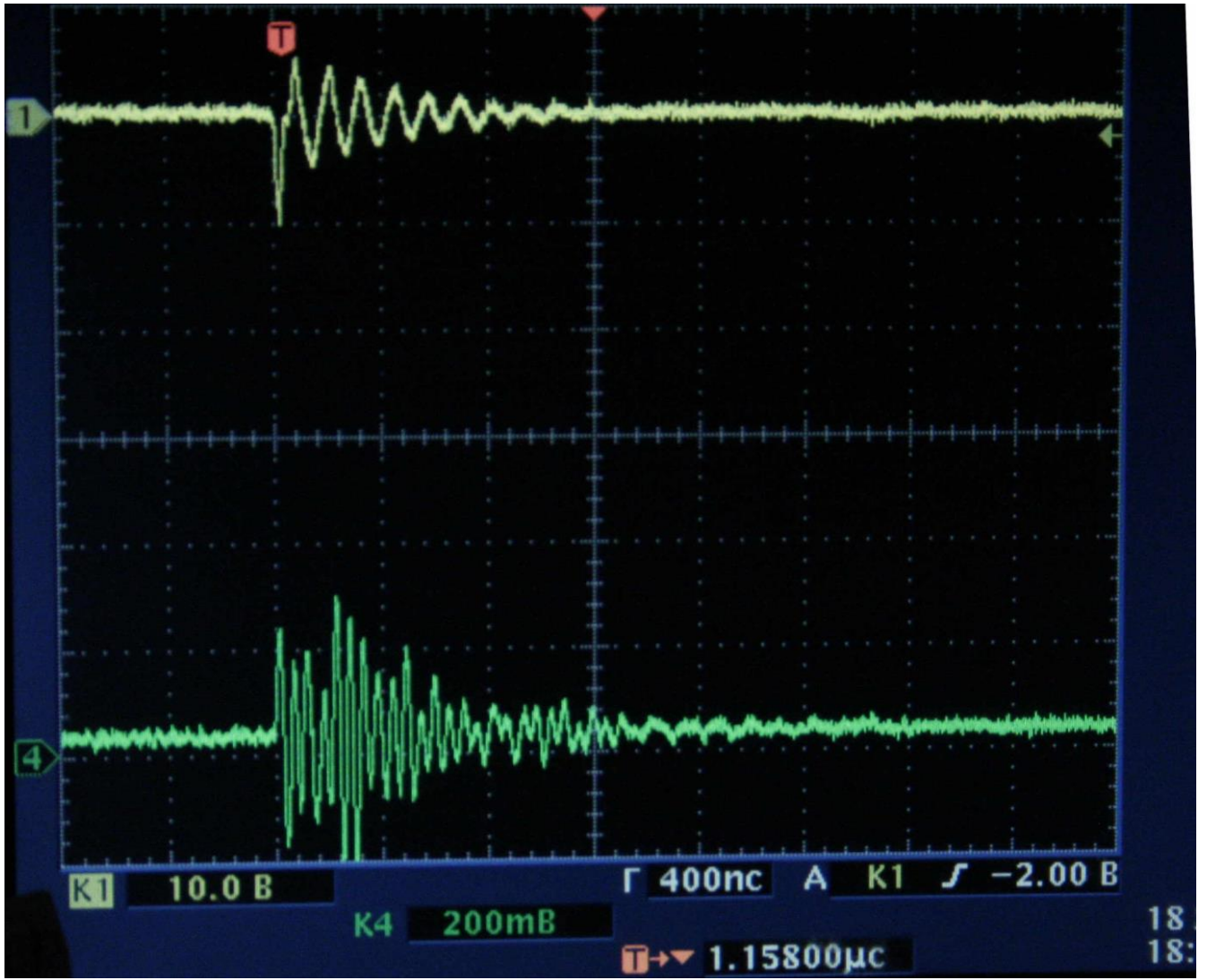

Рисунок 5.18. «Прекурсоры» лидеров (precursor pulses), которые были зафиксированы при полете заземленного болта триггерных разрядов. Верхний луч (желтый) – измерение тока с шунтом 1 Ом. Нижний луч (зеленый) – шар 50 см, который используется для контроля зарядки облака. Большое горизонтальное деление равно 400 нс, а большое вертикальное деление равно 10 А.

синхронизация ИК-камеры осуществлялась по превышению током заданного уровня сигнала, а экспозиция кадра ИК-камеры была в диапазоне 3-9 мс и у ИК-камеры при максимальном пространственном разрешении был небольшой угол зрения, то в этих экспериментах нельзя сказать с высокой уверенностью, что зафиксированное ИК-камерой событие точно соответствует осциллограмме (как многократно было в случаях с токами лидеров, взаимодействующими с заземленной измерительной сферой). Для дальнейшего изучения разрядов, инициированных заземленным болтом в электрическом поле аэрозольного облака, необходимо пользоваться не осциллографами, а компьютерными платами, которые позволят оцифровывать токи продолжительностью около 100 мс с разрешением не хуже, чем 100 МГц.

### 5.3.2.2. Плазменные образования (UPFs), инициированные болтом арбалета внутри отрицательно заряженного аэрозольного облака

ИК-камера FLIR 7700 позволила зафиксировать плазменные образования, которые возникают внутри и около аэрозольного облака, когда в него проникает заземлённый болт. На Рисунке 5.19 можно видеть, как выглядит болт (стрела арбалета) на ИК-кадре.

Большой интерес представляет собой впервые зафиксированное событие (Рисунок 5.20), где можно видеть сложную и яркую иерархическую сеть плазменных каналов, весьма вероятно представляющую из себя сеть UPFs, изображение (1), инициированных заземленным болтом внутри аэрозольного облака, который пролетел по самому нижнему краю кадра (или положительным лидером, который инициирован с поверхности этого болта). Максимальный ток, который зафиксировал осциллограф, достигает большой величины, около 40 А (2). По форме тока осциллограмма похожа на типичную осциллограмму квазиобратного удара при контакте двух каналов, например, Рисунки 3.9-3.10, и, возможно, болт арбалета до момента вхождения в облако контактировал с каким-нибудь плазменным каналом в течение выдержки кадра (9 мс) и этот контакт каналов не попал в кадр ИК-камеры. Но мы также не можем исключить, что на Рисунке 5.20 удалось зафиксировать взаимодействие заземленного болта арбалета с плотной сетью плазменных каналов (UPFs), которые за примерно 1 мкс снабжают болт арбалета большим током

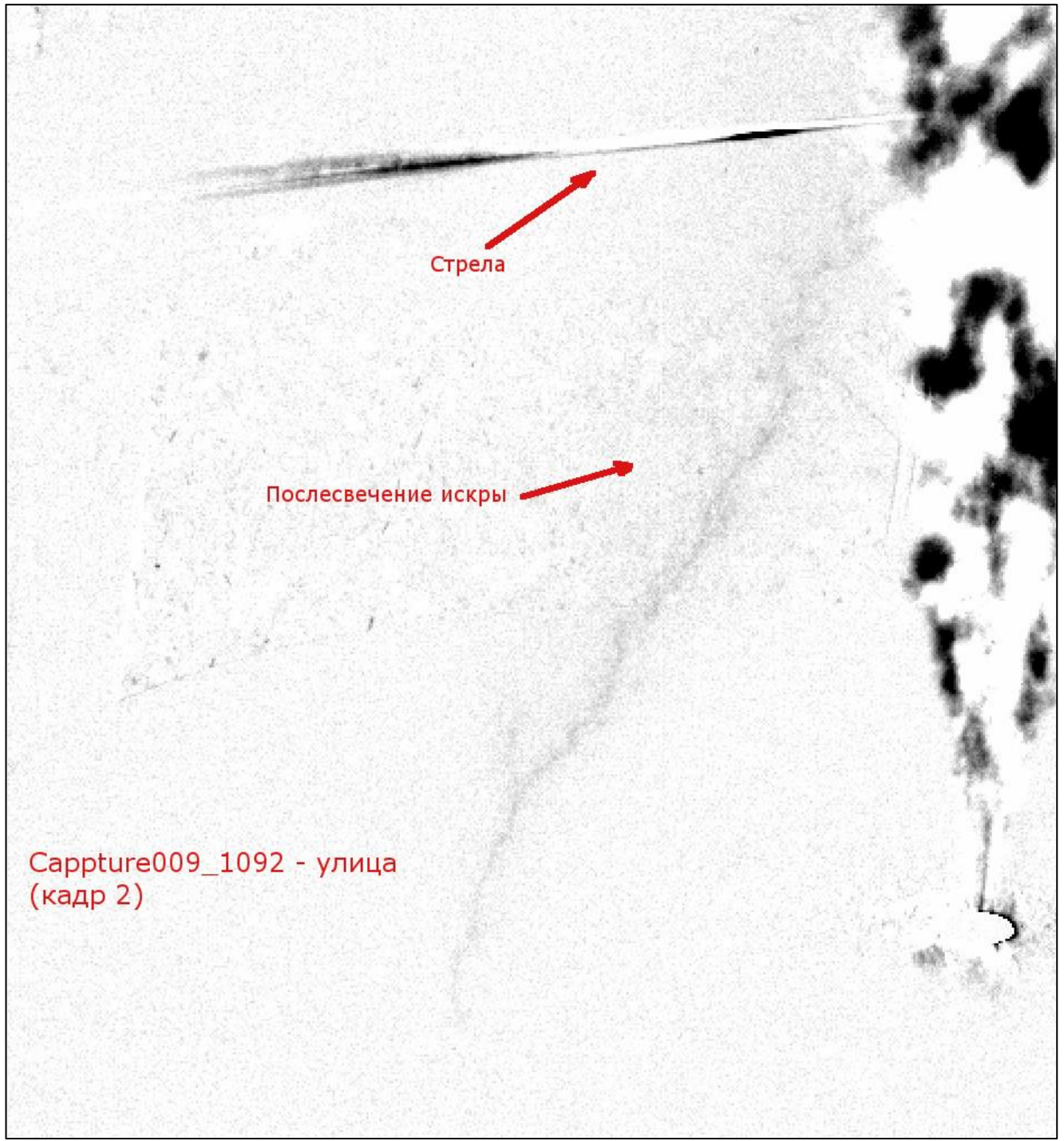

Рисунок 5.19. Болт (стрела), летящая слева-направо, в ИК-изображении (инвертировано).

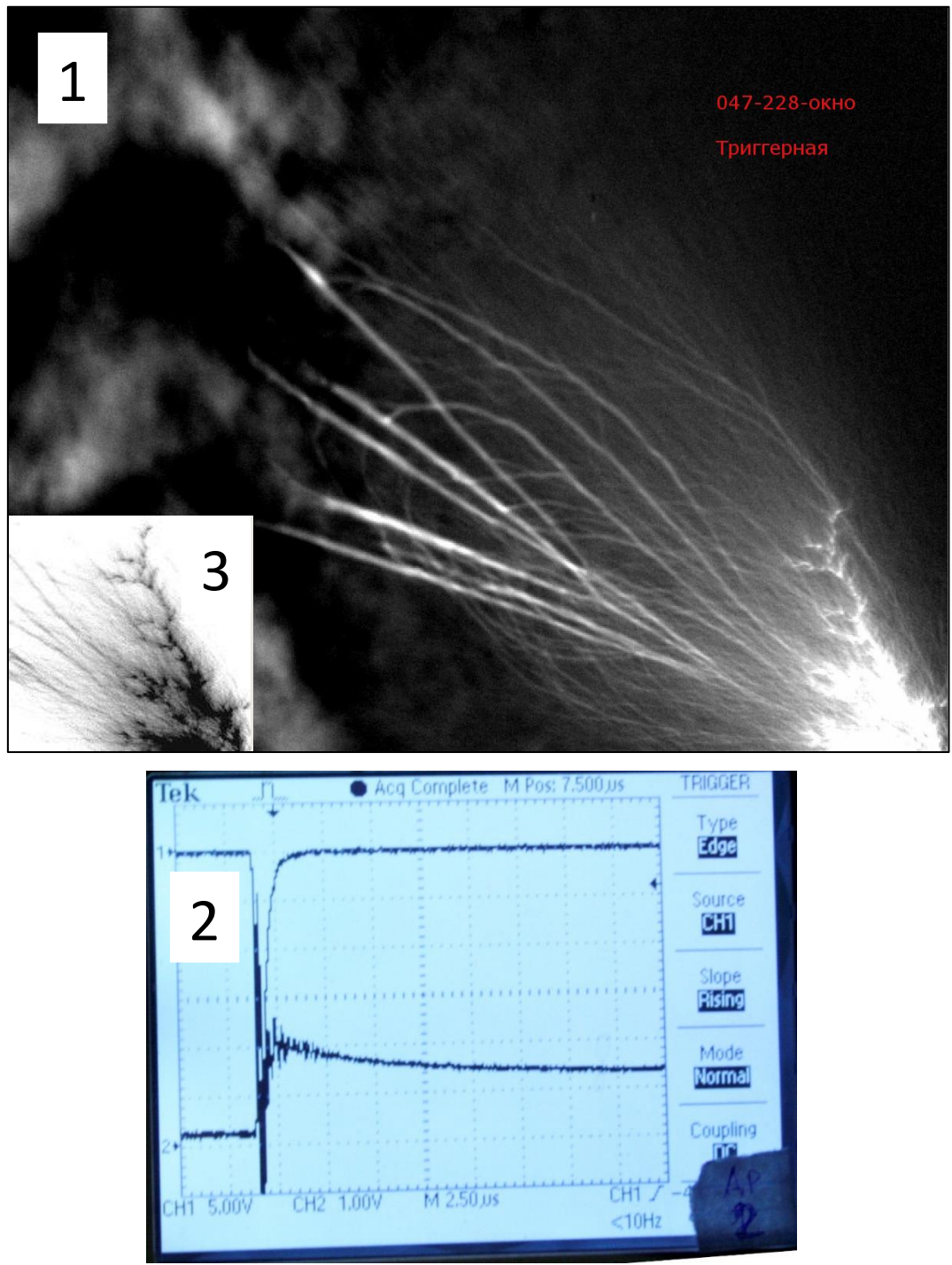

Рисунок 5.20 (событие 047-228_2013-01-18, заряд облака – минус). ИК-изображение (1), фиксирует большую иерархическую сеть плазменных каналов, инициированных заземленным болтом, летящим справа-налево, внутри аэрозольного облака, который пролетел по нижнему краю кадра (или положительным лидером, поднимающимся с болта). Выдержка 3,4 мс, частота кадров 115 Гц, 640х512 пикселей. Осциллограммы тока и контроля заряда облака, которая, возможно, соответствует этому событию (2). Верхний луч осциллограммы – измерение тока с шунтом 1 Ом. Нижний луч – сигнал с шара диаметром 50 см, который используется для контроля заряда облака. Большое горизонтальное деление равно 2.5 мкс, большое вертикальное деление равно 5 А. (3) – увеличенный, осветленный и инвертированный фрагмент правого нижнего угла изображения (1). Интересно, что некоторые яркие UPFs — почти прямые линии.

(40 A). В этой серии измерений многократно наблюдались обычные квазиобратные удары при взаимодействии каналов, но максимумы токов таких взаимодействий не превышали 20 А (Рисунок 3.22). Обращает на себя внимание также форма прямых отрезков, которую имеют шесть центральных, наиболее ярких каналов, с веретенообразными, отходящими от них ветвями. В правом нижнем углу Рисунка 5.20(1) мы видим хорошо идентифицируемый конец восходящего положительного лидера, а обработка этого кадра (осветление) показывает, что яркое светящееся «ядро» в правом нижнем углу, из которого появляется конец положительного лидера, является несколькими близко лежащими ветвями положительного лидера, взаимодействующего с сетью UPFs (Рисунок 5.20(3)). Этот факт говорит о том, что зафиксированная плазменная сеть может взаимодействовать не с самим болтом, а с восходящим положительным лидером (и/или его ветвями), который был инициирован с кончика (верхней части) заземленного болта.

Аналогичная картина зафиксирована и на Рисунке 5.21(I), где также можно видеть большую иерархическую сеть плазменных каналов (но меньшую, чем на Рисунке 5.20, инициированных заземленным болтом внутри аэрозольного облака (или лидером, стартовавшим с заземленного болта). Максимальный ток, которые обеспечивают эти каналы немного меньше и достигает 35 А, также явно меньше, чем на Рисунке 5.20(2) полуширина по полувысоте пика тока (рисунок II), однако ток от 5 до 2 А длится еще 3 мкс (красная стрелка на (II)). Форма осциллограммы тока еще больше напоминает осциллограммы квазиобратных ударов при контакте плазменных каналов, но и здесь на Рисунке 5.21(1) в нижнем углу мы видим хорошо узнаваемый восходящий положительный лидер, окруженные сетью из множества каналов.

### 5.3.2.3. Восходящий положительный лидер, инициированный с заземленного болта, до входа болта в отрицательно заряженное аэрозольное облако

На Рисунке 5.23 мы можем видеть ИК-изображение (I), которое фиксирует небольшой (около 7 см) положительный восходящий лидер (I.2), стартовавший с заземленного болта (I.1), хорошо видна восходящая к облаку стримерная корона положительного лидера (I.3),

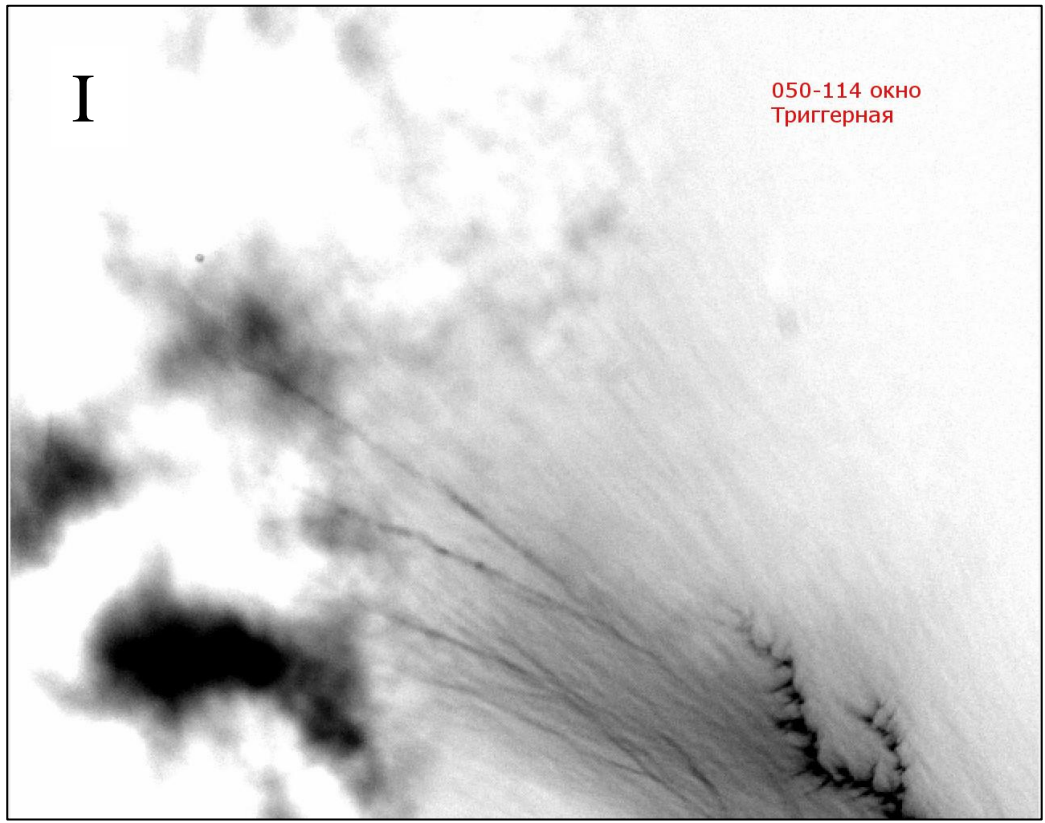

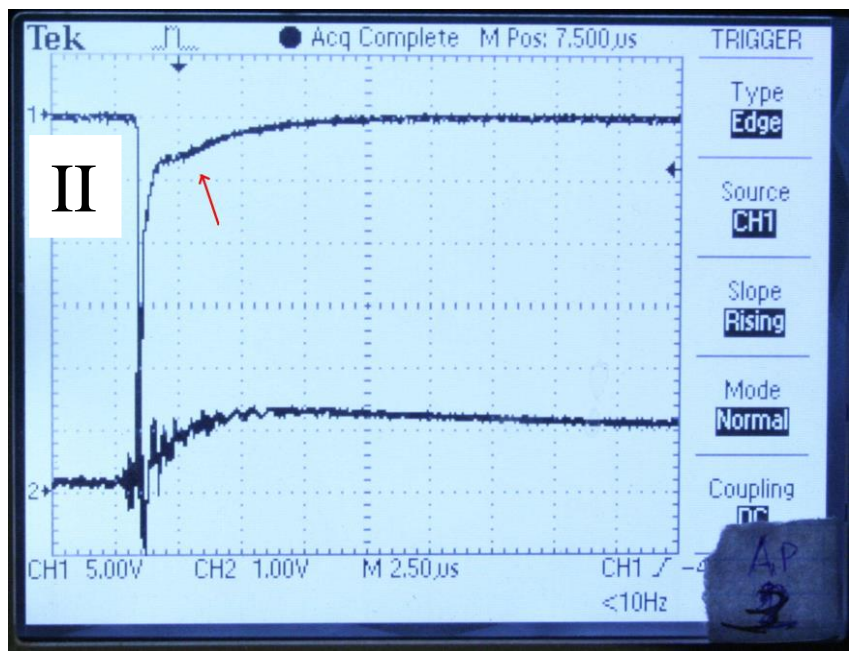

Рисунок 5.21 (событие 050-114_2013-01-18, заряд облака – минус). ИК-изображении (1), фиксирует иерархическую сеть плазменных каналов, инициированных заземленным болтом, летящим справа-налево, внутри аэрозольного облака (или положительным лидером, поднимающимся с болта). Выдержка 5 мс, частота кадров 115 Гц, 640х512 пикселей. Осциллограммы тока и контроля заряда облака (2). Верхний луч осциллограммы – измерение тока с шунтом 1 Ом. Нижний луч – сигнал с шара диаметром 50 см, который используется для контроля зарядки облака. Большое горизонтальное деление равно 2.5 мкс, большое вертикальное деление – 5 А.

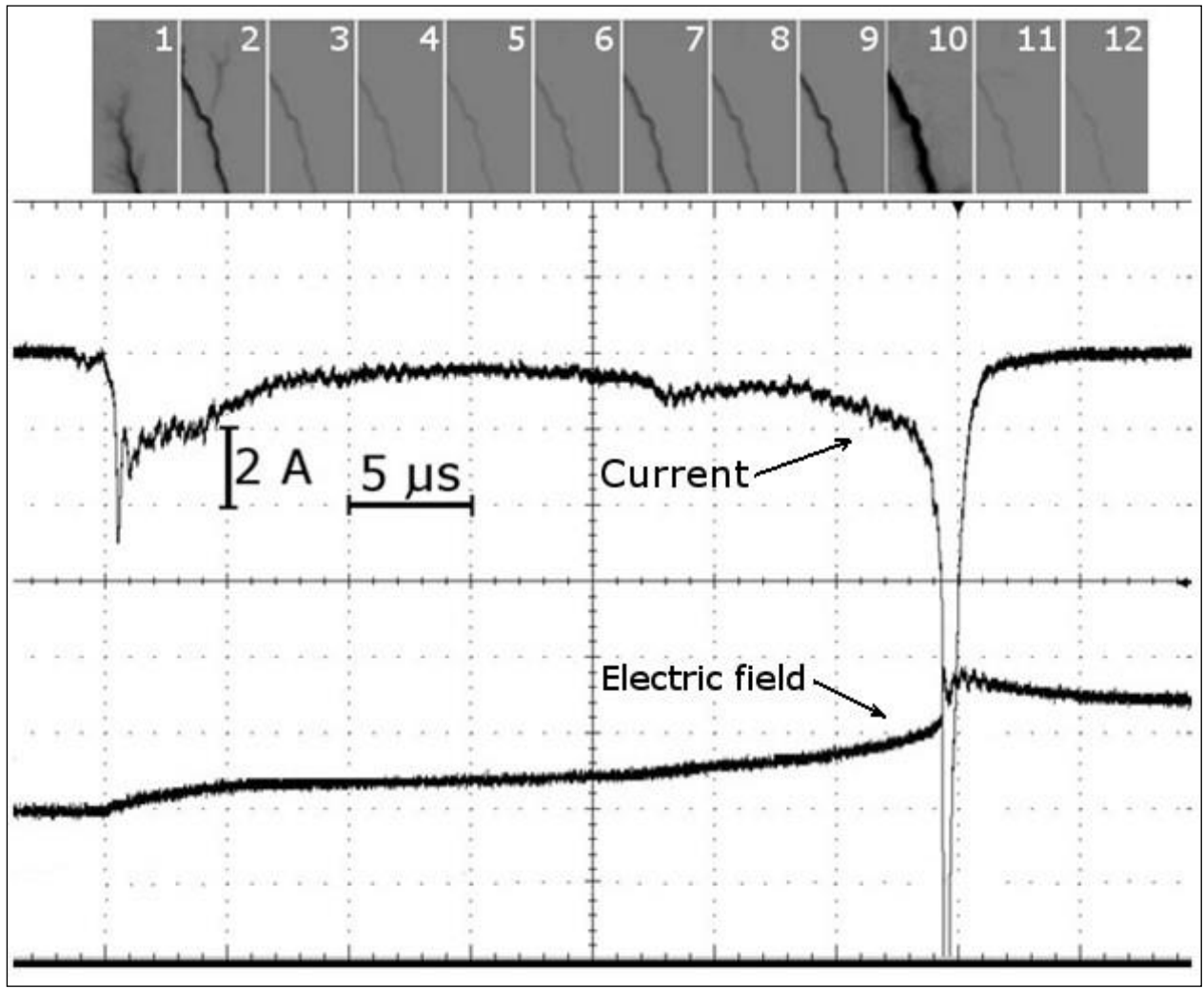

Рисунок 5.22 (адаптировано из [Kostinskiy et al., 2015c]). Двенадцать видеокадров, включающих фрагмент стадии движения вверх восходящего положительного лидера (кадры 1-8), сквозную фазу и квазиобратный удар (кадры 9-10) взаимодействия восходящего положительного лидера и нисходящего отрицательного лидера, а также распад плазмы единого канала (кадры 11-12). Кадры с экспозицией 4,44 мкс (225 000 кадров в секунду) получены камерой видимого диапазона FASTCAM SA4. В данных экспериментах (в тот же день) болт арбалета не использовался и стримеры, и лидеры инициировались самопроизвольно в электрическом поле отрицательно заряженного аэрозольного облака. Все кадры точно синхронизированы с осциллограммами тока и динамики заряда, расположенными ниже кадров, то есть, ширина и местоположение кадра соответствуют току и динамике заряда на осциллограммах в момент времени записи кадра. Соответственно, резкое усиление тока на осциллограммах, например, максимум тока в момент квазиобратного удара, соответствует резкому увеличению свечения на кадре 10.

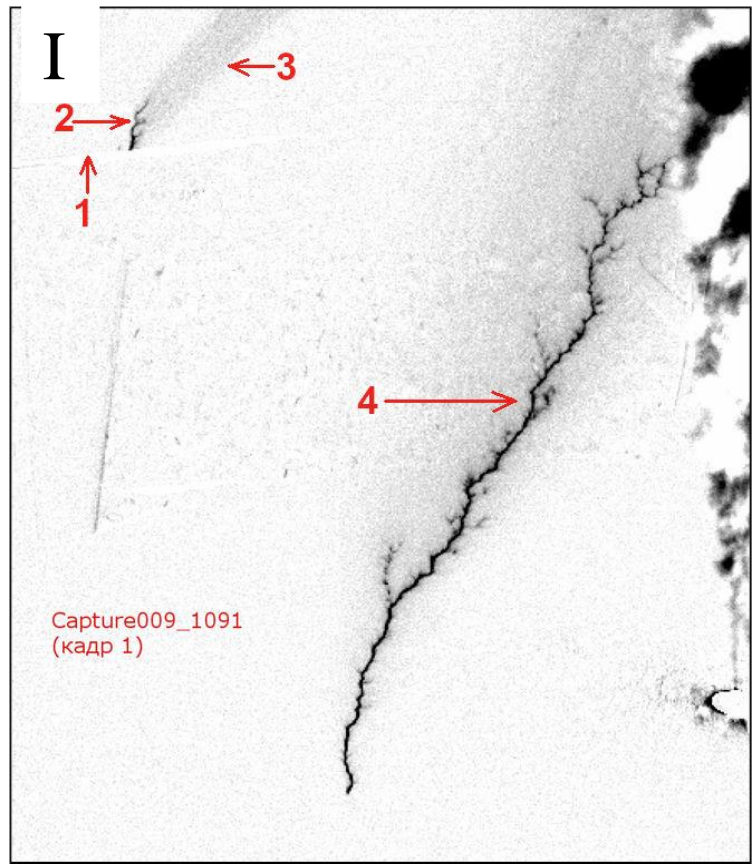

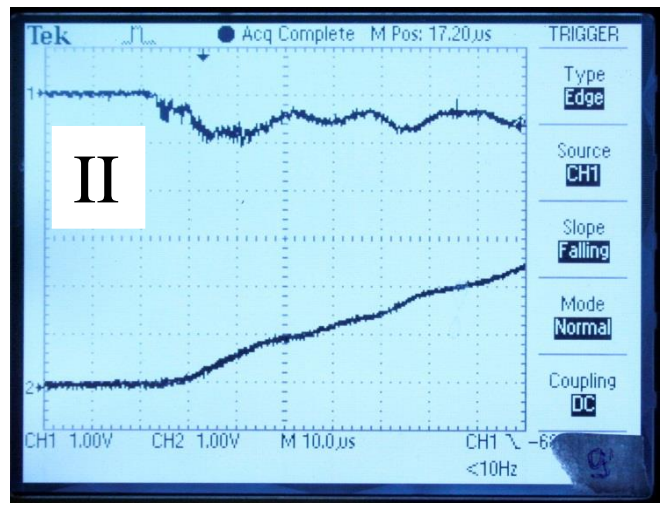

**Рисунок 5.23** (событие 009-1091_2013-01-18, заряд облака – минус). ИК-изображение (I), фиксирует положительный восходящий лидер (2), стартовавший с заземленного болта, который летит слева-направо (1), хорошо видна восходящая к облаку стримерная зона положительного лидера (3), также зафиксирован восходящий с заземленной сферы положительный лидер (4). Выдержка 4.3 мс, частота кадров 115 Гц, 640x512 пикселей. Осциллограммы тока и контроля заряда облака (II). Верхний луч осциллограммы – измерение тока с шунтом 1 Ом. Нижний луч – сигнал с шара диаметром 50 см, который используется для контроля зарядки облака. Большое горизонтальное деление равно 20 мкс, большое вертикальное деление – 1 А.

а также на кадре зафиксирован восходящий с заземленной сферы положительный лидер (I.4). Осциллограмма тока (II) фиксирует ток от 0.3 до 1 А в течение 80 мкс (зафиксирован не весь временной ход тока во время этого события). Так как и провод болта (I.1) и положительный лидер (I.4), восходящий со сферы напрямую связаны с системой измерения тока, то мы не можем исключить, что измеряем не только ток, текущий через заземленный болт (I.1), но и ток, одновременно текущий через положительный лидер (I.4). Длина (около 1 м) восходящего положительного лидера (I.4) и большое время протекания тока (больше 80 мкс) говорит в пользу того, что основной вклад в ток дает именно восходящий положительный лидер (I.4), а не лидер длиной всего 7 см. Но мы можем корректно сделать следующее заключение: ток, текущий через заземленный болт, когда он приближается к заряженному облаку (в каком месте осциллограммы тока он бы не суммировался с током положительного лидера (I.4)), не превышает 0.3 – 1 А. Это обычные токи для положительных восходящих лидеров, инициируемых в поле заряженного аэрозольного облака и, следовательно, восходящий с заземленного болта положительный лидер имеет те же параметры, что и обычный положительный лидер длинной искры.

### 5.3.2.4. Плазменные образования (UPFs), инициированные болтом арбалета внутри *положительно* заряженного аэрозольного облака

Как отмечалось выше, инициировать разряды в электрическом поле положительно заряженного аэрозольного облака гораздо сложнее, чем в поле отрицательного облака. Тем важнее инициировать разряды в положительно заряженном облаке.

На Рисунке 5.24(I) справа мы видим инициацию ярких веретенообразных плазменных образований, возможно UPFs, прямо на заземленном болте, который летит справа-налево. «Веретено» плазменных каналов сходится к плазменным образованиям внутри облака. ИК-изображение (II), фиксирует через 9 мс после первого кадра более слабые плазменные образования на заземленном болте, который сдвинулся влево. Хочется подчеркнуть, что в положительно заряженном облаке яркие плазменные каналы,

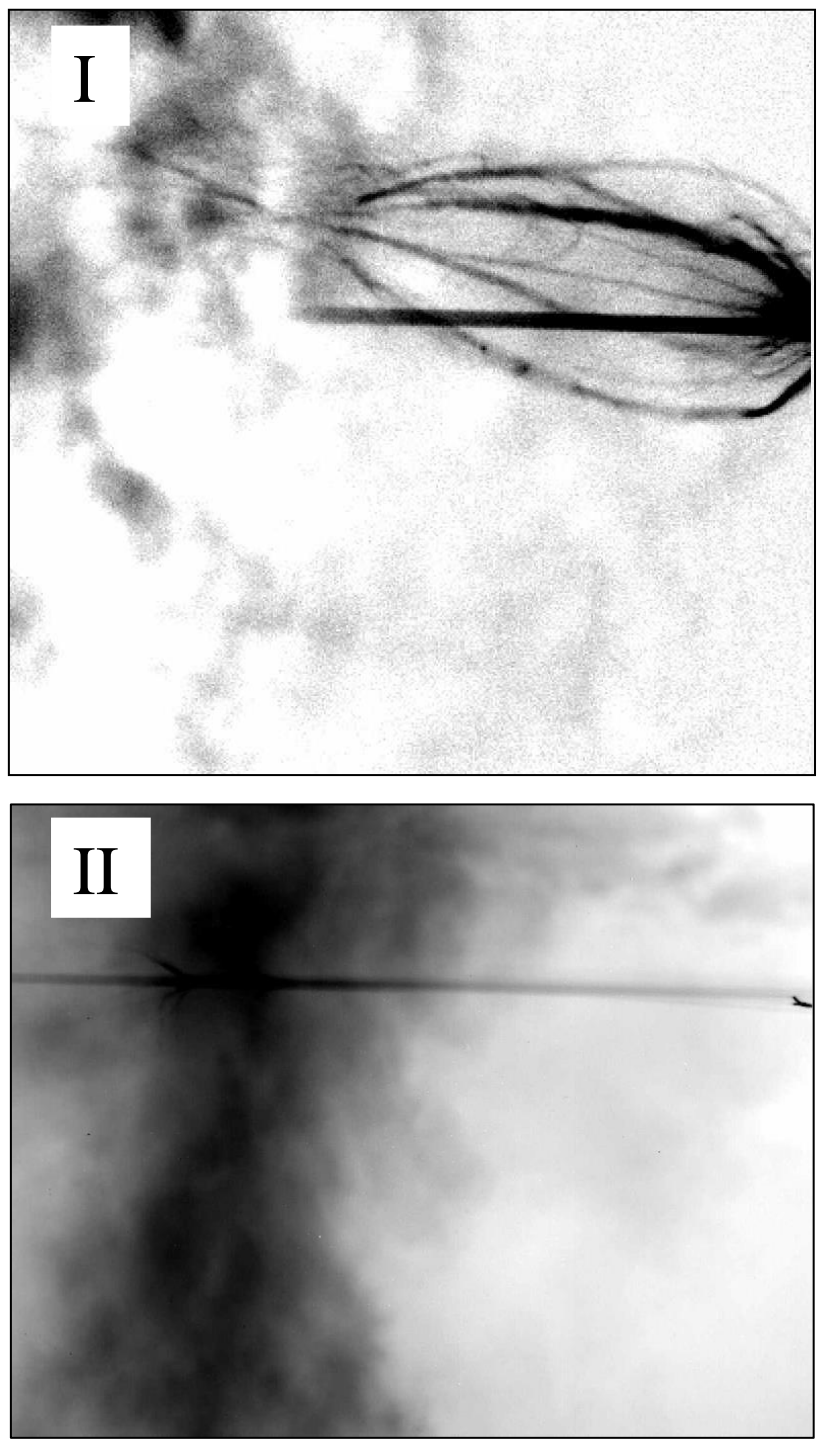

Рисунок 5.24 (Событие 113-182+_2013-01-30, заряд облака – Плюс). ИК-изображение (I), инвертированное, фиксирует образование ярких веретенообразных плазменных образований, возможно UPFs, прямо на заземленном болте, который летит справа-налево; ИК-изображение (II), фиксирует более слабые плазменные образования на заземленном болте, который сдвинулся влево.  Выдержка кадров 7 мс, частота кадров 115 Гц, 640х512 пикселей (полное изображение кадра).

возможно UPFs, взаимодействуют не с отрицательным лидером, который инициирован болтом, а с самим болтом. Эта ситуация не соответствует ситуации взаимодействия заземленного болта с плазменными образованиями внутри отрицательно заряженного облака (Рисунки 5.20-5.21). Это обстоятельство еще раз подтверждает большую разницу между отрицательно и положительно заряженным аэрозольным облаком с точки зрения гораздо большей легкости инициирования стримеров и лидеров в поле отрицательно заряженного облака. Рисунки 5.25-5.26, где плазменные сети также возникают на движущемся заземленном болте, подтверждают этот вывод.

Очень интересное изображение сложной плазменной сети внутри положительно заряженного облака можно видеть на Рисунке 5.27. ИК-изображение (I) фиксирует сложную и мощную иерархическую сеть плазменных образований (скорее всего UPFs), расположенных сверху и снизу от движущегося заземленного болта и инициированного болтом. Около поверхности болта все каналы имеют яркие отрезки. Болт движется справа-налево. На кончике болта также видны плазменные каналы. Увеличенный фрагмент изображения плазменной сети (II), показывает, насколько иерархична и сложна ее структура, которая совершенно непохожа на известные плазменные объекты, такие как стримеры, лидеры, спейс-лидеры, спейс-стемы. Причем многие из этих каналов, как и описанные в предыдущих главах 1-4, имеют яркое свечение в ИК-диапазоне, которое позволяет предположить, что это горячие, высокопроводящие каналы, которые были ранее названы необычными плазменными образованиями (UPFs).

Исходя из анализа Рисунков 5.24-5.27, мы можем предположить, что проникновение отрицательных лидеров (по аналогии с поляризованными отрицательными концами заземленных болтов) в положительно заряженное грозовое облако может приводить к инициации сложной сети плазменных каналов, которые будут взаимодействовать с отрицательным лидером. Этот процесс может быть подготовительным процессом для импульсного развития канала молнии и инициации следующих ударов.

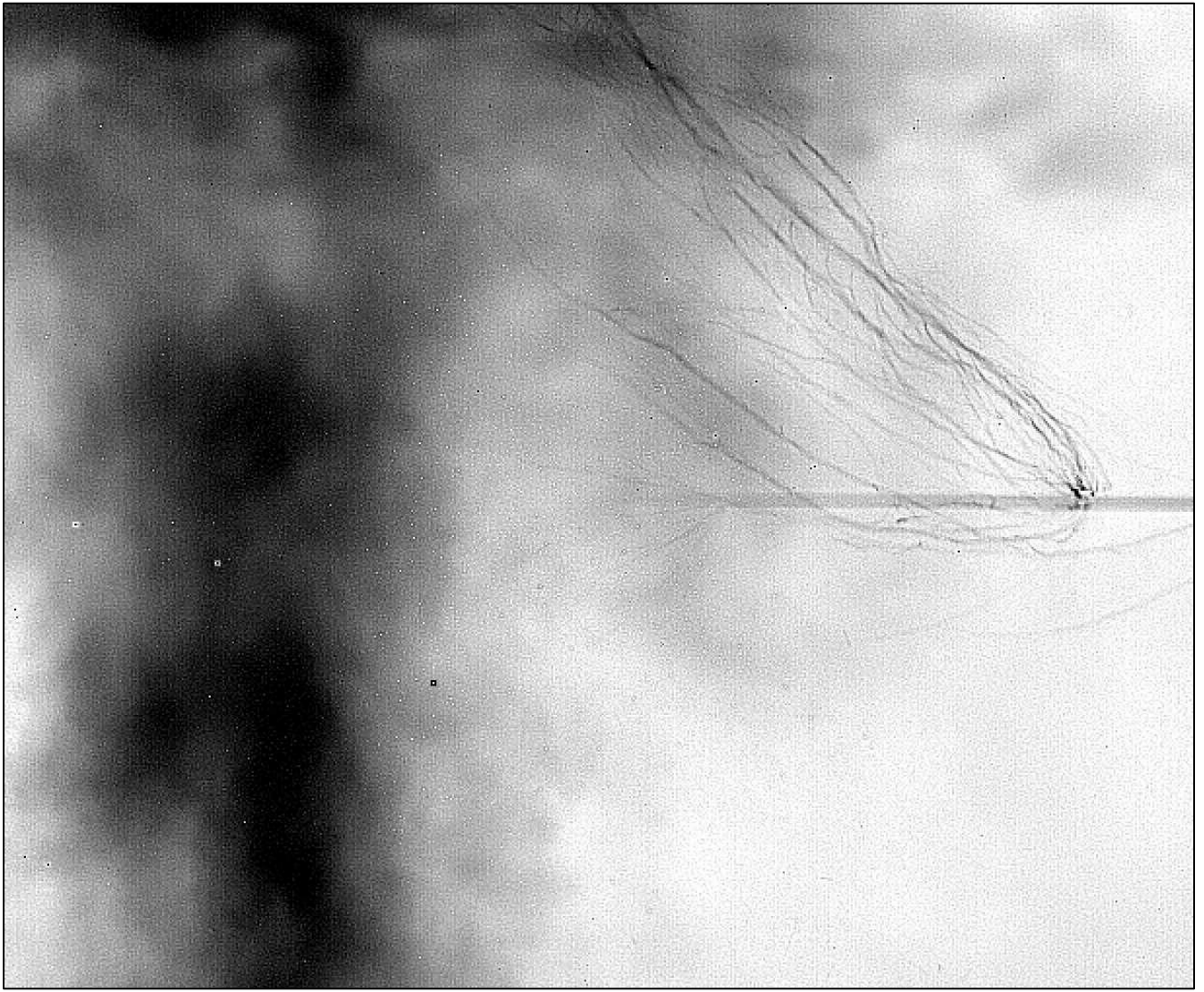

Рисунок 5.25 (Событие 119-170+_2013-01-30, заряд облака – Плюс). ИК-изображение, которое фиксирует образование сети плазменных образований, поднимающихся вверх с середины болта. Также слабые каналы инициируются с кончика болта. Болт движется справа-налево. Выдержка кадров 7 мс, частота кадров 115 Гц, 640х512 пикселей.

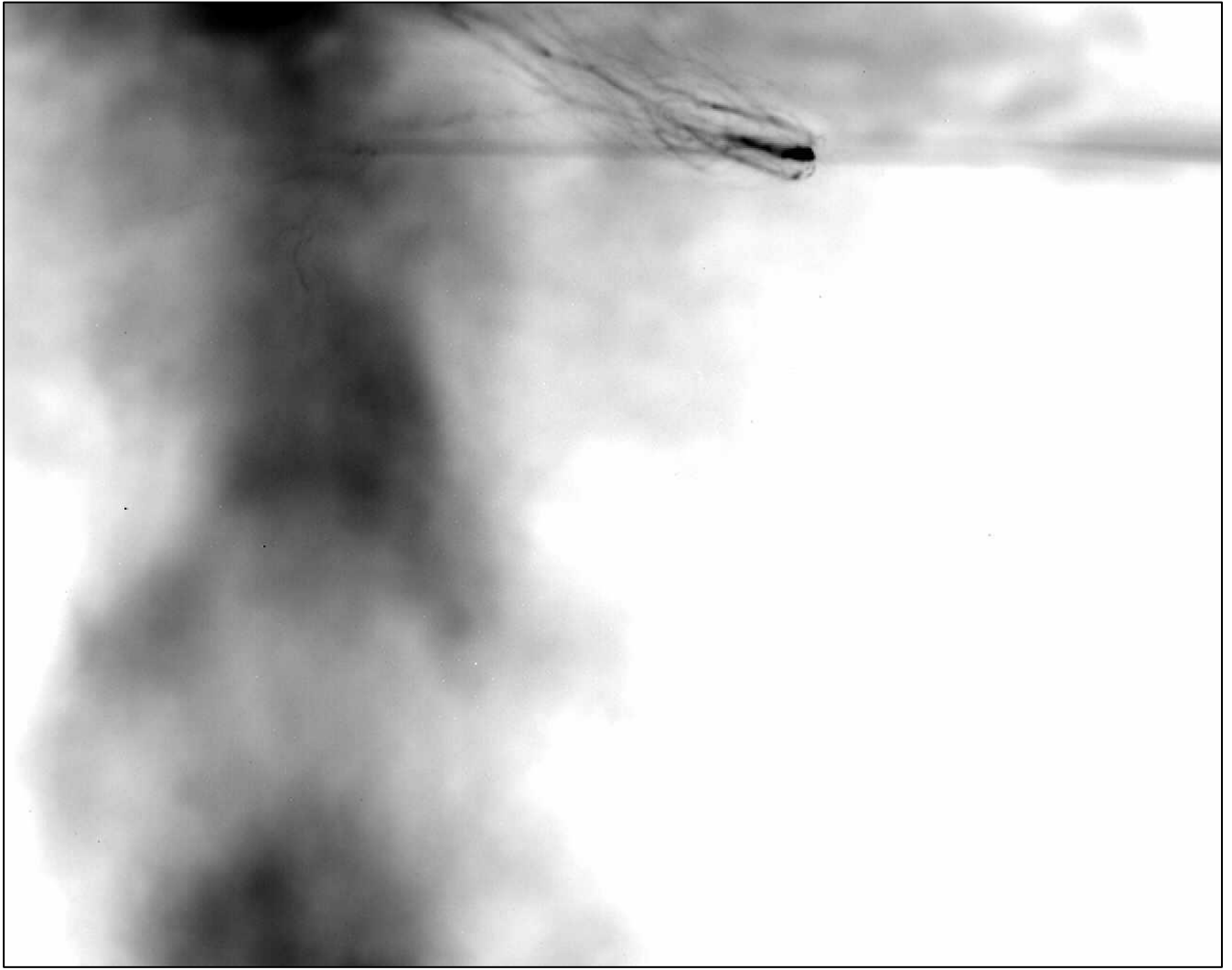

Рисунок 5.26 (Событие 121-304 +_2013-01-30, заряд облака – Плюс). ИК-изображение, которое фиксирует образование сети плазменных образований, поднимающихся вверх с середины болта. Болт движется справа-налево. Выдержка кадров 7 мс, частота кадров 115 Гц, 640х512 пикселей.

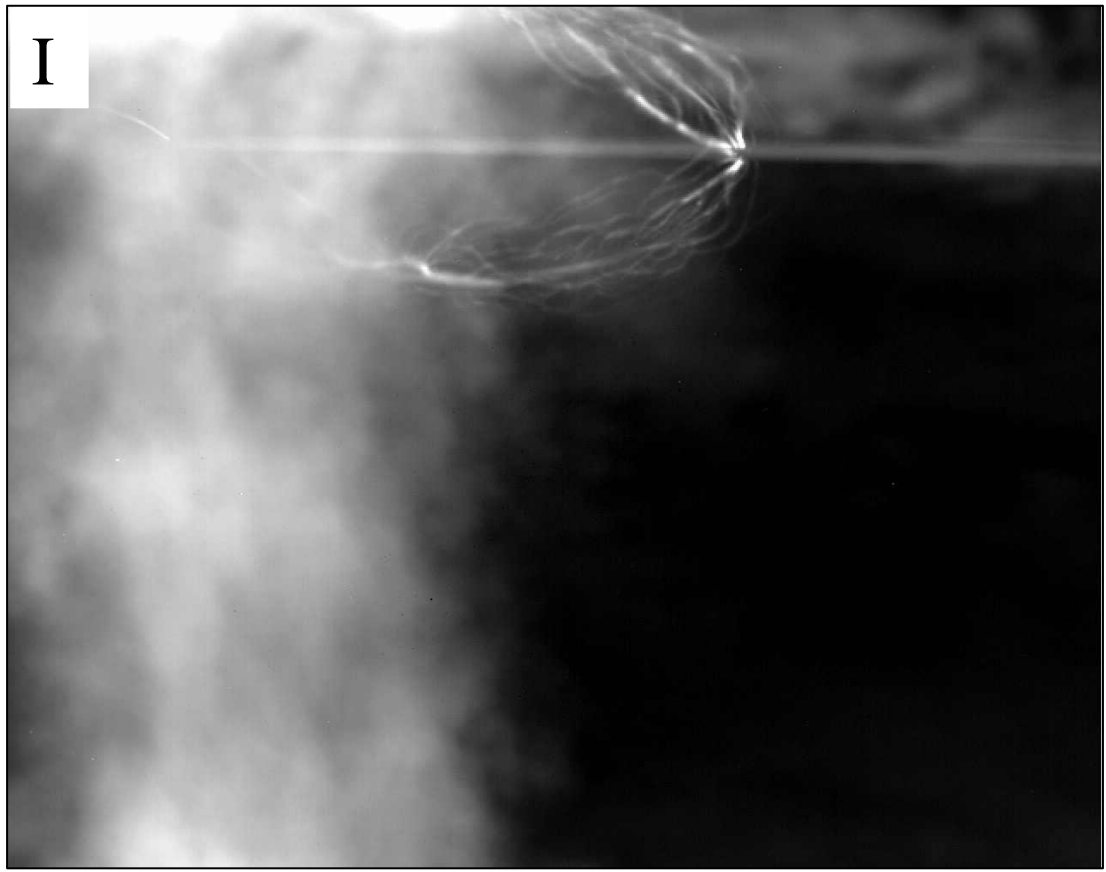

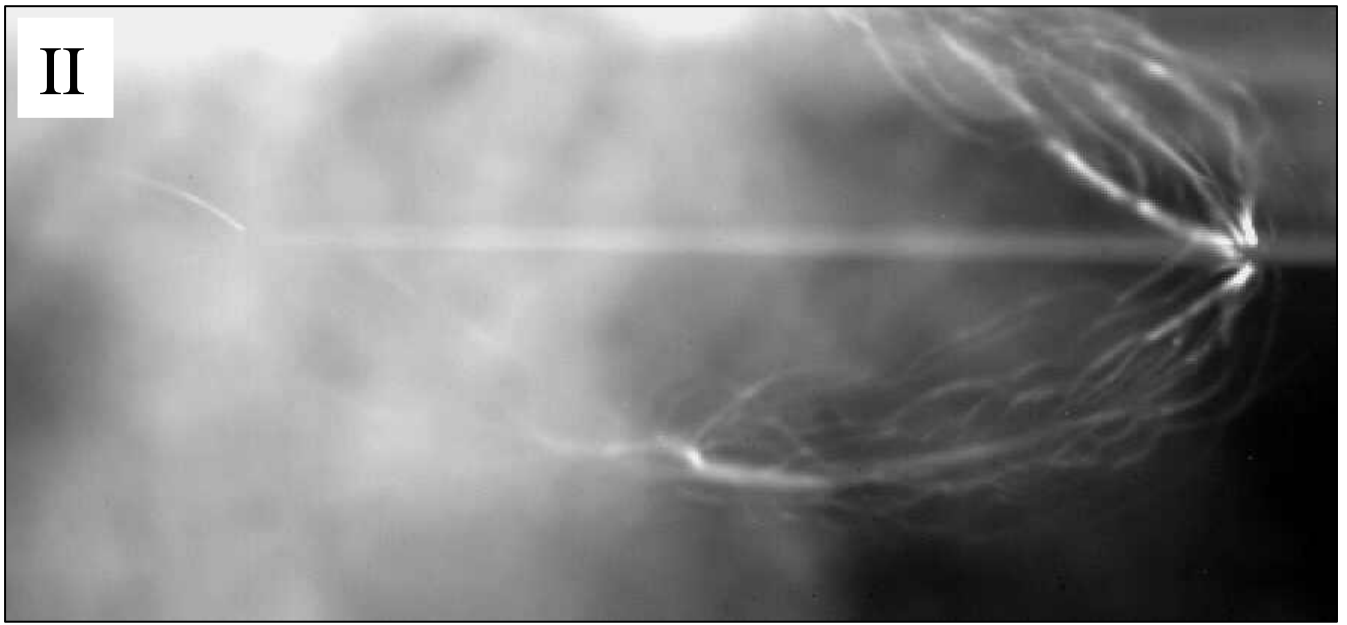

Рисунок 5.27 (Событие 120-84+_2013-01-30, заряд облака – Плюс). ИК-изображение (I), которое фиксирует сложную и мощную иерархическую сеть плазменных образований, расположенных сверху и снизу от движущегося заземленного болта. Болт движется справа-налево. На кончике болта также видны плазменные каналы. Увеличенный фрагмент изображения плазменной сети (II), показывает, насколько иерархична и сложна ее структура. Выдержка кадров 7 мс, частота кадров 115 Гц, 640х512 пикселей.

## 5.4. Выводы главы 5

1. Впервые продемонстрирована возможность моделирования в лабораторных условиях высотно-инициированных триггерных молний (altitude-triggered lightning) и триггерных молний с помощью инициации болтом арбалета (моделирующего летательный аппарат) разрядов, в электрическом поле облака отрицательно заряженного водного аэрозоля.

2. Обнаружена высокая степень подобия разряда, инициированного болтом в электрическом поле искусственного облака заряженного аэрозоля и высотно-инициированной молнии в природных условиях, а также инициированного заземленным болтом и триггерной молнией в поле грозового облака.

3. Параметры токов прекурсоров (стримерных вспышек) восходящих положительных лидеров в электрическом поле искусственного отрицательного облака заряженного аэрозоля оказались очень близкими к токам прекурсоров триггерных положительных лидеров в поле грозового облака, отличаясь по периоду колебаний из-за небольшой длины провода, прикрепленного к болту.

4. Благодаря использованным новым методикам ИК-измерений впервые удалось зафиксировать необычные плазменные образования (UPFs), инициированные длинными проводящими предметами (болтом арбалета с заземленным металлическим проводом) или восходящими с них лидерами, внутри отрицательного и положительного облака заряженного аэрозоля, причем токи, фиксируемые благодаря проводу, прикрепленному к болту арбалета, оказались в районе 30-40 А. Эти UPFs также кардинально отличаются по морфологии каналов (сетей) от стримерных вспышек и отрицательных и положительных лидеров длинной искры, как и ранее обнаруженные UPFs (см. главы 1-4).

5. UPFs, обнаруженные в отрицательно заряженном облаке образуют мощные разветвленные иерархические сети каналов, похожие на зафиксированные внутри облака без заземленного болта, но большие по числу каналов и более нагретые, судя по ИК-изображениям и общим текущим токам. Особенностью инициированных болтом UPFs являются также UPFs, имеющие почти прямые центральные каналы (от которых отходят в форме веретена более мелкие каналы) также или боле нагретые, чем положительные лидеры, зафиксированные на тех же изображениях.

6. UPFs, обнаруженные в положительно заряженном облаке также образуют мощные разветвленные иерархические сети каналов, направленные в сторону оси облака, но они инициируются непосредственно с заземленного болта и не похожи на UPFs, возникающие самопроизвольно (без болта) внутри положительно заряженного облака аэрозоля (глава 4).

**ГЛАВА 6. Ступенчатое развитие отрицательного и положительного лидера, приводящее к мощной вспышке стримерной короны: исследования длинных искр, инициированных генераторами импульсных напряжений (ГИН), в целях моделирования развития отрицательных и положительных каналов молний**

В данном разделе представлены результаты исследований [Kostinskiy et al., 2018], где изучалась физика ступенчатого развития положительных и отрицательных лидеров, развивающихся в воздушных промежутках от 4 до 10 м, в электрическом поле генераторов импульсных напряжений (ГИН) которые подавали на разрядный промежуток высокие напряжения с фронтом 100 мкс и длительностью импульса 7500 мкс. Ступенчатое развитие лидеров, где каждая ступень завершилась яркой вспышкой стримерной короны, наблюдалось как для отрицательных (ожидаемых для «классического» ступенчатого процесса), так и для положительных (ожидаемых в случае так называемого «рестрайков» лидеров, restrike process). Мы будем называть рестрайки (restrikes), резкие удлинения лидерных каналов на десятки сантиметров, — «ступенями положительного лидера» или «положительными ступенями». В разделе представлены изображения ступеней отрицательных и положительных лидеров высокого пространственного и временного разрешения с яркими вспышками стримерной короны. Морфология вспышек стримерной короны во время ступеней практически не зависела от полярности лидера. Наблюдались стримерные вспышки с симметрией, близкой к сферической. Для положительных лидеров каналы ступеней были почти прямыми (что ранее не отмечалось в других исследованиях) и имели длину от 50 до более чем 120 см. Для отрицательных лидеров большинство ступеней были изогнутыми, а их длина в плоскости кадра составляла несколько десятков сантиметров. Результаты исследований говорят в пользу предположения, что физический механизм ступенчатого развития положительных лидеров, несмотря на его не изученность, принципиально отличается от теоретического механизма квазинепрерывного («оптически непрерывного») развития положительного лидера как в длинных искрах [Bazelyan and Raizer, 1998], [Bazelyan et al., 2007], [Popov, 2009], так и, возможно, в молнии (см. подробное изложение ниже).

Под квазинепрерывным движением положительного лидера длинной искры, понимается движение оптически мало разрешимыми ступеньками, длина которых сопоставима с размером головки лидера (∼1 см или меньше) [Горин и Шкилев, 1974], [Базелян и др., 1978, с. 61], [Базелян и Райзер, 1997, с. 232–233]. По современным представлениям, движение положительного лидера является принципиально прерывистым движением мелкими ступеньками (∼1 см или меньше) и не может быть непрерывным, так как преобразование стримерной короны в лидерный канал (стримерно-лидерный переход) происходит в стеме (месте истечения стримеров размером ∼1 см или меньше) на электроде или в головке положительного лидера [Bazelyan and Raizer, 1998], [Bazelyan et al., 2007], [Popov, 2009].

Первый механизм образования больших ступеней положительного лидера, не имеет в данный момент теоретического объяснения. Большие «внезапные, резкие удлинения» (скачки, ступени) положительного лидера длиной в десятки сантиметров возникают, по современным преставлениям, в условиях относительно высокой абсолютной влажности ( > 10 г/м$^3$) [Les Renardieres Group, 1977].

Второй механизм (режим) больших удлинений положительного лидера известен давно (например, [Les Renardieres Group, 1974], [Горин и Шкилев, 1974]). Если приложенное к промежутку напряжение нарастает достаточно медленно (меньше, чем 5 кВ/мкс) [Базелян и др., 1978, с. 53], то положительный лидер *всегда* начинает свое движение в этом режиме. В русскоязычной литературе второй режим называется «вспышечной фазой» развития положительного лидера (например, [Горин и Шкилев, 1974], [Базелян и др., 1978, с. 53]), так как на фотохронограммах канал лидера неоднократно вспыхивает и удлиняется, а потом следует темновая пауза, когда лидер не виден. После того, как напряжение вырастет до определенной величины, «вспышечный» режим распространения лидера сменяется квазинепрерывным ([Les Renardieres Group, 1974], [Горин и Шкилев, 1974]).

В этом исследовании мы сосредоточимся на первом, малоисследованном и малопонятном в данный момент механизме внезапного возникновения ступеней положительного лидера длиной в десятки сантиметров, которые возникают при высокой влажности (>10 г/м$^3$) после того, как отчетливо сформировался квазинепрерывно развивающийся положительный лидер.

## 6.1. Введение в главу 6.

Экспериментально установлено группой сэра Базиля Шонланда (sir Basil Ferdinand Jamieson Schonland) в ЮАР в 1930-х годах, что молния распространяется благодаря предшествующему обратному удару плазменному процессу, который они назвали лидером. Группа Шонланда также установила, что отрицательные лидеры молний, развивающиеся в чистом воздухе, распространяются ступенчато [Schonland, 1933,1938, 1956], [Schonland et al, 1935, 1938a, 1938b]. Также Шонланд предположил, что перед каналом лидера молнии движется некоторая плазменная структура, испускающая стримеры или являющаяся некоторой модификацией стримеров (pilot streamer) [Schonland, 1953]. К сожалению, в то время экспериментальное и теоретическое представление о стримерах только формировалось и взгляды Шонланда на pilot streamer сейчас трудно однозначно интерпретировать. Современные представления о тонкой структуре стримерной зоны отрицательного лидера, где происходит процесс формирования ступеней, включают в себя такие основные элементы стримерной зоны отрицательного лидера (разной плазменной природы), как спейс-стем и спейс лидер, что было впервые было установлено Стекольниковым и Шкилевым [Stekolnikov and Shkilyov, 1963] при исследовании длиной искры с помощью фотохронограмм электронно-оптических преобразователей, Рисунок 6.1 (в публикации [Stekolnikov and Shkilyov, 1963] они спейс-стем еще называют «ступенчатым лидером»). Более развернуто и подробно стримерная зона отрицательных лидеров длинной искры была исследована Гориным и Шкилевым [Горин и Шкилев, 1976], но они называли спейс-стем (самое непонятное плазменное образование стримерной зоны отрицательного лидера) —«стримером второго рода». Наиболее ясное изложение их схемы в почти современных терминах формирования ступени отрицательного лидера приведено на Рисунке 6.2 из [Rakov and

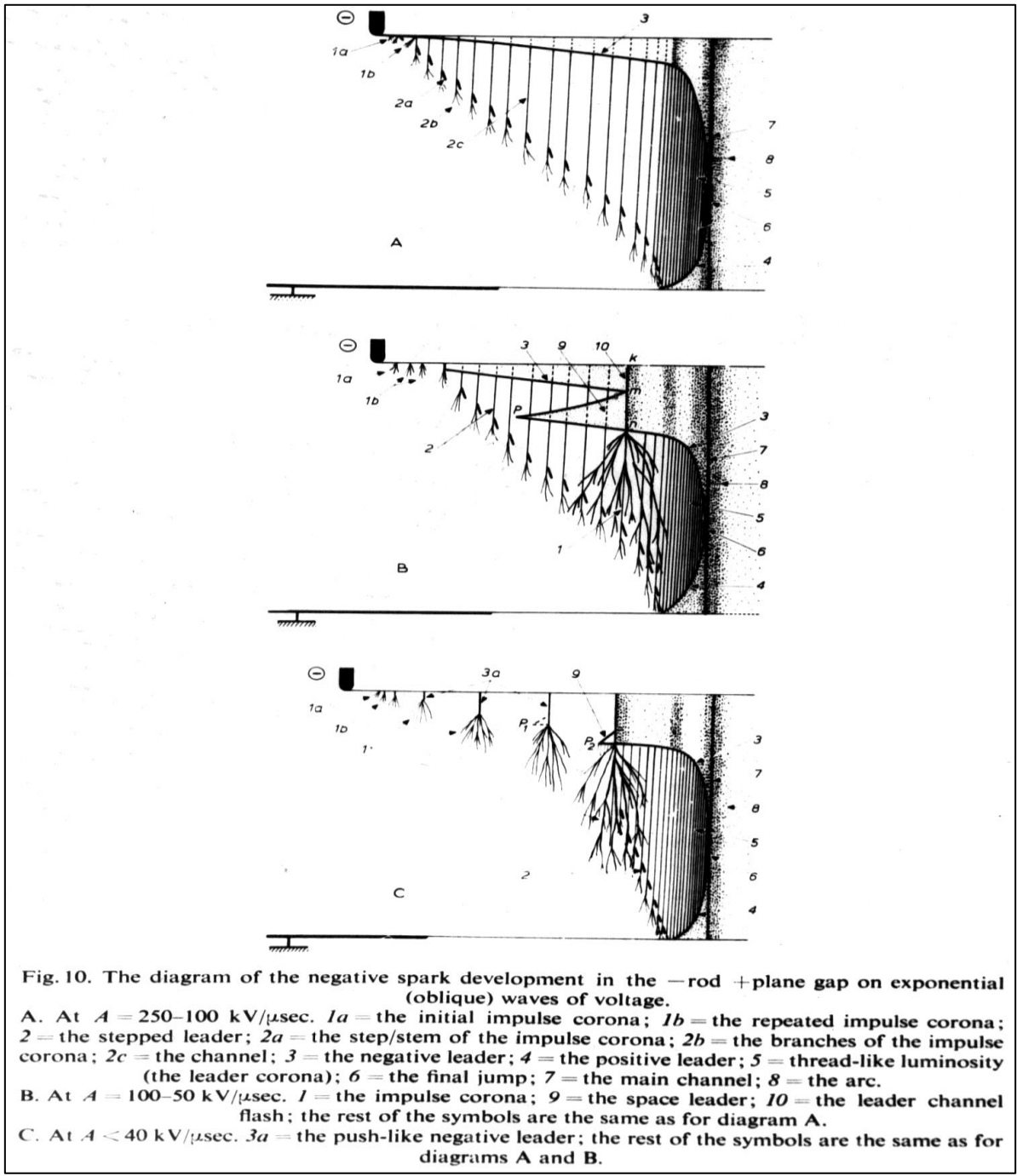

Fig. 10. The diagram of the negative spark development in the —rod +plane gap on exponential (oblique) waves of voltage.

A. At A = 250–100 kV/μsec. *1a* = the initial impulse corona; *1b* = the repeated impulse corona; *2* = the stepped leader; *2a* = the step/stem of the impulse corona; *2b* = the branches of the impulse corona; *2c* = the channel; *3* = the negative leader; *4* = the positive leader; *5* = thread-like luminosity (the leader corona); *6* = the final jump; *7* = the main channel; *8* = the arc.
B. At A = 100–50 kV/μsec. *1* = the impulse corona; *9* = the space leader; *10* = the leader channel flash; the rest of the symbols are the same as for diagram A.
C. At A < 40 kV/μsec. *3a* = the push-like negative leader; the rest of the symbols are the same as for diagrams A and B.

Рисунок 6.1 (Fig.10 из [Stekolnikov and Shkilyov, 1963]). Скетч разных режимов распространения отрицательных лидеров. (А) — без образования спейс-лидера, (В) — «обычный» режим с образованием спейс-лидера, который обозначается буквой Р, (С) — вспышечный режим с образованием спейс-лидеров, обозначенных $P_1$ и $P_2$. На всех трех рисунках спейс-стем ошибочно называется «ступенчатым лидером». В будущих исследованиях было показано, что 2а – спейс-стем, 2b – отрицательные стримеры спейс-стема, 2с – положительные стримеры спейс-стема. Разрядные промежутки 1-3 м, напряжение 1.5/1000 мкс.

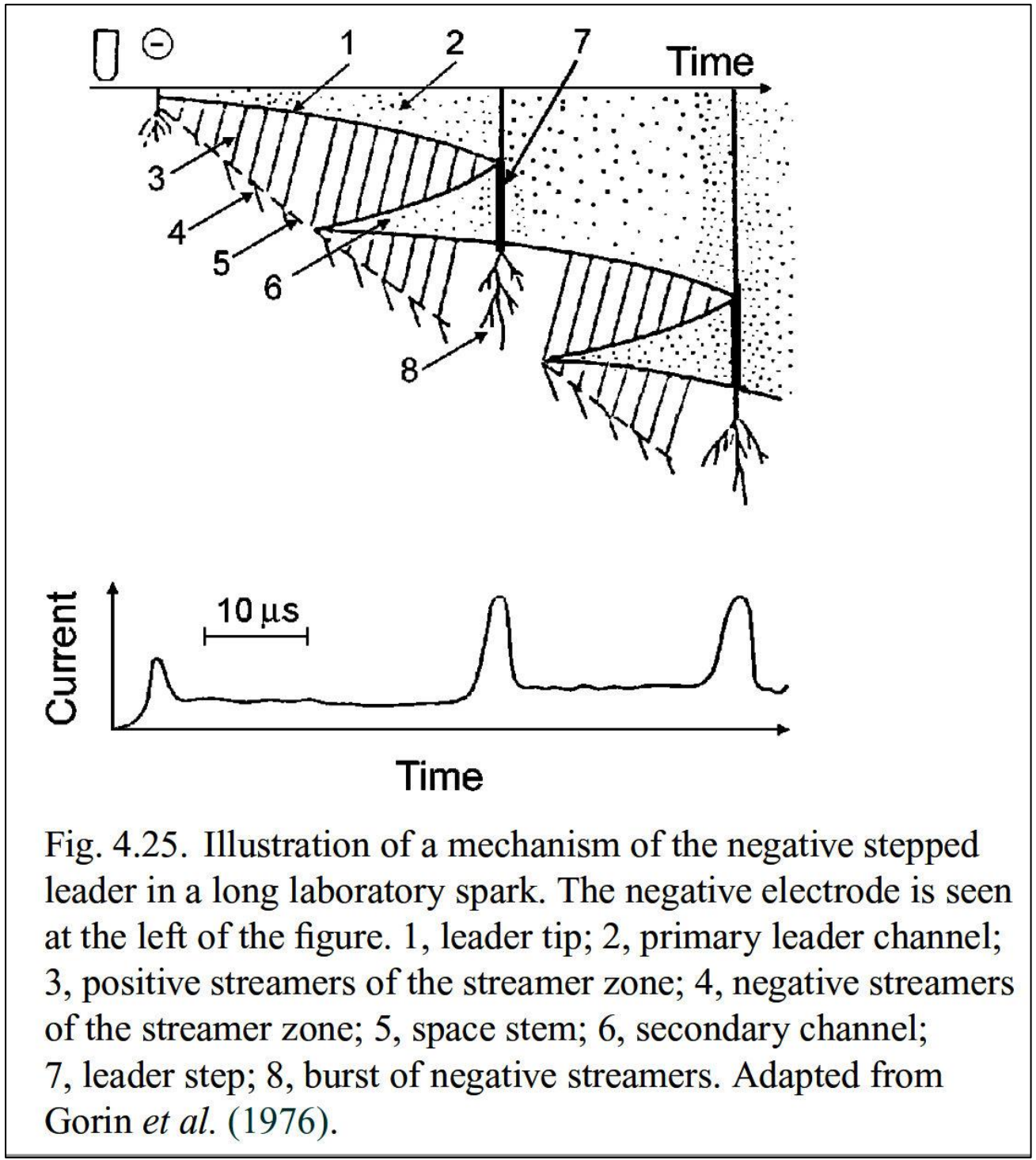

Fig. 4.25. Illustration of a mechanism of the negative stepped leader in a long laboratory spark. The negative electrode is seen at the left of the figure. 1, leader tip; 2, primary leader channel; 3, positive streamers of the streamer zone; 4, negative streamers of the streamer zone; 5, space stem; 6, secondary channel; 7, leader step; 8, burst of negative streamers. Adapted from Gorin *et al.* (1976).

Рисунок 6.2 (Любезно предоставлен Владимиром Раковым). Схема механизма образования ступеней отрицательного лидера длинной искры, которая также является предполагаемым механизмом развития ступеней отрицательного лидера молнии [Rakov and Uman, 2003]. Эту схему Rakov и Uman адаптировали из работ [Горин и Шкилев, 1976], [Gorin et al., 1976].

Uman, 2003], но здесь спейс-лидер называют «вторичным каналом». Мы будем придерживаться терминов спейс-стем и спейс-лидер во всем тексте. Ступень отрицательного лидера, образуется во время контакта спейс-лидера и основного отрицательного канала лидера, завершаясь мощной вспышкой стримерной короны, начинающейся из области близкой к головке лидера.

Элементы тонкой структуры стримерной зоны отрицательных лидеров в естественных и триггерных молниях стали наблюдать только недавно [Biagi et al., 2010, 2014], [Hill et al., 2011], [Petersen and Beasley, 2013], [Gamerota et al., 2014, 2015], [Jiang et al., 2017] и пока экспериментальные данные не дают возможности надежно отличить в стримерной зоне, спейс-лидер от спейс-стема на этих изображениях, хотя сам факт фиксирования структуры стримерной зоны отрицательного лидера молнии является чрезвычайно важным и полезным. В настоящее время можно отчетливо установить только то, что некоторые плазменные образования отделены от основного канала (например, Рисунок 6.21 [Petersen and Beasley, 2013], Рисунок 6.22 [Biagi et al., 2010]), на концах отрицательных лидеров наблюдаются вспышки, делающие более яркими области около конца лидера (например, Рисунок 6.20 [Hill et al., 2011]) и обнаруживаются структуры «волосяного типа», идущие от каналов лидера, которые можно интерпретировать, как стримеры (например, Рисунок 6.21 [Petersen and Beasley, 2013], Рисунок 6.22 [Biagi et al., 2010]), но эти же структуры могут быть и ветвями лидерных каналов. Вывод в упомянутых выше статьях относительно присутствия спейс-стемов и/или спейс-лидеров делается на основании того, что между основным каналом молнии и плазменными образованиями в стримерной короне существуют более темные места. Поэтому в литературе (например, [Biagi et al., 2010], [Jiang et al., 2017]) часто используется такое выражение как спейс стем/лидер (space leaders/stems), по нашему мнению, без достаточных объяснений, что усложняет процесс понимания. Может создаться впечатление, что спейс-стем и спейс-лидер — это почти одно и тоже плазменное образование. Корректно понимать, по нашему мнению, такое обозначения (space leaders/stems) стоит в том смысле, что «перед основным каналом отрицательной молнии видны какие-то плазменные образования точную природу которых мы не можем идентифицировать и поэтому мы считаем, что это либо спейс-стем, либо спейс-лидер». Эксперименты с длинной искрой прямо указывают, что это совершенно два разных плазменных образования: спейс-стем в длинной искре существует всегда, а спейс-лидер появляется в достаточно длинных промежутках. Ступени отрицательного лидера появляются только тогда, когда существует спейс-лидер. Спейс-лидер является тонким плазменным каналом той же природы, что и основной лидер (преимущественно термическая ионизация плазмы с температурой нейтралов выше 3000-5000 К) с двумя выраженными головками, которые снабжаются положительными стримерами. Спейс-стем является гораздо более холодным плазменным образованием, поэтому его и назвали Горин и Шкилев, [Горин и Шкилев, 1976] стримерами второго рода, то есть относительно холодными плазменными образованиями. Состояние плазмы спейс-стема даже длинной искры, не говоря уже о молнии, пока изучено крайне слабо, но уже сейчас понятно, что оно отличается от плазмы лидеров и спейс-лидеров Хорошо известно, что спейс-стем «живет» благодаря не только положительным стримерам, которые связывают его с основным каналом (или со спейс-лидером), но и благодаря отрицательным стримерам, которые стартуют с другого его конца и обеспечивают движение спейс-стема.

В отличие от отрицательных лидеров, для которых ступенчатый процесс является основным механизмом движения, длинные положительные лидеры (в промежутках более 2-3 метров при скорости нарастания напряжения > 5 кВ/мкс) в лабораторных условиях обычно распространяются квазинепрерывно, то есть движутся небольшими ступеньками, которые часто оптически неразрешимы и поэтому движение канала кажется и часто называется) непрерывным. Длина этих ступенек сопоставима с размером головки лидера (~1 см или меньше), а физическим механизмом образования этих ступенек, который является и механизмом, обеспечивающим движение, считается механизм ионизационно-перегревной неустойчивости [Bazelyan and Raizer, 1998], [Bazelyan et al., 2007], [Popov, 2009]. Для этого процесса распространения положительного лидера мы будем использовать термин «квазинепрерывный» по отношению к удлинению лидера ступеньками размером около диаметра головки лидера (0.3-1 см), которое выглядит «оптически непрерывным процессом». Для так называемых коммутационных импульсных напряжений (фронт импульса напряжения 100-400 мкс, длительность импульса 2.5-10 мс) квазинепрерывно распространяющиеся лидеры положительной искры обычно (например, [Горин и Шкилев, 1974], [Les Renardieres Group, 1977], [Domens et al., 1991], [Gallimberti et al., 2002]) характеризуются относительно постоянным диапазоном тока 0.2-2 А и относительно постоянной средней скоростью движения от 1 до $3 \times 10^4$ м/с (1-3 см/мкс), а также высокой извилистостью канала. Ступенчатое развитие положительных лидеров в длинных искрах, как писалось выше, наблюдалось при двух условиях: (1) относительно длительное время нарастания напряжения (скорость поднятия напряжения меньше, чем 5 кВ/мкс, так называемый вспышечный режим распространения лидера [Bazelyan and Raizer, 1998]) или (2) при относительно высокой абсолютной влажности (> 10 г/м$^3$ [Les Renardieres Group, 1977]) при скорости подъема напряжения >10 кВ/мкс.

## 6.2. Ступенчатое развитие положительного лидера длинных искр

На наш взгляд, необходимо более подробно остановиться на состоянии исследований больших ступеней положительного лидера, которые возникают во время квазинепрерывного развития и прерывают его на время (что будет предметом нашего исследования). Наше исследование не касалось вспышечной формы развития положительного лидера до момента старта квазинепрерывного лидера при небольших скоростях поднятия напряжения (меньше, чем 5 кВ/мкс) [Bazelyan and Raizer, 1998], [Bazelyan and Popov, 2020], но мы ниже рассмотрим некоторые особенности этого режима.

Сначала сделаем краткий обзор исследований для случая высокой абсолютной влажности и скоростей поднятия напряжения >10 кВ/мкс. Этот случай мы будем исследовать ниже. Коллаборация европейских ученых Les Renardieres [Les Renardieres, 1972, 1974, 1977] на основании своих исследований положительных лабораторных искр в промежутках 5-10 м сообщала о резком увеличении длины положительного лидерного канала на десятки сантиметров на фоне квазинепрерывного развития. Каждое из этих удлинений положительного лидера сопровождалось сильным импульсом тока и мощной вспышкой стримерной короны, выходящей из области, близкой к головке лидера. В то же время падение напряжения вдоль канала лидера уменьшалось из-за резкого увеличения проводимости канала. Группа Les Renardieres назвала эти события «restrikes» или «reillumination» (для этих явлений не выработаны устоявшиеся русскоязычные термины, так как они не были подробно исследованы в Советском Союзе). Мы будем далее называть это явление («restrikes») «ступенями положительного лидера» или

«положительными ступенями». Точное определение рестрайков (restrikes), данное Les Renardieres Group [Les Renardieres, 1977, стр. 134], звучит так: «Это быстрое прерывистое увеличение длины лидера, сопровождающееся кратковременным сильным увеличением его светимости и резким увеличением тока в разряде». В таком виде определение применимо и к ступеням отрицательного лидера, так как [Ortega et al., 1994] и [Reess et al., 1995] также называют «restrikes» процесс образования ступеней отрицательного лидера, что означает, что каждый из терминов просто указывает на прерывистое (скачкообразное) расширение лидерных каналов, независимо от конкретного физического механизма для положительного или отрицательного лидера. Группа Les Renardieres обнаружила, что вероятность наблюдения ступеней положительного лидера быстро увеличивается с увеличением абсолютной влажности, от очень низкой при < 8 г/м$^3$ до практически 100% при > 12 г/м$^3$ (Рисунок 6.3). Ступеней положительного лидера может быть несколько. При высокой влажности количество положительных ступеней увеличивалось с увеличением длины разрядного промежутка, при этом было зарегистрировано в среднем 3-4 положительных ступени в 10-метровом разрядном промежутке и не менее 1 ступени в промежутке 5 м (Рисунок 6.4). Длина первой положительной ступени для промежутка 5 м обычно составляла 0,6–1,0 м, а соответствующее ей импульсное внедрение заряда в объем составляло 12–60 мкКл (для промежутка 10 м длина первой ступени составляла в среднем 0.55-0.65 м, а внедренный заряд составлял 15-25 мкКл). Средняя длина ступени для промежутка 10 м увеличивалось с номером ступени, достигая 1,5 м для 3-ей и 4-ой положительной ступени, а соответствующее им импульсное внедрение заряда в объем составляло 35–40 мкКл (Рисунок 6.5). Положительные ступени вызывают резкое увеличение общей средней скорости лидера, когда влажность превышает 10 г/м$^3$ (фронт напряжения составлял 390 мкс). Амплитуда тока во время ступеней положительного лидера оценивалась в пределах от 10 А до сотен ампер [Les Renardieres Group, 1972, стр. 132]. Для положительных ступеней, характеризующихся более короткими импульсами тока и света с быстрым нарастанием (доли микросекунды), Baldo и Rea [Baldo and Rea, 1973, 1974] сообщили, что средняя скорость движения вперед положительных ступеней оценивается в $10^6$ м/с (пока эта методика измерений дает наиболее достоверные цифры), что на 1-2 порядка больше, чем наблюдаемая скорость квазинепрерывно расширяющихся положительных лидеров длинной искры 1–3x$10^4$ м/с. Для положительных ступеней, характеризующихся более или

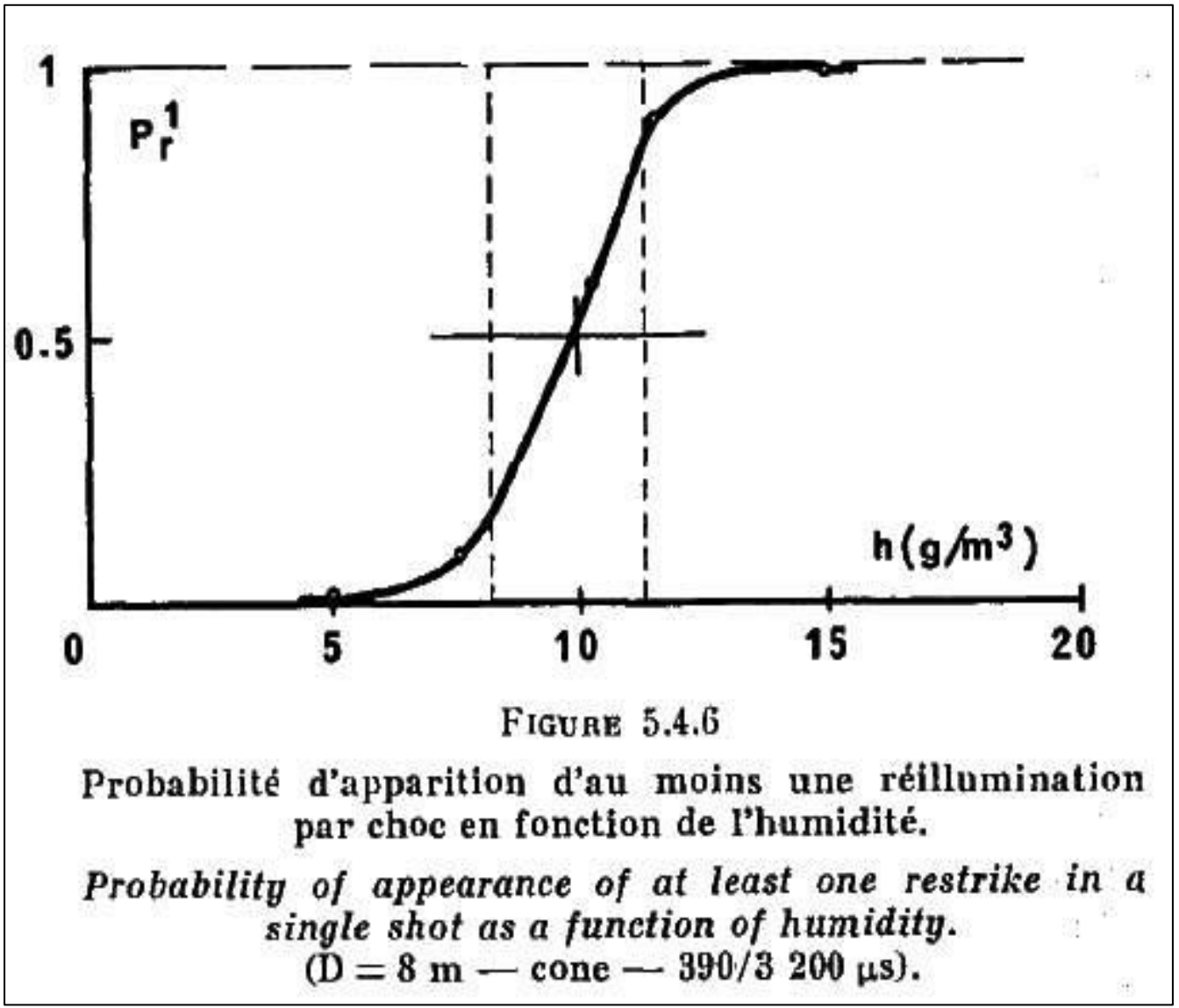

FIGURE 5.4.6

Probabilité d'apparition d'au moins une réillumination
par choc en fonction de l'humidité.

*Probability of appearance of at least one restrike in a
single shot as a function of humidity.*
(D = 8 m — cone — 390/3 200 μs).

Рисунок 6.3 (Fig. 5.4.6 из [Les Renardieres Group, 1977]). Вероятность появления хотя бы одной ступени (restrike) положительного лидера в одном разряде, как функция абсолютной влажности воздуха. Промежуток стержень-плоскость длиной 8 м, стержень оканчивался конусом с минимальным радиусом в 1 см. Данная кривая снята для конкретного импульса напряжения 390/3200 мкс, что означает, что для других импульсов она может немного меняться.

TABLEAU 3.5.1 — TABLE 3.5.1

*Caractéristiques des réilluminations*     Restrike Characteristics

| $T_{cr}$ (μs) | | | 130 | | | | 260 | | | | 500 | | | |
|---|---|---|---|---|---|---|---|---|---|---|---|---|---|---|
| Numero des réilluminations. *Restrike number* | | | 1° *1st* | 2° *2nd* | 3° *3rd* | 4° *4th* | 1° *1st* | 2° *2nd* | 3° *3rd* | 4° *4th* | 1° *1st* | 2° *2nd* | 3° *3rd* | 4° *4th* |
| Cône *Cone* | 5 m | $\bar{L}_s$(m) | 0,7* | | | | 0,6* | 0,4* | | | 0,6* | | | |
| | | $\bar{L}_r$(m) | 2,4 | | | | 2,6 | 2,9 | | | 2,7 | | | |
| | | $\bar{Q}_s$(μC) | 12 | | | | 15 | 13 | | | | | | |
| | | $\bar{Q}_r$(μC) | 50 | | | | 85 | 110 | | | | | | |
| | | $\bar{T}_r$(μs) | 120 | | | | 165 | 225 | | | 260 | | | |
| | 10 m | $\bar{L}_s$(m) | 0,6 | 0,95 | 1,2 | | 0,8 | 1,1 | 1,2 | | 0,6 | 1,1 | 1,3 | 1,6* |
| | | $\bar{L}_r$(m) | 1,5 | 4,2 | 5,0 | | 2,4 | 3,9 | 5,3 | | 2,7 | 4,8 | 5,5 | 6,5 |
| | | $\bar{Q}_s$(μC) | 20 | 30 | 40 | | 25 | 38 | 40 | | 25 | 40 | 40 | 45 |
| | | $\bar{Q}_r$(μC) | 45 | 126 | 150 | | 60 | 100 | 140 | | 50 | 85 | 105 | 120 |
| | | $\bar{T}_r$(μs) | 85 | 201 | 205 | | 190 | 200 | 245 | | 240 | 305 | 325 | 370 |
| Hyperboloïde *Hyperboloid* | 5 m | $\bar{L}_s$(m) | 0,8* | | | | 0,9* | | | | 0,8* | | | |
| | | $\bar{L}_r$(m) | 2,0 | | | | 2,05 | | | | 2,3 | | | |
| | | $\bar{Q}_s$(μC) | - | | | | 60 | | | | . | | | |
| | | $\bar{Q}_r$(μC) | - | | | | 80 | | | | . | | | |
| | | $\bar{T}_r$(μs) | 185 | | | | 225 | | | | 300 | | | |
| | 10 m | $\bar{L}_s$(m) | 0,55 | 1,2 | | | 0,5 | 0,9 | 1,25 | 1,1* | 0,5 | 0,6 | 1,0 | |
| | | $\bar{L}_r$(m) | 1,35 | 3,6 | | | 1,5 | 3,2 | 5,0 | 5,5 | | 2,9 | 5,5 | |
| | | $\bar{Q}_s$(μC) | 20 | 40 | | | 16 | 33 | 39 | 40 | 15 | 20 | 35 | |
| | | $\bar{Q}_r$(μC) | 50 | 120 | | | 40 | 93 | 140 | 160 | 30 | 50 | 100 | |
| | | $\bar{T}_r$(μs) | 110 | 170 | | | 113 | 180 | 223 | 246 | 200 | 290 | 320 | |
| Demi-sphère *Hemisphere* | 5 m | $\bar{L}_s$(m) | 1,0* | | | | 0,9* | 0,5* | | | 1,0* | | | |
| | | $\bar{L}_r$(m) | 1,8 | | | | 2,05 | 3,3 | | | 2,2 | | | |
| | | $\bar{Q}_s$(μC) | - | | | | 60 | | | | 40 | | | |
| | | $\bar{Q}_r$(μC) | 110 | | | | 140 | 230 | | | 130 | | | |
| | | $\bar{T}_r$(μs) | 200 | | | | 280 | 320 | | | 385 | | | |
| | 10 m | $\bar{L}_s$(m) | 0,6 | 0,75 | 0,9* | | 0,6 | 0,7 | 0,7* | | 0,65 | 0,7 | - | |
| | | $\bar{L}_r$(m) | 2,0 | 3,90 | 4,8 | | 2,3 | 3,95 | 3,9 | | 2,5 | 4,5 | | |
| | | $\bar{Q}_s$(μC) | 24 | 33 | 40 | | 25 | 30 | 30 | | 25 | 25 | : | |
| | | $\bar{Q}_r$(μC) | 60 | 120 | 145 | | 85 | 140 | 130 | | 75 | 115 | | |
| | | $\bar{T}_r$(μs) | 115 | 210 | 220 | | 157 | 224 | 260 | | 295 | 390 | 470 | |

\* Petit nombre de chocs disponibles.
\* *Only few shots available.*

Рисунок 6.4 (Fig. 3.5.1 из [Les Renardieres Group, 1972]). Характеристики ступеней положительного лидера (restrikes) в зависимости от длительности фронта 130, 260, 500 мкс для промежутков стержень-плоскость 5 и 10 м и полусферических, эллипсоидных и конических электродов (стержень имел диаметр 60 см, чтобы уменьшить корону с боковых поверхностей электрода). $T_{cr}(\mu s)$ – длительность фронта напряжения в мкс, $T_r(\mu s)$ – момент времени, когда появляется ступень в мкс, $\bar{L}_s(m)$ – средняя длина ступени в метрах, $\bar{L}_\text{к}(m)$ – средняя длина канала лидера перед образованием ступени в метрах, $\bar{Q}_s(\mu C)$ – средний заряд, внедренный ступенью в разрядный объем в мкКл, $\bar{Q}_r(\mu C)$ – средний заряд, внедренный лидером в разрядный объем, до образования ступени в мкКл.

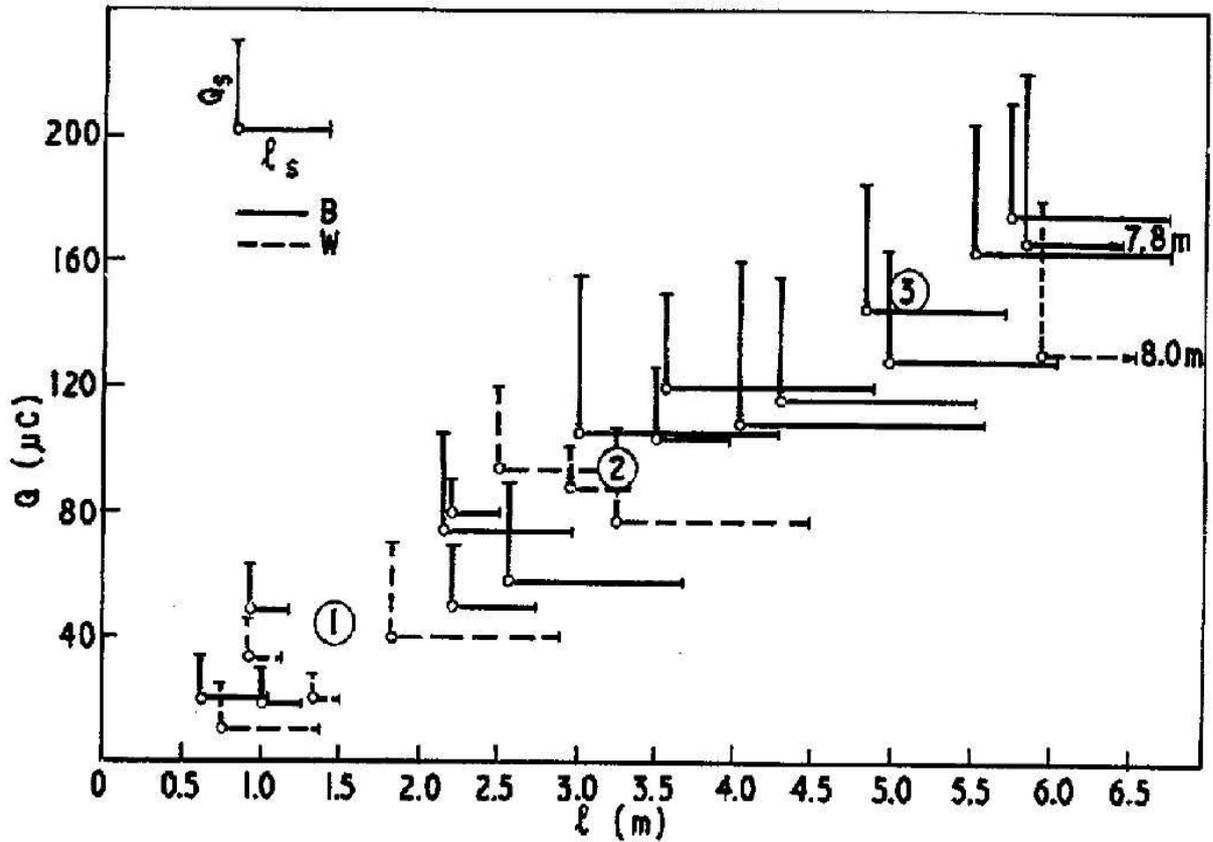

**Discontinuités dans la charge et la longueur du « Leader » qui se produisent pendant une réillumination en fonction de la longueur du « Leader » et de la charge au moment d'une réillumination. Pointe hyperbolique, intervalle de 10 m ($T_{cr} = 260$ μs).**

*Step increases in charge and leader length that occur at a restrike as a function of the leader length and charge at the moment before the restrike. Hyperboloid tip 10 m gap ($T_{cr} = 260$ μs).*

Рисунок 6.5 (Fig. 3.5.2 из [Les Renardieres Group, 1972]). Возрастание длины ступени и величины внедренного заряда в зависимости от номера ступени. Сплошные линии – завершенный разряд, пунктирные линии – незавершенный разряд (то есть, разряд, остановившийся в разрядном промежутке и не достигший заземленной поверхности). Разряд стержень-плоскость в промежутке – 10 м, стержень диаметром 60 см с кончиком в форме гиперболоида. Фронт напряжения равен 260 мкс.

менее симметричными по фронту и спаду импульсами тока и света (с временем нарастания в несколько микросекунд), невозможно было определить направление распространения и измерить скорость лидера, так как весь канал освещался практически одновременно [Gallimberti, 1979]. [Yue et al., 2015] также сообщили о двух типах токов ступеней положительных лидеров с более коротким (обычно 0,97 мкс) и большим (обычно 3,8 мкс) временем нарастания. Интересно, что в обоих исследованиях токам с более длительным временем нарастания, в отличие от токов с более коротким временем нарастания, предшествовал некоторый фоновый ток (background current), что указывает на то, что лидер поддерживал некоторую разрядную активность непосредственно перед началом положительной ступени. Ток положительной ступени имеет тенденцию увеличиваться с уменьшением тока квазинепрерывного лидера [Les Renardieres Group, 1977, стр. 141]. После положительной ступени, которая инжектирует большой положительный объемный заряд в промежуток (вызывающий уменьшение электрического поля в головке лидера), развитие канала лидера обычно приостанавливается, а затем начинается снова по тому же пути. Такое поведение положительных ступеней длиной в десятки сантиметров нельзя описать классическими моделями развития положительного лидера, базирующимися на идее развития ионизационно-перегревной неустойчивости в головке положительного лидера размером около 1 см (например, [Bazelyan & Raizer, 1998]; [Popov, 2003, 2009]). Недавние наблюдения ступеней положительного лидера в длинных искрах в условиях высокой влажности (15,5–16,5 г/м$^3$) были опубликованы [Chen et al., 2016]. Они обнаружили, что двухмерная (в плоскости кадра) межкадровая скорость положительных ступеней составляет от 0,5 до 2,0 × 10$^5$ м/с по сравнению с 2,2 × 10$^4$ м/с во время «стабильной стадии ("stable stage") развития лидера» (то есть, во время квазинепрерывного развития без образования ступеней). В их исследовании было описано, что ряд положительных ступеней демонстрирует некоторую структуру в своих стримерных зонах. Дополнительные изображения ступеней положительного лидера в длинных искрах, полученные с помощью высокоскоростной видеокамеры, полученные той же группой, можно видеть на рисунке 9 в статье [Zhou et al., 2015]. Они обнаружили, что скорость развития положительных ступеней составляла от 0,5 до 3,0×10$^5$ м/с (в среднем 10$^5$ м/с) по сравнению с 1,7 до 2,2×10$^4$ м/с (в среднем 1,88×10$^4$ м/с) во время квазинепрерывного развития положительного лидера. В обоих экспериментах точность измерения скорости ограничивались временным интервалом между кадрами камеры, который был около 8 мкс. Время нарастания напряжения в обоих исследованиях составляло 185 мкс. Максимумы тока положительных ступеней, измеренные [Zhou et al., 2015] и [Chen et al., 2016], по-видимому, порядка ампер, а в одном случае наблюдалось превышение 15 А, что аналогично наблюдениям, опубликованным [Yue et al., 2015], которые обсуждаются ниже. [Xie et al., 2014, рисунки 2 и 3] сообщили о положительной ступени, максимальный ток которой составил около 3 А (разрядный промежуток 3 м, импульс напряжения 250/2500 мкс и абсолютная влажность 8,2 г/м$^3$). Измерения скоростей движения лидеров при помощи скоростных камер имеет большие погрешности, которые мы оцениваем в 2-3 раза, из-за большой величины выдержки кадра по сравнению с временем процесса образования ступени положительного лидера. Поэтому, эти более низкие скорости движения ступеней лидера нельзя считать противоречащими более ранним измерениям, которые имели гораздо лучшие методики измерений, основанные на фотохронограммах стрик-камер с усилением изображения с непрерывной временной разверткой [Les Renardieres Group, 1972, 1974, 1977] или измеряющих движение плазменного фронта непрерывно благодаря линейке фотоумножителей [Baldo and Rea, 1973, 1974].

Далее мы обсудим резкое расширение положительных лидеров, которые, с высокой вероятностью, не были связаны с условиями высокой абсолютной влажности (вспышечный режим). В большинстве случаев время нарастания напряжения было относительно большим, около 1 мс или более и это развитие попадает под условия возникновения вспышечного режима развития лидера. [Yue et al., 2015] получили изображения с помощью высокоскоростной видеокамеры как положительных, так и отрицательных длинных искр, при этом измерялись токи на высоковольтном электроде. Для положительных искр при длине промежутка 6 м и импульсах напряжения 250/2500 мкс наблюдалось квазинепрерывное распространение лидера, а для импульса напряжения 1000/2500 мкс наблюдалось прерывистое (вспышечное) распространение лидера. Импульсы тока вспышек лидера имели максимумы в диапазоне от менее 1 А до более 10 А (см. [Yue et al., 2015, рисунки 12 и 13]). Информации об абсолютной влажности, к сожалению, в статье нет. Для отрицательных лидеров длина промежутка составляла 10 м, а импульс напряжения составлял 80/2500 мкс. Максимумы тока отрицательных ступеней составляли от десятков до сотен ампер (см. [Yue et al., 2015, рисунки 15 и 16]), что значительно больше, чем для ступеней положительных лидеров. Средние значения заряда на единицу длины для разных режимов развития лидеров были оценены как 57.9, 84.6 и 107.1 мкКл/м для квазинепрерывного положительного лидера, ступеней положительного лидера и отрицательного лидера, соответственно.

[Domens et al., 1991], которые изучали положительные лидеры в промежутке стержень-плоскость длиной 16,7 м с использованием импульсов напряжения 300/9000 мкс и 750/10000 мкс, идентифицировали три основных типа распространения лидера, которые они обозначили, как: «непрерывный», «колебательный» и «рестрайковый». Даже для «непрерывного» типа отмечены колебания в стримерной зоне. В случае «рестрайков» они написали, что «очень большая стримерная корона (по размеру и яркости) генерирует большой импульс тока с резким фронтом», причем резкий фронт, как особенность положительной ступени, отсутствует в ключевых особенностях, приведенных выше, которые выделяла для положительных ступеней коллаборация Les Renardieres [Les Renardieres Group, 1977]. [Domens et al., 1991] утверждали, что ток всегда резко уменьшается до возобновления разрядной активности (если она потом вообще происходит) в результате экранирующего эффекта, связанного с большим пространственным зарядом, инжектированным стримерной короной. «Колебательный» тип распространения лидера занимает у [Domens et al., 1991] промежуточное положение между «непрерывным» и «рестрайковым». Для него характерны «большие колебания тока от 0,3–0,4 А до 2,5–4 А с периодом 150–200 мкс». В отличие от «рестрайка», ток в колебательном типе положительной искры никогда не исчезает, а средний ток аналогичен таковому для «непрерывного» типа развития лидера (порядка 1 А).

О прерывистом движении (вспышечном режиме) положительных лидеров сообщали советские исследователи с 1960-х годов (см., например, [Stekolnikov and Shkilyov, 1963], [Горин и Шкилев, 1974]). На фотохронограммах наблюдалось множество прерывистых свечений, которым были даны разные названия, наиболее употребимыми из которых были «скачок» и «вспышка». Значение термина «вспышка» близко к термину «reillumination», используемому Les Renardieres Group, но с акцентом на вспышку стримерной короны от головки лидера, а не на повышение яркости канала лидера. [Горин и Шкилев, 1974] описали четыре типа ступеней (и мелких ступенек) положительного лидера, развивающегося в промежутках от 2 до 15 метров (наблюдаемых или предполагаемых как происходящие):

1. Небольшие (5 см и менее), возникающие при оптически непрерывном (квазинепрерывном) развитии лидера. Они были наложены на квазинепрерывное удлинение лидера, но показали заметные вспышки стримерной короны. Эти небольшие ступеньки, по-видимому, наблюдаемые в разрядном промежутке 2 м, вероятно, соответствуют колебательному типу распространения лидера, наблюдаемыми [Domens et al., 1991] в разрядном промежутке 16,7 м;

2. «Обычные» вспышки положительного лидера во вспышечном режиме, длина которых варьировалась от примерно 6 см в разрядном промежутке 2 м (скорость нарастания напряжения составляла 2,5 кВ/мкс) до 27–60 см для разрядных промежутков 5–15 м (скорость нарастания напряжения составляла 0,5–0,8 кВ/мкс; продолжительность фронта напряжения равна 4,3 мс). Максимумы импульсов тока составляли 0,5–1,5 А для разрядного промежутка 2 м;

3. «Особые» («специальные») ступени, которые наблюдались во время сквозной фазы (когда у положительного нисходящего лидера уже был контакт с помощью положительных стримеров с заземленной плоскостью или восходящим отрицательным лидером) и характеризовались длинами от 10 до 60 см и токами от 20 до 100 А. Было особо отмечено, что на них влияют длина разрядного промежутка и общая электрическая цепь ГИН (включая тормозное сопротивление);

4. Неразрешенные оптически (гипотетические в то время), длина которых не превышает размера головки лидера (3–10 мм). Этот режим соответствует квазинепрерывному режиму.

Одно потенциально важное различие между результатами [Горина и Шкилева, 1974] и Les Renardieres Group состоит в том, что ступени, изученные [Гориным и Шкилевым, 1974] (по крайней мере, те, которые наблюдались в 2-метровом разрядном промежутке), имели вспышки стримерной короны с относительно низкой интенсивностью, предшествовавшие относительно небольшому удлинению лидерного канала (см. их рисунок 5а), тогда как при ступенях положительного лидера, изученных Les Renardieres Group, относительно большое удлинение канала завершалось мощным всплеском стримерной короны.

В этом разделе мы сосредоточимся на относительно больших (от десятков сантиметров до более метра) ступенях как в отрицательных, так и в положительных лидерах с длинными стримерами, возникающими до начала сквозной фазы, которые были описаны в исследованиях [Les Renardieres Group, 1972, 1974, 1977]. Мы предполагаем, что эти длинные ступени положительного лидера более репрезентативны для процессов, происходящих в молнии.

## 6.3. Развитие положительного канала молнии

При развитии канала молнии положительные лидеры могут двигаться также квазинепрерывно или прерывисто (ступенчато). Восходящие положительные лидеры молний, инициируемые высотными зданиями или инициируемые ракетами, несущими провод (триггерные молнии), часто имеют оптически четко различимые ступени (см., например, [Rakov & Uman, 2003], главы 5–7 и ссылки в них). Восходящие положительные лидеры триггерных молний также могут переходить из ступенчатого режима в режим квазинепрерывного распространения и обратно. Idone ([Idone, 1992, Fig.2], [Rakov & Uman, 2003, Fig.7.7]), представил фотохронограмму движения восходящего положительного лидера, сделанную в диапазоне ближнего УФ, каждая ступень которого состояла из яркого тонкого светящегося объекта, обычно длиной 3–5 м ([Idone, 1992] называл его стемом), и диффузной короны около, которая, казалось, выходила из светящегося объекта в направлении распространения на 5-10 метров.

[Biagi et al., 2011] исследовали 10 ступеней, наблюдаемых на начальных 11 м восходящего положительного лидера при разряде триггерной молнии во Флориде. Длина ступени составляла от 0,4 до 2,2 м, а интервал между ступенями — от 17 до 30 мкс. Соответствующие максимумы тока, измеренные на уровне земли (у основания провода, который несла ракета), находились в диапазоне от 17 до 153 А. [Jiang et al., 2013], также для восходящего положительного лидера при инициации триггерных молний в Китае, сообщили о средних геометрических значениях межступенчатого интервала и максимума тока ступени 20 мкс и 45 А, соответственно.

[Wang et al., 2016] для восходящего положительного лидера, развивающейся с метеорологической башни высотой 325 м в Пекине, Китай, представили, пожалуй, самые подробные данные, полученные высокоскоростной видеокамерой о ступенчатом лидере естественной восходящей положительной молнии. Их время экспозиции составляло 6,18 мкс, а мертвое время — 0,49 мкс. Пространственное разрешение составляло 1,27 м. Они обнаружили, что процесс «скачка» (ступени) проявлялся как удлинение лидерного канала. Слабое свечение перед головкой лидера они посчитали стримерной короной из новообразованного участка лидера. По их мнению, ступени предшествовало усиление (увеличение светимости) стримерной зоны около головки лидера без признаков образования плазменных образований перед головкой лидера. Средние значения длины ступени, интервала между ступенями и средней скорости составили 4,9 м, 62 мкс и $8,1{\times}10^4$ м/с. Средняя нижняя граница скорости удлинения канала на отдельном шаге составила $7,3 \times 10^5$ м/с, что почти на порядок выше средней скорости удлинения лидера. Однако, все эти цифры стоит принимать критично, так как точность их измерений невелика и методики измерений описаны не подробно.

[Rakov et al., 2003] наблюдали высокоинтенсивный восходящий положительный лидер во Флориде, вызванный триггерной молнией, для которой амплитуды ступенчатого тока и межступенчатые интервалы составляли в среднем 1,6 кА (измеренные на уровне земли) и 50 мкс, соответственно. Средняя инжекция заряда одной предполагаемой ступени составляла 31 мКл. Однако в этом эксперименте, по мнению [Rakov et al., 2003], мог наблюдаться разрыв токового канала, образование в верхней части двунаправленного лидера, который мог взаимодействовать с восходящим положительным лидером (который связан с измерительной системой), что могло очень сильно исказить картину восходящего лидера и поэтому к этим данным нужно относиться осторожно (если учитывать их в создании моделей восходящего лидера).  Об очень похожих характеристиках ступеней сообщили [Wada et al., 1996] для восходящего положительного лидера, инициированного 200-метровой дымовой трубой Фукуи в Японии. [Yoshida et al., 2010] сообщили о токах килоамперного масштаба (измеренных на уровне земли), связанных с двумя восходящими положительными лидерами во Флориде, вызванными триггерными молниями.

[Wang and Takagi, 2011], используя фотоэлектрическую систему с временным разрешением 100 нс, сообщили о нисходящем положительном лидере, который показал 20 оптических импульсов, излучаемых при распространении от 299 до 21 м над землей, что они приписали ступенчатому процессу. Эти импульсы светимости во всех отношениях были подобны импульсам, зарегистрированным тем же прибором для отрицательных нисходящих лидеров молний, за исключением времени нарастания импульса (10–90%) и ширины половины пика по полувысоте: средние геометрические значения составляли 2,0 и 3,4 мкс для положительного лидера против 0,4 и 1,1 мкс для отрицательного лидера, исследованного [Lu et al., 2008] соответственно. С другой стороны, [Saba et al., 2015] представили аргументы в пользу того, что импульсы электрического поля перед положительным обратным ударом могут быть "due solely to the upward propagation or a negative connecting leader" («обусловлены исключительно восходящим распространением, то есть отрицательным восходящим лидером», видимо, по мнению [Saba et al., 2015], благодаря влиянию электрического поля ступеней отрицательного лидера). Это означает по мнению [Saba et al., 2015], что то, что может казаться положительной ступенью, может быть вызвано одновременно развивающимся отрицательным ступенчатым лидером, хотя многочисленные эксперименты, представленные Les Renardieres Group и работа [Wang et al., 2016] представляют прямые доказательства истинных (в отличие от предположительно «индуцированных») ступеней положительных лидеров. По-видимому, есть три типа положительных лидеров, которые возникают на наземных сооружениях в ответ на приближение отрицательного ступенчатого лидера, и они могут демонстрировать флуктуации тока и яркости, вызванные ступенями отрицательного лидера. К ним относятся восходящие контактирующие стримерной короной (в сквозной фазе) лидеры (например, [Hill et al., 2016], [Lalande et al., 1998], [Saba et al., 2015], [Visacro et al., 2017]), не контактирующие (или attempted, «пытающие взаимодействовать с нисходящим лидером») восходящие лидеры, которые конкурировали с контактирующим лидером за контакт с нисходящим лидером молнии (например, [Tran & Rakov , 2017]; [Schoene et al., 2008]; [Visacro et al., 2017]), а также восходящие положительные лидеры, индуцированные электрическими полями отрицательных восходящих каналов, инициированных высокими объектами из-за приближающегося положительного заряда облака. В положительных искрах, рассмотренных в настоящей статье, наблюдаемое резкое удлинение положительных лидеров (ступень) не могло быть интерпретировано как вызванное приближением отрицательного ступенчатого лидера, поскольку в промежутке во время резкого удлинения вообще не было отрицательных лидеров.

Стримерная зона в виде кисти на кончике ступени наблюдалась для восходящих отрицательных лидеров Бергером [Berger, 1967], а слабая светимость на нижнем конце ярких ступеней была обнаружена [Schonland et al., 1935] для одного ступенчатого нисходящего лидера. [Edens et al., 2014] представили фотографию, показывающую, по их мнению, ступени отрицательного лидера на высоте около 10 км. Длина ступеней, по их мнению, составляла около 200 м и более (что выглядит мало реалистичным, исходя из их фотографии); то есть значительно длиннее, чем на более низких высотах. Edens et al. объяснили это несоответствие более низкой плотностью воздуха на больших высотах. Каждый шаг, по их интерпретации фотографии, сопровождался «щеткой» короны длиной около 100 м. Этот факт выглядит неправдоподобно, так как матрица используемого ими фотоаппарата со стеклянной оптикой не могла бы фиксировать стримеры даже на расстоянии 300 м, не говоря уже о 10 км, что хорошо было видно по нашим измерениям лидеров и их стримерных корон скоростной камерой в главе 5, где стримеры трудно было обнаружить даже с дистанции 3-6 м. По мнению [Petersen and Beasley, 2013, рисунки 4 и 5], им удалось обнаружить стримерные зоны отрицательных лидеров в естественных нисходящих молниях. Также считают [Qi et al., 2016, рисунок 9], которые снимали молнии на расстояниях 770 и 350 м соответственно. Однако, матрицы скоростных камер, которые они использовали также очень плохо фиксируют УФ-излучение стримеров. Поэтому, пока можно утверждать, что существует вероятность, что они зафиксировали стримерные зоны нисходящих лидеров, но это также могут быть и каналы лидеров с переменной светимостью или ветвлениями и изгибами траектории. Для надежного фиксирования стримерных зон необходимо пользоваться скоростными УФ-камерами или стрик-камерами со специальными матрицами или фотопленками, а также кварцевой оптикой.

[Pu et al., 2017] сообщили о двух восходящих отрицательных лидерах в области разрядов молний, вызванных положительными триггерными молниями. У одного из них была оценена длина ступени лидеров, которая составила от 4,9 до 8 м при среднем геометрическом размере 6,3 м в пределах расстояния 500 м от поверхности земли. Было обнаружено, что типичная инжекция заряда отдельной ступени в двух зафиксированных отрицательных лидерах составляет порядка сотен микрокулонов по сравнению с десятками микрокулонов, о которых ранее сообщалось для отдельных ступеней в восходящем положительном лидере триггерной молнии. [Pu et al., 2017] не представили информации о стримерных коронных, связанных со ступенями отрицательных лидеров, так как их было сложно обнаружить на их кадрах.

В этом разделе мы представляем и интерпретируем оптические изображения ступеней отрицательного лидера и ступеней положительного лидера. Все ступени демонстрируют ярко выраженные стримерные вспышки, и морфология стримерных вспышек, по-видимому, существенно не зависит от полярности лидера. Мы сравниваем два режима распространения положительного лидера (квазинепрерывный и ступенчатый). Результаты экспериментов обсуждаются с целью улучшения нашего понимания распространения лидеров молнии. Основное внимание в этой статье уделяется представлению и сравнению избранных событий положительной и отрицательной полярности, записанных с очень высоким пространственным разрешением, которые четко демонстрируют особенности ступенчатого движения, в отличие от сбора статистики для большого количества событий, которая запланирована для будущего исследования.

## 6.4. Экспериментальная установка

Все экспериментальные исследования, представленные в этой главе, были проведены в Центре высоковольтных исследований Всероссийского научно-исследовательского института технической физики им. ак. Забабахина, расположенного в городе Истра, Московской области. Длинные искры образовывались в электрическом поле генератора импульсных напряжений (ГИН-6МВ, собранный по схеме Маркса) в воздушных промежутках стержень-плоскость, длина которых составляла от 4 до 6 м для отрицательных искр и от 5 до 10 м для положительных искр. В некоторых экспериментах использовался короткий, 30-50 см, вертикальный стержень, который располагался на заземленной плоскости под высоковольтным электродом, чтобы лучше контролировать точку контакта нисходящего лидера с плоскостью.

Фотография экспериментальной установки представлена на Рисунке 6.6. Использовались так называемые коммутационные импульсы напряжения (100/7500 мкс). Ток измерялся с помощью малоиндуктивного шунта, установленного либо на высоковольтном электроде, либо на коротком (около 0,5 м) стержне, установленном на заземленной плоскости. К сожалению, достаточно информативные записи тока доступны только для одного события, представленного в этом разделе. Все изображения, представленные в этом разделе, были получены с помощью высокоскоростной камеры видимого диапазона с функцией усиления изображения (4Picos). Для представленных здесь изображений оптическое усиление встроенного усилителя изображения было установлено на уровне $10^3$–$10^4$. Камера выдавала по два кадра с разрешением 1360×1024 пикселей каждый. Межкадровый интервал может быть 500 нс или более, а время экспозиции каждого кадра может быть установлено индивидуально в диапазоне от 0,2 нс до 80 с. Диапазон длин волн, которые фиксировал 4Picos в этих экспериментах составлял (315–850 нм), покрывая значительную часть УФ-диапазона (300–440 нм), в котором происходит большая часть оптического излучения стримеров (см., например, [Les Renardieres Group, 1977, стр. 141]). Тем не менее, некоторые более слабые стримерные процессы не могли быть отображены камерой 4Picos. Камера устанавливалась на расстоянии 25, 30 или 90 м от искрового канала. Для одного события (показанного на Рисунке 6.7) дополнительно представлено соответствующее изображение с временным разрешением (фотохронограмма), полученное с помощью электронно-оптического преобразователя с разверткой по времени (ФЭР-14). ФЭР-14 имел чувствительность в диапазоне длин волн от 380 до 750 нм и устанавливался на расстоянии 30 м от искрового канала. Все представленные изображения являются двухмерными в плоскости кадра, и все наши выводы о геометрии отображаемых процессов подпадают под это ограничение.

## 6.5 Вспышки стримеров при образовании ступеней отрицательных лидеров

Начнем анализ с события 2016-10-15_42, которое наблюдалось в эксперименте, проведенном 15 октября 2016 года. Разрядный промежуток составлял 5,3 м (между высоковольтного стержнем и небольшим вертикальным стержнем высотой 0,5 м, расположенным на заземленной плоскости). Были получены два кадра с помощью камеры

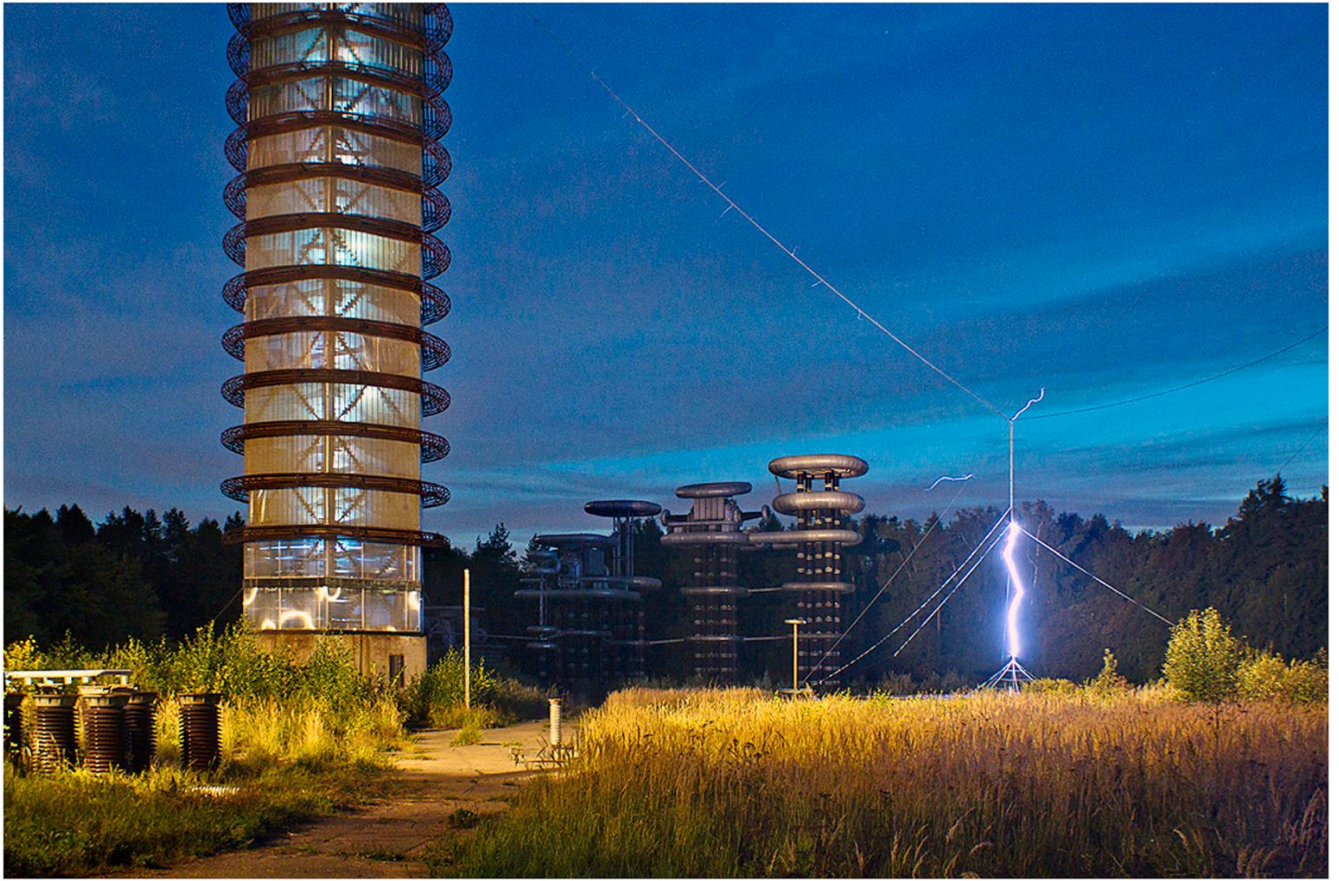

Рисунок 6.6 (адаптировано из [Kostinskiy et al., 2018]). Экспериментальная установка в Центре высоковольтных исследований ВНИИТФ им. Забабахина в Истре, включающая ГИН 6 МВ (слева) и разрядный промежуток (справа). Искра на этой интегральной фотографии была положительной и имела длину около 9 м. Другой генератор, видимый в центре изображения, а также конический объект высотой 3 м, расположенный на плоскости для высоковольтных испытаний, не использовались в экспериментах, описанных в этой статье.

4Picos, которые фиксировали ступень отрицательного лидера с яркой вспышкой стримерной короны. Здесь приведен только первый кадр (см. Рисунок 6.7), поскольку второй кадр содержит менее яркую версию изображения стримерной вспышки, наблюдаемой на первом кадре, которая может быть артефактом системы фиксации изображения камеры (или суперпозицией артефакта и послесвечения разряда). Время экспозиции кадра составляло 200 нс. Соответствующее изображение с временным разрешением, полученное с помощью ФЭР-14, показано на Рисунке 6.8. На изображении видно, что лидеру предшествовала вспышка стримерной короны с отрицательного электрода и за ней последовали три ступени лидера. Изображение, полученное камерой 4Picos (Рисунок 6.7), соответствует второй ступени (Step 2). Интервал времени между началом первой стримерной вспышки (t = 0) и первой ступенью (Step 1) составляет около 21 мкс, а следующие интервалы времени между ступенями составляют 7,8 и 19 мкс. Фотохронограмма показывает, что у каждой из трех ступеней наблюдалась вспышка стримерной короны из района нижнего конца вновь сформированной ступени (район головки), и что третья ступень была ярче, чем вторая и канал ступени двигался перпендикулярно плоскости изображения, чем объясняется сферическая форма конца канала третьей ступени. Этот отрицательный лидер в результате своего развития не дошел до заземленной плоскости и не реализовал квазиобратный удар (незавершенный разряд). Большинство стримерных ветвей, показанных на Рисунке 6.7, берут начало в непосредственной близости от головки лидера, хотя две относительно слабые ветви, кажется, берут начало над вершиной лидера и простираются вверх (от основного разряда). Это поведение стримеров может означать, что некоторые отрицательные стримеры были вынуждены развиваться в обратном направлении из-за быстрой инжекции большого отрицательного пространственного заряда перед головкой лидера.

Далее мы рассмотрим четыре отрицательных разряда, два из которых произошли 12 августа, а два 26 августа 2014 года. 12 августа 2014 года использовался промежуток стержень-плоскость длиной 4 м. 26 августа 2014 года промежуток также составлял 4 м, но расстояние измерялось между высоковольтным стержневым электродом и вертикальным стержнем высотой 0,5 м, который находился на заземленной плоскости.

На Рисунках 6.9 и 6.10 показаны два события, которые наблюдались 12 августа 2014 г. Они демонстрируют мощные вспышки стримерной короны и показывают, что

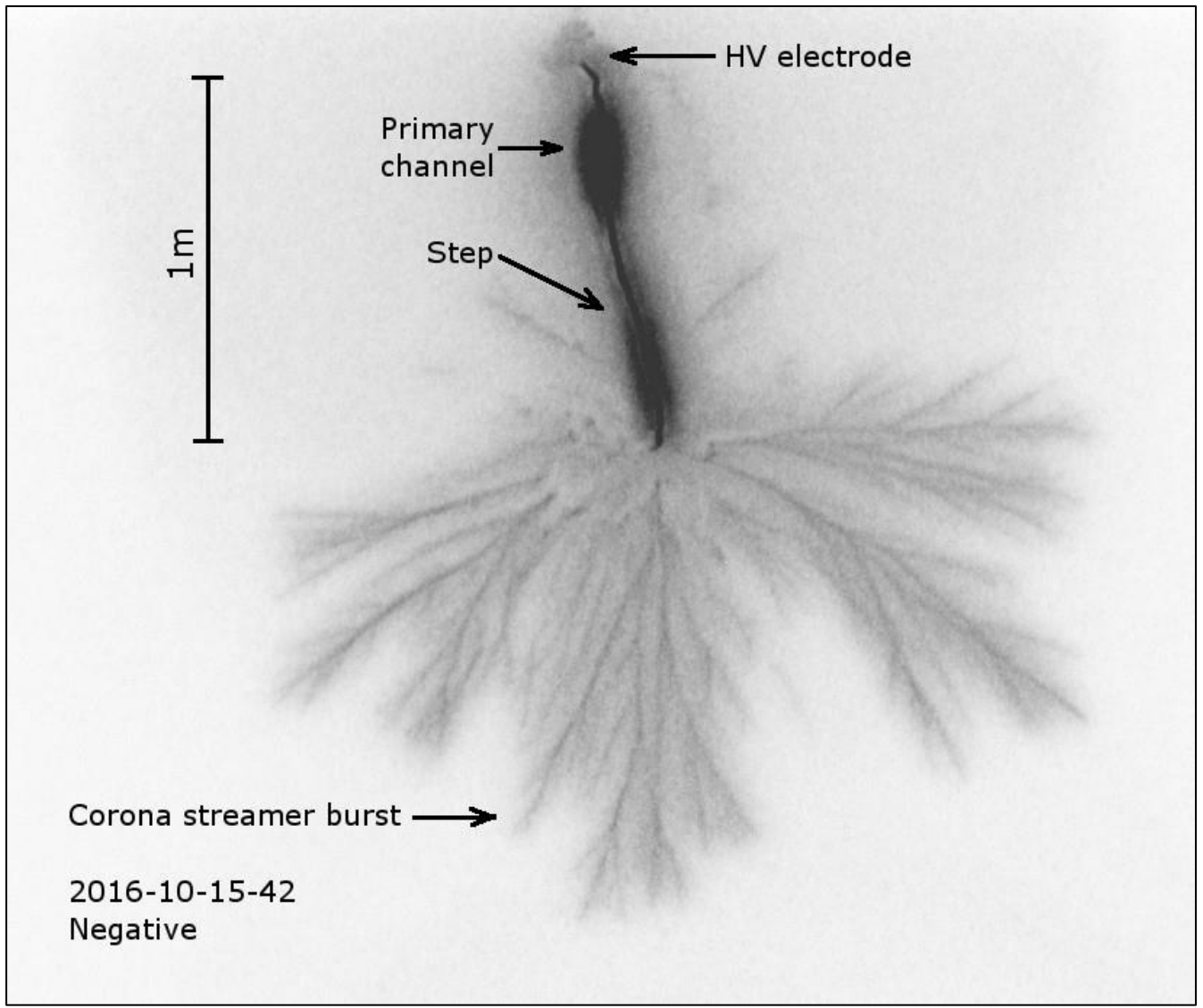

Рисунок 6.7 (адаптировано из [Kostinskiy et al., 2018]), (событие 2016-10-15_42). Изображение высвечивания ступени отрицательного лидера (длиной около 75 см) с ярко выраженной вспышкой стримерной короны, сделанное камерой 4Picos. Изображение инвертированное. Хорошо видно ветвление стримеров. Время экспозиции кадра 200 нс. Фокусное расстояние составляло 50 мм, а значение диафрагмы (относительное отверстие) составляло f/1,4. Размер пикселя изображения в плоскости объекта 4×4 мм2. Температура окружающей среды +2,9 ° С, влажность 91%. (См. также временную развертку этого события на фотохронограмме ФЭР-14, Рисунок 6.8).

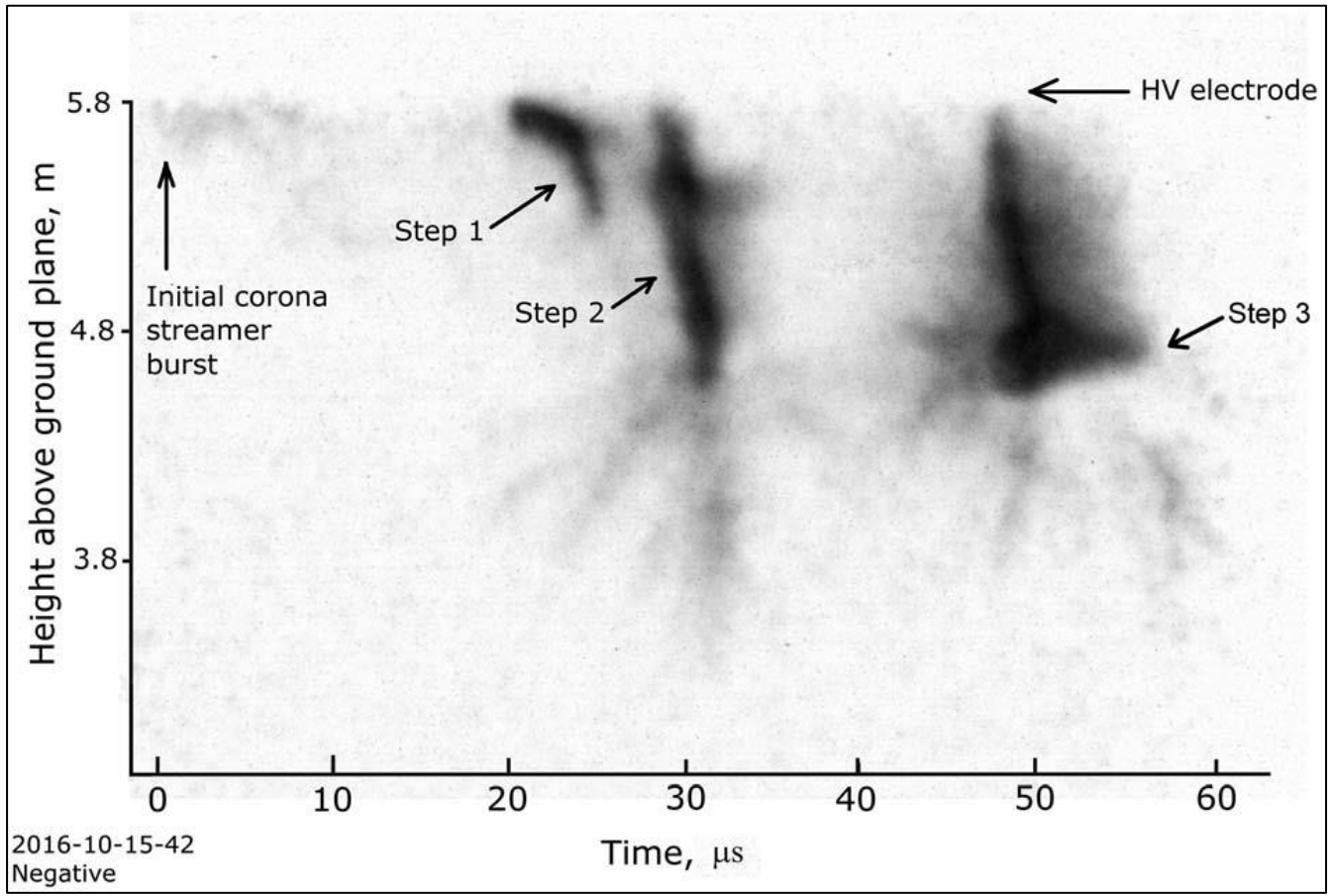

Рисунок 6.8 (адаптировано из [Kostinskiy et al., 2018]), (событие 2016-10-15_42). Изображение с временным разрешением, полученное ЭОП ФЭР-14. Изображение инвертированное. На этом рисунке отмечены начальная вспышка стримерной коронны и три четко различимых ступени. Изображение 4Picos, показанное на Рисунке 6.8, соответствует второй ступени (Step 2). Отрицательный лидер остановился после образования третьей ступени, не дойдя до заземленной плоскости (незавершенный разряд). Начало временной шкалы (t= 0) примерно соответствует началу первой стримерной вспышки на высоковольтном электроде. Вспышки стримерной коронны, во время образования ступеней заметно ярче, чем начальные стримерные вспышки.

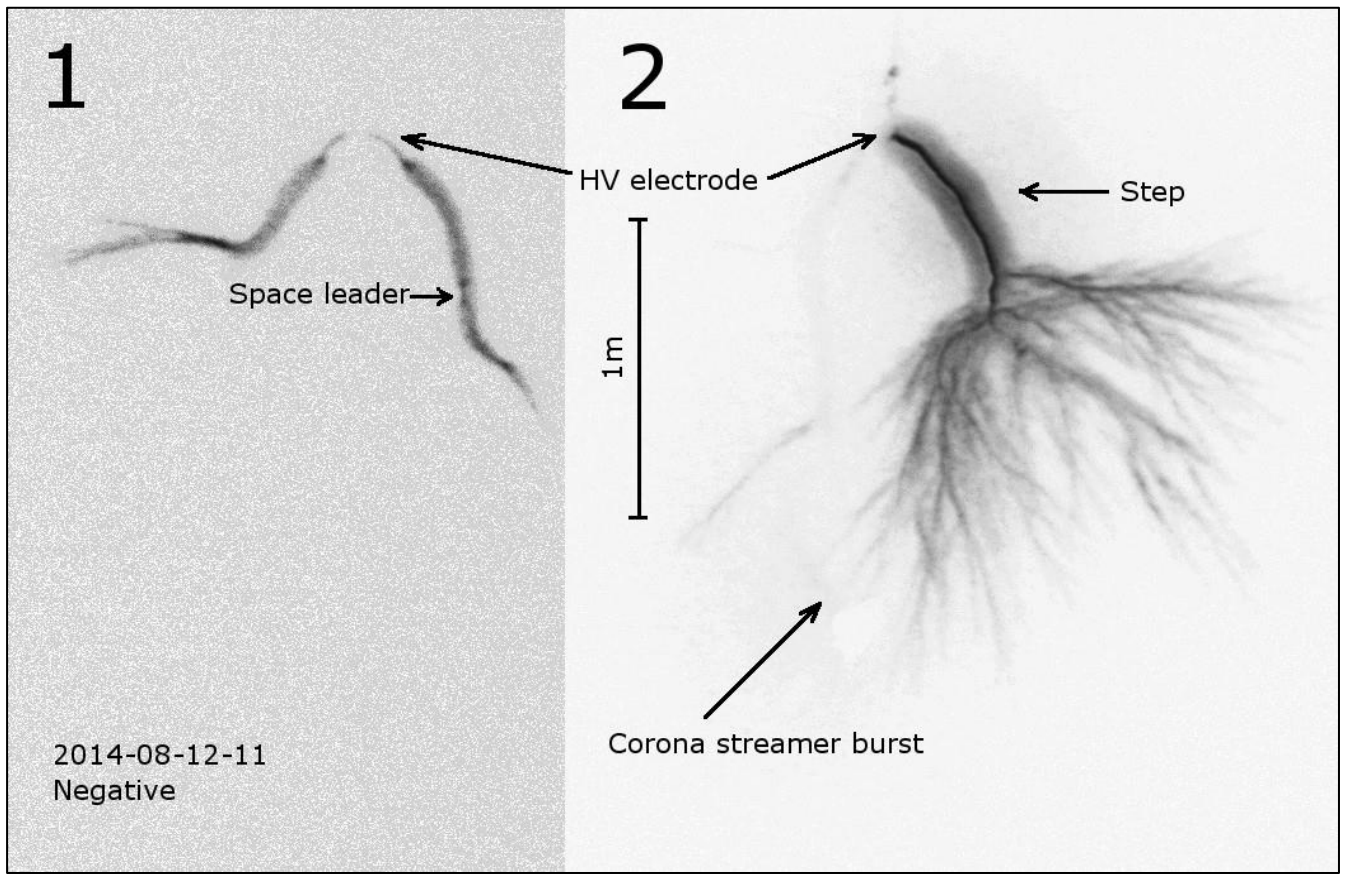

Рисунок 6.9 (адаптировано из [Kostinskiy et al., 2018]), (событие 2014-08-12_11). Ступень отрицательного лидера с яркой вспышкой стримерной короны показана на панели (2), а предшествующие ей в стримерной короне спейс-стем и спейс-лидер (правый канал) показаны (с гораздо большим усилением контраста) на панели (1). Изображение инвертированное. Спейс-лидер, благодаря положительным стримерам с его обоих концов контактирует с коротким правым отрицательным каналом, движущимся от высоковольтного электрода, а также со спейс-стемом перед ним. Несмотря на то, что спейс-лидер длиной всего в несколько сантиметров, хорошо видны обе головки на концах спейс-лидера. Время экспозиции для каждого из кадров 4Picos составляло 1 мкс, а межкадровый интервал — 7 мкс. Обратите внимание, что одна мощная стримерная ветвь, по-видимому, берет начало от боковой поверхности канала ступени; то есть выше нижнего конца ступени. Левая ветвь лидера на панели (1) состоит из короткого отрицательного канала, движущегося от высоковольтного электрода, и спейс-стема, который взаимодействует с каналом лидера благодаря положительным стримерам (спейс-лидера в этой ветви нет). Фокусное расстояние объектива составляло 50 мм, а значение диафрагмы (относительное отверстие) было f/4. Размер пикселя изображения в плоскости кадра составляет 4,4 × 4,4 мм$^2$.

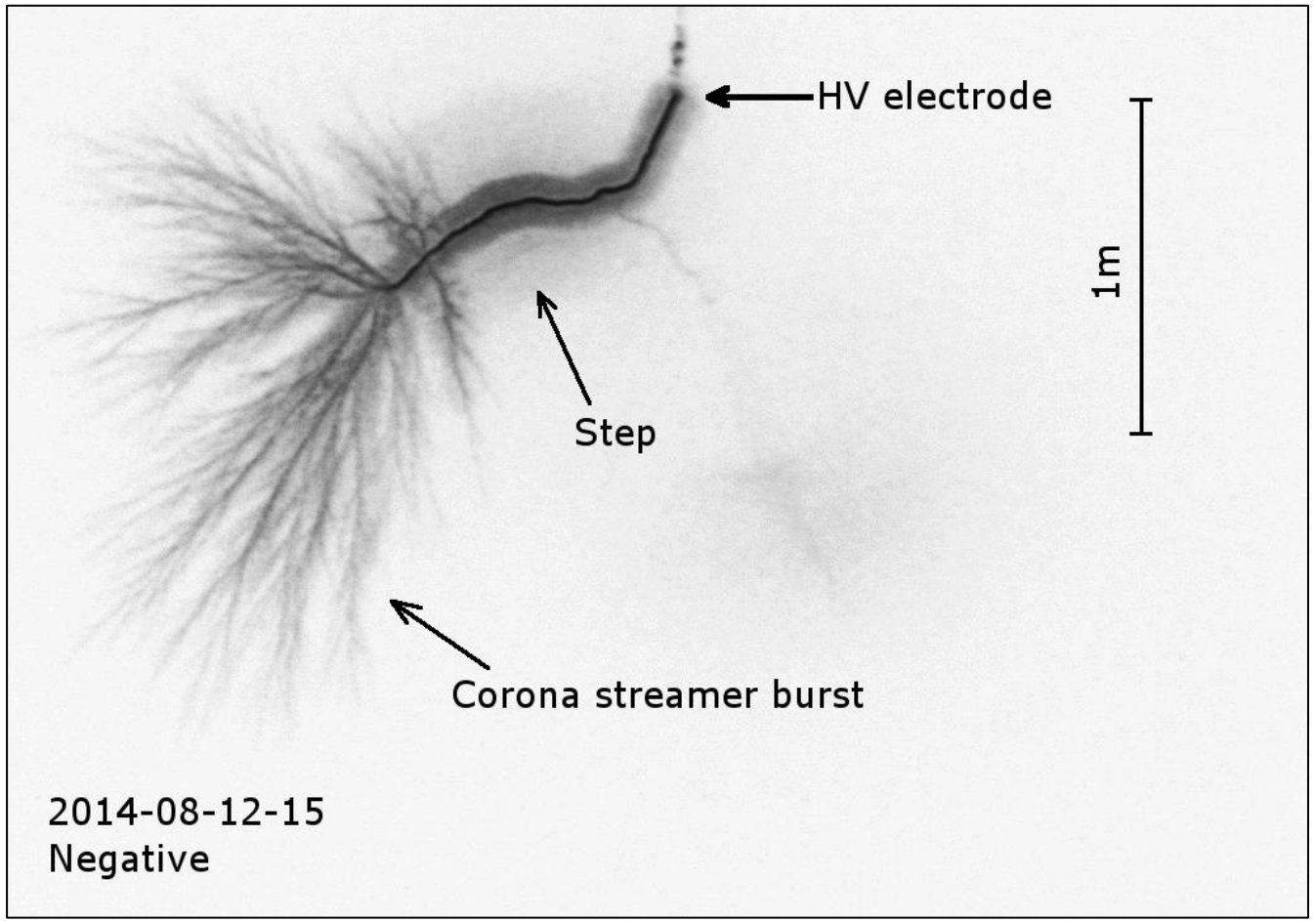

Рисунок 6.10 (адаптировано из [Kostinskiy et al., 2018]), (событие 2014-08-12_15). Ступень отрицательного лидера с яркой вспышкой стримерной короны. Отображается только первый из двух кадров 4Picos (второй может содержать артефакт фиксации события камерой). Время экспозиции кадра равно 1 мкс. Изображение инвертированное. По крайней мере две яркие ветви стримеров исходят от боковой поверхности ступени, то есть выше нижнего конца канала ступени. Фокусное расстояние составляло 50 мм, а значение диафрагмы (относительное отверстие) было f/4. Размер пикселя изображения в плоскости объекта 4×4 мм2.

одна или несколько ярких стримерных ветвей могут исходить от боковой поверхности канала образовавшейся ступени, то есть выше конца ступени (или вновь образовавшаяся головки лидера).

На Рисунках 6.11 и 6.12 показаны два события, которые наблюдались 26 августа 2014 года и чьи вспышки стримерной короны демонстрировали близкую к сферической симметрию, хотя для события, показанного на Рисунке 6.12, это могло быть отчасти из-за необычной геометрии канала. Отчетливо видна разница в морфологии вспышек стримерной короны на Рисунках 6.11 и 6.12 и на Рисунках 6.7 и 6.9, которая связана с разницей во времени экспозиции (50 нс на Рисунках 6.11 и 6.12 против 1 мкс на Рисунках 6.7 и 6.9). Более продолжительное время экспозиции (в 20 раз) приводит к значительному интегрирующему эффекту изображения, когда стримеры реальной видимой длиной 5-10 см со скоростью 100 см/мкс «прочерчивают» линии, которые принято называть «стримерными каналами». В реальности, стримеры, при атмосферном давлении и электрических полях стримерной зоны отрицательного лидера, имеют плазменные каналы длиной около 5-10 см, что мы и видим на Рисунках 6.11 и 6.12, когда была уменьшена до 50 нс экспозиция кадра.

Для отрицательных лидеров, представленных выше, все ступени были изогнутыми и извилистыми, а их двумерная длина составляла несколько десятков сантиметров и доходила до 1 м.

## 6.6. Вспышки стримеров при образовании ступеней положительных лидеров

4Picos-изображения распространения положительных лидеров, полученные в эксперименте, проведенном 15 июля 2014 г., представлены на Рисунках 6.13–6.16. Длина разрядного промежутка составляла 5 м. Как отмечалось во введении, ранее было установлено, что появление ступеней положительных лидеров (restrikes) сильно зависит от абсолютной влажности воздуха [Les Renardieres Group, 1977]. В этом эксперименте температура воздуха составляла 17°C, относительная влажность составляла 75%, а абсолютная влажность была около 11 г/м$^3$. Большинство (примерно 80%) положительных лидеров в тот день сопровождались четко выраженными ступенями, что отвечает

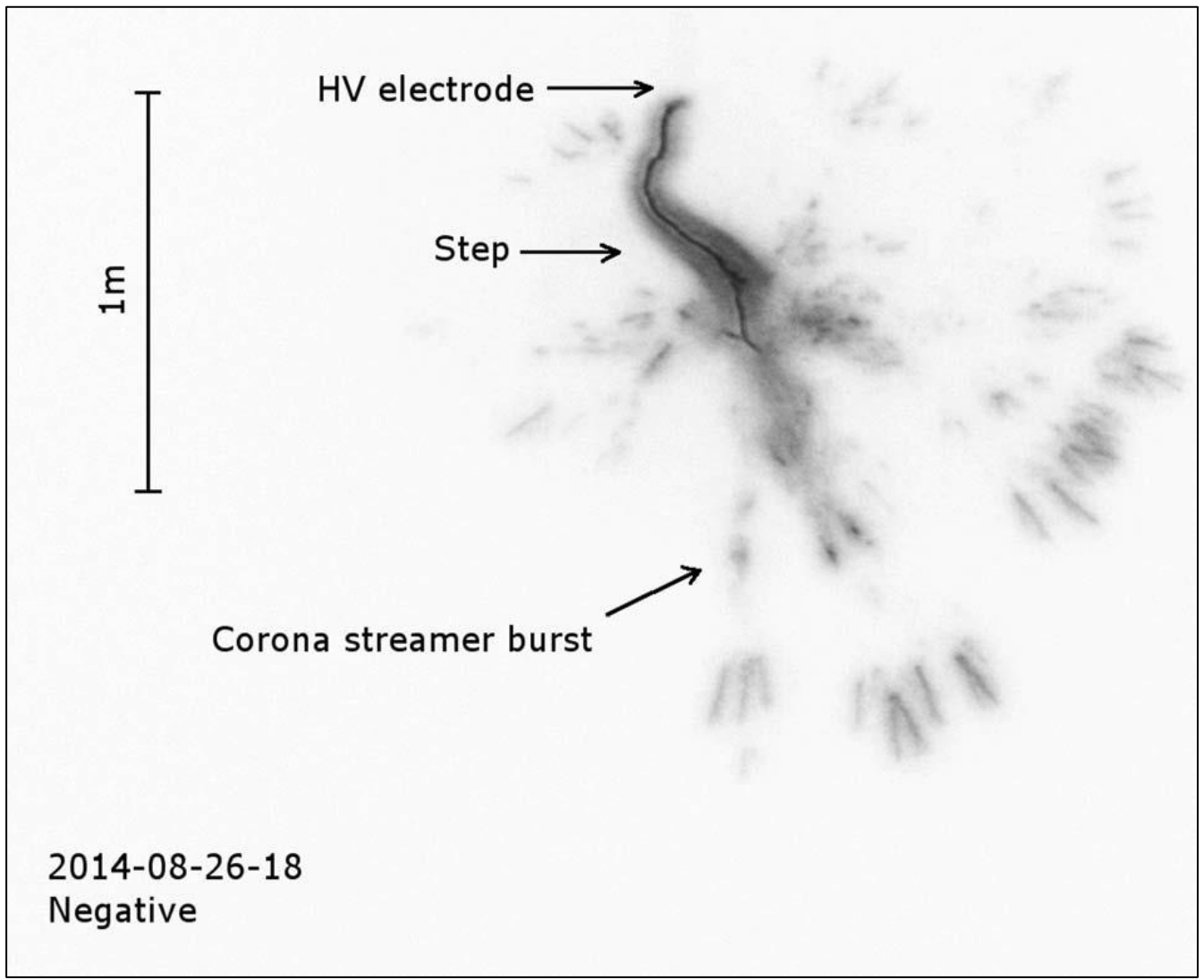

Рисунок 6.11 (адаптировано из [Kostinskiy et al., 2018]), (событие 2014-08-26_18). Ступень отрицательного лидера с яркой вспышкой стримерной короны. Отображается только первый из двух кадров 4Picos (второй может содержать артефакт фиксации события камерой). Время выдержки кадра равно 50 нс. Изображение инвертированное. Стримеры вспышки, кажется, распространяются во всех возможных направлениях, что делает общую геометрию вспышки почти сферической. Также в отличие от Рисунков 6.7 и 6.9, отрицательные стримеры выглядят прерывистыми, что связано с гораздо более коротким временем экспозиции, 50 нс, против 1 мкс на Рисунках 6.7 и 6.9, более продолжительное время экспозиции приводит к более значительному интегрирующему эффекту изображения (в реальности, двигаясь, стримеры длиной около 5-10 см «прочерчивают» линии). Фокусное расстояние составляло 50 мм, а значение диафрагмы (относительное отверстие) составляло f/0,95. Размер пикселя изображения в плоскости кадра составляет 3,1×3,1 мм². Температура окружающей среды 11°С, влажность 81%.

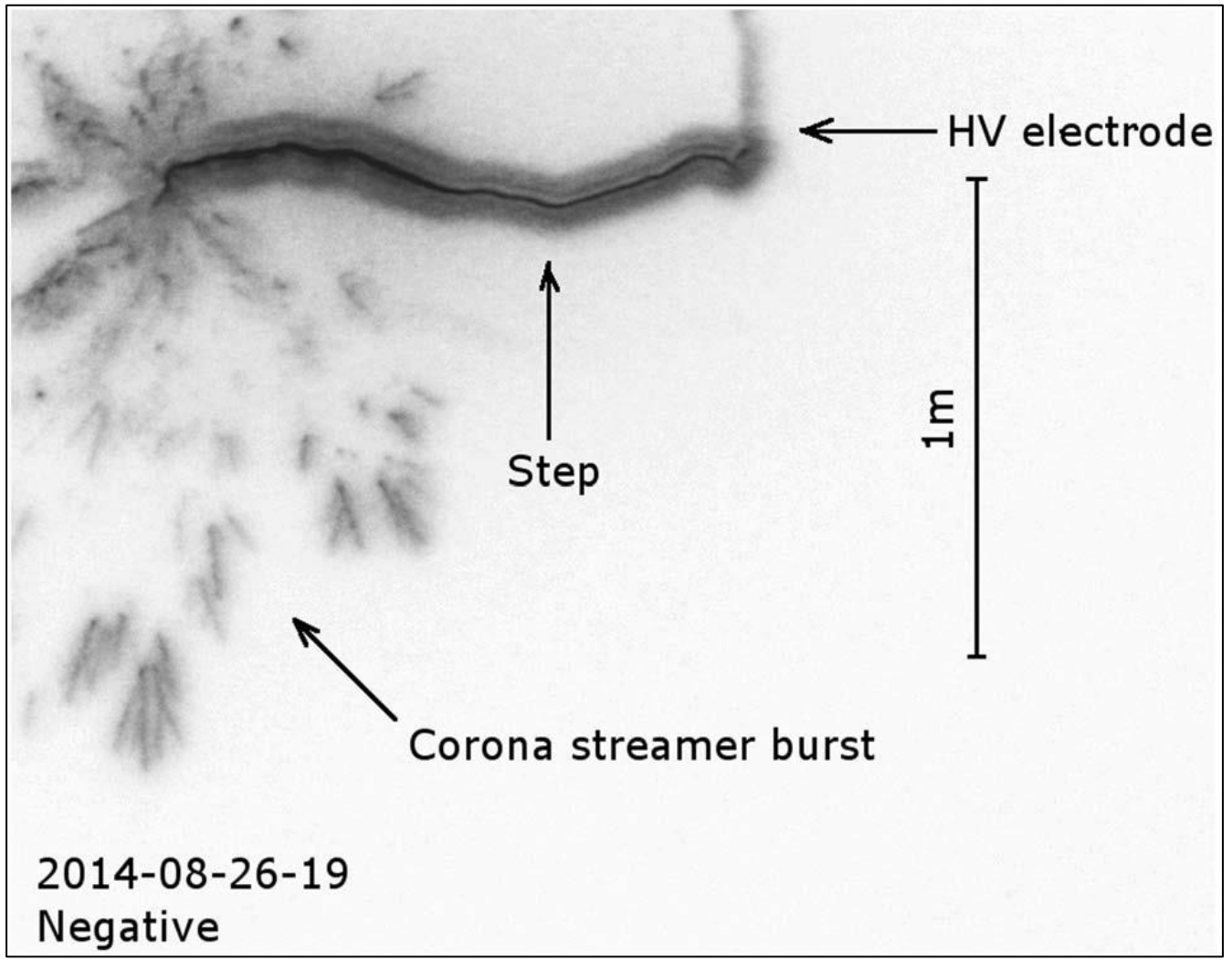

Рисунок 6.12 (адаптировано из [Kostinskiy et al., 2018]), (событие 2014-08-26_19). Ступень отрицательного лидера с яркой вспышкой стримерной короны. Отображается только первый из двух кадров 4Picos (второй может содержать артефакт фиксации события камерой). Изображение инвертированное. Стримеры вспышки, кажется, распространяются во всех возможных направлениях, хотя отчасти это может быть связано с необычной геометрией канала, идущего перпендикулярно первоначальному направлению электрического поля. Этот лидер не дошел до противоположного электрода (незавершенный разряд). Также в отличие от Рисунков 6.7 и 6.9, отрицательные стримеры выглядят прерывистыми, что связано с гораздо более коротким временем экспозиции, 50 нс, против 1 мкс на Рисунках 6.7 и 6.9, более продолжительное время экспозиции приводит к более значительному интегрирующему эффекту изображения. Время выдержки кадра равно 50 нс. Фокусное расстояние составляло 50 мм, а значение диафрагмы (относительное отверстие) составляло f/0,95. Размер пикселя изображения в плоскости кадра составляет 3,1×3,1 мм². Температура окружающей среды 11°C, влажность 81%.

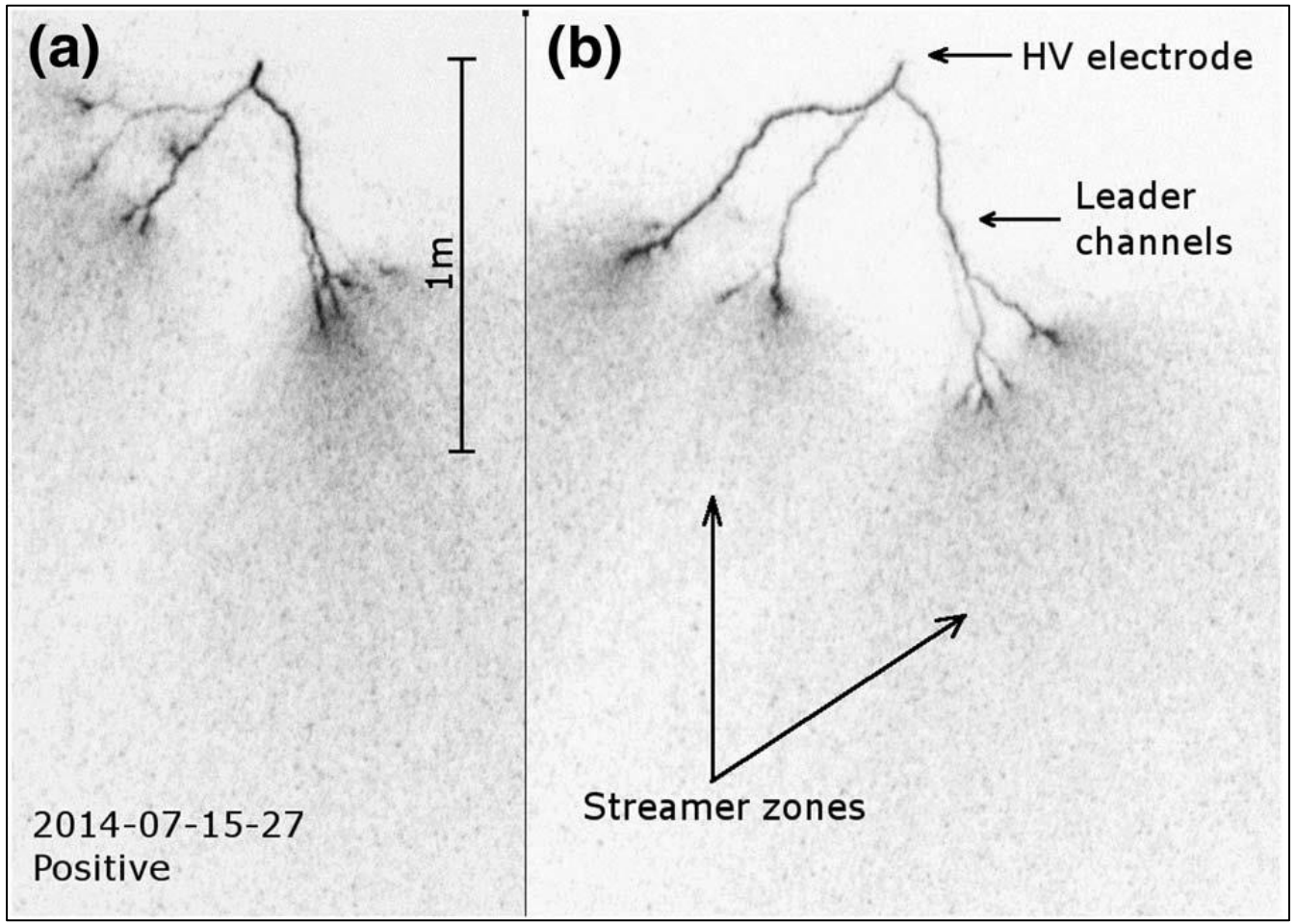

Рисунок 6.13 (адаптировано из [Kostinskiy et al., 2018]), (событие 2014-07-15_27). Пример квазинепрерывного распространения положительного лидера без ступеней. Кадры 1 и 2 показаны на панелях (a) и (b) соответственно (изображения инвертированные). Время экспозиции для каждого из двух кадров составляло 2 мкс, а межкадровый интервал составлял 5 мкс. Фокусное расстояние составляло 50 мм, а значение диафрагмы (относительное отверстие) составляло f/0,95. Размер пикселя изображения в плоскости объекта составляет 3,7×3,7 мм$^2$.

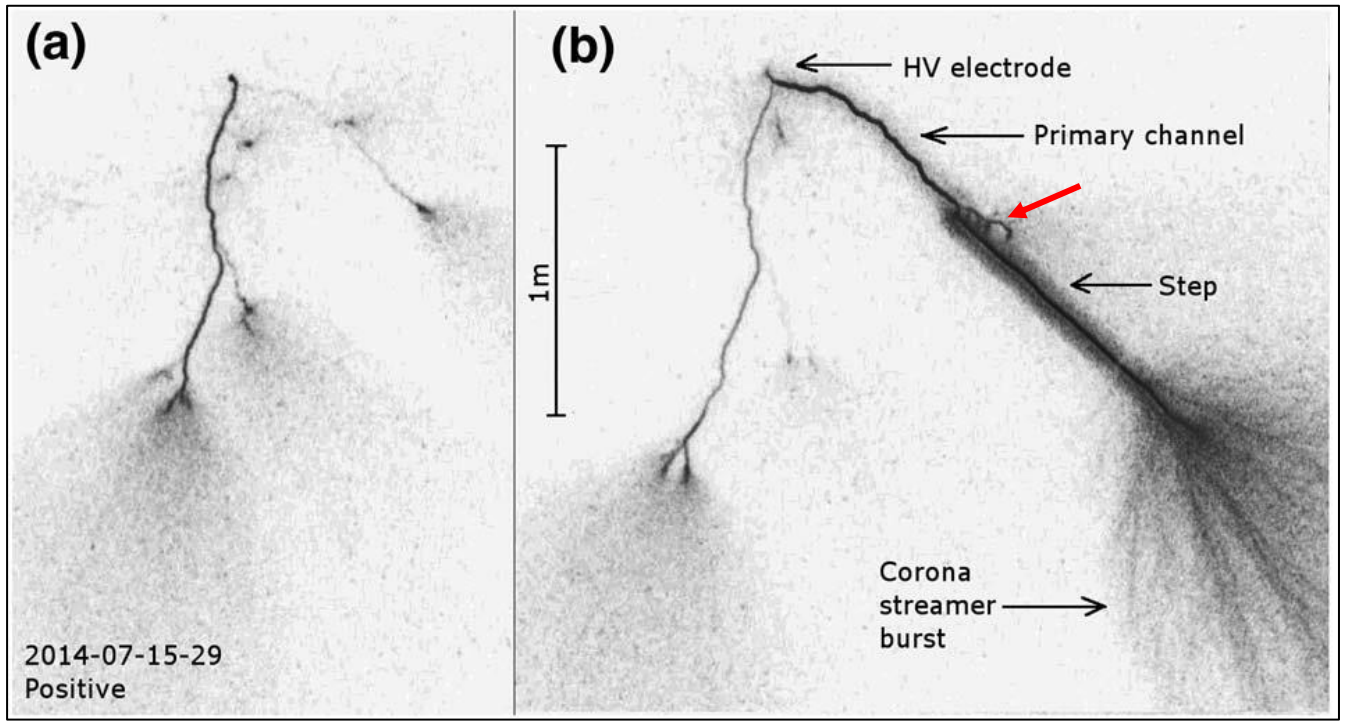

Рисунок 6.14 (адаптировано из [Kostinskiy et al., 2018]), (событие 2014-07-15_29). Пример положительного лидера, ветви которого демонстрируют разные режимы распространения: квазинепрерывные для левой ветви и квазинепрерывный, сменяющийся длинной ступенью для правой. Кадры 1 и 2 показаны на панелях (a) и (b) соответственно (изображения инвертированные). Длина ступени в плоскости кадра около 95 см. Время экспозиции для каждого из двух кадров составляло 2 мкс, а межкадровый интервал составлял 5 мкс. Фокусное расстояние составляло 50 мм, а значение диафрагмы (относительное отверстие) составляло f/0,95. Размер пикселя изображения в плоскости объекта составляет 3,7×3,7 мм².

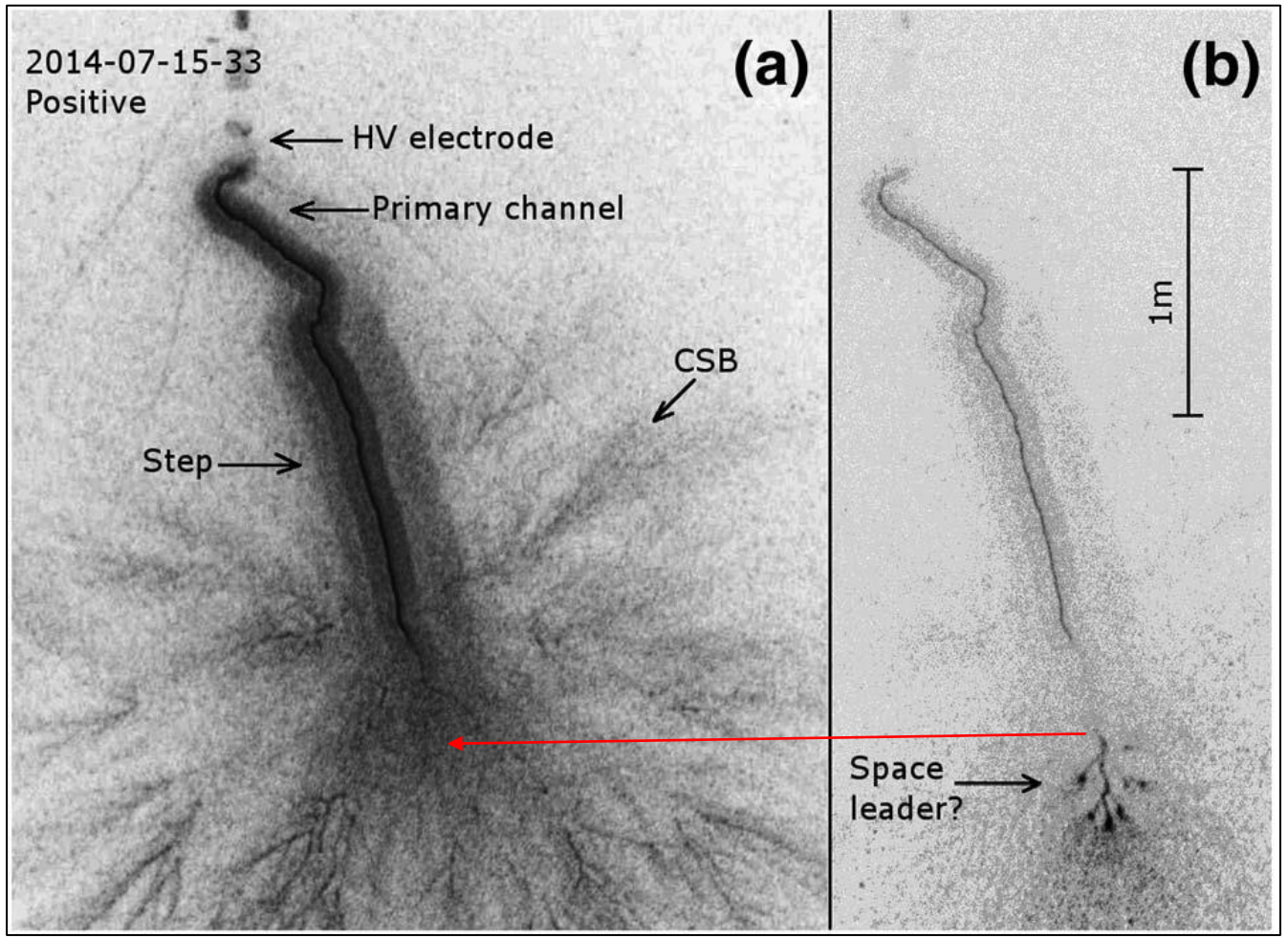

Рисунок 6.15 (адаптировано из [Kostinskiy et al., 2018]), (событие 2014-07-15_29). Ступень положительного лидера с наиболее сильно структурированной вспышкой стримерной короны в нашем наборе данных видна на первом кадре камеры 4Picos, показанном на панели (а). Изображения инвертированные. Длина ступени в плоскости кадра около 123 см. CSB означает вспышку коронного стримера. Второй кадр, показан на панели (b), более контрастной, и он показывает то, что выглядит как «спейс-лидер» с шестью или семью нисходящими ветвями на заметном расстоянии от конца канала положительного лидера. Время экспозиции первого кадра составляло 1 мкс, второго — 0,5 мкс. Межкадровый интервал составлял 5 мкс. Фокусное расстояние составляло 50 мм, а значение диафрагмы (относительное отверстие) составляло f/0,95. Размер пикселя изображения в плоскости объекта составляет 3,7×3,7 мм².

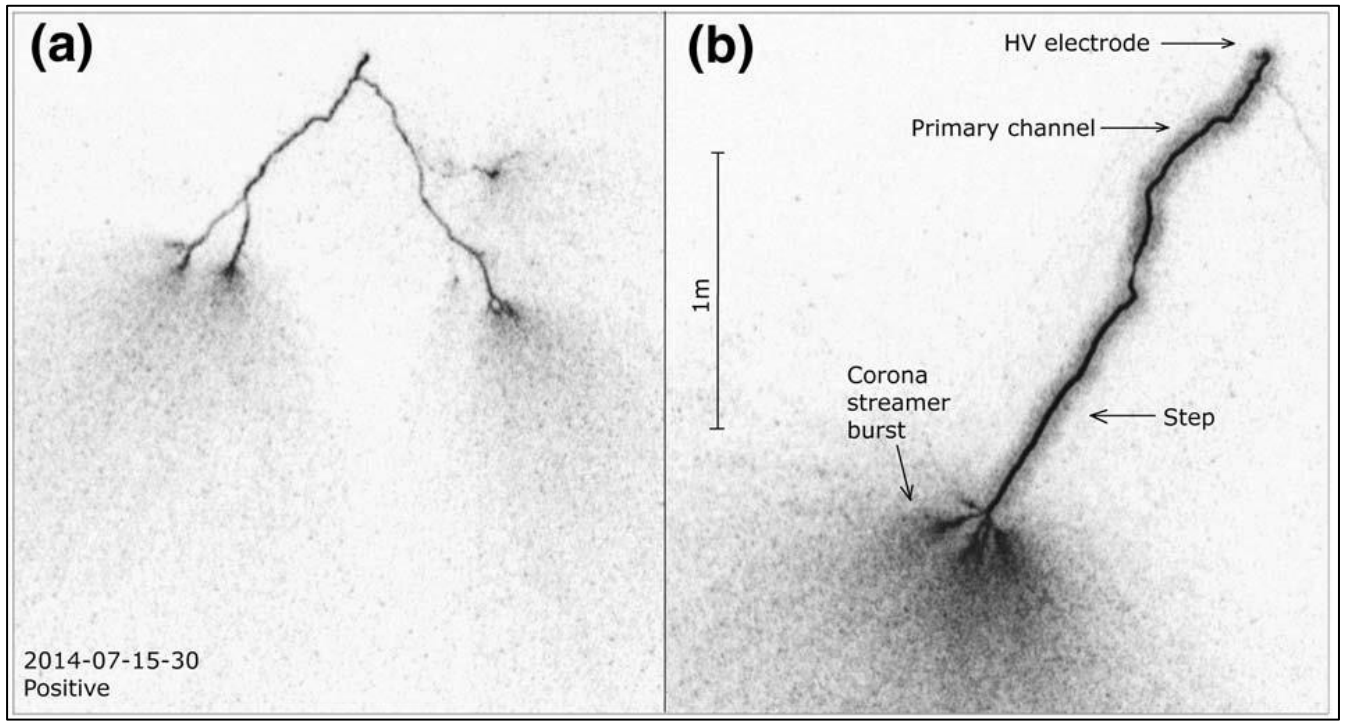

Рисунок 6.16 (адаптировано из [Kostinskiy et al., 2018]), (событие 2014-07-15_30). Ступень положительного лидера длиной 75 см, которая вызвала значительное ослабление свечения других ветвей лидера, наблюдаемых на предыдущем кадре. Кадры 1 и 2 показаны на панелях (a) и (b) соответственно (инвертированные). Время экспозиции каждого из двух кадров составляло 2 мкс, а межкадровый интервал - 7 мкс. Фокусное расстояние составляло 50 мм, а значение диафрагмы (относительное отверстие) составляло f/0,95. Размер пикселя изображения в плоскости объекта составляет 5,2×5,2 мм$^2$.

тенденции, показанной на Рисунке 6.3 (Fig. 5.4.6 из [Les Renardieres Group, 1977]). На Рисунке 6.13 показано событие без ступеней положительного лидера, а на Рисунках 6.14–6.16 показаны три события со ступенями положительного лидера. Еще один положительный лидер со ступенью показан на Рисунке 6.17. Последний был получен в эксперименте, проведенном 12 октября 2014 г. Длина промежутка составила 5 м. Ток лидера измеряли на высоковольтном электроде. Температура составляла около 11°C, относительная влажность 97–98%, абсолютная влажность 9,6–10 г/м3. В отличие от ступеней отрицательного лидера, описанных в разделе 6.5 выше, все четыре ступени положительного лидера (restrikes), представленные здесь, были относительно прямыми и легко отличимыми от более извилистого первичного канала.

На Рисунке 6.13 показан пример положительного лидера, который все время движется квазинепрерывно, без ступеней. Время экспозиции для каждого из двух кадров составляло 2 мкс, а межкадровый интервал составлял 5 мкс. Лидер имеет несколько развивающихся ветвей (типичное поведение в наших экспериментах) со средней скоростью от 2,7 до $4 \times 10^4$ м/с. Такие скорости типичны для квазинепрерывно распространяющихся положительных лидеров.

На Рисунке 6.14 показан пример положительного лидера, ветви которого демонстрируют два разных режима распространения. Как и на Рисунке 6.13, время экспозиции для каждого из двух кадров составляло 2 мкс, а межкадровый интервал составлял 5 мкс. На панели (b) видны две основные ветви. Левая ветвь развивалась непрерывно со средней скоростью $3{,}7 \times 10^4$ м/с, а правая ветвь после квазинепрерывного движения образовала почти прямую ступень длиной около 0.95 м, яркость которой заметно (в 1,6 раза) выше яркости извилистой части лидера, распространявшейся квазинепрерывно. Нижняя граница средней скорости правой ветви составляла $1{,}4 \times 10^5$ м/с. Фактическая скорость, как мы предполагаем, будет намного выше. [Baldo and Rea, 1973, 1974], сообщили о скорости движения волны свечения $\sim 10^6$ м/с во время образования ступеней, а такое большое расхождение объясняется большой величиной экспозиции кадра и большим межкадровым временем. Вспышка стримерной короны правой ветви структурирована, в отличие от более или менее однородной (диффузной) стримерной зоны левой ветви квазинепрерывного развития лидера. Некоторые из стримеров ступени, по-видимому, исходят из боковой поверхности прямого сегмента канала выше конца лидера. Аналогичное поведение наблюдалось у отрицательных лидеров (см. Рисунки 6.9-6.10). Вероятный различный механизм распространения положительного лидера в квазинепрерывной и ступенчатой фазах хорошо виден в месте контакта извилистой и прямой части правого лидера (указано красной стрелкой на панели (b)). Извилистая (квазинепрерывная) и прямая (ступень) части положительного лидера взаимодействуют в этой области через боковую поверхность благодаря 5 точкам контакта, что является крайне нехарактерным для известного ранее развития положительного лидера. Создается впечатление, что прямая часть (ступень), родившаяся в стримерной зоне квазинепрерывного лидера, начала развиваться не из головки квазинепрерывного лидера (если это движение вообще было сверху вниз, а не изнутри объема, подобно развитию спейс-лидера). Нельзя исключить, что прямой плазменный канал (ступень), распространялась подобно ступени отрицательного лидера, — вверх и вниз из области возникновения в объеме перед головкой квазинепрерывного лидера.

На Рисунке 6.15 показан момент высвечивания положительной ступени лидера с наиболее сильно структурированной вспышкой положительной стримерной короны в нашем наборе данных. Обращает на себя внимание, что помимо направленных вниз стримеров, несколько разветвленных стримеров, по-видимому, возникают выше конца лидерного канала и движутся в обратном направлении по отношению к направлению движения канала, что делает общую геометрию вспышки почти сферической. Это похоже на геометрию стримерной вспышки отрицательного лидера, показанную на Рисунке 6.7. Длина ступени составляла около 123 см (самая длинная в этом наборе данных). На втором кадре показано «плавающее» (находящееся на заметном удалении от конца основного канала) разветвленное плазменное образование стримерного или лидерного типа. Больше всего это плазменное образование похоже на спейс-лидер (существующий в стримерной зоне отрицательного лидера) внутри стримерной вспышки положительного лидера. Отличие этого плазменного образования от спейс-лидера, распространяющегося в стримерной зоне отрицательного лидера в том, что отрицательный (верхний) конец не имеет головки, в то время как положительные концы плазменного образования имеют 5 хорошо различимых головок. Неясно, привело ли это событие в стиле спейс-лидера к новой ступени или нет. Если да, то эта ступень была бы короче предыдущей, показанной на Рисунке 6.15а. Возможно, «плавающее» плазменное образование в действительности связано с первичным каналом очень слабо светящемся каналом, яркость которого ниже, чем чувствительность камеры, но это нам кажется маловероятным. Это событие может являться первым свидетельством существования спейс-лидера (или спейс-стема, что нам кажется менее вероятным) внутри стримерной зоны положительного лидера. Поэтому оно требует более подробного обсуждения. Можно предложить три возможных объяснения этого наблюдения:

1. Промежуток между плавающим плазменным образованием и каналом над ним является артефактом камеры. Один из известных артефактов камеры 4Picos (и других подобных камер) проявляется в уменьшении интенсивности света во втором кадре в местах, где были очень яркие пиксели на предыдущем кадре (место, где был на предыдущем кадре настолько яркий плазменный канал, что люминофор не успел восстановить свою излучательную способность к моменту фиксации следующего кадра). Однако положение верхнего конца плавающего образования соответствует на Рисунке 6.15a (конец красной стрелки) области рассеянной светимости, удаленной от канала лидера, а не области очень яркого канала ступени, что свидетельствует против этого конкретного артефакта камеры. Мы не можем исключить существование другого артефакта камеры как причину разрыва канала в месте между плавающим плазменным образованием и каналом над ним на Рисунке 2.2.15b, но нам в результате нескольких лет работы с камерой 4Picos с каналами самой разной яркости не удалось обнаружить другой артефакт, могущий привести к этому эффекту.

2. Разрыв между плазменным каналом и кандидатом в спейс-лидеры реален, и плавающее плазменное образование на самом деле является спейс-лидером (или аналогом UPFs), отрицательная часть которого слишком тусклая, чтобы мы могли ее зафиксировать. Известно, что положительные стримеры (и лидеры) начинают распространяться раньше от «плавающего» проводящего объекта внутри объема [Rakov and Uman, 2003, стр. 269, Figure 7.4], чем отрицательные стримеры и лидеры (из-за в 2 раза большего электрического поля, необходимого для их поддержания), что потенциально может объяснить появление плазменного образования, у которого нисходящие положительные концы намного ярче, чем отрицательные. Расширенная обработка изображения указывает на возможность образования нового контакта между плазменным образованием и основным каналом (не виден на Рисунке 6.15a). В таком случае плавающее плазменное образование фактически начинает взаимодействовать с основным каналом лидера, и в этом случае на Рисунке 6.15b показан переход от вспышки стримерной короны, показанной на Рисунке 6.15а, к зоне вторичной стримерной короны между основным каналом и кандидатом в спейс-лидеры, как на Рисунке 6.9(1).

На Рисунке 6.16(b) показана ступень положительного лидера длиной 75 см, которая вызвала значительное ослабление других ветвей лидера, видных на предыдущем кадре (а). Средняя скорость движения нижней границы составляла $1{,}1$–$1{,}2 \times 10^5$ м/с. Морфология стримерной короны в случае панели (а), по-видимому, не отличается от стримерных зон квазинепрерывно расширяющихся ветвей. Экспозиция второго кадра, вероятно, началась после завершения высвечивания ступени, так как уже начала формироваться «обычная» не структурированная стримерная зона. Обращает на себя внимание пучок из пяти относительно ярких ветвей на нижнем конце лидера, особенность, которая несколько похожа на кажущийся плавающим пучок ветвей, показанный на втором кадре на Рисунке 6.15(b).

На Рисунке 6.17 на панелях (а) и (b) показаны два кадра разветвленного положительного лидера с соответствующей осциллограммой тока, показанной на панели (с). Ступень длиной 50 см на панели (б) возникла в правой ветви. Это привело к значительному ослаблению других ветвей, которые видны в предыдущем кадре (а). Ток во время квазинепрерывного развития лидера и непосредственно перед образованием ступени составлял около 4 А. Ток во время ступени составлял около 21 А. Форма импульса тока при образовании ступени была сложной и имела резкий фронт с длительностью около 1 мкс. Вспышка стримерной короны в этом случае не структурирована, вероятно, по крайней мере частично, из-за более длительного времени экспозиции (8 мкс по сравнению с 2 мкс или меньше для других событий, представленных в этом разделе).

Быстро возникающие участки положительного лидера (ступени) для положительных лидеров, представленных в этом разделе, были почти прямыми (в отличие от ступеней отрицательных лидеров, описанных выше) и имели двумерную длину в плоскости кадра в диапазоне от 50 до более 120 см. Данное свойство ступеней положительного лидера настолько существенное и явное, что оно хорошо видно не только на изображениях камеры с усилением изображения и короткими экспозициями (4Picos), но и на обычных интегральных фотографиях. На Рисунке 6.18 (разрядный промежуток 5

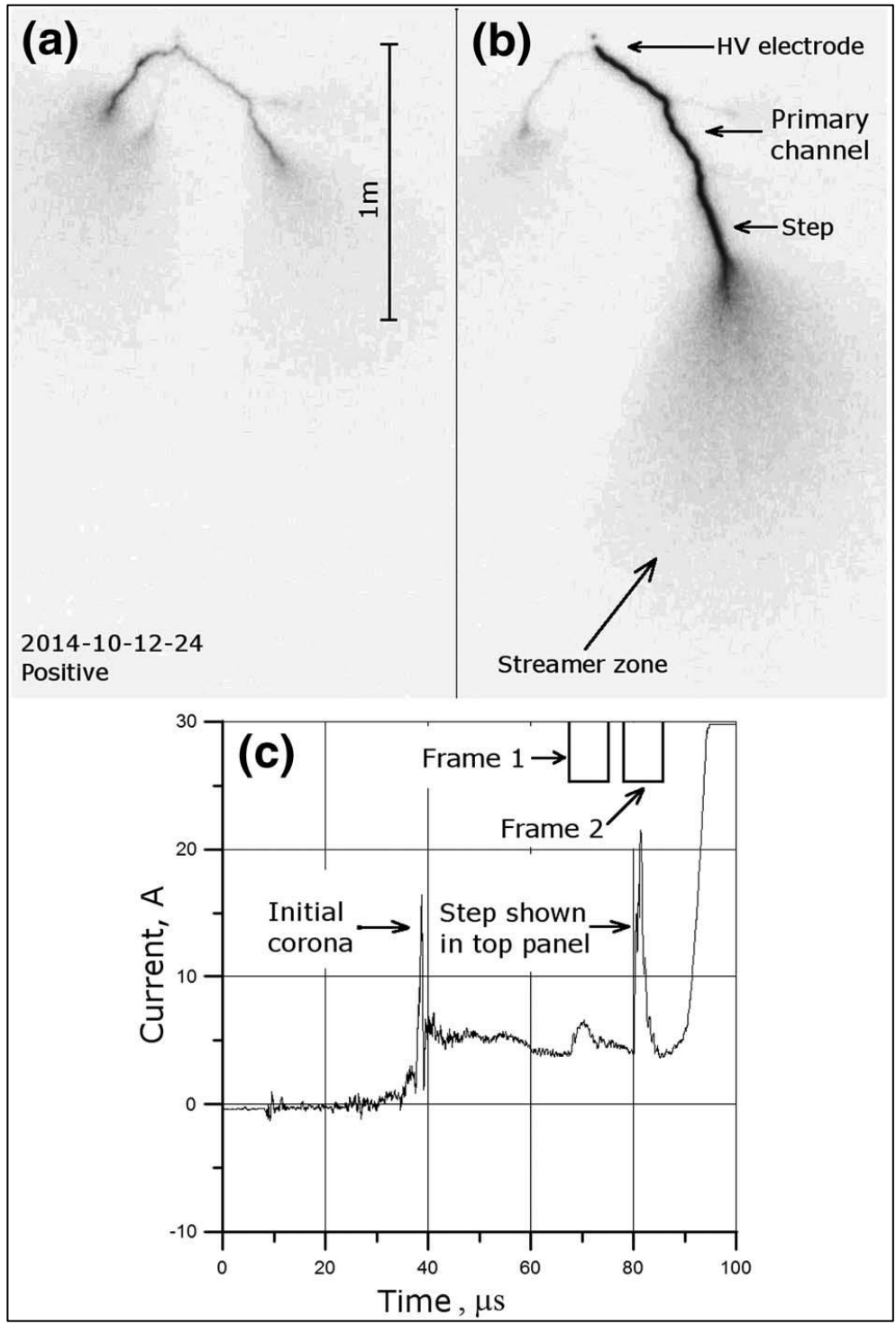

Рисунок 6.17 (адаптировано из [Kostinskiy et al., 2018]), (событие 2014-10-12_24). Кадры (инвертированные), полученные камерой 4Picos показаны на панелях (a) и (b), а соответствующая осциллограмма тока показана на панели (c). На панели (b) видна ступень длиной 50 см в правой ветви и значительное ослабление других ветвей, видимых на панели (a). Время экспозиции каждого из двух кадров составляло 8 мкс, а межкадровый интервал составлял 3 мкс. Осциллограмма тока была сглажена с помощью скользящего окна усреднения длительностью 0,42 мкс. Фокусное расстояние объектива составляло 100 мм, значение было f/4.5. Размер пикселя изображения в плоскости объекта составляет 5,2×5,2 мм². Температура воздуха составляла 1,3°C, а относительная влажность составляла 94%.

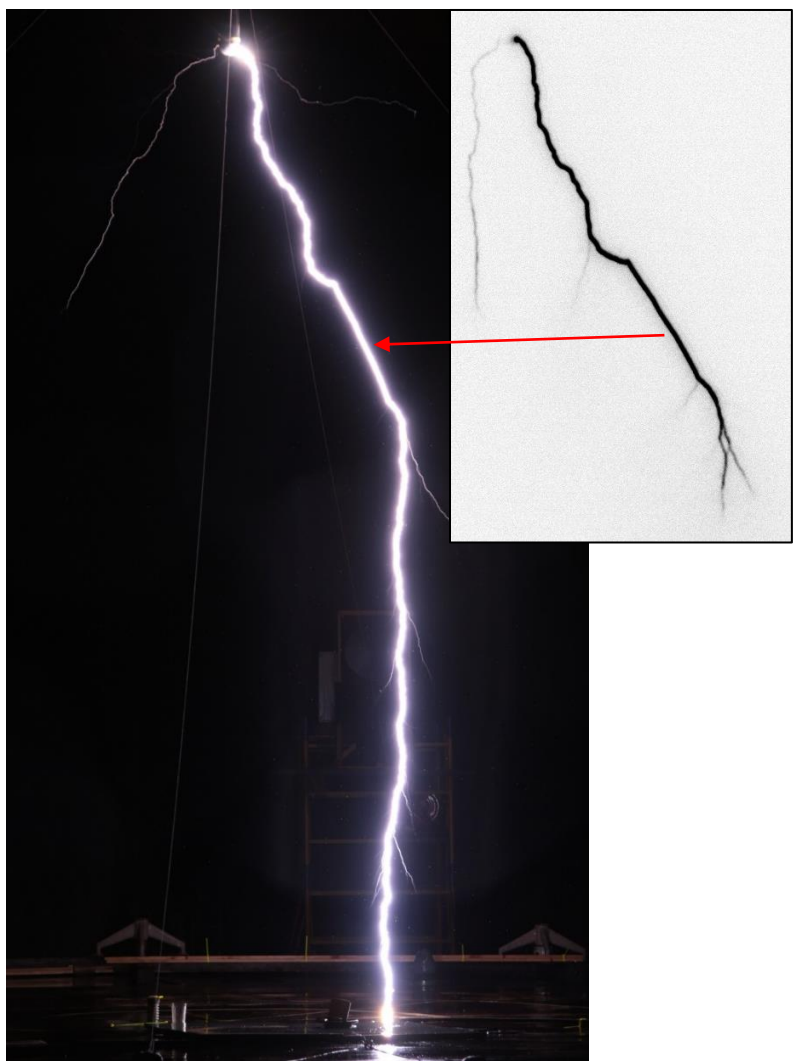

Рисунок 6.18 (событие 2014-09-23_12). Ступень положительного лидера имеет тот же прямолинейный вид на интегральной цветной фотографии (выдержка 4 с, f/8), что и на кадре камеры с усилением изображения 4Picos (выдержка кадра 10 мкс, изображение инвертировано). Разрядный промежуток имеет длину 5 м.

м) ступень положительного лидера, зафиксированная камерой с усилением изображения 4Picos (черно-белое, инвертированное изображение, выдержка 10 мкс), в точности повторяет почти прямолинейную форму, зафиксированную фотоаппаратом на интегральной цветной фотографии (выдержка 4 с). На Рисунке 6.19 (разрядный промежуток — 6 м) ступень положительного лидера (красная стрелка) имеет прямолинейный вид на интегральной цветной фотографии (выдержка 4 с). На черно-белом кадре камеры с усилением изображения 4Picos можно видеть движение положительных лидеров в квазинепрерывном режиме до начала образования ступени (выдержка кадра 1 мкс). Оба Рисунка 6.18 и 6.19 фиксируют, что и до и после образования ступени положительный лидер двигался по извилистой траектории и только во время образования ступени положительный лидер двигался практически прямолинейно. В местах начала и конца ступени на обоих рисунках видны четкие изгибы канала положительного лидера, что также говорит в пользу догадки, что физический механизм распространения лидера во время образования ступени принципиально отличается от механизма квазинепрерывного развития.

## 6.7. Обсуждение результатов, представленных в главе 6

Отрицательные и положительные лидеры молнии различаются по ряду аспектов (см. [Rakov and Uman, 2003, раздел 5.3.2 и ссылки в нем], [Williams, 2006]. Например, в отличие от отрицательных лидеров молнии, которые всегда оптически ступенчатые, положительные лидеры, кажется, оптически могут двигаться либо непрерывно, либо ступенчато [Bazelyan and Raizer, 1998]. Хорошо известно (например, [Горин и Шкилев, 1974, 1976], [Allibone, 1977, рис. 4], [Golde, 1977, рис. 5]), что, при той же длине разрядного промежутка, напряжение пробоя для отрицательной полярности примерно в 2 раза выше, чем для положительной полярности. Кроме того, в случае плавающего (незаземленного) проводящего объекта, помещенного в разрядный промежуток, положительный стример от этого объекта запускается раньше, чем отрицательный (например, [Castellani et al., 1998a, 1998b]), что также справедливо для высотно инициированной триггерной молнии [Rakov and Uman, 2003, стр.269]. Все эти связанные с полярностью различия обычно объясняются тем фактом, что отрицательные стримеры

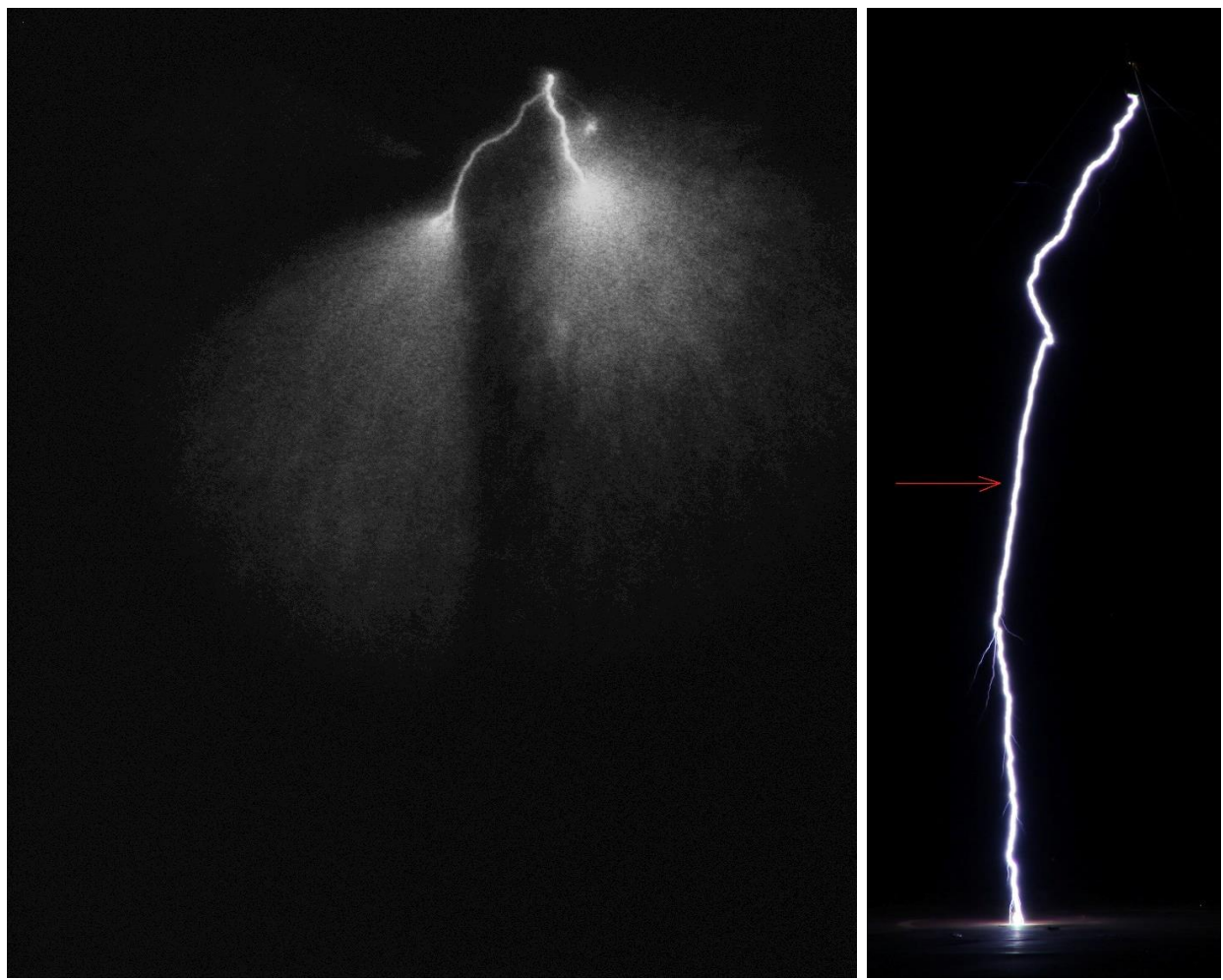

Рисунок 6.19 (событие 2014-07-22_05). Ступень положительного лидера (красная стрелка) имеет прямолинейный вид на интегральной цветной фотографии (выдержка 4 с, f/8). На черно-белом кадре камеры с усилением изображения 4Picos можно видеть движение положительных лидеров в квазинепрерывном режиме до начала образования ступени (выдержка кадра 1 мкс). Разрядный промежуток имеет длину 6 м.

требуют для своего распространения примерно в 2 раза более высоких электрических полей, чем положительные стримеры (при атмосферном давлении, около 10-12 кВ/см для отрицательных, по сравнению с 4.5-5 кВ/см для положительных стримеров [Базелян и Райзер, 2001, стр. 91]). В настоящее время нет однозначного ответа на вопрос, почему положительные стримеры требуют гораздо меньших электрических полей для поддержания своего движения. Однако существует популярное объяснение, основанное на ожидаемом различном поведении электронов в электрическом поле положительного и отрицательного электродов (см., например, [Стекольников, 1960, с. 98], [Ogawa & Brook, 1964], [Rakov and Uman, 2003, глава 5], [Райзер, 2009, стр. 588–589]). Упрощенно можно описать объяснение, приведенное в этих работах так. В случае положительного лидера электроны, присутствующие или созданные перед головкой лидера (или внутри его стримерной зоны), движутся к головке, поскольку они притягиваются к положительному заряду на ней, и результирующая ионизация происходит в усиливающемся электрическом поле (электроны дрейфуют в нем в сторону усиливающегося поля головки). Напротив, в случае отрицательного лидера электроны имеют тенденцию «дрейфовать» от движущейся головки лидера туда, где электрическое поле снижается, поскольку электроны отталкиваются отрицательным зарядом головки лидера. Таким образом, ионизация и формирование стримеров для отрицательного лидера происходит в менее благоприятных условиях, чем для положительного лидера. Другими словами, для поддержания движения стримера требуется более сильное электрическое поле, когда электроны должны перемещаться в область с более низким полем (от головки лидера), что имеет место в случае отрицательных лидеров.

Одним из следствий более низкого поля, необходимого для распространения положительных стримеров, является то, что гораздо легче инициировать положительные стримеры со спейс-стема в сторону отрицательного лидера, чем отрицательные стримеры, в то время как для положительных лидеров в квазинепрерывном режиме не требуется спейс-стем для их развития. С этой точки зрения, разница между квазинепрерывным (положительная полярность) и ступенчатым (отрицательная полярность) режимами распространения лидера является результатом разницы между критическими полями распространения, необходимыми для положительных и отрицательных стримеров. Другими словами, лидер любой полярности в первую очередь приводится в движение «более эффективными» положительными стримерами, которые в случае отрицательного лидера должны инициироваться из пространства перед головкой лидера. Характеристики развитых (горячих) положительных и отрицательных лидерных каналов, такие как температура и проводимость, по-видимому, схожи [Базелян и Райзер, 2001].

Как отмечалось выше, в зависимости от скорости нарастания напряжения на промежутке, положительный лидер длинной искры может перемещаться либо квазинепрерывно, либо периодически (вспышечная фаза, например, [Базелян и Райзер, 1997], [Gorin et al., 1976, рисунок 1]. Вспышечная форма развития положительного лидера (см., например, рисунок 6.2b в [Базелян и Райзер, 1997]) и рисунок 3.2b в [Базелян и др, 1978], которая происходит при относительно низких скоростях нарастания напряжения в промежутке (< 5 кВ/мкс [Базелян и др., 1978, с. 53]), можно рассматривать как своего рода тип ступенчатого (прерывистого) распространения, хотя физический механизм вспышечного распространения [Bazelyan and Popov, 2020] принципиально отличается от механизма распространения отрицательных ступенчатых лидеров. Для положительных искр [Yue et al., 2015] наблюдали квазинепрерывное развитие лидера для импульса напряжения 250/2500 мкс и прерывистое распространение (вспышечный механизм) для импульса напряжения 1000/2500 мкс. Другим важным фактором, способствующим ступенчатому развитию положительных лидеров, является относительно высокая абсолютная влажность [Les Renardieres Group, 1977].

[Горин и Шкилев, 1974], [Базелян и др., 1978, с. 61], [Базелян и Райзер, 1997, с. 232–233] на основании фотохронограмм длинной искры, сделанных на ЭОП с высоким пространственным и временным разрешением утверждали, что непрерывно движущийся положительный лидер будет двигаться маленькими ступеньками размером с головку лидера. Возможно, самый ранний вывод о том, что оптически непрерывное развитие положительных лидеров может быть ступенчатым, был сделан МакИкроном [McEachron, 1939, стр. 191], который предположил, что временные интервалы между ступенями могут стать достаточно короткими, чтобы обеспечить положительному лидеру почти непрерывное развитие.

Представленные здесь результаты экспериментов позволяют напрямую сравнить (1) различные режимы распространения положительных лидеров и (2) тонкую структуру стримерных вспышек, сопровождающих процесс образования ступеней в положительных и отрицательных лидерах.

Наблюдаемые различия в морфологии стримерных вспышек положительной короны (степень их структурирования) могут быть связаны с тем, простиралась ли вспышка за пределы стримерной зоны положительного лидера, существовавшей непосредственно перед образованием ступени. Если новообразованная головка лидера расположена внутри стримерной зоны, то стримерная вспышка из района этой головки будет ослаблена, но она будет гораздо более сильной, если новообразованная головка лидера достигнет области, свободной от пространственного заряда, существующей стримерной зоны. То есть, в первом случае электрическое поле существующей стримерной зоны снижает потенциал головки лидера, а во втором — нет, что ведет к мощной вспышке стримерной короны в свободном от зарядов воздухе. С этой точки зрения, чем длиннее ступень положительного лидера, тем более вероятно ее распространение за пределы ранее существовавшей стримерной зоны и тем более структурированной должна быть вспышка ее стримерной короны, что согласуется с наблюдениями: наиболее структурированных вспышек связанных с самой длинной ступенью (123 см; Рисунок 6.15) и наименее структурированная стримерная вспышка с самым коротким шагом (50 см; Рисунок 6.17). Ситуация с отрицательными лидерными ступенями менее ясна: ярко выраженная структура всплеска в этом случае присутствует всегда (см. Рисунки 6.7 и 6.9–6.12), и, похоже, нет зависимости степени структурированности отрицательной стримерной вспышки от длины ступени (сравните, например, Рисунки 6.11 и 6.12, предполагая, что общая длина отображаемого канала в каждом случае в значительной степени определяется вновь сформированной ступенью).

Обычно считается, что (1) стримерные зоны положительных и отрицательных лидеров существенно различаются (например, [Базелян и др., 1978, стр.67] и (2) стримеры простираются в прямом направлении (в пределах угла разветвления 60–90° или около того) [Горин и Шкилев, 1974]. Наши наблюдения показывают, что эти два предположения не выполняются для больших ступеней положительного лидера, которые возникают после начала квазинепрерывной фазы развития лидера (и прерывают ее). Сравнение Рисунков 6.7 (положительная вспышка стримерной короны) и 6.15 (отрицательная вспышка стримерной короны) ясно показывают, что вспышки стримерной короны при образовании ступеней положительных и отрицательных лидеров могут быть практически одинаковыми по форме и что некоторые стримеры вспышки могут распространяться в обратном направлении по отношению к траектории канала лидера, придавая общей форме вспышке вид, близкий к сферическому.

Возможно, что ступени отрицательного лидера и ступени положительного лидера («restrikes» по терминологии Les Renardieres Group [Les Renardieres Group, 1972, 1974, 1977]) имеют гораздо больше общего, чем считалось ранее. Отличительной особенностью ступеней отрицательного лидера являются спейс-стем и спейс-лидер, которые наблюдаются в стримерной зоне перед головкой канала отрицательного лидера. Спейс-лидер движется в двух направлениях и в конечном итоге сливается с основным каналом. Резкая передача высокого потенциала основного канала к нижнему концу новообразованной ступени (новая головка лидера) вызывает мощный всплеск стримерной короны, который мы и наблюдали на Рисунках 6.7-6.12. Может ли благодаря подобному механизму слияния квази-спейс-лидера (горячего плазменного образования, расположенного перед основным каналом, как на Рисунке 6.15) произойти вспышка стримерной короны, в момент образования ступени положительного лидера? До этого нашего исследования спейс-лидеры (и даже спейс-стемы) никогда не наблюдались в стримерных зонах положительных лидеров. Кроме того, [Baldo and Rea, 1973, 1974] для длинных искр, и [Wang et al., 2016] для восходящего положительного лидера молнии, сообщали, что ступень положительного лидера распространяется в прямом направлении (от основного канала). С другой стороны, [Pilkey, 2014, рисунки А-1 — А-3] сообщил о «плавающем» сегменте, похожем на спейс-лидер, развивающемся за сотни микросекунд, и сливающемся с боковой частью поверхности положительного канала, приводя к резкому увеличению свечения основного канала. Если этот результат говорит в пользу существования спейс-лидера (или подобного горячего, высокопроводящего плазменного образования) в стримерной зоне положительного лидера, то в этом случае можно предположить, что нет концептуальной разницы между ступенями отрицательного лидера и ступенями положительного лидера («restrikes» по терминологии Les Renardieres Group [Les Renardieres Group, 1972, 1974, 1977]), и что ступени положительного лидера, инициированные контактом основного канала положительного лидера с квази-спейс-лидером, просто гораздо труднее наблюдать, но они могут играть важную роль в развитии положительных каналов молнии.

Вполне вероятно, что электрический заряд вспышки стримерной короны как при образовании отрицательной, так и положительной ступени оказывает одинаковое влияние на новообразованную головку лидера: инжектируемый объемный заряд той же полярности, что и у головки, снижает электрический потенциал головки чем снижает способность головки генерировать стримеры. Дополнительный приток заряда к головке лидера необходим для увеличения электрического потенциала головки, чтобы преодолеть влияние пространственного заряда перед головкой (выброшенного в объем перед головкой во время образования ступени). Этот процесс требует времени и, по-видимому, частично отвечает за паузу между двумя последовательными ступенями, как отрицательного, так и положительного лидера.

Важно отметить, что наблюдения [Baldo и Rea, 1973, 1974] и [Wang et al., 2016] указывают, что ступень положительного лидера — это процесс движения вперед (от головки вглубь разрядного объема), развивающийся со скоростью порядка $10^6$ м/с. Интересно, что в отличие от ступеней положительного лидера, которые являются удивительно прямыми (Рисунки 6.14–6.19), ступени отрицательного лидера, которые, как известно, образованы движением спейс-лидеров и основных лидеров, обычно изогнуты (см. Рисунки 6.9–6.12). Наиболее вероятной кажется гипотеза, что положительная ступень лидера развивается вдоль более или менее прямой траектории стримеров (такой, как наблюдаемая во вспышке стримерной короны в правом нижнем углу Рисунка 6.14b), в то время как траектория отрицательной ступени определяется криволинейной траекторией положительного конца спейс-лидера и основного лидера (см. Рисунок 6.9a). Ясно, что необходимы дополнительные данные и теоретическое моделирование этого процесса, чтобы улучшить наше понимание ступеней позитивных лидеров («restrikes» по терминологии Les Renardieres Group [Les Renardieres Group, 1972, 1974, 1977]).

[Tran and Rakov, 2016] с помощью высокоскоростной видеокамеры наблюдали разряд грозовой молнии, который начинался, по их мнению, с двунаправленного лидера. Отрицательный конец двунаправленного лидера распространялся к земле (и в конечном итоге привел к обратному удару), в то время как его положительный конец, развивающийся в основном горизонтально, начинался на высоте около 4 км над поверхностью земли и демонстрировал резкое удлинение, напоминающее гигантскую ступень. Эта «ступень» была относительно прямой и имела двумерную длину около 1 км.

(Для сравнения, [Pilkey, 2014, рис. A-1 и A-2], сообщал о том, что, похожее событие было инициировано «спейс-лидером» примерно в 84 м от канала положительного лидера на высоте 3 км и через 364 мкс (время неточное, так как определяется числом кадров видеосъемки) после контакта с боковой поверхностью этого канала). Похоже, это событие происходило в чистом воздухе, распространялся плазменный канал с очень высокой скоростью, порядка $10^6$ м/с, и вызвал вспышку стримерной короны на его дальнем конце. На наш взгляд, эти события нельзя отнести к процессу распространения положительного и отрицательного лидера из-за очень больших размеров предполагаемых [Tran and Rakov, 2016] и [Pilkey, 2014] ступеней и нереально большими скоростями распространения лидеров. Скорее всего, в этих работах наблюдался контакт уже существовавших двунаправленных лидеров, чья длина принималась за длину ступеней, а скорость вычислялась исходя их прибавления к каналу уже существовавшей длины присоединяющегося двунаправленного лидера. [Tran and Rakov, 2016] предварительно интерпретировали наблюдаемое ими событие, как гигантскую ступень положительного лидера километрового масштаба, который развился из спейс-стема/лидера (понятие которое нам кажется некорректным) и соединился с ранее существующим каналом, создав большую ветвь, которая в конечном итоге стала частью основного канала. Событие могло инжектировать значительное количество положительного заряда около новообразованного положительного конца (вероятно, это была самая интенсивная вспышка стримерной короны в этом развитии молнии). Этот инжектированный объемный заряд мог препятствовать дальнейшему расширению положительного конца, как видно на их кадрах скоростной камеры. Подобный эффект блокировки удлинения лидера при вспышке стримерной короны наблюдался в длинных положительных искрах, при этом вероятность преодоления эффекта пространственного заряда была выше при более высокой скорости нарастания напряжения в промежутке [Les Renardieres Group, 1972, 1977]. Однако точно такая же блокировка движения положительного конца двунаправленного лидера была бы и при контакте двух двунаправленных лидеров, а не контакте канала и спейс-стема.

В обеих этих работах ([Tran and Rakov, 2016], [Pilkey, 2014, рис. A-1 и A-2]) авторы имели дело с внутриоблачными процессами, когда облака могли закрывать часть или всю длину уже существовавших двунаправленных лидеров и какую-то часть двунаправленного лидера можно было принять за чрезвычайно большой «спейс-лидер».

Нам кажется, что эта замена двунаправленного лидера длиной сотни метров или даже километр «спейс-лидером», который должен иметь длину всего в несколько метров, приводит к спорной интерпретации экспериментальных данных. На наш взгляд, в физике молнии пришло время смены парадигмы, где один гигантский двунаправленный лидер Каземира необходимо заменить на множество взаимодействующих двунаправленных лидеров и других плазменных образований, по крайней мере, на стадии начальных импульсных пробоев (стадии IBPs).

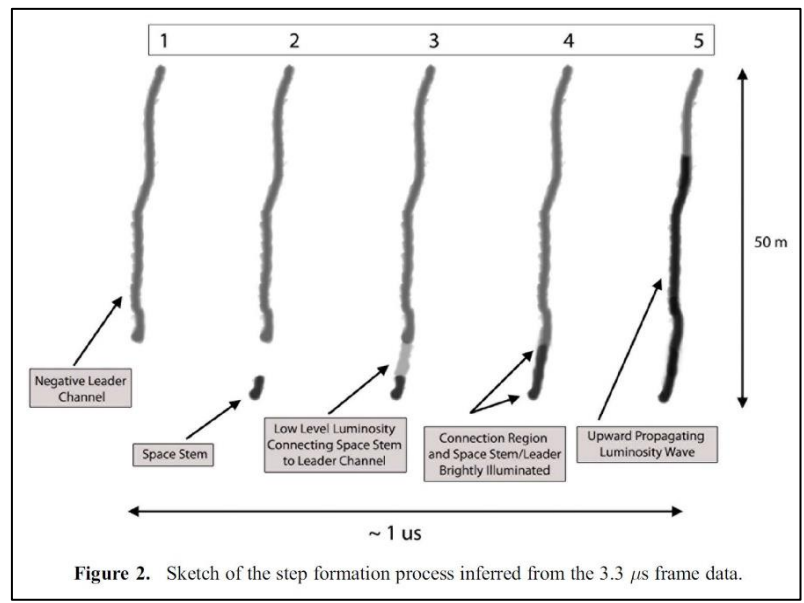

**Figure 2.** Sketch of the step formation process inferred from the 3.3 μs frame data.

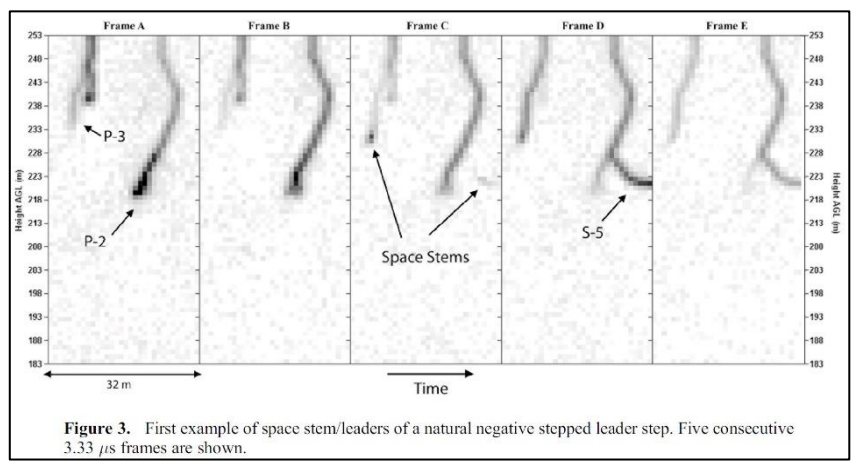

**Figure 3.** First example of space stem/leaders of a natural negative stepped leader step. Five consecutive 3.33 μs frames are shown.

Рисунок 6.20. (адаптированные Figure 2, 3 из [Hill et al., 2011]). Схема (Figure 2) и видеокадры скоростной камеры (Figure 3) образования ступени нисходящего отрицательного лидера молнии. Кадры имеют низкое пространственное разрешение, матрица камеры мало чувствительна в УФ-области и, поэтому, кадры не позволяют сделать вывод о типе плазменных образований в стримерной зоне отрицательного лидера (стримерная корона, спейс-стем, спейс-лидер или более темные места обычного лидера?). Видимо поэтому [Hill et al., 2011] пишут через косую черту — «спейс стем/лидер».

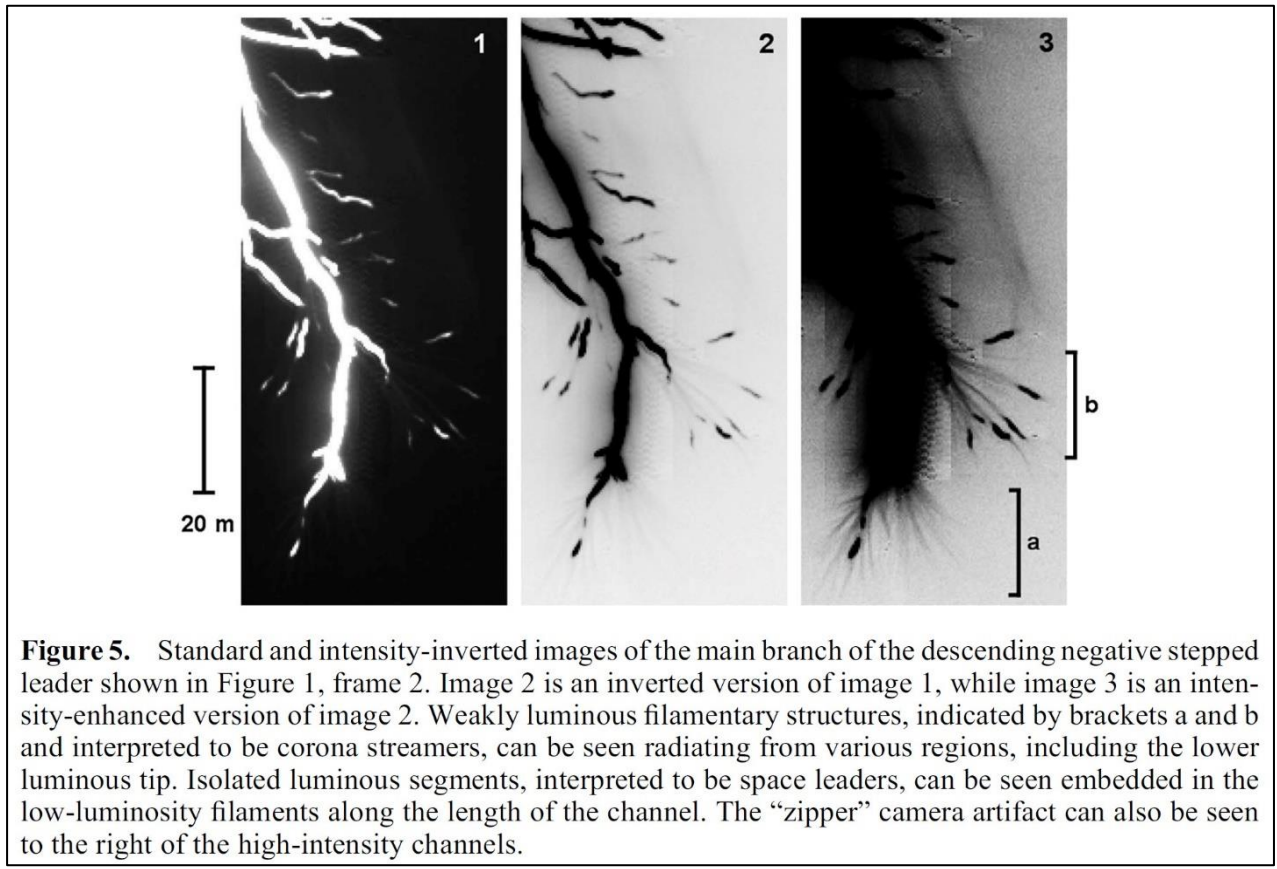

**Figure 5.** Standard and intensity-inverted images of the main branch of the descending negative stepped leader shown in Figure 1, frame 2. Image 2 is an inverted version of image 1, while image 3 is an intensity-enhanced version of image 2. Weakly luminous filamentary structures, indicated by brackets a and b and interpreted to be corona streamers, can be seen radiating from various regions, including the lower luminous tip. Isolated luminous segments, interpreted to be space leaders, can be seen embedded in the low-luminosity filaments along the length of the channel. The "zipper" camera artifact can also be seen to the right of the high-intensity channels.

Рисунок 6.21. (адаптированные Figure 5 из [Petersen and Beasley, 2013]). Видеокадры скоростной камеры стримерной зоны нисходящего отрицательного лидера молнии. [Petersen and Beasley, 2013] претендуют на то, что они идентифицируют стримерную зону и спейс-лидеры на своих кадрах. Но кадры имеют недостаточное пространственное разрешение, матрица камеры мало чувствительна в УФ-области и, поэтому, кадры не позволяют сделать надежный вывод о типе плазменных образований в стримерной зоне отрицательного лидера (стримерная корона, спейс-стем, спейс-лидер или более темные места обычного лидера?). Обычная стримерная зона между удлиненными плазменными образованиями имеет форму веретена, а на кадрах более темные светлые участки находятся внутри «канальных структур» и на каждый канал имеется одна структура, в стримерной зоне длинной искры всегда есть спейс-стемы и спейс-лидеры, то есть структура должна иметь, по крайней мере три уровня.

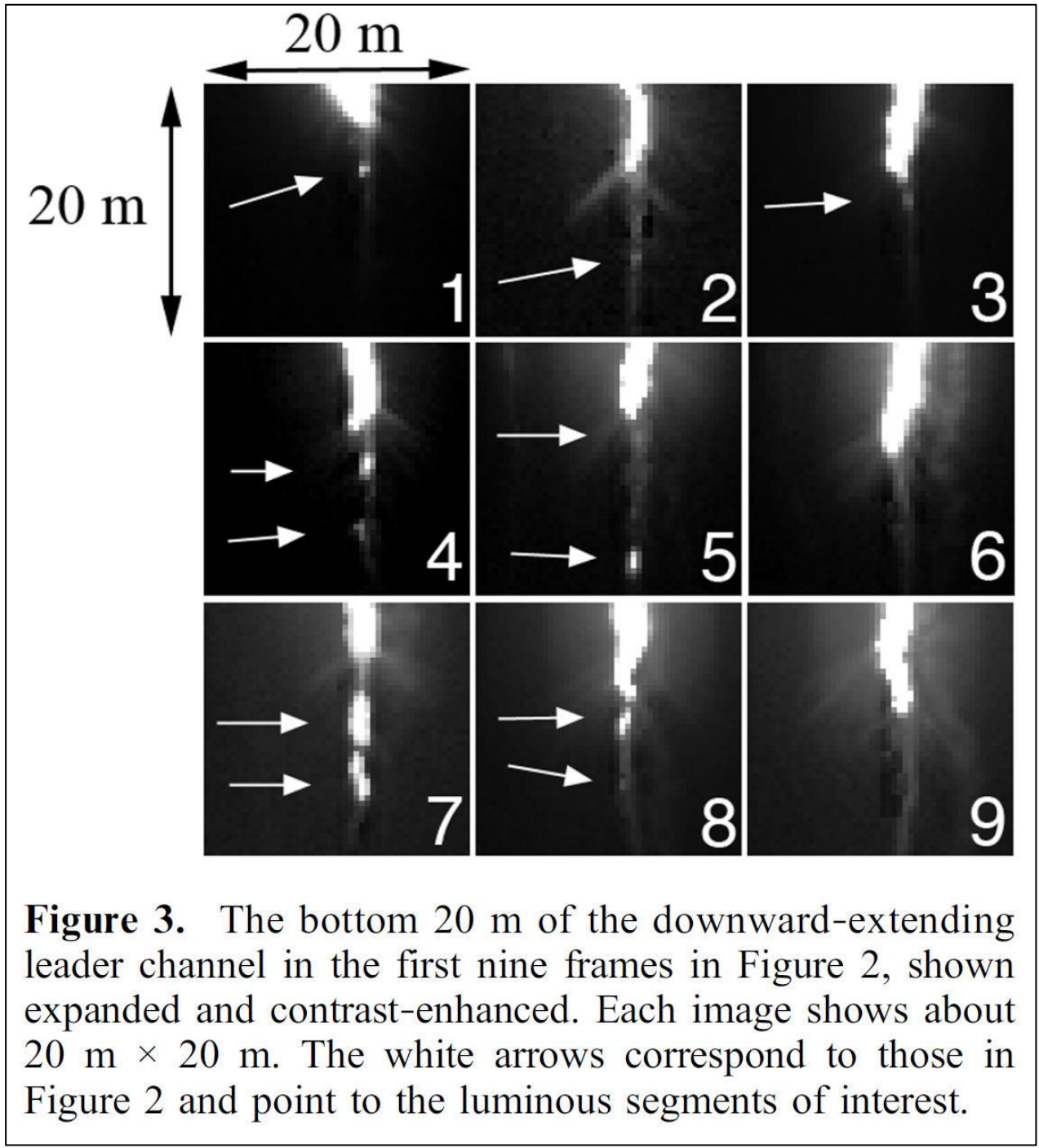

**Figure 3.** The bottom 20 m of the downward-extending leader channel in the first nine frames in Figure 2, shown expanded and contrast-enhanced. Each image shows about 20 m × 20 m. The white arrows correspond to those in Figure 2 and point to the luminous segments of interest.

Рисунок 6.22. (адаптированные Figure 3 из [Biagi et al., 2010]). Видеокадры скоростной камеры стримерной зоны нисходящего отрицательного лидера молнии. [Biagi et al., 2010] считают, что они идентифицировали стримерную зону, спейс-стемы и спейс-лидеры на своих кадрах. Но кадры имеют недостаточное пространственное разрешение, матрица камеры мало чувствительна в УФ-области и, поэтому, кадры не позволяют сделать надежный вывод о типе плазменных образований в стримерной зоне отрицательного лидера, хотя и представляют большой интерес.

## 6.8. Выводы главы 6

1. В этом исследовании наблюдались как квазинепрерывный (оптически непрерывный, «обычный»), так и ступенчатый режимы распространения, проявляемые разными ветвями одного и того же положительного лидера. Положительные лидеры демонстрировали ступенчатость развития ступенями длиной в десятки сантиметров, при этом морфология вспышки стримерной короны была аналогичной наблюдаемой у отрицательных лидеров, развивающихся в чистом воздухе.

2. Для любой полярности вспышки стримерной короны могут иметь почти сферический вид, что предполагает, что быстрая инжекция значительного объемного заряда перед головкой лидера может создать электрическое поле, которое снизит потенциал головки лидера.

3. Представлены свидетельства существования плазменного образования плавающего типа (удаленного от основного канала лидера), которые можно попытаться интерпретировать в терминах спейс-лидера (более вероятно) или спейс-стема, которые находятся перед концом положительного лидера. Это плазменное образование похоже на спейс-лидер внутри положительной стримерной зоны, с разветвленной, идущей вперед, положительной частью с несколькими головками и отсутствием структуры отрицательной части плазменного образования, лежащей напротив основного положительного лидерного канала.

4. Сравнение с характеристиками положительных и отрицательных лидеров, наблюдаемых при развитии молнии, показывает, что для любой полярности ступени длинных искр качественно похожи на ступени молнии.

5. Наши результаты по изучению ступеней положительного лидера вместе с результатами, опубликованными Les Renardieres Group для положительных искр и [Wang et al., 2016] для восходящего положительного лидера природной молнии, выступают против представления (например, [Saba et al., 2015], [Visacro et al., 2017]) о том, что положительный лидер не может порождать собственные ступени, а возможны только колебания тока и светимости, индуцированные на канале положительного лидера электрическим полем или стримерной короной одновременно развивающегося отрицательного ступенчатого лидера.

# ГЛАВА 7. Механизм инициации молнии от инициирующего события (IE) через стадию изменения электрического поля (IEC) до стадии первых импульсов начального пробоя (IBPs)

В данном разделе изложены результаты исследований [Kostinskiy et al., 2020a], где, основываясь на экспериментальных результатах последних лет, дано качественное описание возможного механизма (называемого далее для краткости Механизмом, чтобы отделить от других физических механизмов, обсуждаемых в данном разделе) инициации молнии. Механизм является комплексным (составным) процессом и охватывает основные этапы инициации и развития молнии, начиная с инициирующего молнию самого первого события (IE — the initiating event), за которым следует изменение начального электрического поля (IEC — the initial electric field change), с последующими несколькими начальными импульсами пробоя (IBPs — the initial breakdown pulses) вплоть до старта большого отрицательного лидера, надежно регистрируемого радиофизическими системами и общепризнано считающегося «молнией». Механизм предполагает, что инициирование происходит в области ~0.1-1 км$^3$ со средним электрическим полем E > 0,3 МВ/(м·атм), которая содержит из-за статистического движения заряженных гидрометеоров и турбулентности многофазной среды многочисленные небольшие «$E_{th}$-объемы», под которыми мы будем понимать объемы размером ~$10^{-4}$-$10^{-3}$ м$^3$, где электрическое поле E $\geq$ 3 МВ/(м·атм) обеспечивает размножение «классических» разрядных лавин, то есть, в этих областях частота ионизации выше частоты прилипания $\nu_i > \nu_a$. Механизм описывает инициацию и развитие молнии в результате любого из двух наблюдаемых типов инициирующих событий (IE): сильного УКВ (VHF)-события, такого как компактный внутриоблачный разряд (compact intracloud discharge — CID, NBE — narrow bipolar event), или слабого УКВ (VHF)-события. Согласно Механизму, оба типа инициирующих событий вызваны большой группой, широко распределенных в объеме с сильным электрическим полем грозового облака, лавин релятивистских убегающих электронов и позитронов, где начальные электроны и позитроны являются вторичными частицами широкого атмосферного ливня (ШАЛ, extensive air shower — EAS). Электроны и позитроны проходят через множество $E_{th}$-объемов (которые мы также будем называть «воздушными электродами»), тем самым вызывая почти одновременную (синхронизованную) инициацию множества положительных стримерных вспышек. Из-за ионизационно-перегревной неустойчивости вдоль траекторий стримерных вспышек появляются необычные плазменные образования (unusual plasma formations — UPFs). Эти UPFs во время протекания IEC-стадии, благодаря небольшим сквозным фазам между ними и образовавшимися в результате их взаимодействия двунаправленными лидерами, объединяются в трехмерные (3D) сети горячих высокопроводящих плазменных каналов, что приводит к наблюдаемому слабому току и изменению электрического тока с небольшими импульсами. Последующее развитие этих 3D плазменных сетей приводит к контакту двух (или более) сетей, благодаря взаимодействию (сквозным фазам) их плазменных каналов, что вызывает первый IBP. Каждый последующий IBP возникает, когда еще одна трехмерная сеть каналов горячей плазмы объединяется с цепочкой сетей, созданной более ранними IBPs. Цепочка IBPs приводит к появлению «большого» ступенчатого отрицательного лидера, надежно фиксируемого наземными радиосистемами, который общепризнано считается молнией [Rakov and Uman, 2003]. Таким образом, мы делаем попытку создать качественный Механизм инициации молнии, как целостного процесса (цепочки превращений плазмы из одной формы в другую), в отличие от прежних критериев инициации молнии, за который принимался, например, инициация первого стримера (например, [Dawson & Duff, 1970], [Gurevich et al., 1999], [Babich et al., 2016]), инициация лавины убегающих электронов [Gurevich et al., 1992] или первой стримерной вспышки [Rison et al., 2016].

## 7.1. Введение в главу 7

Несмотря на большие усилия научного сообщества, до сих пор нет общепринятого, даже качественного последовательного механизма инициирования молнии от первого инициирующего события (IE), за которым следует последующее развитие, вплоть до начала отрицательного ступенчатого лидера (например, [Rakov and Uman, 2003], [Gurevich and Zybin, 2001], [Dwyer & Uman, 2014]). Эта ситуация частично объясняется исключительной сложностью и комплексностью явления «молния», которое требует как экспериментальных, так и теоретических знаний о самой молнии, а также информации, как минимум из атмосферной физики высоких энергий, радиофизики атмосферных разрядов, физики турбулентных многофазных заряженных аэрозолей, физики газового разряда высокого давления и физики длинных искр. Этот список легко можно было бы расширить. Однако после недавнего значительного прогресса в экспериментальных и теоретических работах в настоящее время существует острая необходимость хотя бы в качественном построении единого механизма, описывающего в пространстве и времени происхождение и развитие молнии.

Насколько нам известно, только [Petersen et al., 2008] попытались описать возникновение молнии от момента инициации первого стримера до появления отрицательного ступенчатого лидера (см. Рисунок В.38). До [Petersen et al., 2008] проблема зарождения молнии в первую очередь была связана с появлением первой лавины или первого стримера и этой стадией ограничивалась. Основываясь на измерениях с помощью УКВ-интерферометра (VHF), работающего на частоте 20–80 МГц, [Rison et al., 2016] «предварительно» пришли к выводу, что исходным событием всех молний является узкое биполярное событие (NBE/CID или компактный внутриоблачный разряд — КВР), вызванное «быстрым положительным пробоем» (FPB), новым процессом, который они предложили. Исследуемые КВР (NBE/CID) имели кажущуюся скорость $4–10 \cdot 10^7$ м/с. [Attanasio et al., 2019] недавно развили механизм распространения FPB, основанный на модернизации модели Гриффитса и Фелпса [Griffiths & Phelps, 1976], описывающей возникновение молнии благодаря гигантской стримерной вспышке, стартующей с гидрометеоров (подробный анализ недостатков данной модели во Введении). По данным с датчиков электрического поля (называемых «быстрой антенной» или «FA» с типичной полосой пропускания 0,1-2500 кГц), типичные изолированные КВР (NBE) имеют характерную биполярную форму волны с длительностью 10-30 мкс и большими амплитудами импульсов электрического поля (например, [Willett et al., 1989], [Nag et al., 2010], [Karunarathne et al., 2015]. Типичные КВР (NBE) также имеют большую мощность в полосе частот HF/VHF от 3 до 300 МГц [Le Vine, 1980]. Для десяти положительных NBE, инициирующих внутриоблачные молнии (IC), [Rison et al., 2016] обнаружили пиковые мощности от 1 до 274000 Вт в диапазоне VHF (30-300 МГц), а для 5 отрицательных NBE, инициирующих молнии облако-земля (CG), они обнаружили пиковые мощности NBE в диапазоне от 1 до 600 Вт. [Tilles et al., 2019] далее сообщили, что некоторые положительные NBE вызваны

«быстрым отрицательным пробоем» с кажущейся скоростью распространения центроидов интерферометра $4 \cdot 10^7$ м/с.

Недавние результаты показывают, что большинство вспышек молний не инициируются «классическими» NBE, как предполагали [Rison et al., 2016]; гораздо чаще вспышки инициируются гораздо более короткими и гораздо более слабыми событиями. [Marshall et al., 2019] сообщили о первых примерах вспышек с гораздо более слабыми инициирующими событиями (IE). Две IC-молнии были инициированы VHF-событиями с длительностью 1 мкс и пиковой VHF-мощностью 0,09 Вт и 0,54 Вт; в данных FA этих двух вспышек не было совпадающих импульсов. Обращает на себя внимание, что данные FA в основном измеряют движение зарядов в масштабах длин волн > 50 м, в то время как данные VHF в основном измеряют движение зарядов с длинами волн <5 м, поэтому отсутствие импульса FA во время инициирующего события (IE) предполагает, что заряды перемещались на расстояние порядка 5 м, а не 50 м. Две CG-молнии, исследованные [Marshall et al., 2019] были инициированы VHF-событиями длительностью 1 и 2 мкс и VHF-мощностью 0,14 Вт и 0,64 Вт; у этих событий были слабые короткие импульсы быстрой антенны (FA), совпадающие с одним из VHF событий, инициирующих CG-молнии. [Lyu et al., 2019] изучили 26 IC-молний, которые произошли в пределах 10 км от их VHF-интерферометра, и обнаружили, что NBE инициировали только 3 из 26 IC-молний; остальные 23 IC-молнии были вызваны слабыми VHF-событиями длительностью менее 0,5 мкс. [Bandara et al., 2019] исследовали 868 отрицательных CG-молний, происходивших на расстояниях 17–125 км от измерительной системы, и обнаружили, что только 33(4%) из них были инициированы отрицательными NBE; эти относительно слабые отрицательные NBE имели VHF-мощность в диапазоне 1-1300 Вт.

В этой статье мы описываем в качестве первой попытки основные этапы возможного механизма инициирования и развития молнии от инициирующего события (IE) до нескольких первых классических начальных импульсов пробоя (IBPs). Мы с самого начала осознаем рискованность такой попытки, поскольку многие процессы и явления, которые составляют основу предложенного нами Механизма, в настоящее время в достаточной степени не изучены или еще не рассматривались в такой тесной взаимосвязи друг с другом. Эта позиция ограниченных знаний дает значительное пространство для теоретических спекуляций. Но, на наш взгляд, построение единого

Механизма, состоящего из последовательной цепочки плазменных превращений, также имеет преимущества, поскольку позволяет будущим исследованиям сосредоточиться на количественном анализе и улучшении (или существенных изменениях) каждого звена этой цепочки. Таким образом, наша цель — улучшить понимание не только отдельных ключевых стадий развития молнии, но и всего процесса, который объединяет эти ключевые стадии.

Так как наш Механизм базируется на результатах исследований последних нескольких лет, когда зарубежными учеными было введено в оборот много понятий, не имеющих пока однозначного перевода на русский язык, то мы приводим частичный список важных терминов и сокращений, используемых в этом разделе:

t1. *Инициирующее молнию событие (the initiating event — IE)* молнии является первым электромагнитным проявлением инициирования молнии и может быть слабым (weak) NBE, описанным [Marshall et al., 2014a, 2019] и [Lyu et al., 2019] или более сильным, «классическим» NBE, как описано [Rison et al., 2016] и [Lyu et al., 2019]. Как было сказано выше, слабые IE имеют VHF-мощность <1 Вт и длительность ≤ 1 мкс, в то время как KBP(NBE/CID) имеют на порядки большую VHF-мощность и длительность 10–30 мкс.

t2. *Узкое биполярное событие (narrow bipolar event — NBE) или compact intracloud discharges (CIDs) или компактный внутриоблачный разряд (КВР).* Мы будем пользоваться в основном термином NBE, который сейчас наиболее употребим. Все эти термины являются синонимами, которые обозначаю особый тип электрического события, которое происходит во время протекания или вблизи грозы, но не является ни внутриоблачной молнией (IC), ни молнией типа облако-земля (CG). Это важнейшее явление открыл Давид Ле Вайн в 1980 году [Le Vine, 1980]. NBE в данных FA (быстрой антенны) имеет биполярную форму волны длительностью 10-30 мкс; в диапазоне VHF 60–66 МГц сильные NBE имеют большую мощность (30 000–300 000 Вт или 45–55 дБВт) [Rison et al., 2016]. Сильные NBE являются самыми мощными VHF-явлениями в атмосфере, превосходя «классические молнии на порядок» [Smith et al. 1999]. Данные FA для слабых NBE имеют меньшие амплитуды, чем NBE, и могут иметь либо биполярные, либо «в основном монополярные» формы волны; слабые NBE также имеют меньшую VHF-мощность, составляющую 3–300 Вт или 5–25 дБВт [Rison et al., 2016].

t3. **Начальное изменение электрического поля (an initial electric-field change — IEC)**. Важнейший этап развития молнии, который относительно недавно выделил Томас Маршалл с соавторами [Marshall et al., 2014a]. Как описали этот этап [Marshall et al., 2014a] и [Chapman et al., 2017], представляет собой относительно длительный период (40-9800 мкс), который начинается с IE и заканчивается первым классическим начальным импульсом пробоя (IBP). [Marshall et al., 2019] показали, что во время проведения IEC было много импульсов VHF с длительностью 1–7 мкс и что некоторые совпадающие пары импульсов быстрой антенны (FA) и импульсов VHF, похоже, увеличивают IEC (как «усиливающие события» («enhancing events»)).

t4. **Начальный импульс пробоя (an initial breakdown pulse — IB-импульс или IBP)** — это очень сильный биполярный электрический импульс, возникающий в первые несколько миллисекунд после инициации молнии (для CG-молний в первые сотни микросекунд), обычно обнаруживаемый с помощью FA (например, [Weidman & Krider, 1979], [Nag et al., 2009]. Самые большие IBP называются «классическими IBP» и они систематически сопровождаются VHF-импульсами во вспышках CG [Kolmašová et al., 2019]. По нашему определению классические IBP имеют длительность $\geq$ 10 мкс, амплитуду $\geq$ 25% от наибольшего IBP и часто имеют субимпульсы. Амплитуды классических IBPs могут быть 30-160 кА (это параметры мощных обратных ударов молнии) несмотря на то, что IBPs могут возникать через чрезвычайно короткие интервалы времени после инициации молнии (40 мкс-3 мс). По существу, все молнии имеют серию (серии) IBPs, которые происходят в течение нескольких мс после IEC ([Mäkelä et al., 2008], [Marshall et al., 2014b]). По данным [Mäkelä et al., 2008] на большой выборке, не менее 90% всех молний в Северной Европе имеют стадию начальных импульсов пробоя (IBPs), которая предшествует появлению отрицательного лидера, причем для остальных 10% также может присутствовать эта стадия, но экспериментальные данные не давали возможности сделать однозначный вывод. Мы называем период, в течение которого возникают IBPs, «IB-стадия» молнии. Во время протекания IB-стадии существуют биполярные импульсы, меньшие по амплитуде или более короткие по длительности, чем классические IBPs, но они также являются IBPs. Классические и меньшие IBPs, в принципе, могут быть вызваны разными плазменными процессами.

t5. **Стример** — это холодная плазма быстрой волны ионизации, которая поддерживается собственным электрическим полем головки, благодаря поляризации во внешнем электрическом поле, как описано, например, [Райзер, 1992, стр. 423]. В этой статье термин «стример» означает только «обычные» стримеры, которые наблюдались в течение многих десятилетий в газовых разрядах и длинных искрах при давлениях 0.1-1 атм и имеют длину от сантиметров до нескольких метров [Райзер, 1992, стр. 423].

t6. **Необычное плазменное образование (an unusual plasma formation — UPF)** представляет собой короткий канал (каналы) горячей высокопроводящей плазмы, как описано [Kostinskiy et al., 2015a, 2015b], см. главы 1-5. UPFs часто представляют собой иерархическую сеть каналов горячей плазмы длиной в десятки сантиметров.

t7. **Положительный лидер** — это канал горячей плазмы, который удовлетворяет определенным условиям длины в окружающем электрическом поле, так что лидер при определенных условиях будет самораспространяющимся, как описано [Bazelyan et al., 2007a]. При электрическом поле 0,45 - 0,50 МВ/(м атм) (в зависимости от влажности) горячий плазменный канал любой длины будет самораспространяющимся, благодаря положительным стримерам.

t8. ***EE-объем (EE-volume), Estr₊-объем (Estr₊-volume), Eth-объем (Eth-volume):***

*EE-объем (0,1-1 км³)* — это область в грозовом облаке со средней величиной электрического поля $E > 0,28$-$0,35$ МВ/(м·атм) и с большим числом заряженных гидрометеоров разных размеров. Гидрометеоры могут быть жидкими или твердыми, большими или маленькими по размеру, при условии, что их много и они несут значительные электрические заряды, так что турбулентные и статистические движения могут приводить к появлению небольших областей EE-объема с существенно большими электрическими полями. EE-объем может иметь сильно неоднородные электрические поля (в масштабе сотен метров) и состоять из множества близко расположенных турбулентных областей, которые могут быть образованы одинаково или противоположно заряженными встречными многофазными потоками воздуха (например, [Karunarathna et al., 2015]; [Yuter & Houze, 1995]).

$E_{str\,+}$ *-объем (может быть повторяющимися областями с характерным размером порядка 100 м)* — это области в грозовом облаке с $E \geq 0,45$-$0,5$ МВ/(м·атм). Столь большие величины E достаточны для поддержания движения положительных стримеров в воздухе [Базелян и Райзер, 1997].

$E_{th}$-объем ($10^{-3}$-$10^{-4}$ $м^3$) или «воздушный электрод» — это небольшая область в грозовом облаке с E > 3 МВ/(м·атм); столь большие величины E достаточны для образования «классических» электронных лавин, которые при выполнении критерия Мика [Райзер, 1992, стр. 425] могут трансформироваться в классические газоразрядные стримеры.

t9. **EAS-RREA (широкий атмосферный ливень (ШАЛ) — лавина убегающих релятивистских электронов, extensive air shower — relativistic runaway electron avalanche)**, например, [Gurevich & Zybin, 2001]; [Dwyer, 2003], возникает, когда поток вторичных заряженных частиц ШАЛ входит в область в сотни метров с электрическим полем E > 280 кВ/(м·атм). Для проблемы инициирования молнии важны ШАЛ с энергией первичных частиц в диапазоне $\varepsilon_0 \geq 5 \cdot 10^{14}$-$10^{15}$ эВ (как описано ниже).

## 7.2. Экспериментальные и теоретические основания Механизма

Предлагаемый механизм, несмотря на его сложность, определяется и регламентируется достоверно установленными экспериментальными и теоретическими работами. В этом разделе мы перечисляем (i1, i2…) основные наблюдения и теоретические идеи, которые учитываются при разработке нашего Механизма.

i1. Как указывалось выше, [Rison et al., 2016] использовали интерферометр для обнаружения VHF-излучения во время возникновения молнии. Их прибор фиксировал источники VHF-излучения в облаках со скоростью около одного источника во временном окне 1 мкс и программное обеспечение в результате вычислений ставило «точку» на карте в трехмерном пространстве и времени. Эту «точку» называют центроидом (the centroid). Три положительных NBEs, которые были инициирующими событиями (IE) для трех IC молний, имели длительность 10-20 мкс и очень короткие, экспоненциально растущие фронты увеличения активности VHF с длительностью 1-3 мкс. Источники VHF-излучения (центроиды), ассоциированные с NBEs продвигались вниз с видимой скоростью 4-10·$10^7$ м/с на расстояния 500-600 м. [Bandara et al., 2019] обнаружили, что 33 из 868 CG-молний были инициированы слабыми отрицательными NBE с мощностью VHF от 1 до 1300 Вт или от 0 до 31 дБВт.

i2. Также, как указано выше, [Marshall et al, 2019] показали, что IE двух отрицательных CG-молний и двух IC-молний были связаны со слабым импульсом VHF, а не NBE, длительностью около 1 мкс и мощностью <1 Вт. [Lyu et al., 2019] показали, что 23 IC-молнии имели IE, который был связан со слабым импульсом VHF, длительностью ≤ 0,5 мкс, а не NBE.

i3. [Marshall et al., 2014b] изучили инициирование 18 CG-молний и 18 IC-молний и показали, что для каждого из 36 IE не было какой-либо измеренной электрической активности в течение 100-300 мс перед инициацией каждой из этих молний. После IE в каждой молнии возникает IEC-стадия. [Chapman et al., 2017] обнаружили, что продолжительность IEC-стадии развития молнии составляет в среднем 230 мкс для 17 CG-молний (диапазон 80–540 мкс) и 2700 мкс для 55 обычных IC-молний (диапазон 40–9800 мкс), а у некоторых молний было несколько IEC-стадий. Физический процесс, вызывающий IEC-стадию развития молнии, пока неизвестен, но, очевидно, результатом является разделение и накопление заряда, достаточного для возникновения первого классического IBP.

i4. Классические IBPs имеют нормированные по дальности (на 100 км) амплитуды в среднем около 1 В/м [Smith et al., 2018], расчетные пиковые токи 1–165 кА, переносимый во время каждого импульса большой заряд находился в диапазоне 0.12-1.7 Кл (средний 0.44 ± 0.40 Кл) [Betz et al., 2008], [Karunarathne et al., 2014], [N. Karunarathne et al., 2020]. Высокоскоростные видеокамеры показывают, что при каждом классическом IBPs есть яркие вспышки в видимом диапазоне (400-900 нм) [Stolzenburg et al., 2013, 2014], [Campos & Saba, 2013], [Wilkes et al., 2016]. [Stolzenburg et al., 2013] показали и описали серии световых вспышек, совпадающие с несколькими сериями IBPs в CG-молниях (Рисунок 1.18), следующим образом: «линейные сегменты заметно удаляются от первой световой вспышки на 55–200 мкс, затем тускнеют по всей длине, затем последовательность светимости повторяется вдоль того же пути» общей протяженностью 300-1500 м в течении IB-стадии. Эти вспышки света указывают на быстрое появление (не менее, чем за 20 мкс, что является выдержкой кадров скоростной камеры) горячих каналов с высокой проводимостью, которые в большинстве случаев исчезают через 40–100 мкс; через 2-5 мс IBPs переходят в отрицательный ступенчатый лидер с гораздо меньшей светимостью, чем IBPs, Рисунок 1.19 [Stolzenburg et al., 2013].

i5. [Gurevich et al., 1992] и [Gurevich & Zybin, 2001] теоретически предсказали важную роль космических лучей в возникновении молний. Они предположили, что в электрическом поле E> 218 кВ/(м·атм) космические лучи могут вызывать лавины убегающих электронов. Учет упругого рассеяния уточнил значение порогового поля убегающих электронов E> 284 кВ/(м·атм) ([Dwyer, 2003], [Babich et al., 2004], [Lehtinen & Østgaard, 2018]. [Gurevich et al., 1999] были первыми, кто предположил, что совместное действие ШАЛ и убегающих в поле грозового облака электронов может играть значительную роль в инициировании первого стримера в грозовом облаке (они прировняли инициацию стримера к инициации молнии). [Dwyer, 2003, 2007] представил другой механизм генерации убегающих электронов. Для этого последнего механизма важны не только убегающие электроны, но и позитроны и энергичные фотоны, создающие эффект положительной обратной связи, который экспоненциально увеличивает количество лавин убегающих электронов [Dwyer, 2003, 2007]. Механизм, предложенный Двайером [Dwyer, 2003, 2007] требует гораздо более высоких средних электрических полей, чем механизм, предложенный Гуревичем с соавторами [Gurevich et al., 1992] и [Gurevich & Zybin, 2001].

i6. Используя устанавливаемый на воздушном шаре измеритель электрического поля во время активной грозы, [Marshall et al., 2005] подсчитали, что область, где возникли три CG-молнии, имела среднее электрическое поле E> 284-350 кВ/(м·атм) и занимала объем 1-4 км$^3$ с вертикальной и горизонтальной протяженностью 300-1000 м. Этот объем является экспериментальным примером EE-объема, как определено выше. Исходя из первого обнаруженного VHF-источника каждой молнии, три инициирования произошли в пределах 1,1 км от аэростата. [Marshall et al., 2005] обнаружили, что электрическое поле E в облаке превысило порог распространения лавины релятивистских убегающих электронов в E>284 кВ/(м·атм) примерно за 100 секунд до одной из вспышек молнии.

i7. [Базелян и Райзер, 1997, 2001], [Bazelyan et al., 2007a], и [Popov, 2009] теоретически показали ключевую роль «ионизационно-перегревной неустойчивости» в переходе холодной плазмы положительных стримеров в стеме и головке лидера в горячий плазменный канал лидера длинной искры (стримерно-лидерный переход); этот переход (в зависимости от силы тока) происходит менее чем за ~0,2-0,5 мкс при давлении 1 атм ([Popov, 2009], [da Silva & Pasko, 2013]). Если концентрация нейтральных молекул в атмосфере n уменьшается с высотой, то, согласно недавним теоретическим расчетам, время развития ионизационно-перегревной неустойчивости изменяется пропорционально $n^{-\alpha}$, где α находится в диапазоне от 1 до 2 ([Bazelyan et al., 2007b], [Riousset et al., 2010], [da Silva & Pasko, 2012, 2013]. Мы будем использовать значение $\alpha \approx$ 1 ([Velikhov et. al, 1977], [Raizer, 1991, p. 223-227], [Riousset et al., 2010]) в наших оценках, поскольку нет экспериментального подтверждения таких высоких значений, как $\alpha \approx 2$. Здесь мы будем использовать термин «ионизационно-перегревная неустойчивость», хотя этот физический процесс называется по-разному в разных областях физики, включая «thermal instability» («тепловую неустойчивость») ([Raizer, 1991], [Nighan, 1977], и «thermal ionizational instability» («термоионизационная нестабильность») [da Silva & Pasko, 2012].

i8. [Kostinskiy et al., 2015a, 2015b] экспериментально показали, что в электрических полях 500-1000 кВ/(м·атм) внутри искусственно заряженных аэрозольных облаков активно инициируются необычные плазменные образования (UPFs), а также двунаправленные лидеры длиной 1-3 м. Также экспериментально было показано, что в электрических полях 500-1000 кВ/(м·атм) UPFs возникают внутри плазмы положительных стримерных вспышек за счет ионизационно-перегревной неустойчивости [Kostinskiy et al., 2019], глава 2.

i9. [Colgate, 1967] предположил, что турбулентность в грозовом облаке может значительно увеличить локальное электрическое поле E на масштабах около 100 м. Трахтенгерц с соавторами теоретически показали, что из-за гидродинамической неустойчивости электрическое поле E в грозовом облаке может колебаться на размерах порядка 100 м ([Trakhtengerts, 1989], [Trakhtengertz et al., 1997], [Mareev et al., 1999], [Iudin et al., 2003]. [Trakhtengerts and Iudin, 2005], [Iudin, 2017] и [Iudin et al., 2019] теоретически оценили, что более значительные усиления E в меньшем масштабе (10-100 см) возможны благодаря статистическому движению облаке гидрометеоров разных размеров и зарядов. Многие исследования связывают возникновение молнии с турбулентным движением воздуха во время грозы. Например, [Dye et al., 1986] изучали одну небольшую грозовую ячейку с одним главным восходящим потоком и соседними нисходящими потоками. [Dye et al., 1986] обнаружили, что формирование грозового заряда происходит в основном на границе восходящего и нисходящего потоков грозовой ячейки. [Yuter and Houze, 1995] обнаружили, что грозы со многими грозовыми ячейками имеют "a three-dimensional multicellular radar reflectivity pattern that contained a rapidly evolving, complex jumble of updrafts and downdrafts" («трехмерную многоячеистую диаграмму отражательной способности радара, которая содержит быстро развивающуюся сложную мешанину восходящих и нисходящих потоков»). [Karunarathna et al., 2015] изучили местоположения положительных NBEs во время гроз и обнаружили, что многие положительные NBEs произошли в регионах гроз, где ожидаемое крупномасштабное электрическое поле E было направлено вверх (согласно физическому консенсусу), в то время как положительные NBEs требовали поля, направленного вниз. [Karunarathna et al., 2015] пришли к выводу, что эти NBEs, вероятно, происходят "in highly dynamic regions (such as contiguous updrafts and downdrafts in and beside the overshooting cloud top) where turbulent motions can stir, fold, and invert oppositely charged regions, thereby leading to a strong negative E (needed for positive NBE initiation) over a very limited distance." («в высокодинамичных регионах (таких как рядом протекающие восходящие и нисходящие потоки в верхней части облака и рядом с ним), где турбулентные движения могут перемешивать, складывать и инвертировать противоположно заряженные области, что приводит к сильному отрицательному E (необходимому для инициирования положительных NBEs) на очень ограниченном расстоянии». [Bruning and MacGorman, 2013] изучали характеристики молний на основе данных LMA (Lightning Mapping Array, системы картирования молнии) от двух гроз суперячейки и обнаружили, что "shape of lightning flash energy spectrum is similar to that expected of turbulent kinetic energy spectra in thunderstorms" («форма энергетического спектра молнии подобна ожидаемой для спектров турбулентной кинетической энергии во время грозы»), что позволяет предположить, что "advection of charge-bearing precipitation by the storm's flow, including in turbulent eddies, couples the electrical and kinematic properties of a thunderstorm." («адвекция заряженных осадков грозовых потоков, в том числе в турбулентных водоворотах, сочетает электрические и кинематические свойства грозы»). [Brothers et al., 2018] смоделировали электрификацию многоячеистой грозы и грозы суперячейки, используя модель с пространственным разрешением больших вихрей (сетка 125 м). Они обнаружили для обоих гроз, что турбулентность с большими вихрями вызвала больше областей, где мог произойти пробой из-за более близко расположенных небольших карманов противоположных зарядов.

## 7.3. Условия и явления, которым должен удовлетворять и объяснять Механизм

Основываясь на приведенных выше экспериментальных и теоретических результатах, Механизм должен удовлетворять следующим условиям (c1, c2…) и последовательно объяснять следующие явления:

c1. В целом Механизм должен объяснить, как работает инициация молнии. Развитие молнии должно начаться сразу после инициирующего события (IE). В частности, Механизм должен объяснить последовательность стадий инициирования молнии: IE, IEC и несколько первых классических IBPs на IB-стадии. За этими тремя начальными стадиями молнии следует хорошо изученная стадия большого отрицательного ступенчатого лидера, который общепризнано считается молнией.

c2. Оптическое излучение IE кажется довольно слабым на изображениях скоростных камер с матрицей, которые чувствительны в диапазоне длин волн 400-900 нм [Stolzenburg et al., 2014, 2020]. Таким образом, Механизм не должен содержать начальную мощную вспышку видимого света, подобную мощной вспышке света, которая возникает во время протекания IBPs или обратных ударов молнии.

c3. Для обоих типов IE (NBEs или Weak (слабых) NBEs): во время протекания IEC-стадии Механизм должен обеспечить создание длинных плазменных каналов (или сетей каналов), длиной в несколько километров, чтобы значительный заряд мог сохраниться в чехлах этих каналов, и чтобы заряд мог течь, тем самым обеспечивая IEC-стадию с последующей IB-стадией.

c4. Для IE, которые представляют из себя NBEs (например, [Rison et al., 2016]): Механизм должен объяснять создание самого NBE, включая короткую мощную вспышку положительных стримеров с экспоненциально увеличивающимся временем нарастания

VHF-излучения в течение нескольких микросекунд, общей продолжительностью 10–30 мкс и очень сильный VHF-излучением.

с5. Для IEs, которые являются NBEs: до и во время мощного VHF-излучения, производимого NBEs, Механизм не должен содержать уже существующий сильно излучающий длинный канал (каналы) горячей проводящей плазмы [Rison et al., 2016].

с6. Для IEs, которые являются NBEs: Механизм должен содержать физический процесс, который движется со скоростью, близкой к скорости света, чтобы соответствовать экспериментальным данным [Rison et al., 2016], а также обеспечивать биполярный импульс короткой длительности в данных FA (быстрой антенны).

с7. Механизм не должен противоречить общеизвестным и хорошо проверенным данным по физике газового разряда и физике длинной искры: например, скорость распространения положительных стримеров в зависимости от величины электрического поля; лавинно-стримерный переход; стримерно-лидерный переход; быстрое прилипание электронов к молекулам кислорода и т. д. В частности, лабораторные измерения показывают, что скорость положительных стримеров обычно составляет $1\text{-}10{\cdot}10^5$ м/с с максимумом $5{\cdot}10^6$ м/с для E, равного 3-4 МВ/(м·атм) [Les Renardieres Group, 1977].

с8. Механизм должен объяснить, как первый классический IBP (с током ≥ 10 кА и яркой световой вспышкой) возникает за такое удивительно короткое время после IE (первого инициирующего молнию события).

## 7.4. Некоторые основные компоненты Механизма

### 7.4.1. IE, EAS-RREA и лавинно-стримерный переход

Механизм предполагает, что IE состоит из большой группы классических электронных лавин, распределенных в пространстве объемом ~0.1-1 км³, а не из одной, пусть гигантской лавины. «Классическая» электронная лавина развивается в электрическом поле E ≥ 3 МВ/(м·атм). Механизм также предполагает, что большинство классических электронных лавин в группе запускаются электроном (или несколькими электронами), инициированными благодаря релятивистским электронам, позитронам или фотонам высокой энергии релятивистской лавины убегающих электронов [Gurevich & Zybin, 2001], [Dwyer, 2003], которые возникают благодаря ШАЛ ($\varepsilon_0 > 10^{15}$ эВ) в области сильного электрического поля E > 0,4 МВ/(м·атм). В разделе 7.6.0 более подробно обсуждаются необходимые и достаточные условия, необходимые для создания классических электронных лавин.

Чтобы лавина превратилась в положительный стример (т.е. претерпела лавинно-стримерный переход), лавина должна произвести около $10^8$-$10^9$ электронов в объеме около 0,3-0,5 мм$^{-3}$. Это критерий Мика [Raizer, 1991]. Когда эта электронная плотность достигается (2-3·$10^8$-$10^9$ мм$^{-3}$), сильное электрическое поле E, создаваемое поляризацией головки лавины в электрическом поле грозового облака, запускает самоподдерживающуюся ионизацию перед головкой (движущуюся волну ионизации). Соответственно, E головки стримера начинает в несколько раз превышать 3 МВ/(м·атм). После этого стример начинает двигаться самостоятельно. На этот процесс указывает грубая оценка электрического поля шара таких размеров $E = \frac{1}{4\pi\varepsilon_0}\frac{Q}{R^2}$, E≈9·$10^9$·(1.6·$10^{-19}$)·$10^9$/(0.4-0.2·$10^{-3}$)$^2$≈9-36 МВ/м. Но для того, чтобы стример полностью сформировался благодаря лавинному размножению, начинающегося с одного или нескольких начальных электронов, необходимо E > 3 МВ/(м·атм) на всем протяжении роста лавины. Эта длина определяется критерием Мика $\alpha_{eff} \cdot d \approx 18 - 20$, где $\alpha_{eff}$ — коэффициент Таунсенда, который составляет 10-12 см$^{-1}$ для воздуха атмосферного давления при E ≈ 3,0-3,2 МВ/(м·атм) [Raizer, 1991]. Это означает, что длина лавины, когда она перейдет в стример, будет около 2 см при атмосферном давлении и таком электрическом поле [Raizer, 1991], и эта длина будет экспоненциально увеличиваться с высотой. См. раздел 7.6.1 для более подробного обсуждения лавинно-стримерного перехода.

## 7.4.2. Положительные стримерные вспышки

Положительные стримерные вспышки состоят из большого количества положительных стримеров. Положительные стримерные вспышки, начинающиеся с металлических электродов, хорошо изучены. Задача определения параметров стримерной вспышки в грозовом облаке во время возникновения молнии ограничивает соответствующие экспериментальные данные следующими параметрами: напряжение и время нарастания Е (или фронты) должны быть > 100 мкс, а размер «воздушных электродов» в облаке должны быть ≥ 2-5 см. Мы дадим подробное объяснение этим ограничениям в следующем разделе.

### 7.4.2.1 Фронт и продолжительность типичной индивидуальной положительной стримерной вспышки

На Рисунке 7.1 показана типичная осциллограмма тока стримерной вспышки на электродах диаметром ~5-25 см в разрядных промежутках 4-20 м [Les Renardieres Group, 1977, Bazelyan & Raizer, 1998]. На панели (b) показана фотохронограмма (временная развертка) движения стримерной вспышки от электрода. Импульс напряжения, вызывающий положительные стримерные вспышки, имел довольно медленный фронт ~ 200-500 мкс и длительность импульса 2,5-10 мс.

В таких стримерных вспышках, стартующих в высоковольтного электрода (Рисунок 7.1) передний фронт нарастания тока составляет ~ 25–40 нс, и он нарастает до первого разветвления стримеров (1). В момент (2) начинается второе ветвление стримеров (для второй стримерной вспышки) с фронтом ~ 30-60 нс (3). Вся стримерная вспышка длится около 300-500 нс. Напряжение, при котором инициируется стримерная вспышка, сильно зависит от размеров электрода и его емкости. Например, для стержня с полусферическим концом диаметром 100 мм напряжение инициирования стримеров составляет $\cong$ 509 кВ (а электрическое поле Е на поверхности электрода достигает 6 МВ/м), а для сферы диаметром 1 м напряжение достигает 1855 кВ (а Е на поверхности электрода 3,2 МВ / м), [Les Renardieres Group, 1977]. Также размер электрода, из-за разной емкости, существенно влияет на общий заряд стримерной вспышки: для полусферы диаметром 100 мм общий заряд вспышки составляет 6,8 ± 3,4 мкКл, а для сферы диаметром 1 м общий заряд вспышки составляет 62 ± 1 мкКл. В облаках из-за относительно длительного

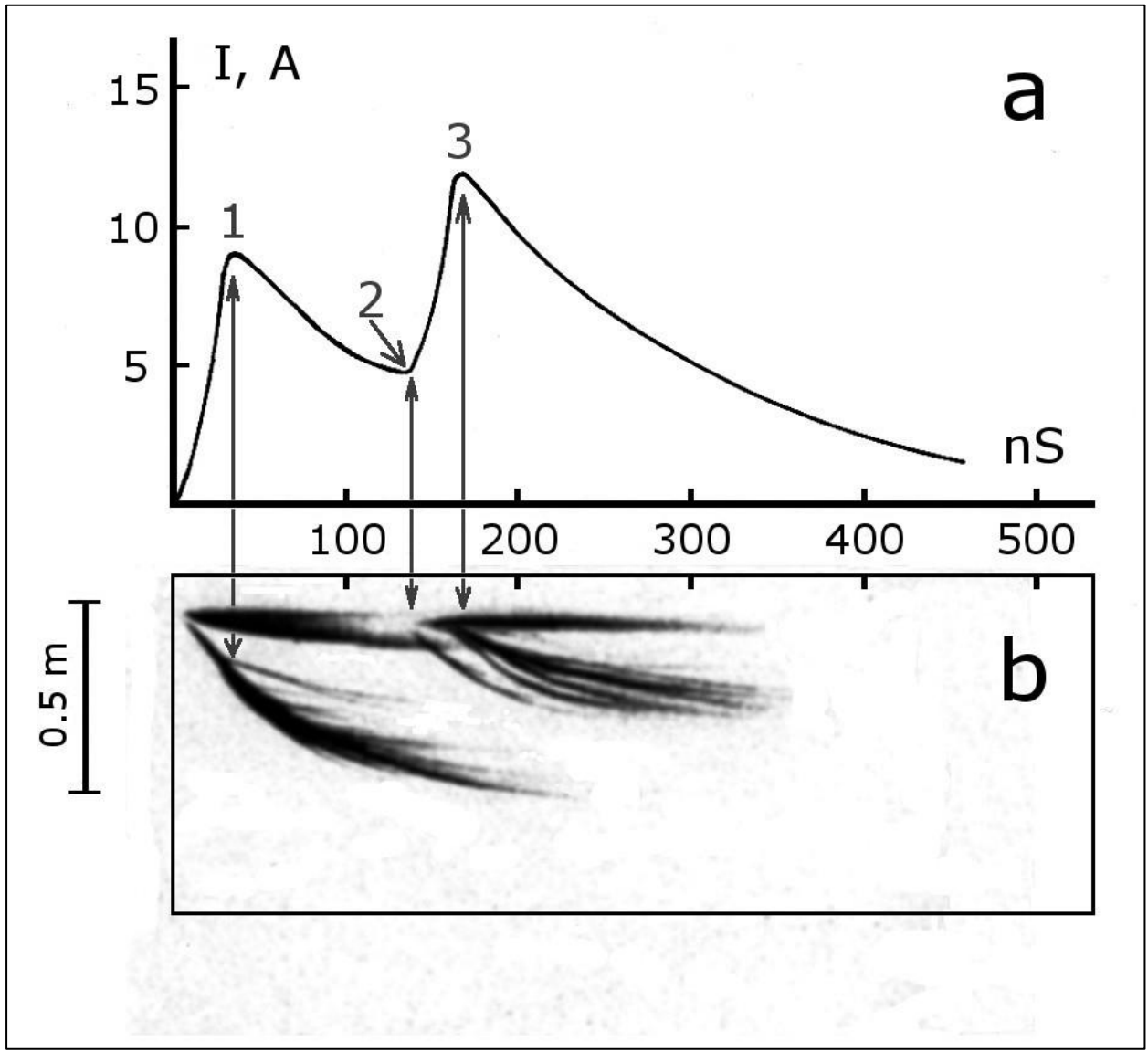

Рисунок 7.1 (адаптировано из [Kostinskiy et al., 2020a]). Типичная стримерная вспышка от металлического электрода. Диаметр электрода ~5-25 см. Фронт напряжения ~200-500 мкс, разрядный промежуток — стержень-плоскость 4-20 м. (a) ток на высоковольтном электроде; (b) одновременная фотохронограмма (временная развертка) движения стримерной вспышки. Шкала времени применима к рисункам (a) и (b).

времени жизни электрических полей генерация стримеров, скорее всего, будет происходить в E, близком к полю пробоя 3-4 МВ/(м·атм), без большого перенапряжения и с очень медленным повышением напряжения по сравнению с разрядами на электродах. [Kostinskiy et al., 2015a, 2015b] изучали положительные стримерные вспышки в электрическом поле заряженного аэрозольного облака, и эти эксперименты, на наш взгляд, полезны для нашего понимания положительных стримерных вспышек в грозовом облаке. При генерации разрядов заряженным аэрозольным облаком нарастание фронта напряжения на заземленной сфере диаметром 5 см составляло 300-500 мс, а продолжительность приложенного напряжения составляла несколько десятков минут, что ближе к реальным условиям в грозовом облаке, чем параметры приложенного напряжения генераторами импульсных напряжений.

Поэтому мы более подробно опишем положительные стримерные вспышки в электрическом поле аэрозольного облака и покажем их развитие на Рисунке 7.2, так как понимание их развития полезно при построении Механизма (более подробно это событие рассматривается в главе 2). Важно, что даже в условиях очень медленно меняющейся E (например, более 300-500 мс), создаваемой заряженным аэрозольным облаком, текущие фронты (времена нарастания) стримерных вспышек не изменились и имели длительность около 30 нс, как в эксперименте, описанный выше с фронтами 200-500 мкс, и показанный на 2.3.1.1. Первая стримерная вспышка на Рисунке 7.2a (1) имела пик тока 1,1 A, фронт тока составлял $30 \pm 5$ нс, полуширина пика на полувысоте составляла $90 \pm 10$ нс, а время спада — $147 \pm 10$ нс. Изображение этой стримерной вспышки показано на рисунке 2.3.1.2b.I(1). Общая длительность тока вспышки составляла около 200 нс. Стримеры продолжали существовать и двигаться даже после того, как ток на электроде упал до нуля, поскольку E во всей области от сферы (2) до облака (3) превышал порог, необходимый для движения положительных стримеров в воздухе $E_{str+} \geq 0{,}45\text{-}0{,}5$ МВ/ (м·атм) [Базелян и Райзер, 1997]. Стримеры долетели до центра облака за 1,7 мкс. Следовательно, длительность тока вспышки стримера, измеряемая на электроде ($\approx$ 200 нс), определяется временем окончания гальванической (токовой) связи стримеров с металлической сферой. Сами стримеры продолжали перемещаться в объеме и просуществовали не менее 1,7 мкс. Вспышка имела длину не менее 1,2 метра, и стримеры переместились от поверхности сферы, Рисунок 7.2b.I(2), в область прохождения СВЧ-пучка, Рисунок 7.2b.I(5). Поскольку стримеры двигались и не переходили в положительные лидеры, мы знаем, что

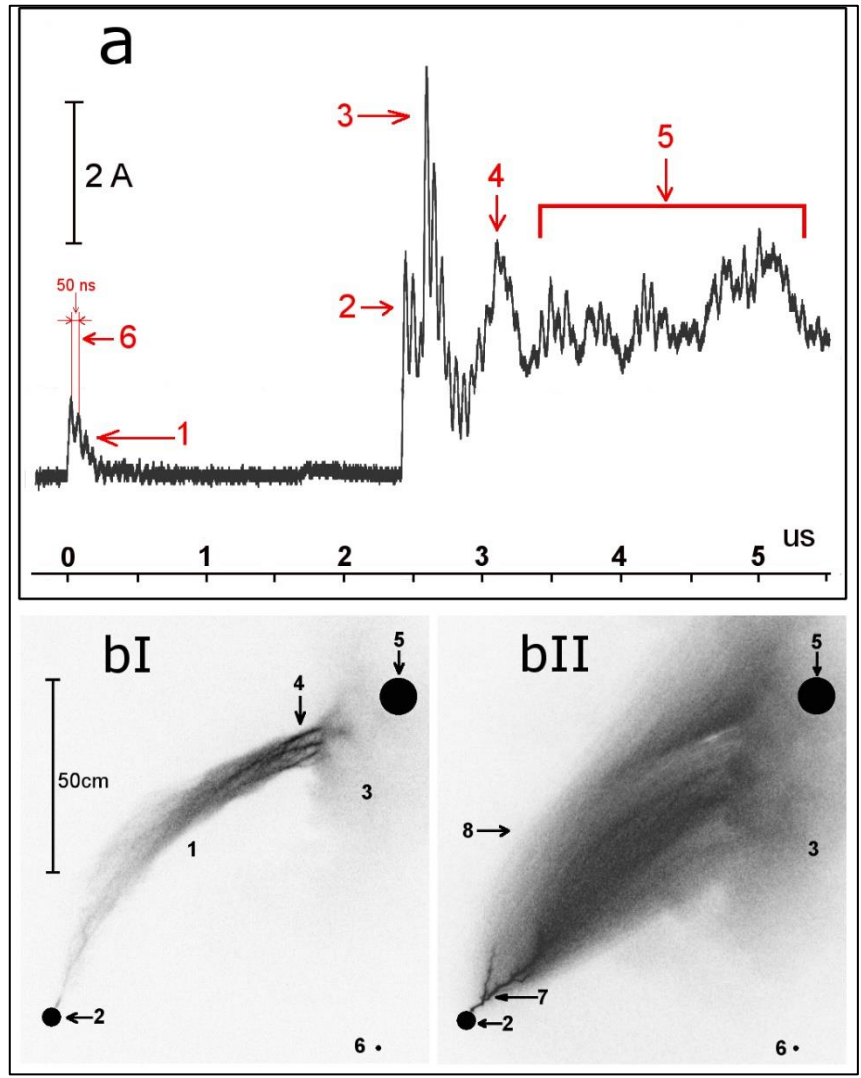

Рисунок 7.2 (адаптировано из [Kostinskiy et al., 2020a]). (а). Осциллограмма типичной положительной стримерной вспышки, стартующей с электрода без перенапряжения. Рост фронта напряжения на заземленной сфере диаметром 5 см составил 300-500 мс, а общая продолжительность приложенного напряжения - 5 минут. Электрическое поле на поверхности электрода ≈3 МВ/м.: 1, 2, 3 - первая, вторая и третья вспышки, 4 - ток инициирования лидера, 5 - ток движущегося лидера. (б). Два последовательных изображения стримерной вспышки, соответствующие осциллограмме (а). Вспышка стартовала от заземленного электрода (2) в электрическом поле заряженного аэрозольного облака. Два изображения были сделаны камерой 4Picos с усилением изображения: bI - первый кадр с выдержкой 2 мкс; bII - второй кадр с выдержкой 10 мкс; между кадрами 1 мкс, оба кадра инвертированы. 1 - первая вспышка положительных стримеров; 2 - заземленная сфера диаметром 5 см с токоизмерительным шунтом; 3 - облако заряженных водного аэрозоля; 4 - UPFs; 5 - зона прохождения диагностического СВЧ-пучка; 6 - центр заземленной плоскости, в которой находится сопло, из которого истекает пар; 7 - восходящий положительный лидер; 8 - стримерная корона положительного лидера (более подробно это событие рассматривается в главе 2) значения электрического поля Е на их пути было 0,5 МВ/м ≤ Е ≤ 1 МВ/м. (Верхнее значение Е <1 МВ/м было получено следующим образом: искусственное аэрозольное облако в этих экспериментах было отрицательно заряжено. Если бы электрическое поле в промежутке от облака до земли превысила ~ 1 МВ/м (порог поддержания движения отрицательных стримеров), то отрицательные стримеры двигались бы от отрицательного облака к заземленной плоскости. Поскольку во всех экспериментах с отрицательно заряженным облаком не наблюдались вспышки отрицательных, движущиеся от облака к заземленной плоскости, но регулярно наблюдали, положительные стримеры, пересекающие весь промежуток от заземленной плоскости к облаку, то мы предполагаем, что электрическое поле Е находится в диапазоне 0,5 МВ/м ≤Е≤1 МВ/м). Средняя скорость положительных стримеров первой вспышки в плоскости изображения оказалась равной ≈ $7 \cdot 10^5$ м/с. Небольшие максимумы и минимумы тока на осциллограмме первой стримерной вспышки Рисунок 7.2а(6), а также на всей осциллограмме тока являются артефактами измерительной схемы и не должны приниматься во внимание. Два других максимума тока, которые соответствуют второй и третьей стримерной вспышке на Рисунке 7.2а (2,3), имели такое же время нарастания тока (≈ 30 нс), что и первая стримерная вспышка, несмотря на гораздо более высокие токи (3.14 А, 5.8 А). Таким образом, время нарастания стримерной вспышки при таких значениях Е не меняется и характеризует физические параметры отдельных стримеров. Полный ток отдельной стримерной вспышки от электрода большого диаметра может достигать 100–200 А, но длительность переднего фронта стримерной вспышки изменяется в небольших пределах 25–35 нс. Такой быстрый фронт дает мощное VHF-излучение с максимумом в диапазоне 30-40 МГц. Максимальный ток на Рисунке 7.2а(4), скорее всего, связан с ионизационно-перегревной неустойчивостью, которая привела к появлению лидера, Рисунок 7.2b(7). Характеристики тока лидера отличаются от токов стримерных вспышек и имеет следующие параметры: время нарастания 195 ± 10 нс (в 6 раз медленнее первых трех положительных стримерных вспышек (1,2,3)), полуширина на полувысоте 180 ± 10 нс и время спада 210 ± 10 нс. На более поздних этапах развития небольшие пики с аналогичными параметрами соответствуют току положительных восходящих лидеров (Рисунок 7.2а(5)), которые перемещаются небольшими ступеньками длиной порядка размеров головки лидера 0.3-1 см (квазинепрерывный режим распространения лидера).

### 7.4.2.2. Длина и проводимость длинных стримеров

Для длинных стримеров движение головки стримера поддерживается собственным сильным Е, который ионизирует воздух перед головкой (поле перед головкой возникает из-за поляризации плазмы электронных лавин во внешнем электрическом поле). Поле перед головкой стримера Е значительно превышает пробойное поле $E_{th}$, и по разным оценкам оно составляет 10–30 МВ/(м·атм), [Базелян и Райзер, 1997, 2001]. Поле Е за головкой стримера быстро падает до значения, близкого к напряженности окружающего электрического поля. В этом случае частота ионизации очень быстро уменьшается, а частота прилипания электронов к молекулам кислорода начинает играть основную роль [Bazelyan & Raizer, 1998, с.25]. При Е < 0,5 МВ/(м·атм) наблюдается сильная потеря проводимости плазмы за головкой стримера всего за 100–200 нс при давлении 0,3–1 атм. При типичных значениях электрического поля для грозового облака основным механизмом потерь является трехчастичное прилипание электронов к молекулам кислорода [Kossyi et al., 1992], [Bazelyan & Raizer, 2000]:

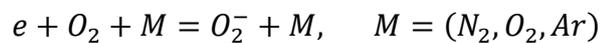

$$e + O_2 + M = O_2^- + M, \qquad M = (N_2, O_2, Ar)$$

со скоростью прилипания, падающей пропорционально квадрату давления. Таким образом, если скорость распространения стримеров $2\text{-}5 \cdot 10^7$ см/с, а характерное время прилипания $1\text{-}2 \cdot 10^{-7}$ с, то длина проводящего канала стримера при атмосферном давлении будет 2-10 см. Концентрация электронов в головке стримера достигает $1\text{-}3 \cdot 10^{14}$ см$^{-3}$, а за головкой стримера концентрация в плазме экспоненциально уменьшается с увеличением расстояния до головки [Базелян и Райзер, 1997, 2001]. Практически реальная длина стримеров при атмосферном давлении видна на Рисунке 7.3 [Kostinskiy et al., 2018] и составляет 10–20 см. Физическая природа прилипания электронов к кислороду не меняется для положительных и отрицательных стримеров. На высоте 6 км над Землей, где давление снизилось примерно в 2 раза, длина токопроводящего канала стримеров увеличится в четыре раза и достигнет длины 8-40 см (в зависимости от Е), и на высоте 9-10 км длина стримеров будет 18-90 см с таким же экспоненциальным падением проводимости за головкой стримеров.

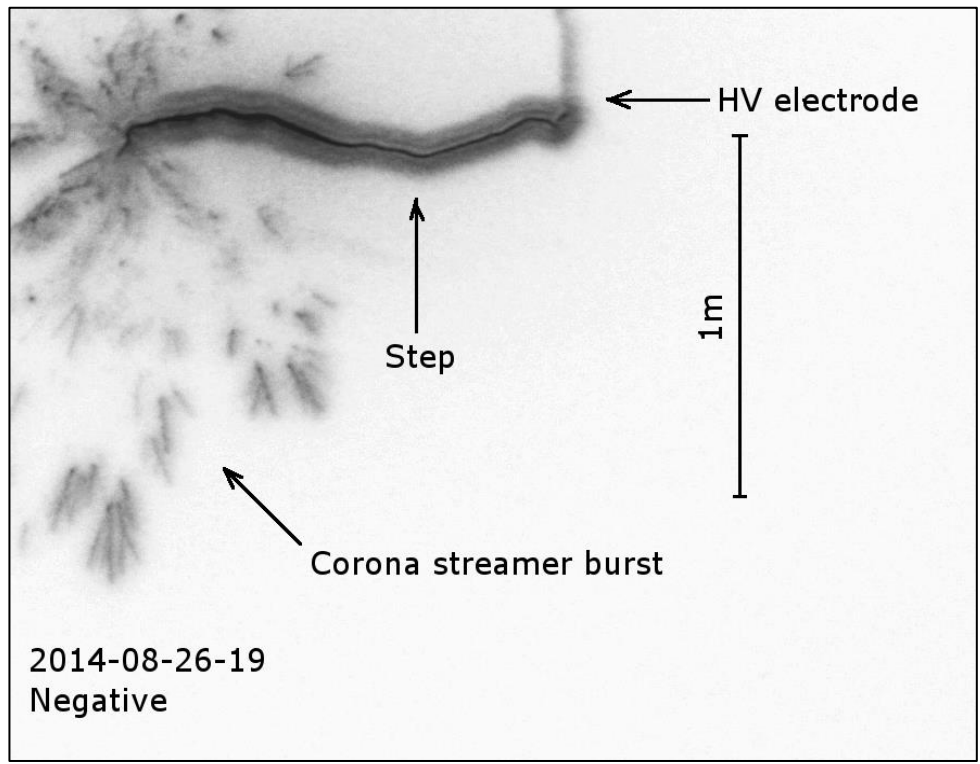

Рисунок 7.3 (адаптировано из [Kostinskiy et al., 2020a]), (событие 2014_08-26-19, адаптировано из [Kostinskiy et al., 2018]). Отрицательная ступень лидера с яркой вспышкой стримерной короны. Отображается только первый из двух кадров камеры с усилением изображения 4Picos (второй может содержать артефакт камеры). Выдержка кадра — 50 нс. Обращает на себя внимание, что стримеры простираются во всех возможных направлениях, хотя отчасти это может быть связано с необычной геометрией канала. Этот лидер не дошел до противоположного электрода (незавершенное событие). Фокусное расстояние объектива составляло 50 мм, а значение диафрагмы (относительное отверстие) составляло f 0,95. Размер пикселя изображения в плоскости объекта составляет 3,1×3,1 мм2. Температура окружающей среды составляла 11°C, а влажность — 81% (это событие подробно рассмотрено в главе 6).

### 7.4.3. Инициирующее событие (IE), как почти одновременное (синхронизованное) инициирование большого числа положительных стримерных вспышек в грозовом облаке

В предыдущем разделе мы описали отдельные стримерные вспышки, возникающие на металлическом электроде, но (почти во всех случаях) в грозовом облаке металлических электродов нет (мы не рассматриваем здесь случай длинных металлических объектов типа ракет и самолетов). В грозовом облаке есть различные гидрометеоры, которые статистически распределены внутри облака и несут разные заряды. Мы предполагаем, что инициирующее событие молнии создает систему положительных стримеров, распределенных в пространстве и времени.

### 7.4.3.1. Стримерная корона, стартующая с гидрометеоров, создающая множество положительных стримерных вспышек

Одно из возможных событий инициирующих молнию (IE) может заключаться в том, что положительные стримеры инициируются одним или несколькими гидрометеорами (например, [Dawson and Duff, 1970], [Griffiths and Latham, 1974], [Phelps, 1974], [Griffiths and Phelps, 1976], [Petersen et al., 2015]). Например, система заряженных гидрометеоров, распределенных в объеме, может генерировать систему положительных стримерных вспышек, распределенных в объеме. Таким образом, если IE является NBE, суммарная стримерная вспышка может состоять из набора отдельных вспышек, инициированных распределенными в объеме гидрометеорами. Каждая из этих вспышек может быть похожа на стримерные вспышки с металлических электродов, описанные выше. Если IE является слабым событием (например, [Marshall et al., 2019]), это событие также может быть множеством положительных стримерных вспышек, но их может быть меньше. Электромагнитное излучение от распределенных в объеме почти одновременно излучающих положительных стримеров, обнаруженное наземными приборами, может проявляться как точечные источники (центроиды), развивающиеся в трехмерном пространстве и времени. Таким образом, комбинированное событие, состоящее из набора отдельных, но синхронизированных во времени каким-либо процессом стримерных вспышек, может генерировать обычно наблюдаемый одиночный VHF-импульс IE.

Для NBEs [Rison et al., 2016] заявили, что «пробой, по-видимому, вызван пространственно и временно распределенной системой положительных стримеров, в которой полный ток распределяется по некоторой площади поперечного сечения как объемная плотность тока». [Rison et al., 2016] оценили пространственный масштаб NBEs в ~500 м и предположили, что положительные "streamers would be initiated by corona from ice crystals or liquid hydrometeors" («стримеры будут инициированы короной, возникающей на кристаллах льда или жидких гидрометеорах»). Однако для инициирования каждого стримера также требуется наличие теплового свободного электрона (энергия <100 эВ) вблизи инициирующего гидрометеора, чтобы запустить лавину электронов, которая перерастает в стримерную вспышку [Мик и Крэгс, 1960, стр.423], [Dubinova et al., 2015], [Rutjes et al., 2019]. [Dubinova et al., 2015] утверждали, что тепловые свободные электроны в грозовых облаках имеют очень низкую плотность из-за трехчастичного прилипания электронов к молекулам кислорода и воды, поэтому отсутствие свободного электрона препятствует инициированию стримеров гидрометеорами. Кроме того, в некоторых исследованиях сделан вывод, что размер инициирующего гидрометеора должен быть больше, чем можно предположить, 6-20 см (например, [Dubinova et al., 2015], [Babich et al., 2016]). Эти проблемы с инициирующими стримеры гидрометеорами (при условии, что гидрометеоры не участвуют в коллективных процессах, которые могут усилить поля на гидрометеорах) побудили нас рассмотреть альтернативный процесс инициирующего события (IE), как описано ниже.

## 7.4.3.2. Гидродинамические и статистические процессы для усиления электрических полей грозового облака и создания группы положительных стримерных вспышек

Другой возможностью для инициирующего молнию события (IE), а также для изолированного NBEs, могут быть гидродинамические и статистические процессы, которые могут увеличивать E в областях с сильно заряженными гидрометеорами во многих небольших областях облака (создавая «воздушные электроды»), чтобы свободные электроны, попадая в эти области, создали трехмерное множество положительных стримерных вспышек. А эти вспышки будут излучать в VHF-диапазоне, как обсуждалось в предыдущем разделе. Таким образом, этот альтернативный способ создания множества положительных стримерных вспышек в объеме облака также будут отслеживать радиоприборы, которые изучают возникновение и развития молнии. Для мелкомасштабных (диаметр около 2-5 см) усилений E, вероятно, потребуется относительно большая область относительно сильного E, которая, кажется, существует в грозовых облаках, как обсуждается далее.

Основываясь на аэростатных измерениях в активных грозовых облаках, удалось восстановить типичное крупномасштабное распределение заряда облаке. Оно имеет 4-8 горизонтально расположенных слоев заряда, распределенных вертикально с типичным максимальным измеренным вертикальным $E \approx \pm 350$ кВ/(м атм) [Marshall & Rust, 1991], [Stolzenburg & Marshall, 2009]. Локальные средние максимальные значения электрического поля обычно возникают между областями заряда противоположной полярности, поэтому эти максимумы распределены в облаке вертикально. Об объеме больших областей E известно меньше, но [Marshall et al., 2005] благодаря измерениям электрического поля с помощью воздушных шаров, обнаружили объемы 1–4 км$^3$ с вертикальной и горизонтальной протяженностью не менее 300–1000 м, связанные с тремя вспышками молнии. Хочется подчеркнуть, что эти баллонные измерения не могли выявить мелкомасштабные вариации поля размером меньшим 100 м и краткосрочные вариации поля для времен меньших 10 с. Однако [Stolzenburg et al., 2007] в результате исследований 9 запусков баллонов, в которых баллон и/или измерительные приборы были поражены молнией, обнаружили максимальные величины электрического поля непосредственно перед семью ударами молнии в диапазоне 309-626 кВ/(м атм) и предполагаемыми величинами поля 833 и 929 кВ/(м·атм) для двух других ударов. Еще одно наблюдение, сделанное [Stolzenburg et al., 2007] поддерживает идею мелкомасштабного усиления электрического поля, лежащего в основе нашего Механизма, особенно с учетом того, что типичная продолжительность CG-молнии составляет 200-300 мс ([Rakov & Uman, 2003, Таблица 1.1], [Stolzenburg et al., 2007]). Это наблюдение состоит в том, что для 7 из 9 ударов молнии |E| быстро увеличивалось за 2-5 с до удара молнии с величинами dE/dt $\approx$11-100 кВ/(м·с). За несколько секунд до одной из вспышек было ступенчатое увеличение (<1 с) измеренного E на 380 кВ/(м атм), которое длилось всего 1 с, а перед следующей вспышкой было ступенчатое увеличение поля на 505 кВ/(м атм), которое длилось 13 с, и тогда же E снизилось на 15% за последние 2 с перед вспышкой. Таким образом, данные E в [Stolzenburg and Marshall, 2009], [Marshall et al., 2005] и [Stolzenburg et al., 2007] поддерживают идею существования EE-объемов внутри грозовых облаков. Эти работы также согласуются с идеей меньшего масштаба областей, возникающих в течение короткого времени с гораздо большими значениями E, вызванными гидродинамической, турбулентной и статистической природой распределения заряда в грозовом облаке, которая усиливает электрические поля в областях с сильно заряженными и поляризованными гидрометеорами.

Следуя гипотезе [Trakhtengerts, 1989], [Trakhtengerts et al., 1997, 2002], и [Trakhtengerts and Iudin, 2005], мы предполагаем, что разряды в грозовом облаке принципиально отличаются от лабораторных разрядов, возникающих с металлических электродов именно потому, что локальное электрическое поле создается статистически и турбулентно движущимися в облаке заряженными частицами, которые создают мелкомасштабные вариации поля E, которые усиливают поля в областях с сильно заряженными и/или поляризованными гидрометеорами. Эти мелкомасштабные вариации поля не существуют в физике классического электродного газового разряда. В течение нескольких десятилетий разрабатывался подход к возможному локальному увеличению E в грозовом облаке на масштабах от десятков сантиметров до сотен метров. Как указывалось выше (i9), колебания электрического поля на расстояниях около 100 м могут возникать из-за гидродинамических неустойчивостей ([Colgate, 1967], [Trakhtengerts, 1989], [Trakhtengerts et al., 1997], [Mareev et al., 1999], [Iudin et al., 2003]). В EE-объеме эти гидродинамические неустойчивости создают периодические увеличения и уменьшения величины электрического поля. Мы называем объемы этих периодических увеличений и уменьшений поля — «ячейками». Внутри EE-объема существует объемная сеть ячеек. Обнаруженные в этих ячейках усиления электрического поля имеют масштаб в десятки метров. Некоторые ячейки являются $E_{str+}$-объемами, способными поддерживать движение положительных стримеров (как определено в t8). Кроме того, [Trakhtengerts and Iudin, 2005] и [Iudin, 2017], [Iudin et al., 2019] показали, что дополнительные увеличения E в еще меньших масштабах могут быть возможны из-за случайного статистического движения множества заряженных частиц разного размера в облаке, и именно внутри этих очень малых объемов при участии сильно заряженных и/или поляризованных гидрометеоров могут возникать электрические поля пробоя $E_{th}$, а сами эти объемы могут стать «воздушными электродами». К сожалению, эти подходы к средне и мелкомасштабным усилениям электрического поля опираются на теоретические работы, и их пока сложно проверить экспериментально. Расчеты до сих пор не включали полную структуру реальной турбулентности гидродинамических потоков в облаке и полную динамику реальных гидрометеоров из-за очень больших вычислительных сложностей. [Brothers et al., 2018] исследуют эти проблемы с помощью математической модели (large-eddy-resolving model) с вычислительной сеткой 125 м. Их численное моделирование для двух гроз показало «огромное количество текстуры» ("tremendous amounts of texture"), то есть мелкомасштабных пространственных вариаций и неоднородностей в плотности заряда из-за адвекции заряда в турбулентности с большими вихрями. Хотя они не рассчитывали мелкомасштабные электрические поля, [Brothers et al., 2018] отмечают, что должны быть «более благоприятные места для пробоев» из-за наличия большего количества соседних небольших очагов разных зарядов. Кажется разумным предположение, что моделирование такого рода с более высоким пространственным разрешением, включающее меньшие турбулентные вихри вместе с эффектами гидродинамической неустойчивости, приведет к еще более мелким усилениям поля E в метровом и субметровом масштабе.

Гипотеза о мелкомасштабных вариациях E кажется нам многообещающей по двум основным причинам. Во-первых, мы не обнаружили никаких противоречий между этой гипотезой и измерениями параметров молний или электрических полей в реальных грозах. Во-вторых, развитие молнии от IE через IB импульсы сильно варьируется от вспышки к вспышке с широким диапазоном длительностей и амплитуд IEC (например, [Marshall et al., 2014b]), широким диапазоном длительностей IB импульсов, временем между IB импульсами, амплитудами IB импульсов и широким диапазоном количества субимпульсов (0-5) на классических IBPs (например, [Marshall et al., 2013], [Stolzenburg et al., 2014], [Smith et al., 2018]). Особо следует отметить в развитии молнии кажущийся случайным порядок классических IBPs: примерно в 1/3 молний первый классический IBP имеет наибольшую амплитуду, за ним следуют классические IBPs с различными амплитудами, в то время как во многих других молниях наибольший классический IBP инициируется третьим, четвертым или пятым со средним промежутком времени после первого IB-импульса равным 1,4 мс для IC-молний и 0,25 мс для CG-молний [Smith et al., 2018]. Эти вариации в развитии молнии от IE до IB-стадии легче понять в контексте статистического распределения мелкомасштабных областей с большими величинами электрического поля, а не с несколькими большими областями постоянного электрического поля.

### 7.4.3.3. Преимущества процесса гидродинамического и статистического усиления электрического поля E

Во-первых, важно подчеркнуть, что такой процесс увеличения E не зависит от того, в каком фазовом состоянии находятся гидрометеоры, и поэтому он согласуется с тем фактом, что молнии возникают в диапазоне высот от 3 до 15 км (и даже выше). В этом процессе гидрометеоры любого размера (т.е. облачные гидрометеоры размером 5-100 мкм или частицы осадков размером 1-3 мм) могут быть жидкими и иметь высокую проводимость или полностью кристаллизованными с более низкой проводимостью. Основное требование состоит в том, чтобы гидрометеоры несли заряд и статистически двигались в турбулентных потоках, так что гидродинамические неустойчивости в турбулентном облаке будут приводить к изменениям плотности заряда и, таким образом, к увеличению и уменьшению E в масштабах в десятки метров или меньше. То есть, в пределах EE-объема 1000x1000x1000 м или 500x500x500 м и крупномасштабных полях E $\geq$ 284-350 кВ/(м·атм) могут наблюдаться колебания E в масштабах от сантиметров до десятков метров и изменения амплитуды электрического поля E в диапазоне 10-100 кВ/(м·атм) и больше, что мы обсудим более подробно ниже. Конечно, в усилениях поля на масштабах миллиметров и сантиметров могут играть существенную роль отдельные сильно заряженные или поляризованные гидрометеоры, чьи электрические поля будут усилены благодаря турбулентности и коллективному статистическому движению, но без коллективных процессов, по нашему мнению, уже нельзя будет обойтись при описании инициации молнии.

В EE-объемах заряженные гидрометеоры разделены расстояниями от нескольких миллиметров до десятков сантиметров и перемещаются беспорядочно. По аналогии, при броуновском движении множество молекул размером 0,03 мкм могут одновременно ударить большую частицу размером 3 мкм с одной стороны, в то время как меньшее количество молекул ударяет по ней с другой стороны, в результате чего большая частица совершает скачок (а без учета статистики ударов молекул, частицы аэрозоля размером около 0.03 мкм не должны двигаться при равномерных ударах со всех сторон). Одна молекула не может повлиять на гораздо более крупную частицу, которая в миллион раз тяжелее, но в ансамбле всегда есть флуктуации, пропорциональные $\sim\sqrt{n}$ числа молекул, вызывающие броуновское движение больших частиц. Аналогичные процессы должны происходить и в облаке при случайном и турбулентном движении гидрометеоров с разным зарядом [Iudin, 2017; Iudin et al., 2019]. Это движение гидрометеоров может приводить к широкому спектру колебаний электрического поля Е. Редкие, в небольших объемах (1-30 см), но большие по величине поля статистические колебания Е должны быть наложены на более крупномасштабные (около 100-200 м) колебания Е, благодаря гидродинамическим процессам. Эти волны сильных статистических колебаний Е с участием заряженных гидрометеоров по оценкам ([Trakhtengerts & Iudin, 2005], [Iudin, 2017], [Iudin et al., 2019]) имеют масштаб 1-30 см. Обычные металлические электроды, используемые для создания длинных искр в высоковольтных разрядах, имеют подобный масштаб (1-30 см). Гипотетически эти увеличения и уменьшения Е могут складываться и вычитаться с зарядами отдельных сильно заряженных гидрометеоров, и приводить, в небольшом масштабе, к спектру значений Е вплоть до значения пробоя >3 МВ/(м·атм). Объемы с масштабами 1-30 см и $E \geq 3$ МВ/(м·атм) будут играть роль «воздушных электродов», упомянутых ранее. Конечно, такие большие колебания электрического поля в грозовом облаке должны происходить редко (не более одной небольшой области с $E \geq 3$ МВ/(м·атм) на 3-100 кубометров), но, в отличие от высоковольтных промежутков, грозовое облако имеет размеры в кубические километры, а время жизни сильного усредненного электрического поля в облаке может составлять десятки минут. Важно, что вероятность ожидания таких сильных колебаний амплитуды электрических полей Е, как и в других подобных статистических процессах, пропорциональна $\sim\sqrt{t}$. Но облако способно «терпеливо ждать» и, вероятно, в случае NBE и молнии, «успевает дождаться» таких сильных локальных колебаний Е в условиях сильной турбулентности и встречных гидродинамических потоков. Характерное время жизни мелкомасштабного усиления поля может составлять от десятков до сотен миллисекунд ([Iudin, 2017], [Iudin et al., 2019]). Более того, в типичном большом усредненном поле ЕЕ-объема должно быть множество небольших увеличений Е, создающих сеть «горных хребтов» Е с «пиками» и «впадинами» (изображенными на Рисунке 7.4). В предложенном нами Механизме это сеть «вершин» горного хребта электрического поля, превышающая в некоторых местах $E > E_{th} \approx 3$ МВ/(м·атм) на размерах нескольких сантиметров и разделенных расстояниями от сантиметров до десятков метров, в которых могут инициироваться положительные стримеры. В нашем Механизме эти области с большими $E > E_{th}$ («вершины», или $E_{th}$- объемы, или «воздушные электроды») заменяют большие (и/или сильно заряженные) гидрометеоры, предложенные другими исследователями для сильного увеличения E около поверхности гидрометеоров, чтобы возникла возможность инициации положительных стримеров (например, [Dawson & Duff, 1970], [Phelps, 1974], [Griffiths & Phelps, 1976] [Rison et al., 2016]). Другими словами, наш Механизм фокусируется на «ландшафте» E плюс поля сильно заряженных гидрометеоров, а не на распределении

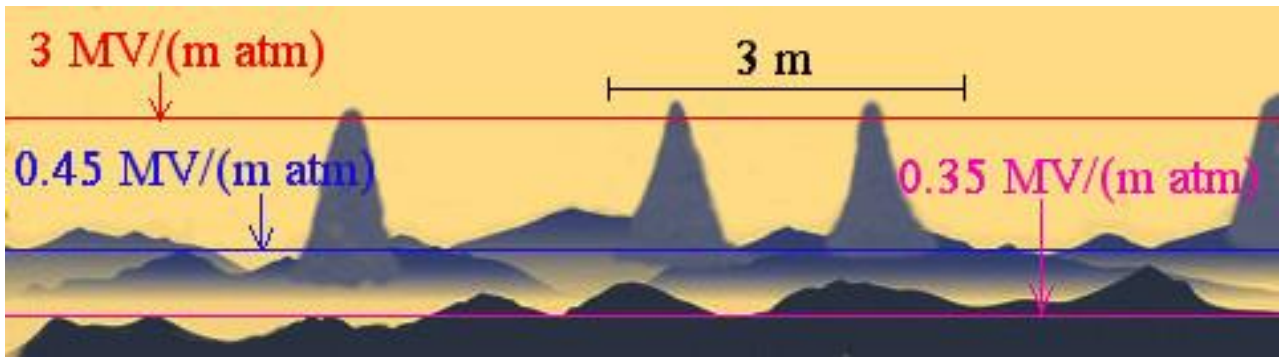

Рисунок 7.4 (адаптировано из [Kostinskiy et al., 2020a]). Возможные вариации электрического поля в турбулентных потоках (струях) грозового облака.

больших электрических полей около индивидуальных гидрометеоров (возможно сталкивающихся).

Второе преимущество этого гидродинамического и статистического подхода заключается в том, что размеры объемов, в которых E превышает $E_{th} \approx 3$-4 МВ/(м·атм), находятся в диапазоне 2-5 сантиметров (см. [Kostinskiy et al., 2020]), и этих размеров при возникновении NBEs и молний на высотах от 3 до 15 км будет достаточно для лавинно-стримерного перехода.

Если начальные стримерные вспышки возникли в областях с очень сильным E, то после этого положительные стримеры смогут поддерживать свое движение, если среднее значение E превышает требуемый минимум $E_{str+} \geq 0{,}45$-0,5 МВ/(м·атм) [Базелян и Райзер, 1997]. Переход стримерных вспышек в токопроводящий канал произойдет, когда стримеры достигнут поля $E \approx 0{,}7$–1,0 МВ/(м·атм). Пройдя расстояние в несколько метров, эти стримеры с большой вероятностью превратятся в короткие (1-30 см) горячие проводящие плазменные образования, подобные UPFs [Kostinskiy et al., 2015 a, b], главы 1-4. Статистически, области с электрическими полями E > 500-1000 кВ/(м·атм) должны быть по размеру гораздо больше, чем области с очень высоким E > 3 МВ/(м·атм).

## 7.5. Механизм инициации молнии от инициирующего события через стадию изменения электрического поля до стадии первых импульсов начального пробоя

В этом разделе мы описываем наш Механизм инициации молнии от инициирующего события IE до появления большого отрицательного ступенчатого лидера, который общепризнанно считается молнией. Наш Механизм непринципиально отличается для двух случаев (1) IE является NBE (сильным или слабым), как в работе [Rison et al., 2016], и (2) IE является гораздо более слабым событием, как в [Marshall et al., 2019]. Сначала мы опишем случай, когда в роли инициирующего события (IE) выступает «классический» NBE.

## 7.5.1. Механизм инициирования молнии благодаря классическому NBE или NBE-IE-Механизм

Важным условием для этого случая Механизма является условие c6: короткое и мощное явление, которое инициирует NBEs, должно двигаться со скоростью, близкой к скорости света [Rison et al., 2016]. Для такого быстрого распространения обычные стримеры не подходят. Даже при относительно большом E положительные стримеры пройдут только 500 м за 500 мкс ($v_{str} \approx 10^6$ м/с), так как для движения на таких больших расстояниях важно только среднее значение E, потому что статистические увеличения поля и ослабления усредняются. Для распространения со скоростью, близкой к скорости света, электромагнитный импульс, по-видимому, должен протекать по хорошо проводящему плазменному каналу, как при обратном ударе молнии или в момент контакта двух двунаправленных лидеров [Rakov and Uman, 2003], [Jerauld et al., 2007]. Даже стреловидные лидеры, движущиеся по распадающейся, но еще хорошо проводящей плазме, каналов предыдущих обратных ударов, развивают только одну десятую скорости света [Rakov and Uman, 2003]. Таким образом, для того, чтобы NBE действительно двигались со скоростью, близкой к скорости света, кажется, что обязательно необходимы высокопроводящие плазменные каналы, но это утверждение прямо противоречит экспериментальным данным условий c5 (до и во время протекания NBE не должен образовываться длинный канал горячей проводящей плазмы) и c2 (во время протекания IE не должно быть мощной вспышки света). Из всех известных нам физических процессов существует единственный процесс, движущийся со скоростью, близкой к скорости света, без помощи высокопроводящего, горячего плазменного канала — это движение релятивистских частиц [Gurevich & Zybin, 2001], [Dwyer, 2003]. Поэтому мы используем релятивистские частицы в нашем Механизме (см. также [Kostinskiy et al., 2020b], глава 8). Последовательность шагов NBE-IE-Механизма для инициации молнии благодаря NBE подробно описаны в следующих подразделах.

### 7.5.1.1. Необходимые условия

NBE-IE будет происходить в EE-объеме грозового облака, как определено ранее: объем около 0,1-1 км$^3$ с высоким усредненным E > 0,28-0,35 МВ/(м·атм) и большим количеством заряженных гидрометеоров различных размеров. Гидрометеоры могут быть жидкими или твердыми, большими или маленькими, и значительное их количество должно нести значительный электрический заряд. Как описано ранее (идея i6), [Marshall et al., 2005] измерили E такой величины в объемах, такой величины.

### 7.5.1.2. Большие электрические поля, возникающие из-за турбулентных и статистических движений гидрометеоров

В EE-объеме облака развиваются гидродинамические неустойчивости, текут турбулентные потоки и происходят статистические движения сильно заряженных гидрометеоров, которые, по-видимому, приводят к сильным локальным статистическим флуктуациям E различных масштабов и размеров на фоне усредненного большого электрического поля: см. Рисунок 7.5.I. Из-за существования широкого спектра колебаний E мы предполагаем, что Et$_h$-объемы («воздушные электроды») относительно малы (2-30 см³) и существуют не менее нескольких десятков миллисекунд [Kostinskiy et

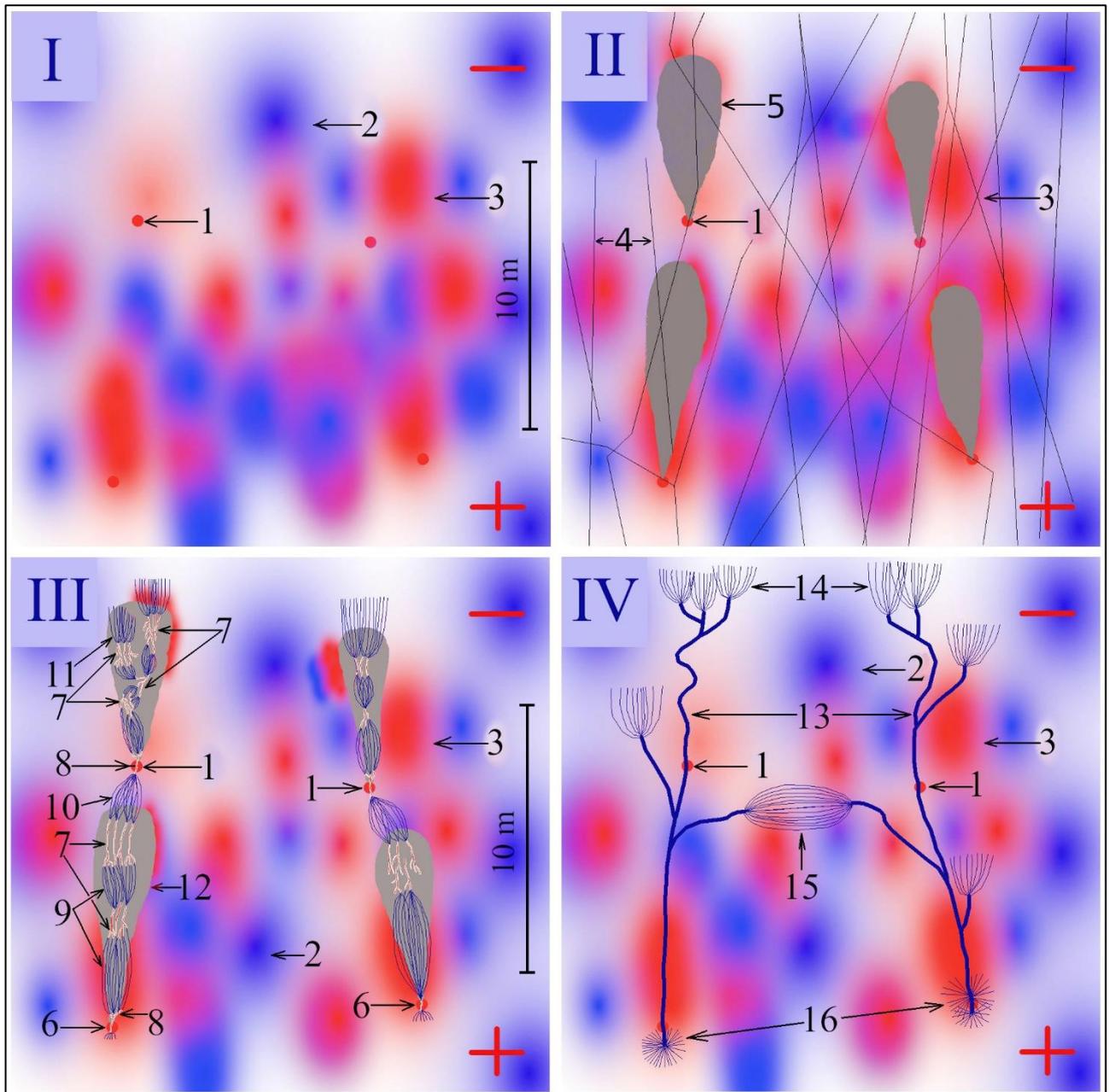

Рисунок 7.5 (адаптировано из [Kostinskiy et al., 2020a]). Схема механизма возникновения NBE, приводящего к инициации молнии. 1,6 — область с электрическим полем E≥ 3 МВ/(м·атм); 2 — область с электрическим полем E <0,28-0,35 МВ/(м·атм); 3 — E$_{str+}$-объем, область с электрическим полем E≥0,45-0,5 МВ/(м·атм); 4 — траектории убегающих электронов; 5 — положительные стримерные вспышки; 7,8 — UPFs; 9 — вторичные положительные стримерные короны, соединяющие UPFs; 10 — вторичные положительные стримерные короны, соединяющие UPFs двух разных стримерных вспышек; 11 — положительная стримерная корона перед UPFs; 12 — траектории (области) первых стримерных вспышек; 13 — горячие высокопроводящие плазменные каналы (двунаправленные лидеры); 14 —стримерные короны положительных лидеров; 15 — стримерная корона двух взаимодействующих больших плазменных каналов (двунаправленных лидеров); 16 — стримерная вспышка отрицательного лидера.

al., 2020b], глава 8. Напомним, что в $E_{th}$-объемах локальное электрическое поле E превышает условный порог пробоя $E_{th} \geq 3$ МВ/(м·атм), необходимый для инициации стримеров (вместе с выполнением критерия Мика). Воздушные электроды отмечены цифрой «1» на Рисунке 7.5.I и цифрой «6» на Рисунках 7.5.II-IV. Простая оценка, приведенная в следующем абзаце, показывает, что для обеспечения классического большого NBE достаточным количеством стримеров для обеспечения общего заряда, перемещаемого в NBE (и тока), в среднем один «воздушный электрод» должен присутствовать в каждом объеме ~5х5х5 $м^3$. Образуются также гораздо большие объемы со средней напряженностью электрического поля, превышающей порог для распространения положительных стримеров $E_{str+} \geq 0,45$-$0,5$ МВ/(м·атм), см. Рисунок 7.5.I (3); эти области представляют собой $E_{str+}$-объемы, определенные ранее (t8). Конечно, многие области в EE-объемах (Рисунок 7.5.I (2)) будут иметь электрические поля, меньшие чем усредненное поле E.

Сделаем простую оценку количества $E_{th}$-объемов («воздушных электродов»), необходимых для обеспечения сильного (классического) NBE. Мы предполагаем, что каждый $E_{th}$-объем может действовать как «воздушный электрод». Пусть заряд, переносимый NBE, составляет около 1 Кл, поскольку расчетные заряды трех NBEs в [Rison et al., 2016] составляли 0.5, 0.7 и 1.0 Кл. Заряд небольшой стримерной вспышки, стартующей с «воздушного электрода» примерно (10 см)$^3$ (= $10^{-3}$-$10^{-4}$ $м^3$ = 0,1-1 $дм^3$), где $E_{th} \geq 3$ МВ/(м·атм) и выполняется критерий Мика (Рисунок 7.5.III (6)), скорее всего будет аналогичен положительному стримерному разряду на Рисунке 7.2.bI(1) с зарядом около $10^{-7}$ Кл. Минимальный заряд отдельной стримерной вспышки определяется зарядом одного стримера. Согласно [Bazelyan and Raizer, 1998, стр. 174-175], измеренный заряд одного положительного стримера при давлении в среднем 0,33 атм (высота ~ 9 км) составляет в среднем $10^{-8}$ Кл. Следовательно, заряд стримерной вспышки, состоящей из 10 стримеров (~$10^{-7}$ Кл), кажется разумным. Таким образом, для обеспечения заряда 1 Кл требуется около $10^7$ воздушных электродов. Мы можем разделить EE-объем 1000х1000х1000 $м^3$ на $10^7$ равных объемов по ~100 $м^3$ ($\approx 4,6 \times 4,6 \times 4,6$ $м^3$). Кажется разумным предположить, что каждый из объемов в 100 $м^3$ в принципе, может содержать одну область размером $10^{-3}$-$10^{-4}$ $м^3$ с большим статистическим увеличением электрического поля до значений $E \geq 3$ МВ/(м·атм), поскольку в нем может быть достаточно много заряженных частиц, включая сильно заряженные гидрометеоры, для статистического усиления поля (например, в наковальне грозового облака [Dye et al., 2007] обнаружили концентрации частиц диаметром 30 мкм - 2 мм в диапазоне $10^4$-$5 \cdot 10^5$ м$^{-3}$), и измерения показывают, что во время грозы многие облачные гидрометеоры и частицы осадков заряжены (например, [Marshall & Stolzenburg, 1998]). Как мы покажем в следующем разделе, NBE может начинаться с синхронного запуска (который ведет к суперпозиции электрических сигналов) большого количества положительных стримерных вспышек, которые происходят в большинстве воздушных электродов ($E_{th}$-объемов).

На рис. 7.5(I) показана схема вертикального поперечного сечения части EE-объема грозового облака. Как предполагалось выше, усредненный вертикальный компонент электрического поля EE-объема составляет E > 0,28-0,35 МВ/(м атм), и направлен вниз, что обозначено большим красным знаком «—» в правом верхнем углу и знаком «+» в правом нижнем углах. Знаки «—» и «+» обозначают относительно удаленные области отрицательного и положительного заряда (размером в километры), которые часто встречаются в горизонтальных слоях, таких как «основной отрицательный» и «нижний положительный» слой заряда грозового облака, изображенные, например, на Рисунке B.25 [Stolzenburg et al., 1998]. Красная заливка области (3) (синяя области (2)) на Рисунке 2.3.1.5(I) указывают области, где электрическое поле E больше (меньше), чем усредненное значение E. Маленькие темно-красные окружности (1) указывают на $E_{th}$-объемы (воздушные электроды, где E> 3 МВ/(м атм)); красные области вокруг $E_{th}$-объемов — $E_{str+}$-объемы (объемы, где могут распространяться на большие расстояния положительные стримеры). Четыре показанных $E_{th}$-объема подразумевают, что приблизительный размер схемы составляет около 20 м по горизонтали и 20 м по вертикали.

Подчеркнем, что схема не показывает детально вектор $\vec{E}$ или величину E; скорее, это примерно указывает относительные величины E в поперечном сечении 20 м x 20 м. На масштабах <1 м мелкомасштабные заряды приведут к очень сложному распределению векторов E, указывающих во многих направлениях, включая восходящие и горизонтальные, а также нисходящие. Это означает, что розовые, красные и синие оттенки также являются средними значениями для затененных областей в масштабе ~ 5 м.

### 7.5.1.3. EAS-RREA-синхронизированный (почти одновременный) запуск большого количества электронных лавин и стримерных вспышек

При выполнении необходимых условий раздела 7.5.1.2, для того чтобы инициировать стримерные вспышки в $E_{th}$-объемах, EE-объем должен быть «засеян» энергичными ионизирующими частицами (электронами, позитронами, фотонами), которые создадут тепловые электроны. Число почти одновременно ионизирующих объем частиц в EE-объеме должно быть настолько большим, чтобы за несколько микросекунд ионизирующие частицы попали в большую часть воздушных электродов ($E_{th}$-объемов): см. Рисунок 7.5.II(6). Ионизирующие частицы будут производить свободные тепловые электроны, необходимые для инициации классических электронных лавин. Для этого идеально подходят широкие атмосферные ливни космических лучей (ШАЛ/EAS) с энергией частиц $\varepsilon_0 > 10^{15}$ эВ, которые при электрическом поле E > 0,284 МВ/(м·атм) будут генерировать релятивистские лавины убегающих электронов (RREA) [Gurevich & Zybin, 2001], [Gurevich et al., 1999], [Dwyer, 2003]. Общая схема процесса синхронизации стримерных вспышек за счет механизма EAS-RREA, например, между верхним отрицательным и нижним положительным зарядом, показана на Рисунке 7.6. Частица космических лучей с энергией $\varepsilon_0 > 10^{15}$ эВ (обозначена (1) на Рисунке 7.6) создает ШАЛ (2). Электроны и позитроны ШАЛ (3) попадают в область сильного электрического поля (5), которое может поддерживать распространение RREA (4) и положительных стримеров (7). Электроны и позитроны EAS-RREA (обозначенные цифрой (6)) пересекают область грозового облака, где есть воздушные электроды и таким образом синхронизируют почти одновременную инициацию множества стримерных вспышек (7), которые стартуют внутри воздушных электродов, через которые проходят электроны (8). Численные оценки этого процесса приведены в работе [Kostinskiy et al., 2020b], глава 8.

Электроны RREA оставляют около 29 тепловых свободных электронов на см длины пути на высоте 8 км и около 18 электронов на см на высоте 12 км [Rutjes et al., 2019]. Лавина релятивистских убегающих электронов (обозначенная цифрой 4 на Рисунке 7.5.II) пересекает весь EE-объем за 1,5–3 мкс. Когда убегающие электроны проходят через $E_{th}$-объем (см. Рисунок 7.5.II(6)), существует высокая вероятность того, что «обычные»

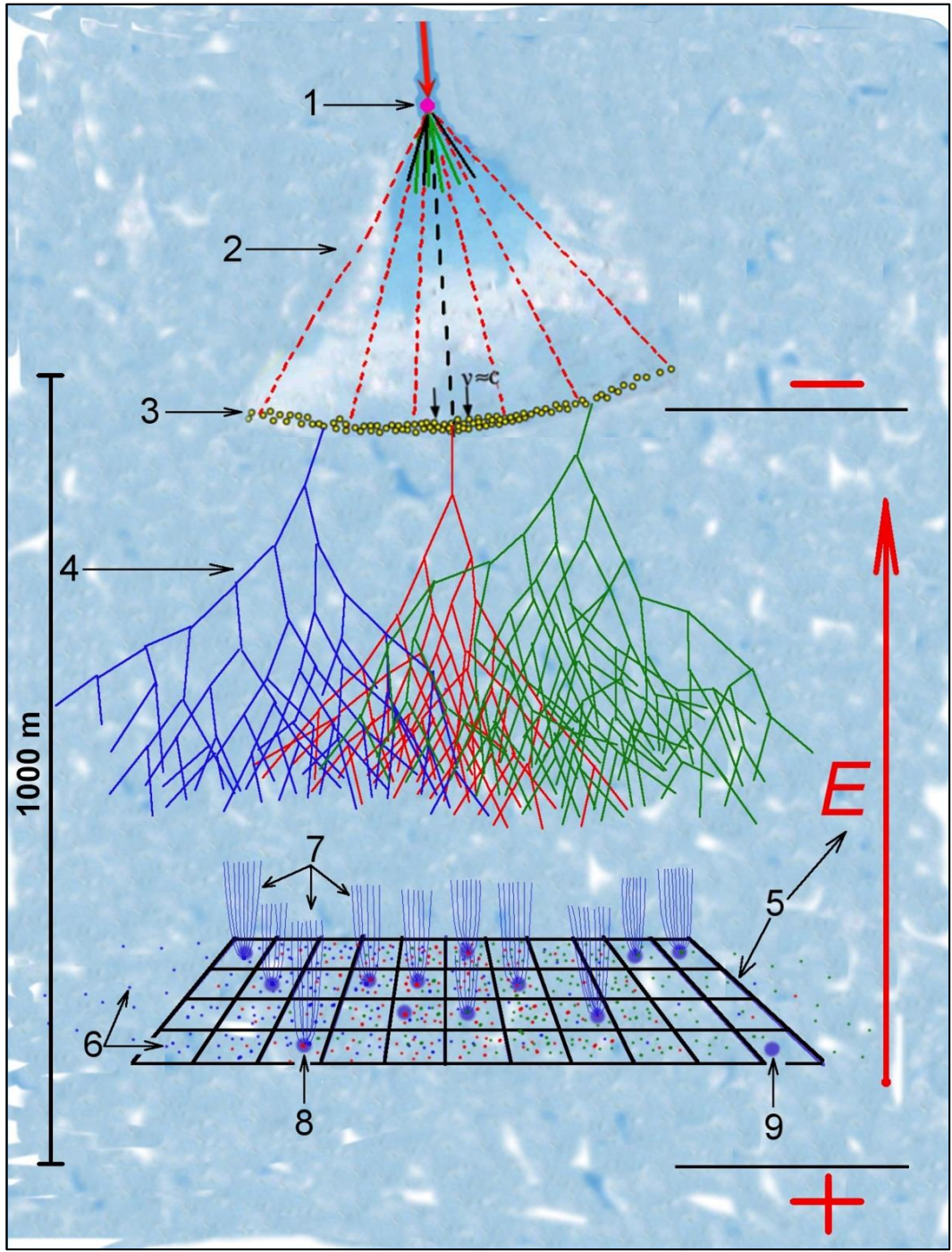

Рисунок 7.6 (адаптировано из [Kostinskiy et al., 2020a]). Схематическое изображение ШАЛ (EAS), который инициирует (синхронизирует) инициацию множества стримерных вспышек в течение нескольких мкс. 1 — первичная частица ШАЛ; 2 — ШАЛ; 3 — вторичные электроны ШАЛ; 4 — RREA (лавина релятивистских убегающих электронов); 5 — область сильного электрического поля; 6 — электроны EAS-RREA пересекают область сильной турбулентности грозового облака, создающей сильные электрические поля; 7 — синхронизированные EAS-RREA механизмом стримерные вспышки; 8 — воздушный электрод (E$_{th}$-volume), через который прошел энергичный электрон; 9 — воздушный электрод, через который не проходили энергичные электроны.

разрядные электронные лавины разовьются настолько, что превратятся в стримеры, потому что большая часть воздушных электродов ($E_{th}$-объемов) находится внутри $E_{str+}$-объема (см. Рисунок 7.5.I (3)). Индивидуальная стримерная вспышка формирует фронт тока примерно за 30 нс, а общая длительность индивидуальной стримерной вспышки будет в диапазоне 150-300 нс (как мы подробно описывали выше). Таким образом, возникают отдельные стримерные вспышки (см. Рисунок 7.5.II(5)), которые генерируют электромагнитные импульсы с максимумом излучения в VHF-диапазоне (30-40 МГц).

Лавины убегающих электронов, засеянные вторичными частицами ШАЛ, действуют как инициирующая волна, движущаяся со скоростью, близкой к скорости света, которая создает в разных точках ЕЕ-объема за 1,5–3 мкс «гигантскую волну» обычных положительных стримерных вспышек, которые мы описали выше (раздел 7.4.2). Можно сказать, что лавина большого количества убегающих электронов «зажигает» почти одновременно многие из $E_{th}$-объемов (см. Рисунок 7.5.II (1)), тем самым формируя фронт излучения NBE в VHF(УКВ)-диапазоне длин волн. Более подробно движение фазовой волны зажигания экспоненциально растущего числа стримерных вспышек показано на Рисунке 7.7 (численная оценка приведена в [Kostinskiy et al., 2020b], глава 8. Таким образом, NBE начинается и является на первом этапе экспоненциально увеличивающейся суперпозицией всех положительных стримерных вспышек, инициированных («зажженных») EAS-RREA вторичными частицами.

Таким образом, NBE-IE Механизм выполняет условия (с4, с5) генерации за несколько микросекунд гигантской вспышки положительных стримеров во время излучения NBEs без предварительного образования горячего плазменного канала с высокой проводимостью. В то же время генерация этой гигантской положительной стримерной вспышки неразрывно связана с лавинами убегающих релятивистских электронов, инициированными вторичными электронами и позитронами ШАЛ ($10^{15}$ эВ $<\varepsilon_0<10^{16}$ эВ). Без ШАЛ вторичные электроны лавин убегающих энергичных электронов, инициированных фоновыми космическими лучами, не могут пересечь большинство небольших воздушных электродов (так как их мало и вероятность попасть в большинство воздушных электродов мала). С другой стороны, только ШАЛ (без создания лавин убегающих в электрическом поле вторичных электронов), даже при начальной энергии частицы $\varepsilon_0 \approx 10^{17}$ эВ, не сможет инициировать достаточное количество стримерных

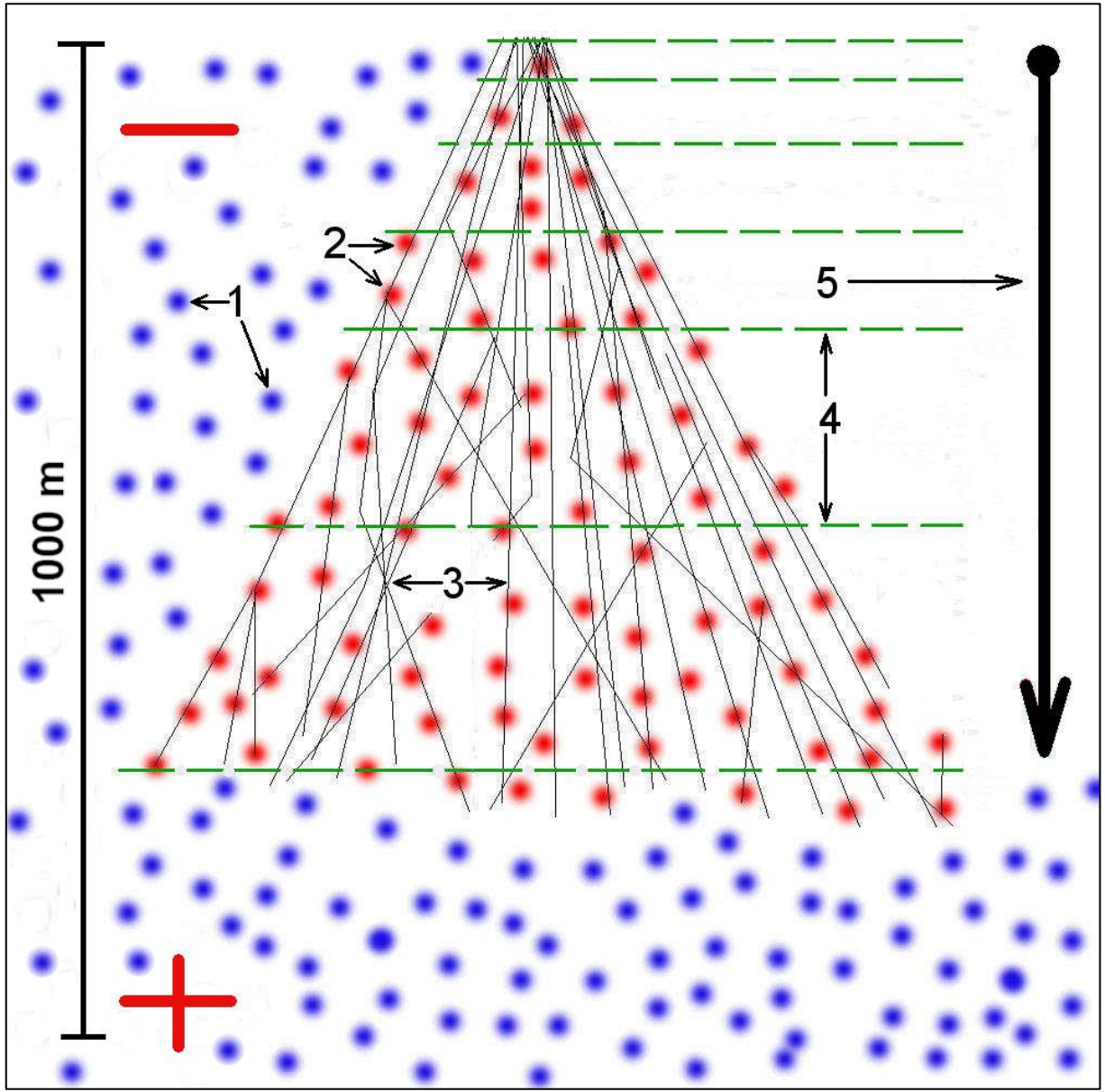

Рисунок 7.7 (адаптировано из [Kostinskiy et al., 2020a]). «Зажигание» фазовой волны стримерных вспышек со скоростью $\sim 10^8$ м/с убегающими релятивистскими частицами (фрагмент). 1 — воздушный электрод, который не пересек энергичный электрон (или позитрон); 2 — воздушный электрод ($E_{th}$-volume), который пересек энергичный электрон; 3 — вторичные электроны лавины EAS-RREA; 4 — шаг лавины релятивистских убегающих электронов (RREA); 5 — фазовая волна стримерных вспышек, которые «зажигаются» в пространстве со скоростью $\approx 10^8$ м/с.

вспышек, чтобы обеспечить мощный УКВ(VHF)-сигнал, сопровождающий NBEs [Kostinskiy et al., 2020b], глава 8.

При построении этой части Механизма мы использовали обычные стримерные вспышки, которые могут двигаться с разумными (экспериментально многократно измеренными) скоростями $v_{str} \approx 2\cdot10^5\text{-}3\cdot10^6$ м/с. Но скорость движения релятивистских убегающих частиц инициирует стримерные разряды в $E_{th}$-объемах (воздушных электродах) вдоль траекторий релятивистских частиц, тем самым придавая процессу развития NBE, фиксируемого по движению центроидов интерферометра [Rison et al., 2016], *фазовую* скорость, близкую к скорости света, удовлетворяющую условиям с6 и с7 (см. выше, раздел 7.3).

Направление движения энергичных вторичных частиц (электронов и позитронов с энергией 0,1-100 МэВ) лавины EAS-RREA не будет определять направление движения стримерных вспышек, поскольку энергичные частицы инициируют только первые затравочные электроны внутри воздушных электродов. Положительные стримеры, как волны ионизации, всегда движутся по силовым линиям электрического поля в направлении отрицательного заряда. Энергичные электроны создают на каждом сантиметре своего пути (при атмосферном давлении) около 74 тепловых электронов со средней энергией около 24 эВ [Rutjes et al., 2019]. Эти тепловые электроны инициируют классические электронные лавины, которые превращаются в стримеры (если выполняется критерий Мика [Базелян и Райзер, 1997]). Головка стримера, поляризованная благодаря окружающему электрическому полю, создает перед собой собственное очень сильное электрическое поле 10-30 МВ/(м·атм). Это сильное электрическое поле определяет характеристики и направление движения отдельного стримера. Лавины энергичных электронов и позитронов EAS-RREA могут двигаться даже перпендикулярно электрическому полю, и это все равно не изменит направление движения стримеров по силовым линиям электрического поля.

### 7.5.1.4. Возникновение и развитие необычных плазменных образований (UPFs)

После инициации стримерных вспышек некоторые стримеры могут двигаться на расстояния ~1-10 м, так как они находятся в более обширных $E_{str+}$-объемах (раздел 7.4.3.2), см. Рисунок 7.5.III(12). Во время движения стримеров в электрических полях из-за ионизационно-перегревной неустойчивости в плазме появляется сеть сильно проводящих плазменных каналов, которые остаются в местах прохождения стримеров (см. Рисунок 7.5.III(7)). Мы будем называть эти сильно проводящие плазменные «необычными плазменными образованиями» (unusual plasma formations — «UPFs»), поскольку мы предполагаем, что они развиваются так же, как «необычные плазменные образования» («UPFs»), которые мы впервые наблюдали в искусственно заряженных аэрозольных облаках (см. главы 1-5, [Kostinskiy et al., 2015a; 2015b]). Важно отметить, что движение одной стримерной вспышки может одновременно генерировать несколько UPFs, Рисунок 7.5.III(7), подробнее глава 1. Эти UPFs через некоторое время начнут взаимодействовать друг с другом благодаря вторичным положительным стримерам, как показано на Рисунке 7.5.III (9) и как показано в экспериментах с заряженным аэрозольным облаком [Andreev et al., 2014], воспроизведенным на Рисунке 7.8, подробнее в главе 1. Время развития ионизационно-перегревной неустойчивости и появления UPFs при атмосферном давлении около 0,2-0,5 мкс [Popov, 2009], а на высотах, где возникает молния, это время может увеличиваться примерно в 2-8 раз [Riousset et al., 2010]. Таким образом, через 5–20 микросекунд (в зависимости от высоты) после начала движения VHF-фронта излучения NBE, весь EE-объем будет состоять из небольших горячих плазменных образований (UPFs), длиной от нескольких сантиметров до десятков сантиметров. См. раздел 7.6.2 для более подробного обсуждения ионизационно-перегревной неустойчивости и стример-UPF перехода.

Чтобы UPFs «выжили» в течение десятков микросекунд, они должны оставаться горячими и проводящими. Следовательно, через UPFs каждого горячего канала должен протекать ток не менее 0,2 А [Базелян и Райзер, 1997]. UPFs, которые расположены близко друг к другу, выживут, если они создают электрическое поле $E \geq 0,45\text{-}0,5$ МВ/(м·атм) в промежутке между собой (то есть, электрическое поле, которое превышает минимально необходимое поле для поддержания распространения положительных стримеров). Скорее всего UPFs обмениваются вторичными положительными стримерами, аналогично способу, которым спей-стем и спейс-лидер обмениваются положительными стримерами между собой и основным каналом отрицательного лидера длинной искры [Горин и Шкилев, 1976], [Les Renardières Group, 1981]. Благодаря вторичным положительным стримерам, между UPFs возникает взаимодействие

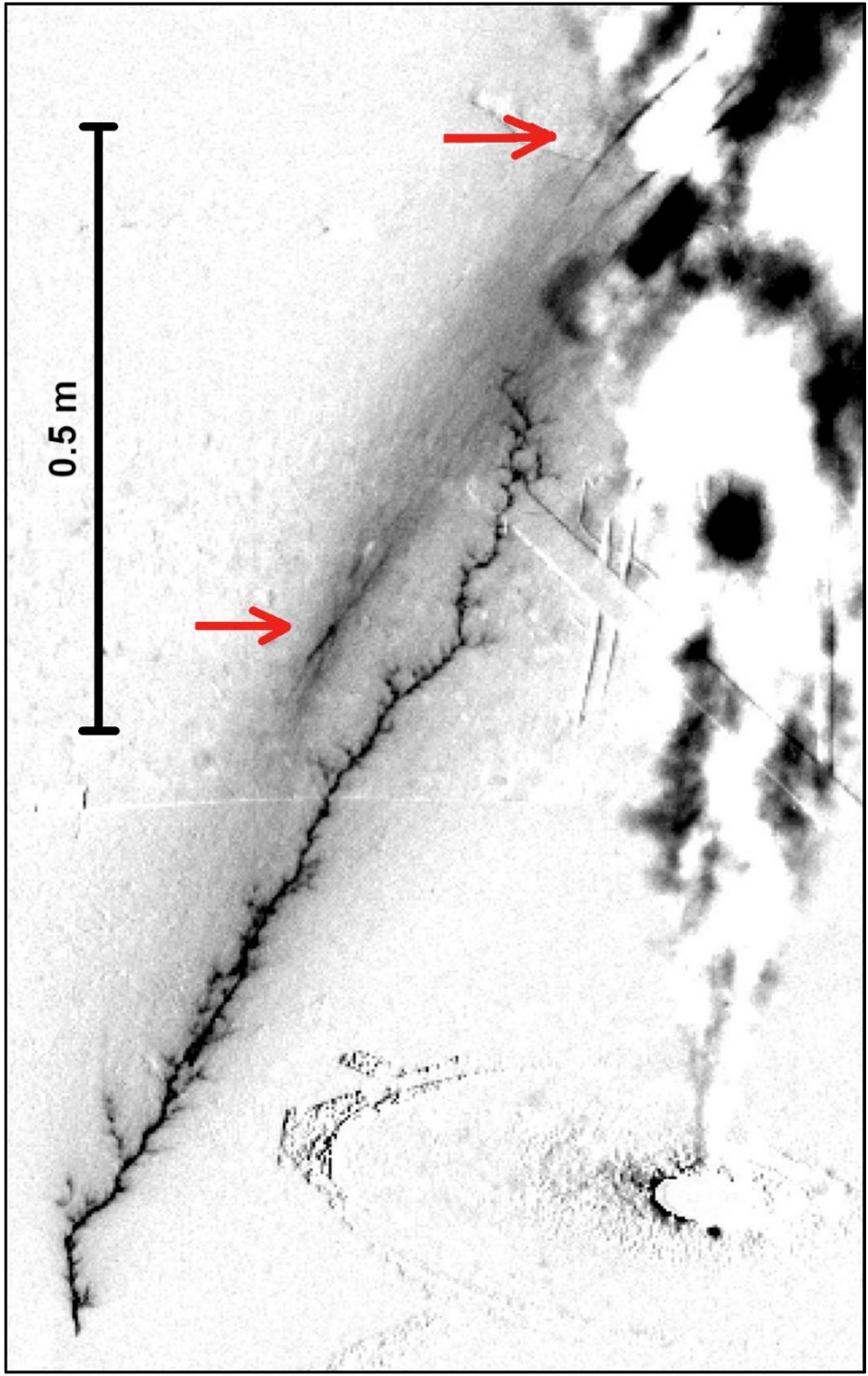

Рисунок 7.8 (адаптировано из [Andreev et al., 2014], [Kostinskiy et al., 2020a]). UPFs, сформированные после прохождения положительной стримерной вспышки, взаимодействуют друг с другом благодаря вторичным положительным стримерам. Отрицательно заряженное аэрозольное облако. Инфракрасное изображение с выдержкой 8 мс. Два UPFs (обозначены стрелками) появились после первой стримерной вспышки. Восходящие положительные лидеры появились и выросли после рождения UPF.

аналогичное небольшим сквозным фазам (например, Рисунок 7.5.III(9,10)), которые заканчиваются объединением UPFs и увеличением общей длины каналов горячей плазмы в процессах, аналогичном небольшим обратным ударам (или ступеням отрицательного лидера в длинной искре) с увеличением тока до 5-15 А за время ~1 мкс. Важно отметить, что индивидуальные UPFs состоят из целой сети каналов, где ток малых каналов питает большие каналы, помогая им выжить дольше: см. Рисунок 7.9 (из [Kostinskiy et al., 2015a], [Andreev et al.,2014]) и Рисунок 7.5.III(7), см. также подробно главы 1-5. Таким образом, внутри ЕЕ-объема многие небольшие каналы объединяются или сливаются в несколько больших каналов, рисунок 7.5.IV(13). Для долгосрочного выживания каждая отдельная цепочка UPFs должна вырасти до такой длины, чтобы потенциал на ее положительном конце достигал примерно 500 кВ, необходимых для инициации положительного лидера [Bazelyan & Raizer, 1998, 2000], [Bazelyan et al., 2007a]. См. раздел 7.6.3 для более подробного обсуждения UPF-положительный лидер переход.

### 7.5.1.5. Развитие отрицательных лидеров

Согласно приведенной выше последовательности плазменных преобразований, многие из UPFs, которые объединились в длинные цепочки и смогли стать «родителями» положительных лидеров, выжили. По мере того, как, возникшие из UPFs, положительные лидеры удлиняются, электрический потенциал их противоположного отрицательного конца будет увеличиваться. Отрицательный лидер будет инициирован с отрицательного конца цепочки UPFs (или конца положительного лидера, который возник в результате объединения UPFs), когда потенциал на отрицательном конце станет примерно в 1,5–2 раза больше, чем потенциал для инициирования положительного лидера [Gorin & Shkilyov, 1974, 1976], [Les Renardières Group, 1977 , 1981]. В этот момент отрицательный лидер на Рисунке 7.5.IV(16) возникает (инициируется) с отрицательного конца сети объединенных UPFs, а цепочка UPFs в результате «слияний и поглощений» превращается в «типичный» двунаправленный лидер, как показано на Рисунке 7.5.IV(13). Количество этих двунаправленных лидеров в ЕЕ-объеме может варьироваться при инициации разных молний от нескольких десятков до нескольких сотен каналов длиной 10–50 м каждый. Количество двунаправленных лидеров будет зависеть от размера начального ЕЕ-объема

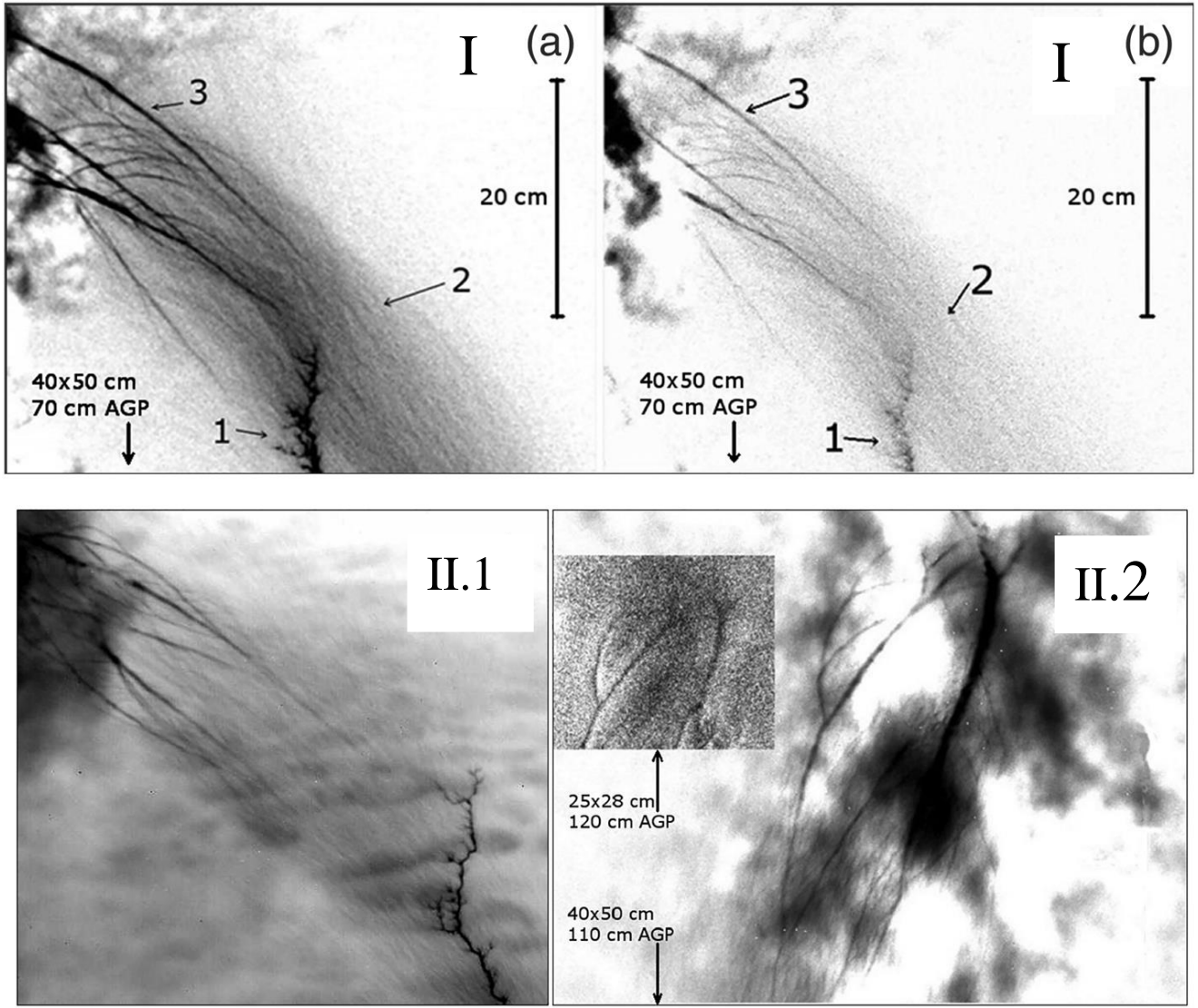

Рисунок 7.9 (Ia, Ib; [Kostinskiy et al., 2015a]) Два последовательных ИК-кадра (инвертированы), с выдержкой 6,7 мс и интервалом 2 мс, которые показывают различные процессы разряда внутри облака. Во время этого события в видимом диапазоне наблюдались только вспышки рассеянного света, а не отдельные каналы. 1 - верхняя часть восходящего положительного лидера (его нижняя часть находится вне поля зрения ИК-камеры); 2 - стримерная зона; и 3 - UPFs. AGP означает «выше заземленной плоскости». (II.1) ИК-кадра (инвертированы), выдержка 6,7 мс. Верхняя часть восходящего положительного лидера (внизу справа) и UPFs (вверху слева), оба кадра внутри аэрозольного облака (адаптировано из [Kostinskiy et al., 2015a]). (II.2) Видимое изображение (вверху слева), полученное камерой 4 Picos, направленной в центр аэрозольного облака на высоте 80–120 см над плоскостью. Камера запускалась током, идущим от шара через шунт через 400 нс после начальной вспышки стримерной короны. Время экспозиции 1 мкс. Изображение размыто, поскольку UPFs находятся внутри облака. Одновременное ИК-изображение (справа). Время экспозиции составляет 8 мс. Видно, что контуры самых ярких UPF на обоих изображениях схожи (адаптировано из [Andreev et al., 2014]) и количества электронных лавин, вызванных прохождением лавиной убегающих электронов через ЕЕ-объем. См. раздел 7.6.4 для более подробного обсуждения перехода положительный лидер-двунаправленный лидер.

Таким образом, NBE-IE механизм инициации молнии позволяет объяснить IEC-стадию развития молнии во время инициирования молнии следующим образом: во-первых, за приблизительно одновременным трехмерным возникновением множества малых UPFs следует слияние небольших UPFs в цепочки или сети UPFs, благодаря вторичным положительным лидерам. Эти небольшие события могут произойти во время протекания IEC-стадии без сильных разрядов. Во-вторых, протекает процесс развития положительных и отрицательных лидеров, возникших из объединения сетей UPFs в фазе подготовки к первому IB-импульсу (IBP). Токи в UPFs, сетях UPFs, а также положительных и отрицательных лидерах вызывают изменение электрического поля во время протекания IEC (стадия начального изменения поля после инициирующего события IE), обнаруживаемое ближайшими быстродействующими антеннами. Мы предполагаем, что слияние двух более крупных цепочек UPFs или небольших двунаправленных лидеров, которые формируются из этих цепочек UPFs, вызывает «усиливающие события» IEC, описанные выше и обсуждаемые в [Marshall et al., 2014a; 2019]. Таким образом, Механизм выполняет часть условий c1 и c3, связанных с IEC-стадией развития молнии.

### 7.5.1.6. Требования к событиям, которые подготавливают IBPs

По определению IEC-стадия начинается с инициирующего молнию события (IE) и заканчивается первым классическим IB-импульсом (IBP). Как указано выше (идея i3), средняя длительность IEC-стадии для CG-молний (молний облако-земля) составляет около 230 мкс (исключительно короткое время), а для IC-молний (внутриоблачных молний) — 2700 мкс (очень короткое время для аккумулирования необходимого для IBP). Моделирование первого и второго классического IBP, которые предшествовали трем CG-молниям показало, что максимальные токи IBP составляли очень большую величину, сравнимую с обратными ударами молний (30–100 кА), они имели длительность 30–50 мкс и заряды в чехлах каналов (расчетные линейные заряды) 0,2–1,2 Кл [Karunarathne et al., 2014]. Таким образом, за чрезвычайно короткое время IE и IEC-стадий электронные лавины, положительные стримеры, UPFs, сети UPFs и двунаправленные лидеры должны были накопить заряд 0,2–1,2 Кл, необходимый для первого классического IB-импульса (IBP).

Важно оценить общую длину сетей UPFs, которые накопят заряд 0,2–1,2 Кл, хотя линейная плотность заряда для UPFs неизвестна, но мы будем исходить из данных по длинной искре. Для длинной искры хорошо известно измеренное значение линейной плотности заряда $\omega$ =0,05–0,07 мКл/м [Les Renardières Group, 1977, 1981], а для хорошо развитого отрицательного лидера молнии $\omega$ = 0,7–1,0 мКл/м [Rakov & Uman, 2003]. Чтобы оценить общую длину сетей UPFs, которые вместе обеспечивают заряд классического IBP, мы, используя линейную плотность заряда в диапазоне 0,07-0,7 мКл/м, находим длины 290-2900 м для 0,2 Кл и 1700-17000 м для 1,2 Кл. Эти длины намного больше, чем наблюдаемые скоростными камерами длины 75-90 м для первых двух IBPs при развитии двух отрицательных CG-молний [Stolzenburg et al., 2013]. Поэтому нам кажется крайне маловероятным, чтобы один двунаправленный лидер длиной 75-90 м мог накопить в своей чехле заряд 0,2-1 Кл. Таким образом, Механизм постулирует, что IEC — это трехмерный процесс с параллельным развитием множества сетей UPFs одновременно, и что каждая сеть UPFs имеет свою собственную трехмерную структуру [Kostinskiy et al., 2015а], см. также главы 1-5. Таким образом, во время протекания IEC-стадии только сети UPFs (а не единичный двунаправленный лидер Каземира) могут накопить требуемый заряд для IB-импульсов с такими большими токами и зарядами. При длине каналов во время протекания IB-импульсов примерно в 100 м, потребуется 3-30 параллельных сетей UPFs для обеспечения заряда IBP в размере 0,2 С и 17-170 параллельных элементов для заряда IBP 1,2 Кл, в зависимости от линейной плотности заряда в цепочках UPFs. Таким образом, трехмерность сетей UPFs является фундаментальной для создания IB-импульсов (IBPs). По мере того, как общая длина и объем каждой сети UPF увеличивается и через них протекают токи, общий заряд в чехлах каждой сети увеличивается. В результате на концах каналов сети UPFs возникает дополнительный заряд и растет потенциал, так как сети поляризуются в электрическом поле грозового облака.

### 7.5.1.7. Первый классический IB-импульс

Согласно NBE-IE-Механизму, первый классический IBP возникает, когда две объемные плазменные системы, каждая из которых состоит из сети соединенных в каналы UPFs, инициирую из сети двунаправленные лидеры, которые встречаются в квази-сквозной фазе и соединяются, как схематично показано на Рисунке 7.10 A,B. Объединение двух сетей включает в себя «сквозную фазу» («4» на Рисунке 7.10A) и «обратный удар» («10» на Рисунке 7.10B) (например, [Jerauld et al., 2007], который продуцирует мощную вспышку света, связанную с первым классическим IB- импульсом [Stolzenburg et al., 2013, 2014], [Campos & Saba, 2013], [Wilkes et al., 2016]. Таким образом, Механизм выполняет условия c1 и c8, объясняя причину первого классического IB- импульса или IBP №1.

### 7.5.1.8. Последующие классические IB-импульсы (IBPs)

По статистическим причинам плазменные сети в некоторых $E_{str+}$- объемах формируются раньше, чем в других $E_{str+}$-объемах, поэтому местоположение первого классического IB-импульса невозможно предсказать. Однако после первого классического IB-импульса, который объединил две плазменных сети Рисунок 7.10.B(1),(2), быстрая поляризация возникающего плазменного комплекса из двух сетей B(1), B(2) в электрическом поле грозового облака создает индуцированные заряды на концах плазменного комплекса в областях B(9) и B(8), где вспыхивают мощные стримерные вспышки (в статьях [Stolzenburg et al., 2013; 2014; 2020] плазменные образования описанные выше, которые создают IBPs и яркие вспышки света, называют «начальным лидером» («initial leader»)). Initial leader *«начальный лидер»* — это ранее введенная в оборот новая сущность (не являющаяся ступенчатым отрицательным лидером или двунаправленным лидером), которая позволяет обозначить прерывистые процессы на стадии инициации молнии, скорее всего связанные с IBPs, которые вызывают сильные изменения электрического поля и яркие вспышки света, похожие на проявления ступенчатого отрицательного лидера, но гораздо более сильные и яркие. Поляризация плазменных образований B(1),

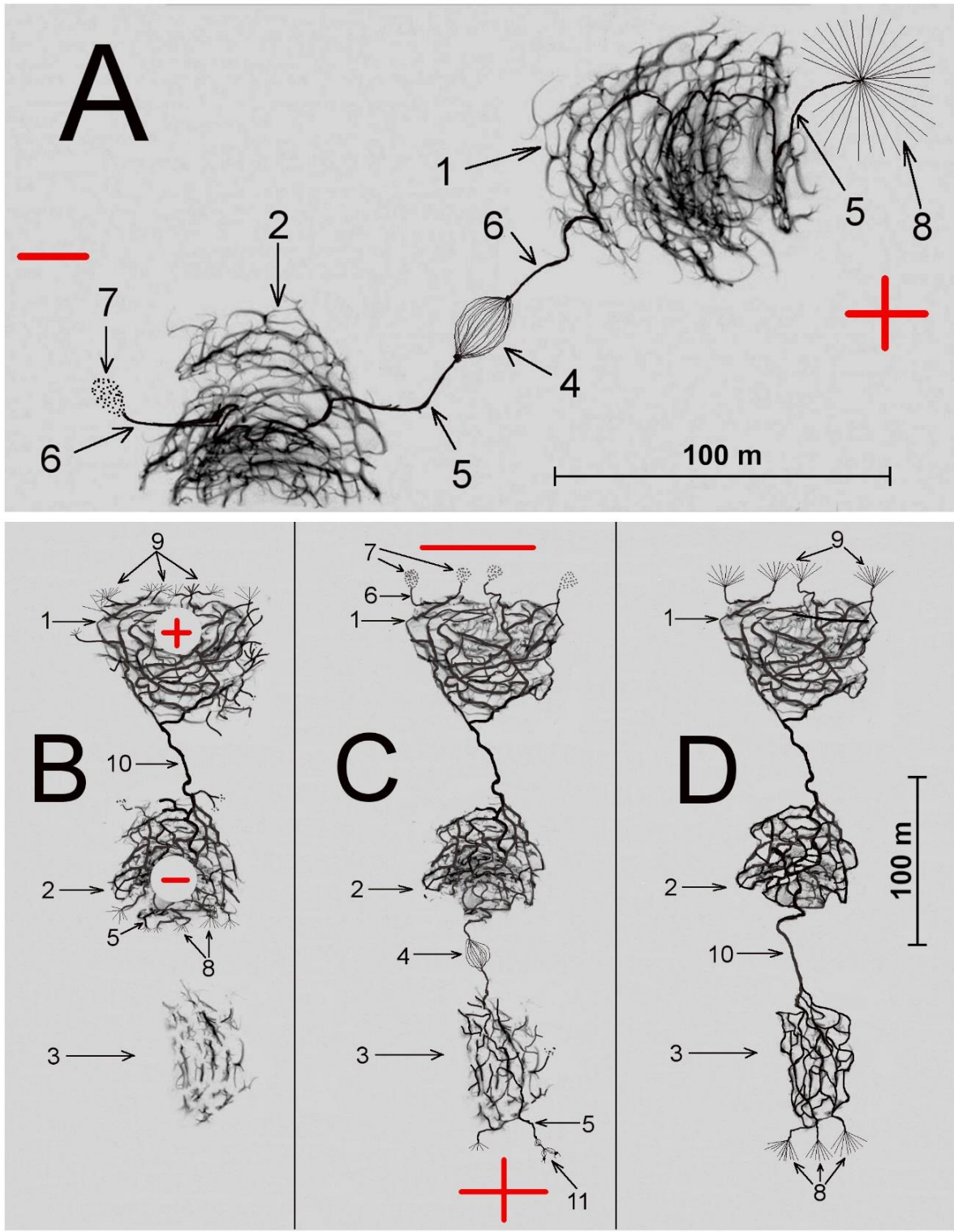

Рисунок 7.10 (схема), (адаптировано из [Kostinskiy et al., 2020a]). (А). Две плазменные сети, образованные после слияния UPFs, взаимодействуют друг с другом (сквозная фаза в IBP №1). (B). Фаза «обратного удара» IBP №1, на которой две плазменные сети объединяются. (C). Сквозная фаза в IBP №2. (D). Фаза «обратного удара» IBP №2. 1 — первая плазменная сеть, образованная слиянием множества UPFs; 2 — вторая плазменная сеть; 3 — третья плазменная сеть; 4 — сквозная фаза взаимодействия плазменных сетей; 5 — отрицательный лидер; 6 — положительный лидер; 7 — стримерная корона положительного лидера; 8 — вспышка стримерной короны отрицательного лидера; 9 — вспышка стримерной короны положительного лидера; 10 — плазменный канал фазы «обратного удара»; 11 — стримерная корона отрицательного лидера.

B(2) создают дополнительное большое электрическое поле на концах этих плазменных сетей и вблизи них. Усиленное электрическое поле в областях B(5),(8) ускоряет развитие существующих сетей UPFs (B(3)), которые находятся ближе всего к отрицательному концу плазменной сети B(2) и разрозненные небольшие сети B(3) превращаются в гораздо более сильно связанную сеть C(3). По мере развития сетей C(2),(3), из них навстречу друг другу начинают двигаться отрицательный и положительный лидеры, которые образуют «квази-сквозную фазу» Рисунок 7.10.C(4) и «квази-обратный удар» Рисунок 7.10.D(10), который будет восприниматься быстрой антенной (FA), как второй классический IB-импульс или IBP № 2.

Последовательность событий, ведущих к IBP№2, соответствует высокоскоростному видео и данным FA [Stolzenburg et al., 2013; 2014; 2020], которые показали, что каждый новый классический IBP «расширял» предыдущий «канал» IBP. Отметим, что в отличие от генерации первого классического IBP, где большие значения электрических полей были вызваны гидродинамикой и статистикой, генерация второго классического IBP может быть вызвана суперпозицией электрических полей из-за: (а) усиленного электрического поля на краях поляризованных плазменных сетей, которые привели к IBP№ 1 и (b) усиленного электрического поля благодаря гидродинамическим и статистическим процессам.

Следующие IB-импульсы могут происходить аналогично образованию IBP№2. Для каждого последующего IB-импульса усиление поля на концах растущего *«начального лидера»* (которым [Stolzenburg et al., 2013; 2014; 2020] называют цепочку вспышек света, причиной которых может быть слияние больших плазменных сетей, схематически изображенную на Рисунке 7.10), в сочетании со статистически распределенными в пространстве возникающими группами плазменных сетей, сильно влияет на линейную траекторию *«начального лидера»*, тем самым учитывая изменения в направлении движения *«начального лидера»* [Stolzenburg et al., 2013] и для «ветвей» *«начального лидера»* [Stolzenburg et al., 2014, 2020]. Хорошо известно, что серии IBPs сильно отличаются одна от другой, включая кажущийся случайным порядок амплитуд IBPs (которые фиксирует быстрая антенна (FA)), длительностей IBPs и интервалов времени между IBPs, как показано при сравнении Рисунков 7.11 и 7.12 ниже (см. также [Wu et al., 2014, рисунок 3], [Bandara et al., 2019, рисунки 4-5], [Smith et al., 2018, рисунки 3-5]).

Предлагаемый Механизм может объяснить эти различия; действительно, гидродинамическое и турбулентное статистическое распределение областей зарядов в облаке и инициация положительных стримерных вспышек в них, благодаря лавинам EAS-RREA, по существу, требует таких сильных вариаций различных параметров IBPs.

Как упоминалось во введении (термин t4), мы определили IB-стадию инициирования молнии, как начинающуюся с первого классического IBP, а предполагаемое окончание стадии наступает после последнего IBP или при переходе к классическому большому отрицательному ступенчатому лидеру. Мы также упомянули, что процессы, вызывающие классические IBPs и небольшие IBPs, неизвестны и могут отличаться. В Механизме физический процесс классических IBP представляет собой соединение двух двунаправленных лидеров, каждый из которых развивается из большой трехмерной сети UPFs (или из трехмерной сети каналов горячей плазмы, созданной из сети UPFs). В Механизме другой физический процесс вызывает как (i) небольшие IBPs, так и импульсы VHF между классическими IBPs и (ii) импульсы фиксируемые быстрой антенной (FA) и импульсы VHF, которые возникают во время IEC (до первого классического IBP): а именно, контакт/слияние двух UPFs, контакт/слияние UPFs или небольшого двунаправленного лидера с цепочкой UPFs или сетью UPFs, а также контакт/слияние двух цепочек (или небольших сетей) из UPFs и двунаправленных лидеров. Мы сгруппируем эти три способа вместе под названием «подготовительные слияния» («preparatory mergers»). Самым крупным из подготовительных слияний, которые происходят во время проведения IEC (начальные усиления электрического поля), являются события по усилению IEC (the IEC enhancing events), то есть события, усиливающие рост электрического поля, о которых говорилось выше.

### 7.5.1.9. Переход к «классическому» отрицательному ступенчатому лидеру

Как описано выше (раздел 7.5.1.5), небольшие отрицательные лидеры будут инициированы с отрицательных концов сетей UPFs (или трехмерной сети каналов горячей плазмы, образованной слиянием сетей UPFs), когда потенциал на отрицательных концах сетей станет в 1,5–2 раз больше, чем потенциал для инициирования положительного лидера. Эти положительные лидеры показаны на Рисунке 7.10(6). Если поблизости есть другие сети UPFs (или другая трехмерная сеть каналов горячей плазмы, образованной слиянием сетей UPFs), то возможно большое число IB-импульсов; в противном случае отрицательный ступенчатый лидер выживет, если цепочка каналов UPFs превратиться в длинный горячий, проводящий плазменный канал, а окружающее электрическое поле облака будет достаточным, чтобы поддержать долгосрочное развитие большого двунаправленного лидера. Когда один из отрицательных лидеров (аналогичный показанным на Рисунке 7.10D(8), в процессе развития всей системы плазменных каналов становится самоподдерживающимся (т.е. становится «классическим» отрицательным ступенчатым лидером), то процесс формирования канала молнии завершается и фаза инициирования молнии заканчивается.

Таким образом, в Механизме движение IB-импульсов электрического поля (и их свечения) в пространстве принципиально отличается от развития отрицательного ступенчатого лидера, поскольку для механизма ступенчатого лидера требуется уже существующий длинный плазменный канал. [Stolzenburg et al., 2014] предположили, что *начальный лидер* (которым они считали цепочку IBPs и их свечение) с каждым последующим IBP, нагревается и что это нагревание в конечном итоге приводит к переходу исходного плазменного образования в классический ступенчатый лидер с горячим проводящим каналом. Недавно [Karunarathne et al., 2020] представили результаты моделирования последовательностей IBPs, которые подтверждают эту гипотезу. Плазменные сети могут также развиваться в других ячейках EE-объема, удаленные от первых IBPs, и они также могут вносить вклад в серии IBPs.

### 7.5.1.10. Сравнение механизма NBE-IE с экспериментальными данными

[Rison et al., 2016] подробно изучили три положительных NBEs, которые инициировали IC-молнии; NBEs двигались вниз, а следующие IB-импульсы двигались вверх (Рисунок B.31. Для каждой вспышки сразу после нисходящего положительного NBE цифровой интерферометр обнаружил серию рассеянных восходящих событий малой мощности, которые [Rison et al., 2016] идентифицировали, как соответствующие началу IEC-стадии (Рисунок B.30). Эти маломощные источники показаны на рисунках 2a, 2b и 3a в [Rison et al., 2016)], Рисунок B.30. Кроме того, их рисунок 3c показывает, что эти маломощные источники продолжались до первого IB-импульса молнии. Мы предполагаем, что маломощные источники были объединением UPFs в цепочки UPFs и UPFs-сети, причем две UPFs-сети, в конце концов, соединились, чтобы произвести первый IB-импульс, как описано выше в разделах 7.5.1.4, 7.5.1.5 и 7.5.1.7 и показано схематично на Рисунках 7.10.A и 7.10.B.

Рисунок 3c из [Rison et al., 2016] также показывает, что маломощные восходящие источники почти непрерывно обнаруживались цифровым интерферометром между первым и вторым IB-импульсами, а также между вторым и третьим IB-импульсами. Мы предполагаем, что источники малой мощности между каждой парой IB-импульсов также были объединением UPFs в цепочки UPFs и, в конечном итоге, в сеть UPFs, которая вызвала следующий IB-импульс, когда образовавшаяся сеть UPFs начала процесс контакта с ранее образованными в единую сеть сетями UPFs (благодаря процессам, обеспечившим предыдущие IB-импульсы), как кратко описано выше в разделе 7.5.1.8. На Рисунках 7.10.B и 7.10.C схематично показан процесс, приводящий ко второму IB-импульсу молнии благодаря слиянию недавно образованных, разрозненных UPFs (Рисунок 7.10.B(3)) в новую сеть UPFs (обозначенную цифрой (3) на рисунках 7.10.B и 7.10.C). На рисунках 7.10.C и 7.10.D схематически показано, что второй IB-импульс возникает, когда новая сеть UPFs (3) присоединяется к объединенной паре сетей UPFs (1,2), которые до этого объединились и произвели первый IB-импульс.

[Bandara et al., 2019] исследовали отрицательные NBE (NNBE), которые инициировали отрицательные CG-молнии (—CG). Определение полярности NBE основано на полярности начального пика осциллограммы быстрой антенны (FA) NBE с использованием физического соглашения о полярности электрического поля. Мы предполагаем, что вспышки положительных стримеров являются основным источником сигнала NBE. Следовательно, для —CG-молний, NNBE перемещают положительный заряд вверх, а следующие IBPs движутся вниз и в конце концов переходят в движущийся вниз отрицательный ступенчатый лидер.

На наш взгляд, NBE-IE Механизм инициации молнии качественно подтверждается измерениями [Bandara et al., 2019], как показано на Рисунках 7.11 и 7.12. Рисунок 7.11a

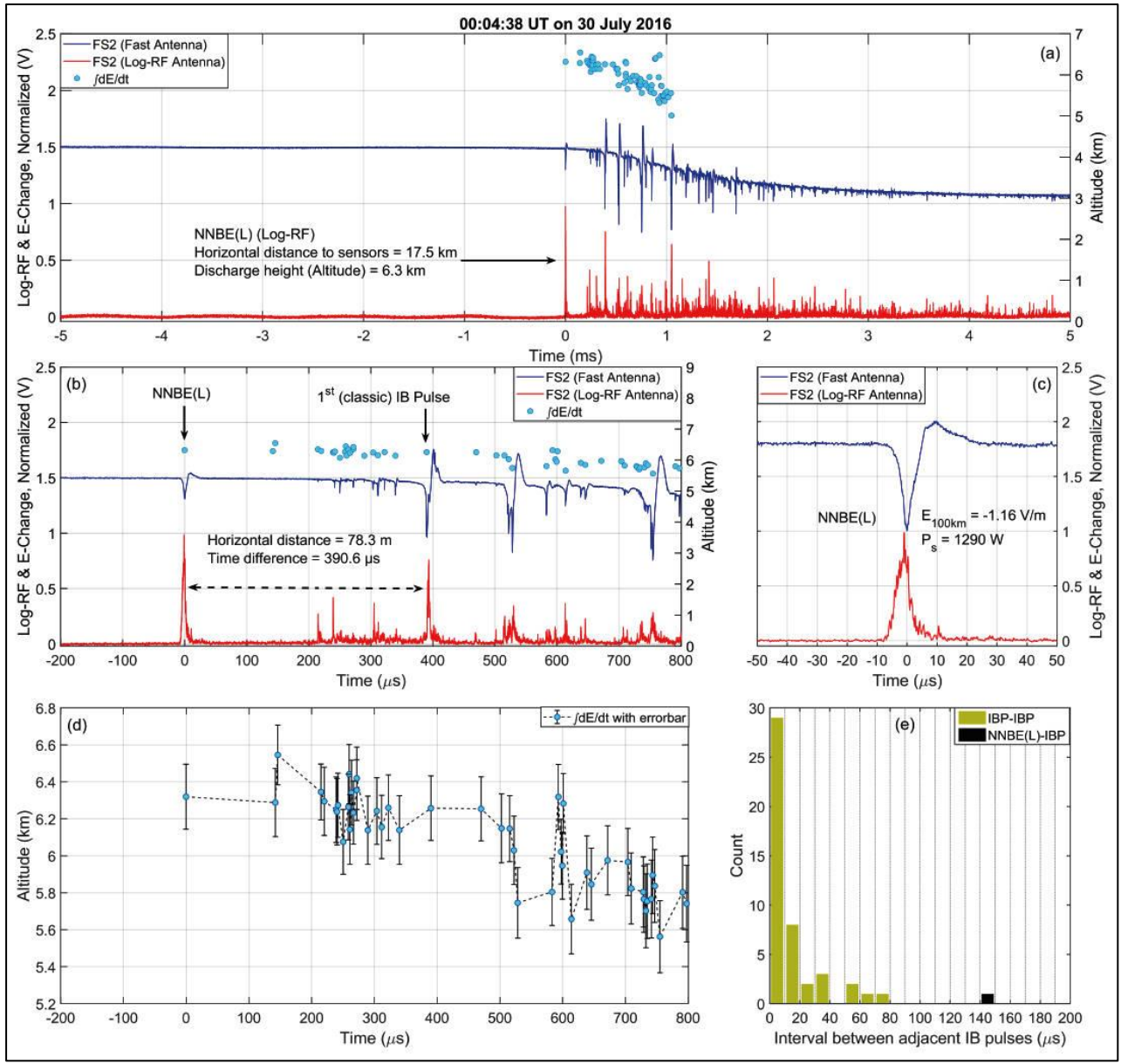

Рисунок 7.11 (адаптировано из [Bandara et al.2019], [Kostinskiy et al., 2020a]). Пример более мощного NNBE (1290 Вт), который, по-видимому, инициировал —CG-молнию (названную NNBE(L) в [Bandara et al., 2019]). Данные FA (синяя линия, неоткалиброванная линейная шкала) и данные Log-RF (красная, неоткалиброванная логарифмическая шкала) нанесены на график как нормализованное напряжение в зависимости от времени (т.е. для каждой кривой наибольшая амплитуда размаха импульса масштабируется до 1,0 В). $E_{100km}$ - это амплитуда от нуля до пика FA (в В/м) NNBE(L), нормированная по дальности до 100 км, а $P_S$ - мощность VHF (в Вт) NNBE(L). (a) Обзор, показывающий 10 мс данных FA и данных Log-RF. Голубые точки представляют собой высоты (правая вертикальная шкала) импульсов FA, определенные с помощью $\int \frac{dE}{dt}$. Высота NNBE(L) составляла 6,3 км. (b) Увеличенный вид (шкала 1 мс) первых событий в (a). (c) Увеличенное изображение (100 мкс) NNBE(L). (d) Высоты импульсов FA с планками ошибок для той же 1 мс, что показана на панели (b). (e) Гистограмма временных интервалов между соседними импульсами FA для той же 1 мс, показанной на (b) и (d). Временной интервал между NNBE(L) и местоположением следующего импульса показан черным цветом на гистограмме.

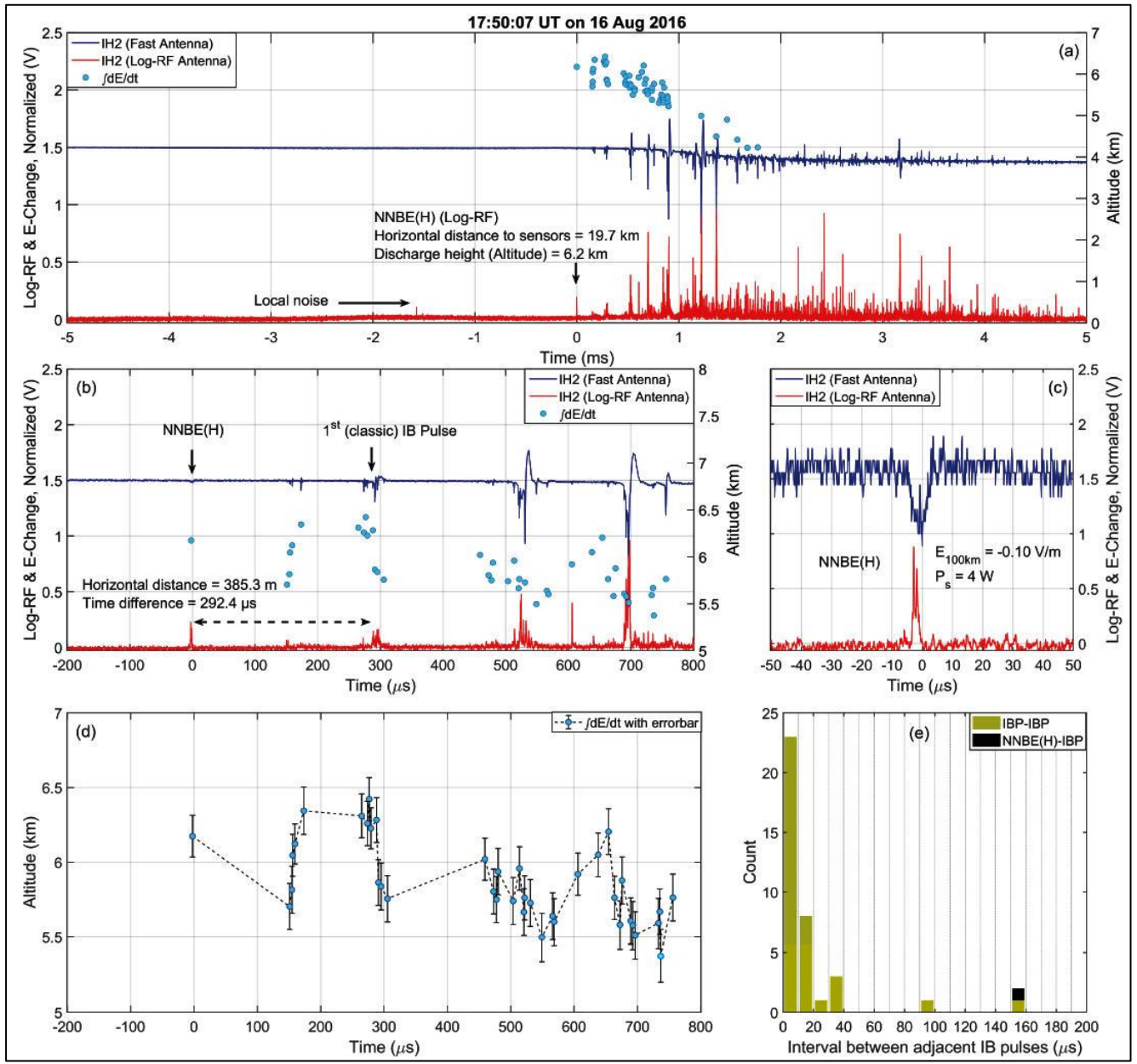

Рисунок 7.12 (адаптировано из (адаптировано из [Kostinskiy et al., 2020a], [Bandara et al., 2019]). Событие, подобное изображенному на Рисунке 7.13, на котором показан пример NNBE с меньшей мощностью (4 Вт), который, по-видимому, инициировал —CG-молнию (называемую NNBE(H).

показывает 10-миллисекундный обзор начальных событий в —CG-молнии, обнаруженной быстрой антенной (FA) и датчиком мощности VHF (называемым Log-RF, с полосой пропускания 186-192 МГц). Обращает на себя внимание, что начальное событие в развитии молнии, NNBE, имело самую большую мощность Log-RF и что мощности, измеренные Log-RF-антенной классических IB-импульсов, также были относительно большими. Синие точки на Рисунке 7.11 представляют собой (z,t) местоположения импульсов быстрой антенны (FA); местоположения источников излучения (x, y, z, t) были определены массивом датчиков, измерявших dE/dt с использованием метода определения координат по запаздыванию времени прибытия (time-of-arrival technique — TOA). Подчеркнем, что из-за используемой шкалы амплитуд некоторые из импульсов FA, где даны положения синих точек, не различимы на Рисунке 7.11. Рисунок 7.11c показывает на шкале 100 мкс развитие NNBE: биполярный импульс FA имел амплитуду -1,16 В/м (дальность, нормированная на 100 км) и длительностью около 20 мкс, тогда как VHF-импульс (Log-RF) имел мощность 1290 Вт и длительность 15-17 мкс. Подчеркнем, что хаотический характер импульсов Log-RF согласуется с гипотезой Механизма о том, что NBEs представляют собой некогерентную суперпозицию множества положительных стримерных вспышек.

На рисунке 7.11b показаны данные FA, данные Log-RF и высоты расположения импульсов FA для первых 800 мкс развития молнии. На рисунке 7.11d показаны высоты импульсов FA с полосами ошибок по z для тех же 800 мкс. Время между NNBE и первым классическим импульсом IB составляло 390 мкс (Рисунок 7.11b). С точки зрения NBE-IE Механизма мы предполагаем на Рисунке 7.11b, что импульсы FA (синие точки) были вызваны слиянием UPFs в относительно длинные цепочки (достаточно длинные, чтобы образовать импульс FA). Соединения UPFs в цепочки начались только через 140 мкс после первого, инициирующего молнию события (IE), на что указывают местоположения очень слабых импульсов FA и VHF. Многие слияния цепочек UPFs произошли во временном диапазоне от 220 до 340 мкс после IE. Мы предполагаем, что некоторые из объединенных сетей UPFs дали начало двунаправленным лидерам. Когда два из этих лидеров (рожденных сетями) встретились и соединились, значительный ток «обратного удара» произвел первый классический IB-импульс. Важно, что «подготовительные слияния» в интервале от 220 до 340 мкс происходили через короткие интервалы времени в диапазоне 2–15 мкс (Рисунок 7.11d,e). Аналогичные подготовительные слияния произошли перед вторым классическим IBP (IBP №2 примерно через 530 мкс после IE) и перед третьим классическим IBP (IBP №3 примерно через 750 мкс после IE). Подготовительные слияния цепочек (или сетей) UPFs перед IBP №2 и IBP №3 могли включать только UPFs, вызванные исходными положительными стримерными вспышками, но могли также включать новые UPFs, вызванные более поздними лавинами/положительными стримерными вспышками. В целом, NBE-IE Механизм качественно согласуется с данными на Рисунке 7.11 о сильном NNBE инициирующих — CG-молнии, о которых сообщили [Bandara et al., 2019].

На Рисунке 7.12 [Bandara et al., 2019] показано инициирование —CG-молнии более слабым NNBE, и это инициирование также согласуется с механизмом NBE-IE. Биполярный импульс IE FA имел амплитуду всего -0,01 В/м (дальность, нормированная на 100 км) и длительность около 10 мкс. Импульс Log-RF имел VHF-мощность 4 Вт и длительность около 5 мкс. Подчеркнем, что импульс FA не был явно биполярным; вместо этого он был «более монополярным по своей природе», как для одного NNBE, о котором сообщили [Rison et al., 2016]. NNBE был первым событием в развитии молнии и имел довольно большую мощность, большую, чем мощность первого классического IBP, но намного меньшую, чем более поздние классические IBP. Первый классический IBP (IBP №1) произошел примерно через 290 мкс после NNBE; очевидны подготовительные слияния, которые начались через 150 мкс после NNBE с большим количеством слияний за 20 мкс до IPB №1. Классический IBP №2 произошел через 530 мкс после IE, а подготовительные слияния начались за 80 мкс до IBP №2. Классический IBP №3 произошел через 700 мкс после IE; подготовительные слияния произошли в период между IBP №2 и IBP №3 и увеличились непосредственно перед IBP №3. Непосредственно перед каждым из первых трех классических IBP слияния происходили с короткими интервалами времени в диапазоне 2-10 мкс (7.12d,e).

Короткое измеренное время (например, 5–150 мкс, Рисунки 7.11 и 7.12) между слияниями можно объяснить близкими расстояниями между взаимодействующими каналами. Мы можем поддержать эту идею, используя простую оценку. Средняя скорость лидера на начальной стадии развития молнии находится в диапазоне $v_L \approx 0.02 - 0.1 \frac{m}{\mu s}$ ([Rakov & Uman, 2003], [Горин, Шкилев, 1976], [Les Renardières Group, 1977, 1981]). Средняя скорость лидеров в сквозной фазе находится в диапазоне $v_{Lbr} \approx 0.1 - 0.3 \frac{m}{\mu s}$ ([Rakov & Uman, 2003], [Горин, Шкилев, 1976], [Les Renardières Group, 1977, 1981]). Следовательно, расстояние между плазменными каналами будет в диапазоне $D_{ch} \approx 0.1 \frac{m}{\mu s} \cdot (5 - 150)\mu s \approx (0.5 - 15)\ m$. Эта общая оценка согласуется с нашими предыдущими оценками (раздел 7.5.1.2) расстояния между воздушными электродами (~4,6 м), с которых начинается весь процесс формирования сети (после прохождения лавины EAS-RREA).

— CG-молния на рисунке 7.12 была инициирована гораздо более слабым NNBE, чем вспышка на Рисунке 7.11, но после IE развитие этих двух вспышек кажется очень похожим. Кажется, что обе инициации достаточно хорошо коррелируют с NBE-IE Механизмом.

### 7.5.1.11. NBE-IE Механизм для событий-прекурсоров и изолированных NBEs

[Rison et al., 2016] описали своего рода «разряд короткой продолжительности» ("short duration discharge"), который они назвали «прекурсорами», поскольку они «иногда происходят до того, как IC-молния начнется в том же месте» («sometimes occur seconds before an IC discharge initiates at the same location»). Два прекурсора, показанные в [Rison et al., 2016], имели длительность 250 мкс и 3 мс. Соответствующие прекурсоры имели IEs с длительностью <1 мкс и 2 мкс и мощностью VHF 10,8 дБВт и 21,6 дБВт. Мощность IE прекурсоров была на 20-30 дБВт больше, чем у других прекурсор-событий. Другой тип кратковременного разряда — это NBE, изолированное во времени и пространстве. Первоначально все NBE считались изолированными разрядами (например, [Willett et al., 1989]). С точки зрения Механизма вполне вероятно, что прекурсоры и изолированные NBE развиваются с помощью NBE-IE Механизма, но их EE-объемы имеют только небольшие $E_{str+}$-объемы (с электрическим полем, достаточным для поддержания движения положительных стримеров), так что развитие больших сетей UPFs и больших двунаправленных лидеров может не произойти, тем самым предотвращая возникновение IBPs. Без IBPs прекурсоры и изолированные NBEs, возможно, не могут развиться в полноценные молнии.

## 7.5.2. Механизм слабого IE, который инициирует молнию

В этом разделе мы описываем, как Механизм описывает возникновение молнии, инициированной событием (IE) короткой продолжительности ($\leq$ 1 мкс) и низкой мощности VHF (<1 Вт), как описано [Marshall et al., 2019] и [Lyu et al., 2019]. Мы назовем эту часть Механизма механизмом слабого IE. Эти IE более слабые, чем все инициирующие события NBE, описанные выше, и явно не являются классическими NBE. Как упоминалось во введении, недавние измерения показывают, что 88% из 26 ближайших IC-молний были инициированы слабыми IE [Lyu et al., 2019], в то время как 96% из 868 CG-молний были инициированы слабыми IE [Bandara et al., 2019]. Поэтому вероятность возникновения молний в 10-25 раз выше в случае слабых IE, чем в случае NBE-IE Механизма. Таким образом, мы можем ожидать, что условия, необходимые для Механизма слабого IE, с большей вероятностью возникнут в грозовом облаке, чем условия для NBE-IE Механизма.

## 7.5.2.1 Первое условие инициирования слабого IE

В соответствии с вышеупомянутыми наблюдениями, мощность VHF для инициирующего молнию события (IE) должна быть меньше, чем 1 Вт. С точки зрения Механизма величина VHF мощности IE существенно зависит от количества $E_{th}$-объемов («воздушных электродов») в ЕЕ-объеме, внутри которых стартуют стримерные вспышки. Если начальных $E_{th}$-volume относительно мало, то стримеров, которые стартуют из этих объемов также мало и VHF-сигнал IE будет слабым. По сравнению с NBE-IE Механизмом, Механизм слабого IE должен иметь меньше вовлеченных в инициацию стримеров воздушных электродов или меньше релятивистских частиц EAS-RREA пересекают воздушные электроды и вызываю стримерные вспышки (может быть реализована и комбинация этих двух процессов).

## 7.5.2.2. Второе условие инициирования «слабого» инициирующего события (Weak IE)

Как описано в разделе 7.5.1.6, начальное изменение электрического поля (IEC) должно накопить заряд 0,2–1,2 Кл [Karunarathne et al., 2014], который измеряется при первом классическом IB-импульсе. Механизм слабого IE (как и NBE-IE Механизм) предполагает, что первый классический IBP вызван контактом двух двунаправленных лидеров, которые возникли внутри больших сетей UPFs (Рисунок 7.10A(4), B(10)). Чтобы иметь достаточный заряд для обеспечения первого классического IBP, общий заряд плазменной системы, накопленный в чехлах двух сливающихся двунаправленных лидеров (и плазменных сетей, которые поддерживают эти лидеры), должен составлять 0,2 – 1,2 Кл. Этот очень большой общий заряд требуется даже для IEC самой малой продолжительности (порядка 100 мкс). Количество заряда, которое будет перемещено во время сквозной фазы и «обратного удара» IBP, должно быть накоплено во время протекания IEC-стадии развития молнии.

## 7.5.2.3. Механизм инициации «слабого» инициирующего события IE (Механизм слабого IE — Weak-IE Mechanism)

Для того, чтобы эти два условия были выполнены, и молния в случае «слабого» IE инициировалась одним и тем же Механизмом, предложенным для мощного инициирующего события IE, которым является NBE-IE Механизм, необходимо, чтобы большая часть EE-объема включала в себя большие пространственные объемы с электрическим полем, достаточным для поддержания положительных стримеров ($E_{str+}$-объемы), см. Рисунок 7.13.I. (причина этого требования будет указана в следующем параграфе). Кроме того, для слабого IE требуется гораздо меньше воздушных электродов (порядка $10^2$-$10^4$ против $10^6$-$10^7$, необходимых для NBE). Поскольку минимальная

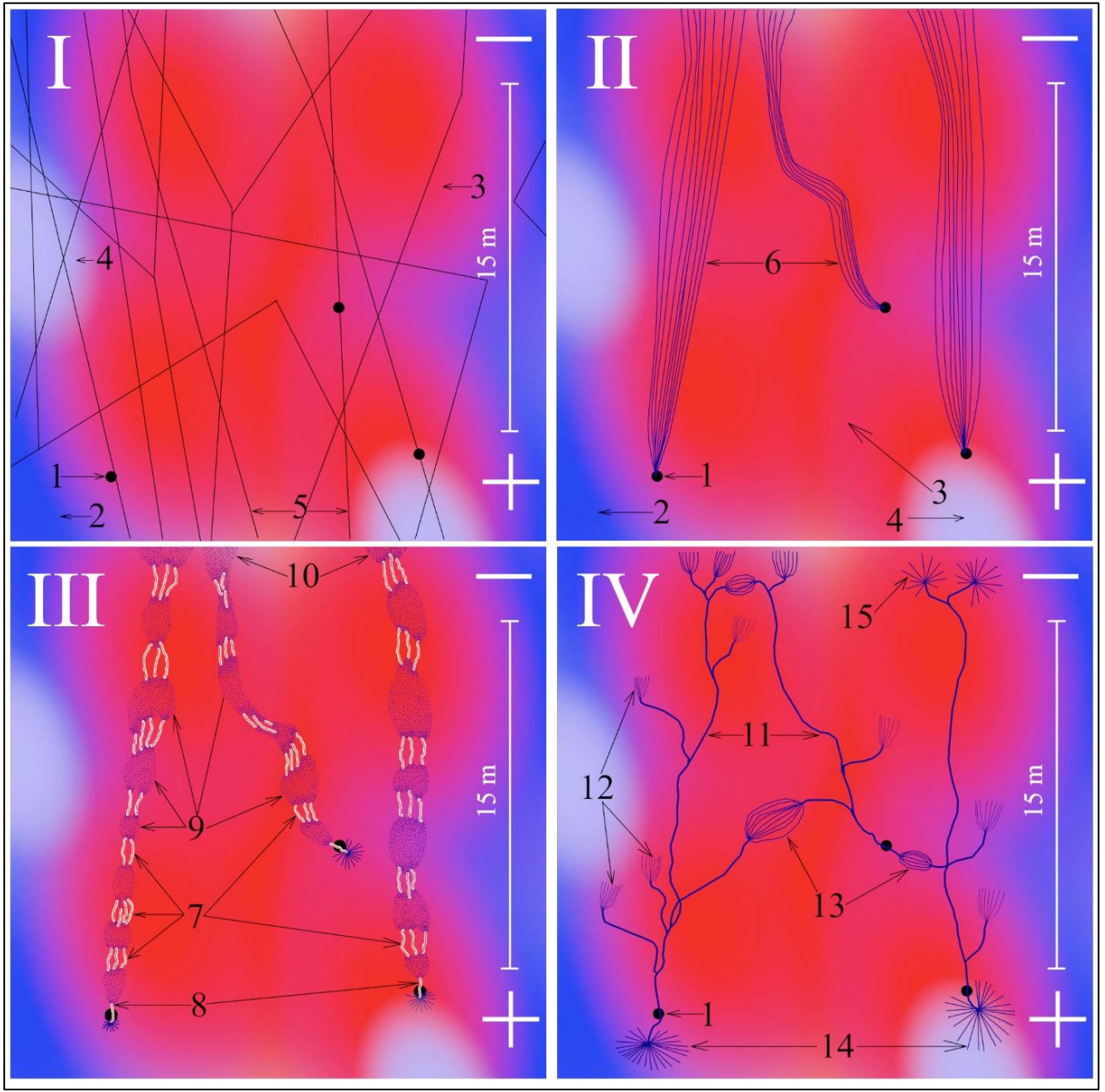

Рисунок 7.13 (адаптировано из [Kostinskiy et al., 2020a]). Возможное инициирование молнии с точки зрения Механизма «слабого» NBE и/или «слабого» инициирующего события (IE). 1 — область с электрическим полем E≥ 3 МВ/(м·атм); 2 — область с электрическим полем E <0,28-0,35 МВ/(м·атм); 3 — область с электрическим полем E ≥ 0,45-0,5 МВ/(м·атм); 4 — область с электрическим полем E≈0,45-0,5 МВ/(м·атм); 5 — траектории убегающих релятивистских электронов (позитронов); 6 — длинные положительные стримерные вспышки; 7,8 - UPFs; 9 — вторичные стримерные короны, соединяющие UPFs; 10 — положительная корона впереди UPFs; 11 — горячие высокопроводящие плазменные каналы; 12 — положительные стримерные короны положительных лидеров; 13 — вторичная положительная стримерная корона двух взаимодействующих в сквозной фазе больших плазменных каналов; 14 — вспышка стримерной короны отрицательного лидера; 15 —вспышка положительной короны положительного лидера.

величина электрического поля в $E_{str+}$-объемах ($\geq 0,45$-$0,5$ МВ/(м·атм)) только примерно на 50% больше, чем среднее электрическое поле в EE-объеме (0,28-0,35 МВ/(м·атм), «ландшафт» электрического поля Weak-IE Mechanism легче (и чаще) реализуется, благодаря статистическим флуктуациям и гидродинамическим процессам, чем ландшафт электрического поля, необходимый для NBE-IE Механизма с его очень большим количеством ($10^7$) $E_{th}$-объемов (воздушных электродов). Этот ландшафт также должен быть насыщен большим количеством лавин релятивистских убегающих электронов и позитронов (Рисунок 7.13.I (5)), чтобы вызвать достаточное число положительных стримерных вспышек, но не большее число, чем требуется в NBE-IE Механизме. Меньшее требование к числу лавин релятивистских убегающих электронов, также увеличивает процент успешных реализаций Механизма слабого IE (Weak-IE Mechanism).

Ключевое различие между Механизмом слабого инициирующего события (Weak-IE Механизмом) и NBE-IE Механизмом основано на различии в характеристиках областей облака, которые могут поддерживать движение стримеров ($E_{str+}$-объемов). Для Механизма слабого IE, стримерные вспышки, начинающиеся в $E_{th}$-объеме, будут иметь очень длинные траектории стримеров (десятки метров) (Рисунок 7.13.II(6)), поскольку траектории будут продолжаться на протяжении всех $E_{str+}$-объемов (Рисунок 7.13.I(3,4)). Из-за длинных траекторий ионизационно-перегревная неустойчивость будет вызывать множество UPFs вдоль траектории каждой стримерной вспышки (Рисунок 7.13.III(7)). Через несколько микросекунд после Weak-IE, UPFs будут соединены в длинные цепочки UPFs с помощью своих собственных вторичных положительных стримерных корон (Рисунок 7.13.III(9)), потому что внутри этих цепочек (между многих UPFs) электрическое поле выше, чем порог распространения положительных стримеров. $E_{str+}$ $\geq 0,45$-$0,5$ МВ/(м·атм). Внутри каждой длинной цепочки UPFs течет ток в диапазоне 5-20 А. Средняя скорость удлинения каждого элемента цепи UPFs или плазменного канала, который выживает и двигается за счет тока соединяющих их положительных стримеров, будет примерно 1-2 см/мкс [Les Renardières Group, 1977]. Когда каналы UPFs сливаются или объединяются в несколько более длинных цепочек, они могут переместиться на 3–6 м друг к другу за 150 мкс. Если UPFs квазиравномерно распределены в пространстве, то объединение UPFs в один большой горячий канал может происходить в виде серии импульсов тока с временем между импульсами 1-3 мкс. Каждая UPFs или небольшая цепочка UPF, которая сливается с основной локальной цепочкой UPFs, будет производить значительный импульс тока в контактирующих высокопроводящих каналах, Рисунок 7.13.IV(11). Мы предполагаем, что эти слияния вызывают VHF-импульсы различной амплитуды в зависимости от длин цепочек UPF, которые сливаются. В конце концов, соединение еще одной цепочки UPFs с существующей длинной цепью UPFs создаст единый плазменный канал длиной в несколько метров, который будет достаточно длинным для того, чтобы на концах объединенного канала возникали сильные отрицательные и положительные стримерные вспышки, Рисунок 7.13.IV(14,15), и родился двунаправленный лидер. Эти стримерные вспышки могут быть похожи на положительные и отрицательные вспышки длинной искры [Kostinskiy et al., 2018, 2015b] (глава 6) и будут производить сильный VHF-сигнал. Объединение длинных цепочек UPFs и/или создание двунаправленного лидера может вызвать события усиливающие начальные изменения электрического поля (IEC), которые производят импульс быстрой антенны (FA) (от длинного тока), совпадающий с VHF- импульсом.

Несколько трехмерных плазменных сетей UPFs, создающих двунаправленные лидеры, должны развиваться близко друг к другу. Затем, как и в механизме NBE-IE, следует первый классический IBP, IBP №1, который возникает, когда два двунаправленных лидера, идущие изнутри плазменных сетей, соединяются друг с другом во время сквозной фазы и фазы обратного удара, Рисунок 7.13.IV(13). Эти события производят мощные VHF-импульсы и большие импульсы фиксируются FA во время первого классического IBP.

Остальная часть Механизма слабого IE идентична NBE-IE Механизму. После IBP №1 электрическое поле в области, расположенной ниже отрицательного конца двух соединенных двунаправленных лидеров (Рисунок 7.13.IV(14)) будет значительно усилено «квазиобратным ударом» при создании IBP №1 и одна или несколько существующих сетей UPFs, находящихся в этой области усиленного электрического поля, также создаст внутри себя двунаправленный лидер, который может войти в контакт с плазменной сетью, которая возникла после IBP №1, тем самым приведя ко второму классическому IBP или IBP №2 и т. д. После того, как произойдет достаточное число IBPs, чтобы возник настолько длинный проводящий канал, что на его концах будет достигнута достаточная разность потенциалов, чтобы стартовал устойчиво само распространяющийся отрицательный ступенчатый лидер, то можно будет говорить об окончании этапа инициирования молнии.

## 7.5.2.4. Сравнение механизма возникновения слабого IE с экспериментальными данными

[Marshall et al., 2019] показывают два примера CG-молний, инициированных Weak-IE событием, воспроизведенных на Рисунке 7.14. По сравнению с NBE-IE Механизмом, показанным на Рисунках 7.11 7.12, инициирующее событие (IE) на Рисунке 7.14а имело гораздо меньшую VHF-мощность и гораздо меньшую продолжительность (-0,14 Вт и 1 мкс, соответственно); Осциллограммы с большим масштабом времени (не показаны на Рисунке) не обнаружили импульса быстрой антенны (FA) с VHF инициирующим импульсом. В течение 130 мкс, которые продолжалась стадия IEC было много VHF-импульсов и только несколько импульсов быстрой антенны (FA). Мы предполагаем, что VHF-импульсы были вызваны слияниями UPFs с существующими цепочками UPFs или каналами лидеров. Присутствовали два события усиления электрического поля (одновременные импульсы FA и VHF: -0.1 В/м и 0.18 Вт и -0.2 В/м и 0.14 Вт), которые произошли в пределах 20 мкс от инициирующего молнию IE; эти первые события после IE произошли намного раньше, чем при инициации молнии NBE-IE событиями, описанными выше. В следующие 90 мкс было 7-10 небольших VHF-импульсов. В следующие (последние на осциллограмме) 20 мкс было много больших VHF-импульсов и одно событие усиления электрического поля (-0.2 В/м и 0.55 Вт), что привело к первому классическому IBP (IBP №1, -2,5 В/м и 3,0 Вт). Эти более сильные VHF-импульсы можно объяснить слиянием более длинных цепочек UPFs, описанных в Механизме слабого IE, и они аналогичны поздним событиям на стадии IEC, во время инициации по NBE-IE Механизму. Изменение момента заряда на стадии IEC составляло 36 Кл·м.

На рисунке 7.14b показано другое инициирование слабым событием (Weak-IE) CG-молнии. Это событие во многом похоже на только что рассмотренное по многим параметрам. Однако IE на рисунке 7.14b был намного сильнее по мощности VHF и длительности (0,64 Вт и 2 мкс, соответственно) и оно совпадало со слабым импульсом

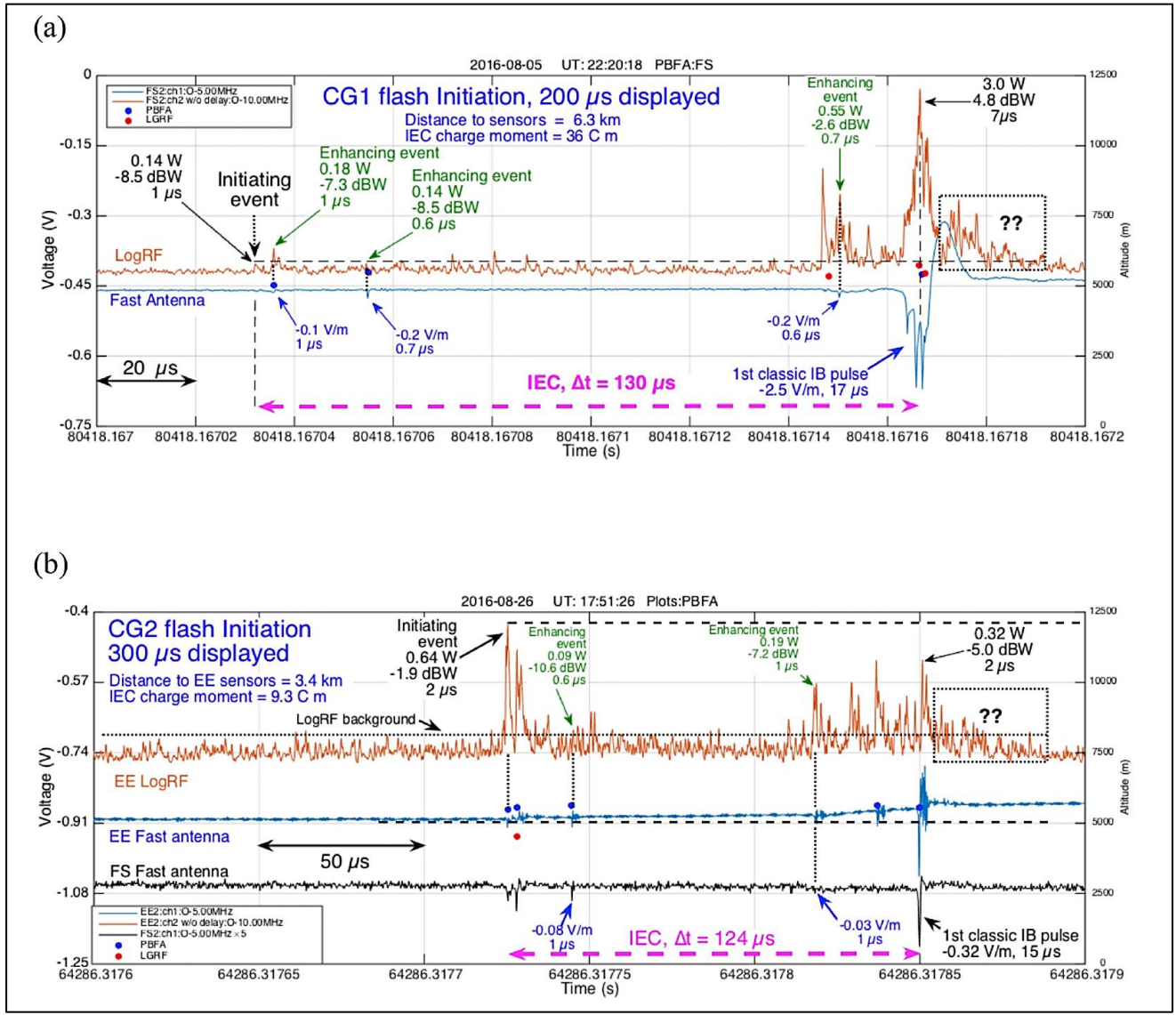

Рисунок 7.14 (адаптировано из [Marshall et al., 2019], [Kostinskiy et al., 2020a])). (a) Первые 200 мкс CG-молнии с Weak-IE (продолжительность — 1 мкс, мощность — 0.14 Вт). Данные быстрой антенны (FA) показаны синим цветом; VHF (LogRF)-сигнал — красным; расстояние от приборов до места инициирования молнии составляло 6,3 км. IEC длилась 130 мкс. (b) Первые 300 мкс другой –CG-молнии, которая тоже была инициирована Weak-IE Механизмом (продолжительность — 2 мкс, мощность — 0.64 Вт); расстояние от приборов до места возникновения молнии составляло 3,4 км. Обозначения осциллограмм такое же, как на панели (a), но включены также данные от дополнительной быстрой антенны FA, которая обозначена FS и показана черным цветом.

FA. IEC-стадия развития молнии длилась 124 мкс, а изменение момента заряда составляло всего 9 Кл·м. В первые 30 мкс после IE было несколько значительных импульсов VHF, за которыми последовало одно слабое усиление поля (-0,08 В/м и 0,09 Вт). В следующие 60 мкс было всего несколько VHF-импульсов. За последние 40 мкс перед первым классическим IBP (IBP №1, -0,32 В/м и 0,32 Вт) было более 20 относительно больших VHF-импульсов.

Оба инициирования молний, изображенное на Рисунке 7.14 кажутся достаточно хорошо соответствующими Механизму Weak-IE, и, за исключением двух различий, они похожи на инициирования по NBE-IE Механизму, показанные на Рисунках 7.11 и 7.12. Одно из различий ожидаемо: характер самого инициирующего события IE (NBE-IE против Weak-IE). Другое отличие состоит в том, что за Weak-IE сразу же следовали VHF-импульсы, в то время как для NBE-IE импульсы VHF не начинались в течение 140–150 мкс. Это различие можно объяснить исходя из представлений, изложенных в Механизме. NBE-IE Механизм порождает большое количество далеко находящихся друг от друга (разделенных расстояниями 5-10 м) коротких UPFs. По-видимому, требуется около 150 мкс, прежде чем эти UPFs сольются в короткие цепочки UPFs, которые затем могут объединиться и произвести наблюдаемые VHF-импульсы. В Механизме Weak-IE длинные положительные стримерные вспышки в $E_{str+}$объемах сразу же могут породить множество близко лежащих (не более 10-20 сантиметров) UPFs на траектории каждой стримерной вспышки. Эти UPFs быстро сливаются, и за 3-5 мкс достигают такой длины, что при их слиянии могут производиться обнаруживаемые в эксперименте VHF-импульсы.

## 7.6. Механизм инициации молнии с точки зрения последовательного перехода и усложнения структуры плазменных образований

### 7.6.0. Условия возникновения классических электронных лавин

Величина электрического поля, необходимого для создания электронных лавин, не зависит от типа поверхности электрода, если в воздухе у поверхности электрода имеется свободный затравочный электрон (или электроны), с которого начинаются электронные лавины. Этот факт многократно подтверждался в экспериментах, в которых электроды и разрядный промежуток облучались ультрафиолетовой лампой или радиоактивным источником (например, [Meek & Craggs, 1953, глава VIII, Irradiation and time lags, pp. 348-373]). Свет ультрафиолетовой лампы или радиоактивного элемента должен обеспечивать постоянное присутствие свободных электронов в разрядном промежутке в то время, когда на разрядный промежуток подается напряжение Е > 3–3,2 МВ/(м·атм). Электрическое поле инициации лавин не равно строго 3 МВ/(м·атм), так как оно зависит от влажности воздуха, газовых примесей, количества твердых и жидких аэрозолей в воздухе, а также от времени, во время которого к промежутку приложено высокое напряжение. Если в воздухе присутствуют свободные электроны, то электрическое поле обеспечивает пробой воздуха, когда частота ионизации $\nu_i$ развивающейся лавины превышает частоту диссоциативного прилипания свободных электронов к молекулам кислорода $\nu_a$ ($O_2 + e \rightarrow O^- + O$) [Коссый и др., 1994]. Если электрическое поле хотя бы на 1% меньше порогового значения, например, 3 МВ/(м·атм), то электроны будут преимущественно прилипать к молекулам кислорода, и лавины перестанут распространяться. Эти фундаментальные процессы в физике газового разряда не зависят от свойств электродов, и пробой газа происходит при электрическом поле ≥ 3 МВ/(м·атм) также в безэлектродных микроволновых разрядах [Gritsinin et al., 1996] и разрядах ВЧ-диапазона [Raizer 1991, pp.378-396], если в области эффективного переменного электрического поля выполняется фундаментальное условие $\nu_i \geq \nu_a$. По нашему мнению, «воздушные электроды» (усиление электрического поля в пространстве Е > 3 МВ/(м·атм)) появляются в облаке из-за статистического движения гидрометеоров (раздел 7.4.3.2). В этом случае воздушный электрод может не содержать ни одного гидрометеора в своем объеме, и в этом смысле он может быть чем-то похож на СВЧ-разряд в свободном пространстве [Gritsinin et al., 1996], только с гораздо меньшей частотой изменения электрического поля. Однако возможно также, что область пространства вблизи очень большого заряженного гидрометеора (или сталкивающихся гидрометеоров), также может быть воздушным электродом, и тогда усиление поля около поверхности гидрометеора также может играть роль в возникновении лавин, особенно, если статистические движения других гидрометеоров также усилят поле около него (для например, [Griffiths &, Latham, 1974], [Petersen et al., 2015], [Cai et al., 2015], [Dubinova et al., 2015], [Sadighi, 2015], [Sadighi et al., 2015]). Условие Е> 3 МВ/(м·атм) является необходимым условием образования лавин, но недостаточным. Достаточным условием для образования лавин (Meek & Craggs, 1953, стр. 348-373) является наличие свободных электронов в объеме сильного поля (3 МВ/(м·атм)). Поэтому, на наш взгляд, порог электрического поля в случае воздушного электрода также будет находиться в диапазоне Е > 3–3,2 МВ/(м·атм).

### 7.6.1. Лавинно-стримерный-переход

Как было отмечено в пионерских работах ([Loeb, 1966], [Phelps, 1974]), ключевую роль в зарождении и развитии молнии должны играть положительные стримеры, так как процесс распространения положительных стримеров в воздухе происходит при меньшем электрическом поле $E_{str+} \approx 450$-$500$ kV/(m·atm), чем любые другие плазменные процессы, включая распространение отрицательных стримеров, требующих $E_{str-} \approx 1000$-$1200$ kV/(m·atm), [Bazelyan and Raizer, 1998, 2000]. Классические стримеры рождаются благодаря классическим разрядным лавинам свободных электронов, которые прошли в электрическом поле $E_{th} \geq 3$ MV/(m·atm) расстояние, необходимое для выполнения критерия Мика [Raizer, 1991]. Критерий Мика подразумевает, что в грозовом облаке должны быть области размером не менее 2-10 см с электрическим полем выше поля конвенциального пробоя $E_{th} \geq 3$ MV/(m·atm), чтобы в некотором объёме внутри этой области $\leq 1$ мм$^3$ были сосредоточены $10^8$-$10^9$ электронов [Райзер, 1992]. При данном числе электронов происходит поляризация плазмы во внешнем электрическом поле $E_{str+} \geq 450$-$500$ kV/(m·atm) и создается условие для развития самостоятельного разряда перед головкой положительного стримера, так как электрическое перед головкой достигает величины $(5$-$10) \cdot E_{th}$. В результате возникает волна ионизации, называемая стримером, движущаяся со скоростями $10^5$-$10^7$ м/с (в зависимости от напряженности электрического поля). Переход лавины в стример является первым и важнейшим преобразованием плазмы из одной формы в другую и его называют — *лавинно-стримерный переход* [Райзер, 1992].

В большинстве исследований считают, что области облака со сверхпробойным полем $E_{th} \geq 3$ MV/(m·atm), где может произойти лавинно-стримерный переход, образуются благодаря заряду гидрометеоров (например, [Babich et al., 2016]), усилению поля на кончиках одного или нескольких заряженных гидрометеоров и/или благодаря гидродинамическим неустойчивостям жидкой фазы этих гидрометеоров ([Loeb, 1966], [Phelps, 1974]). В настоящее время нельзя исключить такие механизмы генерации положительных стримеров, но малое число крупных гидрометеоров в грозовом облаке с большими зарядами и их недостаточный размер резко снижают вероятность того, что такой механизм генерации стримеров является основным (этот вопрос был подробно проанализирован в разделах 7.4.3.1-7.4.3.3). Более крупномасштабный гидродинамический и статистический механизм усиления электрического поля нам кажется более перспективным (возможно при участии электрических полей сильно заряженных гидрометеоров наряду с поляризацией и статистическим усилением), но он в настоящее время требует тщательной экспериментальной проверки ([Colgate, 1967], [Trakhtengerts, 1989], [Trakhtengerts et al., 1997], [Mareev et al., 1999], [Iudin et al., 2003], [Iudin, 2017]).

### 7.6.2. Стримерно-UPFs переход

Плазма стримеров является холодной и после прохождения головки стримера в канале стримера электроны прилипают к молекулам кислорода за 100-200 нс [Kossyi et al., 1992]. Поэтому необходим быстрый механизм нагрева плазмы стримеров, так как нагретая до температур выше 1500-3000 $^0$К плазма имеет шанс «прожить» несколько микросекунд из-за резкого снижения константы прилипания и увеличения константы отлипания электронов [Bazelyan and Raizer, 1998, 2000], [Kossyi et al., 1992]. Единственный известный в настоящее время механизм перехода холодной проводящей плазмы стримеров в горячую плазму небольших каналов (при давлении воздуха 0.1-1 атм) — это достаточно хорошо разработанный теоретически механизм ионизационно-перегревной неустойчивости [Bazelyan and Raizer, 1998, 2000], [Bazelyan et al., 2007], [Popov, 2009]). При ионизационно-перегревной неустойчивости диаметр канала, по которому протекает основная часть тока значительно уменьшается (до размеров 50-100 мкм) по сравнению с исходным диаметром стримера (около 1 мм), что позволяет *тем же самым током* нагреть воздух до необходимых высоких температур [Bazelyan and Raizer, 2000], [Bazelyan et al., 2007], [Popov, 2009]. Для начала развития неустойчивости необходимо, чтобы произошёл либо локальный нагрев воздуха в канале стримера на 10-20% [Raizer, 1992], либо локально увеличилось электрическое поле в районе прохождения стримера [Milikh et al., 2016]. Кандидатом на такой локальный нагрев может быть многократное прохождение стримеров по одному и тому же пути и/или статистическое усиление электрического поля благодаря неравномерности распределения стримеров по объёму стримерной вспышки [Milikh et al., 2016]. В физике длинной искры подобный процесс называют стримерно-лидерным переходом ([Les Renardières Group, 1977, 1981], [Gorin and Shkilyov, 1974, 1976]). Однако, в длинной искре к месту зарождения лидера на электроде (стему) уже изначально приложено необходимое для развития лидера высокое напряжение ([Les Renardières Group, 1977, 1981], [Gorin and Shkilyov, 1974, 1976]), наличие которого трудно предположить в девственном воздухе грозового облака. Поэтому, опираясь на эксперименты [Kostinskiy et al., 2015a, 2015b], мы будем называть этот физический процесс *стримерно-UPFs переходом*, понимая под UPFs один или несколько горячих плазменных каналов длиной ≈5-30 см, образованных после прохождения стримерной вспышки (или нескольких вспышек) в девственном воздухе. При этом, что важно, потенциал поляризации на положительном конце UPF ниже потенциала инициации положительного лидера в данном внешнем электрическом поле (это обстоятельство не дает нам возможность назвать этот процесс стримерно-лидерным переходом, так как после перехода плазмы стримеров в горячее состояние не может стартовать положительный лидер, так как для его инициации в данном электрическом поле длина горячих плазменных образований еще мала).

Есть ещё одна причина, почему мы выделяем UPFs в отдельное плазменное образование, а не говорим, например, об искре (длиной 1-10 см), хотя и UPFs и искра являются горячими высоко проводящими плазменными каналами (или сетями каналов). Важное различие состоит в том, что при классическом искровом разряде весь процесс образования короткой искры проходит в сверхпробойных полях $E_{th} \geq 3$ МВ/(м·атм), неважно, по механизму развития лавин Таунсенда в промежутках длиной несколько миллиметров, или по стримерному механизму в более длинных промежутках [Райзер, 1992]. UPFs же появляется в существенно подпороговых полях $E_{th} \ll 3$ МВ/(м·атм), где, практически всегда, конечным механизмом перехода и необходимого для выживания плазмы нагрева газа, является ионизационно-перегревная неустойчивость. Различие между искрой и UPFs видно и по пороговому электрическому полю. Если для короткой искры порог хорошо известен и понятен, это значение электрического поля $E_{th} \approx 3$ МВ/(м·атм), при котором частота ионизации начинает превосходить частоту прилипания электронов к молекулам кислорода, то для ионизационно-перегревной неустойчивости разумный диапазон электрических полей довольно широк $0.5 \geq E_{th} < 3$ МВ/(м·атм), так как развитие неустойчивости зависит не только и не сколько от величины электрического поля, сколько от масштаба возмущений среды (степени выхода среды из равновесия). Причиной могут быть нагрев, увеличение электрического поля, увеличение концентрации электронов, так как неустойчивость может развиваться, начиная с любого звена цепи: нагрев $\delta T\uparrow$ ведёт к снижению концентрации газа $\delta N\downarrow$, снижение концентрации газа ведёт к увеличению приведённого электрического поля $\delta(E/N)\uparrow$, увеличение приведённого электрического поля $(E/N)$ ведёт к резкому росту энергии электронов и частоты ионизации $\delta v_i\uparrow$, повышение частоты ионизации ведёт к увеличению числа электронов $\delta n_e\uparrow$, увеличение числа электронов ведёт к росту энерговклада в газ $\delta(\sigma E^2)\uparrow$, повышение энерговклада в газ ведёт снова к нагреву $\delta T\uparrow$ и процесс идёт по следующему кругу ([Raizer, 1991], [Bazelyan and Raizer, 1998, 2000], [Bazelyan et al., 2007], [Popov, 2009]). Заметим, что при данном сценарии развития неустойчивости она развивалась при постоянном внешнем, в нашем случае подпороговом, электрическом поле. Процесс может начаться и с любого другого элемента цепочки, например, со скачка всё ещё подпорогового электрического поля:

$$\delta E\uparrow \to \delta(E/N)\uparrow \to \delta v_i\uparrow \to \delta n_e\uparrow \to \delta(\sigma E^2)\uparrow \to \delta T\uparrow \to \delta N\downarrow \to \delta(E/N)\uparrow .$$

Ионизационно-перегревная неустойчивость является универсальным механизмом нагрева газа в подпороговых электрических полях, и она определяет, например, развитие СВЧ-разрядов в существенно подпороговых полях [Бродский и др., 1983], [Богатов и др., 1984], [Батанов и др., 1985], [Аветисов и др.,1990] и при работе газовых лазеров с накачкой несамостоятельными разрядами [Nighan, 1977], [Райзер, 1992, стр. 309-313].

### 7.6.3. Переход UPFs-положительный лидер

Появившиеся благодаря стримерной вспышке горячие высокопроводящие «необычные плазменные образования» (UPFs), мы не можем сразу же считать лидерами (тем более — двунаправленными лидерами), так как для появления положительного лидера из плазмы горячего канала UPF, который находится в данном среднем внешнем электрическом поле

E<450-500 kV/(m·atm), плазменному каналу нужно достичь некоторой минимальной (пороговой) длины, которая благодаря поляризации канала обеспечит необходимый для развития лидера потенциал на головке положительного конца UPF ([Bazelyan and Raizer, 1998, 2000], [Bazelyan et al., 2007]). Единичный UPF, чтобы «выжить», должен породить *устойчиво* развивающийся положительный лидер. Чтобы UPF «дожил» до момента старта лидера с его положительного конца, через UPF с момента возникновения должен постоянно течь ток, иначе электрическое поле будет вытеснено из плазмы. Такой сценарий могут обеспечить цепочки UPFs, настолько близко расположенные друг к другу, чтобы электрическое поле между ними везде превысило порог распространения положительных стримеров $E_{str+} \geq 450-500$ kV/(m·atm). В таком случае между UPFs возникнут вторичные положительные стримеры, концы каналов близлежащих UPFs будут находиться в квази-сквозной фазе взаимодействия и начнут двигаться навстречу друг к другу со скоростями ≈ 2-6 см/мкс до слияния нескольких UPFs в единый канал. Если длина этого, более длинного единого канала, превысит порог возникновения положительного лидера в данном электрическом поле [Bazelyan et al., 2007], то после этого с положительного конца одного из UPFs (или плазменного канала, в который превратятся несколько UPFs) будет инициирован положительный лидер, который сможет самостоятельно удлиняться в данном поле, усиливая поляризацию на противоположном отрицательном конце.

Другой вариант развития UPFs реализуется, если среднее внешнее электрическое поле вокруг UPFs превышает порог распространения положительных стримеров $E_{str+} \geq 450-500$ kV/(m·atm) и протяженность такого поля будет составлять десятки метров. Интересно, что в таком случае через *любое* (начиная, вероятно, с длины 1-5 см) горячее плазменное образование будет течь стримерный ток, поддерживаемый внешним электрическим полем, а, следовательно, плазменный канал будет удлиняться со скоростью 1-3 см/мкс, увеличивая потенциал на отрицательном конце UPFs ([Les Renardières Group, 1981], [Gorin and Shkilyov, 1976]). Данный вариант, по нашему мнению, будет похож на развитие, благодаря положительным стримерам (на обоих концах), горячего короткого вначале (≈1-15 см) *спейс-лидера* в короне отрицательного лидера длинной искры ([Stekolnikov and Shkilyov,1963], [Les Renardières Group, 1981], [Gorin and Shkilyov, 1976]).

Так как процесс развития UPFs статистический и зависит от пространственных размеров и напряженности электрического поля, то UPFs не всегда смогут дорасти до инициирования положительного лидера, а, следовательно, их плазма может распасться. Это очередной пороговый процесс на пути от стримерной вспышки к «классической» молнии и нам кажется, что ему разумно присвоить отдельное название — *переход UPFs-положительный лидер*, подразумевая под этим старт саморазвивающегося во внешнем электрическом поле положительного лидера с положительного конца UPF.

### 7.6.4. Переход положительный лидер — двунаправленный лидер

Когда потенциал на отрицательном конце UPF превысит в 1.5-2 раза потенциал на положительном конце UPF ([Горин и Шкилев, 1974, 1976], [Les Renardières Group, 1977, 1981]), чтобы электрическое поле $E_{str}$—$\geq$1000-1200 kV/(m·atm) смогло обеспечить развитие отрицательных стримеров на расстояние 20-100 см перед UPF, то возникнут отрицательные стримерная вспышки, а потом, скорее всего, стартует небольшой отрицательный лидер. После этого двунаправленный лидер имеет шансы на самостоятельное выживание и развитие.

Картина развития может значительно усложниться, если, как мы и предполагаем в Механизме, одновременно рождаются множество близко лежащих UPFs, связанных между собой положительными стримерами. Если поляризованная трехмерная плазменная сеть UPFs имеет достаточно большие продольные и поперечные размеры, то она сможет формировать перед собой медленно падающее электрическое поле, что повысит длину распространения положительных стримеров. Насколько нам известно, ни теоретически, ни экспериментально такая конфигурация плазмы до сих пор подробно не рассматривалась, поэтому мы не можем детально предсказать эволюцию такой сети UPFs. Однако можно предположить, что по мере слияния UPFs и образования высоко проводящих каналов внутри сети, высоко проводящие каналы сконцентрируют основной ток и именно их поляризационная длина будет определять порог старта положительного лидера из этой сети UPFs. После старта и достаточного развития положительного лидера, необходимый потенциал сформируется на отрицательном конце сети UPFs и там через некоторое время стартует отрицательный лидер.

Подчеркнем, что, согласно представлениям, которые сложились в физике длинной искры ([Горин и Шкилев, 1974, 1976], [Les Renardières Group, 1977, 1981], [Castellani et al., 1998a, 1998b]), а также во время изучения высотно инициированной триггерной молнии (Altitude triggering lightning — ATL, [Rakov and Uman, 2003, стр.269, Figure 7.4.]) первый отрицательный лидер внутри объема, где инициируется молния, может появиться в течение первого десятка микросекунд с момента первого инициирующего молнию события (IE), а не через несколько миллисекунд, как часто по умолчанию предполагается при исследовании молнии с помощью LMA и/или интерференционных систем (например, [Rison et al., 2016]), где отрицательным лидером считается только большой ступенчатый лидер молнии со ступенями не менее 5-10 метров и токами около 1 кА. Однако, на наш взгляд, чрезвычайно трудно предполагать, что такие большие ступени и такой большой ток могут быть в первые микросекунды в момент инициирования молнии. С другой стороны, не видно физических оснований, почему небольшой отрицательный лидер со ступенями меньше метра и токами меньше 100 А не должен возникнуть при достижении потенциала на отрицательных концах плазменных образований порога инициации отрицательного лидера, подобно тому, как это происходит в физике длинной искры или высотно-инициированной триггерной молнии (altitude triggering lightning).

### 7.6.5. IBP-стадия, как переход от цепочек UPFs и небольших двунаправленных лидеров к большим объемным взаимодействующим плазменным сетям

Так как, в соответствие с предложенным нами Механизмом (раздел 7.5), при инициации молнии, благодаря вторичным электронам ШАЛ, которые экспоненциально размножаются по механизму убегающих электронов (EAS-RREA-Механизм, раздел 7.5.1.3, глава 8) инициация стримерных вспышек имеет принципиально объемный характер и приводит к множеству плазменных образований внутри объема 0.1-1 км$^3$, то во всем этом объеме могут возникать различные плазменные образования, которые могут объединятся в цепочки и/или сети. Цепочки UPFs и небольшие положительные и двунаправленные лидеры, могут, сливаясь и объединяясь, переходить в другую форму плазменных образований, которая может включать в себя большие плазменные сети горячих высокопроводящих каналов. Объединение плазменных систем, выросших из сетей UPFs, возможно, приведёт к серии их слияний, которые могут выглядеть на осциллограммах быстрых антенн (FA) как IPBs. IBPs являются, согласно экспериментальным данным ([Stolzenburg et al., 2013, 2014], [Campos and Saba, 2013], [Wilkes et al., 2016]) разрядами высокопроводящих плазменных каналов, которые скорее всего выросли из этих сетей (см. Рисунок 7.10). При этом, каналы и сети, развиваясь, могут значительно увеличивать общую длину, потенциал на концах плазменной системы, скорость движения, общий заряд, но принципиально плазма остаётся той же самой горячей и высокопроводящей. Несмотря на это, такое увеличение параметров каналов, которые выходят из плазменных сетей, может привести к тому, что отрицательный лидер станет отчётливо заметен на больших расстояниях, как в видимом диапазоне, так и в VHF (анализ этих процессов выходит за рамки данной диссертации).

## 7.7. Выводы главы 7

В этой главе мы описали качественную модель физических процессов возникновения молнии от инициирующего разряд молнии события (IE) до нескольких первых IBPs (раздел 7.5). Наш Механизм предполагает, что инициирование молнии развивается следующим образом:

7.7.1. Инициирующее событие (IE) запускает процесс преобразования непроводящего воздуха в проводник. IE может быть либо слабым, либо сильным, называемым здесь Weak-IE или NBE-IE, соответственно. В обоих типах IEs, засеянные широким атмосферным ливнем космических лучей (ШАЛ), релятивистские убегающие электроны и позитроны в сильном электрическом поле, инициируют классические электронные лавины во многих небольших объемах в грозовом облаке, где электрическое поле E > 3 МВ/(м·атм). Трехмерное (3D) множество электронных лавин вызывает 3D-множество, почти одновременных (синхронизированных ШАЛ) обычных положительных стримерных вспышек, которые сильно излучают в VHF-радиодиапазоне. Таким образом, Механизм производит IE и его характерный VHF-импульс.

7.7.2. Начальное изменение электрического поля (Initial E-Change — IEC) следует за IE во всех (состоявшихся, развитых) IC и CG-молниях и, по нашему Механизму, включает возникновение необычных плазменных образований (UPFs). UPFs возникают и развиваются внутри траекторий положительных стримерных вспышек в результате процесса ионизационно-перегревной неустойчивости; затем UPFs сливаются вместе в результате контактов между собой благодаря вторичным положительным стримерам, образуя цепочки или небольшие сети UPFs. Электрические токи в сетях UPFs вызывают относительно медленное изменение электрического поля во время протекания IEC-стадии. Пары цепочек UPFs также сливаются в более длинные и сложные цепи сети. Затем цепи образуют трехмерную сеть каналов горячей плазмы, которая в конечном итоге создает и поддерживает внутри себя развитие двунаправленного лидера. Различные слияния UPFs и двунаправленных лидеров вызывают слабые VHF-импульсы и небольшие импульсы на быстрых антеннах (FA), наблюдаемые во время прохождения IEC-стадии развития молнии.

7.7.3. Первый классический импульс начального пробоя (IBP) завершает IEC и запускает IB-стадию развития молнии. Чтобы создать первый классический IBP, две из трехмерных сетей UPFs (созданные во время прохождения IEC-стадии) должны создать двунаправленные лидеры, выходящие из сетей, а затем сети объединяются, когда их лидеры контактируют друг с другом во время квази-сквозной фазы, а потом квазиобратного удара, которые и формируют первый IBP (начальный импульс пробоя). Каждый последующий классический IBP вызывается слиянием новой трехмерной сети UPFs и/или плазменных каналов с сетью (цепочкой) ранее объединенных сетей UPFs (или плазменных каналов), которые вызвали предыдущие IBPs. Каждое из этих слияний имеет общую стримерную зону, квази-сквозную фазу и квази-обратный удар, который вызывает яркую вспышку света (на видеокадрах), совпадающую с классическим IBP (в данных изменения электрического поля быстрой антенны FA) и импульсом высокой мощности (в данных в VHF-данных). После серии классических IBPs молния переходит в хорошо известную фазу большого отрицательного ступенчатого лидера.

Как описано выше, Механизм согласуется с опубликованными данными инициирования молнии для сильных и слабых NBE-IE событий [Rison et al., 2016], [Lyu et al., 2019], [Bandara et al., 2019] и для вспышек Weak-IE [Marshall et al., 2019], которые не начинаются с классического NBE. Механизм также может быть разумно расширен для объяснения небольших событий типа разрядов-прекурсоров и изолированных NBE, которые не являются инициирующими событиями (IE) молнии, в случаях, когда условия для IEC и/или IBP не создаются. Несмотря на качественный характер предлагаемого механизма, он кажется нам достаточно конкретным, чтобы проверить его основные положения в будущих экспериментах.

В заключение мы рассмотрим несколько важных выводов предлагаемого Механизма. Во-первых, Механизм предполагает, что области, имеющие достаточную величину электрического поля E для инициации обычных положительных стримерных вспышек, возникают из-за мелкомасштабных гидродинамических неустойчивостей и статистических изменений электрического поля, а не из-за взаимодействия электрического поля с только с одними гидрометеорами. Однако, если будет экспериментально доказано, что индивидуальные гидрометеоры или их столкновения (или суперпозиция полей зарядов гидрометеоров, их поляризации и коллективного статистического и гидродинамического усиления поля) способны создавать облачные объемы с полями $E_{th} \geq 3$ МВ/(м·атм) достаточной протяженности для выполнения критерия Мика (воздушные электроды), чтобы инициировать стримерные вспышки, то это обстоятельство существенно не изменит остальной Механизм. Все остальные компоненты Механизма останутся такими же, независимо от физической причины создания воздушных электродов в грозовых облаках.

Во-вторых, в нашем Механизме положительные стримерные вспышки играют ключевую роль в создании инициирующее молнию первое событие (IE), будь то слабый IE или сильный NBE-IE. Хотя в Механизме предполагается, что каждая стримерная вспышка движется с разумной скоростью $< 5 \cdot 10^6$ м/с, видимое на радиоантеннах движение во время IE может быть намного больше, $3\text{-}10 \cdot 10^7$ м/с. Это фундаментальное различие между фактическим физическим движением и кажущейся скоростью состоит в том, что вспышки стримеров инициируются вдоль траектории движения группы вторичных релятивистских заряженных частиц, которые движутся со скоростью, близкой к скорости света. Напротив, [Rison et al., 2016] и [Tilles et al., 2019] постулировали механизмы создания и развития NBEs на основе предложенной ими теории быстрого положительного пробоя (FPB), движущегося вниз, и быстрого отрицательного пробоя (FNB), движущегося вверх со скоростью 4-10·$10^7$ м/с. Они предполагают, что стримерные вспышки могут двигаться с такими большими скоростями на высотах инициации молнии. Нам такие скорости стримеров не кажутся разумными, и, насколько нам известно, нет никаких экспериментальных доказательств столь высоких скоростей стримеров в воздухе при давлениях 0,3–1 атм (см. подробное обсуждение механизма FPB во Введении ).

Механизм предлагает несколько отличающиеся физические процессы для «классических» IBPs с большой амплитудой и большой продолжительностью (например, [Weidman and Krider, 1979]) и более слабых, более коротких IBPs (например, [Nag et al., 2009]). Более слабые и короткие IBPs возникают при слиянии цепочек UPFs или небольших сетей UPFs или сетей уже сформировавшихся плазменных каналов. Классические IBPs возникают в результате слияния двух больших плазменных сетей, каждая из которых произошла от большой сети UPFs. Этот результат может помочь объяснить загадку широкого разброса активности, наблюдаемого на IB-стадии.

Наконец, важной особенностью предложенного Механизма является фундаментальная трехмерность предлагаемых физических процессов создания молнии, включая инициирующее событие (IE), в противовес механизму создания молнии из одного линейного двунаправленного лидера Каземира. Этот подход позволяет объяснить чрезвычайно короткие времена IEC, за которыми очень быстро следуют мощные IBPs. Кроме того, из-за гипотезы о мелкомасштабных и среднемасштабных трехмерных вариациях электрического поля (E), Механизм также непротиворечиво объясняет различное развитие инициирующих событий (IE) в различных молниях, включая широкий диапазон длительностей и амплитуд IEC (например, [Marshall, Schulz, et al., 2014b]), широкий диапазон продолжительности IBPs, времени между IBPs, амплитуды IBPs, количества субимпульсов в классических IBPs (например, [Marshall et al., 2013], [Stolzenburg et al., 2013], [Stolzenburg et al., 2014], [Bandara et al., 2019], и кажущийся случайным порядок амплитуд классических IBPs (например, [Smith et al., 2018]. Эти различия поведения реальных разрядов молнии гораздо труднее понять, если инициирование происходит в единственной гладкой области большого электрического поля E, а результатом инициирования является большой двунаправленный лидер Каземира.

# ГЛАВА 8. Оценка динамики инициирования стримерных вспышек, обеспечивающих пространственно-временной профиль и скорость распространения фазовой волны максимальных VHF-сигналов при развитии КВР (CID/NBE)

В этом разделе с помощью численных методов мы более подробно оцениваем, как широкий атмосферный ливень космических лучей (ШАЛ) может инициировать и синхронизовать почти одновременно (во временном интервале ~1-3 мкс) старт большого числа стримерных вспышек, которые в свою очередь могут обеспечить мощный VHF-сигнал, обычно ассоциируемый с КВР (CID/NBE — далее в этой главе для определенности мы будем пользоваться одним сокращением NBE).

Стримерные вспышки, согласно Механизму изложенному в главе 7 ([Kostinskiy et al., 2020a]), возникают благодаря объемной сети $E_{th}$-областей с электрическим полем выше 30 кВ/(см·атм) («воздушных электродов»), число которых динамически поддерживается турбулентными областями грозового облака, имеющими внутри себя достаточное число сильно заряженных гидрометеоров. Момент синхронизованного старта (в объеме 0.1-1 км$^3$ в рамках промежутка 1-3 мкс) множества стримерных вспышек определяется прохождением через эту область широкого атмосферного ливня космических лучей (ШАЛ), экспоненциально усиленного в электрическом поле грозового облака (разделы 7.5.1.2, 7.5.1.3). В этом разделе приводятся первые численные оценки этого процесса, но в ближайшее время мы планируем на основе данных оценок написать отдельную статью, где основные моменты мы представим более обоснованно и подробно.

Если воздушные электроды с электрическим полем $E \geq 3\,MV\,m^{-1}atm^{-1}$ (в объеме диаметром порядка сантиметров) возникают благодаря сильной турбулентности, статистическому движению сильно заряженных гидрометеоров и/или из-за усиления электрического поля на гидрометеорах, то любая пролетевшая через сечение воздушного электрода энергичная заряженная частица или поглощенный в объеме воздушного электрода фотон должны приводить к обычным плазменным (разрядным) электронным лавинам, которые будут приводить к уходу заряда и снижать электрическое поле внутри воздушного электрода. Мы оценивали (раздел 7.5.1.2), что для мощного NBE нужно накопить одновременно около $10^6$-$10^7$ km$^{-3}$ таких электродов (для слабого $10^3$-$10^4$), чтобы вторичные электроны ШАЛ с энергией первичной частицы $\varepsilon_0 > 10^{15}$ эВ, число которых экспоненциально увеличивается в электрическом поле облака, синхронизовали старт фазовой волны «поджига» обычных стримерных вспышек (раздел 7.5.1.3). При этом всегда существует фоновый уровень облучения атмосферы космическими лучами с гораздо меньшими энергиями $10^4$-$10^{11}$ эВ (например, [Sato, 2015]), которые способны поставлять первые электроны для создания электронных лавин, разряжающих воздушные электроды до прихода ШАЛ.

## 8.1. Расчёт динамики возникновения и гибели «воздушных электродов» в зависимости от высоты над уровнем моря

В настоящее время имеющиеся экспериментальные данные о разрядных микропроцессах внутри грозового облака не позволяют однозначно определить механизм образования небольших областей, с электрическим полем выше пробойного (воздушных электродов), которые обеспечат условия инициации положительных стримеров. Это могут быть большие (от нескольких миллиметров до нескольких сантиметров в длину) поляризованные и сильно заряженные гидрометеоры, как предполагают многие исследователи (например, работы последних лет, [Solomon et al., 2001], [Dubinova et al., 2015], [Sadighi et al., 2015], [Babich et al., 2016, 2017], [Cai et al. 2017]), а могут быть и области высокого электрического поля, которые возникли из-за статистического и гидродинамического движения заряженных частиц [Iudin, 2017], [Babich et al., 2016]. Оба этих подхода встречают значительные трудности. Если воздушный электрод является сильно заряженным и поляризованным гидрометеором, то для инициации стримера даже на высотах 4-6 км требуются очень большие электрические поля, заряды и почти нереальные в природе размеры гидрометеоров ([Solomon et al., 2001], [Dubinova et al., 2015], [Liu et al. 2012], [Богатов, 2013], [Babich et al., 2016]), которые крайне редко наблюдались в грозовом облаке, а с ростом высоты эти проблемы растут пропорционально экспоненциальному падению давления (см. раздел 7.4.3.1, и ниже по тексту), что обычно, не отмечается в приведенных выше работах, чем игнорируется факт, что молнии и CID/NBE успешно инициируются на высотах 12-18 км. Если опираться только на механизм усиления поля из-за статистического и гидродинамического движения заряженных частиц (фактически локального усиления поля из-за сильного сгущения в пространстве заряженных частиц), то для создания воздушных электродов (областей пробойного электрического поля), потребуется на 1-2 порядка больше заряженных частиц в единице объема грозового облака (если опираться на оценки [Trahtenhertz and Iudin, 2005], [Iudin, 2017]), чем удавалось измерить в эксперименте ([Marshall and Winn, 1982], [Weinheimer et al., 1991], [Marshall and Marsh, 1993], [Marsh and Marshall, 1993], [Stolzenburg and Marshall 1998], [Bateman et al., 1999]). Поэтому, скорее всего, воздушные электроды возникают благодаря суммарному электрическому полю, возникающему во время сильной зарядки и поляризации гидрометеоров размером 0.1-2 мм [Solomon et al., 2001] и благодаря механизму усиления электрического поля в процессе статистического и гидродинамического движения ансамбля заряженных частиц [Trahtenhertz and Iudin, 2005], [Iudin, 2017], [Babich et al., 2016]. В наших оценках мы будем оценивать инициацию стримеров на высотах 5.5-16 км, чтобы в результате простых оценок обратить внимание на трудности инициирования стримеров по одному избранному механизму.

Для самых первых оценок параметров воздушных электродов возьмем две наиболее простые формы электрического поля, чтобы определить порядки величин и принципиальную возможность создания воздушных электродов. Это — (1) постоянное электрическое поле и (2) шарообразная область с постоянной плотностью заряда.

Для детального расчета потока фоновых космических лучей, пролетающих через сечение воздушного электрод и инициирующих плазменные процессы, использовалась программа EXPACS (EXcel-based Program for calculating Atmospheric Cosmic-ray Spectrum), [Sato, 2015], которая с заведомо высокой точностью для наших оценок позволяет оценить поток космических лучей на высоте 0-62 км в любой точке Земли, для определенного дня, месяца и года.

### 8.1.1 Оценка зарядов и электрических полей модельных заряженных гидрометеоров, которые могут инициировать стримеры и коронный разряд

Мы можем использовать экспериментальные данные по инициации стримеров и коронного разряда в лабораторных электрических разрядах, так как время растекания заряда по частицам льда является достаточно малым по сравнению с процессами изменения среднемасштабных электрических полей в грозовом облаке. «Благодаря достаточно большой электропроводности льда электрические процессы, например распределение свободных зарядов по поверхности гидрометеоров, протекают на ледяных частицах почти так же, как и на жидких каплях. Для оценки времени релаксации, необходимого для протекания процесса, можно использовать известную формулу $\tau_M \approx \frac{\varepsilon}{\sigma}$. Для льда при T = —10° C, $\varepsilon \approx 2.4 \cdot 10^{-10} F/m$ (для воды 3 раза больше $7.2 \cdot 10^{-10} F/m$) и $\sigma \approx 1.1 \cdot 10^{-7} S/m$, получаем $\tau_M = 2 \cdot 10^{-3}$ с» [Мучник, 1974, стр.168]. Мы, как и подавляющее большинство исследователей, проводивших подобные оценки, не учитывали зависимость $\varepsilon$ льда от частоты, которая может существенно повлиять на момент инициации стримера с поверхности гидрометеора, на что обратили внимание [Dubinova et al., 2015]. При следующей итерации подобных оценок, этот важный фактор необходимо учесть.

Мы будем считать гидрометеоры «сферами» (электрическое поле уменьшается по закону $E = E_0 \left(\frac{r_0}{r}\right)^2$), $r_0$ – радиус границы электрода. Лавины электронов могут развиваться, если электрическое поле $E_0 > E_{th} \approx 3.1 \frac{MV}{m\,atm}$ (пробойным полем мы будем называть значение электрического поля $E_{th}$, при котором частота ионизации больше, чем частота прилипания $\nu_i \geq \nu_a$). Если $E_0 > E_{th}$, то лавины электронов могут развиваться на длине $r_{th}$-$r_0$. $r_{th}^{sph} = r_0 \cdot \left(\frac{E_0}{31\,kV/(cm\,atm)}\right)^{0.5}$, где $r_{th}$ — радиус, на котором частота ионизации $\nu_i$ становится меньше, чем частота прилипания $\nu_a$.

Иногда считается, что разряд с поверхности гидрометеора возникает, если электрическое поле на поверхности гидрометеора превышает $E_{th} \approx 3.1 \frac{MV}{m\,atm}$ или даже $E_{th} \approx 2.4 \frac{MV}{m\,atm}$. Для примера покажем с помощью простой оценки, что это не так. Рассмотрим случай высоты 5.5 км (давление 0.5 атм), заряда Q = 50 пКл на гидрометеоре с диаметром 1.0 mm ($r_0$ = 0.05 см), электрическое поле на поверхности гидрометеора казалось бы выше пробойного $E_Q$ = 1.8 MV/m ($3.6 \frac{MV}{m\,atm}$) > $E_{th}$.

Для старта стримера число электронов лавины в объеме диаметром меньше, чем 1 мм, должно быть n=$10^8$-$10^9$. Число электронов на стадии развития лавины (до момента старта стримера) увеличивается в пространстве по закону

$$n = n_0 exp \left( \int_{r_0}^{r_{th}} \alpha_{eff} \left( \frac{E(r)}{p} \right) dr \right),$$

где $\alpha_{eff} = \alpha - \eta$ является эффективным коэффициентом размножения электронов ($\alpha$ – длина размножения электронов [cm$^{-1}$], $\eta$ – длина прилипания электронов [cm$^{-1}$]. $\alpha_{eff}$, $\alpha$, $\eta$ сильно зависят от $\frac{E}{p}$ или от $\frac{E}{N}$ [Райзер, 1992].

Коэффициенты пересчета приведённых электрических полей будут равны:

$$\frac{E}{p} \left[ \frac{V}{cm\,Torr} \right] = 1.3158 \, \frac{E}{p} \left[ \frac{kV}{cm\,atm} \right]$$

$$\frac{E}{p} \left[ \frac{V}{cm\,Torr} \right] = 3.30 \cdot 10^{16} \frac{E}{N} [V \cdot cm^2] = 0.33 \, \frac{E}{N} [Td]; \; 1\,Td = 10^{-17} [V \cdot cm^2]$$

$$\frac{E}{N} [V \cdot cm^2] = 3.03 \cdot 10^{-17} \frac{E}{p} \left[ \frac{V}{cm\,Torr} \right]$$

$$\frac{E}{p} \left[ \frac{kV}{cm\,atm} \right] = 2.508 \cdot 10^{16} \frac{E}{N} [V \cdot cm^2]$$

Критерий Мика (Meek's criterion) $n_{Meek}$ [Райзер 1992, стр. 424] для изменяющегося электрического поля в случае лавины, начинающейся с одного-двух электронов, будет иметь вид (p – давление в атм, $p_0$ = 1 атм):

$$n_{Meek} = ln \left( \frac{N}{N_0} \right) = 18 - 20 = \int_{r_0}^{r_{th}} \frac{p}{p_0} \alpha_{eff} \left( \frac{E}{p} \right) dr \qquad (8.1)$$

Мы вычислим интеграл (2.3.2.1), с помощью известной эмпирической формулой для $\alpha_{eff} \left( \frac{E}{p} \right)$ [Les Renardieres Group, 1972, стр.54-55]:

$$\frac{\alpha_{eff}}{p} = 106.4 \cdot \left( \left( \frac{E_0}{E_c} \right)^2 - 1 \right), \text{for 31-90 } kV\,cm^{-1}atm^{-1} \; (3.1\text{-}9 \; MV \; m^{-1}atm^{-1}) \quad (8.2)$$

Для сферы это выражение можно переписать в виде:

$$\alpha_{eff}/p = 106.4 \cdot \left( \left( \frac{E_0}{\left( \frac{p}{p_0} \right) E_c} \left( \frac{r_0}{r} \right)^2 \right)^2 - 1 \right); \; \alpha_{eff} [cm^{-1}]; \; E \left[ \frac{kV}{cm} \right]; \; p, p_0 \, [atm]; \; E_c = 31 \left[ \frac{kV}{cm} \right]$$

Для высоты 5.5 км $\frac{p}{p_0} = 0.5$. Число шагов лавины $n_{av}$ (в этом случае для сферы) мы запишем в виде:

$$n_{av} = ln\left(\frac{N}{N_0}\right) = \int_{r_0}^{r_{th}} \alpha_{eff}\left(\frac{E}{p}\right) dr = \int_{r_0}^{r_{th}} \left(\frac{p}{p_0}\right) 106.4 \cdot \left(\left(\frac{E_0}{\left(\frac{p}{p_0}\right)E_c}\left(\frac{r_0}{r}\right)^2\right)^2 - 1\right) dr \quad (8.3)$$

Для контроля правильности полученного значения $n_{av}$, будем использовать также другие, более современные аппроксимационные формулы [Zaengl and Petcharaks, 1994], [Naidis, 2005]

$$\frac{\alpha_{eff}}{\delta} = 0.16053 \cdot \left(\frac{E}{\delta} - 21.65\right)^2 - 2.873 ; \frac{\alpha_{eff}}{\delta} [cm^{-1}]; \quad\quad (8.4)$$

$$E\left[\frac{kV}{cm}\right], \delta = \frac{P}{P_0} = \frac{N}{N_0}, \quad P_0 = 1\,atm \quad \frac{E}{\delta} < 79.4\frac{kV}{cm},$$

$$\frac{\alpha_{eff}}{\delta} = 16.7766 \cdot \frac{E}{\delta} - 800.06 ; \frac{\alpha_{eff}}{\delta} [cm^{-1}]; \quad\quad (8.5)$$

$$E\left[\frac{V}{cm}\right], \delta = \frac{P}{P_0} = \frac{N}{N_0}, \; P_0 = 1\,atm \quad \frac{E}{\delta} \geq 79.4\frac{kV}{cm},$$

В результате вычислений число шагов лавины оказалось в диапазоне $n_{av} \approx 0.034\text{-}0.038$, что значительно меньше, чем даже 1 шаг лавины, то есть, электроны в поле гидрометеора размером 1 мм с довольно большим зарядом 50 пКл не смогут произвести даже одну ионизацию (а стримеру нужно 18-20 шагов лавины). Это происходит потому, что расстояние от поверхности гидрометеора, на которой поле составляет 3.6 MV/(m atm) до значения поля 3.1 MV/(m atm), когда еще существует заметная ионизация, равно всего 40 мкм. Такой гидрометеор даже не сможет зажечь самый слабый коронный разряд, так как для этого нужно 7-9 шагов лавины электронов (см. ниже).

Высота 5.5 км. Оценим электрическое поле и заряд, который необходим для инициации стримера с гидрометеора диаметром 1 мм при давлении 0.5 атм. Нужные $n_{av} \approx 18\text{-}20$ шагов лавины (в оценке получилось $n_{av} \approx 19.7$); может обеспечить приведенное поле 240 $\frac{V}{cm\,Torr}$ на поверхности сферического гидрометеора, которое эквивалентно $E_Q \approx 18.2$ MV/(m atm), а абсолютное значение электрического поля 9.1 MV/m (высота 5.5 км). Это электрическое поле может обеспечить очень большой заряд $Q \approx 253$ пКл. Таким образом, при диаметре гидрометеора 1 мм, на высоте 5.5 км, заряд > 253 пКл может обеспечить старт стримера, а заряды 50 пКл и даже заряд 200 пКл не могут. Чтобы вычислить критерий Мика при таких больших полях, как $\sim 240$ $\frac{V}{cm\,Torr}$ мы использовали еще одну хорошо известную аппроксимационную формулу (2.3.2.6) для $\alpha_{eff}\left(\frac{E}{p}\right)$ ([Браун, 1961], [Brown, 1966]) которая позволяет вычислить $n_{av}$ при электрических полях от 18.2-9 MV/(m atm). При электрических полях 9-3.1 MV/(m atm) мы будем использовать формулу (8.3):

$$\frac{\alpha_{eff}}{p} = 15 \, exp\left(-\frac{365}{\frac{E}{p}}\right); \qquad\qquad (8.6)$$

$$\alpha_{eff}[cm^{-1}]; E\left[\frac{V}{cm}\right], p = Torr, \qquad \frac{E}{p} = (100-800)\frac{V}{cm\,Torr}, (7.6-60.8)\frac{MV}{m\,atm}$$

$$n_{av} = \int_{r_0}^{r_{th}} 760 \left(\frac{p}{p_0}\right) 15 \, exp\left(-\frac{365}{\frac{E_0}{p}\left(\frac{r_0}{r}\right)^2}\right) dr \qquad\qquad (8.7)$$

Контроль расчета по более современным формулам (8.4), (8.5) дает близкий результат $n_{av}$=15.63+3.46=19.1.

Высота 7 км. Стример будет инициирован с гидрометеора ($n_{av}$=16.54+5.04=21.6) диаметром 1 мм на высоте 7 км, согласно формулам (8.2), (8.6), при приведенном электрическом поле на поверхности сферического гидрометеора $\approx 280 \, \frac{V}{cm\,Torr}$ (21.2 MV/(m atm)), абсолютное значение электрического поля будет равно 8.5 MV/m. Контроль расчета по формулам (8.4), (8.5) приводит к близкому значению: $n_{av}$=17.45+2.95=20.4. Это электрическое поле может обеспечить также большой заряд $Q \approx 236$ нКл.

Высота 13 км. Стример может быть инициирован ($n_{av}$=17.457+2.616=20.0) с гидрометеора диаметром 1 мм на высоте 13 км очень большим приведенным электрическим полем на поверхности сферического гидрометеора $450 \, \frac{V}{cm\,Torr}$ (34.2 MV/(m atm)), абсолютное значение электрического поля будет равно 5.57 MV/m. Контроль расчета по формулам (8.4), (8.5) приводит к близкому значению: $n_{av}$=17.33+1.56=18.89. Это электрическое поле обеспечивает заряд $Q \approx 155$ pC.

Высота 16 км. Стример может быть инициирован ($n_{av}$=18.623+1.92=20.54) с гидрометеора диаметром 1 мм на высоте 16 км при огромном приведенном электрическом поле на поверхности сферического гидрометеора $650 \, \frac{V}{cm\,Torr}$ (49.4 MV/(m atm)), абсолютное значение электрического поля 4.93 MV/m. Контроль расчета по формулам (8.4), (8.5) приводит к близкому значению: $n_{av}=18.85+1.14=19.99$. Это электрическое поле на поверхности электрода может обеспечить заряд $Q \approx 137\ pC$.

Отметим, что приведенные электрические поля $49.4 - 34.2$ MV/(m atm), которые нужны гидрометеорам на высоте 13-16 км в десять раз превосходят пробойное поле. Эти электрические поля большинство электронов разряда делают убегающими, и электроны могут достигать на этих расстояниях энергий в сотни эВ). Фактически электроны создают маленький пучок в области действия такого поля. Такие высокие значения электрических полей возникают из-за очень маленьких размеров гидрометеоров, которые требуют для выполнения критерия Мика очень больших электрических полей, подобно тому, как очень большие электрические поля необходимы для инициации стримеров с тонких проводов и иголок. Эта ситуация давно известна в электрофизике инициации коронных разрядов и зафиксирована в критериях зажигания короны Пика для тонкого провода или подобных критериях для других форм тонких коронирующих электродов [Райзер, 2009, стр.626-627].

### 8.1.2. Оценка возможности инициации коронного и стримерного разряда с поверхности заряженных гидрометеоров малых размеров

Физика развития импульсного коронного разряда (до появления стримера) определяется теми же элементарными процессами, что и рождение стримеров, только число свободных электронов лавины (и ступеней ионизации электронной лавины не достигает критерия Мика) и коронный разряд не переходит в стримерный.

Критерий зажигания коронного разряда в воздухе от боковой поверхности провода предложил еще Пик в 1929 г. [Райзер, 1992, стр.435], [Райзер, 2009, стр.626] (в случае гидрометеоров он применим для оценок короны с боковых поверхностей ледяных игл):

$$E_c = 31 \cdot \delta \left(1 + \frac{0.308}{\sqrt{\delta \cdot r_0}}\right), \left[\frac{kV}{cm}\right], \delta = \frac{p}{p_{0=1\ atm}} = 0.1 - 10, r_0 = 0.01 - 1\ cm \qquad (8.8)$$

$E_c$ – абсолютное значение электрического поля в $\left[\frac{kV}{cm}\right]$, которое на графиках приводится по оси $Y$, радиус электрода по оси $X$ в [см]. Графики (ниже) хорошо показывают, как быстро растет критическое поле зажигания короны при уменьшении радиуса провода при атмосферном давлении (1, 2), на высоте 5.5 км (3) и 16 км (4). Из формулы также видно, что даже при радиусе $r_0 = 1\ cm$ ( $p_0 = 1\ atm$) (максимальное значение) $E_c = 40.5\ \frac{kV}{cm}$.

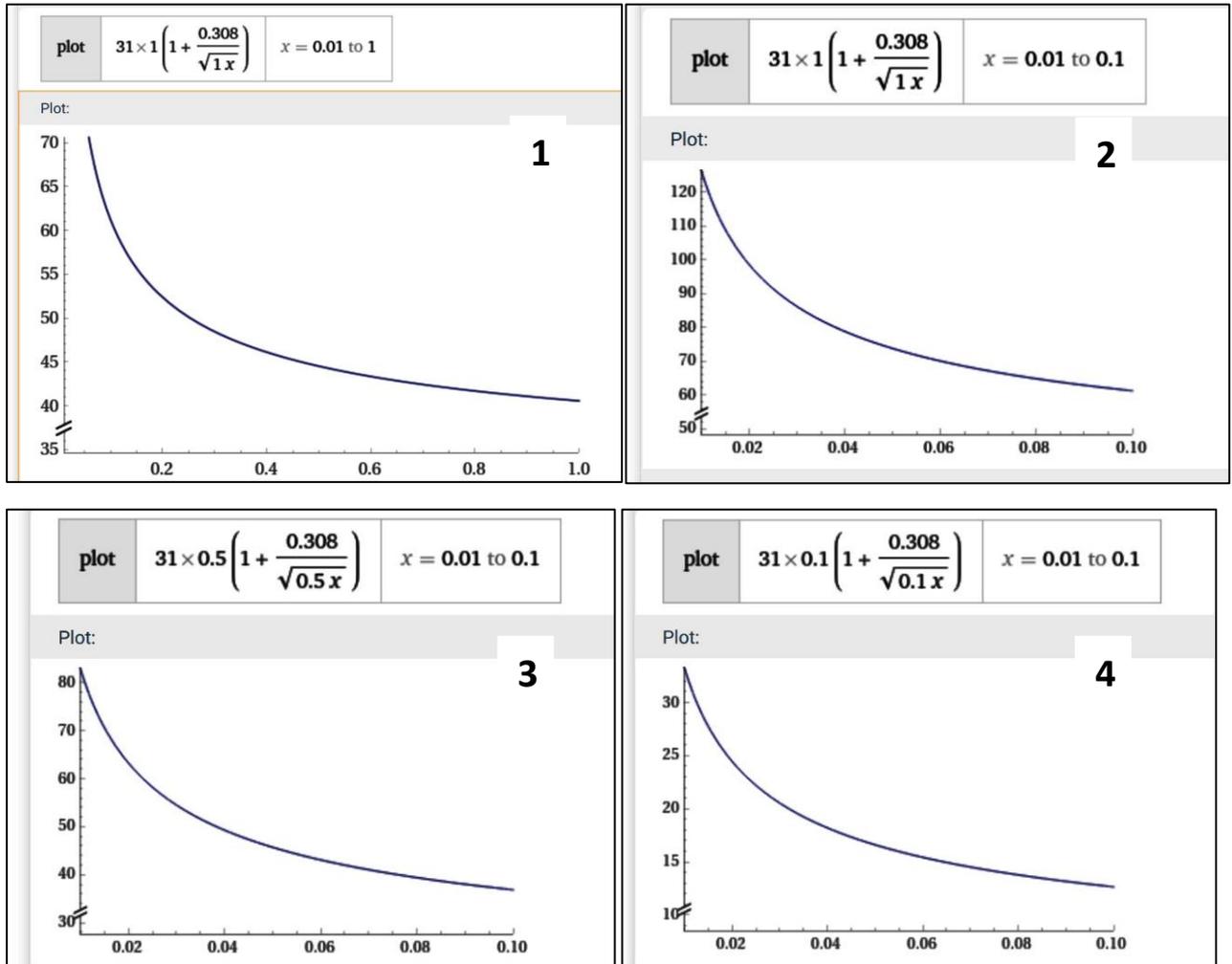

Рисунок 8.1. Расчет абсолютного значения электрического поля $E_c \left[\frac{kV}{cm}\right]$ инициации стримерной короны с боковой поверхности провода по эмпирической формуле Пика (8.8), которое на графиках приводится по оси $Y$, радиус электрода по оси $X$ в [см]. 1, 2 — атмосферном давление; 3 — высота 5.5 км (0.5 атм); 4 — высота 16 км (0.1 атм ).

Для сферического коронирующего электрода используется эмпирическая формула, аналогичная формуле Пика [Райзер, 2009, стр. 627], но с другими коэффициентами

$$E_c = 27.8 \cdot \delta \left(1 + \frac{0.54}{\sqrt{\delta \cdot r_0}}\right),\ \left[\frac{kV}{cm}\right], \delta = \frac{p}{p_0} = 0.1 - 10, \cdot r_0 = 0.01 - 1\ cm, \quad (8.9)$$

$E_c$ – абсолютное значение электрического поля в $\left[\frac{kV}{cm}\right]$, которое на графиках приводится по оси Y, радиус электрода по оси X в см. Ее применяют в случае острия или стержня с закругленным концом, если под $r_0$ понимать радиус закругления. Напряжение зажигания при этом будет примерно $V_c \approx E_c r_0$, если расстояние до противоположного электрода любой формы будет гораздо больше $r_0$. График показывает, что критическое поле зажигания короны при уменьшении радиуса шара растет еще быстрее, чем у провода (так как электрическое поле падает быстрее). Из формулы также видно, что даже при радиусе $r_0 = 1\ cm$ ( $p_0 = 1\ atm$) (максимальное значение) $E_c = 42.8\ \frac{kV}{cm}$.

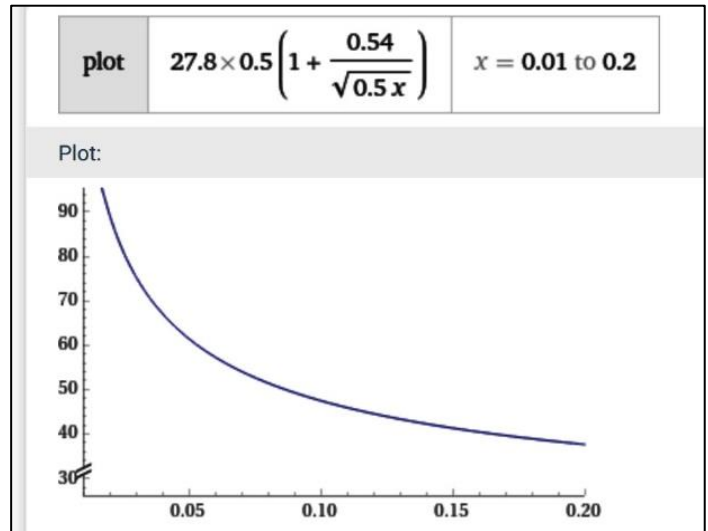

Рисунок 8.2. Расчет абсолютного значения электрического поля $E_c\ \left[\frac{kV}{cm}\right]$ инициации стримерной короны с тонкого сферического электрода по эмпирической формуле (2.3.2.9), которое на графиках приводится по оси $Y$, радиус электрода по оси $X$ в [см]. Атмосферном давление.

    Высота 5.5 км.  Для d = 1 мм ($r_0$ = 0.05 cm), p = 0.5 атм, критическое электрическое поле зажигания короны $E_c$ = 61.4 кВ/см – абсолютная величина поля (по формуле (8.9)). Корона на этом гидрометеоре возникнет, если заряд гидрометеора превысит *170 pC*. При заряде 50 pC и даже заряде 150 pC гидрометеор диаметром d=1 мм на высоте 5.5 км (p=0.5 атм) не сможет зажечь корону (тем более инициировать стример).  Электрическое поле $E_c$=61.4 кВ/см будет соответствовать приведенному электрическому полю 162 $\frac{V}{cm\,Torr}$ , которое гораздо меньше, чем электрическое поле, необходимое для инициации стримера на этой высоте (240 $\frac{V}{cm\,Torr}$).  Интеграл с использованием формулы (8.2), позволяет

вычислить значение числа ступеней лавины $n_{av}$= 9.51. Контрольные формулы (8.4), (8.5) приводят к несколько меньшему, но также разумному значению $n_{av}$= 5.16+2.39=7.55.

[Райзер, 1992, стр. 435] отмечал, что эмпирические формулы Пика несут в себе отпечаток физических критериев типа критерия Мика для стримеров. Найдис [Naidis, 2005] сформулировал эти критерии зажигания короны в явном виде, и они в целом соответствуют нашим оценкам. На Рисунке 8.3 из статьи [Naidis, 2005] $K$ – обозначает число ступеней лавины ($K$= $n_{av}$).

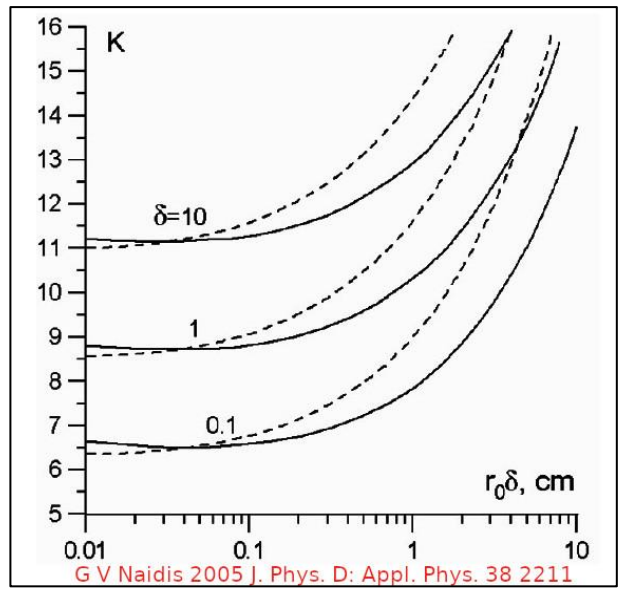

Рисунок 8.3 (адаптировано из [Naidis, 2005]). Число ступеней лавины ($K$= $n_{av}$) необходимых для инициации стримерной короны в зависимости от параметра $r_0\cdot\delta$, где $r_0$ – радиус сферы (сплошная линия) и цилиндра (пунктирная линия). $\delta$ – в единицах атмосферного давления.

Отметим, что теоретический расчет [Naidis, 2005], который суммирован на Рисунке 8.3, не совпадает с экспериментальными результатами [Богатов, 2013], так как в работе [Богатов, 2013] K растет при уменьшении размера частиц. [Богатов, 2013] объясняет это так «Данное противоречие связано, по-видимому, с различными формами зон ионизации в этих двух случаях. Зона ионизации имеет сферически симметричную форму в разряде на сферическом электроде (у [Naidis, 2005] – А.К.), а в случае разряда на сферической частице она локализуется в силовой трубке электрического поля, выходящей из полюса этой частицы. Зоны ионизации вытянутых частиц (т. е. частиц с малым отношением поперечного размера к продольному) локализованы вблизи их торцов и вследствие этого определяются в большей степени их поперечными размерами, чем продольными. Поэтому для вытянутых частиц параметр $K$ в критерии ($\int_{x_0}^{\infty} \alpha_{eff} dx = K$) должен быть ближе к значениям, характерным для сфер с малыми диаметрами, т. е. быть порядка 10». Добавим, что в эксперименте [Богатов, 2013] для сферических частиц измерялся порог зажигания разряда (короны, стримера или искры), так как нельзя было отделить пробой промежутка от зажигания короны, которая приводит к пробою промежутка, инициированному металлическим шариком (моделью «гидрометеора). [Богатов, 2013] на основании своих экспериментов вывел формулу зависимости напряженности минимального внешнего электрического поля (абсолютное значение, при данном давлении окружающего воздуха) от размера гидрометеора

$$E_{cr}^{min}\left(\frac{kV}{cm}\right) = \frac{24 \cdot \delta^{0.1}(atm)}{L^{0.9}(mm)} \quad , \tag{8.10}$$

где $L$ — максимальный размер гидрометеора. В соответствие с формулой (2.3.2.10) для зажигания разряда на высоте 5.5-16 км при больших, но разумных средних внешних электрических полях ($\approx$ 4-2 кВ/см) необходима нереально большая длина гидрометеора 7-14 мм, которые встречаются чрезвычайно редко [Dye et al., 1986, 2007]. Такая большая величина размеров гидрометеоров во многом соответствует более поздним численным расчетам [Dubinova et al., 2015], где, в частности, была учтена зависимость диэлектрической проницаемости льда от частоты, что привело к еще большему увеличению размеров гидрометеоров, которые стали достигать 15-60 мм. Заметим также, что критическое электрическое поле зажигания разряда $E_{cr}^{min}$ в формуле (8.10) [Богатов, 2013] слабо зависит от давления воздуха, а это означает, что приведенное поле (*E/N* или *E/p*), от которого и зависят все плазменные процессы, с высотой растет почти экспоненциально. Требуемые электрические поля и размеры гидрометеоров также говорят о том, что «в одиночку» механизм усиления электрического поля благодаря поляризации гидрометеора даже на очень больших гидрометеорах вряд ли может объяснить синхронную инициацию большого числа стримерных вспышек. Но благодаря этим вычислениям также можно независимо оценить необходимый порядок размера воздушного электрода (не менее 1-2 см в диаметре), который может обеспечить инициацию стримера. Исходя из этих экспериментов и расчетов, по нашему мнению, для обеспечения необходимого размера и уровня электрического поля должны суммироваться вклады в величину поля зарядов больших гидрометеоров, их поляризации, а также статистического усиления поля (см. также обсуждение в разделе 7.4.3.1).

Гидрометеор диаметром 100 мкм (d=0.1 мм, $r_0$=0.005 см, p=0.5 атм), для зажигания короны должен иметь, согласно расчетам, поле на поверхности не менее 200 кВ/см (40 MV/(m atm)), соответствующее критерию зажигания короны Найдиса. При этом, при расчете по формулам (8.4), (8.5) будет сделано $K = n_{av}$= 7.32 (6.81+0.51) шага лавины, что разумно совпадает с другими данными. Заряд этого гидрометеора должен быть не так велик по меркам измеренных зарядов в облаке, не менее, чем *5.7 пКл*. Но сейчас нельзя с полной определенностью сказать существуют ли такие заряды у гидрометеоров размером 100 мкм и меньше ([Gunn, 1949], [Hutchinson and Chalmers, 1951], [Marshall and Winn, 1982], [Weinheimer et al., 1991], [Marshall and Marsh, 1993]).

Для инициации стримера гидрометеор размером 100 мкм должен иметь (при расчете по формулам (8.4), (8.5)) на поверхности гидрометеора очень большое электрическое поле 400 кВ/см (80 MV/(m atm)), которое обеспечит $K = n_{av}$= *19.37*. Это чрезвычайно большие электрические поля, при которых электроны становятся убегающими, но область этого убегания не более 20 мкм. Тем не менее электроны могут разогнаться до примерно 200 эВ. Заряд этого гидрометеора должен быть не менее, чем *11.3 пКл*, то есть в два раза больше, чем необходимо для инициирования короны с гидрометеора того же размера и формы.

<u>Высота 13 км.</u> Для d = 1 мм ($r_0$=0.05 см), p = 0.163 атм, критическое электрическое поле зажигания короны $E_c$=31.7 кВ/см – абсолютная величина поля (по формуле (8.9)). Корона на этом гидрометеоре возникнет, согласно оценке, если заряд гидрометеора превысит 88 пКл. Электрическое поле $E_c$=31.7 кВ/см будет соответствовать приведенному большому электрическому полю 19.4 MV/(m atm) или (256 $\frac{V}{cm\,Torr}$). Интеграл с использованием формулы (8.2), который позволяет получить разумное для заряда 88 пКл значение числа ступеней лавин $n_{av}$= 9.35.

## 8.1.3. Время жизни воздушных электродов с точки зрения инициации электронных лавин фоновыми космическими лучами

Без инициации лавин не может возникнуть стример или коронный разряд. Для инициации лавин всегда необходим первый электрон. В грозовых облаках одним из основных кандидатов на роль поставщика свободных электронов являются космические лучи. Проведем оценку времени жизни воздушных электродов (какова бы ни была их природа) с точки зрения инициации лавин фоновыми космическими лучами.

Для примера мы провели оценку для конкретного события 00:04:38 UT on 30 July 2016 [Bandara et al., 2019], которое произошло на высоте 6.1 км, на широте и долготе университета Миссисипи (г. Оксфорд, США). Если считать, что воздушный электрод может быть образован благодаря статистическому и гидродинамическому механизму [Iudin, 2017], то он является суммарным локальным электрическим полем «сгущения» в пространстве заряженных частиц. Дальнейшие оценки производятся для минимального локально усиленного постоянного поля (идеализация, предположительно созданная зарядами облака), а также для воздушного электрода в форме шара с постоянной плотностью заряда $\rho(\frac{Кл}{м3})$.

Характерный размер идеализированного воздушного электрода мы находили исходя из условий выполнения критерия Мика [Райзер, 1992, стр. 424] для постоянного пробойного электрического поля $E \approx 3\ MV\ m^{-1} atm^{-1}$ на этой высоте, и он составлял около 5 см. Если представить воздушный электрод сгущением заряженных частиц с постоянной плотностью $\rho \left[\frac{Кл}{м^3}\right]$ и общим очень большим избыточным зарядом воздушного электрода около 1 нКл, то «инициировать» стример можно из объема диаметром 1.2 см (S=1.13 cm², V=0.9 cm³, расчеты велись по формулам (8.4), (8.5)). Главную роль в фоновой ионизации атмосферы на этих высотах играют электроны. Также необходимо учитывать вклад позитронов и фотонов [Sato, 2015]. По программе EXPACS мы рассчитали, что за секунду на высоте 6.1 км через см² проходит 1.34 электрона, 0.0565 позитронов и 205 фотонов. Каждый электрон и позитрон, пролетающие через воздушный электрод, приводят к инициации электронных лавин, так как на каждом сантиметре пути порождают на высоте 6.1 км около 35 вторичных электронов [Rutjes et al., 2019], что приводит с высокой вероятностью к инициации лавин. Фотоны высоких энергий имеют слабое поглощение в воздухе и все 205 фотонов создают только 0.13 поглощений в секунду на квадратный сантиметр сечения при диаметре воздушного электрода 5 см. Таким образом, поток всех ионизирующих частиц будет равен 1.53 см$^{-2}$с$^{-1}$.

Следовательно, воздушный электрод (постоянное поле $E \approx 3\ MV\ m^{-1} atm^{-1}$) с диаметром 5 см будет ионизован фоновым космическим излучением на высоте 6 километров в среднем с частотой $\nu_{ae} \approx 30$ с$^{-1}$ ($\tau_{ae} \approx 33$ мс), а воздушный электрод с диаметром 1.2 см (постоянной плотностью $\rho$ и зарядом 1 нКл) с частотой $\nu_{ae} \approx 1.73$ с$^{-1}$ ($\tau_{ae} \approx 578$ мс). Если бы воздушный электрод был сильно заряженным гидрометеором диаметром 1 мм, то придется учитывать не только электроны и позитроны, но и поглощение фотонов во льду, так как им уже нельзя будет пренебречь, как в случае воздуха. Гидрометеор диаметром 1 мм был бы ионизован с частотой $\nu_{ae} \approx 0.037$ с$^{-1}$, то есть мог бы «жить» 27 с, а гидрометеор диаметром 0.1 мм мог бы «жить» целых 45 минут (по отношению к ионизации фоновыми космическими лучами).

Для оценки динамики возникновения и гибели воздушных электродов мы составим простое уравнение для числа воздушных электродов в некотором объёме облака

$$\frac{dN_{ae}}{dt} = a - \nu_{ae} \cdot N_{ae} \qquad , \qquad (8.11)$$

$N_{ae}$ — число воздушных электродов [L$^{-3}$], $t$ — время [S], $a$ — скорость образования воздушных электродов благодаря турбулентности, статистическим флуктуациям электрического поля, усилению электрического поля на гидрометеорах [L$^{-3}$S$^{-1}$], $\nu_{ae}$ — частота гибели воздушных электродов [S$^{-1}$]. В первом приближении мы считаем скорости образования и гибели воздушных электродов постоянными.

Решение этого уравнения будет таким

$$N_{ae} = N_{ae}^0 e^{-\nu_{ae} t} + \frac{a}{\nu_{ae}} (1 - e^{-\nu_{ae} t}) \qquad (8.12)$$

$N_{ae}^0$ — число воздушных электродов в момент прихода в область образования молнии ШАЛ.

Мы оценили в главе 7, что в кубическом километре ($10^9$ м$^3$) для обеспечения VHF-сигнала «классического» NBE должно быть примерно $10^6$-$10^7$ воздушных электродов, то есть нужен один воздушный электрод на 100 м$^3$. Чтобы в среднем хотя бы один воздушный электрод был в этом объеме, необходимо, чтобы гибель воздушных электродов уравновешивалась рождением электродов, т.е. согласно уравнению (8.12) $\frac{a}{\nu_{ae}} \approx 1$. Таким образом, <u>для данной высоты, 6 километров,</u> частота рождения должна составлять не менее 30 воздушных электродов в секунду в объеме 100 м³ (если воздушные электроды будут иметь диаметр 5 см) и не менее 2-х электродов, если воздушные электроды будут иметь диаметр 1.2 см, что потребует очень большого заряда $\approx$ 1 нКл и большого поля на поверхности $E \approx 2.48\ MV\ m^{-1}(5.2 - 5.3\ MV\ m^{-1}atm^{-1})$. Следовательно, для обеспечения классического NBE при минимальном разумном среднем электрическом поле $E \approx 3\ MV\ m^{-1}atm^{-1}$ внутри воздушного электрода, один электрод должен появляться в среднем один раз в секунду в объеме 1.5x1.5x1.5 м³ и «жить» в среднем около 33 мс до момента, когда фоновая частица космических лучей инициирует в нем лавины. Для сильно заряженного электрода диаметром 1.2 см – это время составит около 580 мс.

Для высоты 9 км диаметр воздушного электрода по критерию Мика увеличится до $\geq$ 8 см (при постоянном электрическом поле $E \geq 0.8\ MV\ m^{-1}(3\ MV\ m^{-1}atm^{-1})$), а частота ионизации по аналогичным расчётам с помощью EXPACS возрастет примерно в три раза до $\nu_{cm^2} \approx 4.6\ \frac{1}{cm^2 s}$. Поэтому время жизни такого воздушного электрода уменьшится до 4.3 мс. Для воздушного электрода (который может инициировать стример), образованного сгущением плотности заряда (статистически-гидродинамический механизм), диаметром 1.2 см, зарядом 750 пКл, поле на «поверхности» электрода $E \geq 1.8\ MV\ m^{-1}(6\ MV\ m^{-1}atm^{-1})$), время жизни уменьшится не так сильно, до примерно 192 мс (расчет велся по формулам 8.4, 8.5). Для заряженного гидрометеора диаметром 1 мм частота ионизации составит $\nu_{ae} \approx 0.145$ с$^{-1}$, то есть, его время жизни 6,9 с, а гидрометеор диаметром 0.1 мм мог бы «жить» около 11 минут (по отношению к ионизации фоновыми космическими лучами).

Высота 13 км. Если взять результаты измерений во время гроз в Миссисипи [Karunarathna et al., 2015], то на наиболее частой для образования NBEs высоте 13 км частота ионизации воздуха космическими лучами достигнет $\nu_{cm^2} = 10.6\ \frac{1}{cm^2 s}$. Для высоты 13 км диаметр воздушного электрода по критерию Мика увеличится до 14.7 см (при постоянном электрическом поле $E \approx 3\ MV\ m^{-1}atm^{-1}$). Частота ионизации станет равна $\nu_{ae} \approx 1800$ с$^{-1}$. Время жизни воздушного электрода уменьшится до 0.56 мс. Для воздушного электрода (который может инициировать стример), образованного сгущением плотности заряда (статистически-гидродинамический механизм), диаметром 1.2 см, зарядом 470 пКл, поле на «поверхности» электрода $E \geq$ $1.18\ MV\ m^{-1}(7.2\ MV\ m^{-1}atm^{-1}))$, время жизни уменьшится, до примерно 83 мс. Для заряженного гидрометеора диаметром 1 мм, частота ионизации $\nu_{ae} \approx 0.329$ с$^{-1}$, то есть время ожидания ионизации около 3 с, а гидрометеор диаметром 0.1 мм мог бы «ждать» ионизацию 5 минут (по отношению к ионизации фоновыми космическими лучами).

На высоте 16 км частота ионизации будет равна $\nu_{cm^2} = 13.6\ \frac{1}{cm^2 s}$. Для высоты 16 км диаметр воздушного электрода по критерию Мика увеличится до 23.5 см ($E \approx 0.306\ MV\ m^{-1}$ ($3\ MV\ m^{-1}atm^{-1}$)). Частота ионизации станет равна $\nu_{ae}$=5.9·10$^3$ с$^{-1}$. Время жизни воздушного электрода уменьшится до 0.17 мс. В результате, наши оценки для высот 9-16 км дают очень короткие времена жизни воздушных электродов, в условиях обычных фоновых космических лучей при минимальном для инициации стримера электрическом поле внутри воздушного электрода $E \approx 3\ MV\ m^{-1}atm^{-1}$. Для воздушного электрода (который может инициировать стример), образованного сгущением плотности заряда (статистически-гидродинамический механизм), диаметром 1.2 см, зарядом 350 пКл, поле на «поверхности» электрода $E \geq 0.875\ MV\ m^{-1}(8.6\ MV\ m^{-1}atm^{-1}))$, время жизни уменьшится, до примерно 65 мс. Для заряженного гидрометеора диаметром 1 мм, частота ионизации $\nu_{ae} \approx 0.379$ с$^{-1}$, то есть время ожидания ионизации около 2.6 с, а гидрометеор диаметром 0.1 мм мог бы «ждать» ионизацию 4.4 минут (по отношению к ионизации фоновыми космическими лучами). Интересно, что время жизни гидрометеоров по отношению к ионизации космическими лучами почти не меняется от 13 до 22 км.

### 8.1.4. Выводы раздела 8.1

- Инициация стримеров благодаря заряду сильно заряженных гидрометеоров. Как показали оценки и анализ литературы, гидрометеорам с диаметром около 1 мм (характерный размер осадков) для инициации стримера на высотах 5.5-16 км нужны очень большие заряды 253-137 пКл (фактически предельно измеряемые [Marshall and Winn, 1982], [Marshall and Marsh, 1993], [Bateman et al., 1999]). Подобные же заряды (63–485 пКл) для гидрометеоров, инициирующих стримеры, диаметром 0.5-3 мм при давлении 0.4 атм получили численным счетом ранее [Babich et al, 2016]. Гидрометеоры (частицы осадков) таких размеров и заряды такой величины в принципе измерялись на высотах 5-7 км, но NBEs и молнии возникают вплоть до высот 16-20 км, где гидрометеоры размером более 4 мм измерялись исключительно редко. Если молнии инициируются слабыми и сильными NBEs которые представляют из себя мощные (скорее всего, распределенные стримерные вспышки) [Rizon et al., 2016], [Kostinskiy et al., 2020a], [Shao et al., 2020], то обойтись механизмом инициации стримеров только с поверхности гидрометеоров будет очень трудно, если вообще возможно.

- Инициация стримеров благодаря поляризации гидрометеоров в электрических полях и размер гидрометеоров. В экспериментальной работе [Богатов, 2013] было показано, что для инициации разрядов с гидрометеоров нужны не только очень большие внешние электрические поля, близкие к пробойным, но и размеры гидрометеоров 7-14 мм. Если же пытаться в расчетах «инициировать» стример с гидрометеоров эллиптической формы при измеряемых в облаках электрических полях, то размеры гидрометеоров становятся нереально большими (30-60 мм) [Dubinova et al., 2015]. Таким образом, и механизм инициации стримеров с поверхности гидрометеоров, благодаря только одному процессу поляризации гидрометеора во внешнем электрическом поле, с большим трудом может объяснить инициацию стримеров в грозовом облаке из-за нереально больших размеров гидрометеоров и/или нереально больших значений внешних электрических полей, которые необходимы для осуществления этого механизма.

- Инициация стримеров благодаря образованию воздушных электродов в форме сгущения в пространстве (повышенной концентрации) числа заряженных гидрометеоров (статистический и гидродинамический механизм) с электрическими полями близкими к минимальным пробойным ($3 \ MV \ m^{-1} atm^{-1}$) на больших высотах рождения NBEs вызывает большие сомнения из-за быстрого увеличения диаметра воздушного электрода и резкого уменьшения его времени жизни (не смотря на то, что и поле $E \approx 3 \ MV \ m^{-1} atm^{-1}$ представляется очень большим и пока экспериментально не наблюдалось в грозовых облаках). На это влияют два фактора. Увеличение почти на порядок потока вторичных космических электронов и позитронов (с 6 до 16 км) и, главное, экспоненциальное увеличение размеров воздушного электрода. Поэтому более реалистичным представляется воздушный электрод, с меньшим диаметром, который и на высоте 16 км имеет время жизни более 60 мс. Сократить (сохранить небольшим) диаметр воздушного электрода можно в том случае, если приведенное электрическое поле внутри объема воздушного электрода превышает $E > 3\ MV\ m^{-1} atm^{-1}$ и изменяется в большом диапазоне $5.2 - 8.6\ MV\ m^{-1} atm^{-1}$. Данное положение выглядит почти нереалистичным, но существование самих стримерных вспышек на больших высотах (16-20 км) настойчиво побуждает искать механизмы существования сверхпробойных полей, которые обеспечивают существование воздушных электродов.

- Инициация лавин с гидрометеора требует относительно большого заряда, и он может накопиться пока первый электрон не инициирует лавины (или не произойдет обмен зарядом между другими гидрометеорами). То, что сильно заряженные гидрометеоры имеют относительно маленький размер по сравнению с воздушными электродами, может играть положительную роль для накопления больших зарядов. По отношению к процессу ионизации фоновыми космическими лучами, время ионизации гидрометеора диаметром 1 мм на высотах 5.5-9 км находится в диапазоне 27-6.9 с, а на высотах 13-25 км время жизни изменяется очень мало и составляет 2,5-4 секунды. Для гидрометеора диаметром 0.1 мм (100 мкм) время ионизации (по отношению к процессу ионизации фоновыми космическими лучами) увеличивается примерно на два порядка и на высотах 5.5-9 км находится в диапазоне 45-15 мин., а на высотах 9-25 км время жизни изменяется очень мало и составляет 5-7 минут. Основную роль в ионизации гидрометеоров будут играть фотоны, а не электроны и позитроны (как в случае воздушных электродов), так как вода и лёд из-за высокой плотности поглощают фотоны на три порядка сильнее, чем воздух.

- Формулы Пика для зажигания короны (8.8, 8.9), расчеты [Naidis, 2005], эксперименты [Богатов, 2013] и наши оценки показывают, в случае если поверхность гидрометеора диаметром 1-3 мм сильно неоднородна и имеет множество острых углов (размером 10-100 мкм), как у снежинок или ледяной крупы (Рисунок 8.4), то эти острые углы не будут играть большую роль в коронировании (и создании стримеров), так как, даже при радиусе 100 мкм их площадь будет в 100 раз меньше, чем у частицы в целом и, следовательно, нужно 100 таких острых углов, чтобы вероятность попадания в них энергичного фотона или электрона сравнялась с вероятностью попасть в частицу в целом. Кроме того, электрические поля рядом лежащих острых углов будут сглаживаться из-за сложения электрических полей. Более мелкие острые углы, например 10 мкм, вообще не будут играть существенную роль, так как их площади в $10^4$ раз меньше, чем у частицы в целом и вероятность попадания в настолько малую область их усиленного поля вообще стремится к нулю (и даже частица пролетевшая через эту площадь еще уменьшит вероятность попадания из-за того, что электроны образуются в среднем 1 на 270 мкм, а поле может спадать на десятках микрон). Таким образом, на создание лавин и рождение стримеров (при больших зарядах) гидрометеоров, включая снежинки и крупу будут сильно влиять только общие размеры и форма гидрометеоров, а не их острые концы (углы) [Богатов, 2013].

- Таким образом, для обеспечения уровня VHF-сигнала характерного для NBE, размер воздушных электродов, необходимых для инициации стримеров, ограничивается сверху фоновым космическим излучением и изменяется в пределах от ~ 3-5 см на высоте 5-6 км до ~1-2 см на высоте 16 км. При этом, для обеспечения такого небольшого размера электрода для высот 9-16 км над уровнем моря, приведенное электрическое поле на поверхности однородно заряженных воздушных электродов должно быть высоким 52-86 $kV\ cm^{-1}atm^{-1}$, как и заряд — 1000-350 пКл. Это означает, что сгущение (концентрация зарядов по статистическому механизму) должны быть весьма значительными. Если воздушные электроды имеют меньший размер или электрическое поле внутри воздушных электродов меньше этих значений, но больше 30 $kV\ cm^{-1}atm^{-1}$, то внутри них возникают электронные лавины, которые не превращаются в стример. Такие большие величины зарядов и электрических полей воздушных электродов подчеркивают большие трудности и для статистического и гидродинамического механизма усиления электрического поля. Поэтому, с высокой вероятностью, сильное электрическое поле воздушного электрода, который инициирует стримерные вспышки, формируется одновременным действием нескольких физических механизмов, таких как электрическое поле сильно заряженного гидрометеора, поляризация электрода, статистическое и гидродинамическое усиление электрического поля. При этом, характерный размер воздушного электрода будет в пределах не более 1-3 см и его мы будем использовать при оценках воздействия ШАЛ на грозовое облако.

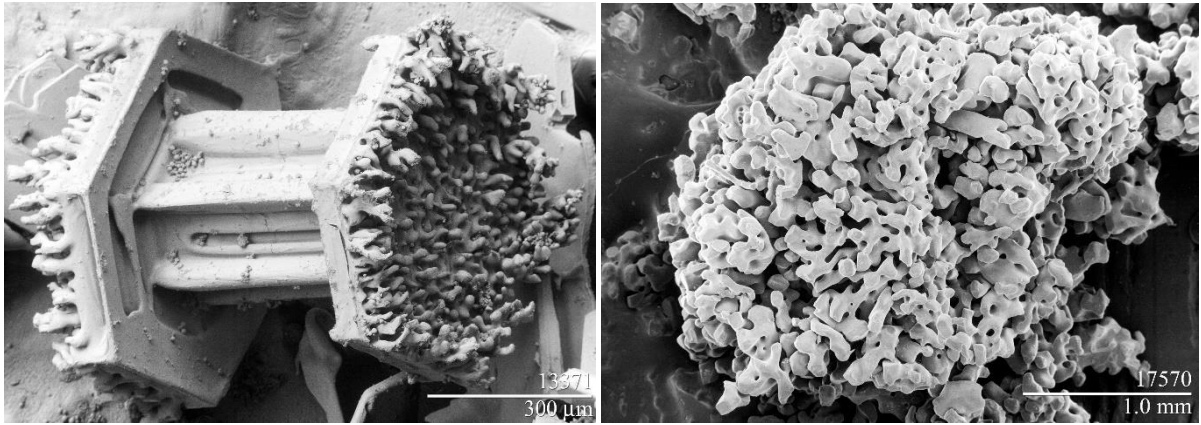

Рисунок 8.4. (слева) Ледяная крупа (graupel) в форме столбчатого снежного кристалла с налипшими на него каплями размера около 10-20 мкм; (справа) ледяная крупа (graupel) в форме эллипсоида с налипшими на него каплями размера около 10-20 мкм. Изображение получено с использованием низкотемпературного сканирующего электронного микроскопа (LT-SEM) в Исследовательском центре сельского хозяйства в Белтсвилле, штат Мериленд, США (Beltsville Agricultural Research Center in the Electron Microscopy).

## 8.2. Синхронная инициация стримерных вспышек благодаря EAS-RREA – механизму

Выше мы рассмотрели некоторые возможные механизмы возникновения и ионизации фоновым излучением воздушных электродов. В этом разделе мы оценим численно механизм синхронизации в объеме 0.1-1 км³ старта стримерных вспышек, число которых экспоненциально увеличивается в электрическом поле благодаря механизму развития лавины релятивистских электронов (RREA), засеянных широким атмосферным ливнем космических лучей (ШАЛ).

Данный расчет является оценкой снизу как минимум по двум причинам. Первая, так как в нашей расчетной модели ШАЛ падает на EE-область вертикально и в данной конфигурации поля (отрицательный заряд находится вверху, положительный заряд находится внизу) вторичные электроны ШАЛ усиливаются в электрическом поле, а позитроны (их всего в несколько раз меньше, чем электронов [Grieder, 2010], [Rutjes et al. 2019]) тормозятся и тоже создают убегающие электроны. При падении ШАЛ под углами большими, чем 45⁰, многие позитроны ШАЛ из-за кулоновского рассеяния начнут двигаться в направлении отрицательного заряда, увеличивая свою энергию и создавая новые электронные лавины, размножающиеся вниз в направлении положительного заряда (своеобразный аналог механизма обратной связи, [Dwyer and Uman, 2014]). Во-вторых, мы учитываем в расчете экспоненциальное размножение только тех вторичных электронов и позитронов, которые упали на границу области сильного электрического поля (далее эти электроны и позитроны мы будем называть посевными электронами). Мы не учитываем то, что в процессе движения ШАЛ через область сильного электрического поля (внутри неё), энергичными частицами ШАЛ создаются дополнительные электроны и позитроны, которые также включаются в процесс размножения.

Согласно Механизму (глава 7) [Kostinskiy et al., 2020a], необходимым условием возникновения классического NBE является быстрое рождение воздушных электродов в ЕЕ-области размером 0.1-1 км$^3$ в каждом объеме $\sim$2x2x2 $-$ 5x5x5 м ($10^{-1}$-$10^{-2}$ m$^{-3}$). Достаточным условием возникновения NBE является почти одновременный старт стримерных вспышек в ЕЕ-объеме. Старт стримерных вспышек, согласно Механизму, инициируется энергичными электронами ($\varepsilon_e$>500 кэВ), который пересекают воздушные электроды. Если выполняется критерий Мика [Raizer,1991], то лавины превращаются в стримеры, так как при движении энергичного электрона через воздушный электрод на каждом сантиметре траектории образуется около 75 тепловых электронов (при атмосферном давлении), что обеспечивает вероятность старта лавин и стримеров близкую к единице.

[Gurevich et al., 1999] первыми обратили внимание на возможную решающую роль ШАЛ в инициировании молнии, так как ШАЛ с энергией в диапазоне $10^{15}$eV $\leq \varepsilon \leq 10^{17}$ eV порождает большое число вторичных энергичных и тепловых электронов и позитронов, число которых может экспоненциально увеличиваться благодаря механизму убегания электронов. [Gurevich et al., 1999] надеялись с помощью вторичных электронов ШАЛ и механизма убегания электронов добиться настолько высокой концентрации электронов в окрестности траектории первичной космической частицы (ствола ШАЛ), что станет возможным старт классического стримера. Дальнейшие численные расчеты показали, что данный механизм инициации стримера очень трудно реализовать, так как плотность электронов, необходимая для старта стримеров, оказалась ниже необходимой на несколько порядков ([Dwyer, 2010], [Babich & Bochkov, 2011], [Rutjes et al., 2019]).

В нашем Механизме роль ШАЛ принципиально отличается от роли ШАЛ в механизме [Gurevich et al., 1999]. Мы предполагаем, что ШАЛ инициирует не один стример около оси ШАЛ, а десятки и сотни тысяч стримерных вспышек, которые расположены в объеме ~0.1-1 км$^3$, то есть, основную роль играют вторичные удаленные на большие расстояния от оси ШАЛ электроны и позитроны, рассеянные по области сильного электрического поля в сильно турбулентной части облака (ЕЕ-области). Сами стримеры инициируются тепловыми электронами (которые возникают вдоль траектории энергичных электронов) в областях, с локальным электрическим полем выше пробойного (воздушные электроды диаметром несколько сантиметров), которые образуются благодаря турбулентному движению, гидродинамическим неустойчивостям, заряду и усилению электрического поля на гидрометеорах.

Рождение классических NBE требует экстремальных значений турбулентности и попадания в эту область облака космических частиц с энергией $\varepsilon_0 > 10^{15}$ эВ, которые относительно редки. Слабые инициирующие молнию события (Weak IE) [Marshall et al., 2014a, 2019] и [Lyu et al., 2019] требуют меньшего числа стримерных вспышек. Образование Weak IE может происходить при меньшей турбулентности или при попадании в сильно турбулентную область менее энергичного ШАЛ. Поэтому, вероятно, Weak IE в среднем случаются гораздо чаще [Marshall et al., 2014a, 2019] и [Lyu et al., 2019].

[Gurevich et al., 1999] оценили, что для рождения стримеров около оси ШАЛ в их модели необходимы ШАЛ с энергией первичной частицы $\varepsilon_0 > 10^{15}$ eV. [Rutjes et al., 2019], [Dubinova et al., 2015] также выделили близкий диапазон энергий ШАЛ ($10^{15}$ eV $\leq \varepsilon_0 \leq 10^{17}$ eV), которые могут играть основную роль (в качестве поставщика первых электронов) в возможной инициации стримеров на гидрометеорах (включая одновременную инициацию нескольких стримеров для обеспечения механизма быстрого положительного стримерного пробоя (FPB), предложенного в [Rison et al., 2016]). По первым оценкам, для реализации нашего Механизма синхронизации стримерных вспышек, наибольшую роль могут играть первичные частицы из близкого диапазона энергий ($5\cdot10^{14}$ eV $\leq \varepsilon_0 \leq 5\cdot10^{15}$ eV).

Частота появления космических лучей на границе атмосферы равна интегралу известного экспериментального распределения, которое в диапазоне $5\cdot10^{14}$ eV $\leq \varepsilon_0 \leq 10^{17}$ eV несколько раз изменяет показатель степени [Tanabashi M. et al., 2018], [Budnev et al, 2020]. Для оценок мы будем использовать приближенную аппроксимацию этого распределения, которая удовлетворяет наши требования к точности оценок:

$$\frac{dN_{\varepsilon_0}}{d\varepsilon_0} \approx a \cdot \varepsilon_0{}^{\mu}, \; km^{-2}s^{-1}sr^{-1}PeV^{-1} , \qquad (8.13)$$

где $a=2.66$, $\mu=-2.73$ для диапазона $10^{14}$eV$\leq \varepsilon_0 \leq 3\cdot10^{15}$eV и $a=4.14$, $\mu \approx -3$ для диапазона $3\cdot10^{15}$eV $\leq \varepsilon_0 \leq 10^{17}$eV. $\varepsilon_0$ в уравнении используется в единицах PeV ($10^{15}$ eV). Интеграл этого распределения будет равен

$$N_{\varepsilon_0}(0.1 \; PeV \leq \varepsilon_0 \leq 100 \; PeV) \approx \frac{a}{\mu+1}\left(\varepsilon_0{}_1^{\mu+1} - \varepsilon_0{}_2^{\mu+1}\right), \; km^{-2}s^{-1}sr^{-1}. \quad (8.14)$$

Частицы с энергией $10^{17}$eV $\leq \varepsilon_0 \leq 10^{19}$eV попадают в атмосферу слишком редко ($\sim 2 \cdot 10^{-4} \; km^{-2}s^{-1}sr^{-1}$), чтобы с их помощью можно было объяснить зарождение молнии и серийных (повторяющихся) NBEs [Bandara et al., 2021]. Частицы с энергией $\varepsilon_0 \leq 10^{13}$ eV попадают в атмосферу часто, но они производят слишком мало посевных электронов, что можно оценить по приближенной формуле $N_e^{EAS} \sim \frac{0.67 \cdot \varepsilon_0(eV)}{10^9}$ [Tanabashi M. et al., 2018, p.429].

Для энергетического интервала $10^{14}$eV $\leq \varepsilon_0 \leq 10^{15}$eV число первичных энергичных частиц ШАЛ на границе атмосферы будет $N_{\varepsilon_0} \approx 82 \; km^{-2}s^{-1}sr^{-1}$, для энергетического интервала $10^{15}$eV $\leq \varepsilon_0 \leq 10^{16}$eV число первичных энергичных частиц ШАЛ на границе атмосферы будет $N_{\varepsilon_0} \approx 1.5 \; km^{-2}s^{-1}sr^{-1}$, для энергетического интервала $10^{16}$eV $\leq \varepsilon_0 \leq 10^{17}$eV число первичных энергичных частиц ШАЛ на границе атмосферы будет $N_{\varepsilon_0} \approx 2 \cdot 10^{-2} \; km^{-2}s^{-1}sr^{-1}$. Таким образом, весь диапазон энергий $10^{14}$eV $\leq \varepsilon_0 \leq 10^{17}$eV требует анализа с точки зрения инициации стримерных вспышек, так как общее число частиц, падающих на границу облака выглядит разумно.

Для предварительных оценок числа всех вторичных частиц в максимуме ШАЛ в диапазоне $10^{14}$eV $\leq \varepsilon_0 \leq 10^{17}$ eV [Dwyer, 2008] используется простую формулу [Gaisser, 1990]

$$N_e^{EAS} \sim 5 \cdot 10^{-2}\varepsilon_0^{1.1} \qquad (8.15)$$

где $\varepsilon_0$ является энергией первичной космической частицы в GeV. Например, $N_e^{EAS} \approx 2\cdot10^5$ для частицы с первичной энергией $10^{15}$ eV. Надо заметить, что в физике космических лучей экспериментальными установками, обычно измеряются электроны и позитроны с энергией не меньше, чем 10-100 MeV, но убегающими в воздухе могут стать электроны с энергией выше 0.5 MeV, число которых в несколько раз больше [Rutjes et al., 2019], чем вычисляется по формуле (8.15). Поэтому первичная частица с энергией $\sim 10^{15}$ eV может производить в максимуме ШАЛ $N_e^{EAS} \sim 10^6$ посевных электронов и позитронов [Rutjes et al., 2019].

## 8.2.1. Радиальное (латеральное) пространственное распределение вторичных электронов, усиленных электрическим полем

Для оценки инициации NBE (Weak IE) благодаря лавинам EAS-RREA, мы должны учитывать радиальное распределение вторичных электронов и позитронов ШАЛ, число которых будет экспоненциально увеличиваться в сильном электрическом поле, благодаря механизму RREA. Вторичные электроны должны почти одновременно попасть в достаточное число имеющихся воздушных электродов, чтобы синхронно инициировать множество стримерных вспышек.

[Dwyer, 2010] и [Babich & Bochkov, 2011] рассчитали методом Монте-Карло радиальное (латеральное) распределения лавины релятивистских электронов (RREA), которая была инициирована в точке одним или несколькими начальными электронами. Их коэффициенты диффузии совпадают с большой точностью [Dwyer et al., 2012]. Они представили свои результаты в виде решений диффузионного уравнения для электронов. Мы будем пользоваться в расчетах уравнением для потока электронов в виде формулы (8.16), которая предложена в статье [Dwyer, 2010]. Эти диффузионные решения являются оценкой снизу, так как они не учитывают рентгеновские кванты. Решение является осесимметричным и определяется расстоянием $z$-$z_0$ по оси распространения RREA и расстоянием до оси $r$. $z_0$ — точка входа посевных электронов в область сильного электрического поля, $N_0$ - число электронов в точке $z_0$, $\lambda$ — шаг лавины убегающих электронов (8.17), рассчитанный в статье [Dwyer, 2003], $D_\perp$ — is the lateral diffusion coefficient (2.3.2.18), $v = 0.89c$, [Coleman & Dwyer, 2006].

$$\Phi_{re}^{c}(r,z) = \frac{N_0}{4\pi\left(\frac{D_\perp}{\nu}\right)(z-z_0)} \cdot exp\left(\frac{z-z_0}{\lambda} - \frac{r^2}{4\left(\frac{D_\perp}{\nu}\right)(z-z_0)}\right)[m^{-2}] \qquad (8.16)$$

$$\lambda = \frac{7300\,[kV]}{\left(E-276\left[\frac{kV}{m}\right]exp\left(\frac{-h}{8.4}\right)\right)} \qquad (8.17)$$

$$\frac{D_\perp}{\nu} = exp\left(\frac{h}{8.4}\right)(5.86\cdot10^4)E^{-1.79}[m],\ E\,[kV/m];\ \ \nu = 0.89c; h\,[km] \quad (8.18)$$

Коэффициент радиальной диффузии (8.18) зависит от напряженности электрического поля $E$ и концентрации молекул воздуха. В нашем случае давление воздуха экспоненциально падает с ростом высоты, что означает, что при одном и том же электрическом поле, коэффициент латеральной диффузии значительно увеличивается на больших высотах. Нас в данном расчете не интересует продольная диффузия, так как она не влияет на общую площадь потока вторичных электронов в электрическом поле.

$$\Phi_{re}^{c}\big(r,(z-z_0)\big) = \frac{N_0}{4\pi\left(exp\left(\frac{h}{8.4}\right)(5.86\cdot10^4)E^{-1.79}\right)(z-z_0)} \cdot exp\left(\frac{z-z_0}{\lambda} - \right.$$

$$\left. \frac{r^2}{4\left(exp\left(\frac{h}{8.4}\right)(5.86\cdot10^4)E^{-1.79}\right)(z-z_0)}\right),[m^{-2}] \qquad (8.19)$$

Распределение (8.16), (8.19) [Dwyer, 2010] и [Babich & Bochkov, 2011] записано в предположении, что все электроны падают в область сильного электрического поля в одной точке (приближение дельта функции Дирака). Но в развитом ШАЛ, интересующего нас диапазона энергий одновременно движется $10^4$-$10^7$ вторичных частиц, которые распределены перпендикулярно оси распространения ШАЛ на сотни метров [Kamata & Nishimura, 1958], [Grieder, 2010]. Поэтому мы не можем просто подставить в уравнение (8.19) общее число $N_0$ частиц ШАЛ. Радиальное распределение частиц ШАЛ часто представляют в виде аппроксимации Nishimura-Kamata-Greizen (NKG). Поэтому корректная модель распространения ШАЛ в сильном электрическом поле грозового облака должна инициировать лавины убегающих электронов по формуле (8.19) в каждой точке поперечного сечения ШАЛ. Уравнение (8.19) должно учитывать число начальных электронов $N_0$ в каждой точке фронта ШАЛ, а не общее число частиц всего ШАЛ. В каждой точке число начальных электронов $N_0$ можно выразить формулой (s11) Nishimura-Kamata-Greizen (NKG).

NKG аппроксимация, используемая для ШАЛ представляется в таком виде [Kamata & Nishimura, 1958]:

$$\rho_e(R) = \frac{N_e^{EAS}}{R_M^2} \cdot C(s) \cdot \left(\frac{R}{R_M}\right)^{s-2} \cdot \left(\frac{R}{R_M} + 1\right)^{s-4.5} \qquad (8.20)$$

где $\rho_e(R)$ — плотность вторичных частиц на расстоянии $r$ от оси ливня, $N_e^{EAS}$ – общее число вторичных частиц ливня, $R_M = 79[m] \cdot exp\left(\frac{h}{8.4}\right)$ – радиус Мольера, $s$ – параметр «возраста» ливня и $C(s) = 0.366 \cdot s^2 \cdot (2.07 - s)^{1.25}$ , [Hayakawa, 1969].

Простые оценки показывают, что на высотах 6-16 км распределение электронов ШАЛ будет широким. Основным механизмом увеличения радиального распределения вторичных электронов ШАЛ является кулоновское рассеяние электронов на ядрах атомов, которые входят в состав молекул воздуха. Кулоновское рассеяние в основном определяет средне-квадратичный радиус ШАЛ $\sqrt{\langle R \rangle^2}$, который включает примерно половину электронов [Мурзин, 1988]: $\sqrt{\langle R \rangle^2} = 0.9 \cdot \left(\frac{E_s}{\varepsilon_{cr}} \cdot t_0\right) = 0.9 \cdot R_M^{\frac{g}{cm^2}} = 0.9 \cdot 9.5 \frac{g}{cm^2} = 8.55 \frac{g}{cm^2}$ , где $E_s = 21\ MeV$ , $t_0\left[\frac{g}{cm^2}\right]$ — радиационная единица длины, $\varepsilon_{cr}$ — критическая энергия электронов, при которой потери на тормозное излучение становятся равными потерям на ионизацию (для воздуха $\varepsilon_{cr} \approx 81\ MeV$). $R_M^{\frac{g}{cm^2}} = 9.5 \frac{g}{cm^2}$ — мольеровский радиус для воздуха в единицах $\frac{g}{cm^2}$. Радиус радиального расширения ШАЛ экспоненциально растет с высотой. Среднеквадратичный радиус ШАЛ $\sqrt{\langle R \rangle^2}$ в метрах будет равен $\sqrt{\langle R \rangle^2}[m] \approx 71.1[m] \cdot exp\left(\frac{h}{8.4}\right)$. Для 6 км среднеквадратичный радиус ШАЛ можно оценить как 144 м; для 9 км — 208 м; для 13 км — 355 м; для 16 км —531 м. Эта оценка показывает, что вторичные электроны хорошо развитого ШАЛ могут засеять значительную часть EE-области. Когда вторичные электроны ШАЛ войдут в область сильного электрического поля, каждый электрон ШАЛ с энергией большей, чем 500 кэВ начнет собственное экспоненциальное размножение, которое в свою очередь также имеет широкое радиальное распределение (8.19).

## 8.2.2. Алгоритм расчета EAS-RREA инициации стримерных вспышек

Общая схема процесса синхронизации стримерных вспышек лавиной EAS-RREA между верхним отрицательным и нижним положительным зарядом изображена на Рисунке 8.5. Космическая частица с энергией $\varepsilon_0 > 10^{15}$ eV (1) создает ШАЛ (2). Электроны и позитроны ШАЛ (3) входят в область сильного электрического поля (5), которое может поддержать распространение RREA (4) и положительных стримеров (7). EAS-RREA электроны и позитроны (6) пересекают область грозового облака с воздушными электродами (9) и синхронизуют почти одновременный запуск множества стримерных вспышек (7), стартующих с воздушных электродов, в объем которых попали электроны (8).

Рисунок 8.6 поясняет алгоритм численного расчета. ШАЛ (EAS) падает на EE-область перпендикулярно и движется из точки $z_0(O_r)$ по оси $z$. Расчет является осесимметричным относительно оси $z$, которая является осью EAS. $R$-радиус является расстоянием до оси EAS в NKG-аппроксимации (формула (11). $R$ изменяется на каждом шаге интегрирования от 5 до конечного значения $R_F$. Данная окружность с радиусом $R$ и шириной $dR$ выбрана таким образом, что она имеет одинаковое число посевных электронов в каждой точке и находится в плоскости $(X\text{-}Y, z{=}z_0)$ проникновения ШАЛ в EE-область. Из этих точек окружности $R$ электроны распространяются в соответствие с уравнением (10) в каждую точку $y_a(z)$. Точка $y_a(z)$ изменяется на каждом шаге интегрирования $(z\text{-}z_0{+}dz)$ от 5 м до конечного расстояния $y_{aF}(z)$, которое определяет границы области, занятой воздушными электродами. Первые 5 метров радиуса ШАЛ требуют отдельного описания из-за сильных флуктуаций энергичных частиц ([Мурзин, 1988], [Grieder, 2010]) и в этом численном приближении мы ими пренебрегаем без большой потери точности общей оценки из-за малой площади этого сегмента и его малого вклада в точки $y_a(z) > 10$ м от оси ШАЛ, которые вносят основной вклад в число электронов, которые ионизируют воздушные электроды.

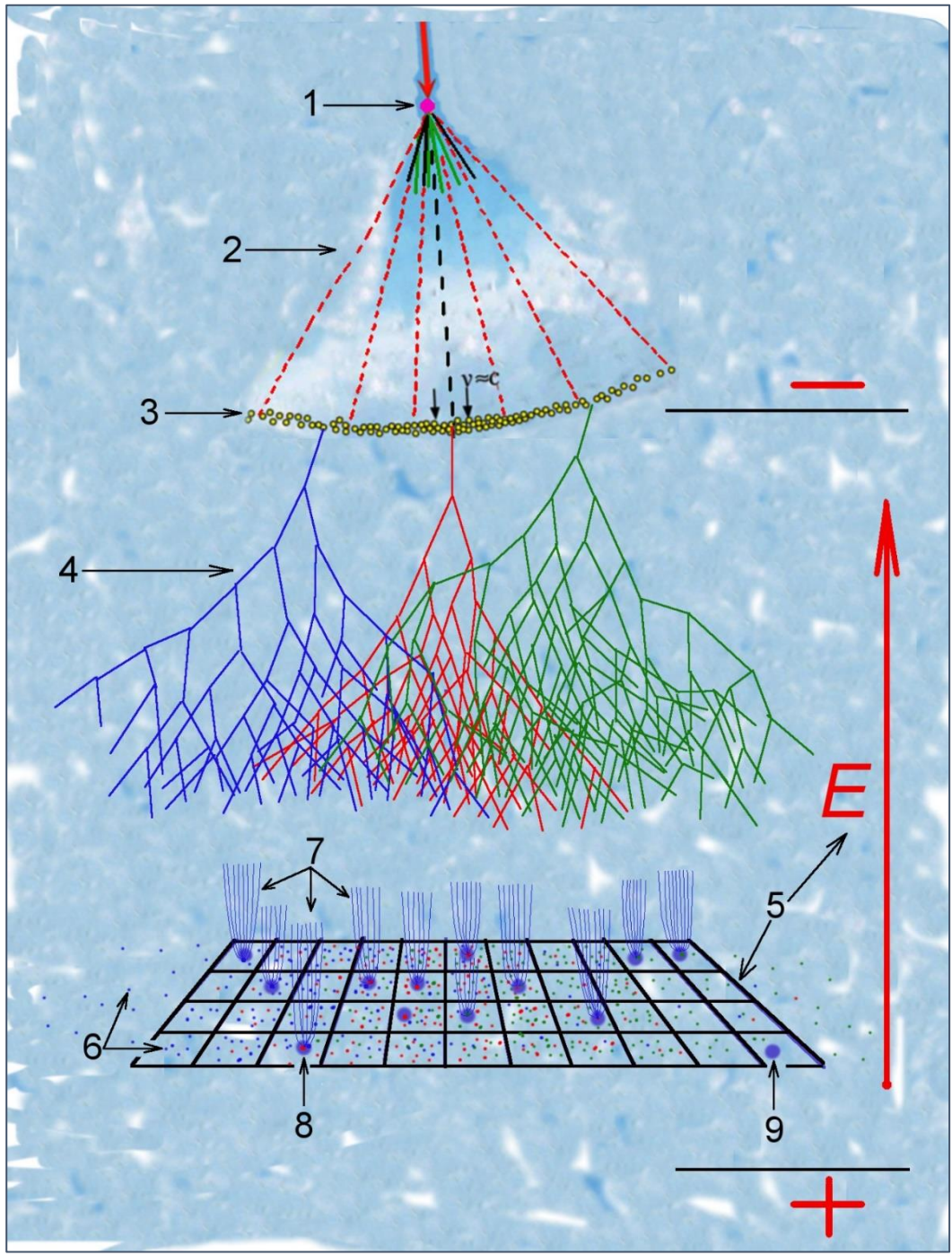

Рисунок 8.5(адаптировано из [Kostinskiy et al., 2020а]): Общая схема процесса синхронизации стримерных вспышек лавиной EAS-RREA между верхним отрицательным и нижним положительным зарядом. 1 - первичная частица ШАЛ; 2 - EAS; 3 - вторичные электроны и позитроны ШАЛ; 4 - RREA; 5 - область сильного электрического поля; 6 - EAS-RREA электроны, пересекающие область сильной турбулентности грозового облака, которая создает сильные электрические поля; 7 - EAS-RREA synchronized streamer flashes; 8 - область сверхпробойного поля, которую пересек энергичный электрон; 9 - область сверхпробойного поля, которую не пересек энергичный электрон

Рисунок 8.6 (адаптировано из [Kostinskiy et al., 2020b]). Схема расчета потока электронов, пересекающих область грозового облака с сильным электрическим полем. Зеленая стрелка и зеленые точки вдоль положительной части оси $x$ показывают порядок варьирования координаты $x_i$ окружности радиуса $R$ (первый цикл), что позволяет вычислить число посевных электронов ШАЛ, посылающих электроны в точку $y_a(z)$; розовая стрелка и розовые точки вдоль отрицательной части оси $x$ показывают порядок варьирования радиуса $R$ (второй цикл), что позволяет вычислить сумму всех электронов в точке $y_a(z)$; красная стрелка и красные точки вдоль положительной части оси $y$ показывают порядок варьирования координаты точки $y_a(z)$, в которых рассчитывается поток электронов (третий цикл); толстые синие стрелки показывают расстояние $r^+$, $r^-$ в уравнении (sl2) от точек $(x_i, y_i^+)$, $(x_i, y_i^-)$ до точки $y_a(z)$; симметричные тонкие синие линии показывают расстояния от точек $(x_i^-, y_i^+)$, $(x_i^-, y_i^-)$ до точки $y_a(z)$.

Из-за осевой симметрии задачи нам достаточно получить значение потока электронов вдоль любого радиуса идущего перпендикулярно от оси $z$ до точки $y_{aF}(z)$ и мы можем выразить его распределением $N_{y_a}(z) = f(y_a(z))$, изменяя положение точки $y_a(z)$(красная стрелка на Рисунке 8.6).

Для данного $R$ определяем плотность посевных электронов в точке $\rho_e(R)$ благодаря NKG-уравнению (8.20) при s=0.9, и получаем окружность с центром в точке $O_r(z_0)$ с постоянной плотностью посевных электронов в каждой точке окружности в месте проникновения ШАЛ в область грозового облака с сильным электрическим полем $E$ (до высоты $z_0$ мы предполагаем в расчете, что электрическое поле было равно нулю). Значение возраста ШАЛ $s=0.9$ отвечает в данном диапазоне энергий $\varepsilon_0$ первичной частицы за развитый в пространстве и времени ШАЛ, имеющий $10^5$-$10^7$ вторичных электронов ([Grieder, 2010], [Rutjes et al., 2019]). Мы называем электрическое поле сильным, если оно способно поддерживать не только убегающие электроны $E > 284$ $kV/(m{\cdot}atm)$, но и может поддерживать движение положительных стримеров $E > 450$ $kV/(m{\cdot}atm)$. Окружность радиусом $R$ является «поставщиком» электронов $N_r$ в точку $y_a(z)$ в соответствие с распределением (8.20), которое получено подстановкой в (8.19) распределения NKG (2.3.2.20). В формуле (8.19) $r$ является расстоянием в плоскости $X$-$Y$ от произвольной точки окружности $R$ до точки $y_a(z)$. Расстояние $r$ мы найдем по известной формуле (8.21), где мы ось ШАЛ поместили в центр окружности, которую в свою очередь поместили в центр системы координат. Точка $y_a(z)$ изменяется по оси $y$, из чего следует, что $x_a = 0$. Также при интегрировании необходимо учитывать два значения $\pm(R^2 - x_i^2)^{0.5}$, так как обе точки $(x_i, y_i^+)$, $(x_i, y_i^-)$ вносят свой вклад в поток электронов $N_{y_a}$ (Рисунок 8.6). Кроме того, решение для каждой точки $(x_i, y_i^+), (x_i, y_i^-)$ должно быть умножено на 2, т.к. точки левой части окружности $R$ (Рисунок 8.6) из-за симметрии вносят точно такой же вклад в $N_{y_a}$. Сумму вклада всех точек окружности $R$ можно получить изменением $x_i$ от 5 до $R$. $R$ изменяется от 5 м до «края» ШАЛ $R_F$. Сумма вклада поток электронов всех окружностей $R$ дает суммарный поток электронов $N_{y_a}$ в точке $y_a(z)$ (в каждом слое $z$-$z_0$+$dz$).

$$N_r(z - zo) = \frac{0.361 \cdot \frac{N_e^{EAS}}{R_M^2} \left(\frac{R}{R_M}\right)^{-1.1} \left(\frac{R}{R_M}+1\right)^{-3.6}}{4\pi \left(exp\left(\frac{h}{8.4}\right)(5.86 \cdot 10^4)E^{-1.79}\right)(z-z_0)} \cdot exp\left(\frac{(z-z_0)}{\lambda} - \frac{x_i^2 + \left(\pm(R^2-x_i^2)^{0.5} - y_a(z)\right)^2}{4\left(exp\left(\frac{h}{8.4}\right)(5.86 \cdot 10^4)E^{-1.79}\right)(z-z_0)}\right)$$

$$(8.20)$$

$$r^{\pm} = ((x_i - x_a)^2 + (y_i - y_a)^2)^{0.5} = ((x_i)^2 + (y_i - y_a)^2)^{0.5} = (x_i^2 + (\pm(R^2 - x_i^2)^{0.5} - y_a)^2)^{0.5}$$

$$(8.21)$$

Вычисляемый поток электронов в точке $(y_a(z), z - z_0)$ осесимметричного радиального распределения в каждом слое EE-области будет равен:

$$N_{y_a}\left(y_a(z), (z-z_0)\right) = \int_{R=5}^{R=R_F} \int_{x=5}^{x=R} \frac{2 \cdot 0.361 \cdot \frac{N_e^{EAS}}{R_M^2} \cdot \left(\frac{R}{R_M}\right)^{-1.1} \left(\frac{R}{R_M}+1\right)^{-3.6}}{4\pi\left(exp\left(\frac{h}{8.4}\right)(5.86 \cdot 10^4)E^{-1.79}\right)(z-z_0)} \cdot exp\left(\frac{(z-z_0)}{\lambda} - \right.$$

$$\left. \frac{x^2 + \left(\pm(R^2-x^2)^{0.5} - y_a(z)\right)^2}{4\left(exp\left(\frac{h}{8.4}\right)(5.86 \cdot 10^4)E^{-1.79}\right)(z-z_0)}\right) dx dR$$

$$(8.22)$$

После проникновения в область сильного электрического поля, число электронов в лавине убегающих электронов увеличивается по экспоненциальному закону $N_e \sim N_e^{EAS} \exp(\frac{(z-z_0)}{\lambda})$ (8.22), где шаг лавины убегающих электронов $\lambda$ определяется формулой (8.17).

Теперь мы можем оценить число стримерных вспышек инициируемых этим потоком электронов с учетом вероятностного характера процессов инициации. Поперечное сечение каждого воздушного электрода в грозовом облаке, в соответствие с предыдущими оценками, находится в диапазоне $10^{-2}$-$10^{-3}$ м$^{-2}$. Вероятность попадания одного случайного электрона в воздушный электрод будет составлять $p = 10^{-2}$-$10^{-3}$ (вероятность равна отношению площади воздушного электрода к одному квадратному метру). Вероятность нескольких подобных независимых событий оценивается по известной формуле Бернулли $P_n^k = C_n^k p^k (1-p)^{n-k}$. Чтобы стримерная вспышка была инициирована достаточно того, чтобы сечение воздушного электрода пересек хотя бы один энергичный электрон. Вероятность такого события вычисляется по упрощенной формуле Бернулли, и она равна $P_{N_{y_a}} = 1 - ((1-p))^{N_{y_a}}$ , где $N_{y_a}$— поток энергичных электронов (по формуле 8.22). Вероятность инициации стримерной вспышки $P_{N_{y_a}}$ при потоке 100 электронов на квадратный метр для воздушного электрода с сечением $10^{-2}$ м$^{-2}$ будет равна 63%, а для воздушного электрода, имеющего сечение $10^{-3}$ м$^{-2}$ та же вероятность инициации будет достигаться при падении 1000 электронов на квадратный метр.

Таким образом, общее число стримерных вспышек во всей ЕЕ-области в зависимости от расстояния (или времени), мы получим интегрированием числа вспышек во всех слоях $dz$ по оси $z$ (Рисунок 8.6):

$$n_{fl} = \int_{z=z_0}^{z=z_F} \int_{y_a=5}^{y_a=y_{aF}} \rho_{E_{th}}(z, y_a(z)) \cdot 2 \cdot \pi \cdot y_a(z) \cdot \left(1 - (1-p)^{N_{y_a}(y_a(z),z)}\right) dy_a dz \quad (8.23)$$

В формуле (8.23) $\rho_{E_{th}}(z, y_a(z))$ является плотностью числа воздушных электродов в кубическом метре. В данных расчетах мы будем считать эту плотность постоянной величиной во всем объеме и равной $\rho_{E_{th}} \approx 10^{-2}\ m^{-3}$, опираясь на оценки мощной вспышке NBE с общим числом воздушных электродов $\sim 10^6$-$10^7$ км$^{-3}$ [Kostinsky et al., 2020a]. Другим аргументом в пользу этого значения плотности воздушных электронов в рамках нашего Механизма является то, что при такой плотности стримерные вспышки со средней скоростью 1-2x$10^6$ м/с пройдут расстояние $\sim$5 м за 2.5-5 мкс. Это позволит создать сеть плазменных каналов и резко повысить их время жизни.

### 8.2.2.1. Результаты оценок

Мы будем производить расчеты числа стримерных вспышек в соответствие с уравнениями (8.22), (8.23). Как говорилось выше, данные расчеты являются оценкой снизу. Оценки проведем для двух высот 13 и 6 км при конфигурации поля и движения ШАЛ в геометрии, показанной на Рисунке 8.5.

На Рисунке 8.7 мы показываем расчет числа стримерных вспышек, который велся для высоты 13 км, число вторичных электронов ШАЛ на границе области сильного электрического поля составляло $N_e^{EAS} = 10^6$, радиус Мольера был R$_M$=328 m. Электрическое поле в этом расчете составляло $E \approx 85$ кВ/м (400 кВ/(м·атм)) и оно приблизительно соответствует минимально возможному электрическому полю, которое может поддержать движение положительных стримеров в условиях чрезвычайно низкой влажности. В оценке использовалась вероятность инициирования воздушных электродов равная p = 0.001, которая соответствует диаметру воздушного электрода ∅≈2 см. Этот диаметр выбирался, исходя из оценок предыдущего раздела. Плотность числа воздушных электродов в этих оценках считалась постоянной внутри области сильного электрического поля и равной $\rho_{E_{th}} = 10^{-2} m^{-3}$ во всем объеме движения ШАЛ. Объем, в котором производились вычисления, имел форму цилиндра с радиусом основания 500 м и переменной высотой z. Радиус цилиндра ограничивался точкой $y_a(z)$=500 м. Если рассмотреть 5 первых шагов лавины убегающих электронов, которые происходят на расстоянии z=1385 м (красная линия на Рисунке 8.7), то общий объем цилиндра будет 1.09 км$^3$, а общее число воздушных электродов около $10^7$ км$^{-3}$. Число всех стримерных вспышек, которые произошли после 5 шагов лавины убегающих электронов в этом случае будет равным 2.4·$10^5$ (Рисунок 8.7). Это означает, что лавине EAS-RREA удалось инициировать вспышки только в 2.2% всех воздушных электродов, которые существовали в объеме. Несмотря на это, даже, если средний ток одной стримерной вспышки был 0.1 А, то общий ток всех стримеров составит около 24 кА. Общее время движения лавины EAS-RREA к этому моменту времени составляет 5.2 мкс. Кривая роста числа стримерных вспышек на Рисунке 8.7 в зависимости от длины z, пройденной лавиной EAS-RREA имеет два выраженных участка. Первый участок кривой соответствует начальному проникновению ШАЛ в область сильного электрического поля, и он составляет 60-70 метров (первые 200-300 нс). Резкий рост числа вспышек от 0 до ~$10^3$ возникает из-за того, что ШАЛ уже состоит из $10^6$ электронов и они сразу же начинают ионизовать воздушные электроды. Этот рост будет соответствовать физике процесса только в случае почти вертикального падения ШАЛ на плоскую широкую область сильного электрического поля, а для других конфигураций EE-области он будет менее резким. Начиная с 70 метров, когда лавина EAS-RREA «полностью вошла» в EE-область, мы на всей траектории движения лавины видим очень близкий к экспоненциальному рост числа вспышек (a, следовательно, и VHF-сигнала) в зависимости от расстояния и времени движения лавины, что качественно соответствует экспериментальным результатам Рисунок B.32b [Rison et al., 2016, Supplementary Figure 1b] в течение первых микросекунд. Мы видим, что оценка дает разумные цифры при

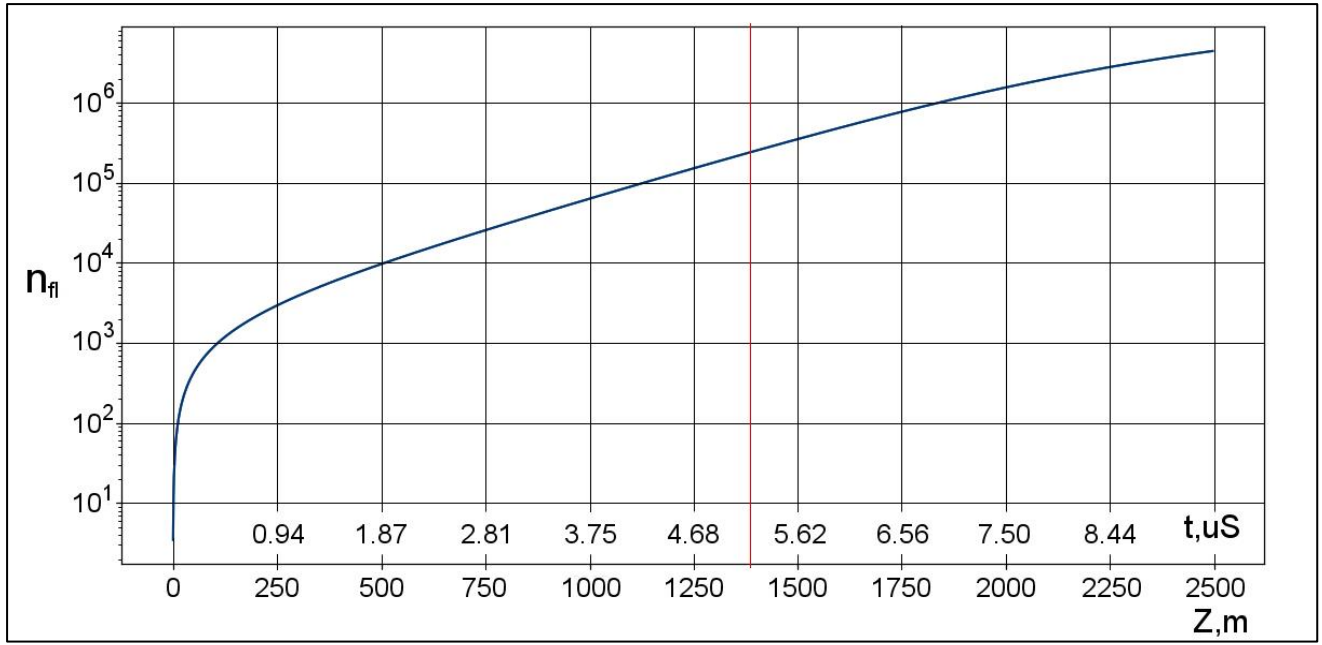

Рисунок 8.7 (адаптировано из [Kostinskiy et al., 2020b]). Оценка числа стримерных вспышек $n_{fl}$ в зависимости от пути $z(t)$, который проходит лавина EAS-RREA внутри EE-области. Расчет велся для высоты 13 км, $N_e^{EAS} = 10^6$, $R_M$=328 m, $400 \frac{kV}{m \cdot atm} \left( 85 \frac{kV}{m} \right)$, $\lambda_{RREA} = 277 \, m$, вероятность инициирования воздушных электродов равна p = 0.001, плотность числа воздушных электродов считалась постоянной и равной $\rho_{E_{th}} = 10^{-2} m^{-3}$. Красная вертикальная линия показывает момент прохождения лавины 1385 м (5.4 мкс с начала движения), что задает объем EE-области ~1.0 км³ ( $y_a(z)$=500 м.)

минимальном поле для распространения стримеров и не очень большой общей длине лавины EAS-RREA.

Рисунок 8.8 показывает, как в данном случае растет с расстоянием $z$ (и временем $t$) число вспышек $n_{fl}$ в каждом поперечном слое воздуха толщиной 1 метр. Хорошо виден также почти мгновенный рост числа вспышек в начальный момент, когда ШАЛ с числом частиц $N_e^{EAS} \approx 10^6$ входит в EE-область. Важно, что даже при таком большом начальном числе электронов, вторичные электроны ШАЛ инициируют не более 5% всех стримерных вспышек EAS-RREA лавины. Это говорит о том, что только вторичным электронам ШАЛ, без механизма усиления лавин убегающих электронов RREA очень сложно инициировать большое число стримерных вспышек за несколько микросекунд, даже с диаметром воздушного электрода ~2 см, не говоря уже о диаметрах воздушных электродов ~0.1-2 мм, которые гипотетически могут быть около самых больших заряженных гидрометеоров. Поэтому, на наш взгляд, даже предварительные оценки показывают, что для инициации значительного числа стримерных вспышек в объеме требуется именно комбинация EAS и RREA.

### 8.2.2.1.1. Влияние напряженности среднего электрического поля на процесс инициации стримерных вспышек, благодаря вторичным электронам ШАЛ

Важную роль может играть зависимость числа стримерных вспышек от напряженности электрического поля, которая прямо определяет как длину шага лавины убегающих электронов $\lambda$ (формула (8.17)) так и радиальное распределение вторичных электронов (формула (8.16)).

Мы продолжаем расчеты числа стримерных вспышек, инициированных EAS-RREA лавиной для высоты 13 км при вертикальном падении ШАЛ. Вероятность инициации случайным электроном принималась за $p = 0.001 (\emptyset \approx 2\ cm), \rho_{E_{th}} = 10^{-2} m^{-3}$, $R_M$=328 m, $N_e^{EAS} = 10^6$. Электрическое поле принимало четыре значения: 85 kV/m (400 kV/(m·atm)), $\lambda = 277$ m; 106 kV/m (500 kV/(m·atm)), $\lambda = 153\ m$; 127.5 kV/m (600 kV/(m·atm)), $\lambda = 106\ m$; 149 kV/m (700 kV/(m·atm)), $\lambda = 81\ m$. Результаты расчетов показаны на Рисунке 8.9. Также, как и для электрического поля 400 kV/(m·atm)

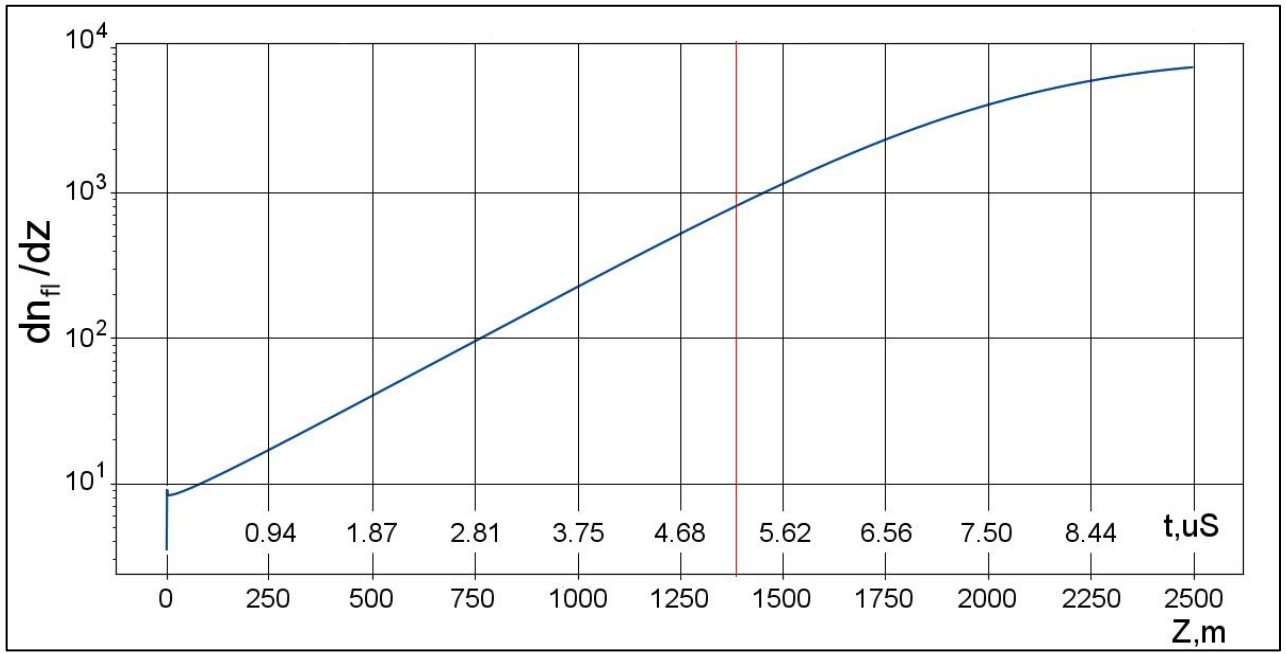

Рисунок 8.8 (адаптировано из [Kostinskiy et al., 2020b]). Число вспышек $dn_{fl}/dz$ в каждом поперечном слое воздуха толщиной 1 метр: высота 13 км, $N_e^{EAS} = 10^6$, $R_M$=328 м, $400\frac{kV}{m \cdot atm}\left(85 \frac{kV}{m}\right)$, $\lambda_{RREA} = 277\ m$, p = 0.001, $\rho_{E_{th}} = 10^{-2}m^{-3}$.

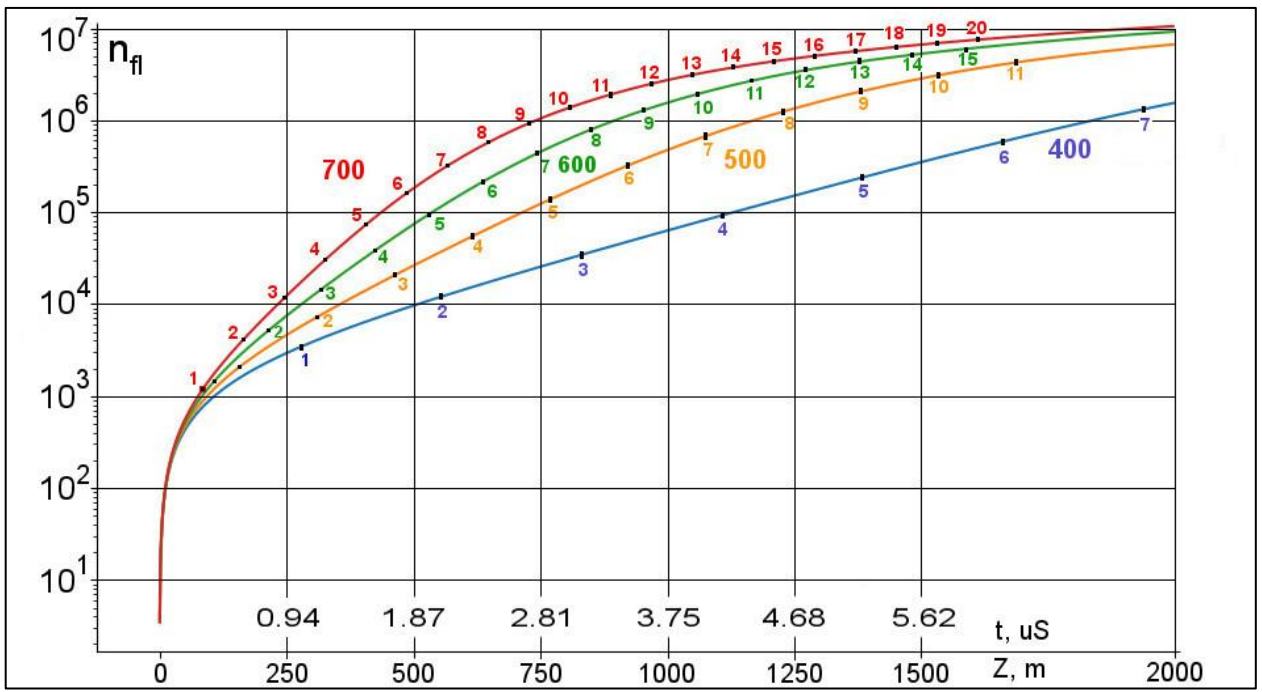

Рисунок 8.9 (адаптировано из [Kostinskiy et al., 2020b]). Зависимость числа стримерных вспышек $n_{fl}(z)$ от напряженности электрического поля. Высота 13 км, вероятность попадания электрона $p = 0.001 (\emptyset \approx 2\ cm), \rho_{E_{th}} = 10^{-2}m^{-3}$, $R_M$=328 м, $N_e^{EAS} = 10^6$. Электрическое поле принимало четыре значения: $400\frac{kV}{m \cdot atm}\left(85 \frac{kV}{m}\right)$, $\lambda_{RREA} = 277\ m$; — синяя линия; $500\frac{kV}{m \cdot atm}\left(106 \frac{kV}{m}\right)$, $\lambda_{RREA} = 153\ m$ — желтая линия; $600\frac{kV}{m \cdot atm}\left(127.5 \frac{kV}{m}\right)$, $\lambda_{RREA} = 106\ m$ — зеленая линия; $700\frac{kV}{m \cdot atm}\left(149 \frac{kV}{m}\right)$, $\lambda_{RREA} = 81\ m$ — красная линия. Цифры около линий обозначают номер шага лавины $\lambda_{RREA}$.

(Рисунок 8.7), кривые роста числа стримерных вспышек в зависимости от длины траектории, пройденной лавиной EAS-RREA имеют два выраженных участка. Первый участок соответствует начальному проникновению ШАЛ в область сильного электрического поля и составляет 60-70 метров. С точностью до толщины линий на графике, линии начинают разделяться через 35 м пути (около 100 нс движения в области сильного поля). Начиная примерно с 70 метров и до 500 м, на всей длине для всех электрических полей можно видеть очень близкий к экспоненциальному рост числа вспышек в зависимости от расстояния (времени движения лавины). Но, начиная с 500 метров, кривая для области облака со средним электрическим полем 700 kV/(m·atm) становится все более пологой, стремясь к насыщению. Также ведут себя кривая для 600 kV/(m·atm), начиная примерно с 800 метров, и кривая для 500 kV/(m·atm), начиная примерно с 1000 метров. Этот переход происходит между 6-м и 7-м шагом лавины для всех кривых. Здесь начинает происходить процесс «исчерпания» имеющихся в объеме воздушных электродов. Зависимость от величины электрического поля является сильной. Так для поля 500 kV/(m·atm) число «стримерных вспышек» равное $10^5$ достигается на первых $\sim$750 метрах, а для 700 kV/(m·atm) всего на $\sim$400 м. Таким образом, оценки числа вспышек при возрастании среднего электрического поля от 400 kV/(m·atm) до 700 kV/(m·atm) еще раньше позволяет достигнуть числа стримерных вспышек, которые могут обеспечить VHF-сигнал, сопровождающий NBE.

## 8.2.2.1.2. Влияние величины объема ЕЕ-области, занятого воздушными электродами, на процесс инициации стримерных вспышек, благодаря вторичным электронам ШАЛ

В следующей серии наших расчетов на высоте 13 км мы оценим влияние на число стримерных вспышек размера ЕЕ-области грозового облака, занятого воздушными электродами, уменьшив радиус ЕЕ-области с 500 м до 250 м и 150 м. Электрическое поле будет изменяться от $400 - 500 \frac{kV}{m \cdot atm} \left( 85 - 106 \ \frac{kV}{m} \right)$. Число посевных электронов ШАЛ будет равно $N_e^{EAS} \approx 10^6$. Объем ЕЕ-области мы будем изменять с помощью предела интегрирования $y_{aF}$. В представленных выше расчетах мы брали $y_{aF} = 500 \ m$ , что задавало общий диаметр цилиндра $\approx 1$ км. Не меняя параметров ШАЛ, мы определим гораздо более узкие области, занятые воздушными электродами $\sim 0.2\ km^3$ ($y_{aF} = 250\ m$, $z = 1000$) и $\sim 0.07\ km^3$ ($y_{aF} = 150\ m,\ z = 1000\ m$). Оценки показывают (Рисунок 8.10), что число стримерных вспышек, несмотря на изменение EE-объема более, чем на порядок, изменяется слабо. Мы рассмотрим различия на значительном расстоянии, пройденном лавиной и равном 900 м. При электрическом поле 400 kV/(m·atm) и $y_{aF} = 250\ m$, объем занятый воздушными электродами уменьшится в 4 раза, что приведет к уменьшению количества стримерных вспышек всего на 23%, а при уменьшении объема в 11 раз ($y_{aF} = 150\ m$) к уменьшению числа вспышек всего в 2 раза. При электрическом поле 500 kV/(m·atm) уменьшение объема занятого воздушными электродами в 4 раза ($y_{aF} = 250\ m$) приведет к уменьшению количества стримерных вспышек на 24%, а при уменьшении объема в 11 раз ($y_{aF} = 150\ m$) к уменьшению числа вспышек в 2.44 раза. Эти оценки позволяют выдвинуть предварительные гипотезы. Основную роль в инициации стримерных вспышек играет область EAS-RREA лавины с радиусом около 250 метров, что можно распространить и на все оценки, приведенные в данном разделе 8.2. Может быть, еще более интересно, что узкие слои грозового облака диаметром 300-500 м, которые могут возникать во время сложных (возможно, встречных) турбулентных воздушных течений в грозовой ячейке или около нее [Karunarathna et al., 2015], [Yuter & Houze, 1995], имеют шанс создать мощные VHF-сигналы, регистрируемые при NBE.

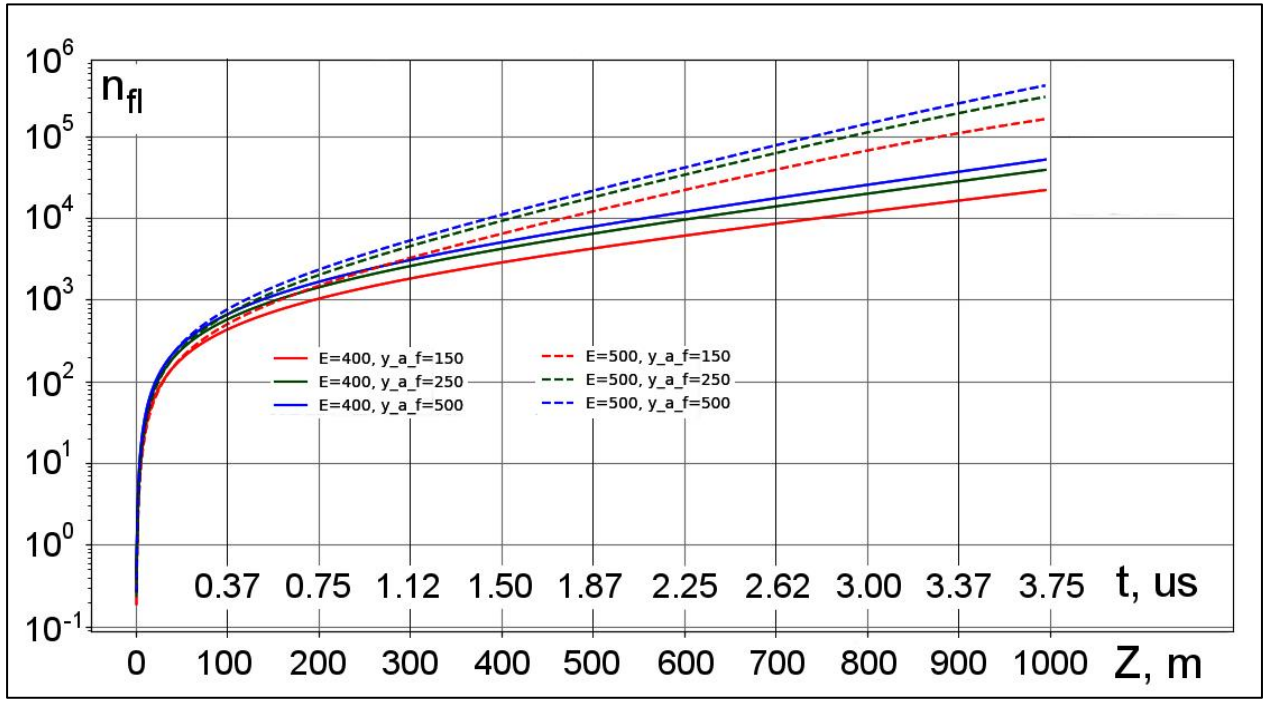

Рисунок 8.10 (адаптировано из [Kostinskiy et al., 2020b]). Влияние на число стримерных вспышек $n_{fl}$ (z) обьемов ЕЕ-области грозового облака (на высоте 13 км), которые задаются радиусами ЕЕ-области: $y_{aF} = 500\ m$ (синие линии), $y_{aF} = 250\ m$ (зеленые линии) и $y_{aF} = 150\ m$ (красные линии), для электрических полей $400\ \frac{kV}{m\cdot atm}\left(85\ \frac{kV}{m}\right)$, $\lambda_{RREA} = 277\ m$ — сплошные линии , $500\ \frac{kV}{m\cdot atm}\left(106\ \frac{kV}{m}\right)$, $\lambda_{RREA} = 153\ m$) — пунктирные линии. Вероятность попадания электрона $p = 0.001(\emptyset \approx 2\ cm), \rho_{E_{th}} = 10^{-2} m^{-3}, R_M{=}328\ m, N_e^{EAS} = 10^6$.

## 8.2.2.1.3. Влияние числа посевных электронов ШАЛ на процесс инициации стримерных вспышек

Нам также нужно оценить роль числа посевных электронов ШАЛ в процессе инициации стримерных вспышек, изменяя число посевных электронов — $N_e^{EAS}$. Высота над уровнем моря была 13 км. На Рисунке 8.9 был показан численный расчет, который велся для четырех электрических полей при числе посевных электронов $N_e^{EAS} = 10^6$. На Рисунке 8.11 показаны расчеты, которые проводились для трех значений числа посевных электронов ШАЛ $N_e^{EAS} \approx 10^3; 10^4;\ 10^5$. Для каждого из этих значений $N_e^{EAS}$ расчет проводился для трех электрических полей: Е= 85 kV/m (400 kV/(m·atm)), 106 kV/m (500 kV/(m·atm)), 127.5 kV/m (600 kV/(m·atm)) (также на Рисунке 8.11 показаны расчеты для $N_e^{EAS} = 10^6$). Результаты, приведенные на Рисунке 8.11 демонстрируют для всех

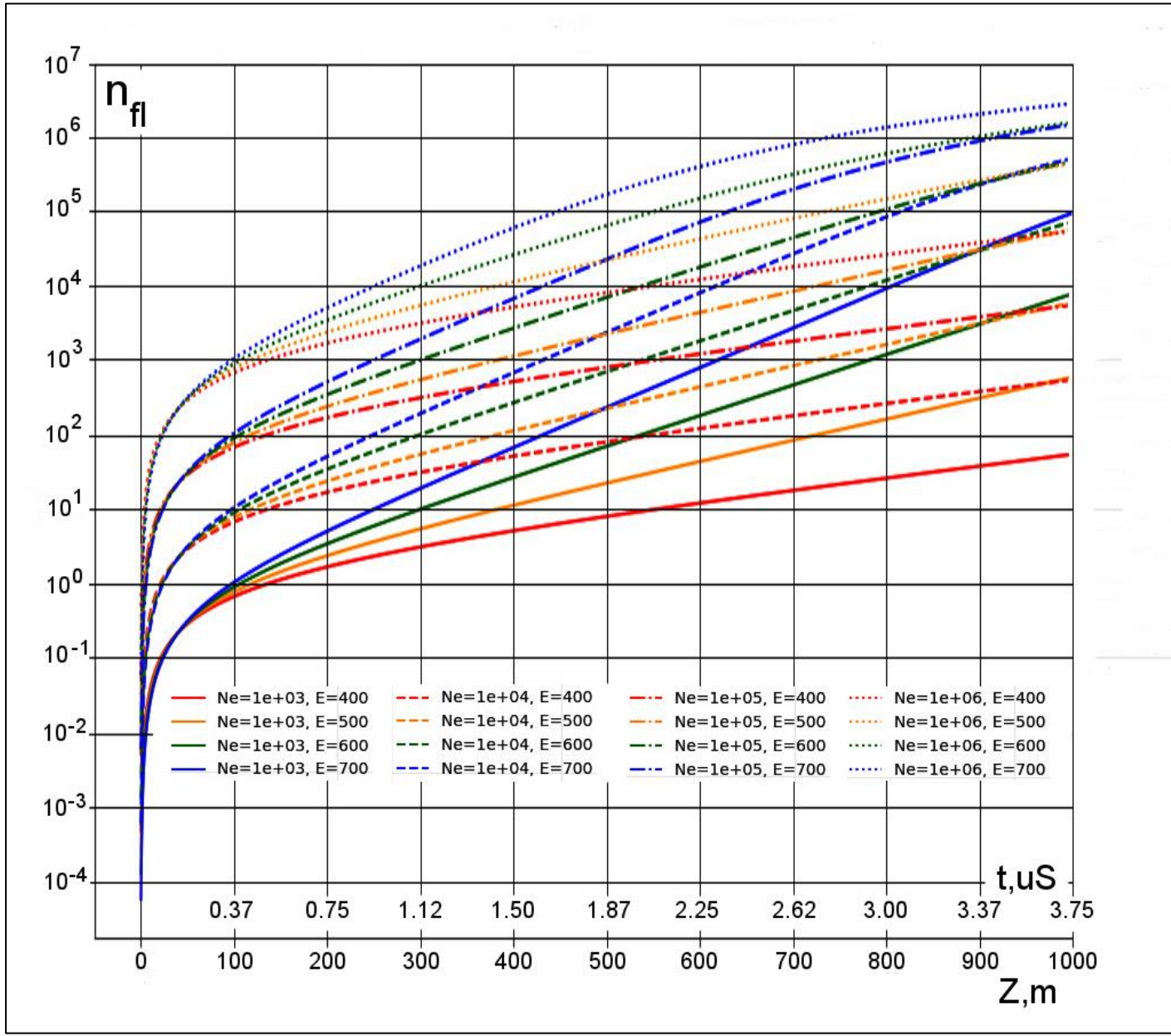

Рисунок 8.11 (адаптировано из [Kostinskiy et al., 2020b]). Число стримерных вспышек $n_{fl}$ (z) на высоте 13 км в зависимости от числа посевных электронов ШАЛ ($N_e^{EAS} = 10^3$(сплошные линии); $10^4$ (пунктирные линии); $10^5$ (штрих-пунктирные линии), $10^6$ (точечные линии). Для каждого из этих значений $N_e^{EAS}$ расчет проводился для четырех электрических полей: $400\frac{kV}{m\cdot atm}\left(85\frac{kV}{m}\right)$, $\lambda_{RREA} = 277\ m$ — красные линии; $500\frac{kV}{m\cdot atm}\left(106\frac{kV}{m}\right)$, $\lambda_{RREA} = 153\ m$ — желтые линии; $600\frac{kV}{m\cdot atm}\left(127.5\frac{kV}{m}\right)$, $\lambda_{RREA} = 106\ m$ — зеленые линии; $700\frac{kV}{m\cdot atm}\left(149\frac{kV}{m}\right)$, $\lambda_{RREA} = 81\ m$ — синие линии. Высота 13 км, вероятность попадания электрона $p = 0.001(\emptyset \approx 2\ cm)$, $\rho_{E_{th}} = 10^{-2}m^{-3}$, $R_M$=328 m.

значений числа посевных электронов $N_e^{EAS}$ похожую экспоненциальную динамику нарастания числа стримерных вспышек, за исключением первых 50-60 метров, как и на всех Рисунках 8.7-8.10. При $N_e^{EAS} \approx 10^3; 10^4; 10^5$, кривые на второй стадии почти не испытывают отклонений от экспоненциального развития. Число вспышек на разных расстояниях лавины EAS-RREA растет в зависимости от $N_e^{EAS}$ почти линейно. Мы можем отчетливо видеть, что при $N_e^{EAS} \approx 10^3$ и электрическом поле 400 kV/(m·atm) число стримерных вспышек даже при пути лавины EAS-RREA 1000 м равно 54. Даже при электрическом поле 600 kV/(m·atm) число стримерных вспышек при $N_e^{EAS} \approx 10^3$ составляет лишь 7.6·$10^3$ (при общем числе воздушных электронов в объеме ~$10^7$). При этом среднее расстояние между стримерными вспышками будет в диапазоне 35-45 м, что не позволит создать впоследствии, согласно предложенному Механизму (глава 7), сеть взаимодействующих горячих плазменных каналов за разумные времена. Эти цифры показывают, что космические лучи с числом посевных частиц $N_e^{EAS} \leq 10^3$ не смогут хоть сколько-нибудь сильно разрядить EE-область, а также создать заметный сигнал NBE. Космические лучи с числом посевных частиц $N_e^{EAS} \approx 10^4$ могут произвести измеряемый сигнал VHF при средних электрических полях 500-600 kV/(m·atm). Возможно, они могут также стать причиной слабого IE, так как, расстояние между вспышками будет находиться в диапазоне 12-15 м. Но оценки показывают, что мощный VHF-сигнал, который сопровождает классический NBE, требует числа посевных электронов большего чем $10^5$ и электрическое поле в диапазоне 500-600 kV/(m·atm).

### 8.2.2.1.4 Влияние давления на процесс инициации стримерных вспышек

При увеличении давления и том же самом среднем электрическом поле быстро уменьшается шаг лавины убегающих электронов (RREA) и рассеяние убегающих электронов на ядрах атомов. На Рисунке 8.12 хорошо видно, что число вспышек $n_{fl}$ быстро растет с длиной z и к 450 метрам движения лавины для всех электрических полей в диапазоне 500-700 kV/(m·atm) $n_{fl}$ превышает $10^5$. Рисунок 8.12 также показывает, что большая часть вспышек на высоте 6 км происходит в радиусе 250 м от оси ШАЛ, причем влияние более удаленных областей лавины начинается только с 600 м (700 kV/(m·atm)), 630 м (600 kV/(m·atm)) и 750 м (500 kV/(m·atm)), когда кривые начинаются разделяться.

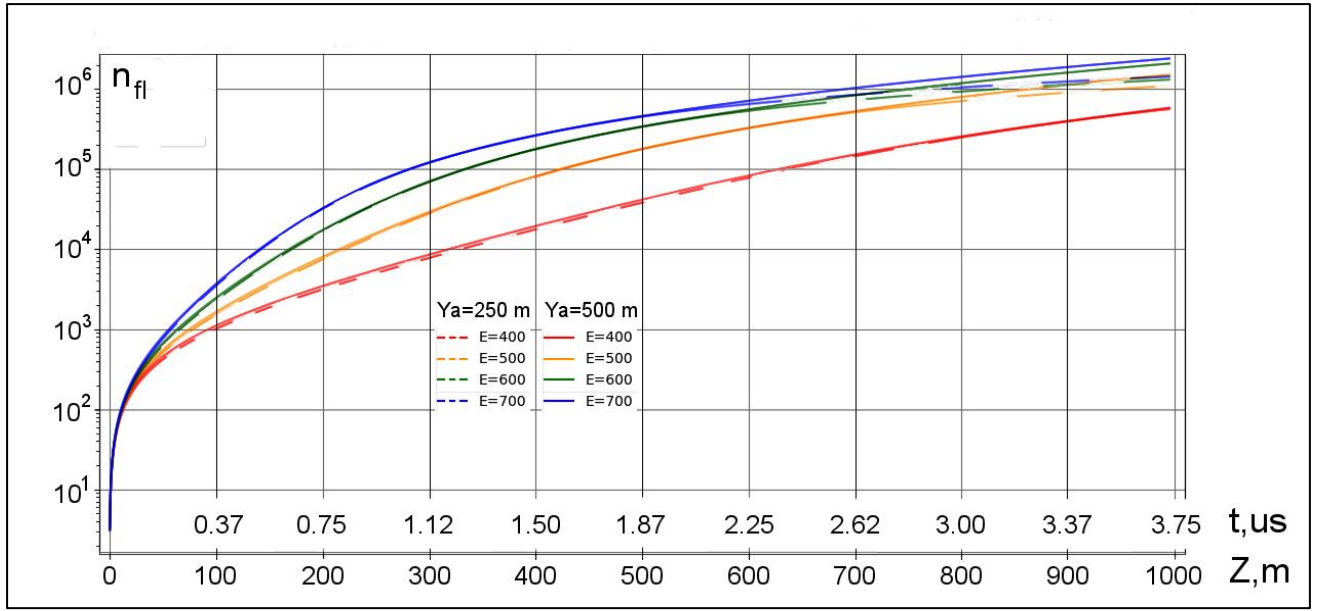

Рисунок 8.12 (адаптировано из [Kostinskiy et al., 2020b]). Влияние на число стримерных вспышек $n_{fl}(z)$ размеров EE-области грозового облака (высота 6 км, $N_e^{EAS} = 10^6$), которые задаются радиусами $y_{aF} = 250\ m$ (пунктирные линии), $y_{aF} = 500\ m$ (сплошные линии): $400\ \frac{kV}{m \cdot atm}\left(196\ \frac{kV}{m}\right)$, $\lambda_{RREA} = 120\ m$ — красные линии; $500\ \frac{kV}{m \cdot atm}\left(244\ \frac{kV}{m}\right)$, $\lambda_{RREA} = 66\ m$ — желтые линии; $600\ \frac{kV}{m \cdot atm}\left(294\ \frac{kV}{m}\right)$, $\lambda_{RREA} = 46\ m$ — зеленые линии; $700\ \frac{kV}{m \cdot atm}\left(343\ \frac{kV}{m}\right)$, $\lambda_{RREA} = 35\ m$ — синие линии. Вероятность попадания электрона $p = 0.001(\emptyset \approx 2\ cm)$, $\rho_{E_{th}} = 10^{-2} m^{-3}$, $R_M = 161$ .

Зависимость числа стримерных вспышек от числа посевных электронов $N_e^{EAS} \approx$ $10^5$ и $10^6$ (Рисунок 8.13) показывает, что на длине 800-900 метров число вспышек становится близким, по-видимому, из-за исчерпания числа воздушных электродов в единице объема (большинство воздушных электродов инициируют вспышки). Также оба Рисунка 8.12, 8.13 отчетливо показывают важную роль длины шага лавины $\lambda_{RREA}$, так как число вспышек, которые были инициированы ШАЛ с меньшим на порядок числом посевных электронов, «догоняет» число вспышек в случаях, которые происходили при меньших электрических полях. Это не удивительно, так как общее число электронов из-за малого шага лавины $\lambda_{RREA}$ для больших электрических полей растет быстро и начинает превышать общее число электронов лавин с меньшим электрическим полем но с бо́льшим начальным числом вторичных электронов ШАЛ ($N_e^{EAS}$)

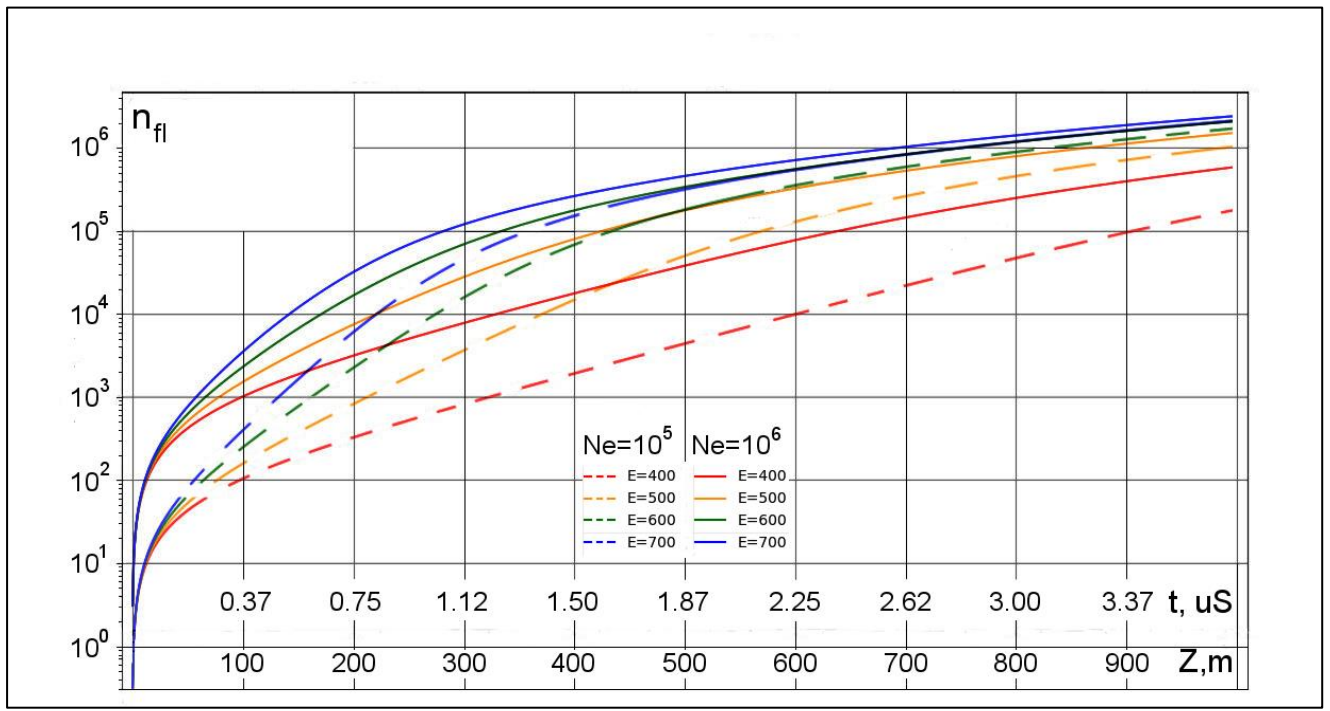

Рисунок 8.13 (адаптировано из [Kostinskiy et al., 2020b]). Число стримерных вспышек $n_{fl}$ (z) на высоте 6 км в зависимости от числа посевных электронов ШАЛ ($N_e^{EAS} = 10^5$(пунктирные линии), $10^6$(сплошные линии). Для каждого из этих значений $N_e^{EAS}$ расчет проводился для четырех электрических полей: $400 \frac{kV}{m \cdot atm} \left(196 \frac{kV}{m}\right)$, $\lambda_{RREA} = 120\ m$ — красные линии; $500 \frac{kV}{m \cdot atm} \left(244 \frac{kV}{m}\right)$, $\lambda_{RREA} = 66\ m$ — желтые линии; $600 \frac{kV}{m \cdot atm} \left(294 \frac{kV}{m}\right)$, $\lambda_{RREA} = 46\ m$ — зеленые линии; $700 \frac{kV}{m \cdot atm} \left(343 \frac{kV}{m}\right)$, $\lambda_{RREA} = 35\ m$ — синие линии. Вероятность попадания электрона $p = 0.001 (\emptyset \approx 2\ cm)$, $\rho_{E_{th}} = 10^{-2} m^{-3}$, $R_M = 161\ m$.

### 8.3. Предварительные выводы главы 8

- В случае справедливости предложенного нами Механизма (глава 7, [Kostinskiy et al., 2020a]) скорее всего только EAS-RREA ливень с числом посевных вторичных электронов ШАЛ $N_e^{EAS} > 10^5$ может обеспечить необходимое число электронов и позитронов для синхронной инициации стримерных вспышек, обеспечивающих VHF-сигнал, сопровождающий NBE во всем диапазоне высот над уровнем моря. ШАЛ обеспечивает не только большое число посевных электронов и позитронов, но и одновременное их широкое начальное радиальное распределение, необходимое для эффективной ионизации воздушных электродов. Только ШАЛ без механизма релятивистских убегающих электронов (RREA) не способны создать достаточное число электронов для синхронизации стримерных вспышек в большом объеме, также как и механизм RREA в реалистичных электрических полях (400-700 kV/(m·atm)) без достаточного числа посевных электронов ШАЛ не сможет создать необходимое число посевных электронов в объеме для инициации большого числа стримерных вспышек. Чтобы RREA механизм (без ШАЛ) достиг в реалистичных сильных электрических полях необходимого числа посевных электронов, которые поставляет ШАЛ, RREA самой необходимо сделать $\ln(10^6) = 13.8 \ \lambda$, но, кроме этого, RREA еще нужны дополнительно 3-7 шагов для того, чтобы электроны лавины RREA инициировали достаточное число стримерных вспышек. Эти оценки показывают ключевую роль ШАЛ в этом процессе.

- Даже при оценке снизу, EAS-RREA лавины с числом электронов $N_e^{EAS} \approx 10^5 - 10^6$ в электрических полях (400-700 kV/(m·atm)) обеспечивают необходимый поток вторичных электронов для одновременной синхронизации в течение нескольких микросекунд множества стримерных вспышек в EE-объеме со средним электрическим полем E> 450 kV/(m·atm).

- EAS-RREA-инициация стримерных вспышек может обеспечить появление множества IE-событий, которые требуют меньшего числа вторичных частиц ШАЛ и меньшего числа инициированных стримерных вспышек по сравнению с NBE (механизм создания Weak NBE).

- Если предложенный Механизм верен, то спектр VHF-излучений различных NBE должен быть непрерывным и переходить в параметры Weak IE, так как статистика взаимодействий ШАЛ и турбулентных заряженных областей будет приводить к большому числу комбинаций между пространственным сечением турбулентных областей и попаданием в них под разными углами ШАЛ различных энергий и «возраста».

- ЕЕ-области могут служить своеобразными стримерными камерами для изучения ШАЛ на разных высотах (эта идея является продолжением идеи [Gurevich et al. 1999, 2004], которые раньше предлагали считать грозовое облако искровой камерой).

- Если предложенный нами Механизм (глава 7, [Kostinskiy et al., 2020a]) верен, то подробное изучение параметров NBE может дать информацию о мощных турбулентных потоках внутри грозового облака.

## Заключение

Полученные в диссертационной работе основные научные результаты заключаются в следующем.

1. Внутри положительно и отрицательно заряженных искусственных водных аэрозольных облаков существует новый класс электрических разрядов «необычные плазменные образования» (unusual plasma formations — UPFs), которые представляют из себя сети плазменных каналов, размеров 10-30 см, некоторые из которых нагреты настолько, насколько нагрет восходящий положительный лидер.

2. Установлен по крайней мере один механизм возникновения UPFs. UPFs инициируется внутри восходящей положительной стримерной вспышки до появления других плазменных образований в электрическом поле аэрозольного облака.

3. Во время сквозной фазы контакта восходящего и нисходящего лидеров длинной искры, возможно ветвление лидеров внутри общей стримерной зоны.

4. Яркость инфракрасного излучения в области контакта восходящего и нисходящих лидеров длинной искры примерно в 5 раз выше, чем для участков образовавшегося единого канала ниже или выше этой области.

5. При движении протяженного проводника (болта арбалета с проводом и без провода) в окрестностях и внутри искусственного облака положительно и отрицательно заряженного водного аэрозоля возникают стримерные вспышки, инициируются длинные искры (лидеры) и UPFs, и возникают квазиобратные удары в заземленные объекты.

6. Во время образования ступеней (длиной 30-120 см) положительного лидера длинной искры (в электрическом поле высоковольтного генератора импульсных напряжений) перед головкой лидера в зоне стримерной короны может существовать плазменное образование, сходное по морфологии со спейс-стемом или спейс-лидером, которые наблюдаются в стримерной зоне отрицательного лидера. Стримерные зоны, возникающие во время образования ступеней положительного и отрицательного лидеров длинной искры морфологическое сходны.

7. Инициацию молнии в грозовых облаках можно описать, как последовательный качественный механизм от инициирующего молнию события (IE) через начальную стадию увеличения электрического поля (IEC), до первых начальных импульсов пробоя (IBPs), которые переходят в ступенчатый отрицательный лидер.

8. Инициирующим молнию событием (IE) могут быть «классические» компактные внутриоблачные разряды (КВР/CID/NBE) или, в гораздо большем числе случаев, — «слабые» КВР (weak NBE), аналогичные по физическому механизму классическим КВР, но меньшие по масштабу. Механизм инициации компактных внутриоблачных разрядов (КВР/CID/NBE), который дает непротиворечивое объяснение спектру оптического излучения КВР, характерному для стримерных вспышек, и скорости распространения источников радиоизлучения внутри грозового облака близкой к скорости света, можно представить, как инициацию вторичными электронами и позитронами ШАЛ в наиболее турбулентных областях грозового облака в небольших объемах со сверхпробойными полями множества стримерных вспышек, внутри которых возникнут цепочки UPFs, взаимодействующие друг с другом посредством вторичных стримерных корон.

9. Инициацию молнии можно представить, как последовательную цепь переходов плазмы из одного состояния в другое (от первых газоразрядных лавин до стримеров, UPFs, положительных, отрицательных, двунаправленных лидеров, сетей плазменных каналов, которые ведут к сквозным фазам взаимодействия длинных плазменных двунаправленных лидеров и/или сетей во время протекания IBPs-стадии развития молнии).

## Словарь терминов

**Быстрый положительный пробой (fast positive breakdown — FPB):** процесс, с помощью которого [Rison et al., 2016], [Attanasio et al., 2019] объясняют физический механизм развития КВР/NBE. По их модели КВР представляет из себя гигантскую положительную стримерную вспышку, развивающуюся в соответствие с моделью Гриффитса-Фелпса [Griffiths & Phelps, 1976] и имеющую скорость 5-10·$10^7$ м/с. В статье [Tilles et al., 2019] эта концепция была распространена и на «быстрый отрицательный пробой» (FNB), который должен был объяснить изменение знака КВР.

**Воздушный электрод:** см. $E_{th}$-объем

**EAS-RREA (широкий атмосферный ливень (ШАЛ) — лавина убегающих релятивистских электронов, extensive air shower — relativistic runaway electron avalanche):** поток вторичных электронов ШАЛ, которые экспоненциально размножаются по физическому механизму убегающих электронов, когда поток вторичных заряженных частиц ШАЛ входит в область в сотни метров с электрическим полем E > 280 кВ/(м·атм). Для проблемы инициирования молнии важны ШАЛ с энергией первичных частиц $\varepsilon_0 \geq$ 5·$10^{14}$-$10^{15}$ эВ (как описано ниже) [Gurevich & Zybin, 2001]; [Dwyer, 2003].

**EE-объем (0,1-1 км$^3$):** область в грозовом облаке со средней величиной электрического поля E > 0,28-0,35 МВ/(м·атм) и с большим числом заряженных гидрометеоров разных размеров. Гидрометеоры могут быть жидкими или твердыми, большими или маленькими по размеру, при условии, что их много и они несут значительные электрические заряды, так что турбулентные и статистические движения могут приводить к появлению небольших областей EE-объема с существенно большими электрическими полями. EE-объем может иметь сильно неоднородные электрические поля (в масштабе сотен метров) и состоять из множества близко расположенных турбулентных областей, которые могут быть образованы одинаково или противоположно заряженными встречными многофазными потоками воздуха (например, [Karunarathna et al., 2015]; [Yuter & Houze, 1995].

**$E_{str +}$ -объем** (могут быть повторяющимися областями с характерным размером порядка 100 м): область в грозовом облаке с E ≥ 0,45-0,5 МВ/(м·атм). Столь большие величины E достаточны для поддержания движения положительных стримеров в воздухе [Базелян и Райзер, 1997].

**E$_{th}$-объем (10$^{-3}$-10$^{-4}$ м$^3$)** или «воздушный электрод»: небольшая область в грозовом облаке с E > 3 МВ/(м·атм); столь большие величины E достаточны для образования «классических» электронных лавин, которые при выполнении критерия Мика [Райзер, 1992, стр. 425] могут трансформироваться в классические газоразрядные стримеры.

**Инициирующее молнию событие (an initiating event — IE)** является первым электромагнитным проявлением инициирования молнии и может быть слабым (weak NBE), описанным [Marshall et al., 2014a, 2019] и [Lyu et al., 2019] или более сильным NBE, как описано [Rison et al., 2016] и [Lyu et al., 2019]. Слабые IE имеют VHF-мощность <1 Вт и длительность ≤ 1 мкс, в то время как КБР(NBE/CID) имеют на порядки большую VHF-мощность и длительность 10–30 мкс.

**Компактный внутриоблачный разряд — КБР (compact intracloud discharges — CIDs)** см. Узкое биполярное событие.

**Начальное изменение электрического поля (an initial electric-field change — IEC)**: важнейший этап развития молнии, который недавно выделил Томас Маршалл с соавторами [Marshall et al., 2014a]. Как описали этот этап [Marshall et al., 2014a] и [Chapman et al., 2017], представляет собой период (40-9800 мкс), который начинается с IE и заканчивается первым классическим начальным импульсом пробоя (IBP). [Marshall et al., 2019] показали, что во время проведения IEC было много импульсов VHF с длительностью 1–7 мкс и что некоторые совпадающие пары импульсов быстрой антенны (FA) и импульсов VHF, похоже, увеличивают IEC (как «усиливающие события» («enhancing events»)).

**Начальный импульс пробоя (an initial breakdown pulse — IB-импульс или IBP)**: очень сильный биполярный электрический импульс, возникающий в первые несколько миллисекунд после инициации молнии, обычно обнаруживаемый с помощью быстрой антенны FA (например, [Weidman & Krider, 1979], [Nag et al., 2009]). Самые большие IBP называются «классическими IBP» и они систематически сопровождаются VHF-импульсами во вспышках CG [Kolmašová et al., 2019]. По нашему определению классические IBP имеют длительность ≥ 10 мкс, амплитуду ≥ 25% от наибольшего IBP и часто имеют субимпульсы. Амплитуды классических IBPs могут быть 30-160 кА (это параметры мощных обратных ударов молнии) несмотря на то, что IBPs могут возникать через чрезвычайно короткие интервалы времени после инициации молнии (40 мкс-3 мс). Практически все молнии имеют серию (серии) IBPs, которые происходят в течение нескольких мс после IEC ([Mäkelä et al., 2008], [Marshall et al., 2014b]). По данным [Mäkelä et al., 2008] на большой выборке, не менее 90% всех молний в Северной Европе имеют стадию начальных импульсов пробоя (IBPs), которая предшествует появлению большого отрицательного лидера, причем для остальных 10% молний также может присутствовать эта стадия, но экспериментальные данные не давали возможности сделать однозначный вывод. Во время протекания IB-стадии биполярные импульсы, меньшие по амплитуде или более короткие по длительности, чем классические IBPs, но также являются IBPs. Классические и меньшие IBPs, в принципе, могут быть вызваны разными физическими процессами.

**Начальный лидер (initial leader — IL):** относительно новая сущность в обширном семействе «лидеров» (не являющаяся ступенчатым отрицательным лидером или двунаправленным лидером), которая позволяет обозначить прерывистые процессы на начальной стадии инициации молнии, скорее всего это процессы, связанные с начальными импульсами пробоя (IBPs), которые вызывают сильные изменения электрического поля и яркие вспышки света, похожие на проявления ступенчатого отрицательного лидера, но гораздо более сильные и яркие. Например, «начальный лидер» ярко светит и может иметь «ветви» [Stolzenburg et al., 2013; 2014; 2020]. [Lyu, Cummer et al., 2019] считают, весь показанный на Рисунке В.44 процесс инициации молнии движением «начального лидера» (IL) и, по их интерпретации, импульсы начального пробоя (IBPs) являются гигантскими ступенями этого IL длиной 200-600 м. Начинается этот гигантский IL с быстрого положительного пробоя (FPB).

**Необычное плазменное образование (an unusual plasma formation — UPF):** короткий канал (или чаще небольшая иерархическая сеть каналов) горячей высокопроводящей плазмы, которые были обнаружены [Kostinskiy et al., 2015a, 2015b, 2015c] внутри или около искусственного заряженного облака водного аэрозоля. UPFs часто представляют собой иерархическую сеть каналов горячей плазмы длиной в десятки сантиметров. UPFs могут взаимодействовать между собой и плазменными лидерами с помощью положительных стримеров. Экспериментально показано, что UPFs возникают внутри длинных стримерных вспышек около границы и внутри отрицательного аэрозольного облака, а также при взаимодействии протяженных металлических предметов с электрическими полями внутри отрицательно и положительно заряженного облака.

**Положительный лидер**: канал горячей высокопроводящей плазмы с положительным зарядом в головке лидера, который удовлетворяет определенным условиям длины в окружающем электрическом поле (или приложенному потенциалу в головке), так что лидер будет самораспространяющимся, как описано [Bazelyan et al., 2007a]. При электрическом поле 0,45 - 0,50 МВ/(м атм) (в зависимости от влажности) горячий плазменный канал любой длины будет самораспространяющимся.

**Стример**: холодная плазма быстрой волны ионизации, которая поддерживается собственным электрическим полем головки, благодаря поляризации во внешнем электрическом поле, как описано, например, [Райзер, 1992, стр. 423]. В этой статье термин «стример» означает только «обычные» стримеры, которые наблюдались в течение многих десятилетий в газовых разрядах и длинных искрах при давлениях 1-0,3 атм и имеют длину от сантиметров до нескольких метров [Райзер, 1992, стр. 423]. Данное пояснение необходимо так как в статьях по физике молнии часто встречаются другие интерпретации термина «стример» более похожие на быстрые волны потенциала при обратных ударах молнии или стреловидных лидеров

**Узкое биполярное событие (narrow bipolar event — NBE) или Компактный внутриоблачный разряд — КВР (compact intracloud discharges — CIDs):** термины являются синонимами, которые обозначаю особый тип электрического события, которое происходит во время протекания грозы, но не является ни внутриоблачной молнией (IC), ни молнией типа облако-земля (CG). Это важнейшее явление открыл Давид Ле Вайн в 1980 году [Le Vine, 1980]. NBE в данных FA (быстрой антенны) имеет биполярную форму волны длительностью 10-30 мкс; в диапазоне ОВЧ 60–66 МГц Сильные NBE имеют большую мощность (30 000–300 000 Вт или 45–55 дБВт) [Rison et al., 2016]. Сильные NBE являются самыми мощными VHF-явлениями в атмосфере, превосходя «классические молнии на порядок» [Smith et al. 1999]. Данные FA для слабых NBE имеют меньшие амплитуды, чем NBE, и могут иметь либо биполярные, либо «в основном монополярные» формы волны; слабые NBE также имеют меньшую VHF-мощность, составляющую 3–300

Вт или 5–25 дБВт [Rison et al., 2016]. NBE может инициировать молнию в нескольких процентах случаев (например, [Rison et al., 2016] и [Lyu et al., 2019])

**Усиливающее событие (enhancing events):** [Marshall et al., 2019] показали, что во время проведения IEC было много импульсов VHF с длительностью 1–7 мкс и что некоторые совпадающие пары импульсов быстрой антенны (FA) и импульсов VHF, похоже, увеличивают изменение электрического поля IEC, как «усиливающие события» (enhancing events).

## Список цитируемой литературы

## Список публикаций по теме диссертации